\documentclass{thesis}
\usepackage[T1]{fontenc}
\usepackage{graphics}

\makeatletter

\providecommand{\LyX}{L\kern-.1667em\lower.25em\hbox{Y}\kern-.125emX\@}
\newcommand{\noun}[1]{\textsc{#1}}
\let\SF@@footnote\footnote
\def\footnote{\ifx\protect\@typeset@protect
    \expandafter\SF@@footnote
  \else
    \expandafter\SF@gobble@opt
  \fi
}
\expandafter\def\csname SF@gobble@opt \endcsname{\@ifnextchar[
  \SF@gobble@twobracket
  \@gobble
}
\edef\SF@gobble@opt{\noexpand\protect
  \expandafter\noexpand\csname SF@gobble@opt \endcsname}
\def\SF@gobble@twobracket[#1]#2{}

 \newenvironment{lyxlist}[1]
   {\begin{list}{}
     {\settowidth{\labelwidth}{#1}
      \setlength{\leftmargin}{\labelwidth}
      \addtolength{\leftmargin}{\labelsep}
      }}
   {\end{list}}

\makeatother

\begin{document}

\prelimpages

\Title{Spin State Detection and Manipulation\\
and Parity Violation in a Single Trapped Ion}

\Author{Michael Schacht}

\Year{2000}

\Program{Physics}

\titlepage

\Chair{E. Norval Fortson}{Professor}{Physics}

\Signature{Blayne Heckel}

\Signature{Aurel Bulgac}

\signaturepage

\doctorialquoteslip

\abstract{Atomic Parity Violation provides the rare opportunity of a low energy window
into possible new fundamental processes at very high mass scales normally investigated
at large high energy accelerators. Precise measurements on atomic systems are
currently the most sensitive probes of many kinds of new physics, and complement
high energy experiments. Present atomic experiments are beginning to reach their
limits of precision due to either sensitivity, systematics or atomic structure
uncertainties. An experiment in a single trapped Barium ion can improve on all
of these difficulties. This experiment uses methods to precisely manipulate
and detect the spin state of a single ion in order to measure a parity induced
splitting of the ground state magnetic sublevels in externally applied laser
fields. The same methods can be used to provide precise measurements of more
conventional atomic structure parameters. Updated and corrected versions of
this document are available from the UW Physics Atomic Physics pages, www.phys.washington.edu/groups/atomic.}

\tableofcontents{}

\listoffigures{}

\listoftables{}

\textpages

\chapter{Introduction and Motivation}

\label{Sec:Introduction}

To the casual outside observer, parity violation might seem to be old news.
Its discovery was made in the 50's in nuclear and high energy experiments, and
even the first observations in atomic systems are decades old by now. The existence
of this chiral asymmetry is well established and well accounted for in the Standard
Model, if a little uncomfortably, and in this context there seems little more
to be learned, yet Parity Non-Conservation(PNC) continues to be the subject
of much attention and effort.

\section{Parity Violation as a tool}

\subsection{A quest for something new}

This idea that the universe is somehow left-handed is still fascinating and
a bit strange, and there remain some fascinating loose ends related to its origin
and precise structure, but the current interest in Parity Violation is not so
much for its own sake, but instead due to more immediately practical motives,
as far as anything can be considered practical here. When its effects are accepted
and in the end understood, Parity Violation becomes a tool, and in particular
a tool for uncovering clues about the current Holy Grail of physics beyond the
Standard Model.

Physicists interested in fundamental processes are, in equal measure, never
satisfied and easily bored. The Standard Model was introduced in the 60's and,
for the moment, shows no compelling signs of failing, yet for a number of pretty
good reasons, physicists are looking for something else. This is not a matter
of any real practical deficiencies as much as a certain lack of elegant internal
consistency, as in every case Standard Model calculations agree with experimental
observations as precisely as anyone has been able to, respectively, calculate
and measure them. The single electron g-2 experiment is a favorite example,
the measured magnetic moment of the electron is found to agree with calculated
predictions using the Standard Model a a part in \( 10^{10} \).

\subsection{Missing Pieces}

Still, there are some nagging rough edges, an embarrassment of adjustable parameters,
a mysterious and not yet well characterized Higgs sector that is particularly
important as it aspires to explain why particles can and do have mass, a failure
to completely account for the lack of anti-matter compared to matter in at least
our local universe, and perhaps its most glaring shortcoming: it completely
fails to account for gravity and in fact can't yet even be made to include it
in a consistent way. Even the simple lack of processes whose natural energy
scales occur between typical electro-weak energies and the characteristic scale
of gravity, the Planck mass, make the structure of the Standard Model as a low
energy limit depend on some uncomfortably careful fine tuning of parameters. 

So motivates the quest for something deeper, something new. It is possible that
these difficulties can be addressed from within the Standard Model through currently
unexplored or misunderstood pieces, phase transitions, other non-perturbative
effects, or more a detailed picture of the Higgs sector. But this is considered
to be significantly less likely and, also calling to mind again the physicists
short attention span, effort is directed at extensions to the Standard Model.
Then in this context, new means anything not explicitly or implicitly already
included in the Standard Model, and in the end this can simply be understood
as new particles.

\section{Heavy particles yield small effects}

These new particles, should they exist, are probably very heavy, simply because
if they weren't, we probably would already have seen them. They are expected
to begin to appear with masses just above the Electro-Weak scale, which is around
the mass of the W and Z bosons, about 100 GeV, though experimental limits have
strongly excluded the possibility of new particles with masses of much less
than about 0.5-1 TeV. This makes finding them particularly challenging since
at the relatively low energies at which we live and presently do experiments,
the effects of physics at large energy scales is heavily suppressed. This is
easily seen even just at tree level, the classical limit.

\subsection{A tree level understanding}

For the exchange of a single heavy particle of mass \( M \) the matrix element
will include a factor of the (renormalized) propagator for this particle and
coupling constants for the vertices. When the exchanged momentum \( q \) is
much less than this mass, the propagator becomes \( q \) independent. 
\[
M\propto \frac{g^{2}}{-q^{2}+m^{2}+im\Gamma }\longrightarrow ^{q^{2}\ll m^{2}}\frac{g^{2}}{m^{2}}\]

This process becomes important when the denominator of the propagator is small,
that is for energies \( q\approx m \). But for the expected new heavy particles
such energies are well above those accessible in most experiments. In this case
the amplitude for this process gets smaller quickly as the mass of the exchanged
particle is increased. This provides a convenient excuse for the absence of
evidence for the existence of either new particles predicted by various possible
new theories such as Supersymmetry or SU(5), and even the Higgs boson of the
Standard Model. But this is convenient only for the theorist, the experimentalist
still faces the challenge of trying to detect these effects and the task is
made increasingly difficult as the search continues to higher masses.

\subsection{Consequences for atoms}

For a comparison, atomic structure is determined almost completely by electromagnetism.
For this case, the tree level interaction is a photon exchange giving a factor
of the photon propagator \( 1/q^{2} \). Typical energies in atomic systems
are around 1eV, so the characteristic size of an effect due to a new heavy particle
with mass of about 1TeV seems to be hopelessly small compared to the natural
size of an electromagnetic process. With coupling constants presumed to be a
similar order,

\[
\frac{g^{2}/m^{2}}{e^{2}/q^{2}}\propto \left( \frac{1}{1TeV}\right) ^{2}=10^{-24}\]

In heavy atoms there is some relief because it turns out that there are enhancements
due to relativistic electrons and nuclear size that increase this considerably.

\[
\frac{\left\langle f\left| H_{New}\right| i\right\rangle }{\left\langle f\left| H_{EM}\right| i\right\rangle }\propto 10^{-14}\]

This would still be very difficult to detect. Even if an experiment was made
sufficiently sensitive for a measurement of this precision, and systematic effects
could be understood and controlled at this level, simply interpreting the result
would require also understanding the background E-M effects to an even higher
precision so they could be separated from the effects due to the new particles.
This would be several orders of magnitude better than even a g-2 experiment,
with a many-body system considerably more complicated than a single electron.
Except in a few very simple cases, it is currently a struggle to understand
even just electromagnetic effects in atoms to a part in \( 10^{3} \).

\section{Parity Violation and new physics}

In practice, these difficulties would make such effects from possible new particles
undetectable if these new processes didn't also happen to have a fundamentally
different character. One difference that can be exploited is parity violation.
The idea of parity and parity violation will be developed more fully in later
sections, but in the end all that is important is that it is a symmetry that
electromagnetic and strong forces conserve exactly and Weak forces violate maximally.

\subsection{A new signature}

Atoms are held together largely by electromagnetism so atomic states have well
defined parities. Perturbations from parity violating interactions can then
appear as a mixing of opposite parity states. In particular, atomic PNC experiments
measure a mixing of \( S \) and \( P \) states, on the order of \( 10^{-10} \)
to \( 10^{-11} \). In a parity symmetric environment E-M won't couple these
states so any mixing must be attributed solely to the parity violating perturbation.
In this way, measuring a parity violating process no longer requires identifying
and understanding a background impossibly many orders of magnitude larger as
the contribution from e-m effects is exactly zero. 

In the same way, just as importantly, the strong force conserves parity. Though
not directly involved in binding an electron to a nucleus, the strong force
is responsible for the structure of the nucleus itself and is not nearly as
easy to deal with as electromagnetism as it effects are largely non-perturbative.
A parity violating strong force would result in a chiral nucleus which creates
an asymmetric e-m potential for the electron, in turn contributing to parity
violating observables. Small corrections due to Weak interactions within the
nucleus do exactly that and must be considered when precision increases. This
turns out to provide valuable information about nuclear structure. But in the
case of effects due to just the strong interactions there is no contribution,
because as with electromagnetism, the Strong force conserves parity, and we
are saved the trouble of precisely understanding another naturally much larger
background that may be even harder to calculate than QED.

\subsection{The Weak Interaction becomes background}

Certainly there are considerable practical difficulties, these will be discussed
in endless detail in later sections. Even excluding that, parity violation isn't
a panacea. For searches for new physics, you still don't get to compare your
result to a background of zero since, of course, the Weak sector of the Standard
Model also violates parity. The dominant effect in atoms is the neutral current
exchange of a single \( Z_{0} \) Weak gauge boson between the nucleus and valence
electrons. Effects due to new particles will appear as corrections to Weak Standard
Model parity violating observables. In this case, however, the Standard Model
effects are already very small since, compared to atomic scale energies the
masses of the Weak boson are already very large. Then with the same simple tree
level estimates the relative sizes of these new effects are given by a ratio
of their masses to the already very high energy Weak scale. Again considering
coupling constants of similar order,

\[
\frac{\left\langle f\left| H_{New}\right| i\right\rangle }{\left\langle f\left| H_{EM}\right| i\right\rangle }\propto \left( \frac{m_{W}}{m_{New}}\right) ^{2}\approx \left( \frac{100\, GeV}{1\, TeV}\right) ^{2}=10^{-2}\]

A measurement of such a parity violating process to 1\% is sensitive to new
physics at mass scales of about 1 TeV. This sort of precision is much more reasonable
and once done, the interpretation becomes far more tractable. This is the real
power of studying parity violation. The electromagnetic and strong forces that
would otherwise mask these much smaller effects become effectively transparent
to them and parity violating effects due to the heavy processes clearly show
through unmodified. In this way parity violation becomes a sensitive differential
test for possible classes of new physics.

\section{High energy experiments and new physics}

These parity violating effects in atomic systems, while now reasonably big enough
to be visible, are still very small and considerable care must be taken in avoiding
systematic errors from external perturbations to measure them precisely. In
addition, detailed atomic and nuclear structure calculations must be made to
interpret them. Meanwhile, much higher energy nuclear or elementary particle
experiments might seem to be a more promising direct route to the same information.

\subsection{Discovery by production}

For sufficiently high energies this would certainly be true. As previously noted,
a given particles effects will dominate near the pole of its propagator, where
the energy scale is equal to the mass of the particle. Perhaps the right way
to find new particles is to search for their resonances when they are directly
produced at continually higher energies. To find a particle with a mass of about
1 TeV in this way requires an accelerator with a beam energy of about a TeV,
in particular a center of momentum energy equal to the new particle mass. Electron-positron
(\( e^{+}e^{-} \)) experiment are the easiest to interpret, but such machines,
like SLAC are limited to 50 GeV. The proton-antiproton (\( p-p \)) collider
at fermilab does reach 2 TeV, and LEP at CERN will shortly reach 4, but a proton
is actually a complicated bag of junk only a very small part of which might
carry a large enough fraction of the total proton energy to produce a heavy
particle themselves, and the probability for all the parts to collective use
their energy to produce a single new heavy particle is unreasonably small. With
these energies the event rate for this kind of search is still too small to
stand out from the background of other processes and the strategy would really
require building accelerators at arbitrarily high energies. This is becoming
increasingly difficult, and in any case is not immediately accessible. So all
experiments are constrained to work at energies much lower than the expected
natural scale of new physics.

\subsection{Sensitivity at low energies}

Energies up to even a few GeV, still below a \( Z \) or \( W \) resonance,
such as in nuclear experiments, are still insufficient for any kind of direct
detection. As before, compared to some generic QED or strong process, 1 TeV
particle effects are typically smaller by \( (1\, GeV/1\, TeV)^{2}=10^{-6} \).
This is a considerably larger fraction than for atomic systems, but still too
small to be untangled from the background, so even in these much higher energy
cases a trick like parity violation must be used. Though the parity violating
effects in nuclei are significantly larger than in atoms by these sorts of factors,
the fraction of that which is due to new heavy physics is the same. Though much
larger relative to general QED effects, the parity violating background due
to QED is already zero and so nothing is gained. As energies increase this fraction
improves, but another sort of difficulty develops. For any kind of direct detection,
these QED effects interpreted as background do go down, but they are also really
the signal which is to be measured precisely enough to detect the small piece
due to new physics. As this small piece becomes a reasonable fraction of the
signal, \( 10^{-2}-10^{-3} \), in colliding beam experiments the entire signal
becomes almost undetectably small and impossible to measure precisely due to
systematic problems signal/noise constraints. This can be partly avoided in
fixed target experiments. In this case the event rate and sensitivity increase
significantly compared to a beam target for the same energy and luminosity simply
because there are many more particles for the probe beam to interact with in
a fixed target than in another beam. The rate is still small compared to typical
accelerator experiments so statistics and systematics will still be a challenge.
Also the effective interaction energy is less since 100GeV colliding beams give
a center of mass energy of 200GeV while yielding only (40 GeV) in the center
of mass frame of a fixed target and the same beam. Still this is a promising
means of attack, though currently only one experiment intends to use such a
strategy.

\subsection{Sensitivity at the \protect\( Z_{0}\protect \) pole}

\label{Sec:Z0PoleSensitivity}

Most experiments avoid these sensitivity problems by looking instead at Weak
processes rather than electro-magnetic. At center of mass energies near the
\( Z \) mass, the \( Z \) propagator finally becomes large and the Weak interaction
dominates. To estimate the relative contribution by TeV scale physics, the previous
tree level approach is still valid, but the result is modified slightly. At
the \( Z \) pole the propagator becomes \( 1/M\Gamma  \), with \( \Gamma  \)
the decay width of the \( Z \). This energy is still presumably much less than
the mass of any new particle so its propagator can still be taken to be the
usual constant and then for any Weak process, not necessarily parity-violating,
the appropriate comparison becomes,

\[
\frac{g'^{2}}{m_{New}^{2}}/\frac{g^{2}}{m_{Z}\Gamma _{Z}}\propto \frac{m_{Z}\Gamma _{Z}}{m_{New}^{2}}\approx \frac{(100\, GeV)(1\, GeV)}{(1\, TeV)^{2}}=10^{-4}\]

This is comparable to the fraction of the contribution of new physics to parity
violating processes at lower energies, in fact even slightly less. This difference
actually makes atomic PNC experiments, more sensitive to new tree level interactions
and able to probe mass scales 5 times higher than in accelerator experiments,
sec.\ref{Sec:NewTree}equivilant. 

In any case there is certainly no advantage here to working at higher energies.
This New to Weak relative ratio could be increased by working slightly off resonance,
but this also decreases sensitivity to the dominant process and so is counter-productive
as it quickly becomes equivilant to the case of mid-range energies previously
discussed, you can't increase sensitivity by lowering your signal. In the end,
there is nothing that can compensate for a very small signal from new physics
because of working at energies below its mass scale. By studying parity violation
at low and mid-range energies the immediate disadvantages of working there,
compared to higher energies, are avoided and in effect all energy scales are
equally bad when sufficient energy for direct production of new heavy particles
is not available.

\subsection{New particles in radiative corrections}

\begin{figure}
{\par\centering \includegraphics{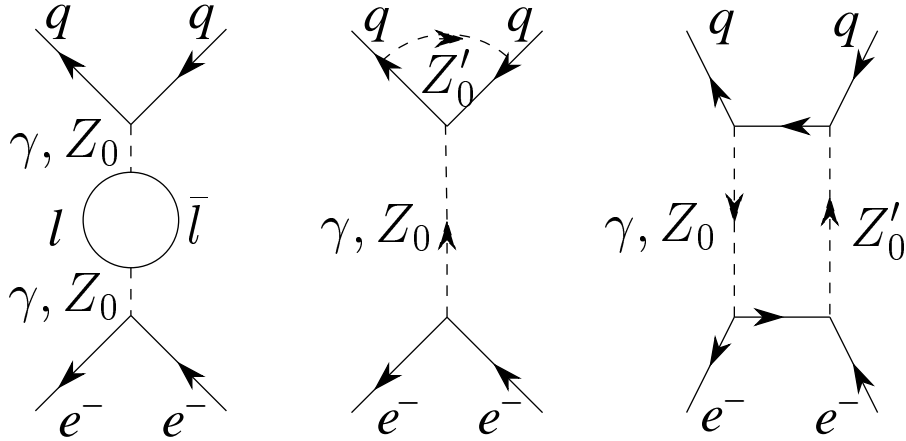} \par}

\caption{Typical possible contributions to Atomic Parity Violation from radiative corrections.\label{fig:FeynmanRadiative}}
\end{figure}

This tree level survey is not a complete picture of possible contributions from
new physics. Observable effects can also appear through radiative corrections
(fig.\ref{fig:FeynmanRadiative}).
This is more complicated to analyze, but will be fully developed in section
\ref{Sec:RadiativeCorrections}. These effects can be distilled and described
with two general parameters, S and T, conventionally defined to be zero if only
Standard Model physics contributes and all these results can be translated into
this universal language for comparison and used to check for consistency or
discrepancies that might indicate the existence of new physics. Different experiments
depend on different linear combinations of these parameters, and sensitivities
are similar though higher energy experiments fare a bit better. A 1\% measurement
of atomic PNC provides a limit on S to about \( \pm 1 \), while the same precision
for a typical collider experiment gives S and T to about \( \pm 0.2 \). 
\begin{figure}
{\par\centering \resizebox*{1\textwidth}{!}{\includegraphics{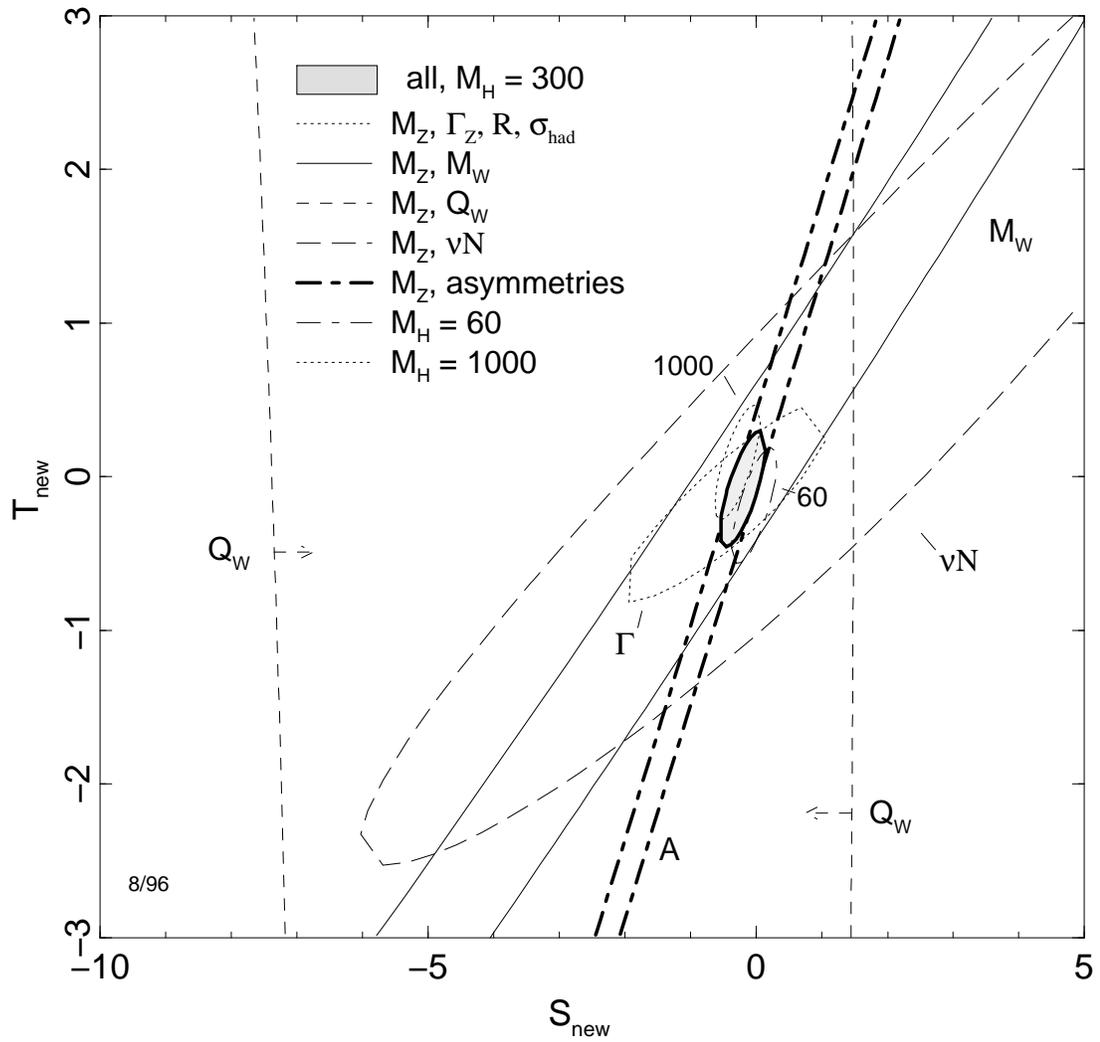}} \par}

\caption{Current constraints on oblique corrections to Standard Model, from \cite{PeskinSTU}\label{fig:S&T}}
\end{figure}
 For the moment the global results are consistent with S and T both zero, fig.\ref{fig:S&T}.
Atomic PNC results are beginning to give a case for S negative, but the precision
is hardly yet enough to be convincing evidence for new physics.

Again, as in the case of new tree level interactions the results for different
experiments are similar and all are about equally sensitive. Higher energy experiments
are a little better for radiative corrections, atomic experiments are a little
better for new tree level interactions. None is inherently significantly better,
so until very high energies can be probed directly all are important for providing
constraints and generating new clues, and all complement each other by providing
this information at a wide range of energy scales. 

To fully appreciate the relative merits of these general classes of measurements,
some details of the effects they are intended to detect must be considered.
This will be treated explicitly later in sections \ref{Sec:NewTree} and \ref{Sec:RadiativeCorrections}.
At the least, Atomic PNC provides just yet another constraint, dependent on
a different combination of these radiative correction parameters, and more importantly
a test of a rather different character since it is at a radically different
energy scale. This alone provides considerable merit. An atomic PNC experiment's
increased sensitivity to new tree level interactions, and the possibility that
the current trend towards negative S makes an even more compelling case for
continued work toward increased precision.

\section{Atomic PNC}

\subsection{The universal quantity}

Precise atomic experiments already exist. The quantity directly measured is
a function of the mixing of opposite parity atomic states due to parity violating
interactions. For any given set of states this mixing is proportional to a quantity
\( Q_{W} \) called the weak charge of the nucleus. This is basically the equivilant
of the nuclear electric charge in electromagnetism, but in this case for the
Weak interaction. So it depends only on the type of atom and is independence
of the particular experiment and independent of the set of states used to measure
it.

\subsection{Precision Goals}

A Stark interference measurement in Cesium currently provides the most precise
result, giving \( Q_{W}\left( Cs\right)  \) to 0.3\%. Prior to that, optical
rotation in Thallium gave \( Q_{W}\left( Th\right)  \) to 1.2\%. These have
already provided important information about the Standard Model, in particular
constraints on possible new heavy gauge bosons. But as nothing new has yet appeared,
even higher precision is required. A useful next target is 0.1\%. This would
give S to \( \pm 0.1 \) and either resolve the inconsistency with high energy
results or provide convincing proof that the discrepancy is real, and along
the way be able to uncover new tree level interactions at energy scales up to
2.5TeV.

It is certainly possible that the existing experiments could be improved to
this level, but they are probably starting to hit their practical limits in
terms of sensitivity and systematics and a new generation of experiments is
required. Optical rotation is the method with the highest sensitivity, but the
best experiments are now limited by an incomplete knowledge of background, non-parity
related rotations, and more importantly, relatively poor atomic theory for the
quantities needed to relate the fundamental theories to the direct experimental
observables. The overall uncertainly is 1.2\% systematics and 2\% atomic theory.
The background systematics could be better understood with a bit more study,
but the theory may tend to prove a tougher challenge as the atoms used in these
experiments, Thallium, Lead, Bismuth, have a rather complicated structure with
many valence electrons or an easily perturbed core which require more elaborate
many-body methods.

In contrast, the Stark interference results with Cs are limited to 0.3\% experimental,
and 0.6\% atomic theory. The atomic structure is relatively simple, being alkali-like
it has a single valence electron outside a tightly bound core, so improving
the theoretical results should be straight-forward with another round of calculations.
Similarly, 7 years of careful study have almost eliminated systematic uncertainties
at this level and the sensitivity is rather lower so the experiment is currently
limited by statistics.

Better statistics on the Cesium experiment, or a modified version of it, and
improved atomic theory could yield the desired 0.1\% precision, but currently
this is the only experiment with even this chance promise, and even if successful,
one result, especially such an anomalous and important result, in a single system,
will always be less than completely satisfying. Just as experiments over a large
range of energy scales is required for to provide constraints and cross checks
from different perspectives and uncover important clues, another, more precise,
atomic experiment is required.

\section{IonPNC}

\begin{figure}
{\par\centering \includegraphics{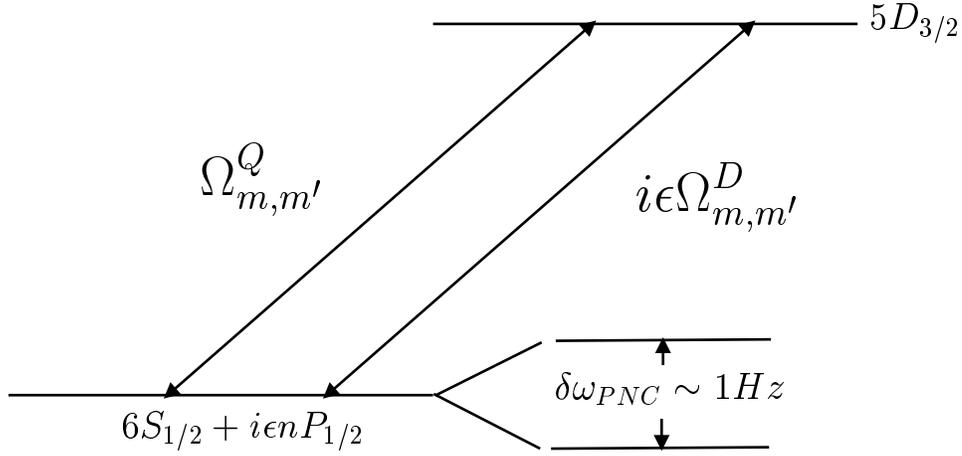} \par}

\caption{\label{fig:ParityCoulingsAndGroundStateSplitting}Energy level diagram of Ba$^+$
and detail of states involved in parity observable.}
\end{figure}

An experiment on a single ion is one possible next-generation project that could
address all of the current difficulties and provide this high precision. The
Barium IonPNC experiment will measure a light shift due to parity violation
that is induced when a particular combination of optical fields is applied,
sec.\ref{Sec:IonPNC}. This light shift splits the ground state \( 6S_{1/2} \)
Zeeman sublevels and this energy shift, or equivilantly the ground state precession
rate it implies is the measured parity violating observable, fig.\ref{fig:ParityCoulingsAndGroundStateSplitting}.

\subsection{Improved Systematics}

The most readily apparent advantage to this technique is improved systematics.
Stark interference experiments use an atomic beam, optical rotation uses an
oven of hot vapor. Both are relatively complicated thermal distributions of
atoms.

For example, with the high densities and temperatures in ovens, collisions and
other interatomic interactions must be considered and accounted for. These can
be fiendishly difficult to understand, and exactly those effects are currently
limiting the experimental precision of the best optical rotation experiment.

For Stark interference measurements the wide spatial extent of the beam make
the uniformity of applied fields very important. Most of the work on systematics
is devoted to understanding and controlling the effects of stray and fringing
or otherwise nonuniform fields.

The ion parity experiment will use the now well developed techniques of the
RF electric quadrupole ion trap. This system, by comparison, is almost trivially
simple. Ideally it is just a single isolated particle at rest in free space
and this is the fundamental reason for its systematic advantages. Of course,
isolated and free is an idealization, an ion trap is not a completely pure environment.
The residual perturbations turn out to be largely unimportant in this case but
there are complications. In particular, the stability and alignment of the applied
optical and magnetic fields is critical \ref{Sec:MisalignmentSystemmatics}
and, as with any precision experiment, complete enumeration of possible difficulties
is elusive. The simple nature of a single ion system, and the small spatial
region over which fields must be carefully controlled make the analysis and
control of these kinds of systematic errors, if not quite trivial, at least
far more tractable.

\subsection{Sensitivity}

This simple feature of a single ion which provides for such advantages in the
analysis of systematic errors might seem to, on the other hand, be detrimental,
if not catastrophic, for sensitivity. Ovens and vapors are macroscopic collections
of atoms and signals are effectively a continuous incoherent average of individual
trials for each of some \( 10^{10} \) atoms. By comparison, one single ion
is an enormous disadvantage. But the same features that ease the systematics
can be exploited to gain back the sensitivity.

The size of parity violating effect for any given atom will depend on the strength
of the applied fields, in all the cases at hand linearly. Generally a dipole
matrix element, \( \Omega _{D} \), is what is directly measured in the experiment.
The size will be given by the mixing \( \varepsilon  \), a dipole matrix element,
\( \left\langle D\right\rangle  \) and the amplitude of an applied electric
field, \( E \), as \( \Omega _{D}\sim \varepsilon \left\langle D\right\rangle E \).
Then the size of the electric field controls the size of the effect. Beams and
ovens are big, requiring, for example, broad laser beams to fill all the interaction
region. In this way field strength is lowered and in the end limited by practical
constraints like available, or usable, laser power. A single ion is very small
and its orbit in the trap is constrained to regions less than 1\( \mu m \).
Low power lasers can be focussed very tightly to easily give much higher fields
strengths.

If a vapor/beam experiment is regarded as many simultaneous trials, the sensitivity
of these and this ion experiment can be discussed in terms of the number of
trials, \( N_{T} \). The precision with which the dipole amplitude can be measured,
\( \delta \Omega _{D} \), generally depends on some line width, \( \Gamma  \),
and the number of trials, and possibly some experimental efficiency factor \( 1/f \)
as \( \delta \Omega _{D}\sim \Gamma /f\sqrt{N_{T}} \). The number of trials
will be given by the number of atoms, \( N \), the total observation time,
\( T \), and the time it takes to make a single trial, \( \Delta t \), as
\( N_{T}=NT/\Delta t \), giving
\[
\delta \Omega _{D}\sim \frac{\Gamma }{f\sqrt{N}}\sqrt{\frac{\Delta t}{T}}\]

This apparently favors small \( \Delta t \) , but at some point as the trial
time is further decreased, the width, \( \Gamma  \), begins to increase. Ideally
the linewidth will be determined by some fundamental structural limit giving
a finite coherence time \( \tau  \) so that \( \Gamma \sim 1/\tau  \), but
it can also have a contribution from the strength of the interaction used to
drive the transition. As the trial time decreased the transition rate must increase
so that something observable happens during the trial time. This leads to a
transition rate contribution to the width of \( \sim 1/\Delta t \). When this
becomes larger than \( 1/\tau  \), \( \delta \Omega _{D} \) begins to increase
again like \( 1/\sqrt{\Delta t} \) with smaller \( \Delta t \) . Then the
uncertainty is minimized for simply, \( \Delta t=\tau  \),

\[
\delta \Omega _{D}\sim \frac{1}{f\sqrt{NT\tau }}\]
and the precision is limited by the coherence time.

In vapors and beams this coherence time is shortened by collisions and transit
times to something less than 1ms. Ion traps are built in ultra-high vacuum systems
where pressures reach below \( 10^{-10}-10^{-11} \) torr giving extremely low
collision rates. As a result in an ion trap,coherence times are limited only
by the natural radiative lifetime of the higher energy state used in the transition.
For the states to be used in the barium ion experiment this is a considerable
longer 50-80s. Since the ion stays in the same place as long as you care to
watch it all of this lifetime is available for observation and there are no
transit times or similar limits.

The overall experimental efficiency,\( f \), is largely dependent on the design
of a particular experiment. Detection efficiency is a measure of how reliably
you can tell that something has happened, in this case the something will be
a spin transition. Any possible conventional method used to detect this would
probably involve looking for a photon emitted either directly by, or as a consequence
of this transition. This kind of system would yield very small detection efficiencies,
less than about 1/1000 because the photon might not come out in the direction
you are looking, and even if it did it might not be seen due to a less than
ideal efficiency of a detector, such as with a photo-multiplier tube of 10\%.
With ion traps, shelving is a common technique that can be used to generate
millions of photons corresponding to single transition from another state. This
kind of light is easy to detect and with variations of these shelving techniques
efficiencies in this experiment will approach 100\%. A detailed analysis of
this experimental efficiency is considered in sec.\ref{Sec:StatisticsAndSensitivity}.

With these experimental conditions the signal to noise ratio is given by,

\[
\frac{\Omega _{D}}{\delta \Omega _{D}}\sim \varepsilon \left\langle D\right\rangle Ef\sqrt{NT\tau }\]
For the IonPNC experiment in particular, with \( \varepsilon \left\langle D\right\rangle E\sim 1Hz \),
\( \tau \sim 100 \) this gives, for \( T=1day \), with \( f\sim 0.4 \), \( \Omega _{D}/\delta \Omega _{D}\sim 1000:1 \).
This turns out to be comparable to the sensitivity of optical rotation experiments,
and larger than the sensitivity for Stark interference measurements, so that
an experiment in a single ion should be at least as sensitive as optical rotation.
Current ion experiments already yield a S/N of 100:1 for measurements of spin
state energy differences. These current results are limited by a large linewidth
from sources of magnetic field noise, but the results are completely consistent
with this statistical analysis so that this expected S/N for a parity measurement
should be realizable with improved technical performance.

\section{Barium}

Besides the general advantages of the ion trap, barium in particular has a number
of advantages as the subject of an atomic parity violation experiment. Most
importantly, barium is a heavy atom. For an atom with atomic number \( Z \),
the amount of mixing of parity eigenstates is increased by \( Z^{3} \) over
the earlier first estimate of the natural size of the effect. The mixing would
be hopelessly small without this enhancement and so the use of a heavy atom
is important. Barium follows cesium in the periodic table and so has a slightly
larger amount of mixing, of similar order compared to any other current atomic
PNC experiment.

\subsection{Atomic Structure}

Barium is also very like cesium in electronic structure. Singly ionized, barium
has a single valence electron outside a tightly bound, 56 electron core. This
is identical to the valence configuration of cesium, beyond that the only important
difference is an extra proton in the nucleus. This makes for relatively straightforward
atomic theory calculations by considering barium to be a single electron system
in a coulomb field modified by the charge distribution of the core electrons.
The largest part of the calculations are given by considering a static core,
but for the precision required for these experiment single, low-level excitations
of the core must be included.

Partly for these reasons, the calculations required to connect an atomic parity
violation experiment to Standard Model predictions are currently most accurate
for cesium. The same techniques can be used for barium and the results should
be at least as accurate.

Similarly, radium may be a good choice for future ion parity experiments. It
is directly below barium in the periodic table and so has the same more easily
calculated electronic structure. It also has a parity violating mixing of atomic
states 50 times larger than barium which would yield tremendous improvements
in sensitivity. An important practical difficulty is that radium has no stable
isotopes, though the longest do last a few days which is plenty long enough
to make a parity measurement in this system possible. But for now the associated
technical problems largely prevent radium from being the ideal first choice
for an ion experiment.

\subsection{Nuclear Structure and Isotope Comparisons}

These precise parity calculations also require detailed knowledge of nuclear
properties. Barium also happens to make this task easier. For theory, the hyperfine
structure and splitting in isotopes with nonzero nuclear spin make them more
difficult to trap and cool. The simplest ion traps will only trap those isotopes
with even-even, spin zero nuclei for which the relevant nuclear structure is
simpler to calculate.

For quantities which must be determined experimentally, there is a miscellaneous
collection of properties that make Barium nuclei good subjects for study. For
example, one important bit of knowledge is the nuclear neutron matter distribution.
To determine this more precisely at smaller length scales, higher energy probes
are needed. In barium these higher energies can be used and still analyzed as
much simpler elastic collisions since the excited states turn out to be widely
separated from the ground state so that they are not excited even at these energies
and as a result don't contribute to the scattering process.

Finally, with atomic PNC experiments there has always been the desire to do
isotope comparisons. In calculating the parity violating mixing of states, atomic
and nuclear contributions can be, approximately, factored into an atomic piece
that is isotope independent and a nuclear piece that is isotope dependent \cite{NuclearTheoryFactoring}.
Then, an experiment on many different isotopes can be used to eliminate, or
at least reduce, the results dependence on precise atomic structure calculations.
The cost is the resulting need for even more precise experimental results and
more accurately known nuclear properties. But as the atomic calculations are
currently the most difficult to do, and the largest sources of uncertainties
in interpreting consequences for the Standard Model, they maybe continue to
be the primary difficult in the future and the trade for a possibly easier nuclear
structure problem could prove worthwhile.

The barium ion parity experiment is the first in which this kind of comparison
would be a practical and promising prospect, in fact is it almost trivial. Different
isotopes can easily be loaded, and identified by the \textasciitilde{}100MHz
scale shifts of their cooling transitions. They already appear regularly during
the standard loading procedure, so far at least three different isotopes have
been seen. For more systematic future work, specific isotopes can be selected
for by adjusting various loading parameters.

The accuracy of this method is improved by studying widely separated isotopes,
those having very different numbers of nuclei. Barium has many stable isotopes,
those with spin zero nuclei that are most easily trapped range from \( Ba^{138} \)
to \( Ba^{146} \) giving a very wide range of \( \Delta N=10 \). This can
even be improved to \( \Delta N=14 \) by included those unstable isotopes with
lifetimes of a few days that could reasonably though to be used for an experiment.
Such a project could be good practice for some future experiment with radium
for which, as mentioned, there are no stable isotopes.

\subsection{Practical Advantages}

Barium is a relatively easy ion to trap and as a result is a popular choice
for trap studies, in fact it was the first single ion to be trapped. The same
practical reasons for this choice still apply and the continued study has made
trapping parameters and procedures well established. 

Many properties such as energies, lifetimes, isotope shifts and branching ratios,
and important effects such as those due to micromotion, magnetic fields and
laser polarizations are well known. The frequencies of cooling transitions are
visible, where lasers are easily available, and stabilization and locking techniques
are abundant and reliable. Sources for neutral barium to ionize and trap exist
and ovens and beam methods have already been well developed. Trapping frequencies
and voltages are well known and accessible with simple equipment. 

None of these pieces make study of a different ion impossible, or even impractical.
If barium wasn't already a good choice for the experiment these practical advantages
wouldn't make it one. Instead all serve to minimize initial practical difficulties
and allow for a quicker route to new areas of interest.

\section{Experiment, Overview, Outline}

The primary purpose of this phase of the IonPNC project was to develop the spin
state manipulation and detection techniques needed to do an experiment of this
kind and determine, and optimize their sensitivity and stability. Effectively
determining precisely how such a measurement will be done in practice and how
well it can be expected to work. A large part of that, as is the case with any
precision measurement, is analyzing systematics, and effort and attention was
equally divided between this sort of theoretical analysis and experiment construction
and operation. 

The following chapters provide a comprehensive summary of the results of these
studies and much of the extensive background required to understand their motivation,
development and analysis. Here also is an attempt to provide a general understanding
of parity violation, and parity violating experiments in general by providing
detailed derivations and discussions of the observable consequences of parity
violation in atoms starting from the parity violation explicitly wired into
The Standard Model. There is a frustratingly large amount of folklore in this
field, as there is in any other, details that are commonly known to be true
but no-one knows why, or who first noticed it was true. In addition there is
extensive recursive referencing such that it is difficult to build the broad
background required to fully understand the developments made and the difficulties
met in this field and the origin of that understanding is spread among dozens
of obscure sources. Here is an attempt to consolidate these results and reduce
the recursion depth.

This document fails, however, to be a completely self contained tutorial on
parity violation due to size , time and ability constraints. Generally, for
results immediately relevant to the experiment are developed in full detail,
but for background tropics there are just enough details provided to connect
results to familiar result from advanced quantum mechanics, field theory and
many-body physics. What is missing is the final details required to connect
background results to elementary quantum mechanics and particle theory. These
details are largely contained in the separately published IonPNC Operators Manual,\cite{Schacht00}.
In addition, only the results that are fundamentally new, or those immediately
required for the same new results, are presented here. There are a great many
peripheral details, including RF Impedance Matching, Optical Resonating Cavities,
Frequency Doubling, Dirac Coulomb Wavefunctions, Many-Body Corrections, Multipole
transitions, Alkali like wavefunctions and matrix elements, two state systems
and resonance and analysis of other old and new methods for studying parity
violation. Some result are necessary for understanding the operation of this
particular apparatus, or understanding the many corrections to the conventional
analysis, but all are existing and well-known. These details are similarly contained
in \cite{Schacht00}.

A final omission that must be mentioned is a colloquial discussion of chiral
symmetry and parity violation. Understanding the structure and origin of parity
violation is not the primary purpose of this project. Instead parity is used
as a tool to provide a sensitive probe for the existence of possible new physics.
But, the idea of parity violation is interesting enough for its own sake that
it deserves some attention. Similarly, to help understand both the parity violating
signal and the associated systematic problems of these experiments, it is productive
to start to build a more intuitive understanding of this chiral symmetry and
the consequences of its conservation or violation in some detail. This includes
an elementary discussion of the discrete parity transformation and the ideal
of scalars and pseudo-scalars and vectors and axial vectors. For background
of this sort, The Ambidextrous Universe by Martin Gardner is highly recommended,
as well as The Feynman lectures for certain details of the physical consequences
of parity violation.

The most glaring omissions and corrections will be included in a revised version
of this thesis available at the IonPNC web site along with the expanded IonPNC
Operators Manuals. These are accessible from the links at the University of
Washington Physics, Atomic Physics web site, www.phys.washington.edu/groups/atomic/.
Currently this groups site is located at www.phys.washington.edu/\textasciitilde{}fortson.
Check there for updated and corrected versions rather than relying on the accuracy
of the archived microfilm version.

\chapter{Atomic Parity Violation and the Standard Model}

The easiest effect to understand of parity violation in atoms, and so far the
only one really exploited or carefully studied, is the mixing of opposite parity
eigenstates by the tree level exchange of a \( Z_{0} \) between a valence electron
and the nucleus. On the atomic side, the general structure of this contribution
is easily understood with simple perturbation theory, the nonrelativistic Schrodinger
equation and a point-like nucleus. For calculations of any accuracy at least
some rudimentary relativistic effects and the finite nuclear size must be included,
and for precision the full Dirac equation, collective effects of the core electrons
and accurate nuclear matter and charge distributions must be included. These
improvements will be introduced and outlined but not extensively pursued. There
are, for present purposes, a practical difficulty as they provide no insight
into new physics. In contrast, the exchange of a single \( Z_{0} \) at tree
level remains a good approximation to a few percent, but corrections to this
will be discussed in considerable details as it is these corrections that contain
the desired information about physics beyond the Standard Model. Treating these
possibilities carefully provides insight into what kinds of new effects atomic
PNC is sensitive to and how they would appear in an experiment.

\section{Tree Level Overview}

\begin{figure}
{\par\centering \includegraphics{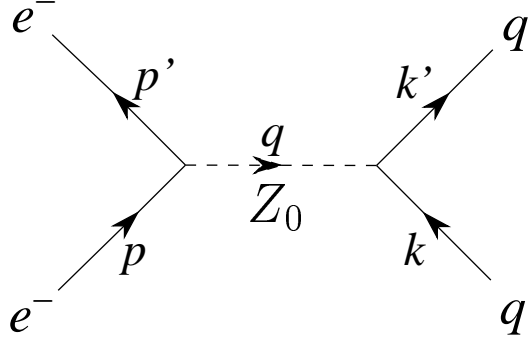} \par}

\caption{\label{fig:Z0Exchange}Tree level exchange of a \protect\( Z_{0}\protect \)}
\end{figure}

A matrix element between arbitrary electronic and nuclear states can be written
as a series of Feynman diagrams. The largest contributions are from the tree
level exchange of a single particle, which correspond to a classical interaction
of fields, fig. \ref{fig:Z0Exchange}.

\subsection{Single Particle Interactions}

The interaction is simplest when the particle states are plane waves, the matrix
element is given immediately by the Feynman rules. For an electron interacting
with another point-like fermion, such as a quark, \( \mathcal{M}=j_{e}^{\mu }(p^{\prime },p)D_{\mu \nu }(p^{\prime },p,k^{\prime },k)j_{q}^{\nu }(k^{\prime },k) \).
The currents are given by \( j^{\mu }=g\bar{u}(p^{\prime })\Gamma ^{\mu }u(p) \)
with \( \Gamma ^{\mu }=(c_{V}\gamma ^{\mu }-c_{A}\gamma ^{\mu }\gamma _{5})/2 \)
and \( g \) the appropriate coupling constant, \( \bar{u}=u^{\dagger }\gamma _{0} \).
The propagator is given by \( D_{\mu \nu }=-ig_{\mu \nu }/(-q^{2}+m^{2}+i\varepsilon ) \)
with \( q=p^{\prime }-p=k-k^{\prime } \). 

For Weak currents, for the exchange of a \( Z_{0} \), \( m=m_{Z} \). The axial
coefficient in the vertex is given by the \( SU(2) \) isospin charge \( c_{A}=T^{3} \).
The vector coefficient includes a contribution from the \( U(1) \) (E+M) charge
appearing from the Higgs mechanism used to break the local gauge symmetry, \( c_{V}=T^{3}-2Q\bar{x} \),
with \( \bar{x} \) given by the weak mixing angle \( \bar{x}=sin^{2}\theta _{W} \)
and the coupling constant is \( g/cos\theta _{W} \) , where it turns out that
\( g=e/sin\theta _{W} \). Also take \( s_{W}=sin\theta _{W} \), \( c_{W}=cos\theta _{W} \).

For atomic wavefunctions the matrix element is best calculated in a spatial
basis.The plane wave solutions provide a momentum space basis which can be integrated
to get the desired form,
\begin{eqnarray*}
\mathcal{M} & = & \int d^{3}r^{\prime }d^{3}rj_{e}^{\mu }(\vec{r}^{\prime })D_{\mu \nu }(\vec{r}^{\prime }-\vec{r})j_{q}^{\nu }(\vec{r})\\
j^{\mu }(\vec{r}) & = & \bar{\psi }_{f}(\vec{r})\Gamma ^{\mu }\psi _{i}(\vec{r})
\end{eqnarray*}
 In this basis the propagator is simply a Yukawa field, which for a point particle
with unit charge is \( D_{\mu \nu }(\vec{r})=V(r)=\frac{1}{4\pi }\frac{e^{-m_{Z}r}}{r} \).
The electron wavefunctions vary on length scales of a few Angstroms. In contrast
the Yukawa field dies away quickly on length scales of a few thousandths of
a fermi, \( c/\hbar m_{Z}\sim 0.002fm \) so on atomic length scales the potential
is effectively a delta function interaction. Fixing the spatial normalization,
\( \int d^{3}rV(r)=1/m_{Z}^{2} \), gives the overall amplitude,

\[
\frac{1}{4\pi }\frac{e^{-m_{Z}r}}{r}\rightarrow \frac{1}{m_{Z}^{2}}\delta ^{3}(\vec{r})\]
 The matrix element is then given by a single integral,
\[
\mathcal{M}_{W}=\frac{g^{2}}{m_{Z}^{2}c_{W}}\int d^{3}rj_{e}^{\mu }(\vec{r})j_{q\mu }(\vec{r})\]
This is equivilant to the limit \( q^{2}<<m_{Z}^{2} \) in the momentum basis
and accurate to \( o(q^{2}/m_{Z}^{2})\sim 10^{-20} \). Errors from this approximation
are completely negligible for calculations accurate to \( 0.1\% \).

Low energy effects are conventionally written in terms of the Fermi coupling
constant \( G_{F} \) where \( G_{F}/\sqrt{2}=g^{2}/8m_{W}^{2} \) with \( m_{Z}^{2}c_{W}=m_{W}^{2} \).
However for comparison with other theories and extensions to the standard model
it is useful to use the actual vertex factors so these coefficients will be
retained explicitly. It will also be convenient to consider and refer to the
vector and axial, parity conserving and parity violating components of the current
independently as
\begin{eqnarray*}
j_{V}^{\mu }(\vec{r}) & = & \bar{\psi }_{f}(\vec{r})\left( c_{V}\gamma ^{\mu }\right) \psi _{i}(\vec{r})\\
j_{A}^{\mu }(\vec{r}) & = & \bar{\psi }_{f}(\vec{r})\left( c_{A}\gamma ^{\mu }\gamma _{5}\right) \psi _{i}(\vec{r})
\end{eqnarray*}
where again
\begin{eqnarray*}
c_{A} & = & T^{3}\\
c_{V} & = & T^{3}-2Q\bar{x}
\end{eqnarray*}

\subsection{Scalar and Pseudoscalar Pieces}

Each \( \Gamma ^{\mu } \) contains a vector, \( c_{V}\gamma ^{\mu } \), a
term whose time component is invariants under a parity transformation, a scalar,
and whose spatial components change sign, a vector, and axial, \( c_{A}\gamma ^{\mu }\gamma _{5} \)
contribution, whose time component is a pseudo-scalar, which changes sign under
parity, and an axial vector which doesn't. The diagonal product of either term
gives a scalar, an effect invariant under a parity transformation. These will
gives effects having a structure identical to electromagnetic processes, a single
photon exchange, and so will likely be invisible in comparison as it is heavily
suppressed by the mass of the exchanged \( Z_{0} \) as discussed in Sec.\ref{Sec:Introduction}.
The cross terms give pseudo-scalars, terms that change sign under parity, which
give effects that violate parity and will be the interesting terms to consider,

\[
\mathcal{M}_{PV}=-\frac{g^{2}}{M_{Z}^{2}c_{W}}\int d^{3}r\left( j_{eA}^{\mu }(\vec{r})j_{qV\mu }(\vec{r})+j_{eV}^{\mu }(\vec{r})j_{qA\mu }(\vec{r})\right) \]

\subsection{Atomic Matrix Elements}

\subsubsection{Electron-Nucleon Interactions}

This gives transition matrix elements for single particles, but in an atom an
electron interacts with a nucleus which is a composite collection of quarks.
The conventional picture of the nucleus is that the quarks are bound into nucleons,
with characteristic sizes given by \( \Lambda ^{QCD} \), and the nucleons are
in turn bound to make the nucleus. For atomic electrons energies are far too
low to resolve any of the quark structure of the nuclei, so the quark currents
\( j_{q}^{\mu } \), can immediately be replaced by nucleon currents \( j_{n}^{\mu } \),
given by the coherent sum of the component quark currents. In the end this just
results in a \( j^{\mu } \) of the same structure with axial and vector coefficients
given by the sum of the quark coefficients. Since these coefficients are written
in terms of \( T^{3} \) and \( Q \), and a nucleon's quark content is chosen
to get the right \( T^{3} \) and \( Q \) for the nucleon, giving \( c_{pA}=1/2 \),
\( c_{pV}=1/2-2\bar{x} \), \( c_{nA}=c_{nV}=-1/2 \). The result still looks
like a current for a pointlike particle and none of the composite quark structure
appears.

\subsubsection{Electron-Nucleus Interactions}

The same is less precisely true for the nucleon as a whole. In most of the atom,
the electron wavefunction changes appreciably on scales of about an angstrom,
which is far longer than the fermi sized length scales of the nucleus,so here
a nucleus could be considered approximately as a pointlike particle with the
appropriate \( T^{3} \) and \( Q \), giving \( c_{NA}=(N_{n}-N_{p})/2=(N-Z)/2 \),
\( c_{NV}=(Z-N)/2-2Z\bar{x}=-(1/2)(N-Z(1-4\bar{x})) \). These can then be identified
as the axial and vector weak charge of the nucleus. But near an atoms own nucleus,
the wavefunction can change very quickly, for a Dirac electron in a Coulomb
field the wavefunction actually diverges at the origin, so that some details
of the nuclear structure may be resolved. It turns out that \( \psi ^{\prime }_{e}(1fm)/\psi _{e}(1fm)\sim 0.01fm \)
so that nuclear structure details begin to contribute at around the few percent
level.

This complicates matters considerably as a nuclear state is generally a complicated
multi-particle wavefunction. But for nuclei in particular, the wavefunction
can be understood well in terms of a product of independent particle wavefunctions
for a well chosen central potential and the end result is that total current
becomes just a sum of the currents of all the nucleons, \( n \), as given by
their single particle wavefunctions  so that effectively the nuclear current
is given by,
\[
j_{N}^{\mu }(\vec{r})=\sum _{n}\bar{\psi }_{nf}(\vec{r})\Gamma ^{\mu }\psi _{ni}(\vec{r})\]

\subsubsection{Multiparticle Electron States}

Similarly, an atom usually consists of many electrons and an arbitrary state
is a multiparticle wavefunction. But for alkali atoms, and alkali-like ions
such as Ba\( ^{+} \), the system is effectively a single electron bound to
a nucleus plus a charge the distribution generated by the remaining core electrons,
which can in term be described, as with nuclei, by a product of single particle
wavefunctions for this core potential. Different atomic states consist of a
changing single particle state for the valence electron and a fixed core and
the current is a sum over the currents of the single particle states, as with
the nucleus.
\[
j_{e}^{\mu }(\vec{r})=\sum _{n}\bar{\psi }_{e_{n}f}(\vec{r})\Gamma ^{\mu }\psi _{e_{n}i}(\vec{r})\]

Notice that these collective currents contain only contributions from changing
one single-particle state, for the electron it is always the valence electron
state. This is a result of using considering only a single interaction, a single
exchange. It is possible that a core electron could be excited to the new state
by the \( Z_{0} \) exchange and the valence electron the drop down to replace
that core electron. This gives the same final state and so contributes to the
total transition amplitude. These processes are slower, the largest effect would
be an electro-magnetic decay of the valence electron which is smaller than the
single valence electron excitation process by \( \alpha  \). But this is a
contribution above the desired \( 0.1\% \) level so precision calculations
must include them. 

This is the beginning of the complicated atomic theory required for these computations.
As mentioned already, these corrections will not be pursued in detail here so
this is also where the results presented here will cease to be rigorously accurate
and are pursued instead to illustrate the general structure of the effects of
the parity violating components of the Weak interaction.

The same corrections would apply to the nuclear currents as well though generally
the nuclear state is unchanged between different atomic energy levels except
for a change of spin. The spin wavefunctions are more easily dealt with, though
for Barium this is irrelevant as its nucleus is spin zero though again that
is only true collectively and when including corrections for the non-trivial
spatial structure of the nucleus this term may appear again.

\subsection{Currents and Dirac Matrix Elements}

Computing these matrix elements precisely then requires detailed knowledge of
the multiparticle wavefunctions, and an even an accurate estimate requires a
good set of single-particle wavefunctions for a reasonable effective potential.
Their general structure, however, can be understood from a simple expansion
of the Dirac wavefunctions in terms of two component spinors.

\subsubsection{Scalar Component}

\( j_{V}^{0} \) is a scalar and can be calculated immediately,
\begin{eqnarray*}
j_{V}^{0} & = & \sum _{n}\bar{\psi }_{nf}c_{V}\gamma ^{0}\psi _{ni}\\
 & = & \sum _{n}c_{Vni}\psi ^{\dagger }_{nf}\left( \gamma ^{0}\right) ^{2}\psi _{ni}\\
 & = & \sum _{n}c_{Vni}\psi ^{\dagger }_{nf}\psi _{ni}
\end{eqnarray*}
When the initial and final wavefunctions are the same this simplifies further.
For example, as discussed above, in general only the spin wavefunctions of a
nucleon will change between atomic energy levels so that, in particular the
spatial wavefunctions will stay the same. Then \( \psi ^{\dagger }\psi \neq 0 \)
requires that the spin wavefunctions also remain unchanged and as a result only
identical initial and final nuclear state contribute to this term which can
then be written in terms of the proton and neutron charge densities,
\begin{eqnarray*}
j_{NV}^{0}(\vec{r}) & = & \sum _{n(ucleon)}c_{Vn}\psi ^{\dagger }_{n}(\vec{r})\psi _{n}(\vec{r})\\
 & = & \sum _{n}c_{Vn}\rho _{n}(\vec{r})\\
 & = & \sum _{n(eutron)}c_{Vn}\rho _{n}(\vec{r})+\sum _{p}c_{Vp}\rho _{p}(\vec{r})\\
 & = & N_{n}c_{Vn}\rho _{n}(\vec{r})+N_{p}c_{Vp}\rho _{p}(\vec{r})\\
 & = & N(-\frac{1}{2})\rho _{n}(\vec{r})+Z(\frac{1}{2}-2\bar{x})\rho _{p}(\vec{r})\\
 & = & -\frac{1}{2}\left( N\rho _{n}(\vec{r})-Z(1-4\bar{x})\rho _{p}(\vec{r})\right) 
\end{eqnarray*}
If the proton and neutron matter distributions are exactly this same this is
further simplified to
\begin{eqnarray*}
j_{NV}^{0}(\vec{r}) & = & Q_{W}\rho (\vec{r})\\
Q_{W} & \equiv  & c_{NV}=-\frac{1}{2}(N-Z(1-4\bar{x}))
\end{eqnarray*}
For Barium, with \( N=82 \), \( Z=56 \) this gives,
\[
Q_{W}\approx -78\]

There is a similar term for the electron, where appropriate which includes instead
the weak charge of the electron, \( c_{eV}=1/2+2\bar{x} \), times the electron
charge density \( \rho _{e} \). Such a term appears only in hight order corrections
to the result being developed here.

\subsection{Pseudoscalar, Vector and Axial Vector Components}

The remaining currents can then be calculated using the appropriate dirac wavefunctions
which solve, 
\[
(\gamma _{\mu }p^{\mu }+V(r))\psi =0\]

The general behavior of the matrix elements by looking at the equations for
the two dimensional spinor components of the full four dimensional wavefunction,
\[
\psi (\vec{r})=\left( \begin{array}{c}
\chi _{+}(\vec{r})\\
\chi _{-}(\vec{r})
\end{array}\right) \]
Now, for explicit calculations a particular representation for the dirac matrices
must be chosen, use,

\[
\begin{array}{ccc}
\gamma _{0}=\left( \begin{array}{cc}
1 & 0\\
0 & -1
\end{array}\right)  & \vec{\gamma }=\left( \begin{array}{cc}
0 & \vec{\sigma }\\
-\vec{\sigma } & 0
\end{array}\right)  & \gamma _{5}=\left( \begin{array}{cc}
0 & 1\\
1 & 0
\end{array}\right) 
\end{array}\]
Substituting these definitions into the Dirac equation gives,

\begin{eqnarray*}
(-e+m+V(r))\chi _{+}+(p_{j}\sigma _{j})\chi _{-} & = & 0\\
(e+m+V(r))\chi _{-}+(p_{j}\sigma _{j})\chi _{+} & = & 0
\end{eqnarray*}
Which simply relates \( \chi _{\pm } \) by, in particular,
\[
\chi _{-}=-\frac{p_{j}\sigma _{j}}{e+m+V(r)}\chi _{+}\equiv -\vec{\sigma }\cdot \vec{\mathcal{P}}\chi _{+}\]
 where \( \vec{\mathcal{P}} \) is used for shorthand as
\[
\vec{\mathcal{P}}=\frac{\vec{p}}{e+m+V(r)}\]
This is complicated in practice if \( V(r) \) is not a simple scalar, but formally
the relation will appear the same as the inverse of the operator that appears
as the denominator should exist since energies will always be non-zero and in
fact positive. Ignoring the possible spin structure of the potential, elementary
manipulation of the spin matrices gives,

\begin{eqnarray*}
j_{V}^{0}=\bar{\psi }_{f}\gamma ^{0}\psi _{i} & = & (1+\vec{\mathcal{P}}_{j}\cdot \vec{\mathcal{P}}_{i})\chi _{f}^{\dagger }\chi _{i}\\
j_{A}^{0}=\bar{\psi }_{f}\gamma ^{0}\gamma _{5}\psi _{i} & = & -\chi _{f}^{\dagger }(\vec{\mathcal{P}}_{j}+\vec{\mathcal{P}}_{i})\cdot \vec{\sigma }\chi _{i}\\
j^{i}_{V}=\bar{\psi }_{f}\gamma ^{i}\psi _{i} & = & -\chi _{f}^{\dagger }(\mathcal{P}_{f,i}+\mathcal{P}_{i,i})\chi _{i}\\
j_{A}^{i}=\bar{\psi }_{f}\gamma ^{i}\gamma _{5}\psi _{i} & = & \chi _{f}^{\dagger }\sigma _{i}(1-\vec{\mathcal{P}}_{j}\cdot \vec{\mathcal{P}}_{i})\\
 & + & (\mathcal{P}_{f,i}(\vec{\mathcal{P}}_{i}\cdot \vec{\sigma })+\mathcal{P}_{i,i}(\vec{\mathcal{P}}_{f}\cdot \vec{\sigma }))\chi _{i}
\end{eqnarray*}

The matrix elements are seen now in terms of the spin and momentum matrix elements
in the more familiar basis of two component spinors.

For this problem it will turn out that much of the motion can be considered
to be approximately non-relativistic. In this case the matrix element simplify
further. Here \( \mathcal{P}^{2}\sim \left( p/m\right) ^{2}<<1 \) so that the
lower spinor component of the dirac spinor is much smaller than the upper component
which then leads in the usual way to the Non-relativistic wavefunction and the
Schrodinger Equations. Rather than start over to get the matrix elements, however,
simply take the non-relativistic limits of these Dirac matrix elements with
\( \mathcal{P}<<1 \) and \( \vec{\mathcal{P}}\approx \vec{p}/m \),

\begin{eqnarray*}
j_{V}^{0}=\bar{\psi }\gamma ^{0}\psi  & \approx  & \chi ^{\dagger }_{+f}\chi _{+i}\\
j_{A}^{0}=\bar{\psi }\gamma ^{0}\gamma _{5}\psi  & \approx  & -(1/m_{f}+1/m_{i})\chi _{+f}^{\dagger }(\vec{p}\cdot \vec{\sigma })\chi _{+i}\\
j_{V}^{i}=\bar{\psi }\gamma ^{i}\psi  & \approx  & -(1/m_{f}+1/m_{i})\chi _{+f}^{\dagger }p_{i}\chi _{+i}\\
j_{A}^{i}=\bar{\psi }\gamma _{5}\gamma ^{i}\psi  & \approx  & \chi _{+f}^{\dagger }\sigma _{i}\chi _{+i}
\end{eqnarray*}

\subsection{Non-Relativistic Nuclei}

The currents that give the matrix elements contain momentum dependent and momentum
independent terms. The momentum dependent terms are generally smaller by a factor
\( v/c \) or \( p/E \). The nucleus, in particular, is approximately fixed.
For hydrogen the electron-nucleus mass ratio is already very small \( m_{e}/m_{N}<1/2000 \)
and for heavy atoms such as Barium with \( N=136 \) the ratio is less than
\( 10^{-5} \). As a result the nucleus, though not precisely stationary, is
certainly moving non-relativistically and its collective motion can be neglected
to a part in \( 10^{5} \).

However, even a fixed nucleus ultimately consists of a confined bag of free
nucleons, and then quarks, which could be moving, in principle, arbitrarily
quickly. It turns out, however, that the nucleon motion is non-relativistic,
though not precisely so, \( v_{n}/c\sim 0.1 \). This is a relatively large
contribution but it is partially masked by the collective motion of all the
nuclei. As mentioned above, electron energies generally aren't sensitive to
sub-nuclear structure and so any momentum-dependent pieces tend to average to
zero. The electron matrix elements are partly sensitive to these details at
the few percent level and the combination results in partial sensitivity to
the motion of the nucleons at about the \( 0.1\% \) level.

This is starting to become important, and again a precision calculation will
then have to consider them more carefully, but here they will be neglected as
the approximation simplifies the matrix elements considerably. Neglecting these
term yields, for the nuclear currents,
\begin{eqnarray*}
j_{NV}^{\mu } & \propto  & (1,\vec{0})\\
j_{NA}^{\mu } & \propto  & (0,\vec{s})
\end{eqnarray*}
so that when combined with the electron currents the spatial components in the
term involving the axial electron current, and the time component in the term
with the vector electron current can be neglected giving,

\begin{eqnarray*}
\mathcal{M}_{PV} & = & \frac{g^{2}}{m_{Z}^{2}c_{W}}\int d^{3}r\left( j_{eA}^{0}j_{NV0}-j_{eV}^{i}j_{NAi}\right) \\
 & \approx  & \frac{g^{2}}{m_{Z}^{2}c_{W}}\int d^{3}r\left( \left( \psi _{e}^{\dagger }\gamma _{5}\psi _{e}\right) \left( \chi ^{\dagger }_{Nf}\chi _{Ni}\right) -\left( \psi _{e}^{\dagger }\gamma _{5}\psi _{e}\right) \left( \chi ^{\dagger }_{Nf}\vec{\sigma }\chi _{Ni}\right) \right) 
\end{eqnarray*}
Again, in the case when the initial and final nucleon wavefunctions are the
same the nucleon wavefunctions simplify to,
\begin{eqnarray*}
\chi ^{\dagger }_{N}\chi _{N} & = & Q_{W}\rho _{N}\\
\chi ^{\dagger }_{N}\vec{\sigma }\chi _{N} & = & \vec{s}_{N}/2
\end{eqnarray*}
Note that the former, \( Q_{W} \) term, grows with the size of the nucleus,
roughly as \( Z \), the vector contributions of each nucleon to the current
add coherently, while the latter term depends on the total spin of the nucleus
which is generally much smaller than \( Z \) since nuclear pairing favors nucleons
pairs of nucleons with opposite spins and the resulting total nuclear spin tends
to be relatively very small. In this way the nuclear spin dependent term is
much smaller then the spin independent term, and of course, for spin zero nuclei,
such as with Barium, it gives zero.

\subsection{Non-relativistic Electrons}

These two component wavefunctions are more familiar and easier to deal with.
For the nucleus they tend to be a good approximation but it is far less accurate
for the electrons as it turns out that the electrons become significantly relativistic
near the nucleus, which is only apparent when studying the detailed solutions
to the dirac equation. Unlike the omissions consider so far, which amount to
corrections of a few percent, this modification changes the result by a factor
of 2-5. Still the structure is unchanged, and even this non-relativistic case
will not be evaluated precisely, so for this survey consider,
\[
\mathcal{j}_{\mathcal{eA}}^{0}\approx \chi ^{\dagger }_{ef}(\vec{\sigma }\cdot \vec{p})\chi _{ei}\]
A simple single particle solution to the Dirac Equation for a Coulomb potential
plus the electron charge distribution and a finite nuclear size provides a reasonable
accurate result.

\section{Parity Mixing}

With the nuclear spin dependent term generally smaller than the remaining spin
independent term, the largest effect comes from the \( j_{eA}^{0}j_{NV0} \),
\[
\mathcal{M}_{PV}=Q_{W}\frac{g^{2}}{m_{Z}^{2}c_{W}}\int d^{3}r\left( \chi ^{\dagger }_{ef}(\vec{\sigma }\cdot \vec{p})\chi _{ei}\rho \right) \]
The parity violating nature of this term is more apparent here in a familiar
form. \( \vec{p} \) is a vector and \( \vec{\sigma } \) is an axial vector
so that the dot product is a pseudo-scalar. It changes sign with a parity transformation,
\( \vec{p} \) changes sign and \( \vec{\sigma } \) doesn't. So that in a left-handed
coordinate system an minus sign would have to be included and the form of the
equation is not independent of the handedness of the coordinate system. The
handedness implicitly defined by this term gives the observable parity violating
processes.

\subsection{Total Spin}

Consider an arbitrary pair of initial and final states, \( \left| n_{i},l_{i},j_{i},m_{i}\right\rangle  \)
and \( \left| n_{f},l_{f},j_{f},m_{f}\right\rangle  \). \( \vec{\sigma }\cdot \vec{\nabla } \)
is a (pseudo)scalar, that is it is invariant under rotation and so commutes
with the total angular momentum operators \( j_{i} \). As a result, it does
not change \( j \) or \( m \) and so the initial and final total angular momentum
must be the same.

\subsection{Point Nucleus}

Further progress now requires details of the nuclear charge distribution and
the electron wavefunctions. As discussed above, corrections from the sensitivity
of the valence electron to nuclear structure is small, around 1\%, so the general
structure can be obtained by considering the nucleus to be uniformly distributed,
and even more simply, since it is so small relative to atomic dimensions, as
a point-like delta function distribution. This results in the matrix element
being given by,
\[
\mathcal{M}_{PV}=Q_{W}\frac{g^{2}}{m_{Z}^{2}c_{W}}\frac{2}{m_{e}}\left( \chi ^{\dagger }_{ef}(0)(\vec{\sigma }\cdot \vec{p})\chi _{ei}(0)\right) \]
The value of the wavefunctions at the origin determine the matrix element. Again,
to compute this accurately the full solution to the dirac equation must be used,
but again, the general structure can be studied with the much simpler non-relativistic
schrodinger equation,
\[
(\frac{p^{2}}{2m}+V(r))\chi =E\chi \]
 This limit also makes the approximation of a point-like nucleus much better
as well but turns out to be poor overall as the solutions to the schrodinger
and dirac equations differ considerably near the nucleus.

\subsection{Angular Momentum}

The behavior of the wavefunctions at the origin is particularly easy to understand
for the schrodinger equation. For a spherically symmetric spin independent potential,
the wavefunction factors the wavefunction into a radial part \( R_{jl}(r) \),
an angular part given by a spherical harmonic \( Y^{l}_{m} \), and a spin \( \chi _{m} \).
For any potential satisfying \( \lim _{r\rightarrow 0}r^{2}V(r)=0 \) the small
\( r \) dependence of the solutions are given by \( R(r)\propto r^{l}. \)
The matrix element requires a gradient of \( \chi _{i} \), so is proportional
to,
\begin{eqnarray*}
\mathcal{M} & \propto  & R_{f}(r)\vec{\nabla }(R_{i}(r)Y^{l_{i}}_{m})|_{r=0}\\
 & \propto  & r^{l_{f}}(r^{l_{i}-1}Y^{l_{i}}_{m}+r^{l_{i}}\vec{\nabla }Y^{l_{i}}_{m})|_{r=0}
\end{eqnarray*}
At \( r=0 \) this is non-zero only for \( l_{f}=0 \) as this gives \( r^{l_{f}}=1 \)
independent of \( r \), while \( l\neq 0 \) gives \( r^{l}=0 \) at \( r=0 \).
In the second factor the first term requires \( l_{i}=1 \) for the same reasons,
while the second term is non-zero for \( l_{i}=0 \) and \( \vec{\nabla }Y^{l_{i}}_{0}\neq 0 \).
But an \( l=0 \), \( S \), wavefunction is spherically symmetric and has zero
gradient so only the first term can contribute. The net result is that this
term is non-zero only for \( l_{f}=0 \) and \( l_{i}=1 \), it only couples
\( S \) and \( P \) states, and again only those having the same \( j \). 

With \( \vec{p}=-i\vec{\nabla } \), the non-zero matrix elements are then given
by,
\[
\mathcal{M}_{PV}=iQ_{W}\frac{g^{2}}{m_{Z}^{2}c_{W}}\frac{2}{m_{e}}\chi ^{\dagger }_{eS}(\vec{\sigma }\cdot \vec{\nabla })\chi _{eP}\]

\subsection{Size}

The overall size of this mixing matrix elements is easy to determine. \( G_{F}=\sqrt{2}g^{2}/8c_{W}m_{Z}^{2}\approx 1.2\times 10^{-5}(GeV)^{-2} \)
, and \( Q_{W}=N-Z(1-4\bar{x})\approx Z/2 \). The \( \chi  \) must have dimensions
of length\( ^{-3/2} \) so that a three dimensional spatial integral gives a
unitless normalization. The derivative gives another factor of length in the
denominator, \( \chi ^{\dagger }_{eS}(\vec{\sigma }\cdot \vec{\nabla })\chi _{eP}\sim 1/l^{4} \).

\subsubsection{\protect\( Z^{3}\protect \) dependence}

For the hydrogen atom length scales are around an angstrom, for a simple electron
bound to a nucleus with charge \( Z \) the lengths scales are reduced linearly
by \( Z \). So apparently \( \chi ^{\dagger }_{eS}(\vec{\sigma }\cdot \vec{\nabla })\chi _{eP}\sim Z^{4}/l^{4} \).
But for a many electron atom the increase is not that extreme. It turns out
that the extra charge results in the electron being more likely to be found
near the nucleus, and in particular at the origin, by a single factor of \( Z \),
since far from the nucleus most the of nuclear charge is shielded by the core
electrons, but the momentum near the nuclear is still increased by \( Z \),
\( \chi ^{\dagger }_{eS}(\vec{\sigma }\cdot \vec{\nabla })\chi _{eP}\sim Z^{2}/l^{4} \).
With \( m_{e}=0.511keV \), This yields the estimate,

, 
\[
\mathcal{M}_{PV}\approx -iZ\frac{\sqrt{2}}{8}1.2\times 10^{-5}\frac{1}{GeV^{2}}\frac{1}{511eV}\frac{Z^{2}}{angstrom^{4}}\]
In these units \( 1/angstrom\sim 1971eV \) giving

\[
\mathcal{M}_{PV}\approx -iZ^{3}6.3\times 10^{-17}eV\]
This gives the typical size of the mixing and some insight into its dependence
on the charge of the nucleus.

\subsubsection{Mixing in Barium}

For Barium with \( Z=56 \) this gives,
\[
\mathcal{M}_{PV}\approx -i1.1\times 10^{-11}eV\]
This is exactly the right order of magnitude. Again an accurate calculation
requires at least Dirac wavefunctions and a finite nuclear size.

\subsection{Phase}

\label{Sec:ImaginaryMixing}

Besides the overall size, this explicit calculation results in a purely imaginary
matrix element. This turns out to be a general result and can be understood
in terms of the parity and time-reversal symmetry of the \( \vec{\sigma }\cdot \vec{p} \)
operator.

Consider any operator \( H \) with well defined symmetry properties under parity
and time reversal, 
\begin{eqnarray*}
T^{\dagger }HT & = & \eta _{T}H\\
P^{\dagger }HP & = & \eta _{P}H
\end{eqnarray*}
 These properties can be used about \( H \) matrix elements of \( H \) between
states particular kinds of states, 
\[
\left\langle f\right| H\left| i\right\rangle \]

For parity eigenstates, matrix elements are easily shown to satisfy a simple
selection rule, 
\[
\left\langle f\right| H\left| i\right\rangle =\eta _{P}\left\langle f\right| P^{\dagger }HP\left| i\right\rangle =\eta _{P}\eta _{f}\eta _{i}\left\langle f\right| H\left| i\right\rangle \]
 This is the basis of the usual selection rules. If the product of all the parity
quantum numbers is negative, this matrix element must be zero. If \( H \) is
\( P \) odd the initial and final states must have opposite parity to give
a nonzero matrix element, similarly \( H \) \( P \) even only couples states
of the same parity. 

A similar transformation using time reversal leads to a relation of this matrix
element to a corresponding matrix element between time reversed states. Time
reversal is weird, eventually you can show, Sakuri 4.4.42\cite{Sakurai}, 
\[
\left\langle f\right| H\left| i\right\rangle =\eta _{T}\left\langle \bar{f}\right| H\left| \bar{i}\right\rangle ^{*}\]
 Where \( \left| \bar{\alpha }\right\rangle  \) are the time reversed states,
and for angular momentum states,
\[
T\left| lm\right\rangle =(-1)^{m}\left| l,-m\right\rangle \]
The time reversal flips the spin of the state. This can be flipped back with
a rotation about an axis perpendicular to \( \hat{z} \). For example, as shown
in \ref{Sec:SpatialWavefunctionRotation},
\[
e^{i\pi J_{y}}\left| l,-m\right\rangle =(-1)^{l+m}\left| l,m\right\rangle \]
This gives 
\[
T\left| lm\right\rangle =(-1)^{l}e^{i\pi J_{y}}\left| l,m\right\rangle \]
where \( (-1)^{l} \) also happens to be the parity, \( \eta  \), of the state.
The time reversed matrix element then becomes
\begin{eqnarray*}
\left\langle f\right| H\left| i\right\rangle  & = & \eta _{T}\left\langle \bar{f}\right| H\left| \bar{i}\right\rangle ^{*}\\
 & = & \eta _{T}\eta _{i}\eta _{f}\left\langle f\right| e^{-i\pi J_{y}}He^{i\pi J_{y}}\left| i\right\rangle ^{*}
\end{eqnarray*}

If the hamiltonian is also rotationally invariant, as it must be if these angular
momentum states are actually eigenstates, the rotation operators leave \( H \)
invariant and yield, 
\[
\left\langle f\right| H\left| i\right\rangle =\eta _{T}\eta _{i}\eta _{f}\left\langle f\right| H\left| i\right\rangle ^{*}\]
The previously derived parity selection rules show that the matrix element is
zero unless \( \eta _{P}\eta _{i}\eta _{f}=1 \) or \( \eta _{i}\eta _{f}=\eta _{P} \)
and the matrix element relation can be written more compactly, 
\[
\left\langle f\right| H\left| i\right\rangle =\eta _{T}\eta _{P}\left\langle f\right| H\left| i\right\rangle ^{*}\]
 Then, if the product of symmetry quantum numbers is odd, the matrix element
must be imaginary, and if the product is even, the matrix element must be real.
\begin{eqnarray*}
PTeven & \Rightarrow  & \left\langle f\right| H\left| i\right\rangle =real\\
PTodd & \Rightarrow  & \left\langle f\right| H\left| i\right\rangle =imaginary
\end{eqnarray*}

\( \vec{\sigma }\cdot \vec{p} \) is \( P \) odd, as already observed, and,
since both vectors change sign with time-reversal, \( T \) even, any non-zero
matrix elements must be purely imaginary, as already seen explicitly in one
case. This phase has important consequences for the resulting observables and
so will generally be written explicitly for emphasis.

\subsection{Perturbation Theory}

In the end the result is simply a mixing of \( S \) and \( P \) states. The
effects of this on the atom are simply understood with first order perturbation
theory. In the usual way, separate the full hamiltonian into a piece that is
easy, which in this case will be taken to include just electromagnetic interactions,
\( H_{EM} \) , and a piece that is small, here Weak, or possible new non Standard
Model interactions, \( H_{W} \) or \( H_{NEW} \). The basis for expanding
solutions to the full hamiltonian is then the usual atomic states and to first
order in the interaction the wavefunctions become,

\[
\left| i\right\rangle =\left| i\right\rangle ^{0}+\sum _{i=f}\left| f\right\rangle ^{0}\frac{\left\langle f\left| H_{I}\right| i\right\rangle ^{0}}{E_{f}^{0}-E_{i}^{0}}+o\left( \frac{H_{I}^{2}}{\Delta E^{2}}\right) \]

As just shown the interaction couples \( S \) and \( P \) states with the
same \( j \) giving, in particular,
\[
\left| nS_{j}\right\rangle =\left| nS_{j}\right\rangle ^{0}+\sum _{n^{\prime }}\left| n^{\prime }P_{j}\right\rangle ^{0}\frac{\left\langle n^{\prime }P_{j}\left| H_{I}\right| nS_{j}\right\rangle ^{0}}{E_{n^{\prime }P_{j}}^{0}-E_{nS_{j}}^{0}}\]
Write the matrix element coefficients in terms of 
\[
i\varepsilon _{nn^{\prime }}=\left\langle n^{\prime }P_{j}\left| H_{I}\right| nS_{j}\right\rangle ^{0}/(E_{n^{\prime }P_{j}}^{0}-E_{nS_{j}}^{0})\]
 Since the matrix element is purely imaginary, \( \varepsilon  \) is real.
This gives,
\[
\left| nS_{j}\right\rangle =\left| nS_{j}\right\rangle ^{0}+i\varepsilon _{nn^{\prime }}\left| n^{\prime }P_{j}\right\rangle ^{0}\]
\( P \) states are similarly modified,
\[
\left| nP_{j}\right\rangle =\left| nP_{j}\right\rangle ^{0}+i\varepsilon _{n^{\prime }n}\left| n^{\prime }S_{j}\right\rangle ^{0}\]
 Energy separations in atoms are of order \( eV \), independent of \( Z \)
since much of the wavefunction is outside the nucleus where most of the nuclear
charge is shielded, though it of course depends on the ionization. With this
observations the general size of the mixing can be determined,
\[
\varepsilon \approx Z^{3}6.3\times 10^{-17}\]
where again for Barium with \( Z=56 \),

\[
\varepsilon \approx 10^{-11}\]

\section{Corrections and Additional Effects}

This component of the tree level exchange account for all of the mixing to within
a few percent. This estimate of the size of the matrix element provides a good
measure of the order of magnitude of the mixing. Again, an accurate calculation
of this term requires a single particle wavefunction solution to the dirac equation
with a finite nuclear size. Precise calculations of the mixing require the nuclear
spin dependent terms where appropriate, and eventually the components of the
current neglected in the approximation of non-relativistic nucleons. Finally,
the many-body effects from the multiple electrons must be included in the initial
and final electron states, as well as the nucleon states when the initial and
final nuclear states are not the same. Even with these contributions only the
tree level \( Z_{0} \) exchange has so far been included. Equally important
at this level of precision are other processes and Standard Model radiative
corrections to this tree level exchange.

\subsection{Electron-Electron Interactions}

A \( Z_{0} \) exchange takes place between any two fermions with a weak charge.
The interaction considered so far is that between the electrons and the nucleus,
electron-electron interactions can also be considered. This is included simply
by contracting the Weak electron currents
\[
\frac{g^{2}}{m_{Z}^{2}c_{W}}\int d^{3}rj_{e}^{\mu }(\vec{r})j_{e\mu }(\vec{r})\]
A similar consideration of the pseudo-scalar pieces of this term, and approximately
non-relativistic electrons gives the largest contribution as a term very similar
to that identified as the primary contribution to the electron-nucleus exchange,
\[
\frac{g^{2}}{m_{Z}^{2}c_{W}}Q_{W}^{e}\int d^{3}rj_{eA}^{\mu }(\vec{r})\rho _{e}(r)\]
Here \( \rho _{e}=\sum _{e}\psi ^{\dagger }(r)\psi (r) \) is the core electron
mass density, normalized to have unit volume so the \( Q_{W}^{e} \) includes
all the information about the number of electrons, with \( Q^{e}=c^{e}_{A}=T^{3}-2Q\bar{x}\approx Z/2(1+4\bar{x}) \).
This ends up being small just due to atomic structure.

\subsection{\protect\( Z_{0}\protect \) Vertex Renormalization}

\begin{figure}
{\par\centering \includegraphics{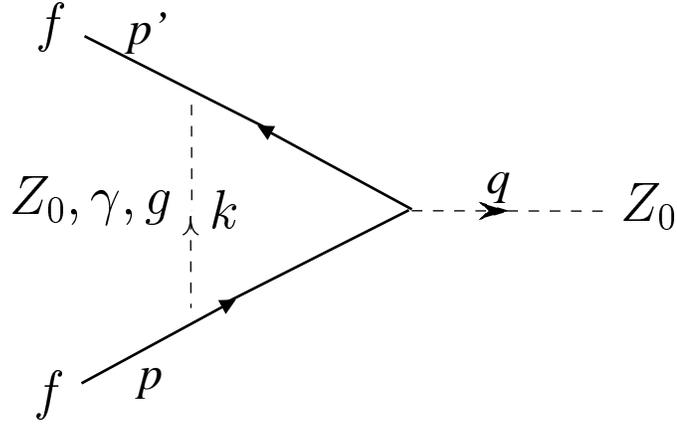} \par}

\caption{\label{Fig:Z0VertexCorrections}General \protect\( Z_{0}\protect \) vertex
corrections for electron or nucleon from \protect\( Z_{0}\protect \), \protect\( \gamma \protect \)
or \protect\( g\protect \) loop.}
\end{figure}

The structure of either the electron or nucleon vertex is altered by vertex
corrections, figure\ref{Fig:Z0VertexCorrections}.

A photon in the vertex loop changes the structure of the vertex by \( o(\alpha ) \),
just as the electromagnetic vertex of an electron acquires an anomalous magnetic
moment, and both the vector and axial components of the \( Z_{0} \) vertex
are adjusted slightly. However, the analogous contribution for a gluon in a
nucleon vertex on the axial term is relatively very large. This effectively
renormalized the axial charge of a nucleon, which has so far appeared only in
the nuclear spin dependent term of the tree level exchange, and is included
by modifying the weak coupling constant \( g \) to \( g\rightarrow \sim 1.25g \).

\subsection{\protect\( \gamma \protect \) Vertex Renormalization}

\begin{figure}
{\par\centering \includegraphics{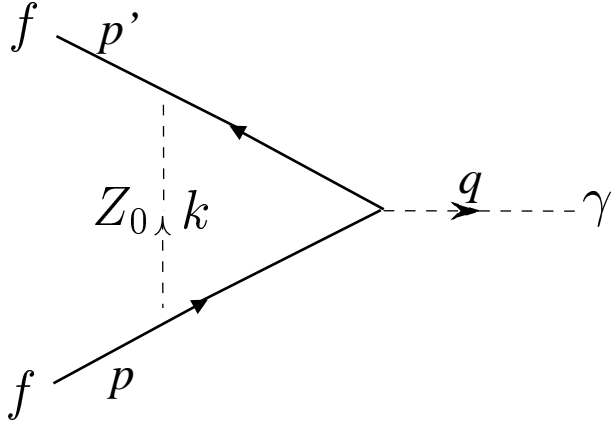} \par}

\caption{\label{Fig:PhotonVertexCorrection}\protect\( \gamma \protect \) axial vertex
correction from \protect\( Z_{0}\protect \) loop.}
\end{figure}

An additional \( Z_{0} \) in the loop adds a contribution \( o(m_{f}^{2}/M_{Z}^{2}) \)
where \( m_{f} \) is the mass of the fermion whose vertex is being modified.
This is a small correction to the \( Z_{0} \) vertex, but it will also modify
the \( \gamma  \) vertex, figure \ref{Fig:PhotonVertexCorrection}.
 The modification to a nucleon's photon vertex is the largest since it has the
largest mass. It effectively gives a small axial component to the photon vertex,
\[
\Gamma ^{\mu }_{n,EM}=Ze\gamma ^{\mu }\rightarrow Ze\gamma ^{\mu }+o(m_{n}^{2}/M_{Z}^{2})Ze\gamma ^{\mu }\gamma _{5}\]
Then a simple photon exchange between an electron and a nucleus contains a pseudo-scalar
term. In effect, the nucleus aquires a chiral electro-magnetic current distribution
and the electron is no longer bound by a spherically symmetric potential and
the same kind of parity mixing occurs. This term involves the axial current
of the nucleon, which as pointed out previously is further suppressed by the
spin of the nucleus. In nuclei with non-zero spin this amounts to another \( \sim 1\% \)
adjustment to the atomic state parity mixing.

\section{New Tree Level Physics}

\label{Sec:NewTree}

The parity violating effects considered so far are those due to Standard Model
processes, in particular, the largest effect is given by a single \( Z_{0} \)
exchange between an atomic electron and the nucleus. Possible extensions to
the Standard Model also naturally violate parity and these new processes can
alter the size and structure of the resulting atomic observables. New physics
can be understood to mean new particles. The effects of new physics can be studied
by considering effects of the addition of various kinds of new particles to
the electron-nucleus interaction.

\subsection{General New Tree Level Interactions}

\begin{figure}
{\par\centering \includegraphics{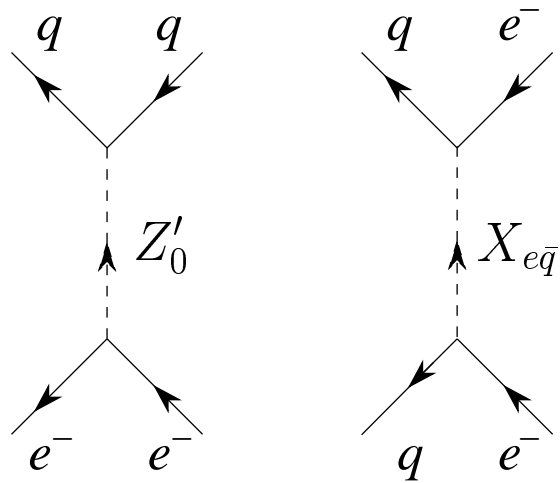} \par}

\caption{\label{fig:NewTreeDiagrams}Typical tree level contributions to Atomic Parity
Violation from new physics.}
\end{figure}

The most straight-forward effects to consider are those due to new tree-level
exchanges, figure \ref{fig:NewTreeDiagrams}.

These examples include possible new particles that occur in extended models
like \( SO(10) \) or \( SU(5) \), \cite{CMS}. These include new \( Z_{0}^{\prime } \)'s
similar to the Standard Model \( Z_{0} \) and leptoquarks. Additional \( Z_{0}^{\prime } \)'s
give contributions with identical structure to the Standard Model \( Z_{0} \)
exchange, which in the limit of a point nucleus the largest contribution becomes,

\begin{eqnarray*}
\mathcal{M}_{Z^{\prime }} & = & -Q_{Z^{\prime }}\frac{g^{2}_{Z^{\prime }}}{4m_{Z^{\prime }}^{2}}\int d^{3}r(\bar{\psi }_{ef}\Gamma ^{\mu }\psi _{ei})(\bar{\psi }_{Nf}\Gamma _{\mu }\psi _{Ni})\\
 & \rightarrow  & -Q_{Z^{\prime }}\frac{g^{2}_{Z^{\prime }}}{4m_{Z^{\prime }}^{2}}(\psi _{ef}^{\dagger }(0)\gamma _{5}\psi _{ei}(0))\\
 & \approx  & -Q_{Z^{\prime }}\frac{g^{2}_{Z^{\prime }}}{4m_{Z^{\prime }}^{2}}\frac{2}{m_{e}}\chi _{fe}^{\dagger }(0)(\vec{\sigma }\cdot \vec{p})\chi _{ie}(0)
\end{eqnarray*}
 The structure of an leptoquark exchange is a little strange since it changes
quarks to electrons, it would look something like,

\[
\mathcal{M}_{X}=-Q_{X}\frac{g^{2}_{X}}{4m_{X}^{2}}\int d^{3}r(\bar{\psi }_{ef}\Gamma ^{\mu }\psi _{Ni})(\bar{\psi }_{Nf}\Gamma _{\mu }\psi _{ei})\]
This terms is generally a bit messy, even for a non-relativistic point nucleus
there are four pseudo-scalar terms involving matrix elements like,
\begin{eqnarray*}
 & \chi ^{\dagger }_{a}\chi _{b}\chi ^{\dagger }_{c}\left( \vec{\sigma }\cdot \vec{p}\right) \chi _{d} & \\
 & \chi ^{\dagger }_{a}\vec{p}\chi _{b}\chi ^{\dagger }_{c}\vec{\sigma }\chi _{d} & 
\end{eqnarray*}
Other models like Technicolor and Supersymmetry, or contributions from one or
more Higgs don't yet appear because they give no new tree level interactions.
Technicolor is intended to account for the Strong interaction so only affect
nucleon interactions, new vertices given by Supersymmetry always involve two
new super-partners where a tree level exchange is just one particle, and similarly
Higgs vertices always involve two Higgs particles. These latter possibilities
then appear first only in loops, radiative corrections. 

The characteristic sizes of these terms can be used to estimate the sensitivity
of a measurement of \( \varepsilon  \) or \( Q_{W} \) to particles with a
certain range of masses. New interactions give a contributions of order \( g^{2}_{X}/m_{X}^{2} \)
to an effect of order \( g^{2}/m_{Z}^{2} \). If the coupling constants are
assumed to be about the same order of magnitude, this is a fractional adjustment
of order \( m_{Z}^{2}/m_{X}^{2} \). For precisions of \( \sim 0.1\% \) this
gives a mass sensitivity of 
\[
m_{X}\sim m_{Z}\sqrt{10^{3}}\approx 30\cdot 80GeV\approx 2.5TeV\]
As discussed in Sec.\ref{Sec:Z0PoleSensitivity} at the \( Z_{0} \) pole this
is generally a correction to a process of order \( g^{2}/m_{Z}\Gamma _{Z} \).
With a similar precision this gives sensitivities of
\[
m_{X}\sim \sqrt{m_{Z}\Gamma _{Z}}\approx \sqrt{80\cdot 0.3}GeV\approx 500GeV\]

These estimates are made assuming the new coupling constants are about the same
as the Weak coupling constant \( g \). For an arbitrary new interaction the
coupling constant could in principle be anything implying that even new particles
with smaller masses may not be detectable in this way because the coupling constant
is very small. But in consistent extensions to the Standard Model, and especially
with Grand Unified Theories there are rigid relations between new and old coupling
constants, in part because part of the point of a Grand Unified Theory is to
reduce the number of fundamental coupling constants. In general the coupling
constants are of the same order of magnitude with freedom for adjustment possible
only be changing the structure of the model. Specific predictions for sensitivity
to new physics then depends on the detailed structure of the model.

\section{Radiative Corrections}

\label{Sec:RadiativeCorrections}

\begin{figure}
{\par\centering \resizebox*{1\textwidth}{!}{\includegraphics{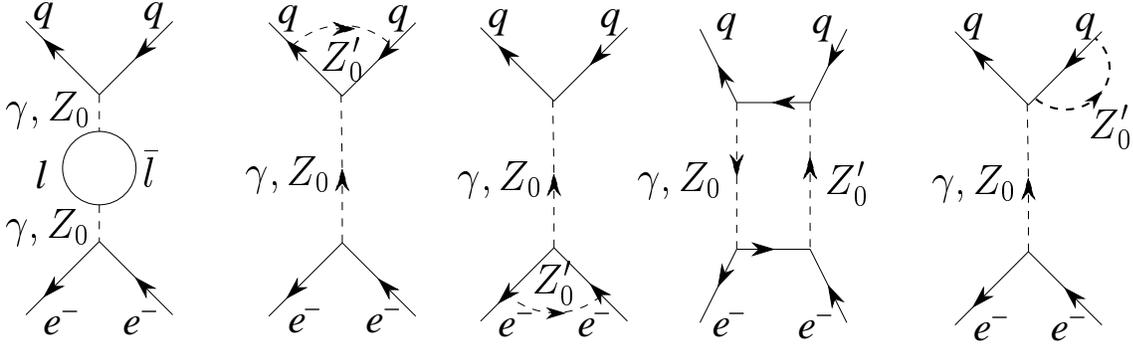}} \par}

\caption{\label{fig:NewLoopDiagrams}Typical radiative correction contributions to Atomic
Parity Violation from new physics.}
\end{figure}

New processes can also appear as corrections to the tree level Weak \( Z_{0} \)
exchange through radiative corrections. The possibilities for the addition of
a single new particle to the tree level diagram are shown in figure \ref{fig:NewLoopDiagrams}.
These largest single-loop corrections all be from new neutral current exchanges.

These are much more complicated to evaluate as they all involve loops which
require careful attention to renormalization details to interpret and generally
change the structure of the effects as well. However, it turns out that the
largest of these radiative corrections can be accounted for very compactly and
result in effects with a structure identical to that the effects already considered
due to the Weak interaction, and so can be included by adjusting the parameters
of these Weak results. The is a slightly specialized and streamlined version
of the methods used in \cite{PeskinSTU}.

\subsection{Oblique Corrections}

These modifications can can be divided into direct corrections which include
vertex corrections and box diagrams and mass renormalization, and oblique corrections,
the vacuum polarizations, which only modify the propagator of the exchanged
particle. For modifications to weak processes, the contributions from new heavy
particles from the direct corrections turns out to be smaller than those from
the oblique corrections by a factor \( m_{f}^{2}/m_{Z^{\prime }}^{2} \),where
\( m_{f} \) is the mass of the lightest fermion attached to the loop and is
in all cases very light relative to the likely \( \sim 1TeV \) mass of a new
\( Z_{0}^{\prime } \). 

The oblique corrections also happen to be the easiest to consider and can be
included simply by modifying the propagator of all the Electro-Weak gauge bosons
from their bare, tree level form \( D^{0} \) to a full, renormalized \( D \).
With only this simple observation, the general structure of the effects of the
oblique corrections can be determined without any further details regarding
the form of the renormalized propagator. Any Electro-Weak vertex will appear
as
\[
\frac{e}{s\sqrt{2}}(j_{+}^{\mu }W_{+}+j_{-}^{\mu }W_{-})+\frac{e}{sc}(j_{3}^{\mu }-s^{2}j_{\gamma }^{\mu })Z+ej_{\gamma }^{\mu }A\]
Matrix elements will be given by contractions of these current through an appropriate
propagator. Omitting explicit Lorentz indices, matrix elements are given by
\begin{eqnarray*}
\mathcal{M} & = & \frac{e^{2}}{2s^{2}}j_{\pm }D_{WW}j_{\mp }\\
 & + & (\frac{e}{sc})^{2}(j_{3}-s^{2}j_{\gamma })D_{ZZ}(j_{3}-s^{2}j_{\gamma })\\
 & + & \frac{e^{2}}{sc}((j_{3}-s^{2}j_{\gamma })D_{Z\gamma }j_{\gamma }+j_{\gamma }D_{\gamma Z}(j_{3}-s^{2}j_{\gamma }))\\
 & + & e^{2}j_{\gamma }D_{\gamma \gamma }j_{\nu }t
\end{eqnarray*}
To determine the modifications to neutral current interaction, adjustments to
the entire Weak sector must analyzed which then require considering affection
on both \( W \) and \( \gamma  \) propagators as well. Note also the presence
of the rather strange looking \( D_{\gamma Z} \) which appears to propagate
a particle that starts as a \( Z_{0} \) and ends as a photon, and vice versa.
No bare, tree level interaction gives this sort of mixing, \( D^{0}_{\gamma Z}=0 \)
giving,
\begin{eqnarray*}
\mathcal{M} & = & \frac{e^{2}}{2s^{2}}j_{\pm }D^{0}_{WW}j_{\mp }\\
 & + & (\frac{e}{sc})^{2}(j_{3}-s^{2}j_{\gamma })D^{0}_{ZZ}(j_{3}-s^{2}j_{\gamma })\\
 & + & e^{2}j_{\gamma }D^{0}_{\gamma \gamma }j_{\gamma }
\end{eqnarray*}
At tree level, at low energy, the unrenormalized propagator yields \( D^{0}(0)=1/m^{2} \),
giving the usual form of the low energy Electro-Weak effective Lagrangian. With
the \( g=e/s \), and \( cm_{Z}=m_{W} \),
\begin{eqnarray*}
\mathcal{M} & = & \frac{g^{2}}{2m_{W}^{2}}(j_{\pm }^{\mu }j_{\mp \mu }+(j_{3}-\bar{x}j_{\gamma })^{2})+\frac{e^{2}}{q^{2}}j_{\gamma }^{2}
\end{eqnarray*}
As usual, if desired, \( g^{2}/2m_{W}^{2}=4G_{F}/\sqrt{2} \), \( \bar{x}=s^{2} \).

\begin{figure}
{\par\centering \includegraphics{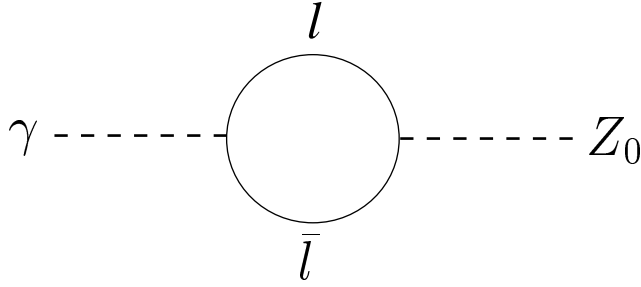} \par}

\caption{\label{Fig:ZGammaLoop}\protect\( Z_{0}-\gamma \protect \) mixing through
loop.}
\end{figure}

With higher order processes it becomes possible to change a \( \gamma  \) into
a \( Z_{0} \) in particular consider any fermion loop with a \( \gamma  \)
and a \( Z_{0} \) attached, fig.\ref{Fig:ZGammaLoop}.
 This gives the first non-zero contribution to \( D_{\gamma Z} \). Rather than
appearing as an additional term this can be included as a modification to the
original form of the matrix element since the Weak neutral current contribution
already contains terms like \( j_{3}j_{\gamma } \). The possible non-trivial
Lorentz structure of the propagators can make this rearrangement a little tricky.
This will be remedied with the detailed development of the structure of the
full propagator. Schematically the result will take the form,

\begin{eqnarray*}
\mathcal{M} & = & \frac{e^{2}}{2s^{2}}j_{\pm }D_{WW}j_{\pm }\\
 & + & (\frac{e}{sc})^{2}(j_{3}-j_{\gamma }(s^{2}-sc(D_{Z\gamma }/D_{ZZ})))D_{ZZ}(j_{3\nu }-(s^{2}-sc(D_{Z\gamma }/D_{ZZ}))j_{\gamma })\\
 & + & e^{2}j_{\gamma }(D_{\gamma \gamma }-2\frac{s}{c}D_{Z\gamma }-D_{\gamma Z}D_{Z\gamma }/D_{ZZ})j_{\gamma }
\end{eqnarray*}
Already the net effect is becoming apparent. With 

\begin{eqnarray*}
D_{ZZ}^{0},D^{0}_{WW} & \rightarrow  & D_{ZZ},D_{WW}\\
s^{2} & \rightarrow  & s^{2}-scD_{Z\gamma }/D_{ZZ}\\
D^{0}_{\gamma \gamma } & \rightarrow  & D_{\gamma \gamma }-2\frac{s}{c}D_{Z\gamma }-D_{\gamma Z}D_{Z\gamma }/D_{ZZ}
\end{eqnarray*}
the effective interaction takes the same form as the original, tree level interaction
with renormalized coupling constants.

\subsection{Full Propagators and Self Energies}

The low energy limit of this result is not immediately apparent. In particular
the structure of \( D_{\gamma Z} \) is not yet known and the explicit form
of the diagonal propagators has not been determined. These require the detailed
forms of the full propagators. 

The oblique corrections are included in the propagator through a self energy
\( \Pi  \). At tree level,
\[
D^{0\mu \nu }=\frac{g^{\mu \nu }-q^{\mu }q^{\nu }/m^{2}}{-q^{2}+m^{2}}\equiv (g^{\mu \nu }-q^{\mu }q^{\nu }/m^{2})D^{0}\]
The momentum dependent term in the numerator doesn't appear in the photon propagator,
and at low energies is negligible in the propagator of a massive particle, but
this Lorentz structure will not appear explicitly again so it is no extra work
to retain it for a more general result. With the addition of the oblique radiative
corrections the full propagator is given by
\[
D^{0}\rightarrow D=\frac{1}{-q^{2}+m^{2}-\Pi (q^{2})}\]
 This form can include Standard Model corrections as well as contributions for
possible new physics. Both these oblique Standard Model corrections and the
previously discussed direct corrections must eventually be included to precisely
interpret any experimental result, but neither will be discussed explicitly
here. The Standard Model oblique corrections can be included implicitly in the
self energies by separately considering contributions from existing Standard
Model particles and new particles as \( \Pi =\Pi _{SM}+\Pi _{NEW} \).

For massive particles, at low energy this modified propagator becomes,
\[
D(0)=\frac{1}{m^{2}-\Pi (0)}\]
 For the photon, with \( m=0 \), the Ward identity requires that \( \Pi _{\gamma \gamma }(0)=0 \)
so that, as seen below, the photon remains massless.This turns out to yield
\( \Pi _{\gamma \gamma }(q^{2})\equiv q^{2}\Pi ^{\prime }_{\gamma \gamma }(q^{2}) \),
where \( \Pi _{\gamma \gamma }^{\prime }(0) \) is finite, this gives a low
energy limit for the photon propagator of, 
\[
D_{\gamma \gamma }\rightarrow \frac{1}{q^{2}}\frac{1}{1-\Pi ^{\prime }_{\gamma \gamma }(0)}\]

Ordinarily, as in the case of the \( W_{\pm } \), the self energies are straight-forward
to calculate and are given as a sum of all 2- point irreducible diagrams with
two external \( W \) legs, call it \( \bar{\Pi }_{WW} \). This is easily seen
from the simple Dyson equation satisfied by the \( W \) propagator,

\[
D_{WW}=D^{0}_{WW}+D^{0}_{WW}\bar{\Pi }_{WW}D_{WW}\]
 giving,
\[
D_{WW}=\frac{D^{0}_{WW}}{1-D^{0}_{WW}\bar{\Pi }_{WW}}=\frac{1}{\left( D^{0}_{WW}\right) ^{-1}-\bar{\Pi }_{WW}}\]
and so \( \Pi _{WW}=\bar{\Pi }_{WW} \).

For the \( Z_{0} \) and the \( \gamma  \) this structure is complicated by
the fact that a \( \bar{\Pi }_{\gamma Z} \) can change a photon to a \( Z_{0} \)
while propagating so the full propagator for a \( Z_{0} \), for example, contains
terms like \( D_{Z}^{0}\bar{\Pi }_{Z\gamma }D_{\gamma }^{0}\bar{\Pi }_{\gamma Z}D_{Z}^{0} \)
as well as the usual \( D_{Z}^{0}\bar{\Pi }_{ZZ}D_{Z}^{0} \). These contributions
are a bit tricky to count correctly. Consider simply modifying the Dyson equations
with the addition of a likely extra term,
\[
D_{ZZ}=D_{ZZ}^{0}+D_{ZZ}^{0}\bar{\Pi }_{ZZ}D_{ZZ}+D_{ZZ}^{0}\bar{\Pi }_{Z\gamma }D^{0}_{\gamma \gamma }\bar{\Pi }_{\gamma Z}D_{ZZ}\]
and similarly for the \( \gamma  \). These give plausible recurrence relations
that can be checked by expanding them a few times. 

To make these kind of calculations a bit tidier omit the explicit appearance
of the \( \Pi  \)'s that glue the propagators together and write the Dyson
equations as,

\begin{eqnarray*}
Z & = & Z^{0}+Z^{0}Z+Z^{0}\gamma ^{0}Z\\
\gamma  & = & \gamma ^{0}+\gamma ^{0}\gamma +\gamma ^{0}Z^{0}\gamma 
\end{eqnarray*}
Expanding the \( Z \) propagator in particular gives,
\begin{eqnarray*}
Z & = & Z^{0}+Z^{0}(Z^{0}+Z^{0}Z+Z^{0}\gamma ^{0}Z)+Z^{0}\gamma ^{0}(Z^{0}+Z^{0}Z+Z^{0}\gamma ^{0}Z)\\
 & = & Z^{0}+Z^{0}Z^{0}+Z^{0}Z^{0}Z+Z^{0}Z^{0}\gamma ^{0}Z+Z^{0}\gamma ^{0}Z^{0}+Z^{0}\gamma ^{0}Z^{0}Z+Z^{0}\gamma ^{0}Z^{0}\gamma ^{0}Z\\
 & = & \cdots 
\end{eqnarray*}
This gives the desired terms involving sequences of an arbitrary number of \( Z^{0} \)
and \( Z^{0}\gamma ^{0} \) terms, but leaves out any terms with repeated \( \gamma ^{0} \)'s
which are surely a valid contribution. This is partly remedied by replacing
the middle, bare propagator in the additional term of each Dyson equation with
a full propagator,
\begin{eqnarray*}
Z & = & Z^{0}+Z^{0}Z+Z^{0}\gamma Z\\
\gamma  & = & \gamma ^{0}+\gamma ^{0}\gamma +\gamma ^{0}Z\gamma 
\end{eqnarray*}
Expanding this as before it quickly becomes apparent that this set of equations
over-counts many contributions. For example, two iterations already gives the
term \( Z^{0}\gamma ^{0}Z^{0}\gamma ^{0}Z^{0} \). 

The trouble in this case is that since the full \( \gamma  \) propagator contains
\( \gamma -Z \) mixing that has already been included in the full \( Z \)
propagator. All that is really needed is a \( \gamma  \) propagator with no
additional mixing to a \( Z \) to generate the appropriate intermediate \( \gamma  \)
propagators. To that end define a modified propagators, \( \bar{D} \), for
the \( Z \) and \( \gamma  \) that don't include mixing. These are easily
defined in terms of the previous \( \bar{\Pi } \). The Dyson equations the
same as for the \( W \),

\[
\bar{D}=D^{0}+D^{0}\bar{\Pi }\bar{D}\]
giving usual,
\[
\bar{D}=\frac{1}{(D^{0})^{-1}-\bar{\Pi }}\]
The Dyson equation for the full \( Z \) propagator, in particular, can then
be written,

\begin{eqnarray*}
D_{ZZ} & = & \bar{D}_{ZZ}+\bar{D}_{ZZ}\bar{\Pi }_{Z\gamma }\bar{D}_{\gamma \gamma }\bar{\Pi }_{\gamma Z}D_{ZZ}
\end{eqnarray*}
A similar expansion of the recurrence gives,

\begin{eqnarray*}
Z & = & \bar{Z}+\bar{Z}\bar{\gamma }Z\\
 & = & \bar{Z}+\bar{Z}\bar{\gamma }\bar{Z}+\bar{Z}\bar{\gamma }\bar{Z}\bar{\gamma }Z\\
 & = & \cdots 
\end{eqnarray*}
as desired. The mixing now appears explicitly and the \( \bar{D} \)'s in each
term can then be expanded to give the correct full series.

Solving for the full propagator,
\begin{eqnarray*}
D_{ZZ} & = & \frac{1}{\bar{D}_{ZZ}^{-1}-\bar{\Pi }_{Z\gamma }\bar{D}_{\gamma \gamma }\bar{\Pi }_{\gamma Z}}\\
 & = & \frac{1}{(D^{0}_{ZZ})^{-1}-\bar{\Pi }_{ZZ}-\bar{\Pi }_{Z\gamma }\bar{D}_{\gamma \gamma }\bar{\Pi }_{\gamma Z}}\\
 & \equiv  & \frac{1}{(D^{0}_{ZZ})^{-1}-\Pi _{ZZ}}
\end{eqnarray*}
This can also be seen from modifying the original candidate for the Dyson equation
using \( \bar{D}_{\gamma \gamma } \) as it was originally motivated,

\begin{eqnarray*}
D_{ZZ} & = & D_{ZZ}^{0}+D_{ZZ}^{0}\bar{\Pi }_{ZZ}D_{ZZ}+D_{ZZ}^{0}\bar{\Pi }_{Z\gamma }\bar{D}_{\gamma \gamma }\bar{\Pi }_{\gamma Z}D_{ZZ}\\
 & \equiv  & D_{ZZ}^{0}+D_{ZZ}^{0}\Pi _{ZZ}D_{ZZ}
\end{eqnarray*}
Either form gives the correct expression for the self energy,
\[
\Pi _{ZZ}=\bar{\Pi }_{ZZ}+\bar{\Pi }_{\gamma Z}\bar{D}_{\gamma \gamma }\bar{\Pi }_{\gamma Z}\]
with the analogous result for the photon. Substituting for \( \bar{D} \) gives
\begin{eqnarray*}
\Pi _{ZZ} & = & \bar{\Pi }_{ZZ}+\frac{\bar{\Pi }_{\gamma Z}^{2}}{(D_{\gamma \gamma }^{0})^{-1}-\bar{\Pi }_{\gamma \gamma }}=\bar{\Pi }_{ZZ}+\frac{\bar{\Pi }_{\gamma Z}^{2}}{q^{2}-\bar{\Pi }_{\gamma \gamma }}\\
\Pi _{\gamma \gamma } & = & \bar{\Pi }_{\gamma \gamma }+\frac{\bar{\Pi }_{\gamma Z}^{2}}{(D_{ZZ}^{0})^{-1}-\bar{\Pi }_{ZZ}}=\bar{\Pi }_{\gamma \gamma }+\frac{\bar{\Pi }_{\gamma Z}^{2}}{q^{2}-m_{Z}^{2}-\bar{\Pi }_{ZZ}}
\end{eqnarray*}
The \( \bar{\Pi } \) are the usual irreducible 2-point Green's Functions. 

With these definitions the structure of the off-diagonal \( \gamma -Z \) propagator
can be determined. An expansion of the full propagator should give,
\begin{eqnarray*}
D_{\gamma Z} & = & \bar{\gamma }\bar{Z}+\bar{\gamma }\bar{Z}\bar{\gamma }\bar{Z}+\bar{\gamma }\bar{Z}\bar{\gamma }\bar{Z}\bar{\gamma }\bar{Z}+\cdots \\
 & = & \bar{\gamma }Z=\gamma \bar{Z}\\
 & = & \bar{D}_{\gamma \gamma }\bar{\Pi }_{\gamma Z}D_{ZZ}=D_{\gamma \gamma }\bar{\Pi }_{\gamma Z}\bar{D}_{ZZ}
\end{eqnarray*}

These result again provides the proper result of a massless photon. The Ward
identity will also give \( \bar{\Pi }_{\gamma Z}(0)=0 \), as well as \( \bar{\Pi }_{\gamma \gamma }(0)=0 \)
as previously discussed, so that \( \Pi _{\gamma \gamma }(0)=0 \). This also
yields \( \Pi _{ZZ}(0)=\bar{\Pi }_{ZZ}(0) \). It will be convenient to be able
to explicitly represent the low energy limit for either case. As before define
\( \bar{\Pi }(q^{2})\equiv q^{2}\bar{\Pi }^{\prime }(q^{2}) \) for \( \bar{\Pi }_{\gamma \gamma } \)
and \( \bar{\Pi }_{\gamma Z} \), where generally \( \bar{\Pi }^{\prime }(0) \)
will be finite. This gives, for \( q^{2}\rightarrow 0 \),
\begin{eqnarray*}
\Pi _{ZZ}=\bar{\Pi }_{ZZ}+q^{2}\frac{\bar{\Pi }_{\gamma Z}^{\prime 2}}{1-\bar{\Pi }^{\prime }_{\gamma \gamma }} & \rightarrow  & \bar{\Pi }_{ZZ}\\
\Pi _{\gamma \gamma }=q^{2}\bar{\Pi }^{\prime }_{\gamma \gamma }+\frac{q^{4}\bar{\Pi }_{\gamma Z}^{\prime 2}}{q^{2}-m_{Z}^{2}-\bar{\Pi }_{ZZ}} & \rightarrow  & q^{2}\bar{\Pi }^{\prime }_{\gamma \gamma }
\end{eqnarray*}
so that
\begin{eqnarray*}
D_{ZZ}(0) & = & \bar{D}_{ZZ}(0)=\frac{1}{m_{Z}^{2}-\bar{\Pi }_{ZZ}}\\
D_{\gamma \gamma } & \rightarrow  & \bar{D}_{\gamma \gamma }=\frac{1}{q^{2}}\frac{1}{1-\bar{\Pi }^{\prime }_{\gamma \gamma }}
\end{eqnarray*}
Notice that for the \( Z \) propagator the \( \gamma -Z \) mixing turns out
to not contribute to the low energy limit.

\subsection{Low Energy Effective Theory}

With these details about the form of \( D_{\gamma Z} \) in particular the structure
of the interaction with these oblique corrections can be determined more carefully,
and more explicitly. Using \( D_{\gamma Z}=\bar{D}_{\gamma \gamma }\bar{\Pi }_{\gamma Z}D_{ZZ} \)
or \( D_{\gamma Z}=D_{\gamma \gamma }\bar{\Pi }_{\gamma Z}\bar{D}_{ZZ} \) as
convenient, the interaction becomes
\begin{eqnarray*}
\mathcal{M} & = & \frac{e^{2}}{2s^{2}}j_{\pm }D_{WW}j_{\mp }\\
 & + & (\frac{e}{sc})^{2}(j_{3}-s^{2}j_{\gamma })D_{ZZ}(j_{3}-s^{2}j_{\gamma })\\
 & + & \frac{e^{2}}{sc}((j_{3}-s^{2}j_{\gamma })\bar{D}_{\gamma \gamma }\bar{\Pi }_{\gamma Z}D_{ZZ}j_{\gamma }+j_{\gamma }\bar{D}_{\gamma \gamma }\bar{\Pi }_{\gamma Z}D_{ZZ}(j_{3}-s^{2}j_{\gamma }))\\
 & + & e^{2}j_{\gamma }D_{\gamma \gamma }j_{\nu }t
\end{eqnarray*}
which can finally be written in the form,

\begin{eqnarray*}
\mathcal{M} & = & \frac{e^{2}}{2s^{2}}j_{\pm }D_{WW}j_{\pm }\\
 & + & (\frac{e}{sc})^{2}(j_{3}-j_{\gamma }(s^{2}-sc\bar{D}_{\gamma \gamma }\bar{\Pi }_{\gamma Z}))D_{ZZ}(j_{3\nu }-(s^{2}-sc\bar{D}_{\gamma \gamma }\bar{\Pi }_{\gamma Z})j_{\gamma })\\
 & + & e^{2}j_{\gamma }D_{\gamma \gamma }(1-\bar{\Pi }_{\gamma Z}\bar{D}_{ZZ}(2\frac{s}{c}-\bar{D}_{\gamma \gamma }\bar{\Pi }_{\gamma Z}))j_{\gamma }
\end{eqnarray*}

The low energy limit can now be taken immediately. \( \bar{D}_{\gamma \gamma }\bar{\Pi }_{\gamma Z}\rightarrow \bar{\Pi }^{\prime }_{\gamma Z}/1-\bar{\Pi }^{\prime }_{\gamma \gamma } \)
is finite, \( \bar{\Pi }_{\gamma Z}\bar{D}_{ZZ}\rightarrow 0 \) and \( D_{\gamma \gamma }\rightarrow \bar{D}_{\gamma \gamma } \).
Including, for the moment, only the explicit for the photon propagators, the
matrix element gives,
\begin{eqnarray*}
\mathcal{M} & = & \frac{e^{2}}{2s^{2}}\bar{D}_{WW}(0)\left( j_{\pm }^{2}+\frac{1}{c^{2}}\frac{\bar{D}_{ZZ}(0)}{\bar{D}_{WW}(0)}\left( j_{3}-j_{\gamma }\left( s^{2}-sc\frac{\bar{\Pi }^{\prime }_{\gamma Z}(0)}{1-\bar{\Pi }^{\prime }_{\gamma \gamma }(0)}\right) \right) ^{2}\right) \\
 & + & \frac{e^{2}}{q^{2}}\frac{1}{1-\Pi _{\gamma \gamma }^{\prime }(0)}j^{2}_{\gamma }
\end{eqnarray*}
This gives the original tree level structure with modified parameters,
\begin{eqnarray*}
\mathcal{M} & = & \frac{4G^{*}_{F}}{\sqrt{2}}\left( j_{\pm }j_{\pm }+\rho ^{*}(j_{3}-s^{*2}j_{\gamma })^{2}\right) +\frac{e^{*2}}{q^{2}}j_{\gamma }^{2}
\end{eqnarray*}
where these renormalized couplings are then given by, 
\begin{eqnarray*}
\frac{4G_{F}^{*}}{\sqrt{2}} & = & \frac{e^{2}}{2s^{2}}\bar{D}_{WW}(0)\\
\rho ^{*} & = & \frac{1}{c^{2}}\frac{\bar{D}_{ZZ}(0)}{\bar{D}_{WW}(0)}\\
s^{*2} & = & s^{2}-sc\frac{\bar{\Pi }^{\prime }_{\gamma Z}(0)}{1-\bar{\Pi }^{\prime }_{\gamma \gamma }(0)}\\
e^{*2} & = & e^{2}\frac{1}{1-\Pi ^{\prime }_{\gamma \gamma }(0)}
\end{eqnarray*}

The tree level forms of the propagators, \( D^{0}_{ZZ}(0)=1/m_{Z}^{2}=1/c^{2}m_{W}^{2} \),
\( D^{0}_{WW}(0)=1/m_{W}^{2} \), \( D_{\gamma Z}=0 \), again gives the original
parameters \( s^{*2}=s^{2} \), \( \rho ^{*}=1 \). The deviation from these
tree level values then contain information about about these oblique corrections.
Weak charged current interactions and electromagnetic processes depend only
on \( G_{F}^{*} \)and \( e^{*2} \) respectively, so they can be determined
independently by experiment, the \( \beta  \) -decay constant and \( \alpha  \)
, and in that sense are fundamental while the bare independent \( G_{F} \)
and \( e^{2} \) are derived quantities. These bare quantities don't appear
here anywhere else so their relation to the corrected values does not need to
be considered in detail. Non-trivial observable effects are still available
through \( \rho ^{*} \) and \( x^{*} \)and with \( G_{F}^{*} \) in particular
as an input, neutral current experiments then can be used to constrain \( \rho ^{*} \)
and \( \bar{x}^{*} \).

This form shows that the low energy effective theory can be completely described
in terms of at most four paramaters. It will turn out that not all of these
are independent and that in fact only three are needed for any energies up to
the Electro-Weak scale. These raw quantities, the 2 point functions at zero
energy are not the most convenient form for these parameters as, among other
things, they are divergent and any sensible use of them requires regularization
and so they become dependent on the renormalization scheme. With a little work
these quantities can be re-written in terms of more well defined, renormalization
and model independent paramaters that can be used to universally describe the
corrections due to contributions from new physics. This can be done in a way
that is valid for a wide range of energy scales, \cite{PeskinSTU},\cite{Schacht00},
allowing an easy mean of comparing the predictions from the results of many
different kinds of experiments, but here the motivation for the definitions
of these paramaters will only be sketched from a low energy perspective.

\subsection{\protect\( \rho \protect \)}

The form of oblique contributions to \( x^{*} \) already appear explicitly.
Adjustments to \( \rho ^{*} \) still appear implicitly in terms of the low
energy limits of the \( Z \) and \( W \) propagators. The previous form for
these limits of the massive propagators could be used here as
\[
\bar{D}\rightarrow \frac{1}{m^{2}-\bar{\Pi }(0)}\]
This gives,

\[
\rho ^{*}=\frac{1}{c^{2}}\frac{m_{W}^{2}-\bar{\Pi }_{WW}(0)}{m_{Z}^{2}-\bar{\Pi }_{ZZ}(0)}=\frac{m_{W}^{2}}{c^{2}m_{Z}^{2}}\frac{1-\bar{\Pi }_{WW}(0)/m_{W}^{2}}{1-\bar{\Pi }_{ZZ}(0)/m_{Z}^{2}}\]
with \( cm_{Z}=m_{W} \) the leading coefficient is just \( 1 \). Now separating
contributions from Standard Model processes and new physics, as discussed previously,
\begin{eqnarray*}
\rho ^{*} & = & \frac{1-\bar{\Pi }_{WW}(0)/m_{W}^{2}}{1-\bar{\Pi }_{ZZ}(0)/m_{Z}^{2}}\\
 & = & \frac{1-(\bar{\Pi }^{SM}_{WW}+\bar{\Pi }^{New}_{WW})/m_{W}^{2}}{1-(\bar{\Pi }_{ZZ}^{SM}-\bar{\Pi }_{ZZ}^{New})/m_{Z}^{2}}\\
 & = & \frac{1-\bar{\Pi }_{WW}^{SM}/m_{W}^{2}}{1-\bar{\Pi }_{ZZ}^{SM}/m_{Z}^{2}}\frac{1-(\bar{\Pi }_{WW}^{New}/m_{W}^{2})/(1-\bar{\Pi }_{WW}^{SM}/m_{W}^{2})}{1-(\bar{\Pi }_{ZZ}^{New}/m_{Z}^{2})/(1-\bar{\Pi }_{ZZ}^{SM}/m_{Z}^{2})}
\end{eqnarray*}
For small \( \Pi /m^{2} \) as these corrections are expected to be, to first
order,
\begin{eqnarray*}
\rho ^{*} & \approx  & \frac{1-\bar{\Pi }_{WW}^{SM}/m_{W}^{2}}{1-\bar{\Pi }_{ZZ}^{SM}/m_{Z}^{2}}\frac{1-\bar{\Pi }_{WW}^{New}/m_{W}^{2}}{1-\bar{\Pi }_{ZZ}^{New}/m_{Z}^{2}}\\
 & \approx  & \frac{1-\bar{\Pi }_{WW}^{SM}/m_{W}^{2}}{1-\bar{\Pi }_{ZZ}^{SM}/m_{Z}^{2}}\left( 1-\frac{\bar{\Pi }_{WW}^{New}}{m_{W}^{2}}+\frac{\bar{\Pi }_{ZZ}^{New}}{m_{Z}^{2}}\right) \\
 & \equiv  & \rho ^{SM}\left( 1-\frac{\bar{\Pi }_{WW}^{New}(0)}{m_{W}^{2}}+\frac{\bar{\Pi }_{ZZ}^{New}(0)}{m_{Z}^{2}}\right) 
\end{eqnarray*}
This gives \( \rho  \) as the value given by Standard Model processes times
a factor that depends on possible new physics given by the difference in the
\( W \) and \( Z \) 2-point functions.

\subsection{Mass and Wavefunction Renormalization}

This particular form of the correction out to be superficially less convenient.
The self energy gives an adjustment to the bare mass that appears in this form
of the propagator which yields the physical, observed mass. Generally it is
convenient to rewrite the propagator in terms of this physical mass since it
is directly accessible by experiment and the bare mass is inherently a derived
quantity. This also requires the introduction of an wavefunction renormalization. 

Rename this bare mass \( m_{0} \) to distinguish it from the physical mass,
\( m \). The physical mass is defined by the pole of the propagator, at \( q^{2}=m^{2} \)
the denominator of the propagator is zero giving \( -m^{2}+m_{0}^{2}-\Pi (m^{2})=0 \).
The propagator then becomes,

\begin{eqnarray*}
\frac{1}{-q^{2}+m_{0}^{2}-\Pi (q^{2})} & = & \frac{1}{-q^{2}+m^{2}+\Pi (m^{2})-\Pi (q^{2})}\\
 & = & \frac{1}{-q^{2}+m^{2}}\frac{1}{1+(\Pi (m^{2})-\Pi (q^{2}))/(-q^{2}+m^{2})}\\
 & \equiv  & \frac{Z(q^{2})}{-q^{2}+m^{2}}
\end{eqnarray*}
with 
\[
Z(q^{2})=\left( 1+\frac{\Pi (m^{2})-\Pi (q^{2})}{-q^{2}+m^{2}}\right) ^{-1}\]

This gives a propagator with the form of a tree level propagator using the real
physical mass with an addition coefficient multiplying the amplitude. This \( Z \)
is the wavefunction renormalization, given something like the resulting net
change in the probability to find a bare particle in a propagating real particle.
In scattering experiments this factor is absorbed in external wavefunctions
and partly compensated by the vertex corrections. For this low energy theory
they will simply be kept explicitly.

Near the mass pole, or for slowly varying \( \Pi  \), \( Z \) can instead
be written more simply in terms of \( \Pi ^{\prime } \), with \( \delta q^{2}=q^{2}-m^{2} \),

\begin{eqnarray*}
\Pi (q^{2}) & = & \Pi (m^{2}+\delta q^{2})\\
 & \approx  & \Pi (m^{2})+\delta q^{2}\Pi ^{\prime }(m^{2})
\end{eqnarray*}
This gives
\[
\frac{\Pi (m^{2})-\Pi (q^{2})}{-q^{2}+m^{2}}\approx \Pi ^{\prime }(m^{2})\]
and 
\[
Z(q^{2})=Z\approx \frac{1}{1+\Pi ^{\prime }(m^{2})}\]
with \( \Pi ^{\prime }(q^{2})=\partial _{q^{2}}\Pi (q^{2}) \). This value for
the wavefunction renormalization is exact at the mass pole, 
\[
Z(m^{2})=\frac{1}{1+\Pi ^{\prime }(m^{2})}\]
for small \( \Pi  \) this becomes simply,
\[
Z(m^{2})\approx 1-\Pi ^{\prime }(m^{2})\]
At low energy, the original, exact form of the renormalization is more appropriate,
\[
Z(0)=\left( 1+\frac{\Pi (m^{2})-\Pi (0)}{m^{2}}\right) ^{-1}\approx 1-\frac{\Pi (m^{2})-\Pi (0)}{m^{2}}\]

\subsection{\protect\( T\protect \)}

With this distinction between bare and physical mass the correction to \( \rho  \)
depends on,

\[
\frac{\bar{\Pi }_{WW}^{New}(0)}{m_{W0}^{2}}-\frac{\bar{\Pi }_{ZZ}^{New}(0)}{m_{Z0}^{2}}=\frac{\bar{\Pi }_{WW}^{New}(0)}{m_{W}^{2}+\Pi _{WW}(m_{W}^{2})}-\frac{\bar{\Pi }_{ZZ}^{New}(0)}{m_{Z}^{2}+\Pi _{ZZ}(m_{Z}^{2})}\]
Since this is already written just to first order in \( \Pi /m^{2} \), the
difference between these real and physical masses can be neglected as it is
a correction of order \( \Pi ^{2} \),
\begin{eqnarray*}
\rho  & \approx  & \rho ^{SM}\left( 1-\frac{\bar{\Pi }_{WW}^{New}(0)}{m_{W}^{2}}+\frac{\bar{\Pi }_{ZZ}^{New}(0)}{m_{Z}^{2}}\right) 
\end{eqnarray*}
The masses that appear are now the real physical masses. This is then conventionally
written in terms of an Electro-Weak correction parameter \( T \) as,
\begin{eqnarray*}
\rho  & \approx  & \rho ^{SM}\left( 1+\alpha T\right) 
\end{eqnarray*}
giving,

\[
-\alpha T=\frac{\bar{\Pi }_{WW}^{New}(0)}{m_{W}^{2}}-\frac{\bar{\Pi }_{ZZ}^{New}(0)}{m_{Z}^{2}}\]
This provides a means of describing the effects of possible new physical processes
on any experiment independent of the particular model of the extension to the
Standard Model. \( T=0 \) for \( \Pi ^{New}=0 \) when there is no new physics.
This particular form for parameterizing the corrections also turns out to have
a great practical benefit because the ultraviolet divergences in a single \( \Pi  \)
will cancel in the difference and the result is independent of the any renormalization
or regularization scheme that would otherwise be needed.

Standard Model corrections are now sufficiently precisely known that \( \rho ^{SM} \)
can be calculated accurately enough that contributions from new physics can
be detected accurately, though there is still a weak dependence on an unknown
Higgs mass. A similar definition of \( T \) could have been made for \( \rho ^{SM} \)as
\[
\rho =\frac{1-\bar{\Pi }_{WW}^{SM}/m_{W}^{2}}{1-\bar{\Pi }_{ZZ}^{SM}/m_{Z}^{2}}\approx \left( 1-\frac{\bar{\Pi }_{WW}^{SM}(0)}{m_{W}^{2}}+\frac{\bar{\Pi }_{ZZ}^{SM}(0)}{m_{Z}^{2}}\right) \equiv 1+\alpha T^{SM}\]
Then giving 
\[
\rho =(1+\alpha T^{SM})(1+\alpha T)\approx 1+\alpha (T^{SM}+T)\]
Measuring \( T \) then requires assuming a Higgs mass to calculate \( T^{SM} \)
to subtract of the Standard Model contribution. An error in that estimate of
the mass of about a factor of 10 changes \( T \) by about \( 0.1 \).

\subsection{Effective Weak Mixing Angle}

The structure of the low energy effective interaction gives an easy result for
\( \rho  \) and a natural definition for a parameter \( T \) to describe the
effects of new physics on \( \rho  \) in a model independent way. The analogous
results for \( s^{2*} \) are not as straight-forward. The effective weak mixing
angle to use a low energy is given by,

\begin{eqnarray*}
s^{*2} & = & s^{2}-sc\frac{\bar{\Pi }^{\prime }_{\gamma Z}(0)}{1-\bar{\Pi }^{\prime }_{\gamma \gamma }(0)}\\
 & = & s^{2}-sc(\bar{\Pi }^{\prime }_{\gamma Z}(0)+\bar{\Pi }^{\prime }_{\gamma \gamma }(0))
\end{eqnarray*}
The derivatives of the 2 point functions are finite so there sum might be a
choice for an additional correction parameter, but that turns out to be cumbersome
to use in other experiments. More importantly, like the bare masses in \( \rho  \),
this is still written in terms of the bare parameter \( s \). \( s \) is not
directly measurable but can be derived from the other coupling constants and
masses. In particular \( G_{F} \) is given in terms of \( s^{2} \) by, 
\[
\frac{4G_{F}^{*}}{\sqrt{2}}=\frac{e^{2}}{2s^{2}}\bar{D}_{WW}(0)=\frac{e^{2}}{2s^{2}}\frac{1}{m_{W0}^{2}+\Pi _{WW}(0)}\]
giving
\begin{eqnarray*}
s^{2} & = & \frac{\sqrt{2}e^{2}}{2\cdot 4G_{F}^{*}}\bar{D}_{WW}(0)\\
 & = & \frac{e^{2}}{4\sqrt{2}G_{F}^{*}}\frac{1}{m_{W0}^{2}-\Pi _{WW}(0)}
\end{eqnarray*}
Here \( e \) is also a bare parameter, but it is similarly given by,

\[
e^{*2}(q^{2})=\frac{e^{2}}{1-\Pi ^{\prime }_{\gamma \gamma }(q^{2})}=4\pi \alpha ^{*}(q^{2})\]
\( \alpha (m_{Z}^{2}) \) is conventionally used for these purposes, \cite{PeskinSTU}.
Also \( m_{Z} \) is used in favor of \( m_{W} \) as it is more accurately
known. The \( s \) can be determined from,

\begin{eqnarray*}
s^{2} & = & \frac{\pi \alpha (m_{Z}^{2})}{\sqrt{2}G_{F}^{*}}\frac{1-\Pi _{\gamma \gamma }^{\prime }(m_{Z}^{2})}{c^{2}m_{Z0}^{2}-\Pi _{WW}(0)}\\
s^{2}c^{2} & = & \frac{\pi \alpha (m_{Z}^{2})}{\sqrt{2}G_{F}^{*}m_{Z}^{2}}\frac{1-\Pi _{\gamma \gamma }^{\prime }(m_{Z}^{2})}{1+\left( \Pi _{ZZ}(m_{Z}^{2})-\Pi _{WW}(0)/c^{2}\right) /m_{Z}^{2}}
\end{eqnarray*}
The leading coefficient is used to define \( s_{0} \). With \( c_{0}^{2}\equiv 1-s_{0}^{2} \),
\begin{eqnarray*}
s^{2}c^{2} & = & s_{0}^{2}c_{0}^{2}\frac{1-\Pi _{\gamma \gamma }^{\prime }(m_{Z}^{2})}{1+\left( \Pi _{ZZ}(m_{Z}^{2})-\Pi _{WW}(0)/c^{2}\right) /m_{Z}^{2}}\\
s_{0}^{2}c_{0}^{2} & = & \frac{\pi \alpha (m_{Z}^{2})}{\sqrt{2}G_{F}^{*}m_{Z}^{2}}\\
sin^{2}(2\theta _{0}) & \equiv  & \frac{4\pi \alpha (m_{Z}^{2})}{\sqrt{2}G_{F}^{*}m_{Z}^{2}}
\end{eqnarray*}
this then provides \( s \) and \( c \) in terms of \( s_{0} \) which can
then be used with \( \bar{\Pi }^{\prime }_{\gamma Z}(0) \) and \( \bar{\Pi }_{\gamma \gamma }(0) \)
to give \( s^{*2} \). This still leads to some gruesome algebra and no natural
definition for any other model and renormalization independent Electro-Weak
correction parameters.

\subsection{\protect\( S\protect \) and \protect\( U\protect \)}

A more natural definition of additional correction paramaters arises from considering
the wavefunction renormalization factors, though this leads to a more complicated
expression for \( s^{*2} \) in terms of these parameters. Near the mass pole
the propagator is given by,
\[
Z(m^{2})=\frac{1}{1+\Pi ^{\prime }(m^{2})}\approx 1-\Pi ^{\prime }(m^{2})\]
The deviation of this propagator from \( 1 \), or the deviation from its Standard
Model value is taken to be \( S \) and can also be used to generally describe
contributions to electro-weak processes from new physics. 
\[
Z(m^{2})\equiv 1-S\]
In this form as the derivative of the self energy it is also finite and so renormalization
scheme independent. 

In practice \( S \) is not defined in terms of these \( Z \) factors, but
in terms of a modified \( Z^{*} \),

For the \( Z \) and \( W \) these are conventionally defined as,

\begin{eqnarray*}
Z^{*}_{Z} & = & 1+\frac{\alpha }{4s^{2}c^{2}}S\\
Z^{*}_{W} & = & 1+\frac{\alpha }{4s^{2}}(S+U)
\end{eqnarray*}

\( U \) doesn't appear at all in neutral current processes at it is related
to the charged \( W \) bosons. Generally it is not consider either as it tends
to be small. With this \( S \) the effective Weak mixing angle can now be determined.
With some algebra,\cite{Schacht00},

\[
s^{*2}=s_{0}^{2}+\frac{\alpha }{c^{2}-s^{2}}(\frac{1}{4}S-s^{2}c^{2}T)\]

\subsection{STU results}

\( S \) , \( T \) and \( U \) provide an easy means of comparing predictions
from the results of any Weak experiment. They are well defined formally in terms
of the various 2-point functions but don't have any immediately obvious physics
interpretation. The explicit result for some specific kind of new particles
provide some insight into their meaning, \cite{PeskinSTU},\cite{Schacht00}.

The contributions to the self energies that determine these parameters will
generally include fermions and scalars. For a scalar, such as a Higgs with mass
\( m_{H} \), in the limit \( m_{H}>>m_{Z} \) , \( S \) , \( T \) and \( U \)
are given by, 
\begin{eqnarray*}
S & \approx  & \frac{1}{12\pi }\ln (\frac{m_{H}^{2}}{m_{0}^{2}})\\
T & \approx  & -\frac{3}{16\pi c^{2}}\ln (\frac{m_{H}^{2}}{m_{0}^{2}})\\
U & \approx  & 0
\end{eqnarray*}
Where \( m_{0} \) is some reference mass scale. 

For a single additional fermion of mass \( m_{f} \),

\begin{eqnarray*}
S & \approx  & -\frac{1}{6\pi }\ln (\frac{m_{f}^{2}}{m_{0}^{2}})\\
T & \approx  & -\frac{3}{16\pi s^{2}c^{2}}\ln (\frac{m_{f}^{2}}{m_{Z}^{2}})
\end{eqnarray*}

A more natural Standard Model extension will include new pair of fermions in
additional \( SU(2) \) doublets. With the particles having masses \( m_{n}\pm \Delta m \),
with \( \Delta m<<m_{n} \),

\begin{eqnarray*}
S & \approx  & \frac{1}{6\pi }\\
T & \approx  & \frac{1}{16\pi s^{2}c^{2}}\frac{\Delta m^{2}}{m_{Z}^{2}}\\
U & \approx  & \frac{2}{15\pi }\frac{\Delta m^{2}}{m_{n}^{2}}
\end{eqnarray*}
In this latter case, \( S \) increases with each new doublet by \( 1/6\pi  \),
but \( T \) changes only if the particle masses are different. This is the
origin of the identification of \( S \) as the isospin conserving parameter,
since its contribution is independent of the masses of each particle in a possible
new doublet, and \( T \) the isospin breaking parameter since with exact isospin
symmetry \( \Delta m=0 \) giving \( T=0 \). In this way \( S \) is sort of
a model independent measure of the existence of some kind of new physics and
\( T \) provides some information about the structure of the addition. Also
\( S \) increases linearly with new particle contributions rather than logarithmically
as for the other cases so that the largest contribution to \( S \) will be
from new isospin doublets. This contribution is positive, for small isospin
symmetry breaking, and so changes to \( S \) from possible new physics are
generally expected to be positive. Also note that \( U \) is very much smaller
than \( S \) or \( T \) as it is suppressed by \( m_{n}^{2} \) rather than
just \( m_{Z}^{2} \).

\subsection{\protect\( Q_{W}\protect \) }

This analysis gives the effects of oblique corrections to any Weak processes.
These processes are accounted for simply by adjusting the tree level coupling
constants. The effects on any observable are then similarly easily determined
by replacing the tree level parameters with the oblique corrected counterparts.
In atomic parity violation, the particle theory details enter through \( G_{F} \)
and \( Q_{W} \). As noted, \( G_{F} \) is conventionally measured with charged
current process and is taken to be externally defined for these atomic processes
and so the only effects to consider explicitly are those contained in \( Q_{W} \)
. At tree level \( Q_{W} \) is given by
\[
Q_{W}\rightarrow Q_{W}^{0}=-(N-(1-4s^{2})Z)\]
The effect of the oblique corrections are to change the overall strength of
the interaction from \( G_{F} \) to \( \rho ^{*}G^{*}_{F} \) and altered effective
mixing angle \( s^{2}\rightarrow s^{*2} \). These changes can then be contained
in \( Q_{W} \),
\[
Q_{W}\rightarrow Q_{W}^{*}=-\rho ^{*}(N-(1-4s^{*2})Z)\]
The \( \rho ^{*} \) and \( \bar{x}^{*} \) can in turn be written in terms
of \( S \) and \( T \) to provide and easy comparison to the results of other
experiments,
\begin{eqnarray*}
\rho ^{*} & = & 1+\alpha T\\
s^{*2} & = & s^{2}+\frac{\alpha }{c^{2}-s^{2}}(\frac{1}{4}S-s^{2}c^{2}T)
\end{eqnarray*}
For the latter \( s=s_{0}=\sqrt{0.2323} \) as defined above, and \( \alpha =\alpha (m_{Z}^{2})=1/129 \).
To first order in \( S \) and \( T \) this gives, 
\[
Q_{W}^{*}=Q_{W}^{0}-(\frac{Z}{c^{2}-s^{2}})\alpha S+(Q^{0}_{W}+\frac{4s^{2}c^{2}Z}{c^{2}-s^{2}})\alpha T\]
For Barium in particular with \( Z=56 \), \( N=82 \),

\begin{eqnarray*}
Q_{W} & = & -78-0.81S-0.03T\\
 & = & -78(1+0.01S+0.0003T)
\end{eqnarray*}

A \( 0.1\% \) measurement of \( Q_{W} \) is effectively insensitive to \( T \)
and provides a constraint on \( S \) to \( \pm 0.1 \).This is comparable to
the sensitivity available with high energy experiments measuring quantities
such as Z and W masses and widths, L/R asymmetries at the \( Z_{0} \) pole,
and deep inelastic neutrino scattering, though atomic experiments are currently
only at precisions of about one percent. Again, the constraints on \( S \)
and \( T \) from all Weak experiments sensitive to these effects are shown
in fig.\ref{fig:S&T}.

\chapter{IonPNC}

\label{Sec:IonPNC}

Information about the Standard Model and new physics, and nuclear structure
is contained in the nuclear Weak charge \( Q_{W} \). Atomic theory gives the
size of the mixing of a given set a states, \( \epsilon  \), which is largest
between \( S \) and \( P \) states. The remaining task at hand is to measure
\( \epsilon  \). In this, all atomic experiments are conceptually the same.
The mixing will yield a nonzero amplitude for a transition previously forbidden
by parity conservation. The presence of this transition amplitude is detected,
and its strength determines the size of \( \epsilon  \). 

Observables generated only by this PNC induced transition amplitude are generally
unmeasurably small so that driving another allowed transition is required. But,
measuring a correction to an existing process from these effects would require
very precise calculuations for interpretation. The ideal arrangement is a differential
measurement, driving both processes but measuring an effect that depends only
on the difference between the parity violating processes and a parity conserving
electromagnetic effect.

\section{Interference and Linearization}

\begin{figure}
{\par\centering \includegraphics{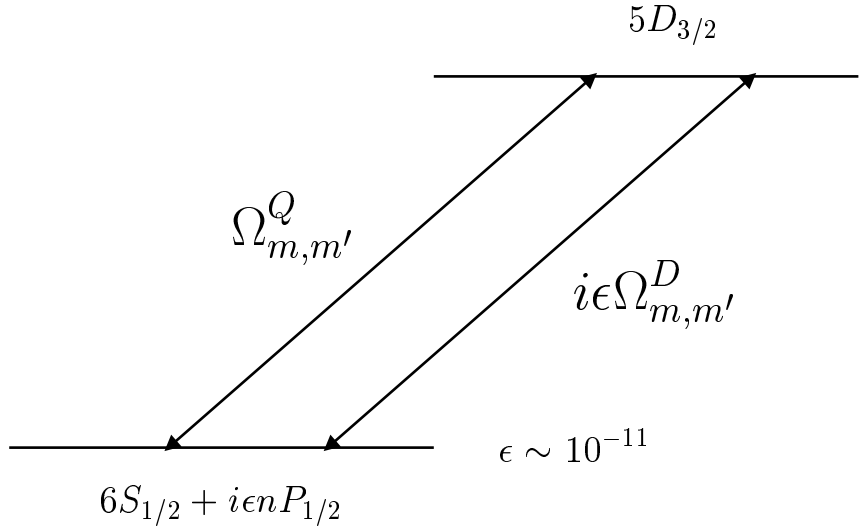} \par}

\caption{A parity violating interaction yields a nonzero amplitude for a transition
previously forbidden by parity.\label{fig:InducedDipole}}
\end{figure}

In all current atomic PNC experiments, the parity induced amplitude is an electric
dipole transition. For pure electromagnetism the atomic states are parity eigenstates
and between two levels with the same parity, an electric dipole transition would
not be allowed as a dipole operator has odd parity and so must change the parity
of a state. With the Weak interaction, atomic states are only mostly parity
eigenstates. With a small amount of an opposite parity state mixed into one
of the original states a dipole transition to the other is now possible, fig.\ref{fig:InducedDipole}.
 Keeping the small size of this induced amplitude, and its relatively imaginary
phase explicit, this coupling will be written as \( \Omega =i\epsilon \Omega ^{D} \).

Detecting this transition directly would be more difficult than it needs to
be. This transition amplitude is proportional to the mixing, but any observable
would involve the square of this amplitude, yielding an effect proportional
to \( \epsilon ^{2} \). \( \epsilon  \) itself is a very small \( 10^{-11} \),
\( \epsilon ^{2} \) would be prohibitively miniscule. Instead a second, parity
allowed transition is also used, \( \Omega ^{EM} \). 

For optical rotation experiments an E-M allowed magnetic dipole transition between
states of the same parity is used. For Stark Interference experiments the nature
of the label finally becomes apparent. A static electric field will also mix
states of opposite parity, a Stark shift, and allow an electric dipole transition
to the second state.

In either case, the total coupling is then given by \( \Omega =\Omega ^{EM}+i\epsilon \Omega ^{D} \).
Expanding to first order in \( \epsilon  \), observables will involve,

\[
\left| \Omega \right| ^{2}=\left| \Omega ^{EM}+i\epsilon \Omega ^{D}\right| ^{2}\approx \left| \Omega ^{EM}\right| ^{2}+2\epsilon Im(\Omega ^{EM*}\Omega ^{D})\]
 Taking the square root, again only to \( o(\epsilon ) \),
\[
\Omega \approx \Omega ^{EM}+\epsilon \Omega ^{EM}Im(\Omega ^{D}/\Omega ^{EM})\]
 Thus the cross term, or interference term, yields an effect linear in \( \epsilon  \).
Notice the effect is independent of the overall magnitude of \( \Omega ^{EM} \)
to \( o(\epsilon ) \), and instead depends only on the magnitude of \( \Omega ^{D} \)
and its phase relative to \( \Omega ^{EM} \). The precise structure and effect
of this cross term depends on the details of the interaction and in general
is slightly complicated by the nontrivial Zeeman structure of the states. But
coarsely, optical rotation experiments look for effects given by \( Im(\Omega ^{D1}/\Omega ^{M1}) \),
and Stark interference measurements, \( Im(\Omega ^{D1}/\Omega ^{Stark}) \).

\section{The Light Shift}

For the Barium ion, the ground state is \( 6S_{1/2} \) and the first excited
state is \( 5D_{3/2} \). Parity selection rules forbid an electric dipole transition
between these two states. Parity violating interactions mix a small amount of
higher energy \( P_{1/2} \) states into the ground state and through this a
transition to the \( D_{3/2} \) state is allowed. To generate the interference
term linear in the mixing, \( \epsilon  \), another coupling between these
two states is needed. In this case an electric quadrupole transition is used. 

The states are now explicitly spin multiplets and the problem must be treated
more generally. For a given set of spin sublevels \( m \) and \( m' \) the
electric dipole and electric quadrupole couplings are given by, 
\begin{eqnarray*}
\Omega _{m'm}^{D} & = & \sum _{i}E_{i}\left\langle 5D_{3/2},m'\left| D_{i}\right| nP_{1/2},m\right\rangle \\
\Omega _{m'm}^{Q} & = & \sum _{i,j}\partial _{i}E_{j}\left\langle 5D_{3/2},m'\left| Q_{ij}\right| 6S_{1/2},m\right\rangle 
\end{eqnarray*}
 The multipole tensor operators are,
\begin{eqnarray*}
D_{i} & = & er_{i}\\
Q_{ij} & = & e(r_{i}r_{j}-\frac{1}{3}\delta _{ij}r^{2})
\end{eqnarray*}

To drive these transitions, the simplest thing to consider is two arbitrary
standing waves. One is placed so that the ion is exactly at the standing wave's
antinode, where the gradient of the electric field is always zero, so it drives
only the dipole transition. The other is placed so that the ion is at its node,
where the amplitude is always zero, this drives only the quadrupole transition.
For a real experiment, this arrangement may be less than optimal due to practical
difficulties or problems with systematic errors, but in exploring the possibilities
it allows the strength of each coupling to be independently controlled and the
effect of each easily identified. The general behavior that emerges is not changed
by any particular implementation.

\subsection{Geometry for a Simplified Case}

\label{Sec:ShiftsForIdealFields}

For completely independent standing waves, with arbitrary directions, polarizations
and phases, this problem is a bit complicated. There are six states with a possible
nonzero coupling between any two spin states from different levels. The choice
of a particular field geometry can be well motivated, but that requires some
formalism that will be developed later. For now, a primitive idea of what can
happen can come from just picking an easy case and blindly working out the consequences.
For a first look consider the special case where neither of the \( S_{1/2} \)
states are connected to the same \( D_{3/2} \) state, in particular suppose
that polarizations and directions are chosen so that only \( \Delta m=\pm 1 \)
transitions are coupled by both the dipole and quadrupole fields.
\begin{figure}
{\par\centering \includegraphics{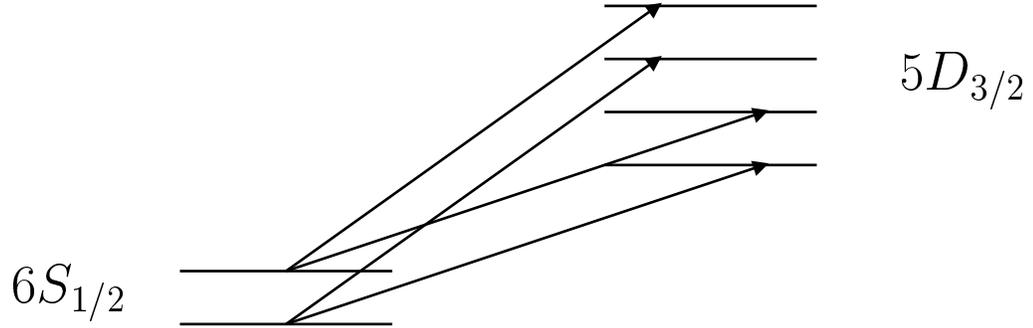} \par}

\caption{\label{Fig:SpinStateCouplingsForIdealFields} Spin state transitions driven
by ideal field geometry.}
\end{figure}
 The effectively factors the problem to two 1+2 state systems which can be solved
easily, rather than the entire 2+4 state system, and includes both interactions
coupling each pair of states so that they can yield a possibly useful interference
term. Initially, and apparently simpler system, a product of two 2-state systems,
would come from simply driving instead the \( \Delta m=0 \) transitions. This
is reasonable for the dipole field with an appropriate choice of coordinates,
but it turns out to put some weird geometric constraints on the quadrupole fields
necessary to drive these transitions. These generalities will be discussed in
complete detail later, for now the simplest configuration to consider is in
fact these two uncoupled \( 1+2 \) state systems resulting from driving only
\( \Delta m=\pm 1 \) transitions.

The fields that generate this can be determined from the explicit expressions
for the spherical form of the tensor operators, but again to build some intuitive
understanding of the problem the general behavior of these operators can be
considered. For \( \hat{z} \) along the angular momentum quantization axis
the operator \( z \) will not change the direction of the angular momentum,
\[
\left\langle jm'\left| z\right| jm\right\rangle \sim \delta _{m'm}\]
 Meanwhile \( x \) and \( y \), which can be written in terms of \( r_{\pm } \)will
change \( m \) by 1,

\( \left\langle jm'\left| x,y\right| jm\right\rangle \sim a\delta _{m',m+1}+b\delta _{m',m-1} \)
Then to get a dipole transition which only drive \( \Delta m=\pm 1 \) transitions
requires,\( D_{z}=0D_{\bot }\neq 0 \)\( D_{\perp } \) is some combination
of \( D_{x} \) and \( D_{y} \). To be explicit, let the dipole field define
the coordinate system and choose the electric field to be in the \( \hat{x} \)
direction and the propagation to be along \( \hat{z} \). The dipole standing
wave will be taken to have the form \( \vec{E}_{D}=\hat{x}E_{D}\cos (kz)\cos (\omega t) \).

Similarly for the quadrupole term, the behavior of the matrix element can be
understood from the way the coordinate operators successively change the angular
momentum, 
\begin{eqnarray*}
\left\langle jm'\left| zz\right| jm\right\rangle  & \sim  & \delta _{m'm}\\
\left\langle jm'\left| r_{\perp }z\right| jm\right\rangle  & \textrm{ }\sim  & a\delta _{m',m+1}+b\delta _{m',m-1}\\
\left\langle jm'\left| r_{\perp }r_{\perp }\right| jm\right\rangle  & \sim  & a\delta _{m',m+2}+b\delta _{m',m-2}+c\delta _{m',m}
\end{eqnarray*}
It is useful to note the particular case of \( xy \). In terms of the spherical
coordinate operators \( xy\sim (r_{+}+r_{-})(r_{+}-r_{-})=r_{+}^{2}+r_{-}^{2} \)
so \( xy \) will only change \( m \) by 2 and not also 0.

Then the quadrupole fields that give \( m=\pm 1 \) are, \( \partial _{z}E_{\perp }\neq 0 \),
or \( \partial _{\perp }E_{z}\neq 0 \). For simple plane waves, the former
corresponds to \( \vec{k}||\hat{z} \), in this case parallel to the dipole
light propagation direction, the latter propagates perpendicularly to the dipole
light but with the polarization fixed parallel. The easiest case to visualize,
though not necessarily to implement, is parallel propagation vectors, the second
case will be considered later and gives the same results. Here depart slightly
from the strategy of just picking some case, looking up the Clebsch-Gordan coefficients
and seeing what happens. There is not yet any reason to fix the polarization
direction, in the \( x-y \) plane, of the quadrupole field, but it will turn
out to be important so allow it to be arbitrary. Also consider it to be fully
linearly polarized so that the polarization direction can be taken to be purely
real, but allow for an overall arbitrary phase in time relative to the dipole
field.
\[
\vec{E}_{Q}=\hat{e}_{Q}\left( \alpha \right) E_{Q}sin(kz)\cos (\omega t+\phi )\]
Where \( \alpha  \) is the angle of the polarization relative to the \( \hat{x} \)
axis so that \( \hat{\epsilon }_{Q}\left( \alpha \right) =cos\left( \alpha \right) \hat{x}+sin\left( \alpha \right) \hat{y} \).
This slightly generalized form will start to give some insight and intuitive
understanding into what happens and why.

After making the rotating wave approximation, \ref{Sec:RWA}, the harmonic time
dependence of the fields can be included instead in the states. Then with these
fields, at the position of the ion \( \vec{r}=0 \), the matrix elements reduce
to,
\begin{eqnarray*}
\Omega _{m'm}^{D} & = & E_{z}\left\langle 5D_{3/2},m'\left| x\right| nP_{1/2},m\right\rangle \\
 & = & E_{D}\left\langle 5D_{3/2},m'\left| x\right| nP_{1/2},m\right\rangle \\
\Omega _{m'm}^{Q} & = & \partial _{\perp }E_{z}\left\langle 5D_{3/2},m'\left| r_{\perp }z\right| 6S_{1/2},m\right\rangle \\
 & = & k\hat{e}_{i}E_{Q}e^{i\phi }\left\langle 5D_{3/2},m'\left| r_{\perp }z\right| 6S_{1/2},m\right\rangle 
\end{eqnarray*}
\( E_{Q} \) and \( E_{D} \) are the amplitudes of the co-rotating terms of
the applied field, the counter-rotating terms are discarded by the rotating
wave approximation. Generally these are half the magnitude of the real field
since the power is equally distributed between the co- and counter-rotating
terms. These factors of two will not be explicitly carried around, in favor
of defining these amplitudes as just described.

Now some more details of the interaction are handy. Writing the operators and
fields in spherical form the Wigner-Eckhart theorem can be used to isolate the
\( m \) dependence into Clebsch-Gordan coefficients. With the summation over
polarizations \( s \) implied, 
\begin{eqnarray*}
\Omega _{m'm}^{D} & = & E_{s}^{(1)}\left\langle 5D_{3/2},m'\left| T_{s}^{(1)}\right| nP_{1/2},m\right\rangle \\
 & = & \frac{\left\langle 5D_{3/2}||D||nP_{1/2}\right\rangle }{\sqrt{2(3/2)+1}}E_{s}^{(1)}\left\langle \frac{3}{2},m'|1,s;\frac{1}{2},m\right\rangle \\
\Omega _{m'm}^{Q} & = & E^{(2)}_{s}\left\langle 5D_{3/2},m'\left| T_{s}^{(2)}\right| 6S_{1/2},m\right\rangle \\
 & = & \frac{\left\langle 5D_{3/2}||Q||6S_{1/2}\right\rangle }{\sqrt{2(3/2)+1}}E_{s}^{(2)}\left\langle \frac{3}{2},m'|2,s;\frac{1}{2},m\right\rangle 
\end{eqnarray*}
 The angular momentum factors \( \sqrt{2(3/2)+1}=2 \) are included for convenience
so that the reduced matrix elements are symmetric under the exchange of initial
and final states. For general fields the amplitudes for a particular dipole
transitions are given by,
\begin{eqnarray*}
E^{D}_{\pm } & = & \frac{\mp E_{x}+iE_{y}}{\sqrt{2}}\\
E_{0}^{D} & = & E_{z}
\end{eqnarray*}
and for quadrupole transitions,
\begin{eqnarray*}
E^{Q}_{\pm 2} & = & (E_{xx}+E_{yy})/2\mp i(E_{xy}+E_{yx})/2\\
E^{Q}_{\pm 1} & = & \mp (E_{xz}+E_{zx})/2+i(E_{yz}+E_{zy})/2\\
E^{Q}_{0} & = & E_{zz}/\sqrt{6}
\end{eqnarray*}
The coefficients \( E_{\pm 1}^{(1,2)} \) are the appropriate spherical tensors
for this case. With these fields, only the \( s=\pm 1 \) terms are non zero,
\begin{eqnarray*}
E^{(1)}_{\pm 1}\equiv E^{D}_{\pm 1} & = & \mp E_{x}+iE_{y}=\mp E_{D}\\
E_{\pm 1}^{(2)}\equiv E_{\pm 1}^{Q} & = & \partial _{z}\left( \mp E_{x}+iE_{y}\right) /2\\
 & = & kE_{Q}\left( \mp \hat{\epsilon }\cdot \hat{x}+i\hat{\epsilon }\cdot \hat{y}\right) e^{i\phi }\\
 & = & \mp kE_{Q}\left( cos\left( \alpha \right) \mp isin\left( \alpha \right) \right) e^{i\phi }\\
 & = & \mp kE_{Q}e^{\mp i\alpha }e^{i\phi }
\end{eqnarray*}

For the alternate choice of geometry having \( \hat{e}^{Q}||\vec{k}^{D} \),
the quadrupole field amplitude is given by,\( E^{\left( 2\right) }_{\pm 1}=\left( \mp \partial _{x}+i\partial _{y}\right) E_{z}/2 \).
At with the polarization for the previous case allow the propagation direction
the the \( x-y \) plane to be arbitrary. This gives the field,

\begin{eqnarray*}
\vec{E}_{Q} & = & \hat{z}E_{Q}sin(\vec{k}\cdot \vec{r})\cos (\omega t+\phi )\\
 & = & \hat{z}E_{Q}sin(cos\left( \alpha \right) x+sin\left( \alpha \right) y)\cos (\omega t+\phi )
\end{eqnarray*}
and the amplitude is again
\begin{eqnarray*}
E_{\pm 1}^{Q} & = & \left( \mp \partial _{x}+i\partial _{y}\right) E_{z}/2\\
 & = & E_{Q}\left( \mp cos\left( \alpha \right) +isin\left( \alpha \right) \right) e^{i\phi }\\
 & = & \mp kE_{Q}\left( cos\left( \alpha \right) \mp isin\left( \alpha \right) \right) e^{i\phi }\\
 & = & \mp kE_{Q}e^{\mp i\alpha }e^{i\phi }
\end{eqnarray*}
identical to the result for the \( \vec{k}_{Q}||\vec{k}_{D} \) geometry.

\subsection{Light Shifts}

With these two transitions now suitably driven, consider the observable consequences,
in particular the resulting energy shifts in the ground state. With these choices
the two ground states are not coupled to each other and the problem factors
into two 1+2 state problems. Exactly on resonance, or for \( \Omega \gg \Delta \omega  \),
the energy shifts for these states is given by, \ref{Sec:1+nState},

\begin{eqnarray*}
\delta \omega _{1/2} & = & \sqrt{\Omega _{3/2,1/2}^{2}+\Omega _{-1/2,1/2}^{2}}\\
\delta \omega _{-1/2} & = & \sqrt{\Omega _{1/2,-1/2}^{2}+\Omega _{-3/2,-1/2}^{2}}
\end{eqnarray*}
simply a incoherent sum of the effects of each individual coupling to the state.
Since the matrix elements coupling the \( \Delta m=0,\pm 2 \) are zero anyways,
this can be written more compactly as,
\[
\delta \omega _{m}=\sqrt{\sum _{m'}\Omega _{mm'}\Omega _{m'm}}=\sqrt{\sum _{m'}\Omega _{m'm}^{*}\Omega _{m'm}}=\sqrt{\left| \sum _{m'}\Omega _{m'm}\right| ^{2}}\]
Or with \( \Omega ^{Q},\Omega ^{D} \)understood to be matrices indexed by the
spin states,

\[
\delta \omega _{m}=\sqrt{\left( \Omega ^{\dagger }\Omega \right) _{mm}}\]
 With \( \Omega =\Omega ^{Q}+i\epsilon \Omega ^{D} \), to linear order in \( \epsilon  \)
\begin{eqnarray*}
\delta \omega _{m} & = & \delta \omega _{m}^{Q}+\delta \omega _{m}^{PNC}\\
\delta \omega _{m}^{Q} & = & \sqrt{\left( \Omega ^{Q\dagger }\Omega ^{Q}\right) _{mm}}\\
\delta \omega _{m}^{PNC} & = & \epsilon Im\left( \left( \Omega ^{Q\dagger }\Omega ^{D}\right) _{mm}\right) /\delta \omega _{m}^{Q}
\end{eqnarray*}
 Recall that in spite of the general appearance of this result, it still requires
that no interaction couples both ground states to the same excited state, though
the general solution to this problem derived later will have almost the same
appearance.

\subsection{Spin Dependence}

Before getting the exact final form, consider the general \( m \) dependence
of the shifts, in particular the \( m\rightarrow -m \) symmetry. In the sums
over matrix elements in each expression all intermediate states \( m' \) are
summed over so the summation variable can be changed to \( -m' \). The matrix
form of the shifts contains this transformation compactly,
\begin{eqnarray*}
\delta \omega _{-m}^{Q} & = & \sqrt{\left| \sum _{m'}\Omega _{m',-m}^{Q}\right| ^{2}}=\sqrt{\left| \sum _{m'}\Omega _{-m',-m}^{Q}\right| ^{2}}\\
 & = & \sqrt{\left( \Omega ^{Q\dagger }\Omega ^{Q}\right) _{-m,-m}}\\
\delta \omega _{-m}^{PNC} & = & \epsilon Im\left( \sum _{m'}\Omega _{m',-m}^{Q*}\Omega _{m',-m}^{D}\right) /\delta \omega _{-m}^{Q}\\
 & = & \epsilon Im\left( \sum _{m'}\Omega _{-m',-m}^{Q*}\Omega _{-m',-m}^{D}\right) /\delta \omega _{-m}^{Q}\\
 & = & \epsilon Im\left( \left( \Omega ^{Q\dagger }\Omega ^{D}\right) _{-m,-m}\right) /\delta \omega _{-m}^{Q}
\end{eqnarray*}

In turn, the \( m \) dependence of the matrix elements is easy to understand
using these particular fields and a simple symmetry property of the Clebsch-Gordan
coefficients,
\[
\left\langle j,-m|k,-s;j',-m'\right\rangle =(-1)^{j-j'+k}\left\langle j,m|k,s;j',m'\right\rangle \]
 The \( k \) dependence is the interesting part here, the others will be fixed
for a comparison, so just write this as
\[
\left\langle j,-m|k,-s;j',-m'\right\rangle =\eta (-1)^{k}\left\langle j,m|k,s;j',m'\right\rangle \]
Here \( \eta =\pm 1 \), so later \( \eta ^{2}=1 \). This can be done safely
since \( k \) is always in integer in these applications and \( j,j' \) are
both either integer or half integer so that \( j-j' \) is always an integer
and \( \left( -1\right) ^{j-j'} \) is well defined.

For either matrix element the sum over polarizations \( s \) can be done in
reverse order, as can the sum over \( m \) in the original expression for the
energy shift, since it is just an index. Attending to only the \( m \) dependence,

\begin{eqnarray*}
\Omega _{-m',-m}^{(k)} & = & \left\langle j,-m\right| T^{(k)}_{s}\left| j',-m'\right\rangle E^{(k)}_{s}\\
 & \propto  & \left\langle j,-m|k,s;j',-m'\right\rangle E_{s}^{(k)}\\
 & = & \left\langle j,-m|k,-s;j',-m'\right\rangle E_{-s}^{(k)}\\
 & = & \eta \left( -1\right) ^{k}\left\langle j,m|k,s;j',m'\right\rangle E_{-s}^{(k)}\\
 & \propto  & \eta \left( -1\right) ^{k}\left\langle j,m\right| T^{(k)}_{s}\left| j',m'\right\rangle E^{(k)}_{-s}
\end{eqnarray*}
 for this dipole field the amplitudes \( E^{D}_{\pm s} \) are related by a
simple sign change. 

\[
E^{D}_{-1}=-E^{D}_{1}\]
For the still slightly more general quadrupole fields the amplitudes can be
written as,

\[
E_{-1}^{Q}=kE_{Q}e^{i\alpha }e^{i\phi }=-e^{2i\alpha }\left( -kE_{Q}e^{-i\alpha }e^{i\phi }\right) =-e^{2i\alpha }E^{Q}_{1}\]
 This may be a bit cumbersome for this introduction, but the generality will
be useful for the quadrupole term. The simple cases of \( \hat{\epsilon }_{Q}||\hat{x},\hat{y} \),
give \( \alpha =0,\pi /2 \) and \( e^{-2i\alpha }=\pm 1 \), so that for the
\( \hat{e}_{Q}||\hat{y} \) polarization the \( s=\pm 1 \) quadrupole amplitudes
are for same, and for \( \hat{e}_{Q}||\hat{x} \) , the amplitudes switch sign.
The transformation of the couplings is then
\begin{eqnarray*}
\Omega _{-m',-m}^{D} & = & \eta \left( -1\right) ^{1}\left\langle j,m\right| D_{s}\left| j',m'\right\rangle E_{-s}^{D}\\
 & = & \eta (-1)^{1}\left\langle j,m\right| D_{s}\left| j',m'\right\rangle \left( -1\right) E_{s}^{D}\\
 & = & \eta \Omega _{m',m}^{D}\\
\Omega _{-m',-m}^{Q} & = & \eta \left( -1\right) ^{2}\left\langle j,m\right| Q_{s}\left| j',m'\right\rangle E_{-s}^{Q}\\
 & = & \eta \left( -1\right) ^{2}\left\langle j,m\right| Q_{s}\left| j',m'\right\rangle \left( -1\right) e^{2i\alpha }E_{s}^{Q}\\
 & = & -\eta e^{2i\alpha }\Omega _{m',m}^{Q}
\end{eqnarray*}

Returning to the energy shifts, the quadrupole shift can be evaluated immediately,
keeping the sum over \( m' \) implicit,

\begin{eqnarray*}
\left( \delta \omega ^{Q}_{-m}\right) ^{2} & = & \Omega _{-m',-m}^{Q}\Omega _{-m',-m}^{Q*}\\
 & = & \left( -\eta e^{2i\alpha }\Omega _{m',m}^{Q}\right) \left( -\eta e^{-2i\alpha }\Omega _{m',m}^{Q*}\right) \\
 & = & \eta ^{2}\Omega _{m',m}^{Q}\Omega _{m',m}^{Q*}\\
 & = & \delta \omega ^{Q}_{m}
\end{eqnarray*}
 For this case, independent of the angle of polarization of the quadrupole field,
the shift is the same for both spin states. Call this shift simply \( \delta \omega ^{Q} \)
since is independent of \( m \). 

For the cross term, finally explicitly consider the more particular quadrupole
fields cases of \( \overrightarrow{E}_{Q} \) along \( \hat{x} \) or \( \hat{y} \),
the quadrupole amplitudes would become,

\begin{eqnarray*}
\Omega _{-m',-m}^{Q_{x,y}} & = & \mp _{x,y}\eta \Omega _{m',m}^{Q_{x,y}}
\end{eqnarray*}
and the change in the shift is given by,

\begin{eqnarray*}
\delta \omega _{-m}^{PNC} & = & \left( \epsilon /\delta \omega ^{Q}\right) Im\left( \Omega _{-m',-m}^{Q_{x,y}*}\Omega _{-m',-m}^{D}\right) \\
 & = & \left( \epsilon /\delta \omega ^{Q}\right) Im\left( \left( \mp _{x,y}\eta \Omega _{m',m}^{Q_{x,y}*}\right) \left( \eta \Omega _{m',m}^{D}\right) \right) \\
 & = & \mp _{x,y}\eta ^{2}\left( \epsilon /\delta \omega ^{Q}\right) Im\left( \Omega _{m',m}^{Q_{x,y}*}\Omega _{m',m}^{D}\right) \\
 & = & \mp _{x,y}\delta \omega _{m}^{PNC}
\end{eqnarray*}
The \( \mp _{x,y} \) denotes a \( - \) for the case of \( \hat{\epsilon }_{Q}||\hat{x} \)
and \( + \) for \( \hat{\epsilon }_{Q}||\hat{y} \). 

For \( \hat{\epsilon }_{Q}||\hat{y} \) this gives \( \delta \omega ^{PNC}_{-m}=\delta \omega _{m}^{PNC} \)
which is a spin independent shift just like that due to the pure quadrupole
term but significantly smaller by the factor \( \epsilon  \) times a ratio
of reduced matrix elements \( \left\langle D\right\rangle /\left\langle Q\right\rangle \sim 10^{3} \).
Somehow measuring this shift to determine \( \epsilon  \) to a part in \( 10^{3} \)
would then require measuring the quadrupole shift to a part in \( 10^{3}/10^{3}\epsilon \sim 10^{11} \).
This is an example of the enormous background from parity allowed interactions
that it was hoped would be avoided by considering this cross term. The interference
generated a shift linear in \( \epsilon  \), but the structure of this parity
induced shift is identical to the quadrupole shift. 

Far more interesting is to consider \( \hat{\epsilon }_{Q}||\hat{x} \). In
this case \( \delta \omega ^{PNC}_{-m}=-\delta \omega _{m}^{PNC} \) so that
the shifts of the two ground state spin levels are in opposite directions and
the total shift will be, 
\[
\delta \omega _{m}=\delta \omega ^{Q}+\delta \omega ^{PNC}_{m}=\delta \omega ^{Q}\pm \delta \omega ^{PNC}\]

\subsection{Parity Violation}

This finally illustrates the fundamental strategy for making an Atomic Parity
Violation measurement using a trapped ion. The spin dependent \( \delta \omega ^{PNC}_{m} \)
shift is now of a fundamentally different character than the quadrupole shift
which is spin independent and moves these levels together. Systematic errors
due to misalignments, polarization impurities and other imperfections can complicate
this simple picture and will be discussed extensively in sec:\ref{Sec:MisalignmentSystemmatics},
but at this level the splitting is a differential signal for parity violation,
an effect that goes to zero as \( \epsilon  \) goes to zero since the quadrupole
term, which is parity allowed, doesn't contribute, so the splitting can then
be unambiguously interpreted as due to parity violation.

It is clear that this shift is zero if parity is conserved since the dipole
amplitude that generates the shift depend on the parity violating mixing of
atomic states. It is not immediately obvious how this result, a spin dependent
shift of magnetic sublevels violates parity. This is more easily seen by interpreting
the splitting to be instead a precession of the spin, the pictures are equivilant,
just as a spin in an external magnetic field can be analyzed statically, by
determining the new eigenstates of the hamiltonian with the addition of the
applied field and their energies, or dynamically in the basis of the original
hamiltonian as a precession of the spin about the applied magnetic field. With
this it is clear that a precession of the ion's spin can be used to define handedness,
such as that the precession is clockwise or counterclockwise when view from
the source of the applied lasers. The direction of the precession with change
sign if the experiment is viewed in a mirror placed parallel to the applied
lasers. 

Many parity conserving, electro-magnetic processes can generate a spin-dependent
shift, but in these cases the fields that give such a shift define a chiral
coordinate system. Circularly polarized electric fields split magnetic sublevels
and the handedness of the circular polarization can be used to define the handedness
of the coordinate systems. Similarly a static magnetic field defines a chiral
coordinate system since magnetic field are generated by current loops or magnetic
moments in the direction of spins. In these cases a mirror image of the experiment
would look like a different physical system, magnetic field in the opposite
direction, circularly polarized light of the opposite handedness. Since the
direction of the precession also changed sign in the mirror image, nothing about
the experiment can be used to define left or right.

The \( \delta \omega ^{PNC} \) splitting is generated by parity symmetric fields.
Both applied lasers propagate in the same direction and polarized the same way.
There is no ambiguity about the handedness of the environment, as the applied
fields can not be used to define a handedness. A mirror image of the lasers
gives the same physical conditions, but the result is still a parity dependent
spin precession. Such an experiment can then be used to unambiguously define
left. Left and right are somehow intrinsically define in the structure of the
Standard Model.

\section{Spin Flip Symmetries}

The potential for a spin dependent shift due to the cross term comes from the
difference in the spin flip symmetry of the matrix elements, one of the quadrupole
or dipole matrix elements changes sign, and the other doesn't. The quadrupole
shift involves the quadrupole coupling twice so the possible sign changes cancel
and the shift is spin independent. The PNC term involves a dipole term and a
quadrupole term. If fields are chosen so that both amplitudes driving these
transitions either do or don't also change sign, the entire coupling changes
sign and the results is a splitting of spin state energies which can be easily
distinguished from the quadrupole shift by its spin dependence.

The result is straightforward and easy to understand, but the analysis was a
bit abstract. It involved the plausible, put possibly mysterious spin flip symmetry
properties of the Clebsch-Gordan coefficients, and the behavior, under the same,
of the spherical tensor amplitudes, which are themselves less than intuitive
compared to simple cartesian vectors. The former can be understood easily by
considering instead the spin flip symmetry of the states, (Sec.Clebsch-Gordan
Relations), and the latter avoided by returning to a cartesian basis. 

The preference for a cartesian basis for this problem is also a practical advantage
rather than just a philosophical bias. The couplings are given by the sum of
the products of matrix elements and field amplitudes over all polarization and
propagation components. These fields have been chosen to be linearly polarized
so that with a proper chose of coordinates, this sum can be given by a single
term in a cartesian basis so that the relationship between couplings is given
easily by the relationship between the possible terms in the sum. At the same
time, linear polarization requires that at least two spherical tensor amplitudes
are non zero, for example \( E_{x}\neq 0 \), \( E_{y},E_{z}=0 \) gives \( E^{\left( 1\right) }_{\pm 1}\neq 0 \),
\( E^{\left( 1\right) }_{0}=0 \), and the coupling requires a sum over many
terms which may change differently when spins are flipping making the relationships
between entire couplings less transparent. If the fields were chosen so that
only one particular spherical tensor amplitude was nonzero, the sum giving the
coupling in a spherical basis would require only one term, while the same result
using a cartesian basis would require many terms, again for example \( E^{\left( 1\right) }_{+1}\neq 0 \),
\( E^{\left( 1\right) }_{-1},E^{\left( 0\right) }_{0}=0 \) requires \( E_{x},E_{y}\neq 0 \),
\( E_{z}=0 \). In addition the states are conventionally indexed in terms of
a spherical basis and it will be convenient later to consider mechanisms for
restricting the interactions to specific subsets of state, independent of polarization,
so that a spherical basis again becomes the more natural choice. So it is valuable
to construct tools for use in either basis.

Finally, the entire discussion of these spin flip symmetries can be cast in
terms of parity transformations and so return the problem full circle to its
origin. In addition, this spherical tensor analysis is somewhat cumbersome and
must be repeated endlessly when considering systematic errors. When some simple
properties of the matrix elements and operators under rotations or reflections
are understood, the general structure and origin of the PNC signal and perturbative
systematic problems are evident by inspection.

\subsection{Transformed States}

The overall coupling for either transition is given by,

\[
\Omega ^{\left( k\right) }_{mm'}=\left\langle j,m\left| T^{(k)}_{s}\right| j',m'\right\rangle E^{\left( k\right) }_{s}\]
The mechanical method for computing this was just presented, the Wigner-Eckhart
Theorem is invoked to write the \( m \) dependence in terms of Clebsch-Gordan
coefficients and the spin flip symmetry then comes from those coefficients,
\[
\Omega _{-m',-m}^{(k)}=\eta \left( -1\right) ^{k}\left\langle j,m\right| T^{(k)}_{s}\left| j',m'\right\rangle E^{(k)}_{-s}\]
Now instead, consider transforming the states in the defining expression,
\[
\Omega ^{\left( k\right) }_{-m,-m'}=\left\langle j,-m\left| T^{(k)}_{s}\right| j',-m'\right\rangle E^{\left( k\right) }_{s}\]
This kind of transformation was considered when deriving the phase of the mixing
matrix element, sec.\ref{Sec:ImaginaryMixing}. In that case time reversal was
used to flip the spin. But time reversal is weird and parity is the general
topic, so try to construct some other transformation \( F \) to flip the spin,
\[
F\left| j,m\right\rangle \propto \left| j,-m\right\rangle \]
Also just consider unitary representations of the spin flip operator, \( F^{\dagger }=F^{-1} \),
so that amplitude of the states are preserved, 
\[
\left\langle j,m\right| F^{\dagger }F\left| j,m\right\rangle =1\]
In this case the constant of proportionality must be just a phase and will depend
on the particular transformation used to represent the spin flip. 

\[
F\left| j,m\right\rangle =\left| j,-m\right\rangle e^{if\left( j,m\right) }\]
It will turn out that, for this particular problem, only this property of unitarity
will be required in considering the action of a spin flip operation on a state.
The action of the spin flip on an operator, however, will always be important,
and when later considering off diagonal elements of these matrix products of
the coupling the detailed transformation of the states must be known, so particular
representations of a spin flip must be considered. Also in briefly considering
the particular effects of these representations on states, some insight is gained
into their workings and, as a by-product, provides an easy means of deriving
the spin flip symmetry of the Clebsch-Gordan coefficients that was used above.

\subsection{Geometrical Representations}

As already noted, parity itself doesn't change spin orientations as a spin is
an axial vector, but the geometric explanation for this reveals the pieces that
can be used to flip spin. Recall the parity transformation can be understood
as a mirror reflection, \( M_{\hat{n}} \), followed by a rotation by \( \pi  \),
\( R_{\hat{n}}\left( \pi \right)  \). The axis of rotation is the vector that
was first reflected so that \( P=R_{\hat{n}}\left( \pi \right) M_{\hat{n}} \)When
the direction is a unit vector it is easily seen that the reflection changes
the sign of one of the coordinates and the rotation changes the signs of the
other two. 

With this axis in the \( \hat{z} \) directions, neither operation changes the
spin, looking down at the \( x-y \) plane a counter-clockwise rotation about
the \( \hat{z} \) axis remains counter-clockwise after a reflection of the
\( \hat{z} \) axis, so spin up remains spin up, the rotation about the \( \hat{z} \)
axis similarly doesn't change the spin orientation. With the transformation
axis perpendicular to \( \hat{z} \), anywhere in the \( x-y \) plane, the
mirror reversal changes counter-clockwise to clockwise, spin up to spin down,
and the rotation reorients to again point up. 

Both cases illustrate how the total parity transformation preserves spin orientation
and the latter suggests two operations that can be used to flip spin giving
the possible representations
\[
F=R_{\hat{n}}\left( \pi \right) \]
or
\[
F=M_{\hat{n}}\]
where \( \hat{n} \) is contained in the \( x-y \) plane. The mirror reflection
is more consistent with the spirit of parity, but the rotation is more straightforward
to implement so the action of both on angular momentum states and coordinate
operators will be considered. Either representation could be picked, along with
a particular choice of orientation axis, and the initial analysis simplified,
but for various particular calculations the variety will be convenient and by
keeping thing general the representation independence of the results will be
apparent explicitly.

\subsubsection{Spatial Wavefuntions}

\label{Sec:SpatialWavefunctionRotation}

A general angular momentum state is a combination of orbital angular momentum
and spin. To rigorously determine the transformation of a total angular momentum
state, the behavior of both pieces must be considered. The action of the rotation
and mirror reflection operations on spatial wavefunctions is the easiest to
obtain as they are immediate consequences of the transformation of the coordinate
operators which are easily understood geometrically. For a given angular momentum
\( l \) and orientation \( m \), the angular part of the wavefunction is given
by a Spherical Harmonic that is a Legendre polynomial, a phase and a real normalization
coefficient,
\[
Y^{\left( l\right) }_{m}\left( \theta ,\phi \right) =f\left( l,m\right) P^{\left( l\right) }_{m}\left( cos\left( \theta \right) \right) e^{im\phi }\]
The Spherical Harmonics have the symmetry,
\[
Y^{\left( l\right) }_{-m}=\left( -1\right) ^{m}Y^{\left( l\right) *}_{m}\]
or 
\[
f\left( l,-m\right) P^{\left( l\right) }_{-m}=\left( -1\right) ^{m}f\left( l,m\right) P^{\left( l\right) }_{m}\]
and
\[
P^{\left( l\right) }_{m}\left( -s\right) =\left( -1\right) ^{l+m}P^{\left( l\right) }_{m}\left( s\right) \]

The reflection of the \( \hat{n} \) direction leaves \( \theta  \) unchanged,
but a rotation about the same axis results in \( z\rightarrow -z \) or \( cos\left( \theta \right) \rightarrow -cos\left( \theta \right)  \)
so that \( R_{\hat{n}}P^{\left( l\right) }_{m}R_{\hat{n}}^{-1}=\left( -1\right) ^{l+m}P^{\left( l\right) }_{m} \).
The transformation of \( \phi  \) can also be worked out geometrically, with
\( \phi _{0} \) be the angle of \( \hat{n} \) in the \( x-y \) plane relative
to the \( \hat{x} \) axis, 
\begin{eqnarray*}
R_{\hat{n}}\phi R_{\hat{n}}^{\dagger } & = & 2\phi _{0}-\phi \\
M_{\hat{n}}\phi M_{\hat{n}}^{\dagger } & = & 2\phi _{0}-\phi +\pi 
\end{eqnarray*}
For the particular cases of \( \hat{n}=\hat{x},\hat{y} \) where \( \phi _{0}=0,\pi /2 \)
this gives, 
\begin{eqnarray*}
R_{\hat{x}}\phi R_{\hat{x}}^{\dagger } & = & -\phi \\
M_{\hat{x}}\phi M_{\hat{n}}^{\dagger } & = & \pi -\phi \\
R_{\hat{y}}\phi R_{\hat{x}}^{\dagger } & = & \pi -\phi \\
M_{\hat{y}}\phi M_{\hat{n}}^{\dagger } & = & 2\pi -\phi \sim -\phi 
\end{eqnarray*}
With \( x=cos\left( \phi \right)  \), \( y=sin\left( \phi \right)  \) and
\( cos\left( -\phi \right) =cos\left( \phi \right)  \), \( sin\left( -\phi \right) =-sin\left( \phi \right)  \),
\( cos\left( \pi -\phi \right) =-cos\left( \phi \right)  \), \( sin\left( \pi -\phi \right) =sin\left( \phi \right)  \)
this yields the correct transformations, 
\begin{eqnarray*}
R^{\dagger }_{\hat{x}}\left( x,y\right) R_{\hat{x}} & = & \left( x,-y\right) \\
M^{\dagger }_{\hat{x}}\left( x,y\right) M_{\hat{x}} & = & \left( -x,y\right) \\
R^{\dagger }_{\hat{y}}\left( x,y\right) R_{\hat{y}} & = & \left( -x,y\right) \\
M^{\dagger }_{\hat{y}}\left( x,y\right) M_{\hat{y}} & = & \left( x,-y\right) 
\end{eqnarray*}

With the action of these operators on the coordinates established, the transformation
of the wavefunctions follow easily.

\begin{eqnarray*}
R_{\hat{n}}Y^{\left( l\right) }_{m}\left( \theta ,\phi \right) R_{\hat{x}}^{\dagger } & = & \left( -1\right) ^{l+m}Y^{\left( l\right) }_{m}\left( \theta ,2\phi _{0}-\phi \right) \\
 & = & \left( -1\right) ^{l+m}f\left( l,m\right) P^{\left( l\right) }_{m}\left( cos\left( \theta \right) \right) e^{im\left( 2\phi _{0}-\phi \right) }\\
 & = & \left( -1\right) ^{l+m}e^{2im\phi _{0}}f\left( l,m\right) P^{\left( l\right) }_{m}\left( cos\left( \theta \right) \right) e^{-im\phi }\\
 & = & \left( -1\right) ^{l+m}e^{2im\phi _{0}}Y^{\left( l\right) *}_{m}\left( \theta ,\phi \right) \\
 & = & \left( -1\right) ^{l+m}\left( -1\right) ^{m}e^{2im\phi _{0}}Y^{\left( l\right) }_{-m}\left( \theta ,\phi \right) \\
 & = & \left( -1\right) ^{l}e^{2im\phi _{0}}Y^{\left( l\right) }_{-m}\left( \theta ,\phi \right) 
\end{eqnarray*}
For the mirror reflection \( 2\phi _{0}\rightarrow 2\phi _{0}+\pi  \), and
there is no \( \left( -1\right) ^{l+m} \) from changing the sign of \( z \),
\begin{eqnarray*}
M_{\hat{n}}Y^{\left( l\right) }_{m}\left( \theta ,\phi \right) M^{\dagger }_{\hat{n}} & = & \left( -1\right) ^{m}e^{im\left( 2\phi _{0}+\pi \right) }Y^{\left( l\right) }_{-m}\left( \theta ,\phi \right) \\
 & = & \left( -1\right) ^{m}e^{im\pi }e^{2im\phi _{0}}Y^{\left( l\right) }_{-m}\left( \theta ,\phi \right) \\
 & = & \left( -1\right) ^{m}\left( -1\right) ^{m}e^{2im\phi _{0}}Y^{\left( l\right) }_{-m}\left( \theta ,\phi \right) \\
 & = & e^{2im\phi _{0}}Y^{\left( l\right) }_{-m}\left( \theta ,\phi \right) 
\end{eqnarray*}
Then the states transform as,
\begin{eqnarray*}
R_{\hat{n}}\left| l,m\right\rangle  & = & \left| l,-m\right\rangle \left( -1\right) ^{l}e^{2im\phi _{0}}\\
M_{\hat{n}}\left| l,m\right\rangle  & = & \left| l,-m\right\rangle e^{2im\phi _{0}}
\end{eqnarray*}
Note that this is consistent with the usual result for parity,
\begin{eqnarray*}
P\left| l,m\right\rangle  & = & R_{\hat{n}}M_{\hat{n}}\left| l,m\right\rangle =\left( R_{\hat{n}}\left| l,-m\right\rangle \right) e^{2im\phi _{0}}\\
 & = & \left| l,m\right\rangle \left( -1\right) ^{l}e^{-2im\phi _{0}}e^{2im\phi _{0}}\\
 & = & \left| l,m\right\rangle \left( -1\right) ^{l}
\end{eqnarray*}
For completeness, also consider \( \hat{n}=\hat{z} \). \( M_{\hat{z}} \) takes
\( z\rightarrow -z \), or \( cos\left( \theta \right) \rightarrow -cos\left( \theta \right)  \)
and \( Y^{\left( l\right) }_{m}\rightarrow \left( -1\right) ^{l+m}Y^{\left( l\right) }_{m} \).
\( R_{\hat{z}}\left( \pi \right)  \) gives \( \phi \rightarrow \phi +\pi  \),
so that \( e^{im\phi }\rightarrow e^{im\left( \phi +\pi \right) }=e^{im\pi }e^{im\phi }=\left( -1\right) ^{m}e^{im\phi } \)
and \( Y^{\left( l\right) }_{m}\rightarrow \left( -1\right) ^{m}Y^{\left( l\right) }_{m} \).
For the states,

\begin{eqnarray*}
R_{\hat{z}}\left| l,m\right\rangle  & = & \left| l,m\right\rangle \left( -1\right) ^{m}\\
M_{\hat{z}}\left| l,m\right\rangle  & = & \left| l,m\right\rangle \left( -1\right) ^{l+m}
\end{eqnarray*}
yielding the same correct result for parity,
\[
P\left| l,m\right\rangle =R_{\hat{z}}M_{\hat{z}}\left| l,m\right\rangle =\left| l,m\right\rangle \left( -1\right) ^{l+m}\left( -1\right) ^{m}=\left| l,m\right\rangle \left( -1\right) ^{l}\]

Since the mirror and rotation representations of the spin flip are related through
the parity transformation, which is that is reflected in the very similar results
of each transformation, differing only by the sign \( \left( -1\right) ^{l} \)
in all cases which is the parity of the state \( \eta _{l} \), the results
for any representation can be written in a more compact form by defining \( \eta _{F,l} \)
to be the contribution to the parity of the state with angular momentum \( l \)
given by the transformation \( F \). For these spatial wavefunctions, \( \eta _{R,l}=\left( -1\right) ^{l} \),
\( \eta _{M,l}=1 \). With this, the results can be summarized nicely as,
\[
F\left| l,m\right\rangle =\left| l,-m\right\rangle \eta _{F,l}e^{2im\phi _{0}}\]
In this form a parity transformation looks like,
\begin{eqnarray*}
P\left| l,m\right\rangle  & = & R_{\hat{n}}M_{\hat{n}}\left| l,m\right\rangle =\left( R_{\hat{n}}\left| l,-m\right\rangle \right) \eta _{M,l}e^{2im\phi _{0}}\\
 & = & \left| l,m\right\rangle \eta _{R}\eta _{M}e^{-2im\phi _{0}}e^{2im\phi _{0}}\\
 & = & \left| l,m\right\rangle \eta _{R,l}\eta _{M,l}
\end{eqnarray*}
which requires \( \eta _{R,l}\eta _{M,l}=\eta _{l} \) for consistency, which
is satisfied in this simple case. 

Clearly these representations of \( F \) are unitary, as desired. Also, in
this case, the phase is linear in \( m \) so that,
\[
F^{2}\left| l,m\right\rangle =F\left| l,-m\right\rangle \left( -1\right) ^{l}\eta ^{l}_{F}e^{2im\phi _{0}}=\left| l,m\right\rangle \left( -1\right) ^{2l}\eta _{F}^{2l}e^{-2im\phi _{0}}e^{2im\phi _{0}}=\left| l,m\right\rangle \]
and so \( F^{2}=1 \) and \( F=F^{-1}=F^{\dagger } \), so when acting on spatial
wavefunctions \( F \) is hermitian.

\subsubsection{Spin Wavefuntions}

Spins should behave in the same way, angular momentum is angular momentum, but
it isn't immediately clear which branch to use in \( \left( -1\right) ^{l,m} \)
for \( \eta _{F}=1 \) as now \( s \) and \( m \) are half integer. It is
important to get the phases of all the transformations exactly right as they
combine to get the sign change this whole analysis is meant to determine, so
the careful attention to phase is not merely pedantic. For example, expressions
like \( (-1)^{j_{1}}(-1)^{j_{2}} \) appear. For integer valued \( j_{1} \)
and \( j_{2} \) this product is perfectly well defined as \( (-1)^{j_{1}+j_{2}} \).
For either \( j \) a half integer, each root could be represented as \( e^{\pm i\pi /2} \)
giving a result of \( (-1)^{j_{1}\pm j_{2}} \), For \( j_{1} \), \( j_{2} \)
half integer this is an ambiguous result as \( (-1)^{j_{1}+j_{2}}=(-1)^{j_{1}-j_{2}} \).

For Barium, the spin angular momentum will be a single electron, so strictly
only spin 1/2 needs to be considered for this application, but the general case
is not too hard to deal with, with a bit a formalism, and the results provide
an easy derivation for the spin flip symmetry of the Clebsch-Gordan Coefficients.

Rotations can be dealt with mechanically with the usual rotation matrices, \( R_{\hat{n}}\left( \alpha \right) =e^{-i\alpha \hat{n}\cdot \vec{J}} \).
For rotations about the \( \hat{z} \) axis, the results are trivial since \( J_{z} \)
is diagonal,
\[
\left( R_{\hat{z}}\left( \alpha \right) \right) _{mm'}=\left( e^{-i\alpha J_{z}}\right) _{mm'}=\delta _{mm'}e^{-im\alpha }\]
this gives,
\[
R_{\hat{z}}\left| s,m\right\rangle \equiv R_{\hat{z}}\left( \pi \right) \left| s,m\right\rangle =\left| s,m\right\rangle e^{-im\pi }\]
This is identical to the result for orbital angular momentum for \( m=l \)
and integer, and identifies the proper branch to use, \( \sqrt{-1}=-i \) ,
for \( m \) an odd half-integer.

For rotations about an axis in the \( x-y \) plane are more difficulty as exponentiating
\( J_{x,y} \) is non-trivial. Here Wigner's closed form solution for arbitrary
rotations about the \( y \) axis can be used,
\begin{eqnarray*}
\left( e^{-i\alpha J_{x}}\right) _{mm'} & = & d_{mm'}^{\left( j\right) }\left( \alpha \right) \\
d_{mm'}^{\left( j\right) }\left( \alpha \right)  & = & \sum _{k}\left( -1\right) ^{k-m'+m}\frac{\sqrt{\left( j+m'\right) !\left( j-m'\right) !\left( j+m\right) !\left( j-m\right) !}}{\left( j-k+m'\right) !k!\left( j-k-m\right) !\left( k-m'+m\right) !}\\
 &  & \left( cos\frac{\alpha }{2}\right) ^{2j-2k+m'-m}\left( sin\frac{\alpha }{2}\right) ^{2k-m'+m}
\end{eqnarray*}
The sum over \( k \) is for all \( k \) that doesn't give a negative argument
for a factorial in the denominator. For \( \alpha =\pi  \), this has a particularly
simple form. In this case \( cos\left( \alpha /2\right) =cos\left( \pi /2\right) =0 \),
\( sin\left( \alpha /2\right) =sin\left( \pi /2\right) =1 \). Each term in
the sum is zero unless the exponent of the \( cos \) term is zero, \( 2j-2k+m'-m=0 \)
or \( j-k=\left( m-m'\right) /2 \) and at most only one term contributes to
the sum, 
\begin{eqnarray*}
d_{mm'}^{\left( j\right) }\left( \alpha \right)  & = & \left( -1\right) ^{k-m'+m}\frac{\sqrt{\left( j+m'\right) !\left( j-m'\right) !\left( j+m\right) !\left( j-m\right) !}}{\left( j-k+m'\right) !k!\left( j-k-m\right) !\left( k-m'+m\right) !}\\
k & = & j-\left( m-m'\right) /2
\end{eqnarray*}
This term only contributes to the original sum if none of the arguments of the
factorial in the denominator are negative. Consider the terms involving \( j \)
in the denominator,
\begin{eqnarray*}
j-k+m' & = & \left( m-m'\right) /2+m'=\left( m+m'\right) /2\\
j-k-m & = & \left( m-m'\right) /2-m=-\left( m+m'\right) /2
\end{eqnarray*}
The sum is over all \( k \) giving no negative arguments for the factorial
in the denominators, but these arguments will be \( \pm \left( m+m'\right) /2 \)
when the \( cos \) factor is also non-zero. These arguments can both be non-negative
only when they are both zero, requiring \( m'=-m \), only the terms that flip
the spin are non-zero. For this case \( k=j-m=j+m' \) and the rotation matrix
elements are,

\[
d_{mm'}^{\left( j\right) }\left( \alpha \right) =\delta _{m,-m'}\left( -1\right) ^{j+m}\frac{\left( j+m\right) !\left( j-m\right) !}{\left( j+m\right) !\left( j-m\right) !}=\delta _{m,-m'}\left( -1\right) ^{j+m}\]

A rotation about an arbitrary axis in the \( x-y \) plane, defined by the same
\( \phi _{0} \) as above, can be done by a composition of rotations, first
rotate \( \hat{n} \) to the \( \hat{y} \) axis, \( R_{\hat{z}}\left( \pi /2-\phi _{0}\right)  \),
followed by the \( \pi  \) rotation about the \( \hat{y} \) axis given above,
\( R_{\hat{y}}\left( \pi \right)  \), and a final rotation back to the original
orientation of \( x,y \) axis, \( R_{\hat{z}}\left( -\left( \pi /2-\phi _{0}\right) \right)  \).
Again the \( \hat{z} \) rotations are trivial and the composition gives,

\begin{eqnarray*}
 &  & \left( R_{\hat{z}}\left( \frac{\pi }{2}-\phi _{0}\right) R_{\hat{y}}\left( \pi \right) R_{\hat{z}}\left( -\left( \frac{\pi }{2}-\phi _{0}\right) \right) \right) _{mm'}\\
 &  & =R_{\hat{z}}\left( \frac{\pi }{2}-\phi _{0}\right) _{mm_{1}}R_{\hat{y}}m\left( \pi \right) _{m_{1}m_{2}}R_{\hat{z}}\left( -\left( \frac{\pi }{2}-\phi _{0}\right) \right) _{m_{2}m'}\\
 &  & =\delta _{mm_{1}}e^{-im\left( \pi /2-\phi _{0}\right) }\delta _{m_{1},-m_{2}}\left( -1\right) ^{j+m_{1}}\delta _{m_{2},m'}e^{im'\left( \pi /2-\phi _{0}\right) }\\
 &  & =\delta _{m,-m'}\left( -1\right) ^{j+m}e^{2im\left( \pi /2-\phi _{0}\right) }\\
 &  & =\delta _{m,-m'}\left( -1\right) ^{j+m}e^{im\pi }e^{2im\phi _{0}}
\end{eqnarray*}
Here some care must be taken in combining the phases. For any \( j \), \( j+m \)
is an integer and \( \left( -1\right) ^{j+m} \) is well defined, \( e^{im\pi } \)
can also be written as \( \left( -1\right) ^{m} \) but the branch to use for
\( m \) an odd half integer is not obvious in this form as if could be interpreted
as \( \left( \pm i\right) ^{2m} \). To simplify this without ambiguity write
the real, sign only, phase as \( e^{-i\pi \left( j+m\right) } \), then the
phases combine uniquely to give \( e^{-i\pi \left( j+m\right) }e^{i\pi m}=e^{-ij\pi } \).
With the other choice of representation for the sign, the total phase would
give \( e^{i\pi \left( j+2m\right) } \) and the resulting \( e^{2im\pi } \)
can't be unambiguously simplified to \( 1^{m} \) as it is not clear whether
this should be \( \pm 1 \) for \( m \) an odd half integer, the former path
shows that it should be \( -1 \).

With the phase now well defined for all \( j \). The transformations of a spin
state again corresponds to that for orbital angular momentum, as it should since
angular momentum states are defined so that they transform like spherical harmonics,
for \( j=s \),
\[
R_{\hat{n}}\left| s,m\right\rangle =\left| s,-m\right\rangle e^{-is\pi }e^{2im\phi _{0}}\]
 As for the case of orbital angular momentum, the transformed state is modified
by at a most a phase and \( F \) is unitary, as is also apparent from the original
form of the rotation operator. In this case, \( R^{2}\left| s,m\right\rangle =\left| s,m\right\rangle e^{2is\pi } \)
and the phase is 1 only for \( s \) an integer and \( F \) is not, in general,
hermitian. This can be written in terms of the previously defined \( \eta _{F,l} \)
with the modification \( \eta _{R,s}=e^{-is\pi } \) which gives \( \eta _{R,l}=\left( -1\right) ^{l} \)
for \( l \) and integer as before,

\[
R_{\hat{n}}\left| s,m\right\rangle =\left| s,-m\right\rangle \eta _{R,l}e^{2im\phi _{0}}\]

The complementary results for a spin flip based on a mirror reflection are less
straight forward. The desired action on the states is clear, the expectation
value of some component of the spin should change sign,
\begin{eqnarray*}
\left\langle s,m\right| F^{\dagger }\hat{n}\cdot \vec{j}F\left| s,m\right\rangle  & = & -\hat{n}\cdot \left\langle s,m\right| \vec{j}\left| s,m\right\rangle \\
 & = & -\hat{n}\cdot \vec{s}
\end{eqnarray*}
this requires 
\[
F^{\dagger }\hat{n}\cdot \vec{j}F=-\hat{n}\cdot \vec{j}\]
and all perpendicular components should be unchanged. In particular consider
\( \hat{n}=\hat{y} \), the transformation requires,
\[
F^{\dagger }\left( j_{x},j_{y},j_{z}\right) F=\left( j_{x},-j_{y},j_{z}\right) \]

It is easy to show that for at least spin 1/2 representations no such operator
is possible. \( F \) must be unitary, so it can be written as \( e^{if} \)
where then \( f \) is hermitian. For \( j=1/2 \) the only hermitian operators
are \( \vec{j} \), as the algebra quickly closes since \( j_{i}^{2}=j^{2}/3 \).
So \( F \) can only be written as \( e^{i\hat{n}\cdot \vec{j}} \) which is
a rotation. A mirror reflection is a discrete operation not continuously connected
to any three dimensional rotation so \( e^{i\hat{n}\cdot \vec{j}} \) can't
be \( M \) and these are the only possible unitary matrices for \( j=1/2 \)
so there can be no representation for a mirror reflection. 

It is possible that for higher spins there is enough freedom to construct a
spin flip operator since products like \( j_{i}j_{j\neq i} \) are not independent
of \( j_{i} \) and \( j^{2} \), but the procedure is far from straightforward
and completely dependent on a particular representation. It is also possible
that this could be done systematically in more spatial dimensions. A rotation
by \( 180^{\circ } \) always changes the sign of two coordinates, if one of
the coordinates is always a fourth spatial \( w=0 \), the result looks like
a reflection in three dimensions. This is each to implement with the coordinate
operators with an \( SO(4) \) representation, but the resulting action of such
rotations on the angular momentum states isn't as trivial. This is now getting
hopelessly pedantic, though such a strategy will probably be pursued.

The actions of a mirror reflection could be deduced from the action of the rotations
and parity from \( P=M_{\hat{n}}R_{\hat{n}}\left( \pi \right)  \) except that
the parity operator is also not well, or uniquely, defined. For Dirac 4-component
wave functions a parity operator can be constructed but it allows for parity
eigenvalues of \( \pm 1 \). In the end only the relative parity of elementary
particles are important and assignments are made by convention, \( P\left| s,m\right\rangle =\left| s,m\right\rangle \eta _{s} \)
defines \( \eta _{s} \). The given phase by a mirror reflection can then be
determined through 
\begin{eqnarray*}
P\left| s,m\right\rangle  & = & M_{\hat{n}}R_{\hat{n}}\left| s,m\right\rangle \\
 & = & M_{\hat{n}}\left| s,-m\right\rangle \eta _{R,s}e^{2im\phi _{0}}\\
 & = & \left| s,m\right\rangle \eta _{R,s}\eta _{M,s}
\end{eqnarray*}
implying \( \eta _{R,s}\eta _{M,s}=\eta _{s} \), where \( \eta _{R,s}=e^{-is\pi } \)
giving \( \eta _{M,s}=\eta _{s}e^{is\pi } \). For \( s=l \) an orbital angular
momentum wavefunction where \( \eta _{l}=\left( -1\right) ^{l} \) this would
give \( \eta _{M,s}=\left( -1\right) ^{l}\left( -1\right) ^{l}=1 \) as before.
This not completely defined phase from the unknown parity of a spin wavefunction
will not cause any trouble as \( \eta _{F} \) will always appear in complementary
pairs in any physical quantity that is computed for these atomic calculations
and \( \eta _{F}^{*}\eta _{F}=1 \).

\subsubsection{Clebsch-Gordan Coefficients}

The action of any of these spin flip operators is now well defined for any angular
momentum state and they can be immediately used to demonstrate a useful spin
flip symmetry of the Clebsch-Gordan coefficients. The Clebsch-Gordan coefficients
give the transformation of a direct product of two angular momentum wavefunctions
\( \left| j_{1},m_{1}\right\rangle \otimes \left| j_{2},m_{1}\right\rangle =\left| j_{1},m_{1};j_{2},m_{2}\right\rangle  \)
to a combined total angular momentum state, \( \left| j,m\right\rangle  \),
\[
\left| j,m\right\rangle =\sum _{m_{1},m_{2}}\left| j_{1},m_{1};j_{2},m_{2}\right\rangle \left\langle j_{1},m_{1};j_{2},m_{2}|j,m\right\rangle \]
Flipping the spin of each state, and invoking the unitarity of the spin flip,
\begin{eqnarray*}
\left\langle j_{1},-m_{1};j_{2},-m_{2}|j,-m\right\rangle  & = & \eta ^{*}_{F,j_{1}}\eta ^{*}_{F,j_{2}}e^{-2im_{1}\phi _{0}}e^{-2im_{2}\phi _{0}}\\
 & \times  & \left\langle j_{1},m_{1},j_{2},m_{2}\right| R_{\hat{n}}^{\dagger }R_{\hat{n}}\left| j,m\right\rangle \eta _{F,j}e^{2im\phi _{0}}\\
 & = & \eta ^{*}_{F,j_{1}}\eta ^{*}_{F,j_{2}}\eta _{F,j}e^{-2i\left( m_{1}+m_{2}-m\right) \phi _{0}}\left\langle j_{1},m_{1},j_{2},m_{2}|j,m\right\rangle 
\end{eqnarray*}
Now for simplicity pick \( F=R_{\hat{n}}\left( \pi \right)  \) so that \( \eta _{F,j}=\eta _{R,j}=e^{-ij\pi } \)
and the entire product of \( \eta  \)'s is \( e^{i\left( j_{1}+j_{2}-j\right) \pi } \).
The Clebsch-Gordan coefficient is zero expect for \( m=m_{1}+m_{2} \) so the
\( \phi _{0} \) dependent part of the phase never contributes to a non-trivial
coefficient. The triangle selection rule requires that \( \left| j_{1}-j_{2}\right| \leq j\leq j_{1}+j_{2} \)
where \( j \) can differ by integer values from its largest and smallest possible
values. This requires that if both \( j_{1} \) and \( j_{2} \) are integer
or half integer, \( j \) must be an integer, while if only one of \( j_{1},j_{2} \)
is half integer, \( j \) is also half integer. Then at most two of \( j_{1},j_{2},j \)
are half integer values, and never only one so \( j_{1}+j_{2}-j \) is always
an integer and the \( j \) dependent phase can be simplified \( e^{-i\left( j_{1}+j_{2}-j\right) \pi }=\left( -1\right) ^{j_{1}+j_{2}-j} \),
finally yielding,
\[
\left\langle j_{1},-m_{1};j_{2},-m_{2}|j,-m\right\rangle =\left( -1\right) ^{j_{1}+j_{2}-j}\left\langle j_{1},m_{1},j_{2},m_{2}|j,m\right\rangle \]

The result should be independent of the representation used, and this rotational
spin flip has trivial \( \eta  \), though it is a bit mysterious how this holds
for \( F=M_{\hat{n}} \) where \( \eta _{M,j}=\eta _{j}e^{ij\pi } \). This
gives \( \eta ^{*}_{F,j_{1}}\eta ^{*}_{F,j_{2}}\eta _{F,j}=\eta _{j_{1}}\eta _{j_{2}}\eta _{j}\left( -1\right) ^{\left( j_{1}+j_{2}-j\right) } \).
The product of intrinsic parities could apparently be anything and if negative
this implies that the corresponding Clebsch-Gordan coefficient is zero because
it would require that both \( \left\langle j_{1},-m_{1};j_{2},-m_{2}|j,-m\right\rangle =\pm \left( -1\right) ^{j_{1}+j_{2}-j}\left\langle j_{1},m_{1},j_{2},m_{2}|j,m\right\rangle  \). 

As an example, consider orbital angular momentum states where all the \( j=l \)
are integers. For this case \( \eta _{M,l}=1 \) which gives\( \left\langle l_{1},-m_{1};l_{2},-m_{2}|l,-m\right\rangle =\left\langle l_{1},m_{1},l_{2},m_{2}|l,m\right\rangle  \)
so that the Clebsch-Gordan coefficient is independent of the sign of the spin
directions. At the same time the first relation derived using rotations still
holds so consistency requires \( \left( -1\right) ^{l_{1}+l_{2}-l}=1 \), which
is the same as the requirements on the parity since \( \eta _{l}=\left( -1\right) ^{l} \).
For this example, this condition is satisfied. For a multipole transition between
states, the parity of the initial state is changed by the parity of the transition
so that the only possible final states that give non-zero matrix elements have
the parity of the product of the parities of the initial state and the transition.
It is easy to show explicitly with the wavefunctions and operators that dipole
transitions only couple states of opposite parity, quadrupole transitions couple
states of the same parity.

Considering these many general representation of the spin flip operator then
yields the desired relation for the Clebsch-Gordan coefficients as well as a
statement about transitions and couplings. Rather than a contradiction, this
constraint on the product of parities actually generates the usual parity selection
rules, but now quite generally for any angular momentum states, 
\[
\eta _{j_{1}}\eta _{j_{2}}\eta _{j}=1\]
 or, since \( \eta ^{2}=1 \), a more natural representation,
\[
\eta _{j}=\eta _{j_{1}}\eta _{j_{2}}\]

\subsubsection{Total Wavefuntions}

With the transformation of the Clebsch-Gordan coefficients determined, the action
of the total wavefunction under such spin flips can be determined. Total angular
momentum includes contributions from orbital angular momentum and spin angular
momentum, 
\[
\left| j,m\right\rangle =\sum _{m_{l},m_{s}}\left| l,m_{l};s,m_{s}\right\rangle \left\langle l,m_{l};s,m_{s}|j,m\right\rangle \]
 Both were shown to transform identically so the total angular momentum state
should transform the same way as is easily shown,

\begin{eqnarray*}
R_{\hat{n}}\left| j,m\right\rangle  & = & \sum _{m_{l},m_{s}}F\left| l,m_{l}\right\rangle \otimes F\left| s,m_{s}\right\rangle \left\langle l,m_{l};s,m_{s}|j,m\right\rangle \\
 & = & \sum _{m_{l},m_{s}}\left| l,-m_{l}\right\rangle \eta _{F,l}e^{2im_{l}\phi _{0}}\otimes \left| s,-m_{s}\right\rangle \eta _{F,s}e^{2im_{s}\phi _{0}}\left\langle l,m_{l};s,m_{s}|j,m\right\rangle \\
 & = & \eta _{F,l}\eta _{F,s}\sum _{m_{l},m_{s}}\left| l,-m_{l}\right\rangle \otimes \left| s,-m_{s}\right\rangle e^{2i\left( m_{l}+m_{s}\right) \phi _{0}}\left\langle l,m_{l};s,m_{s}|j,m\right\rangle 
\end{eqnarray*}
The Clebsch-Gordan coefficient is non-zero only for \( m_{l}+m_{s}=m \) so
that dependence in the phase can be factored out of the sum, and the previous
spin flip transformation of the Clebsch-Gordan coefficient used to do the sum,

\begin{eqnarray*}
R_{\hat{n}}\left| j,m\right\rangle  & = & \eta _{F,l}\eta _{F,s}e^{2im\phi _{0}}\sum _{m_{l},m_{s}}\left| l,-m_{l}\right\rangle \otimes \left| s,-m_{s}\right\rangle \left\langle l,m_{l};s,m_{s}|j,m\right\rangle \\
 & = & \eta _{F,l}\eta _{F,s}\left( -1\right) ^{l+s-j}e^{2im\phi _{0}}\\
 & \times  & \sum _{m_{l},m_{s}}\left| l,-m_{l}\right\rangle \otimes \left| s,-m_{s}\right\rangle \left\langle l,-m_{l};s,-m_{s}|j,-m\right\rangle \\
 & = & \eta _{F,l}\eta _{F,s}\left( -1\right) ^{l+s-j}e^{2im\phi _{0}}\left| j,-m\right\rangle 
\end{eqnarray*}
The form of the transformation is identical. This finally shows that for any
angular momentum state,

\[
F\left| j,m\right\rangle =\eta _{F,j}e^{2im\phi _{0}}\left| j,-m\right\rangle \]
where \( \eta _{F,j}=\eta _{F,l}\eta _{F,s}\left( -1\right) ^{l+s-j} \). For
rotations, \( \eta _{R,s}=e^{-is\pi } \), this becomes \( \eta _{F,j}=e^{-il\pi }e^{-is\pi }\left( -1\right) ^{l+s-j} \),
the phases can be combined unambiguously in complex exponential form, \( \eta _{F,j}=e^{-il\pi }e^{-is\pi }e^{i\left( l+s-j\right) \pi }=e^{-ij\pi } \)
just as for a pure spin state. Similarly for reflections, \( \eta _{R,s}=\eta _{s}e^{is\pi } \)
so that \( \eta _{R,j}=\eta _{l}e^{il\pi }\eta _{s}e^{is\pi }\left( -1\right) ^{l+s-j}=\eta _{l}\eta _{s}e^{i\left( 2l+2s-j\right) \pi } \),
again combine the phases in complex exponential form but here use \( -1=e^{-i\pi } \)
so that \( e^{il\pi }e^{is\pi }\left( -1\right) ^{l+s-j}=e^{il\pi }e^{is\pi }e^{-i\left( l+s-j\right) }=e^{ij\pi } \),
of course just the complex conjugate of the previous result. This gives \( \eta _{R,j}=\eta _{l}\eta _{s}e^{ij\pi } \)
which implies that \( \eta _{j}=\eta _{l}\eta _{s} \) consistent with the parity
selection rules derived immediately above.

\subsection{Operators}

To determine the action on matrix elements, the transformation of the coordinate
operators must also be determined. This is easily done from the transformation
of the azimuthal angle already determined,
\begin{eqnarray*}
R_{\hat{n}}(\theta ,\phi )R^{\dagger }_{\hat{n}} & = & \left( -\theta ,2\phi _{0}-\phi \right) \\
M_{\hat{n}}\left( \theta ,\phi \right) M^{\dagger }_{\hat{n}} & = & \left( \theta ,2\phi _{0}-\phi +\pi \right) 
\end{eqnarray*}
This gives the transformation of \( z \) trivially. For \( x \) and \( y \), 

\begin{eqnarray*}
R_{\hat{n}}\left\{ x,y\right\} R_{\hat{n}}^{\dagger } & = & R_{\hat{n}}\left\{ cos\left( \phi \right) ,sin\left( \phi \right) \right\} R_{\hat{n}}^{\dagger }\\
 & = & \left\{ cos\left( 2\phi _{0}-\phi \right) ,sin\left( 2\phi _{0}-\phi \right) \right\} \\
 & = & \{cos\left( 2\phi _{0}\right) cos\left( \phi \right) +sin\left( 2\phi _{0}\right) sin\left( \phi \right) ,\\
 &  & sin\left( 2\phi _{0}\right) cos\left( \phi \right) -cos\left( 2\phi _{0}\right) sin\left( \phi \right) \}\\
 & = & \left\{ cos\left( 2\phi _{0}\right) x+sin\left( 2\phi _{0}\right) y,sin\left( 2\phi _{0}\right) x-cos\left( 2\phi _{0}\right) y\right\} 
\end{eqnarray*}
For mirror reflections just take \( 2\phi _{0}\rightarrow 2\phi _{0}+\pi  \)
which just changes the signs of all the coefficients so that for either representation,
the results can be written using \( \eta _{F,j} \) which is given, for \( j=k \)
an integer, by \( \eta _{R,k}=\left( -1\right) ^{k} \), and for these spatial
operators \( \eta _{F,k}=1 \),
\[
F\left\{ x,y,z\right\} F^{\dagger }=\eta _{F,1}\left( -cos\left( 2\phi _{0}\right) x-sin\left( 2\phi _{0}\right) y,-sin\left( 2\phi _{0}\right) xcos\left( 2\phi _{0}\right) y,z\right) \]
For completeness, also consider \( \hat{n}=\hat{z} \), this doesn't flip spin,
but for compactness call it \( F_{\hat{z}} \) anyways,
\[
F_{\hat{z}}\left\{ x,y,z\right\} F_{\hat{z}}^{\dagger }=\eta _{F,1}\left\{ x,y,-z\right\} \]
 The transformation of the spherical tensors can then be determined. For the
dipole operators
\begin{eqnarray*}
r^{\left( 1\right) }_{\pm } & = & \left( \mp x+iy\right) /\sqrt{2}\\
r^{\left( 1\right) }_{0} & = & z
\end{eqnarray*}
The transformations give,
\begin{eqnarray*}
R_{\hat{n}}r^{\left( 1\right) }_{\pm }R_{\hat{n}}^{\dagger } & = & R_{\hat{n}}\left( \mp x+iy\right) R_{\hat{n}}^{\dagger }/\sqrt{2}\\
 & = & \left( \mp cos\left( 2\phi _{0}\right) x\mp sin\left( 2\phi _{0}\right) y+isin\left( 2\phi _{0}\right) x-icos\left( 2\phi _{0}\right) y\right) /\sqrt{2}\\
 & = & \left( \mp \left( cos\left( 2\phi _{0}\right) \mp isin\left( 2\phi _{0}\right) \right) x-i\left( cos\left( 2\phi _{0}\right) \mp isin\left( 2\phi _{0}\right) y\right) y\right) /\sqrt{2}\\
 & = & e^{\mp 2i\phi _{0}}\left( \mp x-iy\right) /\sqrt{2}\\
 & = & -e^{\mp 2i\phi _{0}}\left( \pm x+iy\right) /\sqrt{2}\\
 & = & -e^{\mp 2i\phi _{0}}r^{\left( 1\right) }_{\mp }
\end{eqnarray*}
The mirror just gives minus this result,

\begin{eqnarray*}
F\left\{ r^{\left( 1\right) }_{\pm },r^{\left( 1\right) }_{0}\right\} F^{\dagger } & = & \eta _{F,1}\left\{ e^{\mp 2i\phi _{0}}r^{\left( 1\right) }_{\mp },r^{\left( 1\right) }_{0}\right\} \\
F_{\hat{z}}\left( r^{\left( 1\right) }_{\pm },r^{\left( 1\right) }_{0}\right) F_{\hat{z}}^{\dagger } & = & \eta _{F,1}\left( r^{\left( 1\right) }_{\pm },-r^{\left( 1\right) }_{0}\right) 
\end{eqnarray*}

Finally this can be used to quickly get the transformations of the spherical
quadrupole operators,
\begin{eqnarray*}
r^{\left( 2\right) }_{\pm 2} & = & r_{\pm }^{\left( 1\right) 2}\\
r^{\left( 2\right) }_{\pm 1} & = & \sqrt{2}r_{\pm }^{\left( 1\right) }r_{0}^{\left( 1\right) }\\
r^{\left( 2\right) }_{0} & = & \sqrt{2/3}\left( r_{+}^{\left( 1\right) }r_{-}^{\left( 1\right) }+r_{0}^{\left( 1\right) 2}\right) 
\end{eqnarray*}
from the transformation of the dipole operators,
\begin{eqnarray*}
F\left\{ r^{\left( 2\right) }_{\pm 2},r^{\left( 2\right) }_{\pm 1},r^{\left( 2\right) }_{0}\right\} F^{\dagger } & = & \left( e^{\mp 4i\phi _{0}}r^{\left( 2\right) }_{\mp 2},e^{\mp 2i\phi _{0}}r^{\left( 2\right) }_{\mp 1},r^{\left( 2\right) }_{0}\right) \\
F_{\hat{z}}\left( r^{\left( 2\right) }_{\pm 2},r^{\left( 2\right) }_{\pm 1},r^{\left( 2\right) }_{0}\right) F_{\hat{z}}^{\dagger } & = & \left( r^{\left( 2\right) }_{\pm 2},-r^{\left( 2\right) }_{\pm 1},r^{\left( 2\right) }_{0}\right) 
\end{eqnarray*}
Notice the spherical quadrupole operators don't notice the difference between
the mirror reflection and rotation representations of a spin flip.

This generality in the azimuthal angle of the rotation for reflection direction
isn't particularly useful for the immediate purposes of studying the parity
light shift matrix elements, but they will prove useful later when exploited
to implement other transformations and demonstrate other symmetries. For example,
notice that for \( \phi _{0}=\pi /4 \), \( R_{\hat{n}} \) just exchanges \( x \)
and \( y \) coordinates, \( R_{\hat{n}}^{\dagger }\left( x,y\right) R_{\hat{n}}=\left( y,x\right)  \)
allowing for an easy means of relating \( x \) and \( y \) matrix elements.

The transformation of the all the spherical operators can be written very compactly,

\begin{eqnarray*}
Fr^{\left( k\right) }_{\pm s}F^{\dagger } & = & \eta _{F,k}e^{\mp 2si\phi _{0}}r^{\left( k\right) }_{\mp s}\\
F_{\hat{z}}r^{\left( k\right) }_{\pm s}F_{\hat{z}}^{\dagger } & = & \eta _{F,k}\left( -1\right) ^{s}r^{\left( k\right) }_{\pm s}
\end{eqnarray*}

looking very much like the transformation of states,
\[
F\left| j,m\right\rangle =\left| j,-m\right\rangle \eta _{F,j}e^{2im\phi _{0}}\]
Where again \( \eta _{M,j}=\eta _{j}e^{ij\pi } \) and \( \eta _{R,j}=e^{-ij\pi } \)
for the states, and simplifies to \( \eta _{M}=1 \), \( \eta _{R,k}=\left( -1\right) ^{k} \)
for operators. (Notation is little loose, \( j \) is a label as much as it
is and index or argument.)

\subsection{Matrix Elements of Spherical Operators}

These compact results make it very easy to finally determine the spin flip properties
of couplings,
\[
\Omega _{m_{1}m_{2}}^{\left( k\right) }=\left\langle j_{1},m_{1}\right| r^{\left( k\right) }_{s}\left| j_{2},m_{2}\right\rangle E^{\left( k\right) }_{s}\]
through the spin flip properties of the matrix elements, with \( m_{1}=m_{2}+s \)
and \( j_{1}-j_{2},k \) integers,
\begin{eqnarray*}
\left\langle j_{1},m_{1}\right| r^{\left( k\right) }_{s}\left| j_{2},m_{2}\right\rangle  & = & \left\langle j_{1},m_{1}\right| F^{\dagger }Fr^{\left( k\right) }_{s}F^{\dagger }F\left| j_{2},m_{2}\right\rangle \\
 & = & \eta _{F}^{j_{1}*}e^{-2im_{1}\phi _{0}}\left\langle j_{1},-m_{1}\right| \eta _{F}^{k}e^{-2is\phi _{0}}r^{\left( k\right) }_{-s}\left| j_{2},-m_{2}\right\rangle \eta _{F}^{j_{2}}e^{2im_{2}\phi _{0}}\\
 & = & \eta _{F,j_{1}}^{*}\eta _{F,j_{2}}\eta _{F,k}e^{-2i\left( m_{1}-m_{2}+s\right) \phi _{0}}\left\langle j_{1},-m_{1}\right| r^{\left( k\right) }_{-s}\left| j_{2},-m_{2}\right\rangle 
\end{eqnarray*}

This result should be independent of the representation used for \( F \), so
in particular is should be independent of the \( \phi _{0} \) which appears
explicitly in an overall phase. Consistency then requires that either the entire
result is zero, or the phase is zero independent of \( \phi _{0} \) because
\( m_{1}-m_{2}-s=0 \) or \( m_{1}=m_{2}+s \). This gives the usual \( m \)
selection rules. For the rest of the coefficients pick various particular representations
for \( F \). Using \( F=R_{\hat{n}}\left( \pi \right)  \) gives \( \eta _{F,j_{1}}^{*}\eta _{F,j_{2}}\eta _{F,k}=e^{-i\left( j_{1}-j_{2}\right) \pi }\left( -1\right) ^{k} \),
for this application \( j_{1} \) and \( j_{2} \) are both integer or half
integer so the \( j_{1}-j_{2} \) is an integer and the product of \( \eta  \)'s
gives \( \left( -1\right) ^{j_{1}-j_{2}+k} \) reproducing the result obtained
from the Wigner-Eckhart theorem and the spin flip transformation of the Clebsch-Gordan
coefficients. For \( F=M_{\hat{n}} \), \( \eta _{F,j_{1}}^{*}\eta _{F,j_{2}}\eta _{F,k}=\eta _{j_{1}}\eta _{j_{2}}e^{-i\left( j_{1}-j_{2}\right) \pi }=\eta _{j_{1}}\eta _{j_{2}}\left( -1\right) ^{j_{1}-j_{2}} \).
Constancy then requires \( \left( -1\right) ^{j_{1}-j_{2}+k}=\eta _{j_{1}}\eta _{j_{2}}\left( -1\right) ^{j_{1}-j_{2}} \)
yielding the selection rule, \( \eta _{j_{1}}\eta _{j_{2}}=\left( -1\right) ^{k} \)
which becomes the familiar parity selection rule, 

\begin{eqnarray*}
\left\langle j_{1},m_{1}\right| r^{\left( k\right) }_{s}\left| j_{2},m_{2}\right\rangle  & = & \left\langle j_{1},m_{1}\right| P^{\dagger }Pr^{\left( k\right) }_{s}P^{\dagger }P\left| j_{2},m_{2}\right\rangle \\
 & = & \eta _{j_{1}}\eta _{j_{2}}(-1)^{k}\left\langle j_{1},m_{1}\right| r^{\left( k\right) }_{s}\left| j_{2},m_{2}\right\rangle 
\end{eqnarray*}
When considering the composition of these states with \( \eta _{j}=\eta _{l}\eta _{s} \).
The \( s \) is the same for both states so that \( \eta _{j_{1}}\eta _{j_{2}}=\eta _{l_{1}}\eta _{s}\eta _{l_{2}}\eta _{s}=\eta _{l_{1}}\eta _{l_{2}} \)
and requiring \( \eta _{l_{1}}\eta _{l_{2}}=\left( -1\right) ^{k} \).

The entire coupling then changes to,
\begin{eqnarray*}
\Omega _{-m_{1},-m_{2}}^{\left( k\right) } & = & \left\langle j_{1},-m_{1}\right| r^{\left( k\right) }_{s}\left| j_{2},-m_{2}\right\rangle E^{\left( k\right) }_{s}\\
 & = & \left\langle j_{1},-m_{1}\right| r^{\left( k\right) }_{-s}\left| j_{2},-m_{2}\right\rangle E^{\left( k\right) }_{-s}\\
 & = & \left( -1\right) ^{j_{1}-j_{2}+k}\left\langle j_{1},m_{1}\right| r^{\left( k\right) }_{s}\left| j_{2},m_{2}\right\rangle E^{\left( k\right) }_{-s}
\end{eqnarray*}
The \( E_{\pm s}^{\left( k\right) } \) are generally independent, but their
structures are related in a way that allow further simplification. These amplitudes
are constructed out of a symmetric piece, \( E_{sS}^{\left( k\right) } \),
that is the same for both \( \pm s \) and and anti symmetric piece that changes
sign with \( s \), \( E_{sA}^{\left( k\right) } \), as
\begin{eqnarray*}
E_{s>0}^{\left( k\right) } & = & E_{\left| s\right| S}^{\left( k\right) }+E_{\left| s\right| A}^{\left( k\right) }\\
E_{s<0}^{\left( k\right) } & = & E_{\left| s\right| S}^{\left( k\right) }-E_{\left| s\right| A}^{\left( k\right) }
\end{eqnarray*}
For general \( s \) this can be written,

\[
E^{\left( k\right) }_{\pm s}=E^{\left( k\right) }_{\left| s\right| S}+\frac{\left| s\right| }{s}E^{\left( k\right) }_{\left| s\right| A}\]
Dropping the explicit absolute values in the symmetric and antisymmetric parameters,
and understanding them to be implied in this context so that \( E_{s\left\{ S,A\right\} }^{\left( k\right) } \)
is not distinct from \( E_{-s\left\{ S,A\right\} }^{\left( k\right) } \), the
original and spin flipped couplings are given by,
\begin{eqnarray*}
\Omega _{m_{1},m_{2}}^{\left( k\right) } & = & \left\langle j_{1},m_{1}\right| r^{\left( k\right) }_{s}\left| j_{2},m_{2}\right\rangle \left( E^{\left( k\right) }_{sS}+\frac{\left| s\right| }{s}E^{\left( k\right) }_{sA}\right) \\
\Omega _{-m_{1},-m_{2}}^{\left( k\right) } & = & \left( -1\right) ^{j_{1}-j_{2}+k}\left\langle j_{1},m_{1}\right| r^{\left( k\right) }_{s}\left| j_{2},m_{2}\right\rangle \left( E^{\left( k\right) }_{sS}-\frac{\left| s\right| }{s}E^{\left( k\right) }_{sA}\right) 
\end{eqnarray*}
For arbitrary fields, these parts of the field amplitudes are given by,
\[
\begin{array}{cc}
E^{D}_{1S}=\frac{iE_{y}}{\sqrt{2}} & E_{1A}^{D}=-\frac{E_{x}}{\sqrt{2}}\\
E_{0S}^{D}=E_{z} & E_{0A}^{D}=0
\end{array}\]
and
\[
\begin{array}{cc}
E^{Q}_{2S}=\frac{\partial _{x}E_{x}+\partial _{y}E_{y}}{2} & E_{2A}^{Q}=-i\frac{\partial _{x}E_{y}+\partial _{y}E_{x}}{2}\\
E^{Q}_{1S}=i\frac{\partial _{y}E_{z}+\partial _{z}E_{y}}{2} & E_{1A}^{Q}=-\frac{\partial _{x}E_{z}+\partial _{z}E_{x}}{2}\\
E_{0S}^{Q}=\frac{\partial _{z}E_{z}}{\sqrt{6}} & E_{0A}^{Q}=0
\end{array}\]

\subsection{Matrix Elements of Cartesian Operators}

This form isn't any more practically useful than the corresponding transformation
of the Clebsch-Gordan coefficient because the transformation changed the operator.
The spin flipped matrix element is related to a matrix element of a different
operator. To the determine the behavior of the entire coupling then requires
that the field amplitudes and their transformations are also known. A more insightful
picture would emerge if the transformation that changed the state, left the
operator invariant. This can be done by returning to cartesian coordinates.

With \( x=-\left( r_{+}-r_{-}\right) /\sqrt{2} \), \( y=\left( r_{+}+r_{-}\right) /\sqrt{2}i \),
\( z=r_{0} \),
\begin{eqnarray*}
\left\langle j_{1},-m_{1}\right| \left\{ x,y,z\right\} \left| j_{2},-m_{2}\right\rangle  & = & \left\langle j_{1},-m_{1}\right| \left\{ -\frac{r_{+}-r_{-}}{\sqrt{2}},\frac{r_{+}+r_{-}}{\sqrt{2}i},r_{0}\right\} \left| j_{2},-m_{2}\right\rangle \\
 & = & \left( -1\right) ^{j_{1}-j_{2}+1}\\
 & \times  & \left\langle j_{1},m_{1}\right| \left\{ -\frac{r_{-}-r_{+}}{\sqrt{2}},\frac{r_{-}+r_{+}}{\sqrt{2}i},r_{0}\right\} \left| j_{2},m_{2}\right\rangle \\
 & = & \left( -1\right) ^{j_{1}-j_{2}+1}\\
 & \times  & \left\langle j_{1},m_{1}\right| \left\{ -x,y,z\right\} \left| j_{2},m_{2}\right\rangle 
\end{eqnarray*}
Matrix elements of \( x \) and \( y \) operators are simply related to the
spin flipped matrix element of the same operator, and \( x \) and \( y \)
transform the same up to a relative sign. 

This demonstrates this tidy result in way that is independent of the representation
used to the spin flip operator, but still requires the use of the spherical
tensors, and an intuitive understanding of this simple result is slightly obscured.
By picking a few particular spin flip representations a simple geometric picture
emerges and the derivation completely avoids a spherical basis.

For \( \phi _{0}=0 \), or \( \phi _{0}=\pi /2 \), the cartesian coordinate
operators transform very simply,
\begin{eqnarray*}
F_{\phi _{0}=0}\left( x,y,z\right) F^{\dagger }_{\phi _{0}=0} & = & \eta _{F,1}\left( -x,y,z\right) \\
F_{\phi _{0}=\pi /2}\left( x,y,z\right) F^{\dagger }_{\phi _{0}=\pi /2} & = & \eta _{F,1}\left( x,-y,z\right) 
\end{eqnarray*}
For these transformations, the coordinate operators change by, at most, a sign
and it becomes trivial to see that the spin flipped matrix elements involve
the same operator, because the transformation doesn't change the operator. Repeating
the calculation of the spin flipped matrix elements using cartesian coordinates,
\begin{eqnarray*}
\left\langle j_{1},-m_{1}\right| \left\{ x,y,z\right\} \left| j_{2},-m_{2}\right\rangle  & = & \left\langle j_{1},-m_{1}\right| F^{\dagger }F\left\{ x,y,z\right\} F^{\dagger }F\left| j_{2},-m_{2}\right\rangle \\
 & = & \eta _{F,j_{1}}^{*}e^{-2im_{2}\phi _{0}}\\
 & \times  & \left\langle j_{1},m_{1}\right| F\left\{ x,y,z\right\} F^{\dagger }\left| j_{2},m_{2}\right\rangle \eta _{F,j_{2}}e^{2im_{2}\phi _{0}}
\end{eqnarray*}
 If \( \phi _{0} \) is now restricted to \( 0 \) or \( \pi /2 \) , the transformation
is,
\[
F\left\{ x,y,z\right\} F^{\dagger }=\eta _{F}\left\{ -cos\left( 2\phi _{0}\right) x,cos\left( 2\phi _{0}\right) y,z\right\} \]
the matrix element simplifies to,
\begin{eqnarray*}
\left\langle j_{1},-m_{1}\right| \left\{ x,y,z\right\} \left| j_{2},-m_{2}\right\rangle  & = & \eta _{F,j_{1}}^{*}\eta _{F,j_{2}}\eta _{F,1}e^{-2i\left( m_{1}-m_{2}\right) \phi _{0}}\\
 & \times  & \left\langle j_{1},m_{1}\right| \left\{ -cos\left( 2\phi _{0}\right) x,cos\left( 2\phi _{0}\right) y,z\right\} \left| j_{2},m_{2}\right\rangle 
\end{eqnarray*}
For \( \phi _{0}=0 \), the \( m \) dependent phases are all zero, for \( \phi _{0}=\pi /2 \),
\( cos\left( 2\phi _{0}\right) =-1 \), and \( e^{-2i\left( m_{1}-m_{2}\right) \phi }=e^{-i\left( m_{1}-m_{2}\right) \pi }=\left( -1\right) ^{m_{1}-m_{2}} \).
For consistency, \( -\left( -1\right) ^{m_{1}-m_{2}}=1 \) requires \( m_{1}-m_{2} \)
odd. These two cases can also restrict \( m_{1}-m_{2} \) to \( \pm 1 \), the
general transformation would yield the complete familiar \( m \) selection
rules as they appeared when considering the matrix elements of the spherical
tensors.

For the product of \( \eta  \) coefficients, \( F=R_{\hat{n}}\left( \pi \right)  \)
gives the same result as from writing these cartesian operators in terms of
their spherical tensor counterparts,

\[
\left\langle j_{1},-m_{1}\right| \left\{ x,y,z\right\} \left| j_{2},-m_{2}\right\rangle =-\left( -1\right) ^{j_{1}-j_{2}}\left\langle j_{1},m_{1}\right| \left\{ -x,y,z\right\} \left| j_{2},m_{2}\right\rangle \]
As usual, using \( F=M_{\hat{n}} \) gives the same results if the parity selection
rules are satisfied.

This path completely avoids the spherical tensor operators and, as one result,
make it trivial to extend to higher order operators. In particular, quadrupole
operators will get another \( \eta _{F,1} \), use \( \eta _{R,1}=-1 \), and
the final sign just depends on the numbers of \( x \) coordinates, \( n_{x} \),
used in the operator,
\[
\left\langle j_{1},-m_{1}\right| x_{i}x_{j}\left| j_{2},-m_{2}\right\rangle =\left( -1\right) ^{2}\left( -1\right) ^{n_{x}}\left( -1\right) ^{j_{1}-j_{2}}\left\langle j_{1},m_{1}\right| x_{i}x_{j}\left| j_{2},m_{2}\right\rangle \]
Obviously this trivially extends to matrix elements of products of any number
of coordinates,
\begin{eqnarray*}
\left\langle j_{1},-m_{1}\right| x^{n_{x}}y^{n_{y}}z^{n_{z}}\left| j_{2},-m_{2}\right\rangle  & = & \left( -1\right) ^{n_{x}+n_{y}+n_{z}}\left( -1\right) ^{n_{x}}\left( -1\right) ^{j_{1}-j_{2}}\\
 & \times  & \left\langle j_{1},m_{1}\right| x^{n_{x}}y^{n_{y}}z^{n_{z}}\left| j_{2},m_{2}\right\rangle \\
 & = & \left( -1\right) ^{n_{y}+n_{z}}\left( -1\right) ^{j_{1}-j_{2}}\left\langle j_{1},m_{1}\right| x^{n_{x}}y^{n_{y}}z^{n_{z}}\left| j_{2},m_{2}\right\rangle 
\end{eqnarray*}

\subsection{Products of Matrix Elements and Selection Rules}

The intended immediate application of the transformations constructed here is
to be able to quickly, intuitively evaluate the spin flip properties of various
energy shifts. These shifts are the sums of diagonal elements of products of
coupling matrices like
\[
\Omega ^{\left( k_{1}\right) \dagger }_{m,m'}\Omega ^{\left( k_{2}\right) }_{m',m}\]
where a particular coupling is given in a spherical basis by
\[
\Omega _{mm'}^{\left( k\right) }=\left\langle j_{1},m\left| T^{\left( k\right) }_{s}\right| j_{2},m'\right\rangle E^{\left( k\right) }_{s}\]
and in a cartesian basis will involve terms like
\[
\Omega _{mm'}^{\left( k\right) }=\left\langle j_{1},m\left| x^{n_{x}}y^{n_{y}}z^{n_{z}}\right| j_{2},m'\right\rangle E^{\left( k\right) }_{i_{1}i_{2}\cdots i_{k}}\]
with \( n_{x}+n_{y}+n_{z}=k \). For the parity experiment,
\begin{eqnarray*}
\delta \omega ^{Q}_{m} & = & \sqrt{\left( \Omega ^{Q\dagger }\Omega ^{Q}\right) _{mm}}\\
\delta \omega ^{PNC}_{m} & = & \left( \left( \Omega ^{Q\dagger }\Omega ^{D}\right) _{mm}-\left( \Omega ^{D\dagger }\Omega ^{Q}\right) _{mm}\right) /\delta \omega _{m}^{Q}=Im\left( \left( \Omega ^{Q\dagger }\Omega ^{D}\right) _{mm}\right) /\delta \omega _{m}^{Q}
\end{eqnarray*}

For these sorts of quantities, some of the generality considered, the action
of an arbitrary spin flip on a state, becomes moot. For the states, the most
important property of a spin flip operator is that it is unitary, so that the
amplitude of the transformed state differs by at most a phase, \( F\left| j,m\right\rangle =\left| j,-m\right\rangle e^{if\left( j,m\right) } \).
In the shifts, both of the states involved appear once each as initial and final
states so that these possible phases introduced by a particular spin flip operator
exactly cancel. Implicitly summing over the intermediate spin orientations \( m' \),
\begin{eqnarray*}
\left( \Omega ^{\left( k_{1}\right) }\Omega ^{\left( k_{2}\right) }\right) _{-m,-m} & = & \Omega ^{\left( k_{1}\right) }_{-m,m'}\Omega ^{\left( k_{2}\right) }_{m',-m}\\
 & = & \Omega ^{\left( k_{1}\right) }_{-m,-m'}\Omega ^{\left( k_{2}\right) }_{-m',-m}\\
 & = & \left\langle j_{1},-m\left| T^{\left( k_{1}\right) \dagger }_{r}\right| j_{2},-m'\right\rangle \left\langle j_{2},-m'\left| T^{\left( k_{2}\right) }_{s}\right| j_{1},-m\right\rangle E^{\left( k_{1}\right) *}_{r}E^{\left( k_{2}\right) }_{s}\\
 & = & e^{if\left( j_{1},m\right) }\left\langle j_{1},m\left| F^{\dagger }T^{\left( k_{1}\right) \dagger }_{r}F\right| j_{2},m'\right\rangle e^{-if\left( j_{2},m'\right) }\\
 & \times  & e^{if\left( j_{2},m'\right) }\left\langle j_{2},m'\left| F^{\dagger }T^{\left( k_{2}\right) }_{s}F\right| j_{1},m\right\rangle e^{-if\left( j_{1},m\right) }E^{\left( k_{1}\right) *}_{r}E^{\left( k_{2}\right) }_{s}\\
 & = & \left\langle j_{1},m\left| F^{\dagger }T^{\left( k_{1}\right) \dagger }_{r}F\right| j_{2},m'\right\rangle \left\langle j_{2},m'\left| F^{\dagger }T^{\left( k_{2}\right) }_{s}F\right| j_{1},m\right\rangle E^{\left( k_{1}\right) *}_{r}E^{\left( k_{2}\right) }_{s}
\end{eqnarray*}
The transformation symmetries apply term by term to each matrix element, so
the sum over \( r \), \( s \) and even \( m' \) need not be done for these
expressions to hold, except for the initial changing the sign of \( m' \) as
a summation index, so it it generally not necessary to point out when a sum
over a certain index is implied or just not done. Certainly all the sums must
be done to get the desired shift, but the relationships that follow will apply
to each term individually as well.

\subsubsection{Spin Dependence in a Spherical Basis}

For these spherical tensor operators,
\[
Fr^{\left( k\right) }_{\pm s}F^{\dagger }=\eta _{F,k}e^{\mp 2si\phi _{0}}r^{\left( k\right) }_{\mp s}\]
 The matrix elements are then related by,
\begin{eqnarray*}
\left( \Omega ^{k_{1}\dagger }\Omega ^{k_{2}}\right) _{-m,-m} & = & \eta _{F,k_{1}}^{*}\eta _{F,k_{2}}e^{2i\left( r-s\right) \phi _{0}}\\
 & \times  & \left\langle j_{1},m\left| T^{\left( k_{1}\right) \dagger }_{-r}\right| j_{2},m'\right\rangle \left\langle j_{2},m'\left| T^{\left( k_{2}\right) }_{-s}\right| j_{1},m\right\rangle E^{\left( k_{1}\right) *}_{r}E^{\left( k_{2}\right) }_{s}\\
 & = & \eta _{F,k_{1}}^{*}\eta _{F,k_{2}}e^{2i\left( r-s\right) \phi _{0}}\\
 & \times  & \left\langle j_{1},m\left| T^{\left( k_{1}\right) \dagger }_{r}\right| j_{2},m'\right\rangle \left\langle j_{2},m'\left| T^{\left( k_{2}\right) }_{s}\right| j_{1},m\right\rangle E^{\left( k_{1}\right) *}_{-r}E^{\left( k_{2}\right) }_{-s}
\end{eqnarray*}
It was already demonstrated the this result is independent of the representation
used for \( F \) because it was shown that the transformation of each matrix
element is independent of the \( F \) used. In that case the phases generated
from transforming the states cancel the phase generated from transforming the
operator. Here the phases from the states were canceled first, so it must be
that a non zero result here requires the phases from the operators cancel each
other. The \( E_{s}^{\left( k_{1},k_{2}\right) } \) can be be chosen independently
so that these phases must cancel term by term in the \( r \), \( s \) sum.
This simply requires \( r=s \) and the sum collapses to one over a single index.
This is again clear from the familiar \( m \) selection rules. \( T^{\left( k_{1}\right) }_{s} \)
will raise the \( m \) of the state is is operating on by \( s \), so to return
to the same initial state with another transformation requires a \( T^{\left( k_{1}\right) }_{-s} \)
or \( T^{\left( k_{2}\right) \dagger }_{s} \). Using \( F=R_{\hat{n}}\left( \pi \right)  \)
to evaluate the remaining coefficients, \( \eta _{R,k}=\left( -1\right) ^{k} \),
gives \( \eta ^{*}_{R,k_{1}}\eta _{R,k_{2}}=\left( -1\right) ^{k_{1}+k_{2}} \)
and with \( F=M_{\hat{n}} \), \( \eta _{F,k}=1 \) which requires \( \left( -1\right) ^{k_{1}+k_{2}}=1 \)
or \( \left( -1\right) ^{k_{1}}=\left( -1\right) ^{k_{2}} \), the interactions
must have the same parity to couple the same states. This leaves,
\[
\left( \Omega ^{k_{1}\dagger }\Omega ^{k_{2}}\right) _{-m,-m}=\left\langle j_{1},m\left| T^{\left( k_{1}\right) \dagger }_{s}\right| j_{2},m'\right\rangle \left\langle j_{2},m'\left| T^{\left( k_{2}\right) }_{s}\right| j_{1},m\right\rangle E^{\left( k_{1}\right) *}_{-s}E^{\left( k_{2}\right) }_{-s}\]

\subsubsection{Off-diagonal Elements}

This can easily be extended to other matrix elements of products of couplings
which will prove to be useful when considering systematics. Generalizing to
\( \left( \Omega ^{k_{1}\dagger }\Omega ^{k_{2}}\right) _{m_{1},m_{2}}=\sum _{m'}\Omega ^{\left( k_{1}\right) }_{m_{1},m'}\Omega _{m_{2},m'}^{\left( k_{2}\right) } \).In
this case the initial and final states are different to the phases no longer
exactly cancel,
\begin{eqnarray*}
\left( \Omega ^{k_{1}\dagger }\Omega ^{k_{2}}\right) _{-m_{1},-m_{2}} & = & \Omega ^{\left( k_{1}\right) }_{-m_{1},m'}\Omega ^{\left( k_{2}\right) }_{m',-m_{2}}=\Omega ^{\left( k_{1}\right) }_{-m,-m'}\Omega ^{\left( k_{2}\right) }_{-m',-m}\\
 & = & \left\langle j_{1},-m_{1}\left| T^{\left( k_{1}\right) \dagger }_{r}\right| j_{2},-m'\right\rangle \left\langle j_{2},-m'\left| T^{\left( k_{2}\right) }_{s}\right| j_{1},-m_{2}\right\rangle E^{\left( k_{1}\right) *}_{r}E^{\left( k_{2}\right) }_{s}\\
 & = & \eta _{F,j_{1}}e^{2im_{1}\phi _{0}}\eta ^{*}_{F,j_{1}}e^{-2im_{2}\phi _{0}}E^{\left( k_{1}\right) *}_{r}E^{\left( k_{2}\right) }_{s}\\
 & \times  & \left\langle j_{1},m_{1}\left| F^{\dagger }T_{r}^{\left( k_{1}\right) \dagger }F\right| j_{2},m'\right\rangle \left\langle j_{2},m'\left| F^{\dagger }T^{\left( k_{2}\right) }_{s}F\right| j_{1},m_{2}\right\rangle \\
 & \times  & \left| \eta _{F,j_{1}}\right| ^{2}e^{2i\left( m_{1}-m_{2}\right) \phi _{0}}e^{-2i\left( r-s\right) \phi _{0}}E^{\left( k_{1}\right) *}_{r}E^{\left( k_{2}\right) }_{s}\\
 & \times  & \left( \eta _{F,k_{1}}^{*}\eta _{F,k_{2}}\right) \left\langle j_{1},m_{1}\left| T^{\left( k_{1}\right) \dagger }_{-r}\right| j_{2},m'\right\rangle \left\langle j_{2},m'\left| T^{\left( k_{2}\right) }_{-s}\right| j_{1},m_{2}\right\rangle 
\end{eqnarray*}
The \( \left| \eta \right| ^{2} \) gives 1, and \( \phi _{0} \) independence
then requires that \( m_{1}-m_{2}=s-r \) for non-zero terms, consistent with
the usual \( m \) selection rules, and
\begin{eqnarray*}
\left( \Omega ^{k_{1}\dagger }\Omega ^{k_{2}}\right) _{-m_{1},-m_{2}} & = & \eta _{F,k_{1}}^{*}\eta _{F,k_{2}}E^{\left( k_{1}\right) *}_{-r}E^{\left( k_{2}\right) }_{-s}\\
 & \times  & \left\langle j_{1},m_{1}\left| T^{\left( k_{1}\right) \dagger }_{r}\right| j_{2},m'\right\rangle \left\langle j_{2},m'\left| T^{\left( k_{2}\right) }_{s}\right| j_{1},m_{2}\right\rangle 
\end{eqnarray*}

\subsubsection{Different Initial/Final or Intermediate States}

One more generalization must be made to use this for the parity shifts, \( \delta \omega ^{PNC} \)
involves a product of matrix elements between different initial states. The
ground state is a mixture of \( S \) and \( P \) states. For the PNC light
shift term the dipole transition is from the \( P \) component of this state
to the \( D \) state, while the quadrupole transition is from the \( S \)
component. Then for calculations of the ground state shift the initial and final
state are no longer the same, and for calculations of \( D \) state shifts
the intermediate states are no longer the same. The \( j \)'s in each case
are still the same so for \( F=R \) the \( \eta _{R,j}=e^{-ij\pi } \) coefficients
are unchanged and cancel between the pairs, leaving the same \( \eta _{R,k_{1}}\eta _{R,k_{2}}=\left( -1\right) ^{k_{1}+k_{2}} \)
in the end. For \( F=M \), \( \eta _{M,\left\{ j,l\right\} }=\eta _{\left\{ j,l\right\} }e^{ij\pi } \)
and \( \eta _{R,k}=1 \). The parities of each of a pair of states are no longer
equal because the \( l \) of one of the states is different, leaving the factors
\( \eta _{\left\{ j_{11},l_{11}\right\} }\eta _{\left\{ j_{12},l_{12}\right\} } \)
and \( \eta _{\left\{ j_{21},l_{21}\right\} }\eta _{\left\{ j_{22},l_{22}\right\} } \).
For these composite states \( \eta _{\left\{ j,l\right\} }=\eta _{l}\eta _{s} \).
The \( \eta _{s} \) is the same for all state so that what remains is \( \eta _{l_{11}}\eta _{l_{12}}\eta _{l_{21}}\eta _{l_{22}} \)
and the parity selection rules become the same as those for the individual matrix
elements,\( \eta _{l_{11}}\eta _{l_{12}}\eta _{l_{21}}\eta _{l_{22}}=\left( -1\right) ^{k_{1}+k_{2}} \)
which can also be written \( \eta _{l_{1i}}\eta _{l_{2i}}=\left( -1\right) ^{k_{i}} \).
This is just what is needed for a nonzero shift from a product of dipole and
quadrupole couplings. 

With this selection rule satisfied the spin flip relation becomes, in general,
also changing the sign of the \( r \), \( s \) summation indices,

\begin{eqnarray*}
\left( \Omega ^{k_{1}\dagger }\Omega ^{k_{2}}\right) _{-m_{1},-m_{2}} & = & \left( -1\right) ^{k_{1}+k_{2}}E^{\left( k_{1}\right) *}_{-r}E^{\left( k_{2}\right) }_{-s}\\
 & \times  & \left\langle j_{1},l_{11},m_{1}\left| T^{\left( k_{1}\right) \dagger }_{r}\right| j_{2},l_{21},m'\right\rangle \left\langle j_{2},l_{22},m'\left| T^{\left( k_{2}\right) }_{s}\right| j_{1},l_{12},m_{2}\right\rangle 
\end{eqnarray*}
The overall relationship is very simple, the spin flipped shift is given by
the same sum with an overall \( \left( -1\right) ^{k_{1}+k_{2}} \) and the
sign of the spin indices of the tensor field amplitudes changes. This becomes,
for \( m_{1}=m_{2} \), as before,
\begin{eqnarray*}
\left( \Omega ^{k_{1}\dagger }\Omega ^{k_{2}}\right) _{-m,-m} & = & \left( -1\right) ^{k_{1}+k_{2}}E^{\left( k_{1}\right) *}_{-s}E^{\left( k_{2}\right) }_{-s}\\
 & \times  & \left\langle j_{1},l_{11},m\left| T^{\left( k_{1}\right) \dagger }_{s}\right| j_{2},l_{21},m'\right\rangle \left\langle j_{2},l_{22},m'\left| T^{\left( k_{2}\right) }_{s}\right| j_{1},l_{12},m\right\rangle 
\end{eqnarray*}

\subsubsection{Linearly Polarized Light}

These spin flip symmetries apply individually to each term in the sums needed
to get the energy shifts, so in this spherical tensor form there is not yet
an easy way to relate the entire shifts unless the field amplitudes are related
in a particular way, such as \( E^{\left( k\right) }_{-s}=\pm E^{\left( k\right) *}_{s} \)
as is the case for the particular choice of fields discussed for use in this
PNC experiment. As a more general example, this easily simplifies for the special
case of linear polarization. In this case all the polarization components of
\( \vec{E} \) have the same relative phase so that the field can be written
at \( e^{i\phi }\vec{E} \), where the polarization components are now all real.
For a single field this phase can be removed by shifting the time origin, but
for the general case of two independent fields, both phases can not be eliminated
simultaneously and a relative phase factor remains. 

Factoring out the overall relative phase as \( \delta \omega =e^{i\Delta \phi }\delta \bar{\omega } \),
the polarization components of both fields can be made real and for this case
the spherical tensor amplitudes satisfy \( e^{i\phi }E_{-s}^{\left( k\right) }=e^{i\phi }\left( -1\right) ^{s}E_{s}^{\left( k\right) *} \).
The signs of the \( r \), \( s \) spin indices of the field amplitudes \( E^{\left( k\right) }_{-s} \)
in the expression for the spin flipped shift can now be easily changed back
to positive giving,
\begin{eqnarray*}
\delta \bar{\omega }^{\left( k_{1},k_{2}\right) }_{-m_{1},-m_{2}} & = & \left( -1\right) ^{k_{1}+k_{2}}\left( -1\right) ^{r}E^{\left( k_{1}\right) }_{r}\left( -1\right) ^{s}E^{\left( k_{2}\right) *}_{s}\\
 & \times  & \left\langle j_{1},m_{1}\left| T^{\left( k_{1}\right) \dagger }_{r}\right| j_{2},m'\right\rangle \left\langle j_{2},m'\left| T^{\left( k_{2}\right) }_{s}\right| j_{1},m_{2}\right\rangle 
\end{eqnarray*}
The total sign change is \( \left( -1\right) ^{r+s}=\left( -1\right) ^{r-s} \).
The selection rules require that non-zero terms will have \( m_{1}-m_{2}=s-r \)
so the sign can be written \( \left( -1\right) ^{m_{1}-m_{2}} \), 
\begin{eqnarray*}
\delta \bar{\omega }^{\left( k_{1},k_{2}\right) }_{-m_{1},-m_{2}} & = & \left( -1\right) ^{m_{1}-m_{2}}\left( -1\right) ^{k_{1}+k_{2}}E^{\left( k_{1}\right) }_{r}E^{\left( k_{2}\right) *}_{s}\\
 & \times  & \left\langle j_{1},m_{1}\left| T^{\left( k_{1}\right) \dagger }_{r}\right| j_{2},m'\right\rangle \left\langle j_{2},m'\left| T^{\left( k_{2}\right) }_{s}\right| j_{1},m_{2}\right\rangle \\
 & = & \left( -1\right) ^{m_{1}-m_{2}}\left( -1\right) ^{k_{1}+k_{2}}\left( E^{\left( k_{1}\right) *}_{r}E^{\left( k_{2}\right) }_{s}\right) ^{*}\\
 & \times  & \left\langle j_{1},m_{1}\left| T^{\left( k_{1}\right) \dagger }_{r}\right| j_{2},m'\right\rangle \left\langle j_{2},m'\left| T^{\left( k_{2}\right) }_{s}\right| j_{1},m_{2}\right\rangle 
\end{eqnarray*}
The matrix elements of these spherical tensor operators are real, all imaginary
components are contained in the field amplitudes so the complex conjugate can
be taken over the whole term and the sum over \( r \) and \( s \) done again
to recover, simply,
\begin{eqnarray*}
\delta \bar{\omega }^{\left( k_{1},k_{2}\right) }_{-m_{1},-m_{2}} & = & \left( -1\right) ^{m_{1}-m_{2}}\left( -1\right) ^{k_{1}+k_{2}}E^{\left( k_{1}\right) *}_{r}E^{\left( k_{2}\right) }_{s}\\
 & \times  & \left\langle j_{1},m_{1}\left| T^{\left( k_{1}\right) \dagger }_{r}\right| j_{2},m'\right\rangle \left\langle j_{2},m'\left| T^{\left( k_{2}\right) }_{s}\right| j_{1},m_{2}\right\rangle )^{*}\\
 & = & \left( -1\right) ^{m_{1}-m_{2}}\left( -1\right) ^{k_{1}+k_{2}}\delta \bar{\omega }_{m_{1},m_{2}}^{\left( k_{1},k_{2}\right) *}
\end{eqnarray*}
Replacing the relative phases gives,
\[
\delta \omega ^{\left( k_{1},k_{2}\right) }_{-m}=\left( -1\right) ^{k_{1}+k_{2}}\delta \omega _{m}^{\left( k_{1},k_{2}\right) }\]
 for \( m_{1}=m_{2} \) this becomes simply
\[
\delta \omega ^{\left( k_{1},k_{2}\right) }_{-m}=\left( -1\right) ^{k_{1}+k_{2}}\delta \omega _{m}^{\left( k_{1},k_{2}\right) }\]

This gives a trivial demonstration of the existence of the parity light shift,
as discussed later, but only for the ideal case. In considering systematic errors
more general fields must be considered, and one final step yields some powerful
tools that can be used to understand the generalities.

\subsubsection{Spin Dependence in Cartesian Basis}

A bit more can be learned about the spin flip properties of the entire light
shift before having to specify anything about the fields if the analogous transformations
are determined for cartesian operators. For the spherical tensors, the fundamental
complication, as observed before, is that any transformation that flips the
spin of a state also, necessarily, changes the operator. In cartesian coordinates,
however, there are a handful of special transformations that will only, at most,
change the sign of a cartesian operator it acts on. This are simply rotations
or reflections having \( \hat{n} \) parallel to either \( \hat{x} \) or \( \hat{y} \).
These give
\begin{eqnarray*}
F_{\hat{x}}\left\{ x,y,z\right\} F^{\dagger }_{\hat{x}}=F_{\phi _{0}=0}\left\{ x,y,z\right\} F^{\dagger }_{\phi _{0}=0} & = & \eta _{F,1}\left\{ -x,y,z\right\} \\
F_{\hat{y}}\left\{ x,y,z\right\} F^{\dagger }_{\hat{y}}=F_{\phi _{0}=\pi /2}\left\{ x,y,z\right\} F^{\dagger }_{\phi _{0}=\pi /2} & = & \eta _{F,1}\left\{ x,-y,z\right\} 
\end{eqnarray*}
The returning to the shifts, and including the previously discussed generalities,
\begin{eqnarray*}
\delta \omega ^{\left( k_{1},k_{2}\right) }_{-m_{1},-m_{2}} & = & \Omega ^{\left( k_{1}\right) }_{-m_{1},m'}\Omega ^{\left( k_{2}\right) }_{m',-m_{2}}=\Omega ^{\left( k_{1}\right) }_{-m,-m'}\Omega ^{\left( k_{2}\right) }_{-m',-m}\\
 & = & E^{\left( k_{1}\right) *}_{r}E^{\left( k_{2}\right) }_{s}\\
 & \times  & \left\langle j_{1},l_{11},-m_{1}\left| T^{\left( k_{1}\right) \dagger }_{r}\right| j_{2},l_{21},-m'\right\rangle \left\langle j_{2},l_{22},-m'\left| T^{\left( k_{2}\right) }_{s}\right| j_{1},l_{12},-m_{2}\right\rangle \\
 & = & \eta ^{*}_{F,\left\{ j_{1},l_{11}\right\} }\eta _{F,\left\{ j_{1},l_{12}\right\} }\eta ^{*}_{F,\left\{ j_{1},l_{21}\right\} }\eta _{F,\left\{ j_{1},l_{22}\right\} }e^{2i\left( m_{1}-m_{2}\right) \phi _{0}}E^{\left( k_{1}\right) *}_{r}E^{\left( k_{2}\right) }_{s}\\
 & \times  & \left\langle j_{1},l_{11},m_{1}\left| F^{\dagger }T_{r}^{\left( k_{1}\right) \dagger }F\right| j_{2},l_{21},m'\right\rangle \left\langle j_{2},l_{22},m'\left| F^{\dagger }T^{\left( k_{2}\right) }_{s}F\right| j_{1},l_{12},m_{2}\right\rangle 
\end{eqnarray*}
As before the \( m \) dependent phases will cancel with the phases from the
transformation of the operators leaving the result \( \phi _{0} \) independent,
but for now keep these explicit and act ignorant of the selection rules that
guarantee this. For \( \hat{n}=\hat{x} \), \( \phi _{0}=0 \) this phase gives
1, while for \( \hat{n}=\hat{y} \), \( \phi _{0}=\pi /2 \) it yields \( e^{i\left( m_{1}-m_{2}\right) \pi }=\left( -1\right) ^{m_{1}-m_{2}} \)
since \( m_{1}-m_{2} \) must be an integer. This phase can then be written
as \( \left( -1\right) ^{\left( \hat{n}\cdot \hat{y}\right) \left( m_{1}-m_{2}\right) } \). 

As just shown the product of \( \eta  \)'s gives either 1 or \( \eta _{l_{1}}\eta _{l_{2}} \)
depending on the representation used for \( F \). Now use a cartesian representation
for the operators so that \( T^{\left( k\right) }_{s}=x^{n_{x}}y^{n_{y}}z^{n_{z}} \)
and \( n_{x}+n_{y}+n_{z}=1 \) and the field amplitudes become some \( E_{i_{1}i_{2}...i_{k}} \).
Transforming these kinds of operators, with these special spin flip operations,
then gives \( F^{\dagger }x^{n_{x}}y^{n_{y}}z^{n_{z}}F=\eta _{F,1}^{k}\left( -1\right) ^{n_{x}}=\eta _{F,1}^{k}\left( -1\right) ^{n_{y}} \)
where the choice of \( n_{x} \) or \( n_{y} \) comes from the choice of \( \hat{n}=\hat{x} \)
or \( \hat{n}=\hat{y} \) for the axis of reflection or rotation, so write it
as \( \left( -1\right) ^{n_{\hat{n}}} \), and \( \eta _{R,1}=-1 \), \( \eta _{M,1}=1 \),
\begin{eqnarray*}
\delta \omega ^{\left( k_{1},k_{2}\right) }_{-m_{1},-m_{2}} & = & \eta ^{*}_{F,\left\{ j_{1},l_{11}\right\} }\eta _{F,\left\{ j_{1},l_{12}\right\} }\eta ^{*}_{F,\left\{ j_{1},l_{21}\right\} }\eta _{F,\left\{ j_{1},l_{22}\right\} }\eta _{F,1}^{k_{1}+k_{2}}\left( -1\right) ^{n_{1\hat{n}}+n_{2\hat{n}}}\left( -1\right) ^{\left( \hat{n}\cdot \hat{y}\right) \left( m_{1}-m_{2}\right) }\\
 & \times  & \left\langle j_{1},l_{11},m_{1}\left| x^{n_{1x}}y^{n_{1y}}z^{n_{1z}}\right| j_{2},l_{21},m'\right\rangle \left\langle j_{2},l_{22},m'\left| x^{n_{2x}}y^{n_{2y}}z^{n_{2z}}\right| j_{1},l_{12},m_{2}\right\rangle \\
 & \times  & E^{\left( k_{1}\right) *}_{i_{1}i_{2}...i_{k_{1}}}E^{\left( k_{2}\right) }_{i_{1}i_{2}...i_{k_{2}}}
\end{eqnarray*}
The contraction of the matrix elements with the amplitudes is implicitly understood
to look like,
\begin{eqnarray*}
 &  & \left\langle j_{1},m_{1}\left| x^{n_{1x}}y^{n_{1y}}z^{n_{1z}}\right| j_{2},m'\right\rangle E^{\left( k\right) }_{i_{1}i_{2}...i_{k}}\\
 &  & =\left\langle j_{1},m_{1}\left| x_{i_{1}}x_{i_{2}}\cdots x_{i_{n_{x}}}y_{i_{n_{x}+1}}\cdots y_{i_{n_{x}+n_{y}}}z_{i_{n_{x}+n_{y}+1}}\cdots z_{i_{k}}\right| j_{2},m'\right\rangle E^{\left( k\right) }_{i_{1}i_{2}...i_{k}}
\end{eqnarray*}

The \( n_{\hat{n}} \) independent pieces of the coefficient will give the usual
\( \left( -1\right) ^{k_{1}+k_{2}} \) for \( F=R_{\hat{n}}\left( \pi \right)  \),
and \( \eta _{l_{11}}\eta _{l_{12}}\eta _{l_{21}}\eta _{l_{22}} \) for \( F=M_{\hat{n}} \).The
parity selection rules shown before will give \( \eta _{l_{1}}\eta _{l_{2}}=\left( -1\right) ^{k_{1}+k_{2}} \)
when satisfied, making the simplest general representation, with \( n_{\hat{n}}=n_{1\hat{n}}+n_{2\hat{n}} \),

\begin{eqnarray*}
\delta \omega ^{\left( k_{1},k_{2}\right) }_{-m_{1},-m_{2}} & = & \left( -1\right) ^{k_{1}+k_{2}}\left( -1\right) ^{n_{\hat{n}}}\left( -1\right) ^{\left( \hat{n}\cdot \hat{y}\right) \left( m_{1}-m_{2}\right) }\\
 & \times  & \left\langle j_{1},l_{11},m_{1}\left| x^{n_{1x}}y^{n_{1y}}z^{n_{1z}}\right| j_{2},l_{21},m'\right\rangle \left\langle j_{2},l_{22},m'\left| x^{n_{2x}}y^{n_{2y}}z^{n_{2z}}\right| j_{1},l_{12},m_{2}\right\rangle \\
 & \times  & E^{\left( k_{1}\right) *}_{i_{1}i_{2}...i_{k_{1}}}E^{\left( k_{2}\right) }_{i_{1}i_{2}...i_{k_{2}}}
\end{eqnarray*}

The result, as for the relation in the spherical basis, is simply stated each
are of terms contributing to the spin flipped shift are related by a sign given
by the order of each transition \( \left( -1\right) ^{k_{1}+k_{2}} \) and the
number of certain operators used in each matrix element \( \left( -1\right) ^{\hat{n}} \).
This can be written slightly differently using \( n_{x}+n_{y}+n_{z}=k \) so
that \( k \) and \( n_{\hat{n}} \) can be traded instead for \( n_{z} \)
and \( n_{\perp \hat{n}} \) and \( \left( -1\right) ^{k_{1}+k_{2}}\left( -1\right) ^{\hat{n}}=\left( -1\right) ^{n_{z}}\left( -1\right) ^{\perp \hat{n}} \)
which is what would have directly appear writing the \( F=R_{\hat{n}}\left( \pi \right)  \)
explicitly without using \( \eta _{R,1}=-1 \) since \( R_{\hat{n}}\left( \pi \right)  \)
flips the sign of the sign of \( \hat{z} \) and the \( x-y \) component perpendicular
to \( \hat{n} \), while \( M_{\hat{n}} \) just flips the sign of \( \hat{n} \)
and the \( \left( -1\right) ^{k} \) terms come from the products of the parities
of the states.

In this case the derivation and the origin of the various forms the result can
take due to the choice of \( F \), are easily seen since the transformation
of these cartesian operators is easily understood geometrically, but there is
still a strange dependence on the representation of \( F \) through the free
choice of \( n_{x} \) or \( n_{y} \) in the remaining factors that is apparently
ambiguous or inconsistent. Consider, for example, a quadrupole-quadrupole term
involving something like, in particular, \( \left\langle j_{1},m\left| xy\right| j_{2},m'\right\rangle \left\langle j_{2},m'\left| yz\right| j_{1},m\right\rangle  \)
so that \( m_{1}-m_{2}=m \) and \( \left( -1\right) ^{\left( \hat{n}\cdot \hat{y}\right) \left( m_{1}-m_{2}\right) }=1 \).
Using \( \hat{n}=\hat{y} \) implies that this product doesn't change sign for
\( m\rightarrow -m \) as \( n_{y}=2 \) so \( \left( -1\right) ^{n_{y}}=1 \),
while for \( \hat{n}=\hat{x} \), \( \left( -1\right) ^{n_{x}}=-1 \) implying
that the sign does change. This is a contradiction unless the result is actually
zero. For this case that is clear by simple \( m \) selection rules. As pointed
out long ago, sec.\ref{Sec:ShiftsForIdealFields}, \( xy \) will change \( m \)
by 2 or 0, while \( yz \) can change \( m \) by only 1, so that between the
same set of states one must give a zero matrix element. This happens for any
other particular case considered, and one again the initial apparent curse of
generality, becomes a reward. The apparent contradictions generated by these
spin flip transformations turn out to give a set of cartesian selection rules,
of a sort, for determining what terms can be non-zero to begin with. 

These are analogous to the much simpler result from using a spherical basis
where terms were non-zero only if \( m_{1}-m_{2}=s-r \). Here terms will be
nonzero only for \( \left( -1\right) ^{n_{x}}=\left( -1\right) ^{n_{y}}\left( -1\right) ^{m_{1}-m_{2}} \).
In particular, for use in evaluating these ion light shifts where \( m_{1}=m_{2} \),
these rules require \( n_{x}-n_{y} \) to be even.

\subsubsection{Phases}

The spin dependence of the matrix elements has a simply geometric picture in
cartesian coordinates, but one disadvantage of this form is that the origins
of any overall complex phase is slightly obscured. The shifts generally involve
a real or imaginary part of some \( \delta \omega  \), and being able to easily
identify the overall phase of the coupling allows for a quick determination
of the dependence of the shift on the phases of the field driving the transitions.

In the a spherical tensor basis the matrix elements are real, as the are given
explicitly by Clebsch-Gordan coefficient though the Wigner-Eckhart Theorem so
the phase of each term in the total coupling is isolated in the product of the
amplitudes,
\begin{eqnarray*}
Arg\left( \delta \omega ^{\left( k_{1},k_{2}\right) }_{m}\right)  & = & Arg\left( \Omega ^{\left( k_{1}\right) \dagger }_{m,m'}\Omega ^{\left( k_{2}\right) }_{m',m}\right) \\
 & = & Arg\left( \left\langle j_{1},m\left| T^{\left( k_{1}\right) \dagger }_{r}\right| j_{2},m'\right\rangle \left\langle j_{2},m'\left| T^{\left( k_{2}\right) }_{s}\right| j_{1},m\right\rangle E^{\left( k_{1}\right) *}_{r}E^{\left( k_{2}\right) }_{s}\right) \\
 & = & \left\langle j_{1},m\left| T^{\left( k_{1}\right) \dagger }_{r}\right| j_{2},m'\right\rangle \left\langle j_{2},m'\left| T^{\left( k_{2}\right) }_{s}\right| j_{1},m\right\rangle Arg\left( E^{\left( k_{1}\right) *}_{r}E^{\left( k_{2}\right) }_{s}\right) 
\end{eqnarray*}
 In a cartesian basis the matrix elements are no longer generally real so the
total phase of the coupling is distributed between the matrix elements and the
field amplitudes. The demonstration of this separation can be done most easily
by temporarily transforming the cartesian operators to a spherical basis and
this method also shows the simple resolution of the of this difficulty.

Once again the cartesian operators are given by,
\begin{eqnarray*}
x & = & \left( r_{-}-r_{+}\right) /\sqrt{2}\\
y & = & \left( r_{-}+r_{+}\right) /\sqrt{2}i\\
z & = & r_{0}
\end{eqnarray*}
in terms of their spherical counterparts.Then any matrix element of products
of these operators is given by,
\begin{eqnarray*}
\left\langle j_{1},m_{1}\left| x^{n_{x}}y^{n_{y}}z^{n_{z}}\right| j_{2},m_{2}\right\rangle  & = & \left\langle j_{1},m_{1}\left| \left( \frac{r_{-}-r_{+}}{\sqrt{2}}\right) ^{n_{x}}\left( \frac{r_{-}+r_{+}}{\sqrt{2}i}\right) ^{n_{y}}r_{0}^{n_{z}}\right| j_{2},m_{2}\right\rangle \\
 & = & \left( -i\right) ^{n_{y}}\left\langle j_{1},m_{1}\left| \left( \frac{r_{-}-r_{+}}{\sqrt{2}}\right) ^{n_{x}}\left( \frac{r_{-}+r_{+}}{\sqrt{2}}\right) ^{n_{y}}r_{0}^{n_{z}}\right| j_{2},m_{2}\right\rangle 
\end{eqnarray*}
The matrix element is now only of product of the \( r_{s} \)'s with real coefficients,
and so it again is real and the overall phase of the matrix element is just
given by the number of \( y \)'s in the operators up to a real sign since the
sign of the matrix element of this product of spherical tensor operators is
not immediately obvious,
\begin{eqnarray*}
\left\langle j_{1},m_{1}\left| x^{n_{x}}y^{n_{y}}z^{n_{z}}\right| j_{2},m_{2}\right\rangle  & = & \pm i^{n_{y}}\left| \left\langle j_{1},m_{1}\left| x^{n_{x}}y^{n_{y}}z^{n_{z}}\right| j_{2},m_{2}\right\rangle \right| 
\end{eqnarray*}
 This makes the argument of the entire shift again easy to express in cartesian
coordinates as well,
\begin{eqnarray*}
Arg\left( \delta \omega ^{\left( k_{1},k_{2}\right) }_{m}\right)  & = & Arg(\left\langle j_{1},m\left| x^{n_{x_{1}}}y^{n_{y_{1}}}z^{n_{z_{1}}}\right| j_{2},m'\right\rangle \left\langle j_{2},m'\left| x^{n_{x_{2}}}y^{n_{y_{2}}}z^{n_{z_{2}}}\right| j_{1},m\right\rangle \\
 & \times  & E^{\left( k_{1}\right) *}_{i_{1}i_{2}...i_{k_{1}}}E^{\left( k_{2}\right) }_{i_{1}i_{2}...i_{k_{2}}})\\
 & = & \pm Arg(E^{\left( k_{1}\right) *}_{i_{1}i_{2}...i_{k_{1}}}E^{\left( k_{2}\right) }_{i_{1}i_{2}...i_{k_{2}}}\\
 & \times  & i^{n_{y_{1}}}\left| \left\langle j_{1},m\left| x^{n_{x_{1}}}y^{n_{y_{1}}}z^{n_{z_{1}}}\right| j_{2},m'\right\rangle \right| i^{n_{y_{2}}}\left| \left\langle j_{2},m'\left| x^{n_{x_{2}}}y^{n_{y_{2}}}z^{n_{z_{2}}}\right| j_{1},m\right\rangle \right| )\\
 & = & \pm \left| \left\langle j_{1},m\left| x^{n_{x_{1}}}y^{n_{y_{1}}}z^{n_{z_{1}}}\right| j_{2},m'\right\rangle \left\langle j_{2},m'\left| x^{n_{x_{2}}}y^{n_{y_{2}}}z^{n_{z_{2}}}\right| j_{1},m\right\rangle \right| \\
 & \times  & Arg\left( i^{n_{y}}E^{\left( k_{1}\right) *}_{i_{1}i_{2}...i_{k_{1}}}E^{\left( k_{2}\right) }_{i_{1}i_{2}...i_{k_{2}}}\right) 
\end{eqnarray*}

\section{Parity Splitting Revisited}

Finally this can be put to use in trivially deriving the parity light shift
and understanding its origin. The particular choices of the fields used above
can be abandoned temporarily and the general structure of the shifts studied
to determine what can happen and what is required to make useful things happen.
For simplicity the choice of two independent standing waves will be retained,
again positioned so the the ion is at the node of one and the anti-node of the
other, giving a quadrupole coupling from only one of the fields and a dipole
only from the other. The polarizations and propagation directions will now be
completely general.

The quadrupole and parity light shifts were given by,
\begin{eqnarray*}
\delta \omega _{m}^{Q} & = & \sqrt{\left( \Omega ^{Q\dagger }\Omega ^{Q}\right) _{mm}}\\
\delta \omega _{m}^{PNC} & = & \epsilon Im\left( \left( \Omega ^{Q\dagger }\Omega ^{D}\right) _{mm}\right) /\delta \omega _{m}^{Q}
\end{eqnarray*}
with the couplings,
\begin{eqnarray*}
\Omega _{m'm}^{D} & = & \left\langle 5D_{3/2},m'\left| D_{i}\right| nP_{1/2},m\right\rangle E_{i}=\left\langle 5D_{3/2},m'\left| T^{\left( 1\right) }_{s}\right| nP_{1/2},m\right\rangle E^{\left( 1\right) }_{s}\\
\Omega _{m'm}^{Q} & = & \left\langle 5D_{3/2},m'\left| Q_{ij}\right| 6S_{1/2},m\right\rangle \partial _{i}E_{j}=\left\langle 5D_{3/2},m'\left| T^{\left( 2\right) }_{s}\right| nS_{1/2},m\right\rangle E^{\left( 2\right) }_{s}
\end{eqnarray*}

\subsubsection{Formal Results for Linear Polarization}

Again consider the \( m\rightarrow -m \) behavior of the shifts, now using
the spin flip transformations just developed. The fields were chosen to be linearly
polarized, so a quick route is to use the previously derived results for linear
polarization, 
\[
\delta \omega ^{\left( k_{1},k_{2}\right) }_{-m}=\left( -1\right) ^{k_{1}+k_{2}}\delta \omega ^{\left( k_{1},k_{2}\right) }\]
For the quadrupole term \( k_{1}=k_{2}=2 \) so that \( \left( -1\right) ^{k_{1}+k_{2}}=1 \)
and \( \delta \omega ^{Q}_{-m}=\delta \omega ^{Q}_{m} \). The quadrupole shift
is independent, and this shows that this property is completely independent
of the orientation of the field, the only requirement was linear polarization.
Similarly, for the PNC term, \( k_{1}=1 \), \( k_{2}=2 \), giving \( \left( -1\right) ^{k_{1}+k_{2}}=-1 \)
and \( \delta \omega ^{PNC}_{-m}=-\delta \omega _{m}^{PNC} \).

\subsubsection{Off Axis Spin Dependence}

Slightly anticipating pieces of the discussion on systematics, it is productive
to look as the off diagonal terms as well. Later \( \delta \omega _{m_{1},m_{2}} \)
as a matrix will be used in its entirety as a sort of effective hamiltonian
that gives the resulting collective effects of a single state. For the ground
state this will be a \( 2\times 2 \) hermitian matrix so it will look like
the interaction from an effective magnetic field, the diagonal elements give
the \( \hat{z} \) components of the effective field and the off-diagonal elements
the perpendicular components so that non-zero off-diagonal elements would mean
the interaction behaves has an effective spin dependence in a different direction.
Some properties of these terms can be studied from the more general spin flip
relations,

\[
\delta \omega ^{\left( k_{1},k_{2}\right) }_{-m_{1},-m_{2}}=\left( -1\right) ^{m_{1}-m_{2}}\left( -1\right) ^{k_{1}+k_{2}}\delta \omega _{m_{1},m_{2}}^{\left( k_{1},k_{2}\right) *}\]
 The matrix elements in the first off-diagonal row contain any contributions
to an effective vector, or dipole interaction, since these change \( m \) by
at most \( 1 \). For the ground state, there is only one off-diagonal term
having \( m_{1}=-m_{2}=\pm 1/2 \), \( \left| m_{1}-m_{2}\right| =1 \). These
terms must then be related as \( \delta \omega ^{\left( k_{1},k_{2}\right) }_{1/2,-1/2}=-\left( -1\right) ^{k_{1}+k_{2}}\delta \omega _{-1/2,1/2}^{\left( k_{1},k_{2}\right) *} \).
The \( \delta \omega  \) is hermitian giving \( \delta \omega _{m_{1},m_{2}}=\delta \omega ^{*}_{m_{2},m_{1}} \),
so that for these matrix elements of the ground state \( \delta \omega ^{\left( k_{1},k_{2}\right) }_{1/2,-1/2}=-\left( -1\right) ^{k_{1}+k_{2}}\delta \omega _{1/2,-1/2}^{\left( k_{1},k_{2}\right) } \).

For the quadrupole shift, this would give \( \delta \omega ^{Q}_{1/2,-1/2}=-\delta \omega ^{Q}_{1/2,-1/2}=0 \)
so that in fact the quadrupole shift is independent of spin along any axis.
The sole contribution from quadrupole shift is a scalar which is invariant under
rotations. This should be the case, since it was spin independent along the
\( \hat{z} \) axis for any direction and polarization of the applied light.
If there was a spin dependence along a different axis, the entire system could
be rotated to make that dependence along the \( \hat{z} \) axis with the field
in a different directions, inconsistent with the earlier result of independence
along the \( \hat{z} \) axis for any field. For the PNC term this relation
gives no new information as it is just consistent with hermeticity, \( \delta \omega ^{PNC}_{1/2,-1/2}=\delta \omega _{-1/2,1/2}^{PNC*} \).

\subsubsection{General Structure in a Spherical Basis}

This gives the results trivially, but some understanding is lost in the formal
generality, and for the PNC term it only shows the potential for spin dependence,
it doesn't explain when the shift will be non-zero. To study this, continue
to back up a step at a time and look again at the general spin dependence of
each term contributing to the shift in a spherical basis. Consider only the
\( m_{1}=m_{2}=m \) elements of the shift since, as discussed above, any off
diagonal elements are spin defendant shifts in another direction that can be
made to be a spin dependence in the \( \hat{z} \) direction by rotating the
coordinate system. For these elements,

\begin{eqnarray*}
\delta \omega ^{\left( k_{1},k_{2}\right) }_{-m} & = & \left( -1\right) ^{k_{1}+k_{2}}\left\langle j_{1},m\left| T^{\left( k_{1}\right) \dagger }_{s}\right| j_{2},m'\right\rangle \left\langle j_{2},m'\left| T^{\left( k_{2}\right) }_{s}\right| j_{1},m\right\rangle E^{\left( k_{1}\right) *}_{-s}E^{\left( k_{2}\right) }_{-s}
\end{eqnarray*}
Since the spin dependence is the quality of interest consider the difference
in the shifts of the \( \pm m \) spin states, 
\begin{eqnarray*}
\Delta \omega ^{\left( k_{1},k_{2}\right) }_{m} & = & \delta \omega ^{\left( k_{1},k_{2}\right) }_{m}-\delta \omega ^{\left( k_{1},k_{2}\right) }_{-m}\\
 & = & \left\langle j_{1},m\left| T^{\left( k_{1}\right) \dagger }_{s}\right| j_{2},m'\right\rangle \left\langle j_{2},m'\left| T^{\left( k_{2}\right) }_{s}\right| j_{1},m\right\rangle \\
 & \times  & \left( E^{\left( k_{1}\right) *}_{s}E^{\left( k_{2}\right) }_{s}-\left( -1\right) ^{k_{1}+k_{2}}E^{\left( k_{1}\right) *}_{-s}E^{\left( k_{2}\right) }_{-s}\right) 
\end{eqnarray*}
With the sign change of the spin index in the difference many parts of the products
of the amplitudes cancel. Some of that work can be done immediately to provide
a convenient result by writing the amplitudes in terms of symmetric and antisymmetric
pieces,
\begin{eqnarray*}
 &  & E^{\left( k_{1}\right) *}_{s}E^{\left( k_{2}\right) }_{s}-\left( -1\right) ^{k_{1}+k_{2}}E^{\left( k_{1}\right) *}_{-s}E^{\left( k_{2}\right) }_{-s}\\
 &  & =\left( E^{\left( k_{1}\right) *}_{sS}+\frac{\left| s\right| }{s}E^{\left( k_{1}\right) *}_{sA}\right) \left( E^{\left( k_{2}\right) }_{sS}+\frac{\left| s\right| }{s}E^{\left( k_{2}\right) }_{sA}\right) \\
 &  & -\left( -1\right) ^{k_{1}+k_{2}}\left( E^{\left( k_{1}\right) *}_{sS}-\frac{\left| s\right| }{s}E^{\left( k_{1}\right) *}_{sA}\right) \left( E^{\left( k_{2}\right) }_{sS}-\frac{\left| s\right| }{s}E^{\left( k_{2}\right) }_{sA}\right) \\
 &  & =\left( E^{\left( k_{1}\right) *}_{sS}E^{\left( k_{2}\right) }_{sS}+E^{\left( k_{1}\right) *}_{sA}E^{\left( k_{2}\right) }_{sA}\right) \left( 1-\left( -1\right) ^{k_{1}+k_{2}}\right) \\
 &  & +\frac{\left| s\right| }{s}\left( E^{\left( k_{1}\right) *}_{sS}E^{\left( k_{2}\right) }_{sA}+E^{\left( k_{1}\right) *}_{sA}E^{\left( k_{2}\right) }_{sS}\right) \left( 1+\left( -1\right) ^{k_{1}+k_{2}}\right) 
\end{eqnarray*}
For the quadrupole term this becomes, \noun{\( (-1)^{k_{1}+k_{2}}=1 \)},
\begin{eqnarray*}
\Delta \omega ^{Q}_{m} & = & \left\langle j_{1},m\left| Q^{\dagger }_{s}\right| j_{2},m'\right\rangle \left\langle j_{2},m'\left| Q_{s}\right| j_{1},m\right\rangle \left( \left| E^{Q}_{s}\right| ^{2}-\left| E^{Q}_{-s}\right| ^{2}\right) \\
 & = & 2\left\langle j_{1},m\left| Q^{\dagger }_{s}\right| j_{2},m'\right\rangle \left\langle j_{2},m'\left| Q_{s}\right| j_{1},m\right\rangle \frac{\left| s\right| }{s}\left( E^{Q*}_{sS}E^{Q}_{sA}+E^{Q*}_{sA}E^{Q}_{sS}\right) \\
 & = & 2\left\langle j_{1},m\left| Q^{\dagger }_{s}\right| j_{2},m'\right\rangle \left\langle j_{2},m'\left| Q_{s}\right| j_{1},m\right\rangle \frac{\left| s\right| }{s}Re\left( E^{Q*}_{sS}E^{Q}_{sA}\right) 
\end{eqnarray*}
Ideally this splitting is zero. It is possible that a well-concocted interaction
between spin dependent individual terms could cancel out making the sum over
\( s \) zero, but it is easier to understand the simpler case of just making
each term zero. This would require \( \left| E^{Q}_{-s}\right| =\left| E^{Q}_{s}\right|  \),
or \( Re\left( E^{Q*}_{sS}E^{Q}_{sA}\right) =0 \). For the first form one possibility
is immediately clear. As observed above, this is automatically satisfied for
linearly polarized light as that gives \( E^{Q}_{-s}=\left( -1\right) ^{s}E^{Q*}_{s} \).
For the symmetric-antisymmetric component form, this is also seen using the
fact that the symmetric and antisymmetric amplitudes include an explicit factor
of \( i \), except in the case of \( s=0 \) but there \( E^{Q}_{0A}=0 \),
so that for real fields the real part of any product of symmetric and antisymmetric
pieces is always zero. 

The constraints for more general fields are harder to understand in this form,
but the must be considered. The picture is clearer in a cartesian basis. Both
views will be considered in depth when discussing systematics.

The parity term looks similar in this spherical basis, here \( \left( -1\right) ^{k_{1}+k_{2}}=-1 \)
\begin{eqnarray*}
\Delta \omega ^{PNC}_{m} & = & \left\langle j_{1},m\left| Q^{\dagger }_{s}\right| j_{2},m'\right\rangle \left\langle j_{2},m'\left| Q_{s}\right| j_{1},m\right\rangle Im\left( E^{Q*}_{s}E^{D}_{s}+E^{Q*}_{-s}E^{D}_{-s}\right) \\
 & = & 2\left\langle j_{1},m\left| Q^{\dagger }_{s}\right| j_{2},m'\right\rangle \left\langle j_{2},m'\left| Q_{s}\right| j_{1},m\right\rangle Im\left( E^{Q*}_{sS}E^{D}_{sS}+E^{Q*}_{sA}E^{D}_{sA}\right) 
\end{eqnarray*}
There is an automatic sign change from quadrupole-dipole structure of this cross
term, so that for each term to change sign with \( m \) requires \( Im\left( E^{Q*}_{-s}E^{D}_{-s}\right) =Im\left( E^{Q*}_{s}E^{D}_{s}\right)  \).
Again the general geometric requirement for the fields are hard so picture in
this form without some restrictions, but the explicit expression for these spherical
tensor amplitudes in terms of the usual cartesian components provides some insight.

First consider the \( s=\pm 1 \) terms as they correspond to the non-zero amplitudes
for the original choice of fields,

\begin{eqnarray*}
E^{D}_{\pm 1} & = & \left( \mp E^{D}_{x}+iE^{D}_{y}\right) /\sqrt{2}\\
E_{\pm 1}^{Q} & = & \mp \left( \partial _{x}E^{Q}_{z}+\partial _{z}E^{Q}_{x}\right) /2+i\left( \partial _{y}E^{Q}_{z}+\partial _{z}E^{Q}_{y}\right) /2
\end{eqnarray*}
and 

\[
\begin{array}{cc}
E^{D}_{1S}=iE^{D}_{y}/\sqrt{2} & E^{D}_{1A}=-E^{D}_{x}/\sqrt{2}\\
E_{1S}^{Q}=i\left( \partial _{y}E^{Q}_{z}+\partial _{z}E^{Q}_{y}\right) /2 & E_{1A}^{Q}=-\left( \partial _{x}E^{Q}_{z}+\partial _{z}E^{Q}_{x}\right) /2
\end{array}\]
Using these symmetric and anti-symmetric components of the field tensors, the
contribution to the splitting are given by,
\[
E^{Q*}_{1S}E^{D}_{1S}+E^{Q*}_{1A}E^{D}_{1A}=E^{D}_{y}\left( \partial _{y}E^{Q*}_{z}+\partial _{z}E^{Q*}_{y}\right) /2+E^{D}_{x}\left( \partial _{x}E^{Q*}_{z}+\partial _{z}E^{Q*}_{x}\right) /2\]
The second term is exactly what was chosen to be non-zero with the original
fields, the first term is just the second term rotated so that the fields are
contained in the \( y-z \) plane rather than the \( x-z \) plane. The form
of the factors also shows the general freedom on the choice of field configurations.
Two were were originally considered for the quadrupole field, \( \vec{E}^{Q}=\hat{x}cos\left( kz\right)  \)
and \( \vec{E}^{Q}=\hat{z}cos\left( kx\right)  \). These correspond respectively
to nonzero \( \partial _{z}E_{x} \) and \( \partial _{x}E_{z} \), exactly
what is required for a spin independent product term for \( \vec{E}^{D}\propto \hat{x} \).
The full structure of this term shows that in fact the quadrupole field can
be oriented in any way such that \( \hat{k} \) and \( \hat{\epsilon } \) are
in the \( x-z \) plane. Similarly for \( \vec{E}^{Q}\propto \hat{y} \) or
\( \hat{k}_{Q}\propto \hat{y} \) with only \( \vec{E}^{D}\propto \hat{x} \),
the cross term in the product of amplitudes is spin dependent giving no splitting,
as found previously with more mechanical methods. Note that none of this depends
on the direction of propagation of the dipole field, only that it have polarization
components along \( \hat{\epsilon }^{Q} \) and \( \vec{k}^{Q} \). The previous
configurations considered only included cases where \( \hat{\epsilon } \) and
\( \hat{k} \) for both fields were in the same plane.

Now also consider the \( s=0 \) term, clearly this satisfies \( E^{\left( k\right) }_{-0}=E^{\left( k\right) }_{0} \)
so the product of field amplitudes is always spin independent and the resulting
shift spin independent. For the dipole field \( E^{D}_{0}=E^{D}_{z} \), the
transition is driven simply by a polarization component in the \( \hat{z} \)
direction. The quadrupole term is \( E^{Q}_{0}=\partial _{z}E^{Q}_{z}/3 \).
Having both \( k_{z}^{Q} \) and \( E_{z}^{Q} \) non-zero requires \( \vec{E}^{Q} \)to
propagate in some weird direction, such as with \( \vec{k} \) and \( \vec{\epsilon }^{Q} \)
in the \( x-z \) plane but neither parallel to the \( \hat{z} \) axis. This
would gives \( k_{z}\epsilon _{z} \) proportional to \( sin\left( \theta \right) cos\left( \theta \right) =cos\left( 2\theta \right)  \)
which is largest for \( \theta =\pm 45^{\circ } \). 

This seems to conflict with the previous result that gave a non-zero splitting
for parallel quadrupole and dipole polarizations and propagation directions,
because in this basis that would give \( E^{Q}_{0}=0 \) and zero spin dependent
shift even though this is just the first system rotated about the \( \hat{y} \)
axis giving \( \hat{x}\rightarrow \hat{z} \). However, the rotation also changes
the axis of spin dependence from \( \hat{z}\rightarrow -\hat{x} \). This can
be checked explicitly for consistency by making use of the previous results
for off-diagonal matrix elements of \( \delta \omega ^{PNC} \),
\begin{eqnarray*}
\delta \omega ^{\left( k_{1},k_{2}\right) }_{-m_{1},-m_{2}} & = & \left( -1\right) ^{k_{1}+k_{2}}E^{\left( k_{1}\right) *}_{-r}E^{\left( k_{2}\right) }_{-s}\\
 & \times  & \left\langle j_{1},l_{11},m_{1}\left| T^{\left( k_{1}\right) \dagger }_{r}\right| j_{2},l_{21},m'\right\rangle \left\langle j_{2},l_{22},m'\left| T^{\left( k_{2}\right) }_{s}\right| j_{1},l_{12},m_{2}\right\rangle 
\end{eqnarray*}
For this dipole-quadrupole cross term this gives,
\[
\delta \omega ^{PNC}_{-m_{1},-m_{2}}=-\left\langle j_{1},m_{1}\left| D^{\dagger }_{r}\right| j_{2},m'\right\rangle \left\langle j_{2},m'\left| Q_{s}\right| j_{1},m_{2}\right\rangle E^{D*}_{-r}E^{Q}_{-s}\]
Take \( \vec{E}^{D}\propto \hat{z} \), giving only \( E^{D}_{0}=E_{z} \) non-zero,
and \( \vec{\epsilon }^{Q}\propto \hat{z} \), \( \vec{k}^{Q}\propto \hat{x} \)
giving \( E^{Q}_{\pm 1}=\mp E^{Q}/2\sqrt{2} \). This yields,
\[
\delta \omega ^{PNC}_{-m_{1},-m_{2}}=\left\langle j_{1},m_{1}\left| D^{\dagger }_{0}\right| j_{2},m_{1}\right\rangle \left\langle j_{2},m_{1}\left| Q_{s}\right| j_{1},m_{2}\right\rangle E_{s}^{D*}E_{s}^{Q}=\delta \omega _{m_{1},m_{2}}^{PNC}\]
Notice the simplification of the spin indices, the sum over spins can be done
because the dipole term only couple \( \Delta m=0 \) transitions. Similarly
the entire shift is nonzero only for \( \left| m_{1}-m_{2}\right| =1 \) as
the quadrupole term drives only the \( \Delta m=\pm 1 \) transitions.

This term is only a piece of the total light shift, the entire PNC shift is
this term minus its complex conjugate, and as neither changes the total matrix
element doesn't change. For the particular case of the ground state this gives,
\( \delta \omega ^{PNC}_{-1/2,1/2}=\delta \omega ^{PNC}_{1/2,-1/2} \). Since
\( \delta \omega  \) as a matrix is hermitian, \( \delta \omega ^{P}_{-1/2,1/2}=\delta \omega ^{PNC*}_{1/2,-1,2} \)
so that with both requirement these matrix elements must be real.Then if written
in matrix form, this is a term proportional to the usual Pauli spin matrix \( \sigma _{x} \)
representing an effective shift along the \( \hat{x} \) direction. 

The exact form, involving the complex conjugate term was reintroduced to the
the case of \( \vec{k}^{Q}\propto \hat{y} \) could also be considered where
\( E^{Q}_{\pm 1}=iE^{Q}/2 \), giving
\[
\delta \omega ^{PNC}_{-m_{1},-m_{2}}=-\left\langle j_{1},m_{1}\left| D^{\dagger }_{0}\right| j_{2},m_{1}\right\rangle \left\langle j_{2},m_{1}\left| Q_{s}\right| j_{1},m_{2}\right\rangle E_{s}^{D*}E_{s}^{Q}=-\delta \omega _{m_{1},m_{2}}^{PNC}\]
and just as above, for the \( \delta \omega _{-1/2,1/2} \) term, hermeticity
required that this term is purely imaginary, and as it switches sign, the shift
matrix includes a piece proportional to \( \sigma _{y} \).

Again, the existence of the shift is independent of the direction of propagation
of the dipole field, only a polarization component parallel to the polarization
of \( \hat{\epsilon }^{Q} \) is required. Then the propagation direction determined
the direction of the effective shift. The general geometric structure of this
dependence is still hard to see in this form, but will appear easily in the
when the general solution to this problem is developed later. The Pauli spin
matrices can be used as a basis for \( \delta \omega  \) so that an effective
magnetic field \( \delta \vec{\omega } \) can be discussed with \( \delta \omega =\delta \vec{\omega }\cdot \vec{\sigma } \).
The component solutions here are easily seen to be consistent with the general
result that turns out as,
\[
\delta \vec{\omega }\propto \vec{E}^{D}\times \left( \vec{k}^{Q}\times \vec{E}^{Q}\right) +2\left( \vec{E}^{D}\cdot \vec{k}^{Q}\right) \vec{E}^{Q}\]
This form is easiest to use for the first fields considered, when axis were
chosen so that the dipole field drove \( \Delta m=\pm 1 \) transitions, with
propagation and polarization components of both fields in the same plane. \( \vec{k}^{Q}\times \vec{E}^{Q} \)is
then perpendicular to \( \vec{E}^{D} \) and so \( \vec{E}^{D}\times \left( \vec{k}^{Q}\times \vec{E}^{Q}\right)  \)
is back in the original defining plane, perpendicular to \( \vec{E}^{D} \).

For the later cases where the dipole field drives the \( \Delta m=0 \) transitions
it is easiest to expand the triple cross product, 
\[
\delta \vec{\omega }\propto \left( \vec{E}^{D}\cdot \vec{E}^{Q}\right) \vec{k}^{Q}+\left( \vec{E}^{D}\cdot \vec{k}^{Q}\right) \vec{E}^{Q}\]
for the quadrupole fields then chosen \( \vec{E}^{D}\cdot \vec{k}^{Q}=0 \),
and \( \hat{E}^{D}\cdot \hat{E}^{Q}=1 \) so that \( \delta \vec{\omega } \)
simply points along \( \vec{k}^{Q} \).

Hints of this vector structure can be seen from the field amplitudes for this
transition,

\begin{eqnarray*}
E^{Q*}_{\pm 1}E^{D}_{\pm 1} & \rightarrow  & E^{D}_{x}\left( \partial _{x}E^{Q}_{z}+\partial _{z}E^{Q}_{x}\right) +E^{D}_{y}\left( \partial _{y}E^{Q}_{z}+\partial _{z}E^{Q}_{y}\right) \\
 & = & E^{D}_{x}\left( -\left( \partial _{x}E^{Q}_{z}-\partial _{z}E^{Q}_{x}\right) +2\partial _{x}E^{Q}_{z}\right) +E^{D}_{y}\left( -\left( \partial _{y}E^{Q}_{z}-\partial _{z}E^{Q}_{y}\right) +2\partial _{y}E^{Q}_{z}\right) \\
 & = & \left( E^{D}_{x}\left( \vec{\nabla }\times \vec{E}^{Q}\right) _{y}-E^{D}_{y}\left( \vec{\nabla }\times \vec{E}^{Q}\right) _{x}\right) +2\left( E^{D}_{x}\partial _{x}E^{Q}_{z}+E^{D}_{y}\partial _{y}E^{Q}_{z}\right) \\
 & = & \left( \vec{E}^{D}\times \left( \vec{\nabla }\times \vec{E}^{Q}\right) \right) _{z}+2\left( \vec{E}^{D}\cdot \vec{\nabla }\right) E^{Q}_{z}\\
 & \rightarrow  & \left( \vec{E}^{D}\times \left( \vec{k}^{Q}\times \vec{E}^{Q}\right) \right) _{z}+2\left( \vec{E}^{D}\cdot \vec{k}^{Q}\right) E^{Q}_{z}\\
 & = & \left( \vec{E}^{D}\cdot \vec{E}^{Q}\right) \vec{k}_{z}^{Q}+\left( \vec{E}^{D}\cdot \vec{k}^{Q}\right) \vec{E}_{z}^{Q}
\end{eqnarray*}

\subsubsection{General Structure in Cartesian Basis}

This generalized analysis using spherical tensors yielded some new information
by studying the off-diagonal elements of the shift, but for the diagonal elements
it basically exactly reproduced the original analysis, the same expressions
were involved, the only difference was the original of some of the sign changes.
In cartesian coordinates the analysis looks very different, and also becomes
much more intuitively understandable as the geometry is much more straight forward.

In cartesian form, the couplings are given by,

\begin{eqnarray*}
\Omega _{m'm}^{D} & = & \left\langle 5D_{3/2},m'\left| D_{i}\right| nP_{1/2},m\right\rangle E_{i}\\
\Omega _{m'm}^{Q} & = & \left\langle 5D_{3/2},m'\left| Q_{ij}\right| 6S_{1/2},m\right\rangle \partial _{i}E_{j}
\end{eqnarray*}
\( D_{i}=x_{i} \), and \( Q_{ij}=x_{i}x_{j}-x^{2}\delta _{ij}/3 \). \( x^{2} \)
is spherically symmetric, it can't change the angular momentum, so in this case
with \( j_{1}=1/2 \) and \( j_{2}=3/2 \) this term doesn't contribute and
the couplings will generally be given by,
\begin{eqnarray*}
\Omega _{m'm}^{D} & = & \left\langle 5D_{3/2},m'\left| x_{i}\right| nP_{1/2},m\right\rangle E_{i}\\
\Omega _{m'm}^{Q} & = & \left\langle 5D_{3/2},m'\left| x_{i}x_{j}\right| 6S_{1/2},m\right\rangle \partial _{i}E_{j}
\end{eqnarray*}

Once again, recall the general form of the spin flip relations,

\begin{eqnarray*}
\delta \omega ^{\left( k_{1},k_{2}\right) }_{-m_{1},-m_{2}} & = & \eta _{F}\left( -1\right) ^{\left( \hat{n}\cdot \hat{y}\right) \left( m_{1}-m_{2}\right) }E^{\left( k_{1}\right) *}_{r}E^{\left( k_{2}\right) }_{s}\\
 & \times  & \left\langle j_{1},l_{11},m_{1}\left| F^{\dagger }T^{\left( k_{1}\right) \dagger }_{r}F\right| j_{2},l_{21},m'\right\rangle \left\langle j_{2},l_{12},m'\left| F^{\dagger }T^{\left( k_{2}\right) }_{s}F\right| j_{1},l_{22},m_{2}\right\rangle 
\end{eqnarray*}
The \( \eta _{F}\left( -1\right) ^{\left( \hat{n}\cdot \hat{y}\right) \left( m_{1}-m_{2}\right) } \)
comes from flipping the spins of the states and depends on the representation
for the spin flip and the states as shown previously. For rotation the flip
just gives a phase, the \( j \) dependent pieces all cancel since each of the
two \( j \)'s is involved appear twice in complementary places giving \( \eta _{R}=1 \)
and the \( m \) dependence pieces leave the \( \left( -1\right) ^{n_{x}}=\left( -1\right) ^{n_{y}}\left( -1\right) ^{m_{1}-m_{2}} \).
For reflections there are the same phases from the state, giving the same prefactors,
and the parities of all the states which, when written in terms of the parities
of the the fundamental states making up these composites, gives \( \eta _{M}=\eta _{l_{11}}\eta _{l_{12}}\eta _{l_{21}}\eta _{l_{22}} \). 

The general result for any spin flip operator was given above, but for now avoid
any explicit representation of \( F \) other than \( \hat{n} \) parallel to
\( \hat{x} \) or \( \hat{y} \), and assign a parity to each coordinate operator
for a given transformation \( \eta _{F,x_{i}}=\pm 1 \). Then, omitting explicit
mention of the \( l \)'s, the relation can be written,

\begin{eqnarray*}
\delta \omega ^{\left( k_{1},k_{2}\right) }_{-m_{1},-m_{2}} & = & \eta _{F}\eta _{F,x}^{n_{x}}\eta _{F,y}^{n_{y}}\eta _{F,z}^{n_{z}}\left( -1\right) ^{\left( \hat{n}\cdot \hat{y}\right) \left( m_{1}-m_{2}\right) }E^{\left( k_{1}\right) *}_{i_{1}i_{2}...i_{k_{1}}}E^{\left( k_{2}\right) }_{i_{1}i_{2}...i_{k_{2}}}\\
 & \times  & \left\langle j_{1},m_{1}\left| x^{n_{1x}}y^{n_{1y}}z^{n_{1z}}\right| j_{2},m'\right\rangle \left\langle j_{2},m'\left| x^{n_{2x}}y^{n_{2y}}z^{n_{2z}}\right| j_{1},m_{2}\right\rangle 
\end{eqnarray*}
The quadrupole shift is easiest to consider. When this single field is involved
the axis can be rotated so that, for example, \( \vec{E}_{Q}\propto \hat{x}cos\left( kz\right)  \).
This gives,
\begin{eqnarray*}
\delta \omega ^{Q}_{-m_{1},-m_{2}} & = & \eta _{F}\eta _{F,x}^{2}\eta _{F,z}^{2}\left( -1\right) ^{\left( \hat{n}\cdot \hat{y}\right) \left( m_{1}-m_{2}\right) }\left( k_{z}E_{x}^{Q*}\right) \left( k_{z}E^{Q}_{x}\right) \\
 & \times  & \left\langle j_{1},l_{1},m_{1}\left| zx\right| j_{2},l_{2},m'\right\rangle \left\langle j_{2},l_{2},m'\left| zx\right| j_{1},l_{1},m_{2}\right\rangle 
\end{eqnarray*}
The same operator appears twice, so the parities of any coordinate operator
cancel no matter what the representation for \( F \). Also for the quadrupole
transition the initial and final, and intermediate states are the same and \( \eta _{F}=1 \)
even for reflections and as a result the whole shift is spin independent along
any axis since this result is for general \( m_{1},m_{2} \).

This was easy to show when the coordinate system is chosen correctly, but of
course the result should be independent of the coordinate system. Showing this
explicitly for one simple generalization gives some practice in evaluating these
terms than will be useful in considering the PNC term.

Take \( \vec{k}^{Q} \) still along \( \hat{z} \) but now allow \( \hat{\epsilon }^{Q} \)
to be anywhere in the \( x-y \) plane. Then the quadrupole shift will be,
\[
\delta \omega ^{Q}_{m_{1},m_{2}}=\left\langle j_{1},m_{1}\left| zx_{i}\right| j_{2},m'\right\rangle \left\langle j_{2},m'\left| zx_{j}\right| j_{1},m_{2}\right\rangle \left( k_{z}E_{i}^{Q*}\right) \left( k_{z}E^{Q}_{j}\right) \]
The sum will now include terms for \( i,j=1,2 \). The \( i=j \) terms are
spin independent, as before, since the same operators are involved in both matrix
elements and the parities from the transformation of those operators cancel.
But two of the four possible terms involve \( \left\langle xz\right\rangle \left\langle yz\right\rangle  \).
Apparently these terms could change sign with a spin flip, for example \( R_{\hat{y}} \)
changes only the sign of \( y \). Now the phase information derived for these
matrix elements must be used. The phase is given entirely by the number of \( y \)
coordinates as \( i^{n_{y}} \). This then implies that this cross term product
of matrix elements which involves one \( y \) is purely imaginary. The other
term is just the complex conjugate, so since it is also purely imaginary this
gives \( \left\langle zx\right\rangle \left\langle zy\right\rangle =-\left\langle zy\right\rangle \left\langle zx\right\rangle  \).
If the field amplitudes then satisfy \( E_{x}^{Q*}E_{y}^{Q}=E_{x}^{Q}E_{y}^{Q*} \),
or \( Im\left( E_{x}^{Q*}E_{y}\right) =0 \), that is that \( E_{x}^{Q} \)
and \( E_{y}^{Q} \) have the same relative phase these cross terms cancel and
in fact the shift if given by,
\[
\delta \omega ^{Q}_{m_{1},m_{2}}=\left\langle j_{1},m_{1}\left| zx_{i}\right| j_{2},m'\right\rangle \left\langle j_{2},m'\left| zx_{i}\right| j_{1},m_{2}\right\rangle \left( k_{z}E_{i}^{Q*}\right) \left( k_{z}E^{Q}_{i}\right) \]
which is trivially spin independent. Again the condition on the fields implies
that \( \vec{E}^{Q} \) is linearly polarized.

Finally look one last time at the PNC shift. 
\[
\delta \omega ^{PNC}_{m_{1},m_{2}}=\left\langle j_{1},m_{1}\left| x_{i}\right| j_{2},m'\right\rangle \left\langle j_{2},m'\left| x_{j}x_{k}\right| j_{1},m_{2}\right\rangle E_{i}^{D*}\left( k_{j}E^{Q}_{l}\right) \]
Again pick a coordinate system so that \( \hat{\epsilon }^{D}=\hat{x} \), then
\[
\delta \omega ^{PNC}_{m_{1},m_{2}}=\left\langle j_{1},m_{1}\left| x\right| j_{2},m'\right\rangle \left\langle j_{2},m'\left| x_{j}x_{k}\right| j_{1},m_{2}\right\rangle E^{D*}\left( k_{j}E^{Q}_{l}\right) \]
Now the previously derived selection rules can be used, for \( m_{1}=m_{2} \),
a non-zero matrix element requires \( \left( -1\right) ^{n_{x}}=\left( -1\right) ^{n_{y}} \),
or \( n_{x} \) and \( n_{y} \) must differ by an even integer. Here \( n_{x}+n_{y} \)
can be at most \( 3 \), and there is already one \( x \), so these require,
\( \left\{ n_{x},n_{y}\right\}  \) to be \( \left\{ 1,1\right\}  \) or \( \left\{ 2,0\right\}  \).
Both cases leave one coordinate left over which must then be a \( z \) so that,

\[
\delta \omega ^{PNC}_{m}=\left\langle j_{1},m\left| x\right| j_{2},m'\right\rangle \left\langle j_{2},m'\left| x_{a}z\right| j_{1},m\right\rangle E^{D*}\left( k_{a}E^{Q}_{z}+k_{z}E^{Q}_{a}\right) \]
for any non-zero result, where \( x_{a} \) is \( x \) or \( y \). The spin
flipped shift is then given by,
\begin{eqnarray*}
\delta \omega ^{PNC}_{-m} & = & -\eta _{F}\eta _{F,x}\eta _{F,z}\eta _{F,x_{i}}\left\langle j_{1},m\left| x\right| j_{2},m'\right\rangle \left\langle j_{2},m'\left| x_{a}z\right| j_{1},m\right\rangle E^{D*}\left( k_{a}E^{Q}_{z}+k_{z}E^{Q}_{a}\right) \\
 & = & -\eta _{F}\eta _{F,x}\eta _{F,z}\eta _{F,x_{i}}\delta \omega ^{PNC}_{m}
\end{eqnarray*}

Now consider any spin flip operator, for example \( F=M_{\hat{y}} \). This
would give \( \eta _{F,\left\{ x,y,z\right\} }=\left\{ 1,-1,1\right\}  \) and
\( \eta _{F}=\eta _{M}=-1 \) since the initial and final, for the ground state,
or intermediate states, for the \( D \) state, are different and have opposite
parity. A spin dependent shift then required \( x_{a}=x \), as \( \eta _{M}\eta _{M_{\hat{y}},x}^{2}\eta _{M_{\hat{y}}z}=-1 \)
while \( \eta _{M}\eta _{M_{\hat{y}},x}\eta _{M_{\hat{y}}yx}\eta _{M_{\hat{y}}z}=1 \).
Similarly for \( \eta _{M_{\hat{x}}}=\left\{ -1,1,1\right\}  \). Reflections
yield the same results. In this case the sign of two coordinated are flipped,
\( \eta _{R_{\hat{x}}\left( \pi \right) }=\left\{ 1,-1,-1\right\}  \) and \( \eta _{R_{\hat{y}}\left( \pi \right) }=\left\{ -1,1,-1\right\}  \)
but now \( \eta _{F}=\eta _{R}=1 \), and the making product of all the \( \eta  \)'s
negative again requires \( x_{a}=x \). 
\begin{eqnarray*}
\delta \omega ^{PNC}_{-m} & = & -\delta \omega ^{PNC}_{m}
\end{eqnarray*}
 For \( \hat{\epsilon }^{D}=\hat{y} \) the analogous results follow almost
identically. With \( \hat{\epsilon }^{D}=\hat{z} \), selection rules imply
that only \( Q_{xx} \), \( Q_{yy} \), \( Q_{xy} \) or \( Q_{zz} \) give
non-zero matrix elements as these have \( \left| n_{x}-n_{y}\right| =0,2 \).
The same spin flip transformation analysis shows that only \( Q_{zz} \) gives
a spin dependent shift, completely consistent with the results from considering
the \( s=0 \) term in the expression for the shift in a spherical basis.

The off-diagonal terms should be considered as well, the modification to the
previous analysis is to the selection rules, for general \( m_{1} \), \( m_{2} \)
the number of operators must satisfy \( \left( -1\right) ^{n_{x}}=\left( -1\right) ^{n_{y}}\left( -1\right) ^{m_{1}-m_{2}} \),
and the spin flip relation, which for general \( m_{1} \), \( m_{2} \) now
contains a factor \( \left( -1\right) ^{\left( \hat{n}\cdot \hat{y}\right) \left( m_{1}-m_{2}\right) } \).
Again for the particular case of \( m_{1},m_{2}=1/2,-1/2 \) this gives \( \left( -1\right) ^{\left( \hat{n}\cdot \hat{y}\right) \left( m_{1}-m_{2}\right) }=\left( -1\right) ^{\hat{n}\cdot \hat{y}} \)
and \( \left( -1\right) ^{n_{x}}=-\left( -1\right) ^{n_{y}} \), or \( \left| n_{x}-n_{y}\right|  \)
odd. With \( \hat{\epsilon }^{D}=\hat{z} \) this implies nonzero matrix elements
only for \( Q_{xz} \) and \( Q_{yz} \). For either case \( Q_{x_{a}z} \),
any transformation gives, \( \eta _{F}\left( -1\right) ^{\hat{n}\cdot \hat{y}}\eta _{F,x_{a}}\eta _{F,z}^{2}=\eta _{F}\left( -1\right) ^{\hat{n}\cdot \hat{y}}\eta _{F,x_{a}}=\left( -1\right) ^{\hat{y}\cdot \hat{x}_{a}} \).
So the sign of the off-diagonal term for \( Q_{xz} \), and so again real by
hermeticity, corresponding to a shift along the \( \hat{x} \) direction, and
for \( Q_{yz} \) the matrix element changes sign, implying it is imaginary
and that it corresponds to a shift along \( \hat{y} \).

\subsection{General Applications, \protect\( D\protect \) State Shifts}

Collectively these spin flip transformations become a powerful geometrical understanding
of the behavior of matrix elements of coordinate operators, and with that provide
a trivial demonstration of the spin dependence of the PNC light shift. For this
single application these methods are overpowered as in the end they provide
the same result that can be obtained by working out a few Clebsch-Gordan coefficients.
However, the effort was far from wasted. Besides their conceptual advantages,
in building an intuitive understanding of the structure of the interactions,
they will prove to have enormous practical value when considering systematic
errors from polarization and alignment imperfection and will make untangling
the various effect of these kinds of errors as trivial as demonstrating the
spin dependence of the parity shift.

In addition this analysis was independent of the detailed structure of the states
involved in the interaction. In particular, the results hold for completely
general total angular momentum for all of the states. The splitting to be measured
is generally considered to be that of the ground state sublevels, but the information
exists in the \( D_{3/2} \) shifts as well. The details of the systematic problems
are slightly different, as will be discussed extensively later, so that a \( D \)
state parity measurement is general less accurate than an \( S \) state measurement.
But it may still prove useful, for diagnostics or calibration, to additionally
or simultaneously study the \( D \) state shifts and this general spin flip
analysis shows quickly that this information is present and gives its general
size and structure.

\section{Detection }

\label{Sec:Detection}

This general analysis conveniently illustrates how the spin dependence of a
light shift can be used to detect the existence of a parity violating transition.
With the spin dependence for fields arbitrary polarizations and propagation
directions now well developed consider again the initially proposed fields,

\begin{eqnarray*}
\vec{E}^{D} & = & E^{D}cos\left( kz\right) cos(\omega t)\hat{x}\\
\vec{E}^{Q} & = & E^{Q}sin\left( kz\right) cos(\omega t+\phi )\hat{x}
\end{eqnarray*}
The co-rotating pieces of these fields are resonant, the counter-rotating pieces
are neglected in the rotating wave approximation so that the amplitudes that
appear in the couplings will be half of the applied amplitudes,

\begin{eqnarray*}
\vec{E}^{D} & = & \frac{E^{D}}{2}cos\left( kz\right) \hat{x}\\
\vec{E}^{Q} & = & e^{i\phi }\frac{E^{Q}}{2}sin\left( kz\right) \hat{x}
\end{eqnarray*}

\subsection{Amplitudes and Shifts}

These give quadrupole and dipole transition amplitudes of,
\begin{eqnarray*}
\Omega ^{Q}_{m^{\prime }m} & = & e\left\langle 5D_{3/2},m^{\prime }\left| xz\right| 6S_{1/2},m\right\rangle \partial _{z}E_{x}\\
 & = & e\left\langle 5D_{3/2},m^{\prime }\left| xz\right| 6S_{1/2},m\right\rangle e^{i\phi }kE^{Q}/2\\
\Omega ^{D}_{m^{\prime }m} & = & e\left\langle 5D_{3/2},m^{\prime }\left| x\right| 6S_{1/2},m\right\rangle E^{D}/2
\end{eqnarray*}
The quadrupole and PNC light shifts will then be given in terms of,
\begin{eqnarray*}
\left( \Omega ^{Q\dagger }\Omega ^{Q}\right) _{mm} & = & \sum _{m^{\prime }}\left\langle 5D_{3/2},m^{\prime }\left| xz\right| 6S_{1/2},m\right\rangle \left( keE^{Q}/2\right) ^{2}\\
\left( \Omega ^{D\dagger }\Omega ^{Q}\right) _{mm} & = & \sum _{m^{\prime }}\left\langle 6S_{1/2},m\left| x\right| 5D_{3/2},m^{\prime }\right\rangle \left\langle 5D_{3/2},m^{\prime }\left| xz\right| 6S_{1/2},m\right\rangle \\
 & \times  & e^{i\phi }keE^{D}eE^{Q}/4
\end{eqnarray*}
by
\begin{eqnarray*}
\delta \omega _{m}^{Q} & = & \sqrt{\left( \Omega ^{Q\dagger }\Omega ^{Q}\right) _{mm}}\\
\delta \omega _{m}^{PNC} & = & \varepsilon Im(\Omega ^{D\dagger }\Omega ^{Q})_{mm}/\delta \omega _{m}
\end{eqnarray*}
As before, the spin flipped matrix elements of this product are easily related
using \( F=R_{\hat{y}}(\pi ) \) by the number of \( x \)'s and \( z \)'s
in the both operators by \( (-1)^{n_{x}+n_{z}} \). This gives,
\begin{eqnarray*}
\delta \omega _{-m}^{Q} & = & \delta \omega _{m}^{Q}\equiv \delta \omega ^{Q}\\
\delta \omega _{-m}^{PNC} & = & -\delta \omega _{m}^{PNC}
\end{eqnarray*}
the light shift given by the parity violating transition is spin dependent.

\subsection{Relative Phases}

The PNC shift requires the imaginary part of \( \left( \Omega ^{D\dagger }\Omega ^{Q}\right) _{mm} \).
Previously is was shown that the phase of a matrix element of a cartesian operator
is given by the number of \( y \)'s in the operator, by \( i^{n_{y}} \). With
these fields no \( y \) appears so that the matrix elements are real and the
phase of the couplings, \( \Omega  \), are then given by the phases of the
field amplitudes. These phases were chosen so that the dipole field, \( E^{D} \)
is real, and its phase relative to the quadrupole field, \( E^{Q} \) is given
explicitly by \( e^{i\phi } \) so that both \( E^{D} \), and \( E^{Q} \)
are real. Then \( e^{i\phi } \) completely determined the phase of \( \Omega ^{Q} \)
and the PNC shift is proportional to \( Im(e^{i\phi })=sin\phi  \). Then a
non-zero PNC splitting requires \( \phi \neq 0 \) and is maximized for \( \phi =\pi /2 \)
. The splitting is the largest when the dipole and quadrupole fields are exactly
out of phase. 

This is an important constraint on the design of the experiment. A single standing
wave can be used to generate both dipole and quadrupole transitions simply by
positioning the ion in the standing wave at point slightly displaced from the
antinode. But for this configuration the amplitude and gradient of the electric
field are in phase. This would give no splitting and so a more complicated arrangement
must be made.

The phase dependence is an immediate consequence of the \( Im \) operator appearing
in the PNC shift which, in turn, is a consequence of the purely imaginary phase
of the \( S-P \) mixing, recall \( S=S_{0}+i\varepsilon P_{0} \) and this
phase was a consequence of the \( T \)-even symmetry of the parity-violating
interaction \( \vec{\sigma }\cdot \vec{p} \). The \( T \) symmetry of the
parity violating coupling further determines the general requirements for the
measurement.

\subsection{Reduced Matrix Elements}

The cartesian form of the operators most easily shows the spin dependence, but
for explicit calculations a spherical basis is more convenient. For these fields,
\begin{eqnarray*}
E^{D}_{\pm 1} & = & \mp E^{D}_{x}/\sqrt{2}=\mp E^{D}/2\sqrt{2}\\
E^{Q}_{\pm 1} & = & \mp \partial _{z}E^{Q}_{x}/2=\mp e^{i\phi }kE^{Q}/4
\end{eqnarray*}
The Wigner-Eckhart Theorem can then be used to calculate the \( m \) dependence
of the couplings,
\begin{eqnarray*}
\Omega ^{Q}_{m^{\prime }m} & = & \left\langle 5D_{3/2},m^{\prime }\left| -r^{(2)}_{+1}+r^{(2)}_{-1}\right| 6S_{1/2},m\right\rangle e^{i\phi }keE^{Q}/4\\
 & = & -\frac{\left\langle 5D_{3/2}\left| \left| Q\right| \right| 6S_{1/2}\right\rangle }{\sqrt{2(3/2)+1}}\sum _{s=\pm 1}s\left\langle \frac{3}{2},m^{\prime }|2,s;\frac{1}{2}m\right\rangle e^{i\phi }keE^{Q}/4\\
\Omega ^{D}_{m^{\prime }m} & = & \left\langle 5D_{3/2},m^{\prime }\left| -r^{(1)}_{+1}+r^{(1)}_{-1}\right| 6S_{1/2},m\right\rangle eE^{D}/2\sqrt{2}\\
 & = & -\frac{\left\langle 5D_{3/2}\left| \left| D\right| \right| 6S_{1/2}\right\rangle }{\sqrt{2(3/2)+1}}\sum _{s=\pm 1}s\left\langle \frac{3}{2},m^{\prime }|1,s;\frac{1}{2}m\right\rangle eE^{D}/2\sqrt{2}
\end{eqnarray*}
With the Clebsch-Gordan coefficients non-zero only for \( m^{\prime }=m+s \),
the products of couplings are then given by,
\begin{eqnarray*}
\left( \Omega ^{Q\dagger }\Omega ^{Q}\right) _{mm} & = & \frac{\left\langle Q\right\rangle ^{2}}{4}\left( \sum _{s=\pm 1}\left\langle \frac{3}{2},m+s|2,s;\frac{1}{2}m\right\rangle ^{2}\right) \left( \frac{keE^{Q}}{4}\right) ^{2}\\
\left( \Omega ^{D\dagger }\Omega ^{Q}\right) _{mm} & = & \frac{\left\langle Q\right\rangle \left\langle D\right\rangle }{4}\left( \sum _{s=\pm 1}\left\langle \frac{3}{2},m+s|2,s;\frac{1}{2}m\right\rangle \left\langle \frac{3}{2},m+s|1,s;\frac{1}{2}m\right\rangle \right) \\
 & \times  & e^{i\phi }\frac{keE^{Q}}{4}\frac{eE^{D}}{2\sqrt{2}}
\end{eqnarray*}
The spin dependence of these terms has already been established, so just pick
\( m=1/2 \) , for example. The Clebsch-Gordan coefficients give,
\begin{eqnarray*}
\sum _{s=\pm 1}\left\langle \frac{3}{2},\frac{1}{2}+s|2,s;\frac{1}{2}\frac{1}{2}\right\rangle ^{2} & = & \frac{4}{5}\\
\sum _{s=\pm 1}\left\langle \frac{3}{2},m+s|2,s;\frac{1}{2}m\right\rangle \left\langle \frac{3}{2},m+s|1,s;\frac{1}{2}m\right\rangle  & = & \frac{2}{\sqrt{5}}
\end{eqnarray*}
and the product matrix elements finally become,

\begin{eqnarray*}
\left( \Omega ^{Q\dagger }\Omega ^{Q}\right) _{mm} & = & \frac{\left\langle Q\right\rangle ^{2}}{5\cdot 16}\left( keE^{Q}\right) ^{2}\\
\left( \Omega ^{D\dagger }\Omega ^{Q}\right) _{\pm m,\pm m} & = & \pm \frac{\left\langle Q\right\rangle \left\langle D\right\rangle }{16\sqrt{10}}e^{i\phi }keE^{Q}eE^{D}
\end{eqnarray*}
and the shifts are given by
\begin{eqnarray*}
\delta \omega ^{Q} & = & \frac{\left\langle Q\right\rangle }{4\sqrt{5}}keE^{Q}\\
\delta \omega _{m}^{PNC} & = & \pm \varepsilon sin\phi \frac{\left\langle D\right\rangle }{4\sqrt{2}}eE^{D}
\end{eqnarray*}
As before, the parity shift is maximized for \( sin\phi =1 \) so take \( \phi =\pi /2 \).

\subsection{Approximate Sizes}

Precise atomic structure calculations to obtain the reduced matrix elements
are needed to accurately relate the shifts to the field amplitudes and parity
mixing. Reasonable estimates can be obtained from simple approximations such
as by fitting asymptotically to Coulomb wavefunctions, \cite{Schacht00}, or
by using measured quantities such as lifetimes and resonance linewidths. Order
of magnitude estimates can be made easily from typical characteristic dimensional
scales. Atomic length scales are angstroms, so matrix elements of coordinate
operators should have angstrom like sizes,
\begin{eqnarray*}
\left\langle Q\right\rangle  & \sim  & A^{2}\\
\left\langle D\right\rangle  & \sim  & A
\end{eqnarray*}
For practical reasons that are discussed later electric fields of \( 10^{4}V/cm \)
are about the largest that should be used. In terms of a frequency,
\[
1eV/cm\sim 2.4MHz/A\]
With a laser wavelength of \( 2.05\mu  \) this gives a quadrupole shift of,
\begin{eqnarray*}
\delta \omega ^{Q} & = & \frac{\left\langle Q\right\rangle }{A^{2}}\frac{E^{Q}}{10^{4}V/cm}\frac{1}{4\sqrt{5}}\frac{2\pi }{2.05}10^{4}2.4MHz\frac{A}{\mu }\\
 & = & \frac{\left\langle Q\right\rangle }{A^{2}}\frac{E^{Q}}{10^{4}V/cm}0.34MHz
\end{eqnarray*}
and the splitting,
\begin{eqnarray*}
\Delta \omega ^{PNC} & = & \delta \omega _{1/2}^{PNC}-\delta \omega _{-1/2}^{PNC}=2\delta \omega _{1/2}^{PNC}\\
 & = & \varepsilon \frac{\left\langle D\right\rangle }{A}\frac{E^{D}}{10^{4}V/cm}\frac{1}{2\sqrt{2}}24GHz\\
 & = & \varepsilon \frac{\left\langle D\right\rangle }{A}\frac{E^{D}}{10^{4}V/cm}8.5GHz
\end{eqnarray*}
With \( \varepsilon \sim 10^{-11} \) this gives,
\[
\Delta \omega ^{PNC}=\frac{\varepsilon }{10^{-11}}\frac{\left\langle D\right\rangle }{A}\frac{E^{D}}{10^{4}V/cm}0.08Hz\]

With these fields, the spin independent quadrupole shift is of order \( MHz \)
and the parity splitting on the order of \( 0.1Hz \). A more accurate calculation
of the parity mixing and transition matrix elements actually improves this anticipated
splitting considerably, see \cite{KristiThesis} for reference, giving,

\[
\Delta \omega ^{PNC}\approx 2Hz\]

\subsection{Spin Resonance}

Finally, the problem of measuring parity violation reduces to measuring this
small differential shift of the ground state energies. This will be done by
using techniques that make conventional ion state manipulation and detection
techniques spin sensitive. Full details about the development, implementation
and preliminary applications of these techniques are given in sec.\ref{Sec:SpinStuff},
but the general strategy is easy to illustrate.

This small \( \delta \omega ^{PNC}\sim Hz \) would be difficult to measure
directly. If detected as an energy splitting by, for example, measuring the
resonance frequency of the spin transition, the transition would have to be
driven with an applied field of less than \( 1Hz \). With the inevitable environmental
perturbations, in particular for this case noisy broadband sources, it may be
difficult to drive this transition coherently at such a slow rate. Alternately,
the shifts could be measured directly as a precession the the spin by starting
the ion in a spin up state, and measuring how long it takes to get to spin down.
In this mode the precession is affected by stray magnetic fields and so the
size and even the direction of the resulting precession axis would not be well
known, or stable.

\begin{figure}
{\par\centering \resizebox*{1\textwidth}{!}{\includegraphics{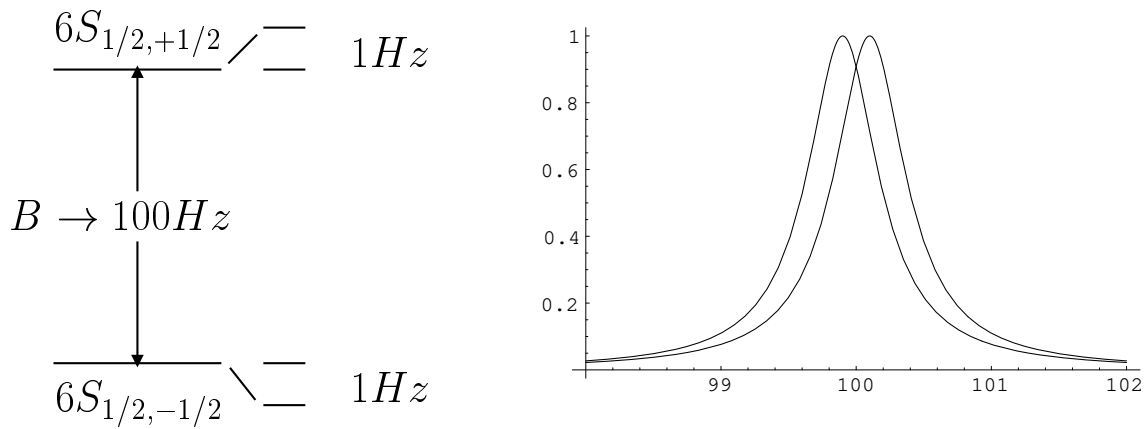}} \par}

\caption{\label{Fix:PNCResonanceShift}PNC generated splitting is detected as shift
in resonance frequency of the spin flip transition from its initial value given
the a Zeeman splitting from an applied static magnetic field.}
\end{figure}

The effects of these kinds of perturbations can be minimized by instead measuring
\( \delta \omega ^{PNC} \) as a change in the splitting of the states due to
an existing applied magnetic field, sec.\ref{Sec:OffDiagonalMatrixElements}.
To maximize sensitivity and minimize systematic problems it will turn out to
be convenient to apply a magnetic field that gives an initial \( 0.1-1kHz \)
splitting of the ground state magnetic sublevels. After setting the ion to a
specific initial spin state, a transition between these spin states can then
be driven with an oscillation magnetic field at the same frequency and checking
for a transition by determining the resulting final spin state. The time averaged
transition probability as a function of the frequency of the applied interaction
will give the usual Lorentzian resonance profile centered around the splitting
frequency. The splitting is then available immediately from the resonance frequency.
When the lasers driving the parity violating transition are applied, the energy
difference of the states is further changed by \( \Delta \omega ^{PNC} \) which
immediately appears as a shift of the resonance frequency. Measuring this shift
then provides \( \Delta \omega ^{PNC} \) directly, \ref{Fix:PNCResonanceShift}.

\subsection{Calibration}

From this resonance frequency shift, \( \varepsilon  \) can finally be determined
and in turn \( Q_{W} \) to provide constraints on \( S \) and \( T \). Generating
\( \varepsilon  \) requires the dipole reduced matrix element, \( \left\langle D\right\rangle  \),
as and the dipole electric field strength, \( E^{D} \). As already discussed,
\( \left\langle D\right\rangle  \) will require precise atomic structure calculations,
\( E^{D} \) will require additional measurements. 

This field would be difficult to accurately predict from the properties of the
applied beam. The power of a laser can be easily monitored external to the UHV
chamber that contains the ion, but before reaching the ion it passes through
windows and past obstacles whose absorption characteristics are not well known.
These could, in principle, be measured independently at another time, before
or after assembly of the system, but it would be difficult to guarantee that
this behavior is stable at the required precision of \( 10^{3} \). Even if
the incident power on the ion was accurately known, the field size also depends
on the beam size, geometry and placement. Placement of the ion relative to the
beam would be particularly difficult to determine independently. 

These practical difficulties make it effectively impossible to accurately predict
the size of the electric field at the ion and for precision measurements the
electric field at the ion must be measured directly using the ion. This could
be done in the same way that the parity shift is measured, using a light shift
that is also dependent on only the dipole field, \( E^{D} \), but now not also
a function of \( \varepsilon  \). There is no immediately obvious candidate
for such a quantity. For the field configuration used to generate the PNC splitting,
the only measurable resonant \( E^{D} \) dependent shift is the \( Q-D \)
PNC shift itself, the \( Q-Q \) shift is \( E^{D} \) independent since ideally
the ion is placed at the antinode of this field, and the previously neglected
\( D-D \) shift is \( o(\varepsilon ^{2}) \). 

Only the quadrupole amplitude is \( \varepsilon  \) independent, so any \( \varepsilon  \)
independent measurement of \( E^{D} \) would require using the quadrupole transition.
The dipole field is ideally placed to that its contribution to the quadrupole
amplitude is zero. Using the quadrupole amplitude to measure the dipole field
strength then require altering the position of the node of the dipole field
so that the dipole field has a non-zero gradient at the ion. With the same linear
polarization the dipole field then gives a spin independent shift of the \( D \)
and \( S \) magnetic sublevels, or more carefully a spin sign independent shift
than can then be measured in the \( D \) state as a change in the separation
of the \( m=\pm 1/2 \) and \( m=\pm 3/2 \) states. Alternately, the beam could
be made to be circularly polarized and the resulting spin dependent shift measured
in the ground state.

Such a strategy is less than ideal because then the amplitude of the field is
not measured under exactly the same conditions that it is used for the parity
shift. Altering the beam could subtly change the amplitude, and then depends
on other parameters, polarization, node position, that must also be precisely
determined and controlled. Also, quadrupole rates are \( \sim MHz \) so that
the resulting shifts are characteristically much larger than the PNC shift and
even larger than the initial splitting given by the magnetic field. Also as
discussed extensively later, \ref{Sec:OffDiagonalMatrixElements}, \ref{Sec:NonresonantLightShifts},
uncertainties in positioning of the beam require that measured shifts be much
smaller than the initial, applied magnetic field generated splitting to reduce
possible systematic errors. 

This would not be a difficulty in an independent measurement, but generally
precludes calibrating the field and measuring the PNC shift simultaneously.
The PNC shift requires a much smaller field and intermittently changing the
size of the applied field is highly undesirable. Precise stability of the magnetic
field during either measurement is required and regularly returning precisely
to two different values is very difficult in practice partly due to non-linear
and hysteresis contributions from ferromagnetic materials, which would also
introduce a long time scale for adjustment if repeatability is possible at all.
Calibration would then have to be done at as a separate measurement, which risks
errors from instabilities of laser power or position, or ion position that alter
the field significantly in the time between parity measurement and calibration. 

The calibration shift could be made intensionally much smaller by adjusting
the beam position only slightly to give a small gradient due to the dipole field
rather than the maximum gradient, or only party circularly polarizing the beam,
but either case then requires precisely measuring these parameters as well,
and these intermediately cases are likely more difficult to determine and control
than some extremal maximum or zero condition.

A final possibility is to measure some non-resonant shift. So far only shifts
due to the \( 6S_{1/2}-5D_{3/2} \) resonance have been considered. The effects
to either state from couplings to non-resonant states are generally very small
and have so far been neglected, but as the PNC shift turns out to be very small
these non-resonant shifts may turn out to be relatively very large, and could
possibly be used for calibration though interpretation would require far more
difficult atomic structure calculations as many transition matrix elements would
be required at the same very high part in \( 10^{3} \) precision, but has the
advantage that it requires no alteration of the beam geometries. In fact, as
seen in \ref{Sec:NonResonantDStateShifts}, these shifts are also \( MHz \)
sized, which, as with the resonant shifts, is obstructively large. 

Any of these methods may be made more practical if the dipole beam can be precisely
attenuated by a factor of \( 10^{3} \) or more to make the resulting shift
more manageable. In either case calibration remains a difficult problem and
must considered as carefully as systematic errors in the measurement of \( \delta \omega ^{PNC} \)
itself. The general possibilities are more easily considered after solving for
the shifts given by arbitrary electric fields. Hints of that structure were
seen from these perturbative calculations, but this route is tedious, and not
obviously correct for general fields. The vector structure can be calculated
easily, and correctly, with the general methods developed in sec.\ref{Sec:LightShiftVectorStructure}.

\chapter{Sensitivity and Systematics}

With the techniques outlined in \ref{Sec:Detection}, and analyzed and implemented
as described in \ref{Sec:SpinStuff}, this few \( Hz \) parity violating splitting
of the ground state magnetic sublevels, though small, would be easily detectable,
provided the system involved only what was so far explicitly described. Of course,
to the contrary, real life barges in with innumerable imperfections, perturbations
and complications which force the consideration of a great many practical constraints. 

These effects can be broadly separated into two general categories. Sensitivity
is the easiest to understand and analyze, and in the end, probably the easiest
to study and optimize. Very simply, the system must not be so noisy that the
shift cannot be detected in a reasonable amount of time, the signal to noise
ratio is high enough that the shift can be detected and precisely measured quickly
enough to be practical and ideally the measurement time is short to minimize
possible systematic problems from drifting rates and alignments.

More challenging, and more sinister, are the systematics errors. This parity
violating shift of a few \( Hz \) is very small compared to the typical \( Mhz \)
rates of atomic processes. Though it will be detectable, there are many other
small effects from parity conserving processes, from impure polarizations, or
non-ideal beam geometries to processes previously neglected in traditional approximations,
that result in a non-parity-symmetric environment and so also can change the
separation of the states by energies that are possibly much larger than the
parity shift. This now affects interpretation. The virtue of the method outlined
so far was that the splitting was generated only by the parity violating transition,
it was proportional to \( \varepsilon  \), and so was an unambiguous signal
of parity violation. With these various potential systematics problems, the
splitting is now polluted by possibly unknown contributions from other processes
and the contribution from true parity violating processes cannot be cleanly
resolved. The possible errors must be exhaustively considered and strategies
developed to detect and minimize or correct for them all.

\section{Sensitivity}

The parity violation induced shift will be measured by a change in the resonance
frequency of a spin flip transition driven by an applied RF magnetic field.
The precision to which the shift can be measured then depends on the width of
this transition resonance and the statistical accuracy to which its position
can be determined. The dependence of the S/N on the linewidths is straightforward.
Clearly, a small shift is better resolved with a narrow linewidth. For a shift
of a few \( Hz \) the linewidth should be no larger than a few \( Hz \), and
for precise determination the width should be many times smaller than the shift.
Larger linewidths can be compensated for with better statistics to more accurately
determine the shape and position of the resonance profile. That typically proves
less efficient than working to reduce the linewidth, as the sensitivity in general
behaves like \( \delta \omega /\omega \propto \Gamma /\sqrt{N} \) where \( \Gamma  \)
is the transition line width and \( N \) is the number of data points taken.
An increase in \( \Gamma  \) by a factor of 2 then requires and increase in
\( N \) by a factor of 4 to yield the same sensitivity.

The detailed dependence of the sensitivity on statistics depends on the particular
techniques used to measure the shift and is fully developed in \ref{Sec:StatisticsAndSensitivity}
where the proper context for the discussion will have already been established.
Understanding and optimizing the linewidth can be done independently. The general
result will always be some lineshape with some characteristic width and generally
a narrower transition profile will give higher sensitivity, though in practice
the linewidth may depend on on other parameters like observation times or transition
rates that also directly effect sensitivity and may have well defined optimal
values or ranges that favor a particular linewidth, rather than just the narrowest
possible. The linewidth will have generally have contributions from external
perturbations, atomic structure related physical limits, and fundamental properties
of the detection technique.

\subsection{Precession and Resonance}

In an ideal environment, with purely monochromatic laser and an absolutely stable
magnetic field the linewidth is determined solely by the transition rate \( \Omega  \)
as the resonance profile is given simply by \( 1/\left( 1+\left( \delta \omega /2\Gamma \right) ^{2}\right)  \).
In principle, \( \Omega  \) can be arbitrarily small, and the transition linewidth
can be made arbitrarily narrow, though a slow transition rate also requires
long measurement times, and unusually long observation times are cumbersome
and patience draining. In practice, the linewidth is influenced by other noisy
processes that either directly change the resonance frequency, and as a result
smear out the resonance, or reduce the effective lifetime of the spin states
generating an intrinsic minimum linewidth from decoherance and giving a practical
maximum for the interaction time for a single measurement trial. 

For the ion experiment this coherence time is a very long 10's of seconds so
that the rate be very slow, and the associated linewidth very narrow. Under
these conditions the character of the resonance profile changes since it is
reasonable to expect to be able to resolve the precession. For a two state system
the time dependent probability to be in the initially unoccupied state is,
\[
p(t)=\frac{sin^{2}\left( \sqrt{1+(\delta \omega /2\Omega )^{2}}\Omega t\right) }{1+(\delta \omega /2\Omega )^{2}}\]
A time average gives the expected Lorentzian signal but when the actual precession
is resolved the profile includes oscillations with the lorentzian envelope.

\begin{figure}
{\par\centering \includegraphics{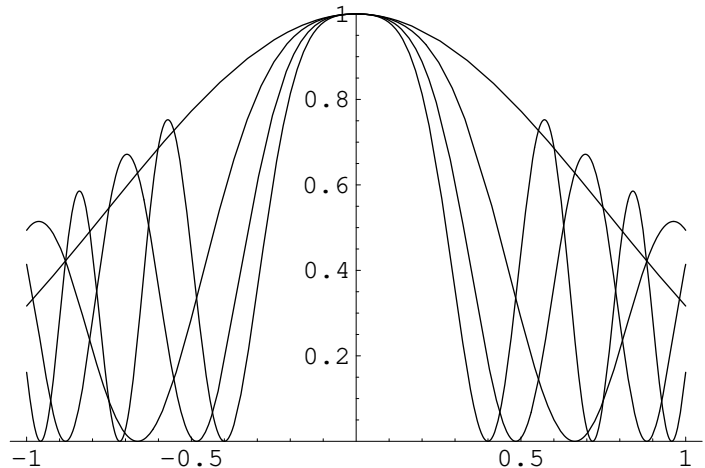} \par}

\caption{\label{Fig:SpinPrecessionProfile}Transition probability as a function of frequency
for resolved spin precession when \protect\( \Omega t=(n+1/2)\pi \protect \)
for \protect\( n=0,2,4,6\protect \).}
\end{figure}

This suggests a possibly much higher sensitivity than suggested by the linewidth
from the transition rate alone. Consider selecting \( \Omega t \) such than
on resonance, at \( \delta \omega =0 \), the spin makes \( n+1/2 \) complete
revolutions, so that it ends up in the opposite state that it was started in.
This require \( \Omega t=(n+1/2)\pi  \). If \( n \) is chosen such that for
\( \delta \omega =\delta \omega ^{PNC} \) the slightly higher transition rate
yields an extra 1/2 revolution in the same time, the ion will end in the same
spin state it started, completely flipped from the on resonance case, fig.\ref{Fig:SpinPrecessionProfile}.
 This gives a 100\% difference in the signal even though the resonance denominator
may have changed very little. Notice that the complementary case of choosing
\( \Omega t=n\pi  \) would have yielded a slightly lower difference as the
on resonance final state would be the initial state, but the off-resonance final
state would be be completely the opposition state as the lorentzian envelope
makes the amplitude of the oscillation slightly less than one. For sufficiently
large \( \Omega  \) this difference would be negligible and the experiment
could be done either way. 

Then, if the profile is sampled sufficiently densely the width should be regarded
as the distance between the zeros. This corresponds to the argument of the sign
changing by \( \pi  \). With \( \Omega t=(n+1/2)\pi  \) the adjacent zeros
on either side will be at \( (n+1)\pi  \), 
\begin{eqnarray*}
\left( 1+\left( \frac{\delta \omega }{2\Omega }\right) ^{2}\right) \left( (n+1/2)\pi \right) ^{2} & = & \left( (n+1)\pi \right) ^{2}\\
\left( \frac{\delta \omega }{2\Omega }\right) ^{2} & = & \left( \frac{n+1}{n+1/2}\right) ^{2}-1\\
 & = & \left( \frac{1+1/n}{1+1/2n}\right) ^{2}-1
\end{eqnarray*}
For moderately large \( n \),
\begin{eqnarray*}
\left( \frac{\delta \omega }{2\Omega }\right) ^{2} & \approx  & \left( 1+\frac{1}{2n}\right) ^{2}-1\\
 & \approx  & \frac{1}{n}\\
\delta \omega  & \approx  & 2\Omega /\sqrt{n}
\end{eqnarray*}
The width is still of order \( \Omega  \) and scales linearly with it, but
is reduced by the factor \( \sqrt{n} \). With only a modest number of revolutions
the linewidth can be reduced significantly, a factor of 2-3, with far less work
than it would take to reduce the linewidth by other means. The full consequences
of this kind of measurement on the sensitivity is discussed in \ref{Sec:StatisticsAndSensitivity}.
Exploiting this requires being able to make coherent transition which in turn
depends on any limits on the coherence time from other sources being much longer
than the observation time. 

Resolving the precession also means that the parity shift could be detected
by measuring the transition probability as a function of time rather than frequency,
looking, for example, at the shift of zeros. The increased rate compared to
exactly on resonance would move cause any non-trivial zeros to occur sooner
by \( \Delta t \). Again with the on resonance pulse giving \( n+1/2 \) rotations,
if the frequency shift is small compared to the lorentzian linewidth the frequency
dependence can be neglected and this time shift is given by,

\begin{eqnarray*}
\Omega (t-\Delta t) & = & n\pi \\
(n+\frac{1}{2})\pi -\Omega \Delta t & = & n\pi \\
\Delta t & = & \frac{\pi }{2\Omega }
\end{eqnarray*}
The shift in the zero get smaller with increasing rates, then favoring slower
rates just as the linewidth as a function of frequency. At least would complete
rotation would have to be mapped out in detail to determine the position of
the zeros. Either method may have additional practical advantages or systematic
problems, but both appear to have improved sensitivity over using simple resonance
profiles.

\subsection{Magnetic Field Noise}

A transition linewidth determined solely by the amplitude of the interaction
driving the transition allows for arbitrarily narrow linewidths. Environmental
and physical constraints generally enter at some level to give a practical lower
bound. Stray magnetic fields give this kind of bound by directly altering the
spin state energy difference and generating a finite observed transition linewidth.

The shift detected, due to the presence of the parity violating transition,
is the shift of a resonance frequency initially determined by an applied magnetic
field. A magnetic field separates the energies of the spin state sublevels by
a few \( MHz \) per \( Gauss \). If this magnetic field fluctuates, the resonance
frequency fluctuates and yields a large spin flip transition linewidth. Sources
of magnetic field noise and drifts abound, from distant motors to the changing
properties of ferromagnetic materials with time and fluctuating temperatures.
These contributions are typically around a few \( mG \), resulting in relatively
very large \( kHz \) sized changes in the resonance frequencies. Linewidths
of a few tenths of a \( Hz \) are necessary to efficiently detect parity shift.
This requires constraining magnetic field fluctuations to less than a \( \mu G \). 

External sources of magnetic field noise must be attenuated by factors of 1000
or more with shielding. This is challenging, but straight-forward. Better performance
is routinely achieved with other projects such as the Mercury EDM experiment.
Local sources of magnetic field fluctuation must be eliminated by a careful
selection of materials. The linewidth in the current system, is dominated by
magnetic field fluctuations. The possible identification of these sources and
means of eliminating them is discussed in more detail in \ref{Sec:SpinFlipTransitions}.

Field fluctuations can also come directly from the applied fields, largely through
the currents used to generate them. Fields are provided by coils and the size
of a field is dependent on coil geometry and placement, and the current in the
coil. Stability of the applied magnetic field then corresponds to stability
of the coil current. Simple electronics are accurate and stable to about a percent
with fluctuations largely temperature related. Some attention to design and
component selection can yield stability to a part in \( 10^{3} \) and careful
attention can increase this a bit more, but significant effort is required to
improve this much beyond a part in \( 10^{4} \). This should be sufficient
for the purposes of this experiment as the initial splitting will be set to
be a few \( kHz \) or less. Current stability of better than a part in \( 10^{3} \)
gives a field stable to less than a \( Hz \) which should reduce this fluctuations
contribution to the line width to less than the limit anticipated from magnetic
field noise.

Generally only fluctuations in the same direction as the applied field significantly
affect sensitivity as those change the magnitude of the total, net magnetic
field linearly. From simple geometry, a perpendicular fluctuation gives a change
of \( \delta B=\sqrt{B^{2}+\delta B_{\perp }^{2}}\approx B(1+\delta B_{\perp }^{2}/B^{2}) \)
which is second order correction in the size of the fluctuation. With the \( mG \)
sized applied fields expected to give a \( kHz \) sized initial shifts, perpendicular
fields with sizes of up to 10's of \( \mu G \) yield changes in the total length
of only a part in \( 10^{4} \) , a few tenths of a \( Hz \) . Typical geometries
for magnetic shields typically shield better along particular directions and
this insensitivity to perpendicular fluctuations eases shield construction and
makes controlling field fluctuations easier in general since only one direction
needs intense attention.

\subsection{\protect\( D\protect \) State Lifetime Shortening}

The initial spin states in both the \( S \) and the \( D \) states are very
long lived. The \( D \) state has a finite lifetime of about \( 80s \) from
possible decay to the ground state, but there is no fast mechanism than changes
the spin. The dominant natural decay process would be some kind of direct magnetic
dipole transition between states. 

A magnetic dipole decay is much slower than an electric dipole from the transition
matrix elements, though typically about the same rate as an electric quadrupole.
Even more significantly, a dipole decay width depends on the cube of the energy
difference between the states. Magnetic sublevels separated by energies of about
a \( kHz \) give transition energies \( 10^{12} \) times smaller than for
transitions at optical wavelengths, such as for the \( D\rightarrow S \) quadrupole
decay, so the lifetimes associated with these kinds of transitions should be
some \( 10^{36} \) times longer, something like \( 10^{38} \) seconds, impossibly
long to observe and an irrelevant perturbation.

However, the spin states involved in the parity measurement are not these pure
stable spin states. The IR laser couples the \( S \) and \( D \) states and
on resonance the occupation probability is equally divided between the \( S \)
and \( D \) states with the detailed distribution among the spin states given
by Clebsch-Gordan coefficients and the particular polarizations. With this mixing,
transitions between spin states in one energy level can be made through the
intermediate states in the other energy level. For example, a transition from
\( S_{-1/2} \) to \( S_{+1/2} \) is now possible from a driven in\( \Delta m=+1 \)
transition to the \( D_{+1/2} \) state and a \( \Delta m=0 \) decay. This
gives an effective short spin state lifetime from the finite lifetime of the
\( D \) state. Since the time average probability to be in the \( D \) state
is \( 1/2 \), the decay rate between spin states is about half the decay rate
of the \( D \) state.

For the unperturbed \( D \) state, this yields a spin lifetime of around \( 150s \),
which is also a minor perturbation. The complication is that the \( D \) state
is not unperturbed. Here is the first re-appearance of a previously neglected
process. So far, only the resonant shifts of the \( S \) and \( D \) state
due to the applied light was considered. This is reasonable since it is certainly
the largest effect, but it is not the only effect. In particular, there is a
strong dipole coupling from the \( D_{3/2} \) state to the \( P_{1/2} \) and
\( P_{3/2} \) states in particular, as well as other \( P \) and \( F \)
and other states in the atom that will be driven by the applied parity laser.
These transitions will be driven far off resonance, yielding small adjustments
to the effects already discussed, but the parity shift itself is small so these
perturbation might not, in this case, be negligible.

These non-resonant dipole couplings will be discussed in more detail later as
they also produce shifts of the spin state energies, but their effects appear
here as well. With the \( D_{3/2} \) state weakly coupled to the \( P \) states
and others, a decay to the ground state through these states to the ground state
is now possible, rather than through a direct quadrupole decay. The amplitude
to be in any of these other states is small since the transition is driven for
far out of resonance, but the decay rate for these states is now the quick \( MHz \)
associated with a typical dipole decay, so the resulting effect can be a noticeable
perturbation. The excitation profile is lorentzian, giving an occupation probability
oscillating with amplitude \( 1/\left( 1+\left( \delta \omega /\Omega \right) ^{2}\right)  \),
with \( \Omega  \) the excitation rate, the time average occupation probability
is just half this. The effective decay rate of the spin states, \( \Gamma  \),
is this probability times the decay rate out of the coupled intermediate state,
\( \Gamma _{i} \). Far off resonance this becomes, \( \Gamma =\left( \Gamma _{i}/2\right) \left( \Omega _{i}/\delta \omega _{i}\right) ^{2} \). 

The total width is the sum of contributions like this from all the states. Increasing
\( \Omega  \) increases the parity splitting, but now apparently also increases
the width of the \( D \) state and, as a result, reduces the lifetime of the
spin states adding this contribution to \( \Gamma _{D}/2 \). The signal to
noise will be proportional to \( \Omega /\left( \Gamma _{D}+\sum _{i}\Gamma _{i}\left( \Omega _{i}/\delta \omega _{i}\right) ^{2}\right)  \).
When this extra width becomes comparable to the width of the \( D \) state,
the sensitivity starts to decrease, even as the transition rate is increased.
Dipole decay rates are around a \( MHz \), the \( D \) state lifetime is something
like \( 0.01Hz \), so the S/N begins to be reduced for \( \Omega =\delta \omega \sqrt{\Gamma _{D}/\Gamma _{P}}\approx \delta \omega 10^{-4} \).
For a \( 6P \) state, \( \delta \omega \approx 3\times 10^{14}Hz \) giving
a maximum rate of \( \Omega \approx 30GHz \). 

There is a similar contribution from both \( P \) states, reducing this practical
maximum by about a half, \( \Omega \approx 15GHz \), and smaller contributions
the reduce the maximum further. A more careful calculation, including crude
calculations of all the dipole matrix elements involved, yields an upper bound
of \( 12GHz \), nicely consistent with this estimate. This is the origin of
the field strength used to estimate the size of the parity splitting, an electric
field of \( 10^{4}V/cm \) gives a \( 6P_{1/2}\rightarrow 6S_{1/2} \) dipole
transition rate of about \( 8.5GHz \). It the the largest field that can be
used before beginning to significantly reduce S/N due to this previously neglected
off-resonant dipole coupling.

\subsection{IR Laser Frequency and Intensity Noise}

The analysis presented so far assumes purely monochromatic light. Real lasers
have a finite linewidth, giving a fluctuating frequency, and variable intensity.
The precise effects will depend on the details of the broadening mechanism and
in general can be very complicated. Their general character can be understood
by considering just adiabatic changes of frequency and linewidth so that the
effects are given approximately simply by making the parameters of the original
static solution time dependent

Intensity fluctuations are straight-forward. Apparently this could be a significant
problem. The absolute shifts of the states are around \( 1MHz \) from the quadrupole
coupling. The parity splitting comes from the small spin dependence of these
large shifts. Even a small change in intensity would change these absolute shifts
by an energy far larger than the parity splitting. However, the quadrupole shifts
of these spin states are correlated and large fluctuations of the shift don't
significantly change the difference between the spin state. As see before, the
splitting is linear in the dipole electric field and independent of the exact
size of the quadrupole field, so a fluctuating intensity gives only a fractional
change of the splitting. For a splitting of about \( 1Hz \), and a desired
linewidth of less than \( 0.1Hz \), the intensity must be stable to \( 10\% \).
This is a very weak requirement, \( 10\% \) stability is achieved by even a
dye laser. For the solid state systems intended to be used here, stabilities
better than \( 1\% \) can be expected. Even if the light source itself is not
sufficiently stable to avoid broadening the spin flip transition, active stabilization
is easily done to \( 1\% \). 

The splitting is even less sensitive to frequency fluctuations. The shifts are
given by \( \omega _{m}^{2}=\delta \omega ^{2}+\Omega _{m}^{2} \). Again the
possible large frequency dependent variations are correlated and the difference
in the shifts is largely independent of frequency. The quadrupole rate dominates
the interaction and runs at about \( 1MHz \). The difference between \( \Omega _{m} \)
and \( \Omega _{-m} \), at about \( 1Hz \), is significantly smaller so that
\( \Omega _{m}=\Omega _{-m}+\Delta \omega ^{PNC}\approx \Omega _{m}+o\left( \Delta \omega ^{PNC}/\Omega _{m}\right)  \),
and the splitting becomes
\begin{eqnarray*}
\Delta \omega  & = & \omega _{m}-\omega _{-m}\\
 & = & \sqrt{\delta \omega ^{2}+\Omega _{m}^{2}}-\sqrt{\delta \omega ^{2}+\Omega _{-m}^{2}}\\
 & = & \Omega _{m}\sqrt{1+\delta \omega ^{2}/\Omega _{m}^{2}}-\Omega _{-m}\sqrt{1+\delta \omega ^{2}/\Omega _{-m}^{2}}\\
 & = & \left( \Omega _{m}-\Omega _{-m}\right) \left( \sqrt{1+\delta \omega ^{2}/\Omega _{m}^{2}}+o(\delta \omega \Delta \omega ^{PNC}/\Omega ^{2}_{m})\right) 
\end{eqnarray*}

The parity laser should have a linewidth of much less than \( 100kHz \), likely
approaching \( 10kHz \), so the \( \Delta \omega ^{PNC} \) independent correctly
is by far the largest and the splitting near resonance is modified by
\[
\Delta \omega ^{PNC}=\left( \Omega _{m}-\Omega _{-m}\right) \left( 1+o(10^{-2})\right) \]
 This correction improves to \( 10^{-4} \) for the better linewidth, and in
either case is sufficient to make the resulting broadening less than \( 0.1Hz \).

Clearly these adiabatic fluctuations do not pollute the measurement. There may
still be some concern with other kinds of laser noise. For example, frequency
components or sidebands of the noise resonant with the parity transition with
the proper polarizations could directly cause spin flip transitions that give
an effective broadening. This possibility requires a more sophisticated analysis
but is also expect to be insignificant.

\subsection{Trapping Fields}

Direct transitions between spin states leading to relaxation can also come from
other fields. Aside from noise sources, one field necessarily present will be
the RF electric fields required for trapping the ion. The details of these fields
will be discussed in \ref{Sec:IonTraps}, the general environment is a RF quadrupole
electric field, \( E \), with sizes of a few tens of \( V/m \), frequencies,
\( \omega  \), of a few 10's of \( Mhz \) and gradients with length scales,
\( \Delta x \), of a few hundred \( \mu m \) that can couple spin states in
the same energy level through a quadrupole transition. These fields would not
be exactly periodic and could lead to some kind of decoherance of the spin.
The exact structure would be hard to calculate but its general size can easily
be estimated as usual. The characteristic size of an electric quadrupole amplitude
was already determined in estimating the \( S-D \) quadrupole rate generated
by the parity lasers. Slightly modifying the size of the terms in that estimate
gives,

\begin{eqnarray*}
\delta \omega ^{Q} & = & e\left\langle Q\right\rangle \frac{E}{\Delta x}\\
 & = & \frac{\left\langle Q\right\rangle }{A^{2}}\frac{E/(10V/m)}{\Delta x/100\mu m}\frac{A^{2}}{100\mu m}\frac{10eV}{cm}\\
 & = & \frac{\left\langle Q\right\rangle }{A^{2}}\frac{E/(10V/cm)}{\Delta x/100\mu m}0.0025Hz
\end{eqnarray*}
This is completely negligible for the time scales considered in this experiment.

\subsection{Study minimization}

The effect considered are all negligible with the proper constraints. There
are very likely other effects not yet considered, but these results are encouraging
enough to pursue studying the further experimentally. The effects of these kinds
of perturbations can be studied independently of a parity measurement. 

Decoherance problems from trapping fields or other fluctuating perturbations
will affect the spin lifetime, so a measurement of the spin lifetime directly
probes for these kinds of problems, and independent of their origin, determined
if the lifetime of these spin states is sufficiently long that a parity experiment
can be done. Initial studies in both the \( S \) and \( D \) states show spin
lifetimes of at least \( 5s \) and \( 30s \) respectively, \ref{Sec:SpinLifetimes}.They
are very likely much longer, though already they are sufficient for a parity
experiment. 

Large spin flip transition linewidths due to fluctuating magnetic fields, or
applied shifting laser intensity can be determined from spin flip transition
profiles, \ref{Sec:SpinFlipTransitions}, and light shifts from other sources
such as off resonance dipole coupling, \ref{Sec:NonresonantLightShifts}. These
show a relatively large \( kHz \) sized spin flip resonance linewidth, that
is perfectly consistent with the currently unshielded dirty magnetic field environment
of the trap, so that appropriate shielding should reduce this sufficiently,
and intensity noise broadening consistent with the applied laser's amplitude
instability so that a sufficiently well regulated laser should eliminate this
trouble as well.

Further work is planned to look for finite limits on the spin lifetime, and
with a sufficiently stable magnetic field, spin precession measurements that
are more sensitive to decoherance effects should uncover any smaller problems
and allow accurate estimates of the size of the effects and clues to their origin.

\section{General Quadrupole Shift Systematics}

For the ideal case, the parity violating shifts are easily distinguished from
the quadrupole shifts because in this case the quadrupole shifts are generally
spin sign independent, so they move sets of states together, while the PNC shifts
only are spin dependent and separate the states. The clean classification disappears
for more general quadrupole fields as arbitrary fields can result in the quadrupole
shift also being spin dependent. 

For the ideal fields the quadrupole shift is spin independent because the fields
are mirror symmetric. The quadrupole shift is a pure electromagnetic, parity
conserving effect, so there can be no result that can be used to define the
handedness of the coordinate system. A spin dependent shift defines a handedness
as it can be considered to lead to a precession in some well defined direction.
More general fields will not be mirror symmetric. Cross products of polarizations
or propagation directions, or circular polarization of the beams can be used
to define a reference handedness so that a resulting spin dependent shift is
not inconsistent. 

Since the quadrupole rate is allow by electromagnetism, the rates are generally
must larger than the parity shift, \( MHz \) to \( Hz \) as estimated earlier.
As a result these kinds of shifts from the quadrupole couplings due to misaligned
fields are characteristically very large. These kinds of effects become the
biggest source of systematic errors in this experiment. It will turn out that
all of these effects will not be as large as this general size and are generally
suppressed by the structure of the states and transitions by what always turns
out to be at least three small parameters given by various errors of spatial
and temporal phase, alignment or polarization. These provide enough freedom
that if these parameters are correct to a precision that can be reasonably expected
to be achieved, these errors can be made negligible.

\subsection{Polarization Impurities}

A simple example to illustrate this is to consider arbitrary polarizations of
the quadrupole fields. Any linear polarization is mirror symmetric, but a small
residual circular polarization can define the handedness of the coordinate system
and give a chiral effect.

Recall the splitting is given easily from the spin flip transformations by,
\begin{eqnarray*}
\Delta \omega ^{Q}_{m} & = & \left\langle j_{1},m\left| Q^{\dagger }_{s}\right| j_{2},m'\right\rangle \left\langle j_{2},m'\left| Q_{s}\right| j_{1},m\right\rangle \left( \left| E^{Q}_{s}\right| ^{2}-\left| E^{Q}_{-s}\right| ^{2}\right) 
\end{eqnarray*}
For linearly polarized fields this was automatically zero since the field amplitudes
in that case are related by \( E^{Q}_{-s}=\left( -1^{s}\right) E_{s}^{Q*} \).
Now allow the polarization to be arbitrary, but keep \( \vec{k}^{Q} \) fixed
so that the only transitions to consider are still \( \Delta m=\pm 1 \). The
transition amplitudes are given by,
\begin{eqnarray*}
E_{\pm 1}^{Q} & = & \mp \left( \partial _{x}E^{Q}_{z}+\partial _{z}E^{Q}_{x}\right) /2+i\left( \partial _{y}E^{Q}_{z}+\partial _{z}E^{Q}_{y}\right) /2
\end{eqnarray*}
For \( \vec{k}||\hat{z} \) these simplify to,
\begin{eqnarray*}
E_{\pm 1}^{Q} & = & \left( \mp \partial _{z}E^{Q}_{x}+i\partial _{z}E^{Q}_{y}\right) /2
\end{eqnarray*}
The spin symmetric pieces of the products of these amplitudes will cancel leaving,
\begin{eqnarray*}
\left| E_{1}^{Q}\right| ^{2}-\left| E_{-1}^{Q}\right| ^{2} & = & \frac{1}{4}(\left( -\partial _{z}E^{Q*}_{x}-i\partial _{z}E^{Q*}_{y}\right) \left( -\partial _{z}E^{Q}_{x}+i\partial _{z}E^{Q}_{y}\right) \\
 & - & \left( \partial _{z}E^{Q*}_{x}-i\partial _{z}E^{Q*}_{y}\right) \left( \partial _{z}E^{Q}_{x}+i\partial _{z}E^{Q}_{y}\right) )\\
 & = & \frac{1}{4}(\left( -\partial _{z}E^{Q*}_{x}\right) \left( i\partial _{z}E^{Q}_{y}\right) +\left( -i\partial _{z}E^{Q*}_{y}\right) \left( -\partial _{z}E^{Q}_{x}\right) \\
 & - & \left( \partial _{z}E^{Q*}_{x}\right) \left( i\partial _{z}E^{Q}_{y}\right) -\left( -i\partial _{z}E^{Q*}_{y}\right) \left( \partial _{z}E^{Q}_{x}\right) )\\
 & = & \frac{i}{2}\left( \left( \partial _{z}E^{Q*}_{y}\right) \left( \partial _{z}E^{Q}_{x}\right) -\left( -\partial _{z}E^{Q*}_{x}\right) \left( \partial _{z}E^{Q}_{y}\right) \right) \\
 & = & Im\left( \left( \partial _{z}E^{Q*}_{y}\right) \left( \partial _{z}E^{Q}_{x}\right) \right) \\
 & = & k^{2}Im\left( E^{Q*}_{y}E^{Q}_{x}\right) \\
 & = & \left( kE^{Q}\right) ^{2}Im\left( \varepsilon _{y}^{Q*}\varepsilon ^{Q}_{x}\right) 
\end{eqnarray*}

With \( \vec{\sigma }=\hat{e}^{*}\times \hat{e} \), for \( \vec{k}||\hat{z} \)
this gives, \( \sigma =\varepsilon ^{*}_{x}\varepsilon _{y}-\varepsilon _{y}^{*}\varepsilon _{x}=2Im\left( \varepsilon ^{*}_{x}\varepsilon _{y}\right)  \).
So the shift becomes proportional to the circular polarization of the quadrupole
beam,
\[
\Delta \omega ^{Q}\sim 2\sigma \left\langle \left\langle Q\right\rangle \right\rangle ^{2}\left( kE^{Q}\right) ^{2}=2\sigma \left| \Omega ^{Q}\right| ^{2}\]

\subsubsection{Quadrupole Polarization}

With a quadrupole field the same size as the dipole field the quadrupole rate
is \( \Omega ^{Q}\approx 1MHz \) so even a very small deviation from perfect
linear polarization \( \sigma \approx 0.1\% \) gives a huge \( kHz \) shift.
Slight better polarizations might be possible, but certainly nowhere near the
\( 10^{-8} \) that would be required to make this negligible compared to the
parity splitting, similarly given by,
\[
\Delta \omega ^{PNV}\sim \varepsilon \left\langle \left\langle D\right\rangle \right\rangle E^{D}\]

The error can be dealt with in a number of ways which nicely illustrate the
general possibilities. First this shift can easily be isolated from the parity
shift with additional measurements. It is independent of the dipole field, so
it will still be present with the dipole beam is removed while the parity shift
will disappear as it depends linearly on the same dipole field. In this way
the presence of this polarization impurity can be detected, and if measured
precisely, corrected. 

For this case, a \( kHz \) sized shift is inconvenient to correct for. The
initial splitting from the applied magnetic field it intended to be a \( kHz \)
or less, so this systematic shift is a large shift relative to that which would
greatly complicate measuring it as the resulting new position of the resonance
would initially be unknown. More importantly, the changing shifts due to fluctuating
laser fields no longer cancel in the difference since the splitting now includes
a contribution directly proportional to the very large quadrupole coupling.
Now a 1\% fluctuation in the rate from frequency or intensity noise gives a
\( 1\% \) fluctuation of the splitting, in the case causing a \( 10Hz \) addition
to the linewidth completely destroying the high S/N of the measurement and making
the parity shift practically undetectable.

These complications require minimizing this shift. This can be done in two ways.
In principle, \( \sigma  \) could be adjusted until this shift is zero. This
could done by measuring the splitting with the dipole field off, so that again
only this residual circular polarization gives a splitting. However this requires
impossibly fine adjustment to a part in \( 10^{8} \) which, even if achieved,
would likely be very unstable and relatively strongly dependent only varying
temperature or mechanical conditions. The only practical possibility is to reduce
\( E^{Q} \). This systematic error gets smaller linearly with the reduction
in this field while the parity splitting is unaffected, as long as the resulting
quadrupole rate, initially around a \( MHz \), is still very much larger than
PNC splitting of a few \( Hz \) so that the splitting is insensitive to fluctuations
in the laser intensity and frequency, which will remain true even for a very
large reduction of the quadrupole field. The polarization can easily be made
linear to a part in \( 10^{2} \) and with some work should be able to made
linear to a part in \( 10^{3} \), so reducing the quadrupole field by a factor
of \( 10^{3} \) keeps that quadrupole rate at a sufficiently quick \( kHz \)
makes up the factors required to reduce this \( MHz \) sized shift by a factor
of \( 10^{6} \) to a less damaging \( 1Hz \) which must then still be corrected
for as described above if not completely eliminated.

Summarizing this general requirement gives.

\[
\sigma _{Q}\left( E^{Q}/E^{D}\right) <10^{-6}\]
with a generously reasonable realization being,
\begin{eqnarray*}
\sigma _{Q} & < & 10^{-3}\\
E^{Q}/E^{D} & < & 10^{-3}
\end{eqnarray*}

\subsubsection{Dipole Polarization}

The effects of a polarization error in the dipole field must also be considered.
If other alignments remain fixed, this effects only the parity term. With 
\begin{eqnarray*}
E^{D}_{\pm 1} & = & \left( \mp E^{D}_{x}+iE^{D}_{y}\right) /\sqrt{2}
\end{eqnarray*}
the splitting is more generally given by,
\begin{eqnarray*}
E_{+1}^{D*}E_{+1}^{Q}-E_{-1}^{D*}E_{-1}^{Q} & = & \frac{1}{2\sqrt{2}}\left( \left( -E^{D*}_{x}-iE^{D*}_{y}\right) \left( -\partial _{z}E^{Q}_{x}\right) -\left( E^{D*}_{x}-iE^{D*}_{y}\right) \left( \partial _{z}E^{Q}_{x}\right) \right) \\
 & = & -\frac{k}{\sqrt{2}}Im\left( E^{D*}_{x}E^{Q}_{x}\right) 
\end{eqnarray*}
The possible components of the field in the \( \hat{y} \) direction from circular
polarization give no contribution to the shift. This can be seen in cartesian
coordinates as well, the coupling will involve a matrix element of \( \left\langle y\right\rangle \left\langle xz\right\rangle  \)
which is non-zero according to the selection rules, but antisymmetric under
a spin flip transformation since, for example \( \left( -1\right) ^{n_{y}}=-1 \)
and since \( \left( -1\right) ^{k_{1}+k_{2}}=\left( -1\right) ^{3}=-1 \), this
coupling doesn't change sign.

With these restrictions circular polarization of the dipole beam gives no systematic
errors, but when more general variations are considered this polarization error
reappears. In particular, if the position of the dipole standing wave is not
placed such that the ion is exactly at its anti-node, if it is off by a phase
\( \delta  \), there will be a small gradient that will couple to the quadrupole
transition and give a splitting with small amounts of circular polarization.
Here the coupling is much smaller with\( \partial _{z}E_{\perp }=sin\left( \delta \right) kE_{\perp } \)

In this case the electric field is still large and must remain so or the parity
splitting will also be reduced so this error can't be minimized by reducing
the field strength. But the coupling now already includes an additional small
parameter \( \delta  \) which appears quadratically since this term will involve
the square of the residual \( E^{D} \) quadrupole amplitude. This makes the
error much less trouble some. If the dipole polarization can also be made linear
to a part in \( 10^{2} \), this error can be made completely negligible with
\( \delta <10^{-3} \).
\[
\sigma _{D}\delta ^{2}<10^{-8}\]
With \( \delta =2\pi \Delta x/\lambda  \), this requires the antinode position
to be set accurately to \( \Delta x<\delta \lambda /2\pi \approx 2\times 10^{-4}\mu m=2A \).
This term can also be detected and minimized independently by adjusting the
polarization and antinode position while monitoring the splitting when only
this dipole field is applied. The parity splitting again disappears since the
quadrupole dipole interference is then zero, the applied quadrupole field is
removed and the residual gradient from this misalignment is in phase with the
amplitude giving \( Im\left( E^{D*}E^{Q}\right) =0 \). The constraint on the
antinode position can be made much less stringent if this (is corrected for)

\subsection{Quadrupole-Quadrupole Error}

\label{Sec:GeneralQQTerm}

These exhaust the possibilities for the influence of a single term, but the
completely general case includes interaction between different couplings from
multiple errors than can give spin dependent shifts. To reliably correctly account
for all of these it is best to start with completely general couplings rather
than add in endless perturbations to the solution of a particular idealized
case.

The field chosen to create the parity violation induced splitting were chosen
to be standing waves. In the optimal case one wave, \( \vec{E}_{1} \), was
positioned so that the ion was at its antinode, and the other, \( \vec{E}_{2} \),
positioned so that the ion was at its node. For this case \( \vec{E}_{1} \)
is uniquely responsible for the dipole coupling and \( \vec{E}_{2} \) only
drives the quadrupole transition. Then the total coupling is 
\begin{eqnarray*}
\Omega  & = & \Omega ^{Q}+i\varepsilon \Omega ^{D}\\
 & = & \Omega _{2}^{Q}+i\varepsilon \Omega _{1}^{D}
\end{eqnarray*}
Now allow for the case where \( \vec{E}_{1} \) also partly drives the quadrupole
transition, and \( \vec{E}_{2} \) partly drives the dipole transitions. Denote
them as small variations to the ideal case by \( \delta \Omega ^{Q}_{1} \)
and \( \delta \Omega ^{D}_{2} \), \( \delta  \) can be interpreted as simply
a label or explicitly as a parameter. These terms will be nonzero only for some
non-zero error in the spatial phase, \( \delta _{1} \), \( \delta _{2} \)
and, as before, will then be proportional to \( sin\left( \delta \right) \approx \delta  \). 

With this generalized case the complete coupling then becomes,
\[
\Omega =\Omega _{2}^{Q}+i\varepsilon \Omega _{1}^{D}+\delta \Omega _{1}^{Q}+i\varepsilon \delta \Omega _{2}^{D}\]

Neglecting the explicit spin substructure for now, the shifts will generally
be given by \( \Omega ^{\dagger }\Omega  \). Neglecting only the \( \varepsilon ^{2} \)
terms as being obviously negligible, this gives the additional contributions
to the shifts

\begin{eqnarray*}
\Omega ^{\dagger }\Omega  & = & \Omega ^{Q\dagger }_{2}\Omega ^{Q}_{2}+\delta \Omega ^{Q\dagger }_{1}\delta \Omega ^{Q}_{1}\\
 & + & i\varepsilon \left( \Omega ^{D\dagger }_{1}\Omega ^{Q}_{2}-\Omega ^{Q\dagger }_{2}\Omega ^{D}_{1}\right) \\
 & + & i\varepsilon \left( \delta \Omega ^{D\dagger }_{2}\Omega ^{Q}_{2}-\Omega ^{Q\dagger }_{2}\delta \Omega ^{D}_{2}\right) \\
 & + & i\varepsilon \left( \Omega ^{D\dagger }_{1}\delta \Omega ^{Q}_{1}-\delta \Omega ^{Q\dagger }_{1}\Omega ^{D}_{1}\right) \\
 & + & i\varepsilon \left( \delta \Omega ^{Q\dagger }_{1}\delta \Omega ^{D}_{2}-\delta \Omega ^{D\dagger }_{2}\delta \Omega ^{Q}_{1}\right) \\
 & + & \left( \Omega _{2}^{Q\dagger }\delta \Omega ^{Q}_{1}+\delta \Omega _{1}^{Q\dagger }\Omega ^{Q}_{2}\right) 
\end{eqnarray*}
The first two terms are exactly those considered above. They are exactly spin
independent for linearly polarized beams and are negligible or correctable when
the constraints previously outlined are satisfied. 

The first term involving \( \varepsilon  \) is the usual term giving the parity
splitting. The remaining terms proportional to \( \varepsilon  \) also include
at least one \( \delta  \)and so will naturally be much smaller than the intended
parity splitting. At the least, for \( \delta <10^{-3} \) these terms give
a \( 0.1\% \) contribution to the splitting, which must be satisfied already
for the dipole field to eliminate problems due to its circular polarization,
and details of the structure of these products actually make these terms even
less of a problem as will be seen in detail shortly.

The remaining term, \( \delta \Omega _{1}^{Q}\Omega _{2}^{Q} \), is the \( E_{1} \),
\( E_{2} \) quadrupole-quadrupole cross term which, as it stands, includes
only one \( \delta  \) and no \( \varepsilon  \) and so is potentially much
larger than the parity term. Estimating the sizes of the matrix elements with
\( \left\langle \left\langle Q\right\rangle \right\rangle k\approx A^{2}/\mu m \),
\( \left\langle \left\langle D\right\rangle \right\rangle \approx A \) the
ratio of these terms can be estimated. Again neglecting the explicit spin structure
and relative phase dependences, with \( \varepsilon \approx 10^{-11} \),
\begin{eqnarray*}
\Omega _{2}^{Q}\delta \Omega ^{Q}_{1}/\varepsilon \Omega ^{D}_{1}\Omega ^{Q}_{2} & \approx  & \delta \Omega _{1}^{Q}/\varepsilon \Omega _{1}^{D}\\
 & \approx  & \left( \delta /\varepsilon \right) \left\langle \left\langle Q\right\rangle \right\rangle kE_{1}/\left\langle \left\langle D\right\rangle \right\rangle E_{1}\\
 & = & \left( \delta /\varepsilon \right) A/\mu m\\
 & = & 10^{-4}\left( \delta /\varepsilon \right) \\
 & = & 10^{7}\delta 
\end{eqnarray*}

This term is naturally very much larger than the parity splitting and, as a
result, is the principle systematic problem with this experiment. If the estimate
is accurate, this error alone makes a parity measurement of this kind impractical.
It turns out that the situation is not nearly as drastic as it appears and the
resolution is hidden in the details of the spin structure of these transitions
that has been, temporarily neglected.

\section{General m+n State Solution}

The solution to the original system considered had a very general appearance,
the shift were given by the diagonal element of the matrix product of the couplings,
\[
\delta \omega _{m}=\left( \Omega ^{\dagger }\Omega \right) _{mm}\]
This form was more a notational convenience than a proper result. It depended
on each ground state sublevel being coupled to different \( D \) state levels
so that the system factors into two trivial problems. This condition was satisfied
for the field configurations considered so far, but for more general conditions
this is not the case, and it is not clear that this solution is appropriate.
In addition, it was argued that the off-diagonal elements could be interpreted
as spin dependent shifts in orthogonal directions and the whole shift took the
form of an effective magnetic field. These ideas require a firmer formal support,
which will be provided by the exact solution to the generally coupled problem.

\subsection{General Couplings and The Rotating Wave Approximation}

\label{Sec:RWA}

The general problem involves two energy levels. Though each level consists of
a number of spin substates, the hamiltonian can still be written will the appearance
of a two state problem using the coupling matrices defined before, \( \Omega  \).
Consider a system with one level having \( m \) substates and the other \( n \).
\( \Omega  \)is then an \( n\times m \) matrix describing the coupling between
any pair of spin states where the pair includes one state from each level and
there are no couplings between states in the same level. \( \Omega  \) is generally
time dependent. Initial this time dependence will be included explicitly. The
hamiltonian will then have the block \( 2\times 2 \) form,

\begin{eqnarray*}
H(t) & = & \left( \begin{array}{cc}
\omega _{0}/2 & \Omega e^{-i\omega t}+\Omega ^{*}e^{i\omega t}\\
\Omega ^{\dagger }e^{i\omega t}+\Omega ^{\dagger *}e^{-i\omega t} & -\omega _{0}/2
\end{array}\right) \\
 & = & \left( \begin{array}{cc}
n\times n & n\times m\\
m\times n & m\times m
\end{array}\right) 
\end{eqnarray*}
The block structure of these matrices allows then to be manipulated just like
trivial \( 2\times 2 \) matrices. Multiplying two matrices with the same block
structure has the same form and when the blocks are just numbers if the elements
are combined using matrix multiplication of the blocks (Reference).

\( \omega _{0} \) is the energy difference between the levels. The diagonal
blocks are then just the energy of the level times the identity matrix of the
appropriate dimension. In this case all the states in each level are taken to
be degenerate, though for the actual experiment they will be separated slightly
be an applied magnetic field. This modification will be considered later, but
in the end, for sufficiently strong couplings, the splitting is negligible.

Solutions to this hamiltonian are complicated by the time dependence of the
interaction. Traditional the Rotating Wave Approximation is made to be able
to transform this to a static problem that is readily solved. To do this, consider
a unitary transformation like
\[
U\left( t\right) =\left( \begin{array}{cc}
e^{-i\left( \omega \prime /2\right) t} & 0\\
0 & e^{i\left( \omega \prime /2\right) t}
\end{array}\right) \]
to transform the states to \( \psi \prime =U\left( t\right) \psi  \). The equation
of motion in terms of the transformed state becomes,
\begin{eqnarray*}
H\psi  & = & i\partial _{t}\psi \\
UHU^{\dagger }U\psi  & = & iU\partial _{t}\left( U^{\dagger }U\psi \right) \\
UHU^{\dagger }\psi \prime  & = & iU\partial _{t}\left( U^{\dagger }\psi \prime \right) \\
 & = & i\left( \left( U\partial _{t}U^{\dagger }\right) \psi \prime +UU^{\dagger }\partial _{t}\psi \prime \right) \\
H^{\prime }\psi ^{\prime } & = & i\partial _{t}\psi ^{\prime }
\end{eqnarray*}
The effective hamiltonian for the transformed state is 
\begin{eqnarray*}
H^{\prime } & = & UHU^{\dagger }-iU\partial _{t}U^{\dagger }\\
 & = & \left( \begin{array}{cc}
e^{-i\left( \omega \prime /2\right) t} & 0\\
0 & e^{i\left( \omega \prime /2\right) t}
\end{array}\right) \\
 & \times  & \left( \begin{array}{cc}
\omega _{0}/2 & \Omega e^{-i\omega t}+\Omega ^{*}e^{i\omega t}\\
\Omega ^{\dagger }e^{i\omega t}+\Omega ^{\dagger *}e^{-i\omega t} & -\omega _{0}/2
\end{array}\right) \left( \begin{array}{cc}
e^{i\left( \omega \prime /2\right) t} & 0\\
0 & e^{-i\left( \omega \prime /2\right) t}
\end{array}\right) \\
 & - & i\left( \begin{array}{cc}
e^{-i\left( \omega \prime /2\right) t} & 0\\
0 & e^{i\left( \omega \prime /2\right) t}
\end{array}\right) \frac{i\omega \prime }{2}\left( \begin{array}{cc}
e^{i\left( \omega \prime /2\right) t} & 0\\
0 & -e^{-i\left( \omega \prime /2\right) t}
\end{array}\right) \\
 & = & \left( \begin{array}{cc}
\omega _{0}/2 & \Omega e^{-i(\omega +\omega \prime )t}+\Omega e^{i\left( \omega -\omega \prime \right) t}\\
\Omega ^{\dagger }e^{i\left( \omega +\omega \prime \right) t}+\Omega ^{\dagger *}e^{-i\left( \omega -\omega \prime \right) t} & -\omega _{0}/2
\end{array}\right) \\
 & + & \frac{\omega \prime }{2}\left( \begin{array}{cc}
1 & 0\\
0 & -1
\end{array}\right) 
\end{eqnarray*}
For \( \omega \prime =\pm \omega  \) one of the time dependent terms of the
interaction becomes static and the other then depends on \( e^{\pm 2i\omega t} \).
Making the Rotating Wave Approximation is to neglect this quickly oscillating
counter-rotating term. This assumes that in the time scale it takes for the
state to change significantly, this term has oscillated many times to that its
effects average to zero. This is valid for \( \partial _{t}\psi <<2\omega  \).
The time derivative of the state will be dominated by its energy, which in this
basis is \( \pm \left( \omega _{0}-\omega \right) /2=\pm \delta \omega /2 \),
\( \delta \omega  \) is the detuning and the requirement for the rotating wave
approximation to be accurate is simply that the interaction is very close to
resonance compared to the frequency of the interaction \( \delta \omega <<\omega  \).

Taking \( \omega \prime =-\omega  \), the hamiltonian for the transformed state
is then simply,

\[
H\prime =\left( \begin{array}{cc}
\delta \omega /2 & \Omega \\
\Omega ^{\dagger } & -\delta \omega /2
\end{array}\right) \]

\subsection{Eigenvalues}

Solving this static system then just requires finding the eigenvalues of \( H \).
This is familiar, and trivial for a simple two state system, but more subtle
for the general case. At least two methods work for finding the eigenvalues
for these \( 2\times 2 \) block matrices with arbitrary dimensions. 

Call the eigenvalues \( \omega  \) temporarily, they are the values that appear
in the eigenvalue equation,
\[
H\psi =\omega \psi \]
The eigenvectors can be computed by writing the states in an explicit component
form. 
\[
\psi =\left( \begin{array}{c}
a\\
b
\end{array}\right) \]
Here \( a \) and \( b \) are respectively \( n \) and \( m \) component
column vectors. The eigenvalue equation becomes,
\begin{eqnarray*}
\left( \begin{array}{cc}
\delta \omega /2 & \Omega \\
\Omega ^{\dagger } & -\delta \omega /2
\end{array}\right) \left( \begin{array}{c}
a\\
b
\end{array}\right)  & = & \omega \left( \begin{array}{c}
a\\
b
\end{array}\right) \\
\left( \begin{array}{c}
\left( \delta \omega /2\right) a+\Omega b\\
\Omega ^{\dagger }a-\left( \delta \omega /2\right) b
\end{array}\right)  & = & \left( \begin{array}{c}
\omega a\\
\omega b
\end{array}\right) 
\end{eqnarray*}
This gives two coupled equations
\begin{eqnarray*}
\left( \delta \omega /2-\omega \right) a & = & -\Omega b\\
\left( \delta \omega /2+\omega \right) b & = & \Omega ^{\dagger }a
\end{eqnarray*}
Substituting each result into the other gives, 
\begin{eqnarray*}
\Omega \Omega ^{\dagger }a & = & \left( \omega ^{2}-\left( \delta \omega /2\right) ^{2}\right) a\\
\Omega ^{\dagger }\Omega b & = & \left( \omega ^{2}-\left( \delta \omega /2\right) ^{2}\right) b
\end{eqnarray*}

Alternately, note that for \( H \) static \( H^{2}\psi =\omega ^{2}\psi  \)
and for this block \( 2\times 2 \) problem \( H^{2} \) turns out to be block
diagonal,
\begin{eqnarray*}
H^{2} & = & \left( \begin{array}{cc}
\delta \omega /2 & \Omega \\
\Omega ^{\dagger } & -\delta \omega /2
\end{array}\right) \left( \begin{array}{cc}
\delta \omega /2 & \Omega \\
\Omega ^{\dagger } & -\delta \omega /2
\end{array}\right) \\
 & = & \left( \begin{array}{cc}
\left( \delta \omega /2\right) ^{2}+\Omega \Omega ^{\dagger } & 0\\
0 & \left( \delta \omega /2\right) ^{2}+\Omega ^{\dagger }\Omega 
\end{array}\right) 
\end{eqnarray*}
Operating this on the component state quickly gives the same constraints for
the eigenvector,
\[
H^{2}\psi =\omega ^{2}\begin{array}{c}
\left( \begin{array}{c}
a\\
b
\end{array}\right) 
\end{array}=\left( \begin{array}{c}
\left( \left( \delta \omega /2\right) ^{2}+\Omega \Omega ^{\dagger }\right) a\\
\left( \left( \delta \omega /2\right) ^{2}+\Omega ^{\dagger }\Omega \right) b
\end{array}\right) \]

\subsubsection{2 State and 1+n State Eigenvalues}

\label{Sec:1+nState}

For a two state system, \( \Omega  \) is simply a scalar, and \( \Omega ^{\dagger }\Omega =\Omega ^{\dagger }\Omega =\Omega \Omega ^{*}=\left| \Omega \right| ^{2} \).
In this case the eigenvector equation for both \( a \) and \( b \) give the
familiar result, \( \omega ^{2}=\left( \delta \omega /2\right) ^{2}+\left| \Omega \right| ^{2} \). 

For \( m=1 \) the solution for \( b \) is similarly trivial as \( \Omega ^{\dagger }\Omega  \)
is again a scalar, \( \Omega ^{\dagger } \) is a \( 1\times n \) matrix and
\( \Omega  \) is \( n\times 1 \), so the product is \( 1\times 1 \) given
by \( \Omega ^{\dagger }\Omega \equiv \left| \Omega \right| ^{2}=\left| \Omega _{1}\right| ^{2}+\left| \Omega _{2}\right| ^{2}+\cdots +\left| \Omega _{n}\right| ^{2} \)and
the energy is again given by \( \omega ^{2}=\left( \delta \omega /2\right) ^{2}+\left| \Omega \right| ^{2} \).
This is the result used in the original analysis of the parity experiment in
the ground state with \( n=2 \) corresponding to the two \( D \) states coupled
to each \( S \) state by the \( \Delta m=\pm 1 \) transitions.

\subsubsection{m+n State Eigenvalues}

For neither \( m \) nor \( n \) equal to one, more work is necessary as the
eigenvector equations for each component are both still matrix equations, but
in this form the components are uncoupled and each can be considered individually.
The components of the eigenvectors of the complete hamiltonian must satisfy

\begin{eqnarray*}
\Omega \Omega ^{\dagger }a & = & (\omega ^{2}-(\delta \omega /2)^{2})a\\
\Omega ^{\dagger }\Omega b & = & (\omega ^{2}-(\delta \omega /2)^{2})b
\end{eqnarray*}

These are themselves eigenvalue equations. \( a \) and \( b \) must be eigenvectors
of \( \Omega \Omega ^{\dagger } \) and \( \Omega ^{\dagger }\Omega  \) respectively,
and they must both have the same eigenvalues since \( \omega  \) is fixed for
the entire state and the complete eigenvector is constructed out of both components.
Call the eigenvectors \( \lambda ^{2} \),
\begin{eqnarray*}
\Omega \Omega ^{\dagger }a & = & \lambda ^{2}a\\
\Omega ^{\dagger }\Omega b & = & \lambda ^{2}b
\end{eqnarray*}

\subsubsection{Positivity}

It is clear that \( \lambda ^{2} \) must be positive and real for consistency
as this would give the energies \( \omega ^{2}=\lambda ^{2}+\left( \delta \omega /2\right) ^{2} \).
This must be true for all \( \delta \omega  \) and in particular for \( \delta \omega =0 \)
which gives \( \omega ^{2}=\lambda ^{2} \). \( \omega  \)must be real since
\( H \) is hermitian, so \( \omega ^{2} \) is real and positive. 

This quality of \( \lambda ^{2} \) can also be seen from the structure of the
operator. \( \Omega ^{\dagger }\Omega  \) and \( \Omega \Omega ^{\dagger } \)
are hermitian so immediately the eigenvalues must be real. For the sign, consider
an arbitrary expectation value,
\begin{eqnarray*}
\psi ^{\dagger }\Omega \Omega ^{\dagger }\psi  & = & (\Omega ^{\dagger }\psi )^{\dagger }(\Omega ^{\dagger }\psi )=\left| \Omega \psi \right| ^{2}>0\\
\psi ^{\dagger }\Omega ^{\dagger }\Omega \psi  & = & (\Omega \psi )^{\dagger }(\Omega \psi )=\left| \Omega \psi \right| ^{2}>0
\end{eqnarray*}
For \( \psi  \) an appropriate eigenvector 
\begin{eqnarray*}
a^{\dagger }(\Omega \Omega ^{\dagger }a) & = & a^{\dagger }a\lambda ^{2}=\lambda ^{2}>0\\
b^{\dagger }\left( \Omega ^{\dagger }\Omega b\right)  & = & b^{\dagger }b\lambda ^{2}=\lambda ^{2}>0
\end{eqnarray*}

\subsubsection{Eigenvectors}

An eigenvector of \( H \) can now be constructed from these components. Consider
first an eigenstate \( a \) of \( \Omega \Omega ^{\dagger } \) with non-zero
eigenvalue. As already seen the eigenvector equation, \( H\psi =\omega \psi  \),
requires
\begin{eqnarray*}
\left( \omega -\delta \omega /2\right) a & = & \Omega b\\
\left( \omega +\delta \omega /2\right) b & = & \Omega ^{\dagger }a
\end{eqnarray*}
 \( \omega  \) is given by \( \omega ^{2}=\left( \delta \omega /2\right) ^{2}+\lambda ^{2} \)
and both positive and negative values of the root can be used. In particular,
this requires that the lower component is given by,
\[
b_{\pm }=\frac{\Omega ^{\dagger }}{\pm \omega +\delta \omega /2}a\]
 Which is easily seen to be an eigenvector of \( \Omega ^{\dagger }\Omega  \)
with eigenvalue \( \lambda ^{2} \) as required,

\begin{eqnarray*}
\Omega ^{\dagger }\Omega b_{\pm } & = & \frac{1}{\pm \omega +\delta \omega /2}(\Omega ^{\dagger }\Omega )\Omega ^{\dagger }a\\
 & = & \frac{\Omega ^{\dagger }}{\pm \omega +\delta \omega /2}(\Omega \Omega ^{\dagger }a)\\
 & = & \lambda ^{2}\frac{\Omega ^{\dagger }}{\pm \omega +\delta \omega /2}a\\
 & = & \lambda ^{2}b_{\pm }
\end{eqnarray*}

These give two eigenvectors of \( H \),
\[
\psi _{\pm }=\left( \begin{array}{c}
a\\
\frac{\Omega ^{\dagger }}{\pm \omega +\delta \omega /2}a
\end{array}\right) \]
These will have energy \( \pm \omega  \), as can be verified by an explicit
calculation,
\begin{eqnarray*}
H\psi  & = & \left( \begin{array}{cc}
\delta \omega /2 & \Omega \\
\Omega ^{\dagger } & -\delta \omega /2
\end{array}\right) \left( \begin{array}{c}
a\\
\frac{\Omega ^{\dagger }}{\omega +\delta \omega /2}a
\end{array}\right) \\
 & = & \left( \begin{array}{c}
\left( \frac{\delta \omega }{2}+\frac{\Omega \Omega ^{\dagger }}{\omega +\delta \omega /2}\right) a\\
\left( \Omega ^{\dagger }-\frac{\delta \omega }{2}\frac{\Omega ^{\dagger }}{\omega +\delta \omega /2}\right) a
\end{array}\right) \\
 & = & \left( \begin{array}{c}
\left( \frac{\delta \omega }{2}+\frac{\lambda ^{2}}{\omega +\delta \omega /2}\right) a\\
\left( 1-\frac{\delta \omega /2}{\omega +\delta \omega /2}\right) \Omega ^{\dagger }a
\end{array}\right) 
\end{eqnarray*}
 With \( \lambda ^{2}=\omega ^{2}+\left( \delta \omega /2\right) ^{2} \) this
becomes,
\begin{eqnarray*}
H\psi  & = & \left( \begin{array}{c}
\left( \frac{\delta \omega }{2}+\frac{\omega ^{2}-\left( \delta \omega /2\right) ^{2}}{\omega +\delta \omega /2}\right) a\\
\frac{\omega +\delta \omega /2-\delta \omega /2}{\omega +\delta \omega /2}\Omega ^{\dagger }a
\end{array}\right) \\
 & = & \left( \begin{array}{c}
\left( \frac{\delta \omega }{2}+\frac{\left( \omega +\delta \omega /2\right) \left( \omega -\delta \omega /2\right) }{\omega +\delta \omega /2}\right) a\\
\frac{\omega }{\omega +\delta \omega /2}\Omega ^{\dagger }a
\end{array}\right) \\
 & = & \left( \begin{array}{c}
\left( \frac{\delta \omega }{2}+\omega -\frac{\delta \omega }{2}\right) a\\
\omega \frac{\Omega ^{\dagger }}{\omega +\delta \omega /2}a
\end{array}\right) \\
 & = & \left( \begin{array}{c}
\omega a\\
\omega \frac{\Omega ^{\dagger }}{\omega +\delta \omega /2}a
\end{array}\right) \\
 & = & \omega \psi 
\end{eqnarray*}

\subsubsection{Dimension and Number of Solutions}

\label{Sec:M+NZeros}

\( \Omega \Omega ^{\dagger } \) is an \( n\times n \) matrix so it must have
\( n \) eigenvector \( a_{i} \). This construction pairs each \( a_{i} \)
with an eigenvector \( b_{i} \) of \( \Omega ^{\dagger }\Omega  \) and uses
each pair to generate two eigenvectors of \( H \). For \( n=m \) this clearly
gives all of the \( n+m=2n \) eigenvectors of \( H \), but for \( n\neq m \)
it is not clear how this can be consistent. In particular, for \( n<m \) this
falls short of the number of eigenvectors of \( H \) by \( n+m-2m=n-m \),
and for \( n>m \) it seems to generate too many \( b_{i} \) for the dimension
of \( \Omega ^{\dagger }\Omega  \). The \( b_{i} \) were constructed from
the \( a_{i} \) from the original eigenvector equation with,
\[
b_{\pm }=\frac{\Omega ^{\dagger }}{\pm \omega +\delta \omega /2}a\]

This gives two eigenvalues for \( H \) only for \( \omega \neq 0 \) and \( \Omega ^{\dagger }a\neq 0 \).
In general \( \omega  \) cannot be zero as even for \( \lambda ^{2}=0 \),
a non-zero \( \delta \omega  \) gives \( \omega ^{2}=\delta \omega ^{2}\neq 0 \).
The more general possibility is that \( \Omega ^{\dagger }a=0 \), for which
it then immediately follows that \( \Omega \Omega ^{\dagger }a=0 \) and \( \lambda ^{2}=0 \).
So the set of \( b_{i} \) generated by this process are not unique and so the
extra ones generated for the case \( n>m \) must be zero and the missing \( n-m \)
eigenvectors of \( H \) are made up of states like
\[
\psi =\left( \begin{array}{c}
a\\
0
\end{array}\right) \]
where \( a \) is an eigenvector of \( \Omega \Omega ^{\dagger } \) with eigenvalue
0. These states will then have an energy given by,
\begin{eqnarray*}
H\psi  & = & \left( \begin{array}{cc}
\delta \omega /2 & \Omega \\
\Omega ^{\dagger } & -\delta \omega /2
\end{array}\right) \left( \begin{array}{c}
a\\
0
\end{array}\right) \\
 & = & \left( \begin{array}{c}
\frac{\delta \omega }{2}a\\
\Omega ^{\dagger }a
\end{array}\right) \\
 & = & \left( \begin{array}{c}
\frac{\delta \omega }{2}a\\
0
\end{array}\right) \\
 & = & \left( \delta \omega /2\right) a
\end{eqnarray*}
Notice that this is the unperturbed energy of all the upper states so this corresponds
to an uncoupled state. The complement is clearly true for \( m>n \), the eigenvectors
of \( H \) not generated from the eigenvectors \( a_{i} \) of \( \Omega \Omega ^{\dagger } \)
are given by \( a=0 \) and \( b \) a zero mode of \( \Omega ^{\dagger }\Omega  \)
as an effectively uncoupled state, and \( \left( \begin{array}{cc}
0 & b
\end{array}\right) ^{T} \) is an eigenstate of \( H \) with energy \( -\delta \omega /2 \). 

These dimensional arguments show that the construction of eigenstates of \( H \)
in this way is consistent and complete and also elegantly show that for \( n\neq m \)
there must always be \( \left| n-m\right|  \) combinations of states in the
energy level with more states that are uncoupled for any fixed polarization,
that is for \( n<m \), \( \Omega ^{\dagger }\Omega  \) must have at least
\( m-n \) zero eigenvalues and for \( m<n \), \( \Omega \Omega ^{\dagger } \)
must have at least \( n-m \) zero eigenvalues. This latter result is not relevant
for these immediate purposes in understanding the systematics of the parity
measurement but will turn out to have great practical consequences for pumping
and spin detection when considered later,\ref{Sec:PolarizationPumping}.

\subsection{The Effective Magnetic Field}

This analysis show that the solutions to these particular classes of otherwise
very complicated \( m+n \) state problems is given simply by the solution to
two much easier \( m \) and \( n \) state problems. As seen, the eigenvectors
and eigenvalues of \( H \) are given the by the eigenvectors and eigenvalues
of \( \Omega ^{\dagger }\Omega  \) and \( \Omega \Omega ^{\dagger } \) . For
every eigenvector of \( \Omega ^{\dagger }\Omega  \) and \( \Omega \Omega ^{\dagger } \)
with eigenvalue \( \lambda ^{2} \) there are eigenvectors of \( H \) with
energy \( \pm \sqrt{\left( \delta \omega /2\right) ^{2}+\lambda ^{2}} \). Each
of these eigenstates corresponds to a particular individual unperturbed spin
state so these eigenvalues of \( H \) can be interpreted as the shifted energies
of the original states and the products of these products of the couplings as
effective hamiltonian. The spin states of the lower and upper energy levels
will evolve as if being respectively acted on by a hamiltonian given by \( \Omega ^{\dagger }\Omega  \)
and \( \Omega \Omega ^{\dagger } \).

For the IonPNC system, one of these effective interactions has a trivial solution.
In this case \( \Omega  \) couples a \( j=1/2 \) level to a \( j=3/2 \) level,
\( n=4 \), \( m=2 \), so that \( \Omega ^{\dagger }\Omega  \) is just a \( 2\times 2 \)
matrix. The Pauli matrices and the identity matrix are a complete basis for
\( 2\times 2 \) hermitian matrices so this must have the general form identical
to that of an interaction with a magnetic field,
\[
\Omega ^{\dagger }\Omega =\delta \omega ^{\left( 0\right) }+\delta \vec{\omega }^{\left( 1\right) }\cdot \vec{\sigma }\]
\( \delta \omega ^{\left( 0\right) } \) gives an overall spin independent shift
of both levels and \( \delta \vec{\omega }^{\left( 1\right) } \) gives a spin
dependent shift. The eigenstates will be exactly the same as those for a magnetic
field \( \vec{B}=\delta \vec{\omega }^{\left( 1\right) } \), spin states pointing
along \( \pm \delta \vec{\omega }^{\left( 1\right) }\vec{B} \) with energies
separated by \( \left| \delta \vec{\omega }^{\left( 1\right) }\right|  \),
and it is clear that the idea of the parity splitting being an effective vector
interaction is an accurate interpretation.

\( \Omega \Omega ^{\dagger } \) is not quite as trivial to analyze as it is
now a \( 4 \) state problem. It can still be understood as an isolated \( j=3/2 \)
spin system, but now there is additional structure to the effective interaction,
it will include more than just a simple dipole. The general structure can be
written in terms of higher order spin operators, \( j^{\left( k\right) }_{s}\rightarrow j^{\left( k\right) }_{i_{1}i_{2}\cdots i_{k}} \),
where \( j^{\left( 1\right) }_{i} \) are just the usual angular momentum operators,
\( j^{\left( 2\right) }_{ij}=\left( j_{i}j_{j}+j_{j}j_{i}\right) /2-\left( \delta _{ij}/3\right) j^{2} \)
and the higher order operators are given similarly. For this case generally
orders up to \( k=4 \) will be included,
\[
\Omega \Omega ^{\dagger }=\delta \omega +\delta \omega _{i}j_{i}+\delta \omega _{ij}j_{ij}+\delta \omega ^{\left( 3\right) }_{s}j^{\left( 3\right) }_{s}+\delta \omega ^{\left( 4\right) }_{s}j^{\left( 4\right) }_{s}\]
The coefficient are given by explicit calculation or, as seen later, using general
symmetry properties of the angular momentum operators through Generalized Pauli
Matrices, \ref{Sec:GeneralizedPauliMatricies}.

\section{Perturbative Misalignment Systematics}

\label{Sec:MisalignmentSystemmatics}

\subsection{The Applied Magnetic Field}

With the solution for general interactions in hand, systematic errors can now
be correctly analyzed by studying the matrix elements of \( \Omega ^{\dagger }\Omega  \)
and \( \Omega \Omega ^{\dagger } \) for which the tools developed using spin
flip transformations can be immediately applied. Generally this requires looking
at all the matrix elements to look for splittings in directions orthogonal to
the intended parity splitting. An intended shift in the \( \hat{z} \) direction
changes only the diagonal matrix elements of these products of couplings, perpendicular
components will be contained in the off-diagonal matrix elements. With the large
applied magnetic field that will be used for this measurement a further simplification
is possible. First, it is necessary to briefly study the effects of this additional
interaction.

With a magnetic field the transformed static hamiltonian becomes,
\[
H=\left( \begin{array}{cc}
\delta \omega /2+\vec{s}_{a}\cdot \vec{B} & \Omega \\
\Omega ^{\dagger } & -\delta \omega /2+\vec{s}_{b}\cdot \vec{B}
\end{array}\right) \]
\( \vec{s}_{0} \) and \( \vec{s}_{1} \) are the appropriate representations
of the spin operator for each level and also implicitly include the relevant
magnetic moments \( \mu _{1} \), \( \mu _{2} \). The constraints for the upper
and lower components of the eigenvectors now become,

\begin{eqnarray*}
\left( \omega -\delta \omega /2-\vec{s}_{a}\cdot \vec{B}\right) a & = & \Omega b\\
\left( \omega +\delta \omega /2-\vec{s}_{b}\cdot \vec{B}\right) b & = & \Omega ^{\dagger }a
\end{eqnarray*}
 Solutions to this system of equations are complicated by the matrix structure
of the left-hand sides, which were previously scalar. However, generally the
operators involves will not have any zero eigenvalues. \( \omega  \) will be
given by \( \lambda  \), the eigenvalues of \( \Omega ^{\dagger }\Omega  \)
and \( \Omega \Omega ^{\dagger } \), and \( \left| \vec{B}\right|  \).

For \( \lambda =0 \), the solutions are trivial as either \( \Omega b \) or
\( \Omega ^{\dagger }a \) will be zero, this is the case where a certain combination
of states in one level is uncoupled by the laser interactions. For that state,
\( \psi  \) the energies are given by \( \left( \omega \pm \delta \omega /2\right) \psi =\left( \vec{s}\cdot \vec{B}\right) \psi  \),
and no further work is necessary, the energies are given completely by the applied
magnetic field and are the same as they would be without the additional laser
interactions. Generally the coordinate system will be chosen so that \( \vec{B}||\hat{z} \)
and so the eigenstates are the usual spin states and there energies are just
given by \( \mu mB \).

For \( \lambda \neq 0 \), \( \lambda  \)is of order \( \Omega  \). \( \Omega  \)
is intended to be on the order of \( MHz \) while the splitting due to the
applied magnetic field will be \( kHz \). So for this case \( \omega \sim \lambda >>\left| \vec{B}\right|  \)
and the eigenvalues are approximately \( \omega \neq 0 \) and the operator
is invertible permitting, with some algebra, a solution of the form,

\begin{eqnarray*}
\left( \omega -\delta \omega /2\right) a & = & \left( \Omega ^{\dagger }\frac{1}{\omega +\delta \omega /2-\vec{s}_{b}\cdot \vec{B}}\Omega +\vec{s}_{a}\cdot \vec{B}\right) a\\
\left( \omega +\delta \omega /2\right) b & = & \left( \Omega \frac{1}{\omega -\delta \omega /2-\vec{s}_{a}\cdot \vec{B}}\Omega ^{\dagger }+\vec{s}_{b}\cdot \vec{B}\right) b
\end{eqnarray*}

The \( \vec{B} \) in the denominator is a correction to \( 1/\left( \omega \pm \delta \omega /2\right)  \)
to second order in \( \left| \vec{B}\right| /\Omega  \) and for these cases
\( \omega \sim \Omega >>\left| \vec{B}\right|  \) and the correction is negligible.
This approximation has a simple interpretation, the width of the transition
profile will be given by \( \Omega  \). If that is narrower then the splitting
given by \( \vec{B} \) the interaction will only couple the spin states exactly
resonant with the interaction. By considering only \( \Omega >>\left| \vec{B}\right|  \)
the magnetic field splitting is assumed to be negligible compared to the width
of the transitions so that the transitions are driven with relative strengths
given only by the couplings \( \Omega  \) and not also the slightly different
detunings. In the same way, the \( \omega  \) that appears in the denominator
is the \( \omega  \) as determined before, \( \omega ^{2}=\lambda ^{2}+\left( \delta \omega /2\right) ^{2} \)
with \( \lambda \sim \Omega  \), plus a shift proportional to the magnetic
field which can then be neglected in the denominator. In all this gives, \( (\omega \pm \delta \omega /2)^{-1}=\left( \sqrt{\lambda ^{2}+\left( \delta \omega /2\right) ^{2}}\pm \delta \omega /2\right) ^{-1}+o(B/\Omega )^{2} \).
This still yields a strange and possible non-trivial lineshape, but exactly
on resonance the form is clear,

\begin{eqnarray*}
\omega a & = & \left( \frac{\Omega ^{\dagger }\Omega }{\lambda }+\vec{s}_{a}\cdot \vec{B}\right) a\\
\omega b & = & \left( \frac{\Omega \Omega ^{\dagger }}{\lambda }+\vec{s}_{b}\cdot \vec{B}\right) b
\end{eqnarray*}
The evolutions of the states in each level is then simply given by the sum of
the initial applied magnetic field and the effective interaction generated by
the laser driven transitions \( \Omega ^{\dagger }\Omega /\lambda  \). \( \lambda  \)
will be given approximately by the scalar piece of the quadrupole shift, \( \lambda \approx \delta \omega _{0}^{Q} \).

\subsection{Off-Diagonal Matrix Elements}

\label{Sec:OffDiagonalMatrixElements}

With the total interaction given simply as the sum of the applied magnetic field
plus the effective interaction generated by the applied laser driven couplings
an important simplification can be made. For the ground state this simplification
has a simple geometric interpretation. 

The couplings give at most an dipole interaction so the total interaction is
given by the vector sum of magnetic field and the vector parameterizing the
effects of the couplings, \( \delta \vec{\omega } \). The energies are given
by \( \left| \vec{B}+\delta \vec{\omega }\right|  \). Expanding this magnitude
\( \left| \vec{B}+\delta \vec{\omega }\right| ^{2}=\left| \vec{B}\right| ^{2}+2\vec{B}\cdot \delta \vec{\omega }+\left| \delta \vec{\omega }\right| ^{2} \).
The couplings are chosen so that \( \Omega \sim MHz>>\left| \vec{B}\right| \sim kHz \)
which implies \( \delta \omega ^{\left( 0\right) }>>\left| \vec{B}\right|  \)
but the difference in the shifts of the spin states, the splitting, is much
smaller than that due to the magnetic field if \( \delta \vec{\omega }\sim Hz<<\left| \vec{B}\right|  \)so
to first order in \( \delta \vec{\omega }/\left| \vec{B}\right|  \) the magnitude
of the vector sum is \( \left| \vec{B}+\delta \vec{\omega }\right| \approx \left| \vec{B}\right| +\hat{B}\cdot \delta \vec{\omega } \).
Only the components of \( \delta \vec{\omega } \) in the direction of \( \hat{B} \)
significantly change the energy. Geometrically this is just the equivilant of
the fact that a small change of vector in a parallel direction changes its length
directly, linearly, while a change in a orthogonal direction modifies the length
only to second order in the magnitude of the shift over the initial magnitude.
In the end this means the effect of components of \( \delta \vec{\omega } \)
perpendicular to \( \vec{B} \) on the splitting can be neglected.

This result can be seen perturbatively directly from the hamiltonian in a way
that easily generalizes to an arbitrary state if the coordinate system is chosen
so that \( \vec{B}||\hat{z} \). In this case the only contribution to off diagonal
elements of the whole interaction is from the effective contribution of the
couplings since \( \vec{\sigma }\cdot \vec{B}=B\sigma _{z} \) which is diagonal.
The diagonal elements, which include contributions from both the magnetic field
and the laser driven couplings, directly modify the energy of the unperturbed
spin states by \( \left( \Omega ^{\dagger }\Omega /\delta \omega ^{\left( 0\right) }\right) _{mm} \).
The effects of the remaining off-diagonal elements is generally very complicated
since it can include quadrupole, octapole and higher order structure, but perturbatively
the results are straightforward if the off-diagonal elements are small compared
to the magnetic field splitting. Their effects will start at second order, for
a given spin state \( m \)
\[
\delta \omega _{m}=\frac{\left( \Omega ^{\dagger }\Omega /\delta \omega ^{\left( 0\right) }\right) _{mm^{\prime }}\left( \Omega ^{\dagger }\Omega /\delta \omega ^{\left( 0\right) }\right) _{m^{\prime }m}}{E_{m}-E_{m^{\prime }}}\]
where \( E_{m}-E_{m^{\prime }}\approx B \) and so these shifts can be neglected
compared to the shifts due to the diagonal matrix elements for sufficiently
small \( \left( \Omega ^{\dagger }\Omega \right) _{m\neq m^{\prime }}<<B \).
\[
\delta \omega _{m}=\frac{\left| \left( \Omega ^{\dagger }\Omega \right) _{mm^{\prime }}/\delta \omega ^{\left( 0\right) }\right| ^{2}}{B}\]

\begin{eqnarray*}
\left( \Omega ^{\dagger }\Omega \right) _{mm} & = & \left( \Omega ^{Q\dagger }_{2}\Omega ^{Q}_{2}\right) _{mm}+\left( \delta \Omega ^{Q\dagger }_{1}\delta \Omega ^{Q}_{1}\right) _{mm}\\
 & + & 2\varepsilon Im(\Omega _{1}^{D\dagger }\Omega _{2}^{Q})_{mm}\\
 & + & 2\varepsilon Im(\delta \Omega _{2}^{D\dagger }\Omega _{2}^{Q}+\Omega _{1}^{D\dagger }\delta \Omega _{1}^{Q}+\delta \Omega _{1}^{Q\dagger }\delta \Omega _{2}^{D})_{mm}\\
 & + & 2\varepsilon Re(\Omega _{2}^{Q\dagger }\delta \Omega _{1}^{Q})_{mm}
\end{eqnarray*}

\subsection{Quadrupole-Quadrupole Misalignment Errors}

\subsubsection{Off-Diagonal Contributions}

This result is intended to be used to analyze the quadrupole-quadrupole cross
term \( 2Re(\Omega ^{Q\dagger }_{2}\delta \Omega ^{Q}_{1}) \), but in this
case it is not immediately clear that this bound on the size of the product
off-diagonal matrix elements is satisfied. The spin independent \( \vec{E}_{2} \)
quadrupole-quadrupole rate still dominates the shift so that \( \delta \omega ^{\left( 0\right) }\approx \Omega ^{Q}_{2} \)
making this error approximately,

\[
\delta \omega _{mm^{\prime }}\sim \left| \left( \delta \Omega ^{Q\dagger }_{1}\Omega _{2}^{Q}\right) _{mm^{\prime }}/\delta \omega ^{\left( 0\right) }\right| ^{2}/B\sim \left| \delta \Omega ^{Q}_{1}\right| ^{2}/B\]

The quadrupole rate for \( \vec{E}_{1} \) is of order \( \delta 1MHz \). If
the position of the antinode of \( \vec{E}_{1} \) can be positioned such that
\( \delta \sim 10^{-3} \), which it must in any case to make the parity term
itself accurate, \ref{Sec:GeneralQQTerm}, \( \delta \Omega ^{Q}_{1}\sim kHz \)
which is comparable to the planned initial magnetic field splitting so that
the effect of these off diagonal terms on the splitting would not be negligible.
However, with no other variations from the ideal geometry, the relative temporal
phases reduce the sizes of these terms further. 

With the fields still polarized exactly in the \( \hat{x} \) direction, and
propagating exactly in the \( \hat{z} \) direction, the matrix element for
this error, \( \Omega ^{Q\dagger }_{2}\delta \Omega ^{Q}_{1}+\delta \Omega ^{Q\dagger }_{1}\Omega ^{Q}_{2} \),
involves only \( \left\langle xz\right\rangle \left\langle xz\right\rangle k^{2}Re\left( E^{*}_{1}E_{2}\right)  \).
Ideally the fields are chosen to be exactly out of phase so that \( Re\left( E_{1}^{*}E_{2}\right) \propto cos\left( \pi /2\right) =0 \).
With a small error in this relative phase, \( \phi  \), this real part becomes,
\( sin\left( \phi \right)  \) and so the general overall size of this matrix
element is then given by two small parameters as \( \delta _{D}\phi  \). For
\( \phi <10^{-3} \) this gives a rate of order \( Hz \). Which then implies
a contribution to the splitting of 
\[
\delta \omega _{mm^{\prime }}\sim \left| \delta \phi \Omega ^{Q}_{1}\right| ^{2}/B\sim (Hz^{2})/kHz\sim 10^{-3}Hz\]
so that these off-diagonal matrix elements can be neglected.

\subsubsection{Diagonal Matrix Elements}

With these off-diagonal matrix elements negligible the remaining possibilities
to be considered are greatly reduced as they are completely contained in the
diagonal matrix elements for which there is the convenient result, in a spherical
basis,
\begin{eqnarray*}
\Delta \Omega ^{(k_{1},k_{2})}_{m} & = & \left( \Omega ^{\left( k_{1}\right) \dagger }\Omega ^{\left( k_{2}\right) }\right) _{mm}-\left( \Omega ^{\left( k_{1}\right) \dagger }\Omega ^{\left( k_{2}\right) }\right) _{-m,-m}\\
 & = & \left\langle j_{1},m\left| T^{\left( k_{1}\right) \dagger }_{s}\right| j_{2},m'\right\rangle \left\langle j_{2},m'\left| T^{\left( k_{2}\right) }_{s}\right| j_{1},m\right\rangle \\
 & \times  & \left( E^{\left( k_{1}\right) *}_{s}E^{\left( k_{2}\right) }_{s}-\left( -1\right) ^{k_{1}+k_{2}}E^{\left( k_{1}\right) *}_{-s}E^{\left( k_{2}\right) }_{-s}\right) 
\end{eqnarray*}
Where the difference in amplitudes can be written,

\begin{eqnarray*}
 &  & E^{\left( k_{1}\right) *}_{s}E^{\left( k_{2}\right) }_{s}-\left( -1\right) ^{k_{1}+k_{2}}E^{\left( k_{1}\right) *}_{-s}E^{\left( k_{2}\right) }_{-s}\\
 &  & =\left( E^{\left( k_{1}\right) *}_{sS}E^{\left( k_{2}\right) }_{sS}+E^{\left( k_{1}\right) *}_{sA}E^{\left( k_{2}\right) }_{sA}\right) \left( 1-\left( -1\right) ^{k_{1}+k_{2}}\right) \\
 &  & +\frac{\left| s\right| }{s}\left( E^{\left( k_{1}\right) *}_{sS}E^{\left( k_{2}\right) }_{sA}+E^{\left( k_{1}\right) *}_{sA}E^{\left( k_{2}\right) }_{sS}\right) \left( 1+\left( -1\right) ^{k_{1}+k_{2}}\right) 
\end{eqnarray*}
For a quadrupole-quadrupole cross term this gives,
\begin{eqnarray*}
\Delta \omega ^{QQ}_{m} & = & \left( \Omega _{1}^{Q\dagger }\Omega _{2}^{Q}+\Omega _{2}^{Q\dagger }\Omega _{1}^{Q}\right) _{mm}\\
 & - & \left( \Omega _{1}^{Q\dagger }\Omega _{2}^{Q}+\Omega _{2}^{Q\dagger }\Omega _{1}^{Q}\right) _{-m,-m}\\
 & = & 2Re\left( (\Omega ^{Q\dagger }_{1}\Omega ^{Q}_{2})_{mm}-(\Omega ^{Q\dagger }_{1}\Omega ^{Q}_{2})_{-m,-m}\right) \\
 & = & 4\left\langle j_{1},m\left| Q^{\dagger }_{s}\right| j_{2},m'\right\rangle \left\langle j_{2},m'\left| Q_{s}\right| j_{1},m\right\rangle \\
 & \times  & \frac{\left| s\right| }{s}Re\left( \delta E^{Q*}_{1sS}E^{Q}_{2sA}+\delta E^{Q*}_{1sA}E^{Q}_{2sS}\right) 
\end{eqnarray*}
The symmetric and antisymmetric pieces of the field amplitudes are given by,
\[
\begin{array}{cc}
E^{Q}_{2S}=\left( \partial _{x}E_{x}+\partial _{y}E_{y}\right) /2\equiv E_{xx}/2 & E_{2A}^{Q}=-i\left( \partial _{x}E_{y}+\partial _{y}E_{x}\right) /2=-iE_{xy}/2\\
E^{Q}_{1S}=i\left( \partial _{y}E_{z}+\partial _{z}E_{y}\right) /2\equiv iE_{yz}/2 & E_{1A}^{Q}=-\left( \partial _{x}E_{z}+\partial _{z}E_{x}\right) /2\equiv -E_{xz}/2
\end{array}\]
As before, the \( \Delta m=0 \) amplitude has no antisymmetric term, \( E_{0A}^{Q} \)
so it will not contribute to this spurious shift.

Notice that for a given transition the symmetric and antisymmetric pieces have
an explicit relative phase factor of \( i \). Since the two applied fields
are ideally exactly out of phase and linearly polarized, this extra factor of
\( i \) allows for a non-zero real part for the products with no relative phase
error. Then there is not yet any additional automatic suppression of this term
but it also means that errors from residual circular polarization of a single
beam, \( \sigma _{1} \) or \( \sigma _{2} \), do not contribute any systematic
shifts, since that would contain an additional \( i \) giving a zero real part,
unless both beams are partly circularly polarized, at which point they give
a contribution proportional to \( \delta _{1}\sigma _{1}\sigma _{2} \) which
is already sufficiently suppressed if \( \sigma <10^{-3} \) without having
to consider the resulting spin structure. 

For still perfectly linearly polarized beams these amplitudes give,

\begin{eqnarray*}
4Re\left( \delta E^{Q*}_{12S}E^{Q}_{22A}+\delta E^{Q*}_{12A}E^{Q}_{22S}\right)  & = & \delta _{D}Re(-i\left( \partial _{x}E^{*}_{1x}+\partial _{y}E^{*}_{1y}\right) \left( \partial _{x}E_{2y}+\partial _{y}E_{2x}\right) \\
 & + & i\left( \partial _{x}E_{1y}^{*}+\partial _{y}E_{1x}^{*}\right) \left( \partial _{x}E_{2x}+\partial _{y}E_{2y}\right) )\\
 & = & \delta _{D}Im(\left( \partial _{x}E^{*}_{1x}+\partial _{y}E^{*}_{1y}\right) \left( \partial _{x}E_{2y}+\partial _{y}E_{2x}\right) \\
 & - & \left( \partial _{x}E_{1y}^{*}+\partial _{y}E_{1x}^{*}\right) \left( \partial _{x}E_{2x}+\partial _{y}E_{2y}\right) )\\
4Re\left( \delta E_{11S}^{Q*}E_{21A}^{Q}+\delta E^{Q*}_{11A}E^{Q}_{21S}\right)  & = & \delta _{D}Re(-i\left( \partial _{y}E_{1z}^{*}+\partial _{z}E_{1y}^{*}\right) \left( \partial _{x}E_{2z}+\partial _{z}E_{2x}\right) \\
 & + & i\left( \partial _{x}E^{*}_{1z}+\partial _{z}E^{*}_{1x}\right) \left( \partial _{y}E_{2z}+\partial _{z}E_{2y}\right) )\\
 & = & \delta _{D}Im(\left( \partial _{y}E_{1z}^{*}+\partial _{z}E_{1y}^{*}\right) \left( \partial _{x}E_{2z}+\partial _{z}E_{2x}\right) \\
 & - & \left( \partial _{x}E^{*}_{1z}+\partial _{z}E^{*}_{1x}\right) \left( \partial _{y}E_{2z}+\partial _{z}E_{2y}\right) )
\end{eqnarray*}
This can be written more compactly as,
\begin{eqnarray*}
4Re\left( \delta E^{Q*}_{12S}E^{Q}_{22A}+\delta E^{Q*}_{12A}E^{Q}_{22S}\right)  & = & \delta _{D}Im\left( E_{1ii}^{*}E_{2xy}-E_{1xy}^{*}E_{2ii}\right) \\
4Re\left( \delta E_{11S}^{Q*}E_{21A}^{Q}+\delta E^{Q*}_{11A}E^{Q}_{21S}\right)  & = & \delta _{D}Im\left( E_{1yz}^{*}E_{2xz}-E_{1xz}^{*}E_{2yz}\right) 
\end{eqnarray*}
For plane waves, \( \partial _{i}\rightarrow k_{i} \). Ideally only \( E_{1x}=E^{D} \),
\( E_{2x}=E^{Q} \) and \( k_{z}=k \) are non zero. Giving only nonzero \( E_{xy} \),

\begin{eqnarray*}
4Re\left( \delta E^{Q*}_{12S}E^{Q}_{22A}+\delta E^{Q*}_{12A}E^{Q}_{22S}\right)  & = & \delta _{D}Im\left( \delta E_{1ii}^{*}\delta E_{2xy}-\delta E_{1xy}^{*}\delta E_{2ii}\right) \\
4Re\left( \delta E_{11S}^{Q*}E_{21A}^{Q}+\delta E^{Q*}_{11A}E^{Q}_{21S}\right)  & = & \delta _{D}Im\left( \delta E_{1yz}^{*}E_{2xz}-E_{1xz}^{*}\delta E_{2yz}\right) 
\end{eqnarray*}
Shifts due to misalignment errors driving the \( \Delta m=\pm 2 \) transitions
are explicitly further suppressed by two small parameters as both of the amplitudes
involved in each term are initially zero. Errors from \( \Delta m=\pm 1 \)
transitions are apparently suppressed by only one small factor giving a potentially
much larger shift.

To illustrate the structure of this problem writing these residual shifts in
terms of perturbations to the ideal geometry, keeping only the largest terms
for each transition,

\begin{eqnarray*}
4Re\left( \delta E^{Q*}_{12S}E^{Q}_{22A}+\delta E^{Q*}_{12A}E^{Q}_{22S}\right)  & = & \delta _{D}Im(\left( \delta k_{1x}E^{D*}+\delta k_{1y}\delta E_{1y}^{*}\right) \\
 & \times  & \left( \delta k_{2x}\delta E_{2y}+\delta k_{2y}E^{Q}\right) \\
 & - & \left( \delta k_{1x}\delta E_{1y}^{*}+\delta k_{1y}E^{D*}\right) \\
 & \times  & \left( \delta k_{2x}E^{Q}+\delta k_{2y}\delta E_{2y}\right) )\\
4Re\left( \delta E_{11S}^{Q*}E_{21A}^{Q}+\delta E^{Q*}_{11A}E^{Q}_{21S}\right)  & = & \delta _{D}Im(\left( \delta k_{1y}\delta E_{1z}^{*}+k\delta E_{1y}^{*}\right) \left( \delta k_{2x}\delta E_{2z}+kE^{Q}\right) \\
 & - & \left( \delta k_{1x}\delta E^{*}_{1z}+kE^{D*}\right) \left( \delta k_{2y}\delta E_{2z}+k\delta E_{2y}\right) )
\end{eqnarray*}
Keeping the largest of these terms,
\begin{eqnarray*}
4Re\left( \delta E^{Q*}_{12S}E^{Q}_{22A}+\delta E^{Q*}_{12A}E^{Q}_{22S}\right)  & = & \delta _{D}Im\left( \delta k_{1x}E^{D*}\delta k_{2y}E^{Q}-\delta k_{1y}E^{D*}\delta k_{2x}E^{Q}\right) \\
 & = & \delta _{D}Im(E^{D*}E^{Q})\left( \delta k_{1x}\delta k_{2y}-\delta k_{1y}\delta k_{2x}\right) \\
4Re\left( \delta E_{11S}^{Q*}E_{21A}^{Q}+\delta E^{Q*}_{11A}E^{Q}_{21S}\right)  & = & \delta _{D}Im\left( k\delta E_{1y}^{*}kE^{Q}-kE^{D}k\delta E_{2y}\right) \\
 & = & \delta _{D}k^{2}Im\left( \delta E_{1y}^{*}E^{Q}-E^{D*}\delta E_{2y}\right) 
\end{eqnarray*}
Note that these can be written quite transparently in vector form, to \( o(\delta \varepsilon ) \)
and \( o(\delta k) \), and with \( \vec{\varepsilon }_{1}\times \vec{\varepsilon }_{2}=0 \),
\( \vec{k}_{1}\times \vec{k}_{2}=0 \),
\begin{eqnarray*}
4Re\left( \delta E^{Q*}_{12S}E^{Q}_{22A}+\delta E^{Q*}_{12A}E^{Q}_{22S}\right)  & = & \delta _{D}Im(E^{D*}E^{Q})\left( \vec{k}_{1}\times \vec{k}_{2}\right) _{z}\\
4Re\left( \delta E_{11S}^{Q*}E_{21A}^{Q}+\delta E^{Q*}_{11A}E^{Q}_{21S}\right)  & = & \delta _{D}k^{2}Im\left( E^{D*}E^{Q}\right) \left( \vec{\varepsilon }_{1}\times \vec{\varepsilon }_{2}\right) _{z}
\end{eqnarray*}

The components driving the \( \Delta m=\pm 2 \) transitions give a possible
error suppressed by two additional small parameters, the misalignments of the
propagation direction of both beams. This provides enough factors that this
term can reasonably be made negligible with \( \delta k<10^{-3} \). The contribution
from the \( \Delta m=\pm 1 \) transition is suppressed by only two small factors,
so that a small, \( 10^{-3} \), error in the position of the dipole fields
node, and the quadrupole fields propagation direction will give \( Hz \) sizes
splittings that can then not be distinguished from the PNC shift. 

The vector form shows more clearly that misaligments of the polarization give
a shift along \( \hat{\varepsilon }_{1}\times \hat{\varepsilon }_{2} \) which
it initially exactly along \( \vec{k}=\hat{z} \), the same direction as the
parity shift. The errors from the \( \Delta m=\pm 2 \) transitions are along
\( \vec{k}_{1}\times \vec{k}_{2} \) which is then perpendicular to both \( \vec{k}_{1} \)
and \( \vec{k}_{2} \) so that initially the resulting shift is perpendicular
to \( \hat{z} \). The characteristic sized of the shifts for both errors are
about the same, but the \( \Delta m=\pm 2 \) shift is perpendicular to the
parity shift so that in the end its resulting effect on the spin state energies
is suppressed by \( \left( \hat{e}_{1}\times \hat{\varepsilon }_{2}\right) \cdot \vec{B} \). 

The vector form also illustrates how this shift is due to fields that are not
mirror symmetric. In each case the cross product gives an axial vector. This
will not change sign under a parity transformation so that the mirror image
could be distinguished from the original geometry by, for example, the orientation
of \( \hat{\varepsilon }_{1}\times \hat{\varepsilon }_{2} \) or \( \vec{k}_{1}\times \vec{k}_{2} \)
relative to any of the other vectors defined by the fields. This chiral arrangement
of fields can then induce a chiral response in the ion.

\subsection{Alternate Geometries}

These results were from perturbations to a particular initial geometry, \( \vec{k}_{1}||\vec{k}_{2} \),
and \( \hat{\varepsilon }_{1}||\hat{\varepsilon }_{2} \). Other beam geometries
can be used to give a parity shift, it is possible that those have less severe
systematic problems. For example, with \( E_{xz}\neq 0 \) for both \( E_{1} \)
and \( E_{2} \) the systematic shift due to \( \Delta m=\pm 1 \) transitions
given by 
\[
4Re\left( \delta E_{11S}^{Q*}E_{21A}^{Q}+\delta E^{Q*}_{11A}E^{Q}_{21S}\right) =\delta _{D}Im\left( E_{1yz}^{*}E_{2xz}-E_{1xz}^{*}E_{2yz}\right) \]
 has contributions from both \( E_{1yz}E_{2xz}=\delta E_{1yz}E_{2xz} \) and
\( E_{1xz}E_{2yz}=E_{1xz}\delta E_{2yz} \) terms. A parity shift is also possible
for \( E_{2xz}\neq 0 \) with \( \hat{\varepsilon }_{1}=\hat{x} \) and \( \vec{k}_{1}||\hat{y} \)
rather than \( \vec{k}_{1}||\hat{z} \), giving \( E_{1xy} \) initially nonzero.
Then the contribution from the shift from the \( E_{1xz}E_{2yz} \) term is
\( \delta E_{1xz}\delta E_{2yz} \) which is suppressed by two small misalignment
parameters. In this case the \( E_{1yz}E_{2xz} \) term is still proportional
to only one small misalignment parameter as \( \delta E_{1yz}E_{2xz} \). In
addition, for the shift from the \( \Delta m=\pm 2 \) term
\begin{eqnarray*}
4Re\left( \delta E^{Q*}_{12S}E^{Q}_{22A}+\delta E^{Q*}_{12A}E^{Q}_{22S}\right)  & = & \delta _{D}Im\left( E_{1ii}^{*}E_{2xy}-E_{1xy}^{*}E_{2ii}\right) 
\end{eqnarray*}
\( E_{1xy} \) is now initially nonzero, and so this shift is nonzero for only
a single misalignment giving nonzero \( \delta E_{2ii} \).

It is possible that other geometries reduce the size of all of these terms.
This requires a general consideration of the possibilities. For general fields
the size of the parity shift is given by,
\begin{eqnarray*}
\Delta \omega ^{PNC}_{m} & = & \varepsilon \left( \Omega _{1}^{D\dagger }\Omega _{2}^{Q}-\Omega _{2}^{Q\dagger }\Omega _{1}^{D}\right) _{mm}\\
 & - & \varepsilon \left( \Omega _{1}^{D\dagger }\Omega _{2}^{Q}-\Omega _{2}^{Q\dagger }\Omega _{1}^{D}\right) _{-m,-m}\\
 & = & 2\varepsilon Im\left( (\Omega ^{D\dagger }_{1}\Omega ^{Q}_{2})_{mm}-(\Omega ^{D\dagger }_{1}\Omega ^{Q}_{2})_{-m,-m}\right) \\
 & = & 4\left\langle j_{1},m\left| D^{\dagger }_{s}\right| j_{2},m'\right\rangle \left\langle j_{2},m'\left| Q_{s}\right| j_{1},m\right\rangle \\
 & \times  & \frac{\left| s\right| }{s}Im\left( E^{D*}_{1sS}E^{Q}_{2sS}+\delta E^{D*}_{1sA}E^{Q}_{2sA}\right) 
\end{eqnarray*}
With 
\[
\begin{array}{cc}
E^{D}_{1S}=\frac{iE_{y}}{\sqrt{2}} & E_{1A}^{D}=-\frac{E_{x}}{\sqrt{2}}\\
E_{0S}^{D}=E_{z} & E_{0A}^{D}=0
\end{array}\]
and the quadrupole amplitudes

\[
\begin{array}{cc}
E^{Q}_{2S}=\left( \partial _{x}E_{x}+\partial _{y}E_{y}\right) /2\equiv E_{ii}/2 & E_{2A}^{Q}=-i\left( \partial _{x}E_{y}+\partial _{y}E_{x}\right) /2\equiv -iE_{xy}/2\\
E^{Q}_{1S}=i\left( \partial _{y}E_{z}+\partial _{z}E_{y}\right) /2\equiv iE_{yz}/2 & E_{1A}^{Q}=-\left( \partial _{x}E_{z}+\partial _{z}E_{x}\right) /2\equiv -E_{zx}/2\\
E_{0S}^{Q}=\partial _{z}E_{z}/\sqrt{6}=E_{zz}/\sqrt{6} & E_{0A}^{Q}=0
\end{array}\]
this gives,

\begin{eqnarray*}
2\sqrt{2}Im\left( E^{D*}_{11S}E^{Q}_{21S}+E^{D*}_{11A}E^{Q}_{21A}\right)  & = & Im\left( E_{1y}^{*}E_{2yz}+E_{1x}^{*}E_{2xz}\right) \\
\sqrt{6}Im\left( E_{10S}^{D*}E_{20S}^{Q}+E^{D*}_{10A}E^{Q}_{20A}\right)  & = & Im\left( E_{1z}^{*}E_{2zz}\right) 
\end{eqnarray*}
The shifts from the quadrupole term were given by,

\begin{eqnarray*}
4Re\left( \delta E^{Q*}_{12S}E^{Q}_{22A}+\delta E^{Q*}_{12A}E^{Q}_{22S}\right)  & = & \delta _{D}Im\left( E_{1ii}^{*}E_{2xy}-E_{1xy}^{*}E_{2ii}\right) \\
4Re\left( \delta E_{11S}^{Q*}E_{21A}^{Q}+\delta E^{Q*}_{11A}E^{Q}_{21S}\right)  & = & \delta _{D}Im\left( E_{1yz}^{*}E_{2xz}-E^{*}_{1xz}E_{2yz}\right) 
\end{eqnarray*}
Any geometries having \( E_{2xz} \) and \( E_{1x} \) non-zero, with \( \vec{k}_{1}||\hat{z} \),
give a parity shift have the same systematic problems, regardless of whether
this amplitude is from \noun{\( \vec{k}_{2}||\hat{x} \)} or \noun{\( \vec{k}_{2}||\hat{z} \).
\( E_{1,2ii} \)} and \( E_{1,2xy} \) are all initially zero, before misalignments,
so error from the \( \Delta m=\pm 2 \) transitions are suppressed by two small
factors, but the \( E_{xz} \) terms in the expression for the shifts from \( \Delta m=\pm 1 \)
transitions are non-zero, so that only an additional misalignment of the polarization
of \( E_{1} \) or \( E_{2} \) in the \( \hat{y} \) direction, giving a nonzero
\( \delta E_{1,2yz} \), gives a nonzero quadrupole splitting. As already noted
using \( \vec{k}_{1}||\hat{y} \) similarly fails to improve the situation,
one contribution from the \( \Delta m=\pm 1 \) transitions is made smaller,
the other remains large, and a previously negligible \( \Delta m\pm 2 \) contribution
is made larger. With nonzero \( E_{1y}E_{2yz} \) giving a parity splitting
results are similar. For \( \vec{k}_{1}||\hat{z} \), only a small \( \delta E_{xz} \)
for \( E_{1} \) or \( E_{2} \), and for \( \vec{k}_{1}||\hat{x} \) only nonzero
\( \delta E_{2,ii} \) are required to give a non-zero shift.

Using the \( E_{1z}E_{2zz} \) term to generate a parity shift, the quadrupole
terms giving shifts due to the \( \Delta m=\pm 1 \) or \( \Delta m=\pm 2 \)
transitions are all initially zero, then requiring two additional misalignments
each for a residual quadrupole splitting. As previously mentioned this field
geometry is not possible without having other field components as \( \vec{\nabla }\cdot \vec{E}=0 \)
requires \( E_{zz}=0 \) if \( E_{xx}=E_{yy}=0 \). Nonzero \( E_{zz} \) then
requires nonzero \( E_{ii} \), which also implies nonzero \( E_{iz} \) giving
initially nonzero terms in the shift due to all transitions. 

More simply a the coordinate system could simply be chosen so that only \( E_{2xz} \)
is not zero, that is \( \vec{E}_{2}\cdot \hat{y}=\vec{k}_{2}\cdot \hat{y}=0 \).
Then a nonzero parity shift requires \( \vec{E}_{1}||\hat{x} \) but allows
either \( \vec{k}_{1}||\hat{z} \) or \( \vec{k}_{2}||\hat{y} \). Both cases
give terms all terms initially zero for the shift due to \( \Delta m=\pm 2 \)
transitions, but a shift due to \( \Delta m=\pm 1 \) transitions given by \( E_{1yz}E_{2xz} \)
which is nonzero for only the single error of a small misalignment of the \( \vec{E}_{1} \)
polarization. This also now accounts for all possible geometries and this and
the former analysis demonstrate clearly that no other possible beam geometry
that gives a nonzero parity splitting is immune to this difficult systematic
shift due to the \( \Delta m=\pm 1 \) transitions that is suppressed by only
two small parameters.

\subsection{Intermediate \protect\( m\protect \) Transition Restrictions}

Amazingly, it turns out that differential shifts of the \( D_{3/2} \) magnetic
sublevels due to off-resonant dipole couplings discussed in sec.\ref{Sec:NonResonantDStateShifts}
can be used to further suppress this potentially troublesome error. It will
turn out that that the off-resonant dipole coupling of the \( D_{3/2} \) state
to other states in the atoms from the strong \( \vec{E}_{1} \) dipole electric
field results in a significant, \( MHz \) sized shift of the \( D_{3/2,\pm 1/2} \)
states relative to the \( D_{3/2,\pm 3/2} \) states, where the quantization
axis is along the electric field. This structure is clear from simply considering
the contributions from \( j=1/2 \) states alone, as they provide no levels
for the \( D_{3/2,\pm 3/2} \) states to couple to and the contribution of the
shift to these states, from these \( j=1/2 \) levels, is zero, while couplings
to \( D_{3/2,\pm 1/2} \) states are nonzero giving finite shifts.

Then with a sufficiently narrow transition width one set of states is shifted
out of resonance and then doesn't participate in the \( S-D \) resonant interaction.
The quadrupole rate for the \( \vec{E}_{1} \) field is also about \( 1MHz \),
about the same size as this \( D \) state spin splitting so this nonresonance
condition is just barely satisfied. This non-resonant shift is proportional
to the square of the dipole electric field, while the resonant rate is linear
in the electric field, so it is possible that the field could be increased slightly
from the value chosen for optimal S/N to allow for this selection of a subset
of \( m \) states. 

Now consider quadrupole transitions to this restricted set of states. This is
most easily evaluated in a coordinate system with \( \vec{E}_{1}||\hat{z} \)
so that the states to which transitions are possible is obvious, the general
structure for arbitrary fields and coordinate systems is developed in sec.\ref{Sec:ShiftVectorStuctureWithmRestrictions},
but the mechanism is easily illustrated with this particular special choice.
Quadrupole amplitudes are given by,
\begin{eqnarray*}
\Omega ^{Q}_{mm^{\prime }} & = & \left\langle 5D_{3/2},m\left| x_{s}\right| 6S_{1/2},m^{\prime }\right\rangle E_{s}
\end{eqnarray*}
If transition to the \( m=\pm 3/2 \) states are used, \( \Delta m=0 \) quadrupole
transitions will not be possible, for transitions to the \( m=\pm 3/2 \) states,
\( \Delta m=\pm 2 \) transitions are not possible. The former case is not particularly
useful. For the latter, the quadrupole amplitude will be given by,
\begin{eqnarray*}
\Omega ^{Q}_{mm^{\prime }} & = & \sum _{s=0,1}\left\langle 5D_{3/2},m\left| x_{s}\right| 6S_{1/2},m^{\prime }\right\rangle E_{s}
\end{eqnarray*}
This is driven by nonzero amplitudes for

\begin{eqnarray*}
E^{Q}_{\pm 1} & = & \mp (E_{xz}+E_{zx})/2+i(E_{yz}+E_{zy})/2\\
E^{Q}_{0} & = & E_{zz}/\sqrt{6}
\end{eqnarray*}
Notice that these amplitudes require some component of \( \hat{\varepsilon } \)
or \( \vec{k} \) in the \( \hat{z} \) direction. For fields completely contained
in the \( x-y \) plane, that is fields with a plane of polarization perpendicular
to the axis of the field generating this differential shift, only \( E^{Q}_{\pm 2} \)
amplitudes are nonzero. These transitions are not driven as they are now out
of resonance, so for these kinds of fields the entire quadrupole coupling is
zero. This result can be expressed in a coordinate system independent form.
With \( \vec{E}_{1} \) giving the direction of the non-resonant shifts, quadrupole
fields with \( \hat{\varepsilon } \) and \( \vec{k} \) perpendicular to \( \vec{E}_{1} \)
give no quadrupole amplitude, so in effect,
\[
\Omega ^{Q}\propto a\hat{\varepsilon }_{2}\cdot \vec{E}_{1}+b\vec{k}_{2}\cdot \vec{E}_{1}\]

The errors causing the larger systematic shifts were due to \( \Delta m=\pm 1 \)
transitions given by,
\begin{eqnarray*}
4Re\left( \delta E_{11S}^{Q*}E_{21A}^{Q}+\delta E^{Q*}_{11A}E^{Q}_{21S}\right)  & = & \delta _{D}Im(E_{1yz}^{*}E_{2xz}-E^{*}_{1xz}E_{2yz})
\end{eqnarray*}
For the usual ideal geometry considered here, \( E_{1xz} \) and \( E_{2xz} \)
are nonzero. For simplicity, take \( \hat{x} \) and \( \hat{z} \) to define
the plane of polarization of \( E_{1} \). The for the possible \( E_{1yz}E_{2xz} \)
contribution to the shift, \( \delta E_{1yz} \) is zero by this definition.
For \( E_{1xz}E_{2yz} \), \( E_{2yz}=\delta E_{2yz}=k_{z}\delta E_{2y} \)
is nonzero for a small misalignment of the \( \vec{E}_{2} \) polarization in
the \( \hat{y} \) direction, but this component has a plane of polarization
perpendicular to \( \hat{\varepsilon }_{1} \) so the restricted transitions
result in zero quadrupole amplitude for this field configuration and this error
gives no shift. In this way the previously troublesome errors from the \( \Delta m=\pm 1 \)
transitions are further suppressed by the factor \( \hat{\varepsilon }_{1}\cdot \delta \hat{\varepsilon }_{2\perp } \)
where \( \delta \hat{\varepsilon }_{2\perp } \) are fluctuations of \( \vec{E}_{2} \)
perpendicular to its initial plane of polarization. 

This result is special to this particular geometry. Again, a parity splitting
is possible with the same \( \hat{\varepsilon }_{1} \) with \( \vec{k}_{1}||\hat{y} \).
Similarly taking \( \hat{x} \) and \( \hat{y} \) to as the plane of polarization
of \( E_{1} \), \( E_{yz} \) and \( E_{xz} \) are zero by definition. For
the contributions from the \( \Delta m=\pm 2 \) transitions,

\begin{eqnarray*}
4Re\left( \delta E^{Q*}_{12S}E^{Q}_{22A}+\delta E^{Q*}_{12A}E^{Q}_{22S}\right)  & = & \delta _{D}Im\left( E_{1ii}^{*}E_{2xy}-E_{1xy}^{*}E_{2ii}\right) 
\end{eqnarray*}
\( E_{1xy} \) is initially non-zero and \( E_{2ii}=\delta E_{2xx}=\delta k_{x}E_{x} \).
This perturbation still has \( \hat{\varepsilon }_{2}||\vec{E}_{1} \) so there
is no additional reduction of this term from the restricted transitions. This
gives the first practical motivation for a preference for beam geometries. Configurations
with \( \vec{k}_{1} \) initially contained in the plane of polarization of
\( \vec{E}_{2} \) can exploit this reduction in systematic shifts from restricted
transitions. With \( \hat{\varepsilon }_{1} \)already required to also be in
the plane of polarization of \( \vec{E}_{2} \) this then corresponds to geometries
whether the plane of both \( \vec{E}_{1} \) and \( \vec{E}_{2} \) are the
same.

\subsection{Cartesian basis analysis}

This analysis was centered around a spherical basis for the field amplitudes.
As usual the origin of the errors obtained from a similar analysis in a cartesian
basis is more immediately intuitive. Though in this case the reduction of certain
errors from the transitions restrictions just discussed is less trivial. In
this basis the quadrupole shifts are given by,

\begin{eqnarray*}
\Delta \omega ^{QQ}_{m} & = & 2Re\left( (\Omega ^{Q\dagger }_{1}\Omega ^{Q}_{2})_{mm}-(\Omega ^{Q\dagger }_{1}\Omega ^{Q}_{2})_{-m,-m}\right) \\
 & = & 2Re(\partial _{i}E_{1j}^{*}\partial _{k}E_{2l}\\
 & \times  & \left\langle j_{1},m\left| x_{i}x_{j}\right| j_{2},m'\right\rangle \left\langle j_{2},m'\left| x_{k}x_{l}\right| j_{1},m\right\rangle \\
 & - & \left\langle j_{1},-m\left| x_{i}x_{j}\right| j_{2},-m'\right\rangle \left\langle j_{2},-m'\left| x_{k}x_{l}\right| j_{1},-m\right\rangle )
\end{eqnarray*}
The scalar pieces of the quadrupole operator will give zero since the initial
and final angular momentums are different. Taking the fields to be plane standing
waves the \( \partial _{i} \) give \( k_{i} \) and the size of the gradient
of the \( \vec{E}_{1} \) field is given by \( \delta _{D} \)

As before a spin flip transformation can be used to change \( -m\rightarrow m \).
The details of the structure of the spin flip are not important here. Since
it must be unitary the result must be the spin flipped state determined up to
a phase. For rotations this phase is only dependent on \( j \) and \( m \).
For mirror reflections the phase also depends on \( l \) through the spatial
parity of the state. In either case, since only diagonal matrix elements are
considered here each state involved in this matrix element appears once as an
initial state and once as a final state these phases will exactly cancel. This
gives,
\begin{eqnarray*}
\Delta \omega ^{QQ}_{m} & = & 2\delta _{D}k_{i}k_{j}Re(E_{1j}^{*}E_{2l}\\
 & \times  & \left\langle j_{1},m\left| x_{i}x_{j}\right| j_{2},m'\right\rangle \left\langle j_{2},m'\left| x_{k}x_{l}\right| j_{1},m\right\rangle \\
 & - & \left\langle j_{1},m\left| F^{\dagger }x_{i}x_{j}F\right| j_{2},m'\right\rangle \left\langle j_{2},m'\left| F^{\dagger }x_{k}x_{l}F\right| j_{1},m\right\rangle )
\end{eqnarray*}
The result is independent of the representation of the spin flip operator so
that any can be used. The simplest choice is a mirror reflection of, or a rotation
about \( \hat{x} \) or \( \hat{y} \) as this leaves all but possibly the sign
of any operator invariant, \( F^{\dagger }x_{i}F=\eta _{i}x_{i} \), so that,
\begin{eqnarray*}
\Delta \omega ^{QQ}_{m} & = & 2\delta _{D}k_{i}k_{j}Re(E_{1j}^{*}E_{2l}(1-\eta _{i}\eta _{j}\eta _{k}\eta _{l})\\
 & \times  & \left\langle j_{1},m\left| x_{i}x_{j}\right| j_{2},m'\right\rangle \left\langle j_{2},m'\left| x_{k}x_{l}\right| j_{1},m\right\rangle )
\end{eqnarray*}
For the various spin flip operators,

\begin{eqnarray*}
M_{\hat{x}} & \rightarrow  & \eta _{i}\eta _{j}\eta _{k}\eta _{l}=(-1)^{n_{x}}\\
M_{\hat{y}} & \rightarrow  & \eta _{i}\eta _{j}\eta _{k}\eta _{l}=(-1)^{n_{y}}\\
R_{\hat{x}} & \rightarrow  & \eta _{i}\eta _{j}\eta _{k}\eta _{l}=(-1)^{n_{y}+n_{z}}\\
R_{\hat{y}} & \rightarrow  & \eta _{i}\eta _{j}\eta _{k}\eta _{l}=(-1)^{n_{x}+n_{z}}
\end{eqnarray*}
Again, this is an apparent inconsistant dependance on the representation of
the spin flip operator, but as usual, the combine to define selection rules,
all must hold for a non-zero matrix element requiring,
\begin{eqnarray*}
(-1)^{n_{z}} & = & 1\\
(-1)^{n_{x}+n_{y}} & = & 1
\end{eqnarray*}
The later also corresponds to \( \left| n_{x}-n_{y}\right|  \) even. Notice
that this is exactly consistant with the results obtained using a spherical
basis.

\begin{eqnarray*}
4Re\left( \delta E^{Q*}_{12S}E^{Q}_{22A}+\delta E^{Q*}_{12A}E^{Q}_{22S}\right)  & = & \delta _{D}Im\left( E_{1ii}^{*}E_{2xy}-E_{1xy}^{*}E_{2ii}\right) \\
4Re\left( \delta E_{11S}^{Q*}E_{21A}^{Q}+\delta E^{Q*}_{11A}E^{Q}_{21S}\right)  & = & \delta _{D}Im\left( E_{1yz}^{*}E_{2xz}-E^{*}_{1xz}E_{2yz}\right) 
\end{eqnarray*}
the terms that appear have exactly the appropriate numbers of each coordinate
to be spin dependant and satisfy the selection rules so that the resulting shifts
are non-zero.

As before, the phase is given by the number of \( y \) operators,
\begin{eqnarray*}
\Delta \omega ^{QQ}_{m} & \propto  & \delta _{D}Re(E_{1j}^{*}E_{2l}i^{n_{y}})(1-\eta _{i}\eta _{j}\eta _{k}\eta _{l})
\end{eqnarray*}
With \( F=M_{\hat{y}} \) it is clear that \( \eta _{i}\eta _{j}\eta _{k}\eta _{l} \)
is odd, giving \( \Delta \omega \neq 0 \), only for \( n_{y} \) odd, which
in turn gives \( i^{n_{y}}=i \), and \( \Delta \omega  \) nonzero by phase
requires,
\begin{eqnarray*}
\Delta \omega ^{QQ}_{m} & \propto  & \pm \delta _{D}Re(iE_{1j}^{*}E_{2l})(1-\eta _{i}\eta _{j}\eta _{k}\eta _{l})=\pm \delta _{D}Im(E_{1j}^{*}E_{2l})(1-\eta _{i}\eta _{j}\eta _{k}\eta _{l})
\end{eqnarray*}
Ideally \( E_{1} \) and \( E_{2} \) are out of phase so that \( Im(E_{1}^{*}E_{2}) \)
is non zero and so any perturbation in alignment or polarization of the same
phase as the original fields gives a nonzero \( \Delta \omega  \), thus is
it not nesseary to consider the effects of circular polarization here as well.

For the usual initial ideal geometry, the most convenient representation is
\( F=R_{\hat{y}} \). For this case it is clear that the quadrupole shift is
spin dependant if \( n_{y}=1,3 \). The selection rules then require that \( n_{z}=1,3 \)
and \( n_{2}=0,2 \) for a non-zero shift. The inital matrix element involved
is \( \left\langle zx\right\rangle \left\langle zx\right\rangle  \). With \( F=R_{\hat{y}} \)
this is trivially spin independant since \( n_{y}=0 \), and the shift can become
spin dependant for any perturbation in the \( \hat{y} \) direction, perpendicular
to the plane of polarization of the ideal beams. Whether the shift is non-zero
depends on the particular pertubation. A single variation of a \( \vec{k} \)
, originally along \( \hat{z} \), in the \( \hat{y} \) direction gives a term
like \( \left\langle yx\right\rangle \left\langle zx\right\rangle  \), this
must be zero since, in particular, \( (-1)^{n_{z}}=-1 \), and \( (-1)^{n_{x}+n_{y}}=1 \).
An addition perturbation of the other \( \vec{k} \) in the \( \hat{x} \) direction
gives a \( \left\langle yx\right\rangle \left\langle xx\right\rangle  \) which
is spin dependant and nonzero. Note that a variation of this second \( \vec{k} \)
along \( \hat{y} \) as well also gives a non-zero coupling as \( \left( -1\right) ^{n_{x}+n_{y}}=1 \),
but now \( (-1)^{n_{y}}=1 \) and the term is not spin dependant. A similar
analysis for perturbations of one \( \vec{k} \) along \( \hat{x} \) requires
an additional pertubation of the other \( \vec{k} \) along \( \hat{y} \),
and for these cases two additional small perturbations in alignment, as well
as the initial dipole anti-node phase error are required for to generate a spurious
quadrupole shift.

Alternately, a single perturbation of either \( \hat{\varepsilon } \) along
\( \hat{y} \), gives a \( \left\langle zy\right\rangle \left\langle zx\right\rangle  \)
which is both spin dependant and non-zero except for the cases of restricted
transitions to the \( m \) sublevels of the excited state, in which case this
is a perturbation with a plane of polarization in the \( y-z \) plane, perpendicular
to the strong dipole electric field in the \( \hat{x} \) direction and the
projection to the \( m=\pm 1/2 \) state in this \( \hat{x} \) direction gives
zero quadrupole amplitude. Then, as before, there is a possible spurious quadrupole
shift depending on only a single geometric misalignment, the polarization, and
the standing wave phase error, that is further supressed when considering only
a restricted set of intermediate state.

This particular example nicely illustrates how this cartesian basis analysis
gives a more geometrically intuitive result. For \( F=M_{\hat{y}} \) any perturbation
along the \( \hat{y} \) direction gives a splitting.  For this geometry, ideally
the plane of polarization is completely contained in the \( x-z \) plane and
it mirror symmetric, but any perturbation in the \( \hat{y} \) direction can
now be used to define left or right relative to the other fields and the fields
are no longer mirror symmetric leading to a possible spurious spin dependant
shift. 

This geometric interpretation is valid for other geometries as well, though
a bit less clear for \( E_{2zx}\neq 0 \), \( E_{1yx}\neq 0 \) and \( \vec{E}_{1}||\hat{x} \).
This is still obviouly mirror symmetric but \( F=M_{\hat{y}} \) suggests that
this term, giving \( \left\langle zx\right\rangle \left\langle yx\right\rangle  \),
is already spin dependant, similarly for \( M=R_{\hat{y}} \). In this case
the selection rules make the shift zero and other representations for \( F \)
make the geometric picture clearer.fv

\subsection{Reduction Strategies}

This analysis accounts for all possible systematic errors due to misalignments
of the parity beams. In all cases the resulting splittings can be suppressed
by at least three small parameters which can be used to reduce the sizes of
these shifts. The overall sizes of these shifts and their general dependence
on the misalignment errors are summarized in tab.\ref{tab:SystemmaticErrorSizes}.
An overall suppression in the order of \( 10^{-9}-10^{-10} \) is required for
these contributions to be negligible compared to the parity splitting. This
is possible with the more reasonable constraints of alignments individually
accurate to a part in \( 10^{-3} \).
\begin{table}

\caption{\label{tab:SystemmaticErrorSizes}Contribution to spin dependent shifts from
perturbations to ideal geometry.}
\vspace{0.3cm}
{\centering \begin{tabular}{|c|c|c|}
\hline 
Term&
Shift&
\( 10^{-7}\delta \omega /\delta \omega _{PNC} \)\\
\hline 
\hline 
\( \varepsilon \Omega _{2}^{Q\dagger }\Omega ^{D}_{1} \)&
\( \varepsilon \Omega _{1}^{D} \)&
- \\
\hline 
\( \Omega _{2}^{Q\dagger }\Omega ^{Q}_{2} \)&
\( \sigma _{Q}\Omega _{2}^{Q} \)&
\( \sigma _{Q}\frac{E^{Q}}{E^{D}} \) \\
\hline 
\( \Omega _{1}^{Q\dagger }\Omega ^{Q}_{1} \)&
\( \sigma _{D}\delta _{1}^{2}\Omega _{1}^{Q} \)&
\( \sigma _{D}\delta ^{2} \) \\
\hline 
\( \left( \Omega _{2}^{Q\dagger }\delta \Omega ^{Q}_{1}\right) _{m\neq m^{\prime }} \)&
\( \phi ^{2}\delta _{1}\frac{\delta _{1}\Omega _{1}^{Q}}{B}\Omega _{1}^{Q} \)&
\( \phi ^{2}\delta _{1}\frac{\delta _{1}\Omega _{1}^{Q}}{B} \)\\
\hline 
\( \left( \Omega _{2}^{Q\dagger }\delta \Omega ^{Q}_{1}\right) _{mm,\Delta m=\pm 2} \)&
\( \delta _{1}\delta k_{1}\delta k_{2}\Omega _{1}^{Q} \)&
\( \delta _{1}\delta k_{1}\delta k_{2} \)\\
\hline 
\( \left( \Omega _{2}^{Q\dagger }\delta \Omega ^{Q}_{1}\right) _{mm,\Delta m=\pm 1} \)&
\( \delta _{1}(\hat{\varepsilon }_{1}\cdot \delta \hat{\varepsilon }_{2\perp })\Omega _{1}^{Q} \)&
\( \delta _{1}(\hat{\varepsilon }_{1}\cdot \delta \hat{\varepsilon }_{2\perp }) \)\\
\hline 
\end{tabular}\par}\vspace{0.3cm}
\end{table}

This is difficult with completely independent fields, but some automatic simplification
may be possible with particular configurations. It was already shown that no
particular field geometry is intrinsically better for systematics, though one
case is unable to take advantage of reductions from intermediate spin state
restrictions. However, the technical details of various  particular ways to
generate and direct these fields may have an advantage in some cases.

For example, somehow using a single beam with guarantee that the polarizations
and directions of propagation are identical, which automatically eliminates
all of the quadrupole-quadrupole systematics if it really can be done precisely.
One way to do this is to simply use a single traveling wave. This gives the
appropriate out of phase amplitude and gradient requires, but it also means
that \( E_{1} \) and \( E_{2} \) are the same, which gives an excessively
high quadrupole rate that can make systematics problems, like shifts due to
residual bits of polarization in the beam, excessively large. Also the strategies
for correcting for this kind of error, by removing the quadrupole beam to eliminate
the PNC shift in order to measure the size of the shift due to this circular
polarization to subtract it off, can not be done because \( E_{1} \) and \( E_{2} \)
are provided from the same beam.

Another means of using a single beam to get a single beam to generate the necessary
fields with the appropriate phases is to add a bit of a running wave to a single
standing wave. In creating a standing wave, a mirror of a finite reflectivity
must be used so that a bit of traveling wave is always present. A larger amount
could be obtained if necessary by using a less reflective mirror. This doesn't
guarantee beam alignment as that now depends on the positioning of the mirror,
and the mirror may alter the polarization of the beam slightly upon reflection
so that the \( E_{1} \) and \( E_{2} \) fields are not precisely identically
polarized. The mirror misalignment errors may not be important, a more expicit
analysis would be required, and polarization alignment could be enforces by
using high quality polarizers inside any standing wave generating cavity, so
this method could turn out to be effective. Again there is the difficulty that,
since the beams are not independent, which it what helps to guarantee their
alignment, there may not be enough freedom to be able to manipulate them enough
to be able to detect and correct problems. This may turn out not to be necessary
if the engineering constraints are sufficient.

These kinds of decisions must be partly put off until more knowledge about the
performance of a mirror and cavity for generating this standing wave is known.
Such considerations will be the central focus of studies involving the mirror
when it becomes operational.

Finally it may turn out that the frequency dependence of the PNC and quadrupole
shifts are slightly difference and could be distinguished. This wasn't immediately
apparent in the analysis considered so far and also required additional theoretical
investigation.

\section{Generalized Pauli Matrices}

\label{Sec:GeneralizedPauliMatricies}

This perturbative approach is straight-forward, but cumbersome. It is based
on variations from a particular ideal geometry so it does not immediately apply
to all geometries. As briefly considered perturbatively for the \( E_{\pm 1}^{Q} \)
amplitude errors, it is possible that for some field configurations some constraints
from systematic errors are eased and with this perturbative approach each case
must be explicitly analyzed, it is difficult to identify general patterns. Also
it will be useful to use these transitions and shifts for other purposes, such
as calibration of the electric fields. The existing final solution can not be
applied to such a problem and the shifts must be calculated again for the different
geometries to be used for this kind of calibration. In addition the geometric
structure of the dependence of the shift on the directions of the applied optical
fields is obscured. These difficulties are remedied by finally solving for the
completely general case.

\subsection{Vector Structure}

For any of the terms, the shifts are given by the matrix elements of a product
like, \( \Omega ^{(k_{1})\dagger }\Omega ^{(k_{2})} \). As already noted, this
is a hermitian matrix. For the ground state, this is a \( 2\times 2 \) matrix,
so it can be written in the form,

\[
\Omega ^{(k_{1})\dagger }\Omega ^{(k_{2})}=\delta \omega +\delta \vec{\omega }\cdot \sigma \]
\( \delta  \)\( \omega  \) is a spin independent shift, and \( \delta \vec{\omega } \)
is a simple spatial vector. This should be given somehow by the vectors that
define the system. For the parity term this should be the polarizations of both
fields, \( \vec{E}^{D} \), \( \vec{E}^{Q} \) and the propagation direction
of the quadrupole field \( \vec{k}^{Q} \). As seen from some of the matrix
elements previously explicitly calculated, the parity vector takes a form something
like,

\[
\delta \vec{\omega }\propto \left( \vec{E}^{D}\cdot \vec{E}^{Q}\right) \vec{k}^{Q}+\left( \vec{E}^{D}\cdot \vec{k}^{Q}\right) \vec{E}^{Q}\]

This structure can be deduced from general symmetry principles, the only vectors
available are the three mentioned, the amplitude of each term must be given
by a scalar which much include the other two terms. A scalar is given from two
vectors by a dot product, \( \vec{E}^{Q}\cdot \vec{k}^{Q}=0 \), so only two
terms remain. Other scalar terms like \( \sqrt{\left| \vec{E}\right| ^{2}} \)
can appear because the sizes appear only linearly in the matrix elements. The
only remaining variables are possible constant coefficients which can be determined
by considering some particular cases, such as \( \vec{E}^{D}\perp \vec{E}^{Q} \)
so that only the term on the direction of \( \vec{E}^{Q} \) remains, and calculating
the shift explicitly. 

This procedure will get the right answers, but it still cumbersome. Instead,
the coefficients and contributing terms can be determined by studying the general
behavior of these kinds of products of matrix elements.

\subsection{Dipole-Dipole Shifts}

To set the stage, consider a simpler problem that is not too awkward to evaluate
by brute force. Take a dipole transition between two \( j=1/2 \) states. The
matrix elements will be given by,

\[
\Omega _{m_{2}m_{1}}=\left\langle \frac{1}{2}m_{2}\right| x_{i}\left| \frac{1}{2}m_{1}\right\rangle E_{i}\]
and, as usual the shifts by,
\[
(\Omega ^{\dagger }\Omega )_{m_{2}m_{1}}=\sum _{m^{\prime }}\Omega ^{\dagger }_{m_{2}m^{\prime }}\Omega _{m^{\prime }m_{1}}\]
The mechanical solution is to use the Wigner-Eckhart theorem and transform to
a spherical basis so that the matrix elements are given by Clebsch-Gordan coefficients,

\[
\Omega _{m_{2}m_{1}}=\left\langle \frac{1}{2}m_{2}\right| T_{s}^{(1)}\left| \frac{1}{2}m_{1}\right\rangle E_{s}=\frac{\left\langle \frac{1}{2}\left| \left| r\right| \right| \frac{1}{2}\right\rangle }{\sqrt{2(1/2)+1}}\left\langle \frac{1}{2}m_{2}|1,s;\frac{1}{2}m_{1}\right\rangle E_{s}\]
where the \( E_{s} \) are given as usual by the \( \Delta m=0 \) and \( \Delta m=\pm 1 \)
transition amplitude. For future use write this transformation explicitly as
\[
E_{s}=U_{s}^{i}E_{i}\]
Where the components of the field amplitudes are,
\begin{eqnarray*}
E_{s} & = & \left( \begin{array}{ccc}
E_{+1} & E_{0} & E_{-1}
\end{array}\right) ^{T}\\
E_{i} & = & \left( \begin{array}{ccc}
E_{x} & E_{y} & E_{z}
\end{array}\right) ^{T}
\end{eqnarray*}
and the transformation is given by
\[
U_{s}^{i}=\left( \begin{array}{ccc}
-1/\sqrt{2} & -i/\sqrt{2} & 0\\
0 & 0 & 1\\
1/\sqrt{2} & -i/\sqrt{2} & 0
\end{array}\right) \]
The complementary transformation is then given by
\[
E_{i}=U_{i}^{s}E_{s}\]
where the transformation matrix is simply,
\[
U_{i}^{s}=U_{s}^{i\dagger }\]
so that 
\begin{eqnarray*}
U^{i\dagger }_{s}U^{s}_{j} & = & \delta _{ij}\\
U^{r\dagger }_{i}U^{i}_{s} & = & \delta _{rs}
\end{eqnarray*}

Computing all the terms explicitly gives,
\begin{eqnarray*}
 &  & \left( \Omega ^{\dagger }\Omega \right) _{m_{2}m_{1}}=\frac{\left\langle r\right\rangle ^{2}}{2}\sum _{m^{\prime }}\left\langle \frac{1}{2}m_{2}\right| x_{r}\left| \frac{1}{2}m^{\prime }\right\rangle E_{r}^{*}\left\langle \frac{1}{2}m^{\prime }\right| x_{s}\left| \frac{1}{2}m_{1}\right\rangle E_{s}\\
 &  & =\frac{\left\langle r\right\rangle ^{2}}{6}\times \\
 &  & \left( \begin{array}{cc}
\vec{E}^{*}\cdot \vec{E}+i(E_{x}E_{y}^{*}-E_{y}E_{x}^{*}) & -i(E_{y}E_{z}^{*}-E_{z}E_{y}^{*})-(E_{z}E_{x}^{*}-E_{x}E_{z}^{*})\\
-i(E_{y}E_{z}^{*}-E_{z}E_{y}^{*})+(E_{z}E_{x}^{*}-E_{x}E_{z}^{*}) & \vec{E}^{*}\cdot \vec{E}-i(E_{x}E_{y}^{*}-E_{y}E_{x}^{*})
\end{array}\right) 
\end{eqnarray*}
Recalling the definition of the pauli matrices \( \vec{\sigma } \), the appropriate
components are easily identified and give as vector form for the matrix of,
\[
\Omega ^{\dagger }\Omega =\frac{\left| \langle r\rangle \right| ^{2}}{6}(\vec{E}\cdot \vec{E}^{*}+i(\vec{E}\times \vec{E}^{*})\cdot \vec{\sigma })\]
 Again, this general structure could easily be guessed. The only parameter is
\( \vec{E} \) and it and its complex conjugate must appear linearly. The spin
independent shift must be a scalar and \( \vec{E}^{*}\cdot \vec{E} \) is the
only possibility. Similarly the vector amplitude can only be \( \vec{E}\times \vec{E}^{*} \). 

This procedure could work for any combination of transitions but quickly begins
to get unwieldy with the rapidly increasing number of possible terms, for example
for a quadrupole-quadrupole term there are four parameters, three possible scalar
combinations and six vectors. With the later added generality of transitions
to only particular states defined along particular directions the situation
is significantly more complicated. Besides this the constant coefficients must
still be obtained from an explicit calculation of particular cases.

These result can be obtained more explicitly. Slightly rearrange the form of
the answer,

\[
\vec{E}\cdot \vec{E}^{*}-i(\vec{E}\times \vec{E}^{*})\cdot \vec{\sigma }=(\delta _{ij}+i\epsilon _{ijk}\sigma _{k})E_{i}E_{j}^{*}\]
 This appears as some operator times \( E_{i}E_{j}^{*} \) which is exactly
the form of the general expression for the shift,
\begin{eqnarray*}
\left( \Omega ^{\dagger }\Omega \right) _{m_{2}m_{1}} & = & \left( \sum _{m^{\prime }}\left\langle \frac{1}{2}m_{2}\right| x_{i}\left| \frac{1}{2}m^{\prime }\right\rangle \left\langle \frac{1}{2}m^{\prime }\right| x_{j}\left| \frac{1}{2}m_{1}\right\rangle \right) (E_{i}^{*}E_{j})
\end{eqnarray*}
Removing the electric field amplitude gives something like,
\[
\sum _{m^{\prime }}\left\langle \frac{1}{2}m_{2}\right| x_{j}\left| \frac{1}{2}m^{\prime }\right\rangle \left\langle \frac{1}{2}m^{\prime }\right| x_{i}\left| \frac{1}{2}m_{1}\right\rangle \propto (\delta _{ij}+i\epsilon _{ijk}\sigma _{k})_{m_{2}m_{1}}\]
 This may be starting to look familiar. 

To make the general structure more apparent use the Wigner-Eckhart theorem to
factor out the radial dependence,

\begin{eqnarray*}
\Omega  & = & \left\langle x_{i}\right\rangle E_{i}=\left\langle x_{s}\right\rangle E_{s}\\
 & = & \left\langle \left\langle r\right\rangle \right\rangle \frac{\left\langle j_{2},m_{2}|1,s;j_{1},m_{1}\right\rangle }{\sqrt{2j_{2}+1}}E_{s}\\
 & = & \left\langle \left\langle r\right\rangle \right\rangle \left( \frac{\left\langle j_{2},m_{2}|1,s;j_{1},m_{1}\right\rangle }{\sqrt{2j_{2}+1}}U_{s}^{i}\right) E_{i}\\
 & \equiv  & \left\langle \left\langle r\right\rangle \right\rangle \frac{\left\langle j_{2},m_{2}|1,i;j_{1},m_{1}\right\rangle }{\sqrt{2j_{2}+1}}E_{i}
\end{eqnarray*}
This defines a sort of cartesian Clebsch-Gordan coefficient,
\[
\left\langle j_{2},m_{2}|1,i;j_{1},m_{1}\right\rangle =\left\langle j_{2},m_{2}|1,s;j_{1},m_{1}\right\rangle U^{i}_{s}\]
where the distinction between this and the usual Clebsch-Gordan coefficient
is made with the index, \( s,r \) compared to \( i,j,k \)

Further defining these with,
\[
(\sigma _{i}(j_{2},j_{1}))_{m_{1}m_{2}}=\left\langle j_{2},m_{2}|1,i;j_{1},m_{1}\right\rangle /\sqrt{2j_{2}+1}\]
or the corresponding relation in a spherical basis,
\[
(\sigma _{s}(j_{2},j_{1}))_{m_{1}m_{2}}=\left\langle j_{2}m_{2}|1,s;j_{1},m_{1}\right\rangle /\sqrt{2j_{2}+1}\]
 gives a coupling that looks like,

\begin{eqnarray*}
\Omega  & = & \left\langle \left\langle r\right\rangle \right\rangle \sigma _{j}(\frac{1}{2}\frac{1}{2})E_{j}\\
\Omega ^{\dagger } & = & \left\langle \left\langle r\right\rangle \right\rangle \sigma _{i}(\frac{1}{2}\frac{1}{2})E_{i}
\end{eqnarray*}
The shifts are then given by
\begin{eqnarray*}
\Omega ^{\dagger }\Omega  & = & \left\langle r\right\rangle ^{2}(\sigma _{i}(\frac{1}{2}\frac{1}{2})\sigma _{j}(\frac{1}{2}\frac{1}{2}))E_{i}E_{j}^{*}\\
 & = & \left\langle r\right\rangle ^{2}\frac{1}{6}(\delta _{ij}+i\epsilon _{ijk}\sigma _{k})E_{i}E_{j}^{*}
\end{eqnarray*}
Suggesting a convenient product rule for these matrices of Clebsch-Gordan coefficients,
\[
\sigma _{i}(\frac{1}{2}\frac{1}{2})\sigma _{j}(\frac{1}{2}\frac{1}{2})=\frac{1}{6}\left( \delta _{ij}+i\epsilon _{ijk}\sigma _{k}\right) \]
These \( \sigma  \) will be referred to as Generalized Pauli Matrices, since
for this case they are, in fact, directly proportional to the usual Pauli matrices.
With a few explicit calculations, or as shown below,
\begin{eqnarray*}
\sigma _{i}(\frac{1}{2},\frac{1}{2}) & = & \frac{1}{\sqrt{1/2(1/2+1)}}\frac{1}{\sqrt{2(1/2)+1}}\left\langle \frac{1}{2}m_{2}\left| j_{i}\right| \frac{1}{2}m_{1}\right\rangle \\
 & = & \frac{2}{\sqrt{3}}\frac{1}{\sqrt{2}}\left\langle \frac{1}{2}m_{2}\left| j_{i}\right| \frac{1}{2}m_{1}\right\rangle \\
 & = & \frac{1}{\sqrt{6}}\sigma _{i}
\end{eqnarray*}
 The Pauli matrices have the well known product rule,
\[
\sigma _{i}\sigma _{j}=\delta _{ij}+i\epsilon _{ijk}\sigma _{k}\]
which yields the previous product for the generalized pauli matrices immediately.

\subsection{General Products}

This multiplication rule provides and immediate result for vector and scalar
contributions to an energy shift for arbitrary fields but these observations
would not be particularly useful if the result applied to only this case of
\( j=1/2 \) initial and intermediate states. In fact, the general structure
of the product is valid for any combination of initial, \( j_{i} \), final,
\( j_{f} \), and intermediate, \( j^{\prime }, \) \( j \)'s. The various
types of multiple components in the result are the same, only the coefficients
of each vary. Varying only \( j^{\prime } \) and leaving \( j_{i}=j_{f}=1/2 \),
the result will still be a \( 2\times 2 \) hermitian matrix which can be written
in terms of the identity and the usual Pauli matrices.
\[
\sigma _{i}(\frac{1}{2},j^{\prime })\sigma _{j}(j^{\prime },\frac{1}{2})=\delta \omega (j^{\prime },\frac{1}{2})\delta _{ij}+i\delta \omega ^{\prime }_{i}(j^{\prime },\frac{1}{2})\varepsilon _{ijk}\sigma _{k}\]
Note that for \( j^{\prime }=1/2 \) the \( \sigma  \) matrices are square
and so the product is immediately sensible. For \( j^{\prime }\neq 1/2 \) the
\( \sigma  \) are now rectangular but \( \sigma (1/2,j^{\prime }) \) has \( 2j^{\prime }+1 \)
columns and \( \sigma (j^{\prime },1/2) \) has \( 2j^{\prime }+1 \) rows so
again the sum over intermediate \( m^{\prime } \) can always be written as
matrix multiplication. For \( j\neq j^{\prime } \), the \( \sigma  \) with
exchanged arguments can also be always traded for a transpose, 
\begin{eqnarray*}
\sigma _{s}(j^{\prime },j)_{m_{2}m_{1}} & = & \left\langle j^{\prime }m_{2}\left| x_{s}\right| j,m_{1}\right\rangle /\left\langle j^{\prime }\left| \left| x\right| \right| j\right\rangle \\
 & = & \left\langle j,m_{1}\left| x_{s}\right| j^{\prime }m_{2}\right\rangle ^{*}/\left\langle j^{\prime }\left| \left| x\right| \right| j\right\rangle \\
 & = & \sigma _{s}(j,j^{\prime })_{m_{1}m_{2}}^{*}\\
\sigma _{s}(j^{\prime },j) & = & \sigma _{s}^{\dagger }(j,j^{\prime })
\end{eqnarray*}
Generally is is easiest to see the flow of the transition without the dagger
as the \( j's \) then are listed in the arguments of the \( \sigma  \) in
the same order as they appear in the matrix elements.

For arbitrary \( j_{i}=j_{f}\neq j^{\prime } \) the product will be convenient
to write in terms of angular momentum operators since there is no conventional
definition of the pauli matrices in anything but a spin 1/2 system, where \( \left\langle 1/2,m_{2}\left| j_{i}\right| 1/2,m_{1}\right\rangle =(1/2)\left( \sigma _{i}\right) _{m_{2}m_{1}} \).
In addition, for \( j_{i}=j_{f}=j=1/2 \) the product can contain quadrupole
structure, though nothing of higher order since this is the product of only
two dipole transitions, and as each dipole can change \( m \) by at most one,
the combination can change \( m \) by at most two, so there is not yet, for
example, any octapole structure in the product and the result takes the form
\[
\sigma _{i}(j,j^{\prime })\sigma _{j}(j^{\prime },j)=\delta \omega (j^{\prime },j)\delta _{ij}+i\delta \omega _{i}(j^{\prime },j)\varepsilon _{ijk}j_{k}+\delta \omega _{ij}(j^{\prime },j)j_{ij}\]
where the quadrupole operator is the usual traceless combination \( j_{ij}=(j_{i}j_{j}+j_{j}j_{i})/2-(\delta _{ij}/3)j^{2} \),
and any \( j \) is just shorthand for \( \left\langle j_{i},m_{2}\left| j_{i,j\cdots }\right| j_{i}m_{1}\right\rangle  \). 

For the previous case of \( j=j^{\prime }=1/2 \) this gives
\[
\sigma _{i}(\frac{1}{2},\frac{1}{2})\sigma _{j}(\frac{1}{2},\frac{1}{2})=\frac{1}{6}\left( \delta _{ij}+i\varepsilon _{ijk}\sigma _{k}\right) =\frac{1}{6}\left( \delta _{ij}+2i\varepsilon _{ijk}j_{k}\right) \]
 For \( j^{\prime }=3/2 \), the case of immediate interest, it turns out,
\[
\sigma _{i}^{\dagger }(\frac{1}{2},\frac{3}{2})\sigma _{j}(\frac{3}{2},\frac{1}{2})=\frac{1}{6}\left( \delta _{ij}-i\varepsilon _{ijk}j_{k}\right) \]
For \( j=1/2 \) and \( j=1 \) it is clear that this form must hold since the
matrix representations of \( j_{i} \) and \( j_{ij} \) for a basis for hermitian
matrices of this \( 2j+1 \) dimension. General Hermitian matrices can have
more structure than can be described by only these dipole and quadrupole operators,
so appeals must be made to the general structure of the transitions to motivate
this form.

This structure also turns out to hold for \( j_{i}\neq j_{j} \). In this case
the final result is not square so the product can not be written in terms of
matrix elements of angular momentum operators, but instead must be written in
terms of other \( \sigma  \)'s. For constancy write all the components in terms
of the original \( \sigma  \)'s but with different dimensions,
\begin{eqnarray*}
\sigma _{i}(j_{f},j^{\prime })\sigma _{j}(j^{\prime },j_{i}) & = & \delta \omega _{0}(j_{f},j^{\prime },j_{i})\sigma (j_{f},j_{i})+i\delta \omega _{1}(j_{f},j^{\prime },j_{i})\varepsilon _{ijk}\sigma (j_{f},j_{i})\\
 & + & \delta \omega _{2}(j_{f},j^{\prime },j_{i})\sigma _{ij}(j_{f},j_{i})
\end{eqnarray*}
These kinds of products don't arise immediately when considering light shifts,
but will appear when considering various restrictions to the \( m \) sublevels
of the intermediate states, and are useful in other kinds of problems not discussed
here. This form is also useful for \( j_{i}=j_{f} \) in intermediate steps
of many calculations where it is convenient to write only the final answers
in terms of angular momentum operators.

\subsection{Products \protect\( j_{i}=j_{f}=j^{\prime }\protect \)}

This general structure for the product is plausible. As discussed above by considering
the net result of a pair of dipole transitions, it is clear the product can
have no \( \Delta m>2 \) components, but it has not been shown explicitly that
the other components can be written in the way shown. In addition, the coefficients
are so far undetermined and must be computed explicitly. For the case of \( j_{i}=j_{j}=j^{\prime } \)
all this can be done using the familiar algebra of the angular momentum operators.

Matrix elements of \( j \) 's can be computed easily using the raising and
lowering operators, \( j_{\pm }=j_{x}\pm ij_{y} \). But, as they are also vector
operators, that is \( [j_{i},v_{k}]=i\varepsilon _{ijk}v_{k} \) for \( v=j \),
they can be computed using the Wigner-Eckhart Theorem.The angular momentum operators
commute with \( j^{2} \) so they don't change the total angular momentum of
a state and only \( j_{f}=j_{i}=j \) must be considered, 
\[
\left\langle j,m_{f}\left| j_{s}\right| j,m_{i}\right\rangle =\left\langle j||j||j\right\rangle \frac{\left\langle j,m_{f}|1,s;jm_{i}\right\rangle }{\sqrt{2j+1}}\]
Here the \( j_{\pm 1} \) are the spherical operators analogous to \( x_{\pm 1}=(-x-iy)/\sqrt{2} \)
rather than the usual angular momentum raising and lowering operators.

\subsubsection{Dipole Angular Momentum Operator Reduced Matrix Elements}

In
\[
\left\langle j,m_{f}\left| j_{s}\right| j,m_{i}\right\rangle =\left\langle j||j||j\right\rangle \sigma _{s}(j,j)\]
the reduced matrix element can be evaluated to make all the terms in the relation
explicit, take the \( j_{0}=j_{z} \) term, this gives,
\[
\left\langle j,m_{f}\left| j_{0}\right| j,m_{i}\right\rangle =m_{i}\delta _{m_{f}m_{i}}=\left\langle j||j||j\right\rangle \frac{\left\langle j,m_{f}|1,0;jm_{i}\right\rangle }{\sqrt{2j+1}}\]
yielding,
\[
\left\langle j||j||j\right\rangle =\frac{m\sqrt{2j+1}}{\left\langle j,m|1,0;j,m\right\rangle }\]
The Clebsch-Gordan coefficient can be computed in the usual way using stretched
states, lowering operators and orthogonality, \cite{Schacht00}, to give
\[
\left\langle j,m|1,0;j,m\right\rangle =\frac{m}{\sqrt{j(j+1)}}\]
so that 
\[
\left\langle j||j||j\right\rangle =\sqrt{j(j+1)}\sqrt{2j+1}\]
and finally,
\[
\left\langle j,m_{f}\left| j_{s}\right| j,m_{i}\right\rangle =\sqrt{j(j+1)}\sqrt{2j+1}\sigma _{s}(j,j)\]

\subsubsection{\protect\( \sigma \protect \) Products using Angular Momentum Matrix Elements}

Now, products of matrix elements of \( x \), products of \( \sigma  \)'s can
be written in terms of products of matrix elements of \( j \). With \( \sigma _{i}\equiv \sigma _{i}(j,j) \)
for convenience,
\[
\sigma _{i}^{\dagger }\sigma _{j}=\frac{1}{2j+1}\frac{1}{j(j+1)}\left\langle j,m_{2}|j_{i}|j,m^{\prime }\right\rangle \left\langle j,m^{\prime }\left| j_{j}\right| j,m_{1}\right\rangle \]
Again the \( j \) doesn't change \( j \), so the intermediate \( j \) can
be summed over without changing the result. With this sum, and the existing
sum over \( m \), completeness gives and identity,
\begin{eqnarray*}
\sigma _{i}^{\dagger }\sigma  & _{j}= & \frac{1}{2j+1}\frac{1}{j(j+1)}\sum _{j^{\prime }m^{\prime }}\left\langle j,m_{2}|j_{i}|j^{\prime },m^{\prime }\right\rangle \left\langle j^{\prime },m^{\prime }\left| j_{j}\right| j,m_{1}\right\rangle \\
 & = & \frac{1}{2j+1}\frac{1}{j(j+1)}\left\langle j,m_{2}\left| j_{i}j_{j}\right| j,m_{1}\right\rangle 
\end{eqnarray*}
Now the familiar angular momentum algebra, \( [j_{i},j_{j}]=i\varepsilon _{ijk}j_{k} \),
can be used to decompose the result into scalar, vector and quadrupole components,
\begin{eqnarray*}
\sigma _{i}^{\dagger }\sigma  & _{j}= & \frac{1}{2j+1}\frac{1}{j(j+1)}\left\langle j,m_{2}\left| \frac{1}{2}(j_{i}j_{j}+j_{j}j_{i})+\frac{1}{2}(j_{i}j_{j}-j_{j}j_{i})\right| j,m_{1}\right\rangle \\
 & = & \frac{1}{2j+1}\frac{1}{j(j+1)}\left\langle j,m_{2}\left| \frac{1}{3}\delta _{ij}j^{2}+\frac{1}{2}[j_{i},j_{j}]+\left( \frac{1}{2}(j_{i}j_{j}+j_{j}j_{i})-\frac{1}{3}\delta _{ij}j^{2}\right) \right| j,m_{1}\right\rangle \\
 & = & \frac{1}{2j+1}\frac{1}{j(j+1)}\left\langle j,m_{2}\left| \frac{j(j+1)}{3}\delta _{ij}+\frac{i}{2}\varepsilon _{ijk}j_{k}+j_{ij}\right| j,m_{1}\right\rangle 
\end{eqnarray*}
The coefficients are then,

\begin{eqnarray*}
\delta \omega _{0}(j,j) & = & \frac{1}{3}\frac{1}{2j+1}\\
\delta \omega _{1}(j,j) & = & \frac{1}{2}\frac{1}{2j+1}\frac{1}{j(j+1)}\\
\delta \omega _{2}(j,j) & = & \frac{1}{2j+1}\frac{1}{j(j+1)}
\end{eqnarray*}
For \( j=1/2 \), \( j(j+1)=3/4 \), \( 2j+1=2 \), this yields the previous
result,
\[
\sigma _{i}(\frac{1}{2},\frac{1}{2})\sigma _{j}(\frac{1}{2},\frac{1}{2})=\frac{1}{6}\delta _{ij}+\frac{1}{3}i\varepsilon _{ijk}j_{k}=\frac{1}{6}\left( \delta _{ij}+2i\varepsilon _{ijk}j_{k}\right) \]
The quadrupole coefficient is irrelevant in this case since the matrix elements
of the quadrupole operator are zero for these \( j=1/2 \) states.

\subsubsection{Products in terms of \protect\( \sigma \protect \)}

In terms of \( \sigma  \) matrices rather than angular momentum operators the
previous relation
\[
\left\langle j,m_{f}\left| j_{s}\right| j,m_{i}\right\rangle =\sqrt{j(j+1)}\sqrt{2j+1}\sigma _{s}(j,j)\]
can be used for the dipole term. Relating the quadrupole components requires
relating \( j_{ij} \) to \( \sigma _{ij}(j,j) \). This can be done in the
same way the \( j_{i} \) was written in terms of \( \sigma _{i}(j,j) \).

\subsubsection{Quadrupole Angular Momentum Operator Reduced Matrix Elements}

The Wigner-Eckhart theorem gives,

\[
\left\langle j,m_{2}\left| j_{ij}\right| j,m_{1}\right\rangle =\left\langle \left\langle j^{2}\right\rangle \right\rangle \sigma _{ij}(j,j)_{m_{2}m_{1}}\]
The reduced matrix element can be determined by evaluating a particular component
of this matrix element, the simplest is \( j_{zz} \),

\begin{eqnarray*}
\left\langle j,j\left| j_{zz}\right| j,j\right\rangle  & = & \left\langle j,j\left| j_{z}^{2}-\frac{1}{3}j^{2}\right| j,j\right\rangle \\
 & = & j^{2}-\frac{1}{3}j(j+1)\\
 & = & \frac{j}{3}(2j-1)\\
 & = & \left\langle \left\langle j^{2}\right\rangle \right\rangle \sigma _{zz}(j,j)_{jj}
\end{eqnarray*}
This component of the \( \sigma  \) matrix is given by 
\[
Q_{zz}=\sqrt{2/3}Q_{0}\]
 and, \cite{Schacht00}, 
\[
\left\langle j,j|2,0;j,j\right\rangle =\sqrt{2j+1}\sqrt{\frac{j}{2j+3}}\sqrt{\frac{2j-1}{j+1}}\]
so that 
\begin{eqnarray*}
\sigma _{zz}(j,j)_{j,j} & = & \sqrt{\frac{2}{3}}\sigma _{0}(j,j)_{j,j}\\
 & = & \sqrt{\frac{2}{3}}\frac{\left\langle j,j|2,0;j,j\right\rangle }{\sqrt{2j+1}}\\
 & = & \sqrt{\frac{2}{3}}\sqrt{\frac{j}{2j+3}}\sqrt{\frac{2j-1}{j+1}}
\end{eqnarray*}
giving

\begin{eqnarray*}
\left\langle \left\langle j^{2}\right\rangle \right\rangle  & = & \frac{j}{3}(2j-1)\sqrt{\frac{3}{2}}\sqrt{\frac{2j+3}{j}}\sqrt{\frac{j+1}{2j-1}}\\
 & = & \frac{\sqrt{j(j+1)(2j-1)(2j+1)(2j+3)}}{\sqrt{6}}
\end{eqnarray*}

\subsection{Projection Operators}

For products with arbitrary \( j_{i}=j_{f} \) and \( j^{\prime } \) the coefficients
of the terms in the product are not as easily determined since the \( x_{i} \)
can not be written immediately in terms on the \( j_{i} \). The coefficient
of the scalar term turns out to remain the same and showing this is fairly straight-forward.

\subsubsection{Scalar Coefficient}

The \( \varepsilon _{ijk}j_{k} \) term in the product is anti-symmetric, so
in particular \( \varepsilon _{iik}j_{k}=0 \). The \( j_{ij} \) term is traceless
over the spatial indices, so computing this trace for the entire relation gives,
\begin{eqnarray*}
\delta _{ij}\left( \sigma _{i}\sigma _{j}\right)  & = & \delta _{ij}(\delta \omega _{0}\delta _{ij}+i\delta \omega _{1}\varepsilon _{ijk}j_{k}+\delta \omega _{2}j_{ij})\\
\delta _{ij}\left\langle j,m_{2}\left| x_{i}\right| j^{\prime }m^{\prime }\right\rangle \left\langle j^{\prime }m^{\prime }\left| x_{j}\right| j,m_{1}\right\rangle  & = & \left\langle j\left| \left| x\right| \right| j^{\prime }\right\rangle ^{2}3\delta \omega _{0}
\end{eqnarray*}
The left-hand side has and implied identity in \( m \) space,
\begin{eqnarray*}
3\delta \omega _{0} & = & \delta _{ij}\left\langle j,m\left| \hat{x}_{i}\right| j^{\prime }m^{\prime }\right\rangle \left\langle j^{\prime }m^{\prime }\left| \hat{x}_{j}\right| j,m\right\rangle /\left\langle j\left| \left| x\right| \right| j^{\prime }\right\rangle ^{2}\\
 & = & \delta _{ij}U^{i}_{s}U^{j}_{r}\left\langle j,m\left| \hat{x}_{s}\right| j^{\prime }m^{\prime }\right\rangle \left\langle j^{\prime }m^{\prime }\left| \hat{x}_{r}\right| j,m\right\rangle /\left\langle j\left| \left| x\right| \right| j^{\prime }\right\rangle ^{2}\\
 & = & \delta _{rs}\sum _{m^{\prime }}\frac{\left\langle j^{\prime },m^{\prime }|1,s;j,m\right\rangle }{\sqrt{2j^{\prime }+1}}\frac{\left\langle j^{\prime },m^{\prime }|1,r;j,m\right\rangle }{\sqrt{2j^{\prime }+1}}\\
 & = & \sum _{s,m^{\prime }}\frac{\left\langle j^{\prime },m^{\prime }|1,s;j,m\right\rangle }{\sqrt{2j^{\prime }+1}}\frac{\left\langle j^{\prime },m^{\prime }|1,s;j,m\right\rangle }{\sqrt{2j^{\prime }+1}}\\
 & = & \frac{1}{2j^{\prime }+1}\sum _{m^{\prime }}\left\langle j^{\prime },m^{\prime }|1,m^{\prime }-m;j,m\right\rangle ^{2}
\end{eqnarray*}
 The general sum over the Clebsch-Gordan coefficients is is independent of \( m \),
 \cite{Schacht00}
\[
\sum _{m^{\prime }}\left\langle j^{\prime },m^{\prime }|1,m^{\prime }-m;j,m\right\rangle ^{2}=\frac{2j^{\prime }+1}{2j+1}\]
 giving,
\[
\delta \omega _{0}=\frac{1}{3}\frac{1}{2j+1}\]
as before.

\subsubsection{Angular Momentum Projection Operators}

The dipole and quadrupole coefficients are not as easily determined. For \( j_{i}=j_{f} \),
algebra similar to that used for computing the \( j_{i}=j_{f}=j^{\prime } \)
products, also using, \( [j_{i},x_{j}]=i\varepsilon _{ijk}x_{k} \), can be
exploited using angular momentum projection operators on the intermediate states
to give the remaining coefficients for the product
\[
\sigma _{i}(j_{f},j^{\prime })\sigma _{j}(j^{\prime },j_{i})=\delta \omega _{0}+i\delta \omega _{1}\varepsilon _{ijk}j_{k}+\delta \omega _{2}j_{ij}\]
Define \( P_{j} \) by \( P_{j}\left| j^{\prime },m\right\rangle =\left| j^{\prime },m\right\rangle \delta _{j^{\prime }j} \).
Then,in an arbitrary products, the intermediate \( j \) states can be summed
over using a projection operator \( P_{j} \) to pick out the correct state,
the sum over all \( j \) and \( m \) in the intermediate state then gives
\( 1 \),
\begin{eqnarray*}
\sigma _{i}(j_{f},j)\sigma _{j}(j,j_{i}) & \propto  & \sum _{m^{\prime }}\left\langle j_{f}m_{2}|x_{i}|jm^{\prime }\right\rangle \left\langle jm^{\prime }|x_{j}|j_{i}m_{i}\right\rangle \\
 & = & \sum _{j^{\prime }m^{\prime }}\left\langle j_{f}m_{2}|x_{i}P_{j}|j^{\prime }m^{\prime }\right\rangle \left\langle j^{\prime }m^{\prime }|x_{j}|j_{i}m_{i}\right\rangle \\
 & = & \left\langle j_{f}m_{2}\left| x_{i}P_{j}x_{j}\right| j_{i}m_{1}\right\rangle 
\end{eqnarray*}
A representation of these projection operators can be built out of \( j^{2} \),
First consider 
\[
K_{jj^{\prime }}\equiv \frac{j^{2}-j^{\prime }(j^{\prime }+1)}{j(j+1)-j^{\prime }(j^{\prime }+1)}\]
 with \( j\neq j^{\prime } \). Operating this on \( \left| j^{\prime \prime },m\right\rangle  \)
gives 
\[
K_{jj\prime }\left| j^{\prime \prime },m\right\rangle =0\]
 for \( j^{\prime \prime }\neq j \) and for \( j^{\prime \prime }=j \), 
\[
K_{jj\prime }\left| j,m\right\rangle =\left| j,m\right\rangle \frac{j(j+1)-j^{\prime }(j^{\prime }+1)}{j(j+1)-j^{\prime }(j^{\prime }+1)}=\left| j,m\right\rangle \]
 \( K_{jj^{\prime }} \) kills the state \( j^{\prime } \) and returns the
state \( j \) with coefficient 1. On any other state it returns the state with
some nontrivial coefficient. Now, consider \( \prod _{j^{\prime \prime }\neq j}K_{jj^{\prime \prime }} \).
Again operating with this on a state \( \left| j^{\prime },m\right\rangle  \)
with \( j\neq j^{\prime } \) gives zero as one of the factors is \( j^{\prime \prime }=j^{\prime } \),
and for \( j^{\prime \prime }=j \) every factors gives 1 times the same state.
So it acts as the desired projection operator for \( j \),
\[
P_{j}=\prod _{j^{\prime }\neq j}K_{jj^{\prime }}=\prod _{j^{\prime }\neq j}\frac{j^{2}-j^{\prime }(j^{\prime }+1)}{j(j+1)-j^{\prime }(j^{\prime }+1)}\]

The \( \sigma  \) products can then be computed with matrix elements like,
\[
\left\langle j_{2}m_{2}\left| x_{i}j^{2n}x_{j}\right| j_{1}m_{1}\right\rangle \]
Generally \( n \) is arbitrarily large, but for this particular application
only a finite number of \( K_{jj^{\prime \prime }} \) are needed to effectively
represent the \( j \) projection operators. A dipole transition can change
the total angular momentum by at most \( \Delta j=1 \). For a non zero matrix
element the intermediate \( j \) must differ by \( j_{1} \) and \( j_{2} \)
by no more than 1. When summing over intermediate states with arbitrary \( j \),
states with \( j \) differing from \( j_{1} \) or \( j_{2} \) by 2 or more
already give zero and don't need to be killed with a \( K \) and so at most
only two \( K \)'s are needed to complete selecting the original intermediate
\( j \) state. For a product of two \( K \) 's the matrix element then involves
only \( j^{4} \) and \( j^{2} \). 

These kinds of matrix elements can be evaluate using the usual angular momentum
operator algebra, \( [j_{i},x_{j}]=i\varepsilon _{ijk}x_{k} \) and the previous
transition from matrix elements of coordinate operates to matrix elements of
angular momentum operators when taken between states with the same \( j \).
As an example,
\[
j^{2}x_{j}=x_{j}(j^{2}+2)-2i\varepsilon _{kjm}x_{m}j_{k}\]
\begin{eqnarray*}
[j^{2},x_{i}] & = & i\varepsilon _{jik}(j_{j}x_{k}+x_{k}j_{j})\\
 & = & i\varepsilon _{jik}(x_{k}j_{j}+i\varepsilon _{kjm}x_{m}+x_{k}j_{j})\\
 & = & 2\varepsilon _{jik}x_{k}j_{j}-\varepsilon _{ijk}\varepsilon _{mjk}x_{m}\\
 & = & 2(x_{i}-\varepsilon _{ijk}x_{k}j_{j})
\end{eqnarray*}
giving, 
\begin{eqnarray*}
\left\langle j,m_{1}\left| x_{i}j^{2}x_{j}\right| j,m_{2}\right\rangle  & = & \left\langle j,m_{1}\left| x_{i}x_{j}(j^{2}+2)-2i\varepsilon _{kjm}x_{i}x_{m}j_{k}\right| j,m_{2}\right\rangle \\
 & = & \left( j(j+1)+2\right) \left\langle j,m_{1}\left| x_{i}x_{j}\right| j,m_{2}\right\rangle \\
 & - & 2i\varepsilon _{kjm}\left\langle j,m_{1}\left| x_{i}x_{m}j_{k}\right| j,m_{2}\right\rangle 
\end{eqnarray*}
The first term is a matrix element of \( x_{i}x_{j}=Q_{ij}+\delta _{ij}r^{2}/3 \)
so it gives a quadrupole term, \( \sigma _{ij}\propto j_{ij} \) and a scalar
term. The \( j \) at the far end of the second term doesn't change angular
momentum to inserting a sum over intermediate state with the same \( j \),
\[
\left\langle j,m_{1}\left| x_{i}x_{m}j_{k}\right| j,m_{2}\right\rangle =\left\langle j,m_{1}\left| x_{i}x_{m}\right| j,m^{\prime }\right\rangle \left\langle j,m^{\prime }\left| j_{k}\right| j,m_{2}\right\rangle \]
Again the quadrupole pieces gives a quadrupole and scalar piece. The scalar
piece combines with the \( j_{k} \) matrix element to give a vector term. The
quadrupole piece is a matrix element between states of the same \( j \) so
it can be written in terms of a matrix element of angular momentum operators
and the product computed as before using the angular momentum operator algebra,

\begin{eqnarray*}
\varepsilon _{kjm}\left\langle j,m_{1}\left| Q_{ij}\right| j,m^{\prime }\right\rangle \left\langle j,m^{\prime }\left| j_{k}\right| j,m_{2}\right\rangle  & \propto  & \varepsilon _{kjm}\left\langle j,m_{1}\left| j_{ij}\right| j,m^{\prime }\right\rangle \left\langle j,m^{\prime }\left| j_{k}\right| j,m_{2}\right\rangle \\
 & = & \varepsilon _{kjm}\left\langle j,m_{1}\left| j_{ij}j_{k}\right| j,m_{2}\right\rangle 
\end{eqnarray*}

This requires the detailed form of the octapole angular momentum operators.
Determining this is straightforward but won't be considered here. The pure octapole
operator is symmetric under an exchange of a pair of indices so that the \( \varepsilon _{kjm} \)
removes this contribution from the final result and only at most quadrupole
operators remain. 

The same must be done for \( \left\langle x_{i}j^{4}x_{j}\right\rangle  \).
This is even more elaborate, but again straight-forward and again yields only
quadrupole, dipole and scalar terms. The results must then be combined to reform
the angular momentum projection operator to finally give the result for the
products of pauli matrices. Note that along the way different kinds of reduced
matrix elements appear in relating matrix elements to the \( \sigma  \) matrices.
In particular \( \left\langle x\right\rangle ^{2} \) and \( \left\langle x^{2}\right\rangle  \),
these must be properly related to get a closed form final result. This is also
done in a straight-forward manner using the Wigner-Eckhart Theorem and closed
form expressions for a handful of Clebsch-Gordan coefficients if desired.

This method is currently cumbersome, though it has the advantage of being rigorous
and general and can be immediately applied to products of an arbitrary number
of \( \sigma  \) matrices of arbitrarily high order, some examples of which
are required for the analysis considered later though not explicitly developed
in this way as more efficient techniques will be developed. It is likely that
this method can be streamlined with a bit more work and used more easily to
quickly provide results for any products, but at this point it is useful for
simply demonstrating the structure of the product of dipole matrices.

\subsubsection{Effective Completeness}

With a bit more work, this ideal of projection operators, and the restricted
set of intermediate states that can contribute to a particular matrix element,
can be used to quickly provide results for dipole products with \( j_{i}=j_{f}=1/2 \).
The \( j^{\prime }=1/2 \) can has already been solved, and the only other possible
intermediate state is \( j^{\prime }=3/2 \) so this method yields only a very
restricted result, but these angular momenta are exactly those of the states
used in the parity transition so the result is practically useful. 

Since dipole transitions from \( j=1/2 \) states are only possible to \( j^{\prime }=1/2 \)
and \( j^{\prime }=3/2 \) states, a sum over intermediate states with give
contributions from only these \( j^{\prime } \)s. 
\begin{eqnarray*}
\left\langle \frac{1}{2}m_{2}\left| x_{i}x_{j}\right| \frac{1}{2}m_{1}\right\rangle  & = & \sum _{j}\left\langle \frac{1}{2}m_{2}\left| x_{i}P_{j}x_{j}\right| \frac{1}{2}m_{1}\right\rangle \\
 & = & \left\langle \frac{1}{2}m_{2}\left| x_{i}(P_{1/2}+P_{3/2})x_{j}\right| \frac{1}{2}m_{1}\right\rangle 
\end{eqnarray*}
With this, a product of operators for intermediate states of one \( j^{\prime } \)
can be written in terms of the product for the other \( j^{\prime } \),

\begin{eqnarray*}
 &  & \left\langle \frac{1}{2}m_{2}\left| x_{i}\right| \frac{3}{2}m^{\prime }\right\rangle \left\langle \frac{3}{2}m^{\prime }\left| x_{j}\right| \frac{1}{2}m_{1}\right\rangle \\
 & = & \left\langle \frac{1}{2}\left| \left| D\right| \right| \frac{3}{2}\right\rangle ^{2}\sigma _{i}(\frac{1}{2}\frac{3}{2})\sigma _{j}(\frac{3}{2}\frac{1}{2})\\
 &  & =\left\langle \frac{1}{2}m_{2}\left| x_{i}P_{3/2}x_{j}\right| \frac{1}{2}m_{1}\right\rangle \\
 &  & =\left\langle \frac{1}{2}m_{2}\left| x_{i}(1-P_{1/2})x_{j}\right| \frac{1}{2}m_{1}\right\rangle \\
 &  & =\left\langle \frac{1}{2}m_{2}\left| x_{i}x_{j}\right| \frac{1}{2}m_{1}\right\rangle -\left\langle \frac{1}{2}m_{2}\left| x_{i}\right| \frac{1}{2}m_{1}\right\rangle \left\langle \frac{1}{2}m_{2}\left| x_{j}\right| \frac{1}{2}m_{1}\right\rangle \\
 &  & =\left\langle \frac{1}{2}m_{2}\left| Q_{ij}+\delta _{ij}\frac{r^{2}}{3}\right| \frac{1}{2}m_{1}\right\rangle -\left\langle \frac{1}{2}m_{2}\left| x_{i}\right| \frac{1}{2}m_{1}\right\rangle \left\langle \frac{1}{2}m_{2}\left| x_{j}\right| \frac{1}{2}m_{1}\right\rangle \\
 &  & =\left\langle \frac{1}{2}\left| \left| Q\right| \right| \frac{1}{2}\right\rangle \left( \sigma _{ij}(\frac{1}{2}\frac{1}{2})+\delta _{ij}\frac{1}{3}\right) -\left\langle \frac{1}{2}\left| \left| D\right| \right| \frac{1}{2}\right\rangle ^{2}\sigma _{i}(\frac{1}{2}\frac{1}{2})\sigma _{j}(\frac{1}{2}\frac{1}{2})\\
 &  & =\left\langle \frac{1}{2}\left| \left| Q\right| \right| \frac{1}{2}\right\rangle \delta _{ij}\frac{1}{3}-\left\langle \frac{1}{2}\left| \left| D\right| \right| \frac{1}{2}\right\rangle ^{2}\frac{1}{6}(\delta _{ij}+2i\varepsilon _{ijk}j_{k})
\end{eqnarray*}
The quadrupole term is zero for \( j=1/2 \). Collecting terms gives,

\begin{eqnarray*}
\sigma _{i}(\frac{1}{2}\frac{3}{2})\sigma _{j}(\frac{3}{2}\frac{1}{2}) & = & \frac{\left\langle \frac{1}{2}\left| \left| D\right| \right| \frac{1}{2}\right\rangle ^{2}}{\left\langle \frac{1}{2}\left| \left| D\right| \right| \frac{3}{2}\right\rangle ^{2}}\left( \frac{\delta _{ij}}{3}\left( \frac{\left\langle \frac{1}{2}\left| \left| Q\right| \right| \frac{1}{2}\right\rangle }{\left\langle \frac{1}{2}\left| \left| D\right| \right| \frac{1}{2}\right\rangle ^{2}}-\frac{1}{2}\right) -\frac{1}{3}i\varepsilon _{ijk}j_{k}\right) 
\end{eqnarray*}

A closed form result requires evaluating all of these reduced matrix elements.
This is not worthwhile just for use in this particular case, though the general
result is required when using angular momentum algebra to simplify the projection
operators. This is at least a quick demonstration of the structure of the product.

\subsection{Explicit Evaluation of Product Expansion Coefficients }

With the structure of the products well established a more practice evaluation
of the coefficients is possible. One method is just to use the structure of
the result.

\[
\sigma _{i}(j_{f},j^{\prime })\sigma _{j}(j^{\prime },j_{i})=\delta \omega _{0}+i\delta \omega _{1}\varepsilon _{ijk}j_{k}+\delta \omega _{2}j_{ij}\]

\subsubsection{Coefficient of Scalar Term}

As already determined \( \delta _{ij}\sigma _{i}\sigma _{j}=3\delta \omega _{0}=1/(2j+1) \).

\subsubsection{Quadrupole Coefficient}

Similarly with \( \varepsilon _{ijk} \) antisymmetric, \( \sigma _{i}\sigma _{i}=\delta \omega _{0}+\delta \omega _{2}j_{ii} \)
and \( \delta \omega _{2} \) can be determined from a particular \( i \),
such as \( i=z \), this gives,

\[
\sigma _{z}(jj^{\prime })\sigma _{z}(j^{\prime }j)=\delta \omega _{0}+\delta \omega _{2}j_{zz}\]
Only one matrix element of this relation is required, take the \( m_{1}=j \),\( m_{2}=j \)
component and is will always a nonzero element of \( j_{zz} \)

\[
\left( \sigma _{z}(jj^{\prime })\sigma _{z}(j^{\prime }j)\right) _{jj}=\delta \omega _{0}+\delta \omega _{2}\left( j_{zz}\right) _{jj}\]
This matrix element of \( j_{zz} \) is easily evaluated and has already computed
for use in relating \( j_{ij} \) to \( \sigma _{ij} \),
\[
\left( j_{zz}\right) _{jj}=\frac{j}{3}(2j-1)\]
The product of the \( \sigma  \) matrices is given by a sum of products of
Clebsch-Gordan coefficients that is greatly simplified by selection rules,

\begin{eqnarray*}
\left( \sigma _{z}(jj^{\prime })\sigma _{z}(j^{\prime }j)\right) _{jj} & = & \left( \sigma _{0}(jj^{\prime })\sigma _{0}(j^{\prime }j)\right) _{jj}\\
 & = & \sum _{m^{\prime }}\frac{\left\langle j,j|1,0;j^{\prime }m^{\prime }\right\rangle }{\sqrt{2j+1}}\frac{\left\langle j,j|1,0;j^{\prime }m^{\prime }\right\rangle }{\sqrt{2j+1}}\\
 & = & \frac{\left\langle j,j|1,0;j^{\prime }j\right\rangle ^{2}}{2j+1}
\end{eqnarray*}
This gives 
\begin{eqnarray*}
\delta \omega _{2} & = & \frac{3}{j(2j-1)}\left( \frac{\left\langle j,j|1,0;j^{\prime }j\right\rangle ^{2}}{2j+1}-\delta \omega _{0}\right) \\
 & = & \frac{3}{j(2j-1)(2j+1)}\left( \left\langle j,j|1,0;j^{\prime }j\right\rangle ^{2}-\frac{1}{3}\right) 
\end{eqnarray*}
For \( j^{\prime }<j \) this happens to be particularly simple since the Clebsch-Gordan
coefficient is zero. For other transitions the particular closed form for the
Clebsch-Gordan coefficients that are involved can be used,

\begin{eqnarray*}
\langle j-1,j|1,0;j,j\rangle  & = & 0\\
\langle j,j|1,0;j,j\rangle  & = & \frac{j}{\sqrt{j(j+1)}}\\
\langle j+1,j|1,0;j,j\rangle  & = & \frac{1}{\sqrt{j+1}}
\end{eqnarray*}
The resulting expressions for the quadrupole coefficients are summarized in
tab.\ref{tab:DipoleProductCoefficients}. Note that this corresponds to \( \Delta m=0 \)
transitions so that naturally, for example, the \( j,j \) matrix element of
the product gives zero for \( j^{\prime }<j \) since there is no excited state
to couple to.

\subsubsection{Dipole Coefficient}

Similarly with \( j_{ij} \) symmetric, \( \varepsilon _{ijk}\sigma _{i}\sigma _{j}=2i\delta \omega _{1}j_{k} \)
and \( \delta \omega _{1} \) can be solved for using a particular term. This
turns out to involves an unwieldy number of Clebsch- Gordan coefficients and
a simpler route is to consider a physical example. Like for the quadrupole coefficient
consider \( \Delta m=+1 \) transitions, this will give zero \( j,j \) matrix
elements for \( j^{\prime }<j \) and \( j^{\prime }=j \) since again there
are no excited states for the \( \left| j,j\right\rangle  \) initial state
to couple to corresponding to, 
\begin{eqnarray*}
\left\langle j-1,j|1,1;j,j\right\rangle  & = & 0\\
\left\langle j,j|1,1;j,j\right\rangle  & = & 0
\end{eqnarray*}
For \( j^{\prime }=j+1 \) a transition is possible and its rate is given by
the Clebsch-Gordan coefficient,
\begin{eqnarray*}
\left\langle j+1,j|1,1;j,j\right\rangle  & = & 1
\end{eqnarray*}
This can be used to evaluate the dipole product coefficients through,

\begin{eqnarray*}
\left( \sigma _{-1}\sigma _{+1}\right) _{jj} & = & \frac{\left\langle j+1,j+1|1,-1;j,j\right\rangle }{\sqrt{2(j+1)+1}}\frac{\left\langle j,j|1,1;j+1,j+1\right\rangle }{\sqrt{2j+1}}\\
 & = & \frac{\left\langle j,j|1,1;j+1,j+1\right\rangle ^{2}}{2j+1}\\
 & = & \left( \frac{\sigma _{x}-i\sigma _{y}}{\sqrt{2}}\frac{\sigma _{x}+i\sigma _{y}}{\sqrt{2}}\right) _{jj}\\
 & = & \frac{1}{2}\left( \sigma _{x}^{2}+\sigma _{y}^{2}+i(\sigma _{x}\sigma _{y}-\sigma _{y}\sigma _{x})\right) _{jj}\\
 & = & \frac{1}{2}\left( (\delta \omega _{0}+\delta \omega _{2}j_{xx})+(\delta \omega _{0}+\delta \omega _{2}j_{yy})+i(i\delta \omega _{1}j_{z}+i\delta \omega _{1}j_{z})\right) _{jj}\\
 & = & \frac{1}{2}\left( 2\delta \omega _{0}-2\delta \omega _{1}j_{z}+\delta \omega _{2}(j_{xx}+j_{yy})\right) _{jj}\\
 & = & \delta \omega _{0}-\delta \omega _{1}\left( j_{z}\right) _{jj}-\frac{\delta \omega _{2}}{2}\left( j_{zz}\right) _{jj}
\end{eqnarray*}
Solving for \( \delta \omega _{1} \),
\[
\delta \omega _{1}=\frac{1}{\left( j_{z}\right) _{jj}}\left( \delta \omega _{0}-\frac{\delta \omega _{2}}{2}\left( j_{zz}\right) _{jj}-\frac{\left\langle j,j|1,1;j+1,j+1\right\rangle ^{2}}{2j+1}\right) \]
The \( j_{zz} \) matrix element is again required, the \( j_{z} \) matrix
elements simply gives \( j \). Substituting these, the Clebsch-Gordan coefficients
and the relevant previously determined product coefficient yields a relatively
simple closed form result for \( \delta \omega _{1} \). These expressions are
also summarized in tab.\ref{tab:DipoleProductCoefficients}.

\begin{table}

\caption{\label{tab:DipoleProductCoefficients}Coefficients of scalar, dipole and quadrupole
component of products of dipole \protect\( \sigma \protect \) matrices for
arbitrary \protect\( j_{f}=j_{i}\protect \) and \protect\( j^{\prime }\protect \)}
\vspace{0.3cm}
{\centering \begin{tabular}{|c||c|c|c|}
\hline 
\( j^{\prime } \)&
\( \delta \omega _{0} \)&
\( \delta \omega _{1} \)&
\( \delta \omega _{2} \)\\
\hline 
\hline 
\( j+1 \)&
\( \frac{1}{3}\frac{1}{2j+1} \)&
\( -\frac{1}{2}\frac{1}{j+1}\frac{1}{2j+1} \)&
 \( -\frac{1}{j+1}\frac{1}{2j+1}\frac{1}{2j+3} \)\\
\hline 
\( j \)&
\( \frac{1}{3}\frac{1}{2j+1} \)&
\( \frac{1}{2}\frac{1}{j}\frac{1}{j+1}\frac{1}{2j+1} \)&
\( \frac{1}{j}\frac{1}{j+1}\frac{1}{2j+1} \)\\
\hline 
\( j-1 \)&
\( \frac{1}{3}\frac{1}{2j+1} \)&
\( \frac{1}{2}\frac{1}{j}\frac{1}{2j+1} \)&
\( -\frac{1}{j}\frac{1}{2j+1}\frac{1}{2j-1} \)\\
\hline 
\end{tabular}\par}\vspace{0.3cm}
\end{table}

\subsection{Higher Order Products and Decomposition}

This gives complete and general result for dipole products for any set of initial
and intermediate states, but for problems involving quadrupole transitions products
of quadrupole matrices with dipole and other quadrupole matrices must be computed.
For \( j_{i}=j_{f}=j^{\prime } \) the products can be computed in the same
way that dipole products for this case were determined, by writing the \( \sigma  \)
in terms of matrix elements of angular momentum operators and using the usual
angular momentum operator algebra once the definitions for the \( j_{ijk} \)
and \( j_{ijkl} \) the will now appear in the result are determined. For \( j_{i}=j_{f}\neq j^{\prime } \)the
methods discussed that make use of angular momentum projection operators written
in terms of \( j^{2} \) can be immediately applied though this process is even
more cumbersome than for the dipole products as the \( j^{2} \) must be maneuvered
around more \noun{\( x \)}'s. 

It turns out that the complications of this latter calculation for general \( j \)
can be avoided by writing the quadrupole operators in terms of a sum of dipole
products. \( \sigma _{ij} \) is symmetric, so consider, \( \sigma _{i}\sigma _{j}+\sigma _{j}\sigma _{i} \).
The antisymmetric dipole part of the products cancel in the sum,
\begin{eqnarray*}
\sigma _{i}\sigma _{j}+\sigma _{j}\sigma _{i} & = & (\delta \omega _{0}+i\varepsilon _{ijk}\delta \omega ^{\prime }_{1}\sigma _{k}+\delta \omega ^{\prime }_{2}\sigma _{ij})+(\delta \omega _{0}+i\varepsilon _{jik}\delta \omega ^{\prime }_{1}\sigma _{k}+\delta \omega ^{\prime }_{2}\sigma _{ij})\\
 & = & 2\left( \delta \omega _{0}+\delta \omega ^{\prime }_{2}\sigma _{ij}\right) 
\end{eqnarray*}
 Since the \( j_{i}=j_{f} \) cases can be dealt with as just described using
angular momentum algebra, the only non trivial cases are for \( j_{i}\neq j_{f} \)
for which the scalar term is zero. Replacing the explicit \( j \) dependence,
\[
\sigma _{ij}(j_{f},j_{i})=\frac{1}{2\delta \omega ^{\prime }_{2}(j_{f},j^{\prime },j_{i})}\left( \sigma _{i}(j_{f},j^{\prime })\sigma _{j}(j^{\prime },j_{i})+\sigma _{j}(j_{f},j^{\prime })\sigma _{i}(j^{\prime },j_{i})\right) \]
This is valid for any intermediate \( j^{\prime } \) that would give a non-zero
transition amplitude. As an example, the \( j=1/2\rightarrow 3/2 \) quadrupole
transition can be represented in two ways,

\begin{eqnarray*}
 &  & \sigma _{ij}(\frac{1}{2}\frac{3}{2})\\
 &  & =\sqrt{\frac{3}{5}}\left( \sigma _{i}(\frac{1}{2}\frac{1}{2})\sigma _{j}(\frac{1}{2}\frac{3}{2})+\sigma _{i}(\frac{1}{2}\frac{1}{2})\sigma _{j}(\frac{1}{2}\frac{3}{2})\right) \\
 &  & =\sqrt{6}\left( \sigma _{i}(\frac{1}{2}\frac{3}{2})\sigma _{j}(\frac{3}{2}\frac{3}{2})+\sigma _{i}(\frac{1}{2}\frac{3}{2})\sigma _{j}(\frac{3}{2}\frac{3}{2})\right) 
\end{eqnarray*}
where the constants of proportionality wore determine from an explicit calculation.
Though a general closed-form result should be possible by computing one component,
such as \( ij=zz \) and using computing any Clebsch-Gordan coefficients involved.

For this case the calculation of products greatly simplifies calculations since
the smallest intermediate \( j^{\prime } \) can be used so that products needed
for intermediate results have less complicated structure. For example, in computing
a product of a dipole and quadrupole matrices, 
\[
\sigma _{ij}(\frac{1}{2}\frac{3}{2})\sigma _{k}(\frac{3}{2}\frac{1}{2})=\left( \sigma _{i}(\frac{1}{2},j^{\prime })\sigma _{j}(j^{\prime },\frac{3}{2})+\sigma _{j}(\frac{1}{2},j^{\prime })\sigma _{i}(,j^{\prime },\frac{3}{2})\right) \sigma _{k}(\frac{3}{2}\frac{1}{2})\]
the two right-most \( \sigma  \)'s can be multiplied first. If \( j^{\prime }=3/2 \)
is used, the product gives another non-square matrix, for which the product
rules have not yet been explicitly determined, and will give another quadrupole
term in addition to the dipole and scalar. For \( j^{\prime }=1/2 \), the resulting
intermediate product is square, and as it is for \( j_{i}=j_{f}=1/2 \) there
is no quadrupole component. This single step reduced the calculation to products
of only \( \sigma _{i}(1/2,1/2) \) terms which can be easily done similarly
for quadrupole products, decomposing both quadrupole operators into products
of dipole matrices with \( j^{\prime }=1/2 \) and multiplying the middle two
dipole \( \delta  \) first similarly reduced the problem to products of \( \sigma _{i}(1/2,1/2) \)
. This trick will be exploited when calculating the ground state shifts in sec.\ref{Sec:LightShiftVectorStructure}.

The same methods simplify dipole-quadrupole, and quadrupole-quadrupole products
for arbitrary \( j \). The simplification is not as dramatic since generally
quadrupole operators will reappear in the products, but with an appropriate
decomposition and a strategically chosen first step the problem can be reduced
to products of square matrices which can be dealt with straightforwardly with
the familiar angular momentum algebra used to compute the product of dipole
matrices for \( j_{i}=j_{f}=j^{\prime } \). Specifically for

\[
\sigma _{ij}(j,j^{\prime })\sigma _{k}(j^{\prime },j)=\left( \sigma _{i}(j,j^{\prime \prime })\sigma _{j}(j^{\prime \prime },j^{\prime })+\sigma _{j}(j,j^{\prime \prime })\sigma _{i}(,j^{\prime \prime },j^{\prime })\right) \sigma _{k}(j^{\prime }j)\]
taking \( j^{\prime \prime }=j \), and multiplying the rightmost \( \sigma  \)
using the already determined dipole product rules resulted in an expression
involving products of only square matrices of dimension \( 2j+1 \) which can
then be multiplied as described earlier in this section. The same can be done
for quadrupole products for arbitrary \( j \) and \( j^{\prime } \). This
procedure will give a general, closed form result for any products of dipole
and quadrupole matrices, for any set of \( j \) and \( j^{\prime } \), fairly
easily, \cite{Schacht00}.

Similar methods should allow products of even higher order matrices as well,
that is, it should be possible to decompose \( n \)th order matrices into sums
of products of dipole matrices and use the dipole product rules already determined
and angular momentum algebra to write the product in terms of an appropriate
set of irreducible \( \sigma  \) matrices, though this has not yet been explicitly
studied for operator of order higher than quadrupole.

\subsection{\protect\( j_{f}\neq j_{i}\protect \)}

Using decomposition arbitrary products can be computed for any set of angular
momentum states with \( j_{f}=j_{i} \). For \( j_{f}\neq j_{i} \) this method
can be used to reduce the result to products of square matrices and one rectangular
matrix. The application of these kinds of products is not immediately apparent
Reduction of all the products of square matrices will finally yield terms like,
\[
\sigma _{i}(j_{f},j_{i})\sigma _{jkl\cdots }(j_{i},j_{i})\propto \left\langle j_{f}m_{2}\left| x_{i}j_{jkl\cdots }\right| j_{i}m_{1}\right\rangle \]
This kind of product of rectangular matrices has not yet been explicitly considered,
and there is no immediately apparent strategy for computing them, though it
should be possible to write the results in terms of matrices of the form \( \sigma _{ijkl\cdots }(j_{f},j_{i}) \).
Further progress will eventually appear in \cite{Schacht00}.

\section{Geometric Structure of the Ground State Light Shifts}

\label{Sec:LightShiftVectorStructure}

\subsection{Full Spin Manifold}

Though a general result for products of dipole and quadrupole matrices for arbitrary
\( j_{f}=j_{i} \) and \( j^{\prime } \) is available, the structure of the
light shifts in the ground state for this experiment when the full set of spin
states in the intermediate state are accessible can be determined easily from
the multiplication rules and decomposition results already explicitly computed.
Shifts in the \( D \) state may also be useful for analyzing systematic problems
but these require the more complicated general results and are not explicitly
considered here. The structure of these \( D \) state shifts will appear in
\cite{Schacht00}

For the ground state, the required products are,
\begin{eqnarray*}
\sigma _{i}(\frac{1}{2}\frac{1}{2})\sigma _{j}(\frac{1}{2}\frac{1}{2}) & = & \frac{1}{6}\left( \delta _{ij}+i\sqrt{6}\varepsilon _{ijk}\sigma _{k}\left( \frac{1}{2}\frac{1}{2}\right) \right) \\
\sigma _{i}(\frac{1}{2}\frac{3}{2})\sigma _{j}(\frac{3}{2}\frac{1}{2}) & = & \frac{1}{6}\left( \delta _{ij}-i\sqrt{\frac{3}{2}}\varepsilon _{ijk}\sigma _{k}\left( \frac{1}{2}\frac{1}{2}\right) \right) 
\end{eqnarray*}
The quadrupole operator can be decomposed as,
\begin{eqnarray*}
 &  & \sigma _{ij}(\frac{1}{2}\frac{3}{2})\\
 &  & =\sqrt{\frac{3}{5}}\left( \sigma _{i}(\frac{1}{2}\frac{1}{2})\sigma _{j}(\frac{1}{2}\frac{3}{2})+\sigma _{i}(\frac{1}{2}\frac{1}{2})\sigma _{j}(\frac{1}{2}\frac{3}{2})\right) 
\end{eqnarray*}
and final results are more convenient in terms of \( j_{i} \) using
\[
\sigma _{i}\left( \frac{1}{2}\frac{1}{2}\right) =\sqrt{\frac{2}{3}}j_{i}\]

\subsubsection{Dipole-Dipole}

The dipole-dipole term is not needed, but for completeness the result obtained
previously is listed here,

\[
\Omega ^{D\dagger }(\frac{1}{2}\frac{3}{2})\Omega ^{D}(\frac{3}{2}\frac{1}{2})=\frac{\left\langle D_{3/2}\left| \left| Q\right| \right| P_{1/2}\right\rangle ^{2}}{6}(\vec{E}^{*}\cdot \vec{E}-(\vec{E}^{*}\times \vec{E})\cdot \vec{j})\]

\subsubsection{Quadrupole-Dipole}

For the PNC term the Dipole-Quadrupole product gives,

\begin{eqnarray*}
 &  & \Omega _{2}^{Q\dagger }(\frac{1}{2}\frac{3}{2})\Omega _{1}^{D}(\frac{3}{2}\frac{1}{2})\\
 &  & =\left\langle Q\right\rangle \left\langle D\right\rangle \sigma _{ij}(\frac{1}{2}\frac{3}{2})\sigma _{k}(\frac{3}{2}\frac{1}{2})\partial _{i}E^{*}_{2j}E_{1k}\\
 &  & =\left\langle Q\right\rangle \left\langle D\right\rangle \sqrt{\frac{3}{5}}(\sigma _{i}(\frac{1}{2}\frac{1}{2})\sigma _{j}(\frac{1}{2}\frac{3}{2})+\sigma _{j}(\frac{1}{2}\frac{1}{2})\sigma _{i}(\frac{1}{2}\frac{3}{2}))\sigma _{k}(\frac{3}{2}\frac{1}{2})\partial _{i}E^{*}_{2j}E_{1k}\\
 &  & =\left\langle Q\right\rangle \left\langle D\right\rangle \sqrt{\frac{3}{5}}\sigma _{i}(\frac{1}{2}\frac{1}{2})\sigma _{j}(\frac{1}{2}\frac{3}{2})\sigma _{k}(\frac{3}{2}\frac{1}{2})(\partial _{i}E^{*}_{2j}+\partial _{j}E^{*}_{2i})E_{1k}
\end{eqnarray*}
The product of the \( \sigma  \) matrices can be simplified using the product
rules just listed,

\begin{eqnarray*}
 &  & \sigma _{i}(\frac{1}{2}\frac{1}{2})\sigma _{j}(\frac{1}{2}\frac{3}{2})\sigma _{k}(\frac{3}{2}\frac{1}{2})\\
 &  & =\frac{1}{6}\sigma _{i}(\frac{1}{2}\frac{1}{2})(\delta _{jk}-i\varepsilon _{jkm}\sqrt{\frac{3}{2}}\sigma _{m}(\frac{1}{2}\frac{1}{2}))\\
 &  & =-\frac{1}{6}(\sigma _{i}\delta _{jk}-i\varepsilon _{jkm}\sqrt{\frac{3}{2}}\sigma _{i}\sigma _{m})\\
 &  & =-\frac{1}{6}\left( \delta _{jk}\sigma _{i}-i\varepsilon _{jkm}\sqrt{\frac{3}{2}}\frac{1}{6}\left( \delta _{im}+i\varepsilon _{imn}\sqrt{6}\sigma _{n}\right) \right) \\
 &  & =-\frac{1}{6}\left( \delta _{jk}\sigma _{i}-i\varepsilon _{ijk}\frac{1}{6}\sqrt{\frac{3}{2}}+\frac{1}{2}\varepsilon _{jkm}\varepsilon _{imn}\sigma _{n}\right) \\
 &  & =-\frac{1}{6}\left( -i\varepsilon _{ijk}\frac{1}{6}\sqrt{\frac{3}{2}}+(\delta _{jk}\delta _{in}+\frac{1}{2}\varepsilon _{jkm}\varepsilon _{imn})\sigma _{n}\right) \\
 &  & =-\frac{1}{6}\left( -i\varepsilon _{ijk}\frac{1}{6}\sqrt{\frac{3}{2}}+\sqrt{\frac{2}{3}}(\delta _{jk}\delta _{in}+\frac{1}{2}\varepsilon _{jkm}\varepsilon _{imn})j_{n}\right) \\
 &  & =-\frac{1}{6}\sqrt{\frac{2}{3}}\left( -i\varepsilon _{ijk}\frac{1}{4}+(\delta _{jk}\delta _{in}+\frac{1}{2}\varepsilon _{jkm}\varepsilon _{imn})j_{n}\right) 
\end{eqnarray*}
Symmetrizing this in \( ij \) to give the original quadrupole operator,

\begin{eqnarray*}
 &  & \sigma _{ij}(\frac{1}{2}\frac{3}{2})\sigma _{k}(\frac{3}{2}\frac{1}{2})\\
 &  & =\sqrt{\frac{3}{5}}\left( \sigma _{i}(\frac{1}{2}\frac{1}{2})\sigma _{j}(\frac{1}{2}\frac{3}{2})+\sigma _{j}(\frac{1}{2}\frac{1}{2})\sigma _{i}(\frac{1}{2}\frac{3}{2})\right) \sigma _{k}(\frac{3}{2}\frac{1}{2})\\
 &  & =-\frac{1}{6}\sqrt{\frac{3}{5}}\sqrt{\frac{2}{3}}\left( \delta _{jk}\delta _{in}+\delta _{ik}\delta _{jn}+\frac{1}{2}\varepsilon _{jkm}\varepsilon _{imn}+\frac{1}{2}\varepsilon _{ikm}\varepsilon _{jmn}\right) j_{n}
\end{eqnarray*}
Note that there is no scalar term, the vector term can be written in a variety
of forms using,

\begin{eqnarray*}
\varepsilon _{jkm}\varepsilon _{nim} & = & \delta _{jn}\delta _{ki}-\delta _{ji}\delta _{kn}\\
\varepsilon _{ikm}\varepsilon _{njm} & = & \delta _{in}\delta _{kj}-\delta _{ji}\delta _{kn}
\end{eqnarray*}
Two possible forms for the result, equivilant up to terms involving \( \vec{\nabla }\cdot \vec{E} \),
which will be zero for these fields, are

\begin{eqnarray*}
 &  & \Omega ^{Q\dagger }(\frac{1}{2}\frac{3}{2})\Omega ^{D}(\frac{3}{2}\frac{1}{2})\\
 &  & =-\frac{\left\langle Q\right\rangle \left\langle D\right\rangle }{2\sqrt{10}}\left( (\vec{E}_{2}^{*}\cdot \vec{E}_{1})\vec{k}_{2}+(\vec{k}_{2}\cdot \vec{E}_{1})\vec{E}_{2}^{*}\right) \cdot \vec{j}\\
 &  & =\frac{\left\langle Q\right\rangle \left\langle D\right\rangle }{2\sqrt{10}}\left( (\vec{E}_{1}\times \vec{\nabla })\times \vec{E}^{*}_{2}-(\vec{E}_{1}\cdot \vec{\nabla })\vec{E}_{2}^{*}\right) \cdot \vec{j}
\end{eqnarray*}
The former is written explicitly for plane waves using \( \vec{\nabla }\vec{E}=\vec{k}\vec{E} \)
since it is hard to write unambiguously in operator form.

This show the general structure hinted at in the perturbative analysis.

\subsubsection{Quadrupole-Quadrupole Term}

The quadrupole term is similarly given by,

\begin{eqnarray*}
 &  & \Omega ^{Q\dagger }(\frac{1}{2}\frac{3}{2})\Omega ^{Q}(\frac{3}{2}\frac{1}{2})\\
 &  & =\left\langle Q\right\rangle ^{2}\sigma ij(\frac{1}{2}\frac{3}{2})\sigma _{kl}(\frac{3}{2}\frac{1}{2})\partial _{i}E_{j}\partial _{k}E_{l}\\
 &  & =\left\langle Q\right\rangle ^{2}\frac{3}{5}\sigma _{i}(\frac{1}{2}\frac{1}{2})\sigma _{j}(\frac{1}{2}\frac{3}{2})\sigma _{k}(\frac{3}{2}\frac{1}{2})\sigma _{l}(\frac{1}{2}\frac{1}{2})(\partial _{i}E_{j}+\partial _{j}E_{k})\left( \partial _{k}E_{l}+\partial _{l}E_{k}\right) 
\end{eqnarray*}
Evaluating the this product of four \( \sigma  \) matrices is a bit cumbersome,
after symmetrizing with respect to \( ij \) and \( kl \) and omitting terms
involving \( \varepsilon _{ijm} \) or \( \varepsilon _{klm} \) because of
symmetry and terms involving \( \delta _{ij} \) or \( \delta _{kl} \) because
\( \vec{\nabla }\cdot \vec{E}=0 \), the product can be written \cite{Schacht00},
\begin{eqnarray*}
\sigma _{ij}(\frac{1}{2}\frac{3}{2})\sigma _{kl}(\frac{3}{2}\frac{1}{2}) & = & -\frac{1}{24}(\delta _{jk}\delta _{il}+\delta _{ik}\delta _{jl}+\delta _{il}\delta _{jk}+\delta _{jl}\delta _{ik})\\
 & + & i(\delta _{jk}\varepsilon _{iln}+\delta _{il}\varepsilon _{jkn}+\delta _{ik}\varepsilon _{jln}+\delta _{jl}\varepsilon _{ikn})j_{n})
\end{eqnarray*}
Note that the scalar term is given by all possible pairs of dot products and
the vector term has component in the direction of the cross products of every
pair with an amplitude given by the dot product of the other two. 

To clean representations of the resulting vector piece are,

\begin{eqnarray*}
\delta \vec{\omega }^{QQ} & = & -i\frac{\left\langle Q\right\rangle ^{2}}{20}((\vec{k}_{1}\cdot \vec{k}_{2})(\vec{E}_{1}\times \vec{E}_{2}^{*})+(\vec{E}_{1}\cdot \vec{E}_{2}^{*})(\vec{k}_{1}\times \vec{k}_{2})\\
 & + & (\vec{k}_{1}\cdot \vec{E}_{2}^{*})(\vec{E}_{1}\times \vec{k}_{2})+(\vec{E}_{1}\cdot \vec{k}_{2})(\vec{k}_{1}\times \vec{E}_{2}^{*}))\\
 & = & -i\frac{\left\langle Q\right\rangle ^{2}}{10}((\vec{k}_{1}\cdot \vec{k}_{2})(\vec{E}_{1}\times \vec{E}_{2}^{*})+(\vec{E}_{1}\cdot \vec{E}_{2}^{*})(\vec{k}_{1}\times \vec{k}_{2})\\
 & - & \frac{1}{2}(\vec{k}_{1}\times \vec{E}_{1})\times (\vec{k}_{2}\times \vec{E}_{2}^{*}))
\end{eqnarray*}
Again the two are equivilant up to terms involving \( \vec{k}\cdot \vec{E}=0 \)
using familiar vector cross product identities. The first two terms in the latter
result were hinted at in the perturbative analysis of the systematic errors
and illustrate the largest problem, from the usual set of ideal fields, a single
misalignments a polarization gives a shift in the direction of \( \vec{\varepsilon }_{1}\times \vec{\varepsilon }_{2} \)
which is principally along the \( \hat{z} \) axis. For perfect alignment, these
cross products, or dot products are all zero.

\subsection{\protect\( m\protect \) Restrictions}

\label{Sec:ShiftVectorStuctureWithmRestrictions}

Previously this error was reduced by restricting the set of intermediate states
to the \( m=\pm 1/2 \) sublevels. The structure of the resulting vector shift
can be determined with these methods using \( m \) projection operators, \( P_{\hat{s},\left\{ m\right\} } \)
constructed much the the angular momentum operators used in evaluating the dipole
\( \sigma  \) products. For states with \( m \) defined along a particular
axis \( \hat{s} \) these projection operators should act to give,
\[
P_{\hat{s},m}\left| j,m^{\prime }\right\rangle _{\hat{s}}=\left| j,m^{\prime }\right\rangle _{\hat{s}}\delta _{mm^{\prime }}\]
\( P_{\hat{s},\left\{ m\right\} } \)is a sum of \( P_{\hat{s},m} \)'s that
give zero for any \( m^{\prime } \) not in \( \left\{ m\right\}  \).

The quadrupole amplitude can then be written as,
\[
\Omega _{m_{2}m_{1}}^{Q}=\left\langle j_{2}m_{2}\left| P_{\hat{s},\left\{ \pm 1/2\right\} }x_{i}x_{j}\right| j_{1}m_{1}\right\rangle \]
This introduces another vector, \( \hat{s} \), into the problem that should
appear in the final result.

With \( \hat{s}=\hat{z} \) these projection operators are easily constructed
using \( j_{z} \). As for the angular momentum projection operators, 
\[
K_{\hat{z},m,m^{\prime }}=\frac{j_{z}-m^{\prime }}{m-m^{\prime }}\]
kills the state \( \left| j,m^{\prime }\right\rangle  \) and gives \( 1 \)
on \( \left| j,m\right\rangle  \), so that these can be combined to build a
spin projection operator,

\[
P_{\hat{z},m}=\prod _{m^{\prime }\neq m}\frac{j_{z}-m^{\prime }}{m-m^{\prime }}\]
Projection on spin state along an arbitrary axis, \( \hat{s} \), are given
simply by rotating \( j_{z} \),
\[
P_{\hat{s},m}=\prod _{m^{\prime }\neq m}\frac{\left( \hat{s}\cdot \vec{j}\right) -m^{\prime }}{m-m^{\prime }}\]

For \( m=\pm 1/2 \) this is explicitly given by

\begin{eqnarray*}
P_{\hat{z},\left\{ \pm 1/2\right\} } & = & P_{1/2}+P_{-1/2}\\
 & = & \frac{j_{z}-3/2}{1/2-3/2}\frac{j_{z}+1/2}{1/2+1/2}\frac{j_{z}+3/2}{1/2+3/2}\\
 & + & \frac{j_{z}-3/2}{-1/2-3/2}\frac{j_{z}-1/2}{-1/2-1/2}\frac{j_{z}+3/2}{-1/2+3/2}\\
 & = & \frac{j_{z}-3/2}{1/2-3/2}\frac{j_{z}+1/2}{1/2+1/2}\frac{j_{z}+3/2}{1/2+3/2}\\
 & - & \frac{j_{z}-3/2}{1/2+3/2}\frac{j_{z}-1/2}{1/2+1/2}\frac{j_{z}+3/2}{1/2-3/2}\\
 & = & \frac{j_{z}^{2}-9/4}{1/4-9/4}\\
 & = & \frac{9/4-j_{z}^{2}}{2}\\
 & = & \frac{9}{8}-\frac{1}{2}j_{z}^{2}
\end{eqnarray*}
or
\[
P_{\hat{s},\left\{ \pm 1/2\right\} }=\frac{9}{8}-\frac{1}{2}(\hat{s}\cdot \vec{j})^{2}\]
as a check, note that for \( m=\pm 3/2 \),

\begin{eqnarray*}
P_{\hat{z},\left\{ \pm 3/2\right\} } & = & P_{3/2}+P_{-3/2}\\
 & = & \frac{1}{2}(\hat{s}\cdot \vec{j})^{2}-\frac{1}{8}
\end{eqnarray*}
so that 
\[
P_{\hat{s},\left\{ \pm 1/2\right\} }+P_{\hat{s},\left\{ \pm 3/2\right\} }=1\]

Now shifts can be determined by evaluating product like
\[
\left\langle \frac{1}{2}m\left| x_{i}\right| \frac{3}{2}m^{\prime }\right\rangle \left\langle \frac{3}{2}m^{\prime }\left| \left( \frac{9}{8}-\frac{1}{2}(\hat{s}\cdot \vec{j})^{2}\right) x_{j}\right| \frac{1}{2}m\right\rangle \]
The quadrupole-quadrupole term involves

\begin{eqnarray*}
 &  & \left\langle \frac{1}{2}m\left| x_{i}x_{j}\right| \frac{3}{2}m^{\prime }\right\rangle \left\langle \frac{3}{2}m^{\prime }\left| \left( \frac{9}{8}-\frac{1}{2}(\hat{s}\cdot \vec{j})^{2}\right) x_{k}x_{l}\right| \frac{1}{2}m\right\rangle \\
 & = & \frac{9}{8}\left\langle \frac{1}{2}m\left| x_{i}x_{j}\right| \frac{3}{2}m^{\prime }\right\rangle \left\langle \frac{3}{2}m^{\prime }\left| x_{k}x_{l}\right| \frac{1}{2}m\right\rangle \\
 & - & \frac{1}{2}\left\langle \frac{1}{2}m\left| x_{i}x_{j}\right| \frac{3}{2}m^{\prime }\right\rangle \left\langle \frac{3}{2}m^{\prime }\left| j_{m}j_{n}x_{k}x_{l}\right| \frac{1}{2}m\right\rangle s_{m}s_{n}
\end{eqnarray*}
which has a contribution from the original term. As a result, for general \( \hat{s} \),
these spin state restriction can reduce the largest quadrupole error. But when
\( \hat{s} \) is set to \( \hat{\varepsilon }_{1} \) the appropriate terms
appear to cancel these largest effects.

Evaluating these matrix elements requires considerably more complicated general
results than those explicitly listed here, but decomposing the quadrupole operator
again works and yields and answer in a straight-forward, though slightly tedious
manner, \cite{Schacht00}. For arbitrary \( \hat{s} \) both scalar and vector
contributions to the various shifts are given by, 

\begin{eqnarray*}
\delta \vec{\omega }^{DD} & = & -i\frac{\left\langle D_{3/2}\left| \left| Q\right| \right| P_{1/2}\right\rangle ^{2}}{6}\left( (\vec{E}_{2}^{*}\times \vec{E}_{1})+\frac{3}{2}((\vec{E}_{1}\times \vec{E}_{2}^{*})\cdot s)s\right) \\
\delta \vec{\omega }^{DQ} & = & -\frac{\left\langle Q\right\rangle \left\langle D\right\rangle }{2\sqrt{10}}(\left( (\vec{E}_{2}^{*}\cdot \vec{E}_{1})-(\vec{E}_{1}\times \hat{s})\cdot (\vec{E}_{2}^{*}\times \hat{s})\right) \vec{k}_{2}\\
 & + & \left( (\vec{k}_{2}\cdot \vec{E}_{1})-(\vec{E}_{1}\times \hat{s})\cdot (\vec{k}_{2}\times \hat{s})\right) \vec{E}_{2}^{*}\\
 & + & \left( \frac{1}{2}(\vec{E}_{1}\times \hat{s})\cdot (\vec{k}_{2}\times \vec{E}_{2}^{*})-(\vec{E}_{1}\times \vec{E}_{2}^{*})\cdot (\vec{k}_{2}\times \hat{s})\right) \hat{s})\\
\delta \vec{\omega }^{QQ} & = & -i\frac{\left\langle Q\right\rangle ^{2}}{10}(\frac{1}{4}(\left( (\vec{k}_{1}\cdot \vec{k}_{2})+(\vec{k}_{1}\cdot \hat{s})(\vec{k}_{2}\cdot \hat{s})\right) (\vec{E}_{1}\times \vec{E}_{2}^{*})\\
 & + & \left( (\vec{E}_{1}\cdot \vec{E}_{2}^{*})+(\vec{E}_{1}\cdot \hat{s})(\vec{E}_{2}^{*}\cdot \hat{s})\right) (\vec{k}_{1}\times \vec{k}_{2})\\
 & + & \left( (\vec{k}_{1}\cdot \vec{E}_{2}^{*})+(\vec{k}_{1}\cdot \hat{s})(\vec{E}_{2}^{*}\cdot \hat{s})\right) (\vec{E}_{1}\times \vec{k}_{2})\\
 & + & \left( (\vec{E}_{1}\cdot \vec{k}_{2})+(\vec{E}_{1}\cdot \hat{s})(\hat{k}_{2}\cdot \hat{s})\right) (\vec{k}_{1}\times \vec{E}_{2}^{*}))\\
 & + & \frac{1}{4}\left( (\vec{E}_{1}\times \vec{E}_{2}^{*})\times (\vec{k}_{1}\times \vec{k}_{2})+(\vec{k}_{1}\times \vec{E}_{2}^{*})\times (\vec{k}_{2}\times \vec{E}_{1})\right) \\
 & + & \frac{1}{2}(\left( \vec{E}_{1}\times \hat{s}\right) \times \left( \vec{k}_{1}\times \vec{k}_{2}\right) (\vec{E}_{2}^{*}\cdot \hat{s})+\left( \vec{E}_{1}\times \hat{s}\right) \times \left( \vec{k}_{1}\times \vec{E}_{2}^{*}\right) (\vec{k}_{2}^{*}\cdot \hat{s})\\
 & + & \left( \vec{E}_{1}\times \hat{s}\right) \times \left( \vec{k}_{2}\times \vec{E}_{2}^{*}\right) (\vec{k}_{1}^{*}\cdot \hat{s}))+\left( \vec{E}_{1}\times \hat{s}\right) \cdot \vec{k}_{2}(\hat{s}\cdot \vec{k}_{1})\vec{E}_{2}^{*}\\
 & - & \frac{1}{2}\left( \vec{E}_{1}\times \hat{s}\right) \cdot \vec{E}_{2}^{*}(\vec{k}_{1}\cdot \vec{k}_{2})\hat{s}-\left( \vec{E}_{1}\times \hat{s}\right) \cdot \vec{k}_{2}(\vec{E}_{2}^{*}\cdot \vec{k}_{1})\hat{s})
\end{eqnarray*}
With \( \hat{s}=\hat{E}_{1} \), these become,

\begin{eqnarray*}
\delta \vec{\omega }^{DD} & = & -i\frac{\left\langle D_{3/2}\left| \left| Q\right| \right| P_{1/2}\right\rangle ^{2}}{6}(\vec{E}_{2}^{*}\times \vec{E}_{1})\\
\delta \vec{\omega }^{DQ} & = & -\frac{\left\langle Q\right\rangle \left\langle D\right\rangle }{2\sqrt{10}}\left( (\vec{E}_{2}^{*}\cdot \vec{E}_{1})\vec{k}_{2}+(\vec{k}_{2}\cdot \vec{E}_{1})\vec{E}_{2}^{*}-(\vec{E}_{1}\times \vec{E}_{2}^{*})\cdot (\vec{k}_{2}\times \hat{E}_{1})\hat{E}_{1}\right) \\
\delta \vec{\omega }^{QQ} & = & -i\frac{\left\langle Q\right\rangle ^{2}}{10}(\frac{1}{4}(\left( \vec{k}_{1}\cdot \vec{k}_{2}\right) (\vec{E}_{1}\times \vec{E}_{2}^{*})+2\left( \vec{E}_{1}\cdot \vec{E}_{2}^{*}\right) (\vec{k}_{1}\times \vec{k}_{2})\\
 & + & \left( \vec{k}_{1}\cdot \vec{E}_{2}^{*}\right) (\vec{E}_{1}\times \vec{k}_{2})+2\left( \vec{E}_{1}\cdot \vec{k}_{2}\right) (\vec{k}_{1}\times \vec{E}_{2}^{*}))\\
 & + & \frac{1}{4}\left( (\vec{E}_{1}\times \vec{E}_{2}^{*})\times (\vec{k}_{1}\times \vec{k}_{2})+(\vec{k}_{1}\times \vec{E}_{2}^{*})\times (\vec{k}_{2}\times \vec{E}_{1})\right) 
\end{eqnarray*}
In the quadrupole shift the original terms appear with slightly different amplitudes,
plus two additional terms. These addition terms remove the polarization error.
This can be seen explicitly by taking \( \vec{k}_{1}=\vec{k}_{2} \), this leaves
two terms,
\begin{eqnarray*}
 &  & \left( \vec{k}_{1}\cdot \vec{k}_{2}\right) (\vec{E}_{1}\times \vec{E}_{2}^{*})+(\vec{k}_{1}\times \vec{E}_{2}^{*})\times (\vec{k}_{2}\times \vec{E}_{1})\\
 &  & =(\vec{E}_{1}\times \vec{E}_{2}^{*})+(\vec{k}\times \vec{E}_{2}^{*})\times (\vec{k}\times \vec{E}_{1})\\
 &  & =(\vec{E}_{1}\times \vec{E}_{2}^{*})+\vec{k}(\vec{k}\times \vec{E}_{2}^{*})\cdot \vec{E}_{1}+\vec{E}_{1}\left( \vec{k}\times \vec{E}_{2}^{*}\right) \cdot \vec{k}\\
 &  & =(\vec{E}_{1}\times \vec{E}_{2}^{*})-\vec{k}\left( \vec{E}_{1}\times \vec{E}_{2}^{*}\right) \cdot \vec{k}
\end{eqnarray*}
\( \vec{E}_{1}\times \vec{E}_{2}^{*} \) is exactly along \( \vec{k} \) so
these terms cancel and, as before, there is no vector contribution to the shifts
from this quadrupole-quadrupole term from only misaligned polarizations.

\section{Other Shifts}

The possible shifts due to general configurations of the applied electric fields
acting on the \( S \) and \( D_{3/2} \) levels has been exhaustively analyzed.
There remain, however, possible effects due to other fields, applied and environmental,
and other processes previously neglected.  

\label{Sec:NonResonantDStateShifts}

An immediate example is the non resonant dipole coupling driven by the parity
fields. Large dipole transition matrix elements exist between the \( S \) and
\( 5D_{3/2} \) states and other \( P \) and \( F \) states in the atom. Though
not resonant with the parity laser they will be driven by it and one result
is additional energy shifts. These shifts will be relatively small, but again,
as the PNC shift is very small, these non-resonant shifts may be a significant
contributions. Their size can be estimated in the usual way from the general
size of the dipole rate determined earlier \( \Omega ^{D}\sim 10GHz \). The
resulting shifts will be of the order of this rate reduced by the detuning,
\[
\delta \omega ^{nonres}\sim \frac{\left| \Omega ^{D}\right| ^{2}}{\Delta E-\omega }\]
In terms of the wavelength of the transitions, \( \omega =2\pi c/\lambda  \).
For the contribution from couplings to the \( 6P_{1/2} \) state this gives
\[
\Delta E-\omega =2\pi c(1/650nm-1/2.05\mu m)\approx 2\times 10^{6}GHz\]
giving 
\[
\delta \omega ^{nonres}\sim \frac{10}{10^{6}}GHz=0.1MHz\]
There are similar contributions from the counter-rotating term \( \left| \Omega ^{D}\right| ^{2}/(\Delta E+\omega ) \)
and all the other states in the atom so that the overall shift is \( MHz \)
sized. The detailed structure of these shifts is calculated in sec.\ref{Sec:NonresonantLightShifts}
to evaluate their utility in making precise measurements of atomic matrix elements,
but the coarse structure of the shifts is also fairly transparent. For the linearly
polarized parity beam, only \( \Delta m=0 \) transitions along \( \hat{z} \)
will be driven. The size of this transition amplitude with be independent of
the sign of \( m \), but rotational invariance, but generally depends on the
magnitude of \( m \) so that couplings of the \( 5D_{3/2,\pm 3/2} \) states
\( m=\pm 3/2 \) states in another energy level will have a different strength
than couplings of the \( 5D_{3/2,\pm 1/2} \) to the \( m=\pm 1/2 \) states
in that same second level. This is obvious for the case of \( j=1/2 \) states
as in such cases there are no \( m=\pm 3/2 \) states to couple to. The end
result is that the \( 5D_{3/2,\pm 3/2} \) states are shifted away from the
\( 5D_{3/2,\pm 1/2} \) states by energies on the order of a \( MHz \). This
shift is what is exploited while driving the parity transition to help reduce
spurious shifts from systematic errors. 

Since the shifts are spin sign independent, a perfectly linearly polarized laser
introduces no additional systematic error due to this process since the shifts
in the ground state will be the same, but bits of circular polarization can
give a shift, just as it does from the resonance quadrupole transitions. These
will have a characteristic size something like \( \sigma \delta \omega ^{nonres} \)
so that for \( \sigma \sim 10^{-3} \) this gives \( kHz \) sized splitting.
Also like the circularly polarization generated resonant splittings, this can
be detected independent of the PNC shift simply by turning off the quadrupole
laser, this spurious shift remains and could be measured and subtracted off,
though this could be slightly complicated in practice since it is comparable
to the intended initial splitting from the applied magnetic field. This shifts
is also largely frequency independent and to could be distinguished from the
parity shift, and even the spurious resonance shifts, but it frequency dependent.
A larger concern is it contribution to the linewidth, a fluctuation intensity
with give a fluctuating shift. The shift is linearly related to the field intensity
so 1 part in \( 10^{3} \) fluctuation of the laser intensity give \( Hz \)
sized contributions to the linewidth. This may be excessive and require improved
regulation or polarization.

Other effects may come from fields not yet considered. Some of these are applied
but neglected, such as the magnetic fields of the laser, others are environmental
perturbations such a fluctuation static electric and magnetic fields, others
are necessary side effects of the details of the apparatus, like the electric
field and associated induced magnetic field of the trap. One amusing possibility
is that the orbit of the ion is a smaller circle of a particular handedness
that results in a spin dependent shift of energy levels from both trapping and
stray fields. Similar perturbations were considered as they affect spin lifetimes
and spin flip transition linewidths and all were found to give negligible negligible
effects. The size of these incoherence effects was estimated from the general
size of the associated matrix elements. Coherent effects would depend on the
same matrix elements and should be similarly negligible. These kind of effects
have not yet been completely considered and exhaustively analyzed but a causal
investigation turns up no significant problems, still the possibilities should,
very shortly, be enumerated and assessed.

\chapter{The IonPNC Apparatus}

The experiment is straightforward in principle, and elegantly simple, but realizing
the concept in practice requires considerable gadgetry. The essential idea is
to place a single ion in a small, stable magnetic field and measure the shift
in the Zeeman resonance when strong optical fields are applied to drive the
parity violating transition. As the central requirement is to hold and manipulate
a single ion, naturally the heart of the system is an RF ion trap.

Complete and explicit construction and operational details are already described
in \cite{KristiThesis}. Only the features and techniques required to understand
the development of the spin state manipulation and detection techniques described
in sec. \ref{Sec:SpinStuff} will be reviewed here. The most directly relevant
are state detection by floresence and shelving. In addition, some further, new
analysis of the trap is also presented as an aid to increasing general understanding
of the trap and to help identify non-ideal details of a trapped ion's environment
that could be a source of systematic errors and practical difficulties for implementing
an optical cavity inside the trap's UHV chamber to provide the stable standing
waves required for the measurement.

\section{The Ion Trap}

\label{Sec:IonTraps}

Ion traps are old news. They are implemented routinely and provide a luxurious
environment for making precision atomic measurements using both traditional
spectroscopic techniques and novel methods unique to traps. Though now common,
they are still fiendishly clever and it is entertaining to study their operation
in some detail to try to build an intuitive understanding of how they work.

\subsection{Static Fields}

The task at hand is to hold a single particle at rest and in free space to provide
an isolated and unperturbed subject for study. For a charged particle it is
natural to consider trying to use electric fields to provide a stable potential
well that confines the particle to a well defined, localized region in space.
It is well known that Maxwell's Equations imply that no configuration of static
electric fields can provide a potential well that is stable in all direction.
With a suitable choice of coordinates the potential in the region of a stationary
point can always be written in the form,
\[
\Phi =\frac{1}{2}(\alpha x^{2}+\beta y^{2}+\gamma z^{2})\]
Giving an electric field,
\[
\vec{E}=-\vec{\nabla }\Phi =-(\alpha x\hat{x}+\beta y\hat{y}+\gamma z\hat{z})\]
For this field to be stable, the force along a given axis must be directed opposite
to the displacement of any perturbation. This corresponds to all of the coefficients
that define the form of the potential being positive, \( \alpha ,\beta ,\gamma >0 \).
However, Maxwell's equation require, in particular, that in charge free space,

\[
\vec{\nabla }\cdot \vec{E}=\alpha +\beta +\gamma =0\]
so that at most two, and also at least one, of these parameters is positive.

This immediately eliminates the possibility of using only static electric fields
to trap a charged particle, but it doesn't yet require the use of other kinds
of fields and may still allow for the use of only electric fields if they are
allowed to become time dependent. Hints to a solution are provided by studying
a generic unstable system in one dimension.

\subsection{Inverted Pendulum}

A simple pendulum, consisting of a mass suspended by rigid rod from a fixed
support, provides straight-forward example of an unstable system. For an arbitrary
displacement from the down position by an angle \( \theta  \), the equation
of motion for \( \theta  \) is the familiar,
\[
\ddot{\theta }-\frac{g}{R}sin\theta =0\]
 Where \( R \) is the length of the rod. For small displacements about the
down position, \( \theta =\delta \theta \approx x/R \), this equation becomes
\( \ddot{x}+(g/R)x=0 \), which yields stable harmonic motion about the equilibrium,
\( \theta =0 \), position. In contrast, small displacements about the inverted
position, \( \theta =\pi +\delta \theta  \), give \( \ddot{x}-(g/R)x=0 \)
which result in the exponential growth of any perturbation that characterizes
an unstable equilibrium. To provide a possible mechanical analogy for an ion
in a time dependent electric field, now consider driving the pendulum by moving
its support. In this case the definition of the inverted position as a stable
or unstable equilibrium is less sharp.

\subsubsection{Equations of Motion for Driven Pendulum}

Let the position of the support be given arbitrarily by \( \vec{R}(t) \). The
resulting equation of motion for \( \theta  \) can be obtained systematically
from the Lagrangian, or by considering the system to be in an accelerated coordinate
system fixed to the support and formally transforming the variables. Alternately,
the modifications to the equation of motion for the simple pendulum can be deduced
from simple transformation properties of the accelerated coordinate system. 

Uniform motion of the support in any direction doesn't change the dynamics of
the system, but accelerations appear as effective forces in the frame of reference
of the pendulum. Note the acceleration of the support by \( \ddot{\vec{R}}=a_{x}\hat{x}+a_{y}\hat{y} \).
Accelerations in the positive \( \hat{y} \) direction appear just like an acceleration
due to gravity and so their effects can be included in the the equations of
motion simply by taking \( g\rightarrow g-a_{y} \). Accelerations in the \( \hat{x} \)
direction give the same sort of effective gravity but in the perpendicular direction
so that it is largest for \( \theta =\pi /2,3\pi /2 \) or \( cos(\theta )=\pm 1 \)
rather than \( sin(\theta )=\pm 1 \). Both possibilities can be included in
an equation of motion of the form,
\[
\ddot{\theta }-\alpha _{x}cos(\theta )-\alpha _{y}sin(\theta )=0\]
with \( \alpha _{x}=a_{x}/R \) and \( \alpha _{y}=(g-a_{y})/R \). For small
displacements about the inverted position this becomes,
\[
\ddot{x}-\alpha _{y}x=\alpha _{x}\]

\subsubsection{X Driving Motion}

The simplest kinds of driving motion to consider are harmonic displacements
of the support. For this case motion in the \( \hat{x} \) direction is the
easiest to consider first. Though it turns out not to provide the desired stable
solution, it is exactly solvable and suggests a useful technique for studying
the non-trivial case of driven motion in the \( \hat{y} \) direction.

For this system take \( a_{y}=0 \) and \( a_{x}=-\omega ^{2}a_{0}cos(\omega t) \).
The sign of this driving term is unimportant as it is equivilant to a phase
change of the argument of \( cos \) or a shift of \( t \), the negative sign
is chosen for convenience. This gives,
\[
\ddot{x}-\omega _{0}^{2}x=-\omega ^{2}a_{0}cos(\omega t)\]
with \( \omega _{0}^{2}=g/R \). One immediately apparent solution is simply
harmonic motion of the form,
\[
x=Acos(\omega t)\]
Substituting this into the equation of motion fixes the amplitude of the oscillation,
\begin{eqnarray*}
-\omega ^{2}A-\omega _{0}^{2}A & = & -\omega ^{2}a_{0}\\
A & = & \frac{a_{0}}{1+\omega _{0}^{2}/\omega ^{2}}
\end{eqnarray*}

This is a valid solution, and represents regular oscillation about the inverted
position, but it is not very general, and provides no information about its
own stability. Formally this is just a particular solution to the equation of
motion. A general solution is given by this plus a solution to the homogeneous
equation, which is seen to give an exponentially diverging solution. A more
physical, intuitive path is to consider small displacements from this solution
and study the behavior of the displacements. 

This particular solution is an oscillation about \( \theta =0 \), a simple
generalization is to add a time-dependent offset.
\[
x=Acos(\omega t)+x_{0}(t)\]
Substituting this into the equation of motion for \( x \) yields an equation
of motion for \( x_{0} \),
\[
\ddot{x}_{0}-\omega _{0}^{2}x_{0}=0\]
 which is now easily solved by exponentials, 
\[
x_{0}(t)=a_{1}e^{\omega _{0}t}+a_{2}e^{-\omega _{0}t}\]
 This provides a complete solution for any initial conditions. 

It is clear from this form that the offset diverges after large times no matter
what the initial condition and so harmonic motion about the inverted equilibrium
position is unstable. This can be understood by constructing a sort of effective
potential for the pendulum in the reference frame of the support,
\begin{eqnarray*}
\ddot{x} & = & \omega _{0}^{2}x-\omega ^{2}a_{0}cos(\omega t)\\
 & = & F/m=-\partial _{x}V/m
\end{eqnarray*}
Integrating yields \( V \),
\[
V=-m\int _{0}^{x}dx(\omega _{0}^{2}x-\omega ^{2}a_{0}cos(\omega t))=-\frac{m}{2}\omega _{0}^{2}x^{2}-m\omega ^{2}a_{0}x\cos (\omega t)\]
With the addition of an \( x \) independent, though time-dependent term, this
can be written as, 
\[
V=\frac{m\omega _{0}^{2}}{2}(x-\frac{\omega ^{2}}{\omega _{0}^{2}}a_{0}\cos (\omega t))^{2}\]
The effective potential is an inverted parabola with a harmonically varying
center position. If the offset is sufficiently displaced from the peak, the
force at all times is repulsive, it is away from the peak, and the average position
also moves further from the peak. For displacements where the offset is occasionally
on both sides of the peak, the force during that time is periodically in both
directions, but it still spends more time on one side than the other and during
those times generally experiences a larger force, so the time average force
is still away from the peak and the offset diverges. This can be summarized
formally by explicitly computing the time averaged force over one period of
the oscillation of the driving motion,
\begin{eqnarray*}
\left\langle F\right\rangle  & = & m\left\langle \ddot{x}\right\rangle \\
 & = & m\left\langle \omega _{0}^{2}x-\omega ^{2}a_{0}\cos (\omega t)\right\rangle \\
 & = & m\omega _{0}^{2}\left\langle x\right\rangle \\
 & = & m\omega _{0}^{2}\left\langle x_{0}\right\rangle 
\end{eqnarray*}
The time average force is always in the same direction as the time average offset
and so the time average offset increases.

This can easily be reduced to an equation of motion for the time average of
\( x_{0} \), \( \left\langle x\right\rangle =\left\langle A\cos (\omega t)+x_{0}\right\rangle =\left\langle x_{0}\right\rangle , \)
and then \( \left\langle \ddot{x}\right\rangle =\left\langle \ddot{x}_{0}\right\rangle  \).
This gives
\[
\left\langle \ddot{x}_{0}\right\rangle =\omega _{0}^{2}\left\langle x_{0}\right\rangle \]
When the driving term is much quick than the natural response of the system
\( \omega >>\omega _{0} \), \( x_{0} \) doesn't change much during one period
of the driving term and \( \left\langle x_{0}\right\rangle \approx x_{0} \)
directly giving an equation of motion for \( x_{0} \),
\[
\ddot{x}_{0}\approx \omega _{0}^{2}x_{0}\]
For this system this relation is exact, but it will also turn out to hold for
more general problems and provides an easy insight into the behavior of the
system by accurately describing the average position.

\subsubsection{Y Driving Motion}

Considering this time average force and an effective potential immediately shows
the case of driven motion in the \( \hat{y} \) direction to be more promising.
For the same kind of harmonic motion for the \( \hat{y} \) direction, the equation
of motion becomes,
\[
\ddot{x}-(\omega _{0}^{2}-\frac{a}{R}\omega ^{2}\cos (\omega t))x=0\]
This looks like motion for in a potential given by,
\[
V=-\frac{m}{2}(\omega _{0}^{2}-\frac{a}{R}\omega ^{2}\cos (\omega t))x^{2}\]
The potential is again parabolic, but now with a harmonically varying amplitude,
for large enough \( a \) or \( \omega  \)the potential is always attractive
for some fraction of the period for any \( x \) allowing for the possibility
that the time averaged force is attractive as well. The seems plausible qualitatively.
While the force is repulsive the pendulum is being pushed away from the equilibrium
position, so that when the force becomes attractive it is more likely further
from equilibrium where the now attractive force is stronger than when it was
closer and repulsive. Then over one period of the driving motion the net attractive
force tends to be stronger than net repulsive force and the overall average
force is attractive. This simple picture, and the general conditions under which
it is valid, emerge from a more detailed quantitative analysis of the motion.

The equation of motion happens to be a Mathieu equation. Approximate solutions
and stability regions are well known but the analysis is traditionally less
than transparent or intuitive. Instead, again consider a solution like that
for the \( \hat{x} \) driving motion, a quick oscillation about a slowly varying
offset from the origin \( x_{0} \),
\[
x=A\cos (\omega t)+x_{0}\]
The equation of motion for \( x_{0} \) is simplified if \( A \) is allowed
to vary in this case. Take \( A \) to be time dependent, but slowly varying
so that time derivatives can be neglected, \( \dot{A}/A<<\omega  \). Applying
this form to the equation of motion gives,
\[
(-A\omega ^{2}\cos (\omega t)+\ddot{x}_{0})-(\omega _{0}^{2}-\frac{a}{R}\omega ^{2}\cos (\omega t))(A\cos (\omega t)+x_{0})=0\]
collecting terms linear in \( \cos (\omega t) \), 
\[
(-A\omega ^{2}-A\omega _{0}^{2}+\frac{a}{R}\omega ^{2}x_{0})\cos (\omega t)+\ddot{x}_{0}-(\omega _{0}^{2}x_{0}-\frac{a}{R}\omega ^{2}A\cos ^{2}(\omega t))=0\]
and fixing the amplitude to eliminate this \( \cos  \) term,
\[
A=\frac{a}{R}\frac{x_{0}}{1+\omega _{0}^{2}/\omega ^{2}}\]
leaves an equation of motion for \( x_{0} \)
\[
\ddot{x}_{0}-(\omega _{0}^{2}-\left( \frac{a}{R}\right) ^{2}\frac{\cos ^{2}(\omega t)}{\omega ^{2}+\omega _{0}^{2}})x_{0}=0\]
time averaging, and as usual taking \( \omega _{0}<<\omega  \) so that \( \left\langle \cos ^{2}(\omega t)x_{0}\right\rangle \approx \left\langle \cos ^{2}(\omega t)\right\rangle \left\langle x_{0}\right\rangle =\left\langle x_{0}\right\rangle /2 \)
and \( \left\langle x_{0}\right\rangle \approx x_{0} \) gives,
\[
\ddot{x}_{0}+(\frac{a^{2}}{2R^{2}}\frac{1}{\omega ^{2}+\omega _{0}^{2}}-\omega _{0}^{2})x_{0}=0\]
For 
\[
\frac{a^{2}}{2R^{2}}\frac{1}{\omega ^{2}+\omega _{0}^{2}}>\omega _{0}^{2}\]
the coefficient of \( x_{0} \) becomes positive, rather than negative, and
the solution for \( x_{0} \) changes from diverging exponentials to bound harmonic
oscillation. This appears in the time averaged force as well,
\begin{eqnarray*}
\left\langle F\right\rangle  & = & m\left\langle \ddot{x}\right\rangle \\
 & = & m\left\langle (\omega _{0}^{2}-\frac{a}{R}\omega ^{2}\cos (\omega t))(A\cos (\omega t)+x_{0})\right\rangle \\
 & = & m\left\langle \omega _{0}^{2}x_{0}-\frac{a}{R}\omega ^{2}\cos ^{2}(\omega t)x_{0}-\frac{aA}{2R}\omega ^{2}\right\rangle \\
 & = & m\left( \omega _{0}^{2}\left\langle x_{0}\right\rangle -\frac{a\omega ^{2}}{R}\left\langle \cos ^{2}(\omega t)x_{0}\right\rangle -\frac{a\omega ^{2}}{2R}\frac{\left\langle x_{0}\right\rangle }{1+\omega _{0}^{2}/\omega ^{2}}\right) \\
 & \approx  & m\left( \omega _{0}^{2}\left\langle x_{0}\right\rangle -\frac{a\omega ^{2}}{R}\left\langle \cos ^{2}(\omega t)\right\rangle \left\langle x_{0}\right\rangle -\frac{a\omega ^{2}}{2R}\frac{\left\langle x_{0}\right\rangle }{1+\omega _{0}^{2}/\omega ^{2}}\right) \\
 & = & -m\left( \frac{a}{2R}\frac{1}{1+\omega _{0}^{2}/\omega ^{2}}-\omega _{0}^{2}\right) \left\langle x_{0}\right\rangle \\
 & \equiv  & -m\omega _{s}^{2}\left\langle x_{0}\right\rangle 
\end{eqnarray*}
with the same requirements as above on \( \omega _{0} \) the coefficient of
\( \left\langle x_{0}\right\rangle  \) becomes negative and implies a stable
restoring force. 

By driving the support vertically with a suitable amplitude and frequency the
initially unstable motion of an exponential diverging with time constant \( \omega _{0} \)
becomes the stable combined motion of an quickly oscillating micromotion at
frequency \( \omega  \) about a slowly oscillating secular motion at a frequency
given by \( \omega _{s} \). The same mechanism can now be applied to the problem
of stably confining a charged particle by changing a partly unstable motion
in a static quadrupole electric field to a effectively stable motion in an oscillating
field.

\subsection{The Pseudopotential}

An explicit analysis, as done with the driven inverted pendulum system, is a
bit cumbersome and in the end slightly redundant. It is also not very general
as the results don't obviously apply to an effective potential that is not quadratic
or a system in more than one dimension. A generalized problem can be treated
more formally to uncover the fundamental behavior of the resulting motion and
the essential requirements for it to be stable and easily understood.

Consider now an oscillating force that varies arbitrarily with position,
\[
\vec{F}(\vec{x},t)=\cos (\omega t)\vec{F}(\vec{x})\]
the equation of motion will be,
\[
m\ddot{\vec{x}}(t)=\vec{F}(\vec{x}(t))=\cos (\omega t)\vec{F}(\vec{x}(t))\]
For constant \( \vec{F} \), solutions are simple harmonic oscillations given
by,
\[
\vec{x}(t)=\vec{A}\cos (\omega t)+\vec{x}_{0}\]
where \( \vec{x}_{0} \) and \( \vec{A} \) are fixed with \( \vec{x}_{0} \)
arbitrary and \( \vec{A} \) given by the equation of motion as,
\[
-m\omega ^{2}\vec{A}=\vec{F}\]

For \( \vec{F} \) arbitrary but slowly varying over the range of motion of
a single oscillation, \( \vec{A}<<\left| \vec{F}^{\prime }\right| /\left| \vec{F}\right|  \),
the short time motion will be the same, with \( \vec{A} \) now also position
dependent as \( \vec{A}(\vec{x}_{0})=-m\omega ^{2}\vec{F}(\vec{x}_{0}) \),
and the long time motion accounted for with a time dependent \( \vec{x}_{0} \),
giving, with all time dependences explicit,
\[
\vec{x}(t)=\vec{A}(\vec{x}_{0}(t))\cos (\omega t)+\vec{x}_{0}(t)\]

Consider again the time averaged force,

\[
\left\langle \vec{F}\right\rangle =m\left\langle \ddot{\vec{x}}\right\rangle =m\left\langle \partial _{t}^{2}\left( \vec{A}(\vec{x}_{0})\cos (\omega t)+\vec{x}_{0}\right) \right\rangle \]
Let \( 1/\omega _{s} \) be the time scale in which \( \vec{x}_{0} \) varies
significantly. Here \( \vec{A} \) depends explicitly only on \( \vec{x}_{0} \)
so it also varies only on time-scales longer than \( 1/\omega _{s} \) and for
\( \omega _{s}<<\omega  \) its time derivatives can be neglected in comparison
to \( \omega  \), \( \dot{A}/A\sim \omega _{s}<<\omega  \) giving,
\[
\ddot{\vec{x}}\approx -\omega ^{2}\vec{A}(\vec{x}_{0})\cos (\omega t)+\ddot{\vec{x}}_{0}\]
Similarly for \( \omega _{s}<<\omega  \), the time average of any product of
a fast \( \omega  \) piece times a slow \( \omega _{s} \) piece approximately
factor, and the time average of a slow piece is approximately its instantaneous
value at any point during a time included in the average. The time average of
this time derivative then gives,
\begin{eqnarray*}
\left\langle \ddot{\vec{x}}\right\rangle  & \approx  & \left\langle -\omega ^{2}\vec{A}(\vec{x}_{0})\cos (\omega t)+\ddot{\vec{x}}_{0}\right\rangle \\
 & = & -\omega ^{2}\left\langle \vec{A}(\vec{x}_{0})\cos (\omega t)\right\rangle +\left\langle \ddot{\vec{x}}_{0}\right\rangle \\
 & \approx  & -\omega ^{2}\left\langle \vec{A}(\vec{x}_{0})\right\rangle \left\langle \cos (\omega t)\right\rangle +\ddot{\vec{x}}_{0}\\
 & = & \ddot{\vec{x}}_{0}
\end{eqnarray*}
So the time averaged force approximately determines the motion of \( \vec{x}_{0} \),
\[
\left\langle \vec{F}\right\rangle \approx m\ddot{\vec{x}}_{0}\]
 For \( \vec{F} \) slowly varying, it be expanded about \( \vec{x}_{0} \),
\begin{eqnarray*}
\vec{F}(\vec{x}(t)) & = & \vec{F}(\vec{A}\cos (\omega t)+\vec{x}_{0}(t))\\
 & \approx  & \vec{F}(\vec{x}_{0}(t))+\cos (\omega t)\vec{A}(\vec{x}_{0}(t))\cdot \vec{\nabla }\vec{F}(\vec{x}_{0}(t))
\end{eqnarray*}

Now, consider again the time averaged force,
\[
\left\langle \vec{F}\right\rangle =\left\langle \cos (\omega t)\vec{F}(\vec{x}(t))\right\rangle \approx \left\langle \cos (\omega t)\vec{F}(\vec{x}_{0}(t))\right\rangle +\left\langle \cos ^{2}(\omega t)\vec{A}(\vec{x}_{0}(t))\cdot \vec{\nabla }\vec{F}(\vec{x}_{0}(t))\right\rangle \]
 In particular since \( \vec{A} \) and \( \vec{F} \) here depend explicitly
only on \( \vec{x}_{0} \), 
\begin{eqnarray*}
\left\langle \vec{F}\right\rangle  & \approx  & \left\langle \cos (\omega t)\vec{F}(\vec{x}_{0}(t))\right\rangle +\left\langle \cos ^{2}(\omega t)\vec{A}(\vec{x}_{0}(t))\cdot \vec{\nabla }\vec{F}(\vec{x}_{0}(t))\right\rangle \\
 & \approx  & \left\langle \cos (\omega t)\right\rangle \left\langle \vec{F}(\vec{x}_{0}(t))\right\rangle +\left\langle \cos ^{2}(\omega t)\right\rangle \left\langle \vec{A}(\vec{x}_{0}(t))\cdot \vec{\nabla }\vec{F}(\vec{x}_{0}(t))\right\rangle \\
 & \approx  & (1/2)\vec{A}(\vec{x}_{0}(t))\cdot \vec{\nabla }\vec{F}(\vec{x}_{0}(t))
\end{eqnarray*}
Substituting the explicit expression for \( \vec{A} \) and suppressing the
explicit position dependence,
\[
\left\langle \vec{F}\right\rangle \approx -\frac{1}{2m\omega ^{2}}\vec{F}\cdot (\vec{\nabla }\vec{F})=-\frac{1}{4m\omega ^{2}}\vec{\nabla }(\vec{F}\cdot \vec{F})\]

The time averaged force is given simply by the gradient of \( U=\left| \vec{F}\right| ^{2}/4m\omega ^{2} \).
Since this time averaged force approximately determined the motion of \( \vec{x}_{0} \),
\( U \) acts as kind of effective potential for \( \vec{x}_{0} \),
\[
m\ddot{\vec{x}}_{0}\approx -\vec{\nabla }U\]
This \( U \) is the pseudo-potential and, as is apparent from the form, it
results in \( \vec{x}_{0} \) being generally driven to regions of small \( \left| \vec{F}\right|  \). 

Again, this picture is valid for \( \omega _{s}<<\omega  \), and \( \vec{A}<<\vec{\nabla }\vec{F}/\left| \vec{F}\right|  \),
where \( \omega _{s}=\dot{x}_{0}/x_{0} \). \( 1/\omega _{s} \) gives the time
\( \tau _{s} \), it takes a particle to travel a significant distance for a
static field with a size equal to the amplitude of the oscillating field, while
\( 1/\omega  \) gives the time, \( \tau  \), it takes for the oscillating
field to change direction. This constraint then just corresponds to, \( \tau _{s}>>\tau  \),
effectively requiring that the field change direction before the ion moves very
far. If the driving frequency is very slow the ion simple leaves the trap before
the electric field can turn around to being it back. But for \( \omega _{s}\sim \omega  \)
the equation of motion would require higher order spatial derivatives and this
constraint in the frequencies, as well as the constraint on the amplitude of
the micromotion just allows for the simple first order form.

For amusement note that \( U \) can in turn be written in terms of the average
kinetic energy of the particle. In terms of the amplitude of the motion \( \vec{F}=-m\omega ^{2}\vec{A} \),
\[
U=\frac{m\omega ^{2}}{4}\left| \vec{A}\right| ^{2}\]
\( \vec{A} \) also approximately gives the velocity of the particle,
\[
\vec{v}=\dot{\vec{x}}\approx -\omega \vec{A}\cos (\omega t)\]
The time averaged kinetic energy of the particle can then be written as
\begin{eqnarray*}
\left\langle T\right\rangle  & = & \frac{m}{2}\left\langle \left| \vec{v}\right| ^{2}\right\rangle \\
 & \approx  & \frac{m\omega ^{2}}{2}\left\langle \left| \vec{A}\cos (\omega t)\right| ^{2}\right\rangle \\
 & \approx  & \frac{m\omega ^{2}}{2}\left\langle \left| \vec{A}\right| ^{2}\right\rangle \left\langle \cos ^{2}(\omega t)\right\rangle \\
 & = & \frac{m\omega ^{2}}{4}\left| \vec{A}\right| ^{2}\\
 & = & U
\end{eqnarray*}
The pseudopotential is simply the time averaged kinetic energy of the particle
and the motion of the offset of the micromotion is to minimize this kinetic
energy,
\[
\left\langle \vec{F}\right\rangle =-\vec{\nabla }\left\langle T\right\rangle \]

\subsection{Quadrupole Fields and Secular Motion}

This can finally be applied to the problem of confining a charged particle with
only an electric field. As before, a static electric field in the region of
a stationary point will take the form,
\[
\vec{E}=\alpha x\hat{x}+\beta y\hat{y}+\gamma z\hat{z}\]
where the coefficients must satisfy \( \alpha +\beta +\gamma =0 \). There will
typically an azimuthal symmetry so that \( \alpha =\beta  \) giving \( \gamma =-2\alpha  \).
Take \( \alpha =2E_{0}/r_{0}=2V/r_{0}^{2} \). The field becomes,
\[
\vec{E}=\frac{2V}{r^{2}_{0}}\left( x\hat{x}+y\hat{y}-2z\hat{z}\right) \]

This field can be given by a potential,
\[
\Phi =\frac{V}{r_{0}^{2}}(r^{2}-2z^{2})\]
The equipotentials are hyperbolas so that generating this field precisely over
some region of space required electrodes that are hyperboloids of revolution,
the surface is given by a hyperbola in the \( r-z \) plane rotated about the
\( z \) axis. In practice any stationary point in a potential has this quadrupole
form and so any stationary point can be used for a trap. The only modification
is a reduction in the size if the resulting electric field compared to a precisely
shaped hyperbolic electrode of the same characteristic size. The potential does
not change as quickly as a function of position for these traps. This could
be accomadated by changing the size of \( r_{0} \) suitably, but it is convenient
to keep this as a measure of the physical size of the trap and introduce \( \alpha <1 \)
to characterize the reduction. The trap in this experiment is made using a circular
gap between two twisted wires, \cite{KristiThesis}. The wire is about \( 100\mu m \)
in diameter and the hole about \( 1/2mm \). This yields reduction factors on
the order of \( \alpha \sim 1/10-1/20 \).

With this reduction factor, making the potential harmonically time dependent
yields a force,
\[
\vec{F}=\cos (\omega t)\frac{2q\alpha V}{r_{0}^{2}}\left( x\hat{x}+y\hat{y}-2z\hat{z}\right) \]
For \( \omega _{s}<<\omega  \) and \( \vec{A}<<\vec{\nabla }\vec{F}/\left| \vec{F}\right|  \),
where \( \omega _{s} \) and \( \vec{A} \) will be identified shortly, this
gives a pseudopotential
\begin{eqnarray*}
U & = & \frac{1}{4m\omega ^{2}}\left| \vec{F}\right| ^{2}=\frac{q^{2}\alpha ^{2}V^{2}}{m\omega ^{2}r_{0}^{4}}(r^{2}+4z^{2})\\
 & \equiv  & \frac{m\omega _{r}^{2}}{2}r^{2}+\frac{m\omega _{z}^{2}}{2}z^{2}
\end{eqnarray*}
The pseudopotential is that of a three-dimensional harmonic oscillator with
frequencies given by
\begin{eqnarray*}
\omega _{r} & = & \sqrt{2}\frac{q\alpha V}{m\omega r_{0}^{2}}\\
\omega _{z} & = & 2\sqrt{2}\frac{q\alpha V}{m\omega r_{0}^{2}}=2\omega _{r}
\end{eqnarray*}
Validity of this pseudopotential approximation then becomes
\[
\frac{\omega _{s}}{\omega }\sim \frac{q\alpha V}{m\omega ^{2}r_{0}^{2}}<<1\longrightarrow \omega ^{2}>>\frac{q\alpha V}{mr_{0}^{2}}\]
For this case, the constraint on the amplitude turns out to be equivilant,
\[
\frac{\vec{A}}{\vec{\nabla }\vec{F}/\left| \vec{F}\right| }=\frac{\vec{F}/m\omega ^{2}}{\vec{\nabla }\vec{F}/\left| \vec{F}\right| }\sim \frac{qV\left| \vec{x}\right| /m\omega ^{2}r_{0}^{2}}{\left| \vec{x}\right| }=\frac{q\alpha V}{m\omega ^{2}r_{0}^{2}}<<1\]

\subsection{Well depth and confinement}

This latter constraint corresponds to requiring that the field change signs
faster than the ion can instantaneously react to it so that the ion is not lost
from the trap during the period. On the other hand if the field oscillates so
quickly that the ion hardly moves during one period it will only effectively
respond to the time average of the field, which is zero. This is also seen in
the dependence of the secular frequencies on \( \omega  \). The secular frequencies
give the effective depth of the well. A well depth can be defined by the value
of the pseudopotential at some reference point, a convenient point is \( z=0 \)
, \( r=r_{0} \) . At this point, 
\[
U=\frac{1}{2}m\omega _{r}^{2}r_{0}^{2}=m\left( \frac{q\alpha V}{m\omega r_{0}}\right) ^{2}\]
The well depth gets weaker with increasing \( \omega  \). 

Confining the ion then requires a well depth sufficiently deep that its kinetic
energy does not allow it to escape from the trap. An ion with a kinetic energy
less than this well depth is then confined to a region of radius \( r \) given
by,
\begin{eqnarray*}
\frac{K}{U} & = & \frac{r^{2}}{r_{0}^{2}}\\
r & = & r_{0}\sqrt{\frac{K}{U}}=\sqrt{\frac{2K}{m\omega _{r}^{2}}}=\frac{\omega r_{0}^{2}}{q\alpha V}\sqrt{mK}
\end{eqnarray*}
For fixed \( U \), confinement is tighter for smaller traps. For well depths
of a few \( eV \) , ions with kinetic energies corresponding to \( mK\sim 10^{-7}eV \)
, will have confinement radii on the order of \( 10^{-3}r_{0} \).

The well depth is a function of the trap size and for fixes secular frequency
confinement is independent of \( r_{0} \). Even further, \( \omega _{r} \)
is generally constrained by stability requirements so that 
\[
\frac{\omega _{s}}{\omega }\equiv \gamma =\frac{q\alpha V}{m\omega ^{2}r_{0}^{2}}=\frac{U}{q\alpha V}<1\]
 If \( \omega  \) only is decreased \( \omega _{r} \) also increases and quickly
moves the trap away from this stability region. Treating instead \( \gamma  \)
as the independent parameter gives,

\[
U=\gamma q\alpha V\]
so that for fixed \( \gamma  \) the voltage must increase to a larger well
depth, which in turn requires either increased \( \omega  \) or \( r_{0} \)
as seen from the expression for \( \gamma  \). The radius of confinement however
becomes,
\[
r=r_{0}\sqrt{\frac{K}{\gamma q\alpha V}}\]
Favoring smaller traps for tighter confinement for a given voltage and secular
to RF frequency ratio.

\subsection{Micromotion}

A deeper will also reduces the amplitude of the micromotion. This is not immediately
obvious. The amplitude of the micro-motion depends on the size of the electric
field,
\[
m\omega ^{2}A=F=eE\]
Certainly a deeper well confined the ion closer to the center of the trap where
the electric field is weaker, but for a tighter trap that field is strong everywhere
than for the weak trap. Writing the form of the electric field,
\[
E=\frac{2\alpha V}{r_{0}^{2}}r\]
gives an amplitude

\[
A=\frac{2q\alpha V}{m\omega ^{2}r_{0}^{2}}r=2\gamma r\]
The amplitude of the micromotion just depends on the displacement and the ratio
of secular to RF frequencies. A tighter trap will certainly reduce \( r \)
for a particle of the same energy, or a particle displaced from the center by
the same force, so for traps with the same \( \gamma  \) the deeper trap has
a reduced micromotion. Note that this arrangement is natural when simply changing
the voltage to the trap and no other parameters.

For thermal kinetic motion this amplitude is given in terms of the confinement
radius,
\[
A=2\gamma r_{0}\sqrt{\frac{K}{\gamma q\alpha V}}\]
An ion may also be displace from the trap by an external force \( F_{e} \).
The pseudopotential gives an effective position dependent force 
\[
F_{pp}=m\omega _{r}^{2}r=\frac{2U}{r_{0}^{2}}r\]
giving,
\[
A=\gamma \frac{F_{pp}r_{0}^{2}}{U}\]
also favoring small traps and large well depths.

\subsection{Design parameters}

An ideal trap has a small confinement region and small micromotion. This is
generally achieved by small traps and large well depths. This can be arranged
in many way using the design parameters involved, \( \omega  \), \( r_{0} \)
and \( V \) . Generally there are practical constants. An immediate example
is the applied voltage, RF power, trap capacitance and feedthrough limit give
a maximum practical voltage of something less than a thousand volts. With this
the possible well depth is immediately available using \( U=\gamma \alpha qV \).
With \( \alpha \sim 1/10 \), \( \gamma \sim 1/10 \). This gives a trap depth
of a few \( eV \). 

Small \( r_{0} \) is desirable, but stability is also required with,
\[
\gamma =\frac{q\alpha V}{m\omega ^{2}r_{0}^{2}}<1\]
Smaller \( r_{0} \) requires higher RF frequencies to maintain stability. A
reasonable range for \( \omega  \) is less than a few dozen \( MHz \) as these
frequencies are easier to work with and produce at high powers than the microwaves
at \( GHz \) frequencies. Taking \( \omega \sim 10MHz \) and using \( m=138GeV/c^{2} \),
\( V=100V/cm \) gives \( r_{0}>\sim250 \mu  \). 

The trap used in this experiment has a diameter of \( 400\mu m \) with \( \alpha \sim 1/20 \),
and is driven at at a frequency of around \( 30MHz \) with voltages around
300-400\( V \). This gives a secular to RF frequency ratio of \( \gamma \sim 0.25 \),
which implies a well depth of \( \sim 10eV \). This gives a confinement radius
of around \( r\sim 20nm \) and a micromotion amplitude \( A\sim 10nm \).

\section{Cooling}

The ion trap provides an effective harmonic potential well. This will confine
the ion in a small, well defined region, but not yet fix it to be motionless
at a specific point. The ion's motion consists of the micromotion along the
quadrupole electric field lines at the RF driving frequency, and the secular
motion of its single RF period time averaged position at the secular frequency.
The micromotion is driven, and can't be eliminated, though it does become zero
at the center of the trap, but the amplitude of the secular motion depends on
the energy is arbitrary. It depends on the energy of the ion and should ideally
be as small as possible.

Further progress toward this ideal is made using the now familiar techniques
of doppler cooling. A laser is tuned to slightly below a strong resonance. When
the ion is moving toward the laser, the laser is shifted into resonance in the
ion's frame of reference and so more strongly absorbed. The re-radiated photon
is emitted isotropically so that the average net change in momentum is opposite
to the ions motion and the ion is slowed. When the ion moves away from the laser,
the laser is doppler shifted further out of resonance and even less strongly
absorbed and its motion is unaffected. 

In neutral atom traps six lasers are required to cool all degrees of freedom
of the motion, a pair of counter-propagating beams along each axis. An ion in
a ion trap when not at rest in the center of the trap, which is when it needs
to be cooled, is moving harmonically, with its velocity periodically in both
directions along all three axis. A single laser can then be used to cool the
ion if it directed in such a way that the motion of the ion always, or at least
frequently, has some component parallel to the beam. This is easily arranged
by, for example, orienting the beam at equal angles to all three principal axis
of the trap, and results in cooling during some part of every period of the
motion. This happens to make a perfectly cylindrically symmetric trap unsuitable
since the motion in the \( x \) and \( y \) directions are degenerate, the
principle axis are not well defined so that motion in the \( x-y \) plane parallel
to the component of the cooling beam in the \( x-y \) plane is quickly damped
out, while the perpendicular motion is unaffected. The traps used in practice
generally have well defined axis.
\begin{figure}
{\par\centering \includegraphics{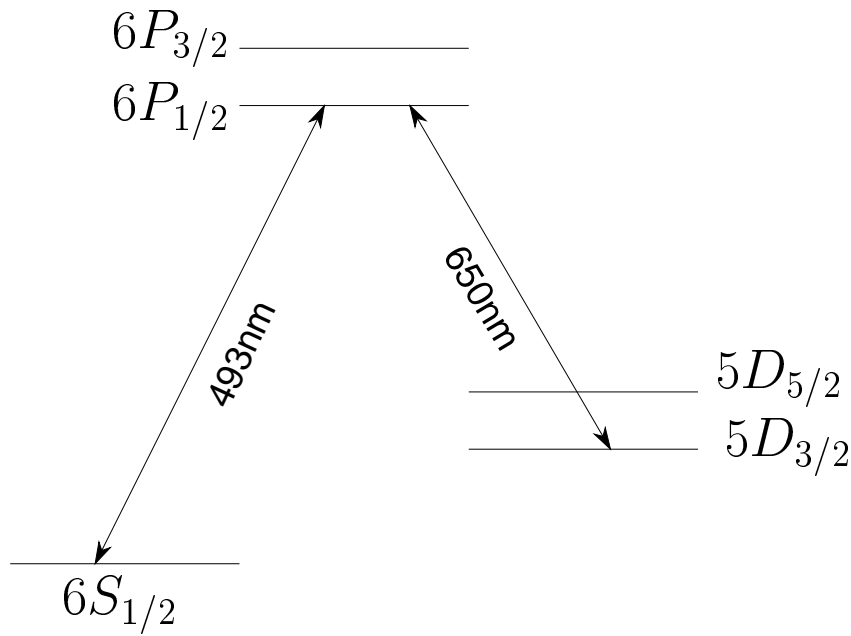} \par}

\caption{\label{fig:CoolingTransitionEnergyLevelDiagram}Cooling and Cleanup transitions
used in trapping Ba\protect\( ^{+}\protect \).}
\end{figure}

For Barium, the \( 6S_{1/2}\rightarrow 6P_{1/2} \) dipole transition at 493nm
is used for cooling, fig.\ref{fig:CoolingTransitionEnergyLevelDiagram}. This
light is provided by a frequency doubled 986nm diode laser. The system typically
yields about a mW of blue light with a linewidth of about 5\( MHz \). The laser
is coarsely tuned to about 100\( Mhz \) to the red of exact resonance for cooling
using the opto-galvonic resonance of a Barium Ion hollow cathode discharge tube
having a linewidth of about a \( GHz \). 

As practical matter, the \( 6P_{1/2} \) state can decay to the \( 5D_{3/2} \)
state as well as the ground state. The \( S:D \) branching ratio is around
2.5. The \( 5D_{3/2} \) state can only decay to the ground state through a
quadrupole transition so is very long lived, about 80-90s. An ion in this state
can not absorb any of the cooling light and so will not be further cooled until
it decays. This makes cooling very inefficient so a second laser is used, tuned
to on resonance with the \( 5D_{3/2}\rightarrow 6P_{1/2} \) transition at 650nm
also using a Barium hollow cathode lamp as a reference. This light is provided
by an external cavity diode laser which also yields about a mW. This is used
as a cleanup beam to kick the ion out of this \( D \) state should it decay
there, and return it to the cooling cycle. Strictly this clean up beam does
no cooling, though it can be made to, but both cooling and cleanup beams are
frequently collectively referred to as the cooling beams.

Both lasers are aimed through the trap and focussed so that they have spot sizes
of about \( 100\mu  \) at the ion. Here just a few 10s of \( \mu W \) are
required to saturate both transitions at a few \( MHz \), as the lifetime of
the \( P_{1/2} \) state is a fraction of a \( \mu s \), giving the maximum
possible cooling rate. This typically yields a final kinetic energy corresponding
to a temperature of a few \( mK \). For traps a few hundred \( \mu  \) in
diameter this confines the ion to a few tens of \( nm \). This is only slightly
higher than the minimum possible temperature corresponding to the ground state
energy of the harmonic oscillator, \( h\omega _{s}/2 \). With \( \omega _{s}\approx 2\pi 10MHz \)
this gives \( T\sim 1/4mK \).

\section{Detection}

The blue photons scattered during cooling also provide a means of detecting
the ion. A PMT intercepts about 1/100th of a solid angle and with its \( \sim 10\% \)
detection efficiency yields a signal of a few 1000cps. With the laser sufficiently
tightly focussed, the PMT correctly positioned and focussed, and the trap suitably
oriented, the background to the PMT from stray scattered light is only a few
dozen counts per second, so the ion signal is easily distinguished from this
background. Fig.\ref{fig:ShelvingData} shows the typical difference in the
PMT signal between a floresing ion and background. 

An narrow-band interference filter, with a width of about 5\( nm \), is used
in front of the PMT to select only the scattered blue light. This helps reduce
the background and ease beam size and quality constraints for the red laser
since in this case stray scattered red will not contribute to the background.
It also provides a means of background subtraction. With the cleanup beam off,
the blue floresence will stop even though the cooling beam is still being applied
since the ion is quickly pumped to the \( 5D_{3/2} \) state where it can no
longer absorb the cooling light. The PMT signal will drop to the background
rate and as with it sensitive to only the blue light, this drop can be attributed
entirely to the loss of floresence and not to the the absence of stray scattered
red light in the background.

\section{Shelving}

\begin{figure}
{\par\centering \resizebox*{1\textwidth}{!}{\includegraphics{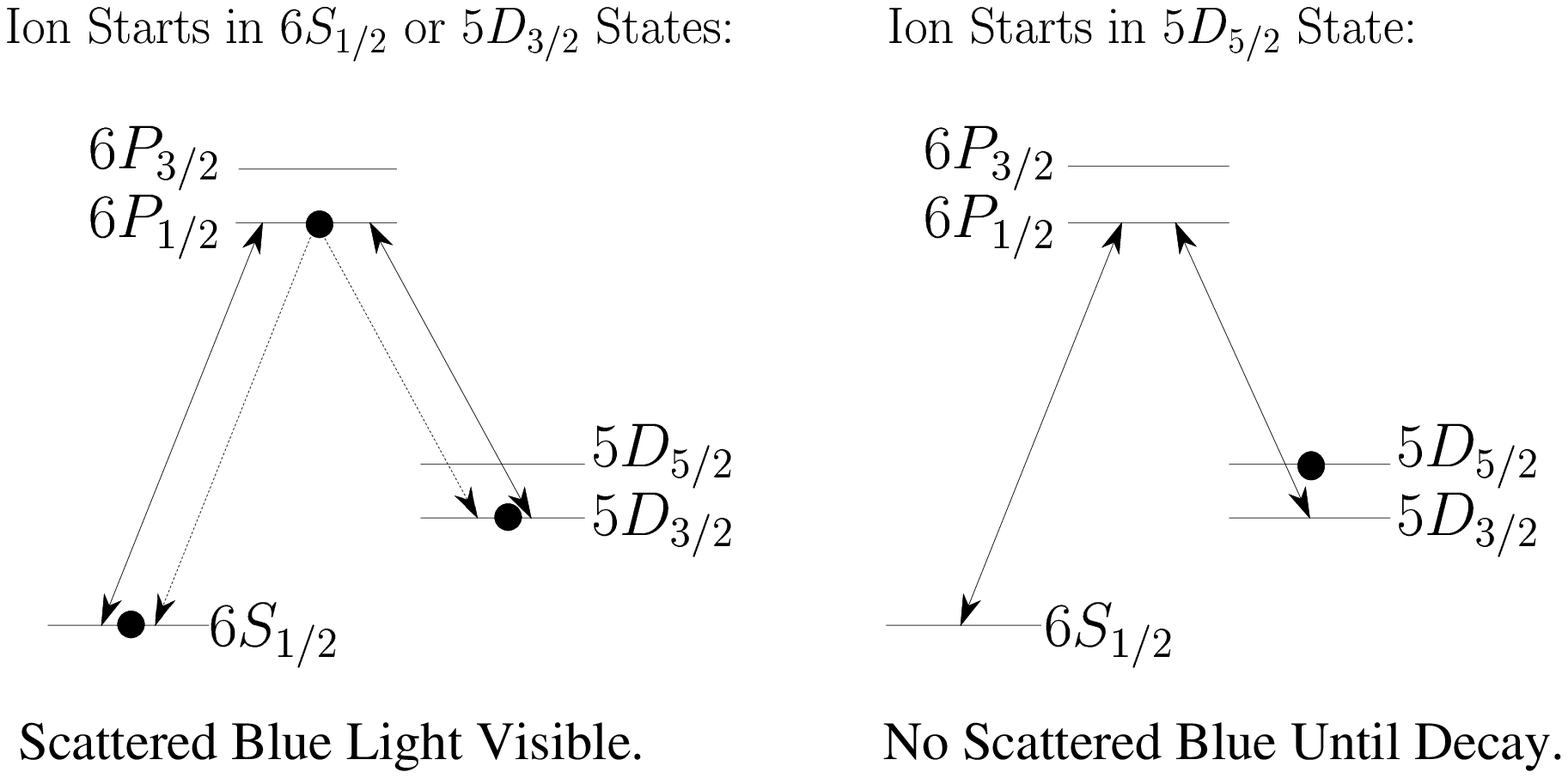}} \par}

\caption{\label{Fig:ShelvingAndFloresence}Detection of shelved ion. }
\end{figure}

While cooling, a decay to the \( 5D_{5/2} \) state from the \( 6P_{1/2} \)
is also possible energetically, and the state is similarly long-lived, 40-50s,
but is too large an angular momentum change for a dipole transition and the
wrong parity for a quadrupole transition so any \( 6P_{1/2}\rightarrow 5D_{5/2} \)
decay would be very slow, and negligible in practice so no additional cleanup
beam is needed for this state. However, driving the ion into this state turns
out to  provide means to a number of valuable results. In this state the ion
will not absorb, and so not scatter, light from the cooling or cleanup beam,
fig\ref{Fig:ShelvingAndFloresence}.

 The ion is transparent to the cooling light when in this state. This makes
it easy to reliably detect when the ion has made this transition, and provides
a means of effectively turning off the ion. This is shelving and the \( 5D_{5/2} \)
state is the shelved state.

\begin{figure}
{\par\centering \includegraphics{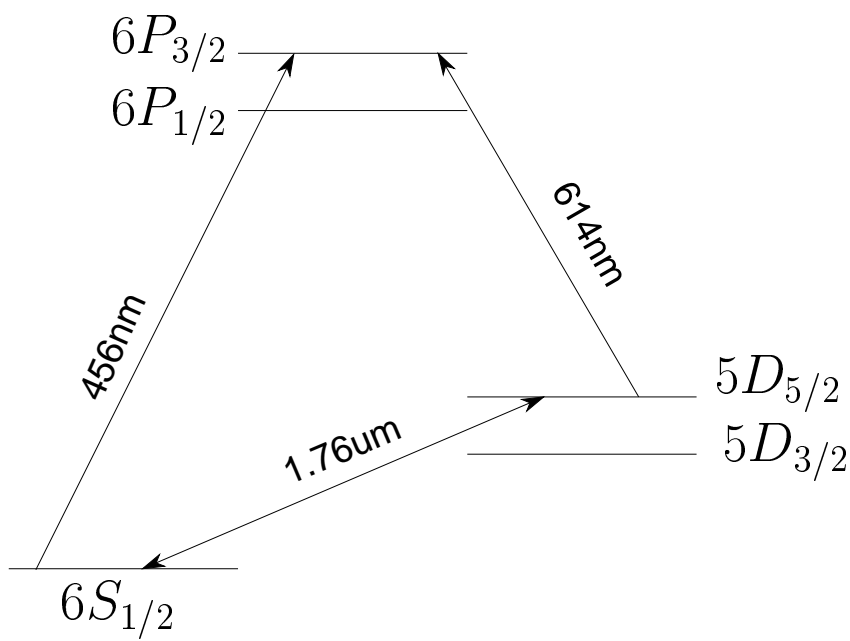} \par}

\caption{\label{fig:ShelvingTransitionsLevelDiagram}Shelving and deshelving transitions.}
\end{figure}

The transition can be made from the ground state directly with a 1.76\( \mu  \)
laser. This is the most efficient and flexible in the end, but requires a carefully
tuned laser with modest power. In this project, the shelving transition is made
indirectly through the \( P_{3/2} \) state, fig.\ref{fig:ShelvingTransitionsLevelDiagram}. 
A Barium ion discharge lamp, identical to the lamps used to tune both cooling
beams, provides an incoherent source of light for all the transitions in Ba\( ^{+} \).
The light corresponding to the \( 6S_{1/2}\rightarrow 6P_{3/2} \) transition
is selected with an interference filter and focussed onto the ion, driving the
transition at a rate of a few per second. The \( P \) state is very short-lived
and from it, the ion effectively immediately decays back the the \( S \) state
70-80\% of the time or to either of the two \( 5D \) states otherwise. When
decaying to the \( D \) states the \( 6P_{3/2}\rightarrow 5D_{5/2} \) transition
to the shelved state is preferred 10:1 to the transition to the \( 5D_{3/2} \).
In the latter case, if the cooling cleanup beam is being applied simultaneously
the ion will quickly appear back in the ground state from which another shelving
attempt can be made, otherwise shelving fails giving a maximum shelving probability
of about 90\%. 

Once in the shelved state the ion must will decay back to the ground state radiatively
after some 40-50s. Often the only information needed is whether the ion is shelved
and one that has been determined the ion is no longer needed in the shelved
state. Waiting for a decay out of the shelved state to resume measurements will
be tediously slow in this case so it is also convenient to be able to deshelved
the ion. Once again this can be done directly with a \( 1.76\mu m \) laser
or indirectly through the \( 6P_{3/2} \) state. The latter procedure is used
in this system. In this case the \( 614nm \) light from the same discharge
tube is selected with an interference filter do excite the ion to the \( 6P_{3/2} \)
state from the \( 5D_{5/2} \) state where it can the decay back to the ground
state.

\begin{figure}
{\par\centering \resizebox*{1\textwidth}{!}{\includegraphics{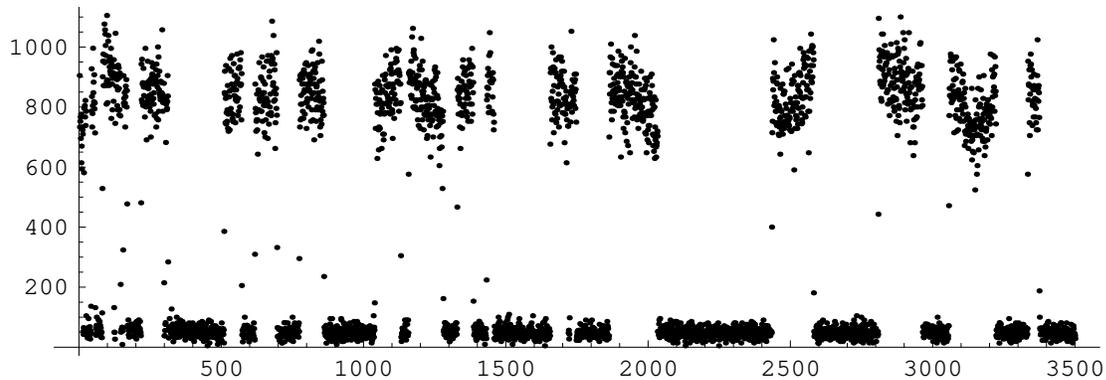}} \par}

\caption{\label{fig:ShelvingData}PMT signal log showing shelving transitions.}
\end{figure}

Again, once in the shelved state absorption of the blue laser is no longer possible
and so if the cooling beams are being applied the transition to the shelved
state can be detected with almost 100\% efficiency. Fig.\ref{fig:ShelvingData}
shows the PMT signal while the ion is periodically shelved and deshelved. 

\section{Loading}

With a stable, deep trap and suitably tuned cooling and cleanup beams trapping
is now possible. An ability to shelve the ion is not strictly necessary, but
turns out to be very useful while attempting to load a single ion. An oven loaded
with pure barium is heated to a dull glow and a small opening in the oven provides
a source for a thermal beam of neutral barium atoms. This beam is aimed in the
general direction of the trap. In this system the hot oven is also used as an
electron source, though typically an electron beam is provided independently
by a separately heated and biased filament. The relatively very light electrons
are rapidly accelerated and steered by the large RF electric field generating
the trapping potential, an electron initially at rest would move many hundreds
of cm during one half cycle of the RF. Then careful alignment is not generally
required, any electrons are accelerated towards the trap during the attractive
phase of the RF electric field. The electrons ionize the Barium atoms to provide
Barium ions to trap.

\subsection{Incident Ion Trajectories}

It is not immediately clear if ions must be created inside the trapping region
to be trapped since ions created outside the trap are repelled by the trap electrode.
Far from the center of the trap the electric field is basically that of a line
charge \( E=\alpha /r \). The pseudo-potential directs the ions towards weaker
electric fields, so outside the trap, the net average force is further from
the trap. The approximate electric field from the wire of the trap electrode
can be used to estimate the size of this repulsive potential relative to the
trap's well depth. The electric potential in this case is \( V(r)=Vln(r/r_{0})+V_{0} \)
giving \( E=V/r \), where \( r_{0} \) is the size of the trap. \( V \) and
\( V_{0} \) are determined by taking the voltage at the trap electrode, \( r\approx r_{0} \)
to be the applied RF voltage, \( V(r_{0})=V_{0}\equiv V_{RF} \), and the potential
at some typical reference point,\( r_{ref} \), a few centimeters away where
the oven is located to be ground, \( V(r_{ref})=0=Vln(r_{ref}/r_{0})+V_{RF} \)
giving \( V=-V_{RF}/ln(r_{ref}/r_{0}) \). This gives \( E=V_{RF}/ln(r_{ref}/r_{0})/r \). 

\begin{figure}
{\par\centering \includegraphics{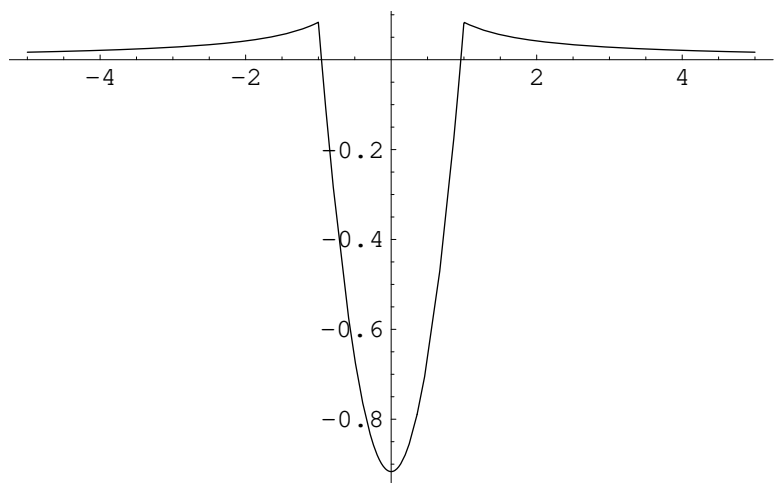} \par}

\caption{\label{Fig:TrapPseudoPotential}Approximate trap pseudo-potential inside and
outside of trapping region.}
\end{figure}

The pseudopotential generated by the RF is then given by,

\[
U=\frac{1}{4m\omega ^{2}}\left| \vec{F}\right| ^{2}=\frac{q^{2}V_{RF}^{2}}{4m\omega ^{2}ln(r_{ref}/r_{0})^{2}}\frac{1}{r^{2}}\]
This can be written in terms of the trap secular frequency and, in turn, the
well depth,

\begin{eqnarray*}
\omega _{r}^{2} & = & 2\frac{q^{2}V^{2}}{m^{2}\omega ^{2}r_{0}^{4}}\\
U_{0} & = & \frac{1}{2}m\omega _{r}^{2}r_{0}^{2}
\end{eqnarray*}
as
\begin{eqnarray*}
U & = & \frac{1}{4}\left( \frac{1}{2}m\omega _{r}^{2}r_{0}^{2}\right) \frac{1}{ln(r_{ref}/r_{0})^{2}}\frac{r_{0}^{2}}{r^{2}}\\
 & = & \frac{U_{0}}{4ln(r_{ref}/r_{0})^{2}}\frac{r_{0}^{2}}{r^{2}}
\end{eqnarray*}
With \( r_{ref}/r_{0}\approx 30-50 \), the pseudopotential at \( r\approx r_{0} \)
is \( U\approx U_{0}/100 \) and at \( r\approx r_{ref} \), \( U\approx 0 \),
fig.\ref{Fig:TrapPseudoPotential}. 

For \( U_{0} \) around a few \( eV \) the potential at the trap entrance at
\( r\approx r_{0} \) is just barely low enough that an ion will thermal energies
of \( 1/40-1/20eV \) could enter the trap. To clearly settle this case the
potential near the transition region would have to be determined more accurately.
Also near the electrode the electric field is strong enough that a pseudopotential
may no longer be a good approximations. Even if ions created outside the trap
can enter the trapping region, clearly they must be cooled more than an ion
created in the trap, since an entering ion must lose a few eV of kinetic energy
while a local ion must only lose its thermal energy.

In either case, an ion appears in the trap and is rapidly cooled by the cooling
lasers. Typically this takes only as long as required for oven to heat up, which
is usually less than a minute. This is immediately detected as a large floresence
signal in the PMT. At this point the oven and electron beam may be shut off
and the laser tuning and alignment and PMT position can be adjusted to maximize
the detected floresence.

\subsection{Counting}

The floresence signal indicates a trapped ion, but doesn't unambiguously indicate
the number of trapped ions. More ions should scatter more light, but only if
there is sufficient power in the cooling laser to saturate many ions. Typically
multiple ions will result in some ions being positions further from the beam
axis where the intensity is weaker and often the result is that many ions have
a single similarly to a single ions, though usually a few cooled ions do give
approximately a simple multiple of the single ion signal, but this is not consistently
reliable. 

Instead shelving can be used to count the ions. A shelved ion will not absorb
cooling light so it will not contribute to the floresence signal until it decays
from the \( 5D_{5/2} \) state. The decay time is not well defined but has an
exponential distribution so that even if all ions are shelved at the same time,
they will most likely decay at different times. Then the ions can be counted
by applying the shelving lamp until all ions are shelved, which it indicated
by the complete disappearance of floresence, then blocking the shelving beam
and waiting for the ions to decay. There will be discrete jumps in the signal
as each ion decays back to the ground state and begins to again be cooled. Counting
the number of jumps needed to restore the full signal gives the number of ions.
Similarly the shelving rate can be slowed down enough that individual shelving
transitions can be seen by discrete drops in the floresence. 

Both methods can require a few attempts to get an accurate count since in some
cases two or more ions can react closely enough in time that their individual
decays or shelving transitions can't be resolved, but more importantly for these
purposes the presence of more than one ion can be determined quickly, and a
single ion easily verified.

\subsection{Multiple Ions}

If more than one ion is initially present, shelving can also be used to reduce
the count to just one. Shelved ions absorb no blue light, so they can't be cooled,
neither then could they be heated. Changing the frequency of the blue laser
so that it is slightly bluer than resonance heats the ions just as down-tuning
cools them. Ions moving away from the laser are more likely to absorb and get
an average net extra momentum kick in the same direction as their initial velocity
and so their energy increases. This will eventually add enough kinetic energy
to an ion that it can escape the trap. 

It would do no good to boil out all the ions in this way, but if one is shelved
it will not be affected immediately by this re-tuning and will be unaffected
while the others are ejected from the trap. Then to reduce the number of ions
in the trap, the shelving light can be applied until the floresence signal indicates
that one ion has been shelved, or all can be shelved and the decays waited out
until only one is left shelved. At this point the cooling laser can be re-tuned
to the heating side of resonance. The PMT signal drops as the ions are heated
and their orbit increases so that they interact less often with the laser and
in weaker parts of the beam. After a few seconds the ions are usually gone from
the trap. The laser can be re-tuned to cooling frequencies and the shelved ion
deshelved resulting in the desired single ion. This also can take a few tries,
with many ions it is hard to be sure that just one is shelved so typically this
cleaning out is done in stages, removing a few at a time until only a few remain
that can be more easily individually manipulated. Also there is the risk that
the shelved ion that is intended to be kept decays while heating the others
and is itself heated and lost. In practice these complications are not much
of an obstacle and a single trapped ion can be obtained in about a minute after
a few of the shelving and heating cycles, only very occasionally losing all
the ions.

A safer, but slower and less exciting method of reducing the number of ions
is simply to shut off the cooling beams for a while. Collisions between ions
will occasionally result in one being knocked out of the trap and after a long
enough time letting them many ions fight it out, a single ion remains which
is then not further heated by collisions. This process can take anywhere from
a few seconds to 10 minutes, but always eventually results in a single ion and
seldom fails by losing all the ions.

\subsection{Isotopes}

The particular isotope of the ion can be determined at this point. Different
isotopes have slightly difference cooling frequencies, the shift is of order
\( 100MHz \). The OG signal correspond to the most frequent \( A=138 \) so
that by maximizing the count rate with the tuning of the laser, any shift in
the transition frequency is easily detected as an offset from the peak of the
tuning curve. At least three different isotopes have already been trapped and
detected in this way. In addition, when the shifts are known, the cooling lasers
can be tuned to preferentially cool a certain target isotope making that more
likely to be trapped. This has not been systematically studied, but does seem
to qualitatively change the rate at which various isotopes are trapped. This
could be enormously useful in practice for doing isotope comparisons.

\subsection{Ion Lifetime}

Immediately after loading, there is a relatively high concentration of Barium
in the system even with the oven off. This seems to lead to a reduces lifetime
for individual ions in the trap as the are frequently replaced by ions from
the background. The trap is too deep for a trapped ion to be knocked out, and
the pseudopotential far from the trap repels ions, so a likely exchange mechanism
is for a neutral atom that passed through to exchange an electron with the trapped
ion and be trapped itself while the originally trapped, and now neutral ion
is free to leave. 

Usually this exchange would not be detectable expect as some, probably unmeasurable
slow, loss of coherence. The exchange can be detected if the new ion is a different
isotope as the floresence will change significantly and retuning will yield
much different frequencies. This happen regularly a few minutes after loading,
especially if the oven happened to be needed to be on for a long period of time
while loading. It is never observed more than a few minutes after loading, and
rarely when the oven was needed hot for only a few seconds to load, so it is
likely that this exchange occurs only when there is much background barium in
the system and afterward the ion in the trap after many hours is the same one
originally loaded. 

Except for this effect there is no limit to the time an ion remains trapped
if it is being regularly cooled. When this system is operating ions are lost
only due do power failures and tuning errors in the blue laser. Barring these
events an ion stays trapped for many days, typically up to a week and on one
occasion 6 weeks.

\section{State Manipulation and Detection}

\label{sec:StateManipulationAndDetection}

Trapping and loading and cooling finally yield a single ion that is very well
localized and relatively motionless. Now a number of precision measurements
become possible. The ones that will be particularly relevant in understanding
the methods developed for the parity measurement are those that exploit the
ability to efficiently detect individual transitions between energy levels of
an ion. For a measurement of the PNC lightshift with spin resonance, the initial
spin state must set, the spin flip transition driven, and the final state detected.
The same coarse sequence involving different energy levels, rather than different
spin states in the same level, is followed for various measurements in single
ions. The general ideas are best illustrated with a few examples.

\subsection{\protect\( 5D_{5/2}\protect \) Lifetime}

The simplest experiment to implement is a measurement of the \( 5D_{5/2} \)
lifetime. The most conceptually clear way to determine the lifetime is to start
an ion in the \( 5D_{5/2} \) state, watch for the decay and record the decay
time. Averaging the decay times gives the lifetime. With an ion trap this ideal
can be realized.

A conventional atomic experiment can effectively do this by preparing a large
number of atoms in the initial \( D_{5/2} \) state and  recording the floresence
from the emitted photon in the decay as a function of time. This method would
be impractical in an ion trap because the detection efficiency of the decay
is very low. Though the initial state can be set and detected easily using the
shelving lamp and floresence signal, each decay produces only one photon. The
solid angle and quantum efficiency of the PMT yield a combined detection efficiency
of \( 1/1000 \) so only 1 in a thousand decays is detected. An experiment in
a vapor, or even a large trapped cloud, compensates for this with a large number
of atoms but for a single trapped ion an accurate measurement would instead
require excessively long observation times, especially if the background of
the detector is considered. An emitted photon would be detected, on average,
every \( 1000*50s=5\times 10^{4}s \). With background from, for example, a
dark count of even 1 every few seconds, detecting this occasional extra photon
would require that the rate be determined to a part in \( 10^{4} \) which then
requires detecting some \( 10^{8} \) decays, then requiring \( 10^{12}s \)
of observation time to detect a single transition. This is completely unreasonable
and detecting a decay in this way is not practical, but with the tools already
discussed this transition can be detected reliably with other methods.

The \( 5D_{5/2} \) state is the shelved state. In this state the ion will not
absorb or scatter the cooling light. Then a transition to this state, driven
by the shelving lamp, is detected simply by the disappearance of the blue floresence
detected by the PMT and a decay to the ground state is indicated by the reappearance
of the PMT signal. The decay time is given by the time between these two large
discrete changes in the floresence. Repeating this sequence a number of times
quickly yields enough data to reliably determine the average decay time.

\begin{figure}
{\par\centering \resizebox*{!}{0.9\textheight}{\includegraphics{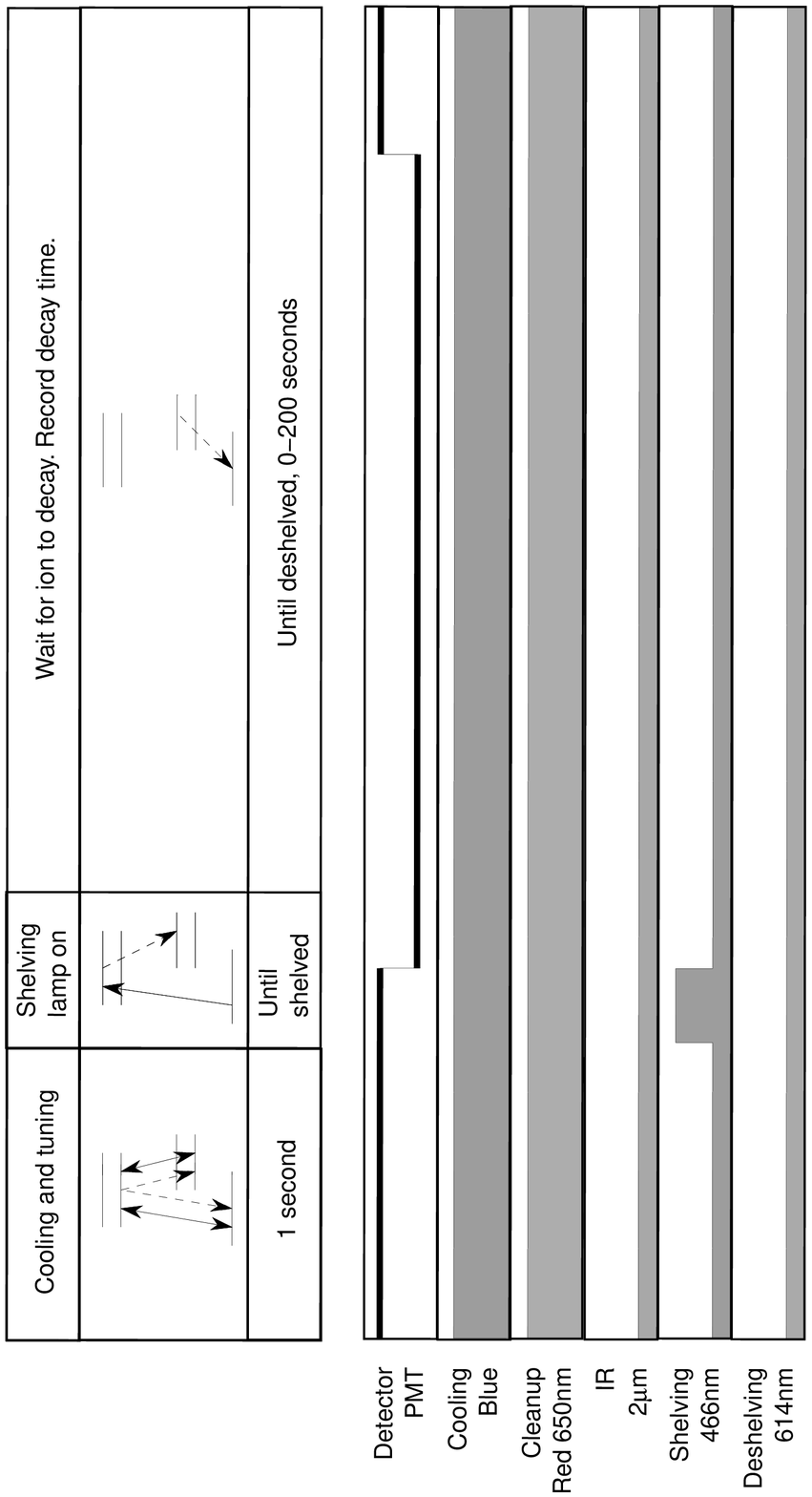}} \par}

\caption{\label{Fig:5D_5/2LifetimeMeasurementSequence}\protect\( 5D_{5/2}\protect \)
lifetime measurement sequence.}
\end{figure}

\begin{figure}
{\par\centering \includegraphics{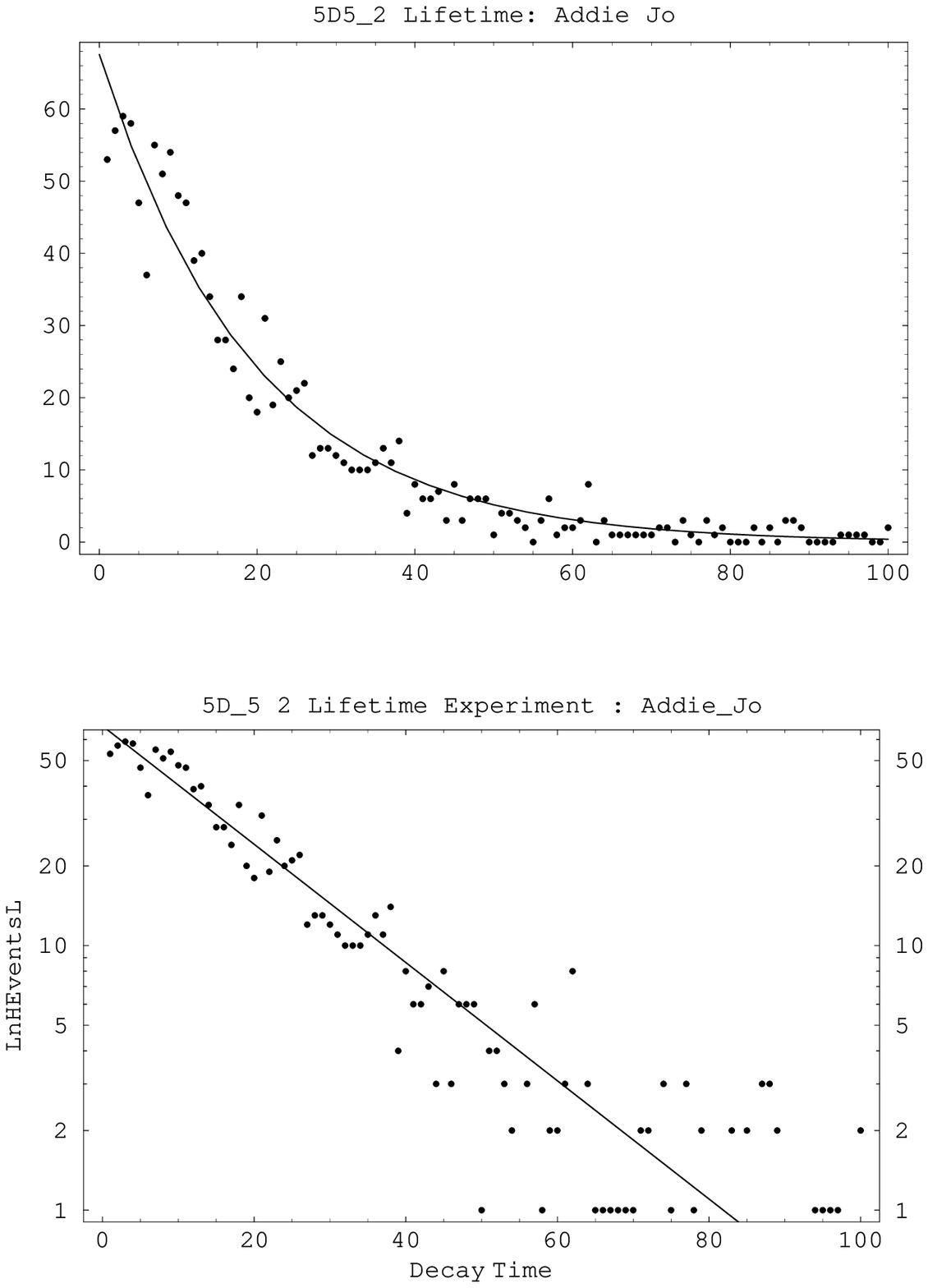} \par}

\caption{\label{Fig:5D_5/2LifetimeData}\protect\( 5D_{5/2}\protect \) lifetime data.
Number of events as a function of decay time.}
\end{figure}

This measurement sequence is shown schematically in fig.\ref{Fig:5D_5/2LifetimeMeasurementSequence}. 
This kind of diagram will be used throughout the remaining discussions of experimental
procedures. It provides information about the state of all the light sources
used in the experiment at every point in the measurement, a brief description
of the purpose of each step, a simple level diagram indicating the transitions
being driven or detected and the time required for each step. This measurement
sequence yield data like that shown in fig.\ref{Fig:5D_5/2LifetimeData}.

This kind of measurement in an ion trap now avoids complications from effects
like Doppler Broadening, collisional quenching and radiation trapping, but this
particular method has the subtle disadvantage that the decay is measured while
the ion is interacting with the relatively strong cooling lasers. With perfect
lasers this shouldn't have a large effect on the quantity being measured, though
non-resonant dipole coupling of the \( D_{5/2} \) state to other states does
slightly reduce its lifetime. In this system the diode laser used for the cleanup
transition has some small broadband background that has some non-negligible
component at the 614nm \( 5D_{5/2}\rightarrow 6P_{3/2} \) deshelving transition
so that the time spent in the \( D_{5/2} \) state is reduced by direct excitations
out of it. This could be easily remedied with a simple interference filter,
but just as easily it can be avoided with a slightly different measurement sequence
that will also avoid the inherent broadening from the off-resonant dipole couplings
and be more generally useful for other transitions that can't be seen directly
like this shelving transition.

Instead of continuously monitoring the ion with the cooling beams consider blocking
them immediately after the ion has been shelved. The ion now waits to decay
in complete darkness, in the absence of any applied interactions. The decay
is still not directly detectable, but at any time the ion can be checked to
see if it has decayed by applying the cooling beams. If floresence is detected,
the ion has decayed during the time waited, if no floresence is seen, the ion
is still shelved and so has not decayed. The difference in the floresence in
the two cases is some \( 1000cps \) so the ion's state can be determined with
almost 100\% certainty in even as little as \( 0.1s \). With a background of
40-70cps this gives a signal of \( 4\pm 2 \) counts if the ion is shelved and
about \( 100\pm 10 \) if it has decayed, these cases are easily distinguished. 

\begin{figure}
{\par\centering \resizebox*{!}{0.9\textheight}{\rotatebox{90}{\includegraphics{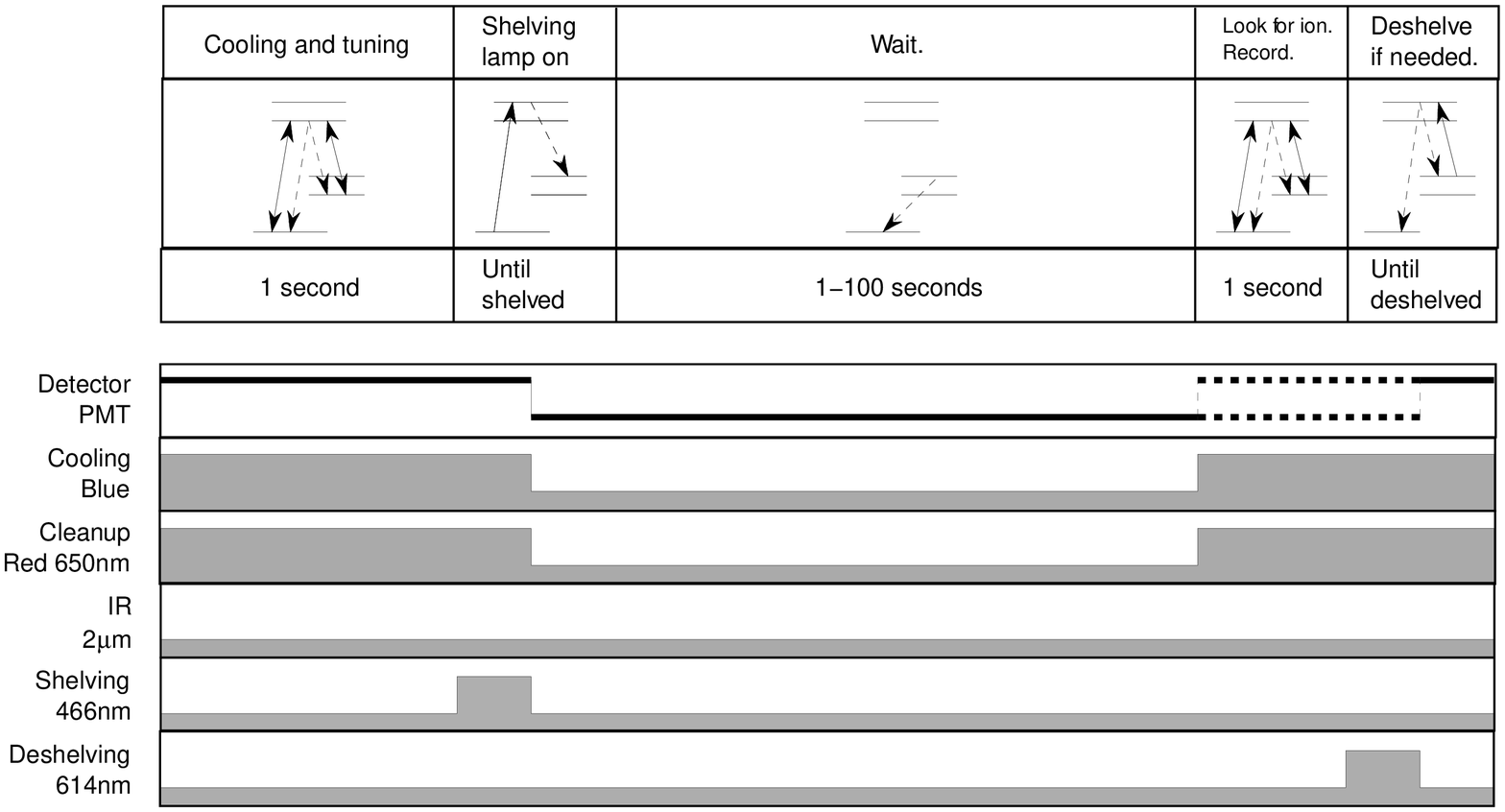}}} \par}

\caption{\label{Fig:5D5_2BlindLifetimeMeasurementSequence}Blind shelving \protect\( 5D_{5/2}\protect \)
lifetime measurement sequence.}
\end{figure}

\begin{figure}
{\par\centering \resizebox*{0.8\textwidth}{!}{\includegraphics{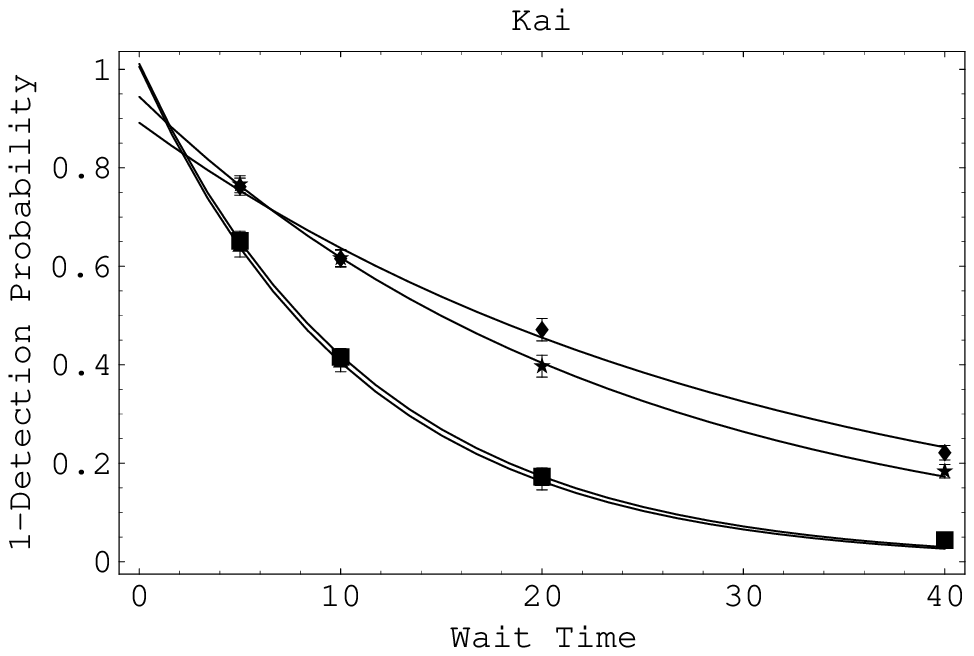}} \par}

\caption{\label{Fig:5D_5/2BlindLifetimeData}\protect\( 5D_{5/2}\protect \) lifetime
data using blind shelving. Normalized shelving probability as a function of
wait time after pumping to \protect\( D_{5/2}\protect \) state with and without
red and blue cooling lasers.}
\end{figure}

Rather than the precise decay time, repeating this sequence many times, for
many difference times, fig.\ref{Fig:5D5_2BlindLifetimeMeasurementSequence},
gives the decay probability as a function of time which also yields the lifetime
fig.\ref{Fig:5D_5/2BlindLifetimeData}.


\subsection{\protect\( 5D_{3/2}\protect \) Lifetime}

A small variation on this latter method of determining the \( 5D_{5/2} \) lifetime
can also be used for the \( 5D_{3/2} \) state. Again the ideal measurement
would be to start the ion in this state and measure the decay time, or at least,
the decay probability as a function of time. Setting the initial state is easily
done with the cooling and cleanup beam. The purpose of the cleanup beam is to
prevent the ion prevent the ion from getting stuck in the \( 5D_{3/2} \) state
by a decay from the \( 6P_{1/2} \) state during cooling. So simply shutting
off the cleanup beam quickly pumps the ion to the desired \( D \) state. 

\begin{figure}
{\par\centering \resizebox*{1\textwidth}{!}{\includegraphics{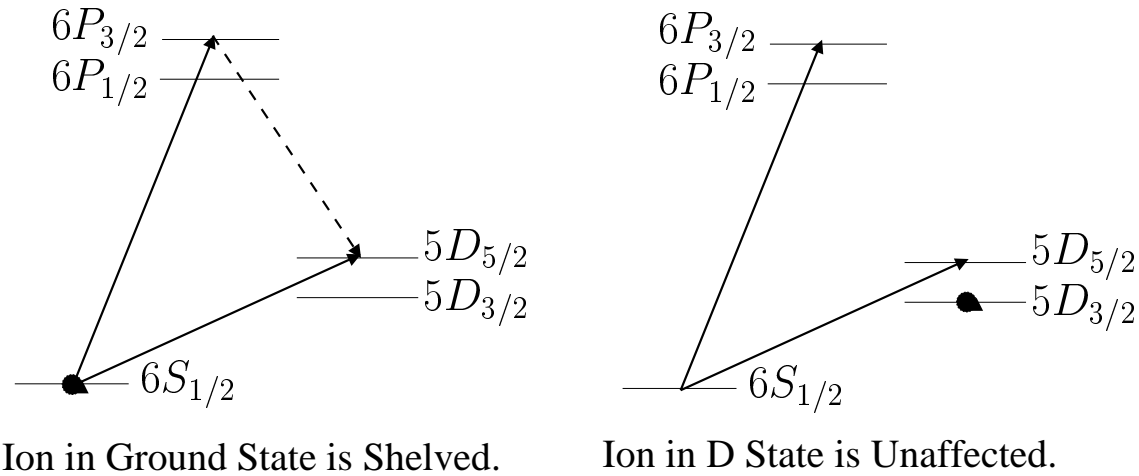}} \par}

\caption{\label{Fig:S-DStateDetection}Resolving \protect\( 6S_{1/2}\protect \) and
\protect\( 5D_{3/2}\protect \) levels using shelving transitions from the \protect\( 6S_{1/2}\protect \)
state.}
\end{figure}

At this point the beams can be shut of and the ion allowed to decay back to
the ground state. Checking for the decay after some time is no longer immediately
possible with the cooling beams as both states are part of the cooling cycle,
so the ion would florese whether it had decayed or not, however a single extra
step restores this easy determination. The shelving lamp shelves drives the
\( 6S\rightarrow 6P_{3/2} \) transition so it will shelve the ion only if the
ion starts in the ground state. Similarly a shelving transition driven directly
by a \( 1.76\mu m \) laser couples only the ground state to the shelved state.
In either case an ion in the \( 5D_{3/2} \) state will be unaffected, fig.\ref{Fig:S-DStateDetection}
 A decay from the \( 5D_{3/2} \) state can then be detected by attempting to
shelve the ion by applying the shelving lamp for a brief period of time and
then checking to see if the shelving transition was made. If the ion has decayed,
it can be shelved and when the cooling beams are applied this state is indicated
by the absence of blue floresence. An ion still in the \( 5D_{3/2} \) state
can not be shelved and will immediately florese when the cooling beams are applied. 

\begin{figure}
{\par\centering \resizebox*{!}{0.9\textheight}{\includegraphics{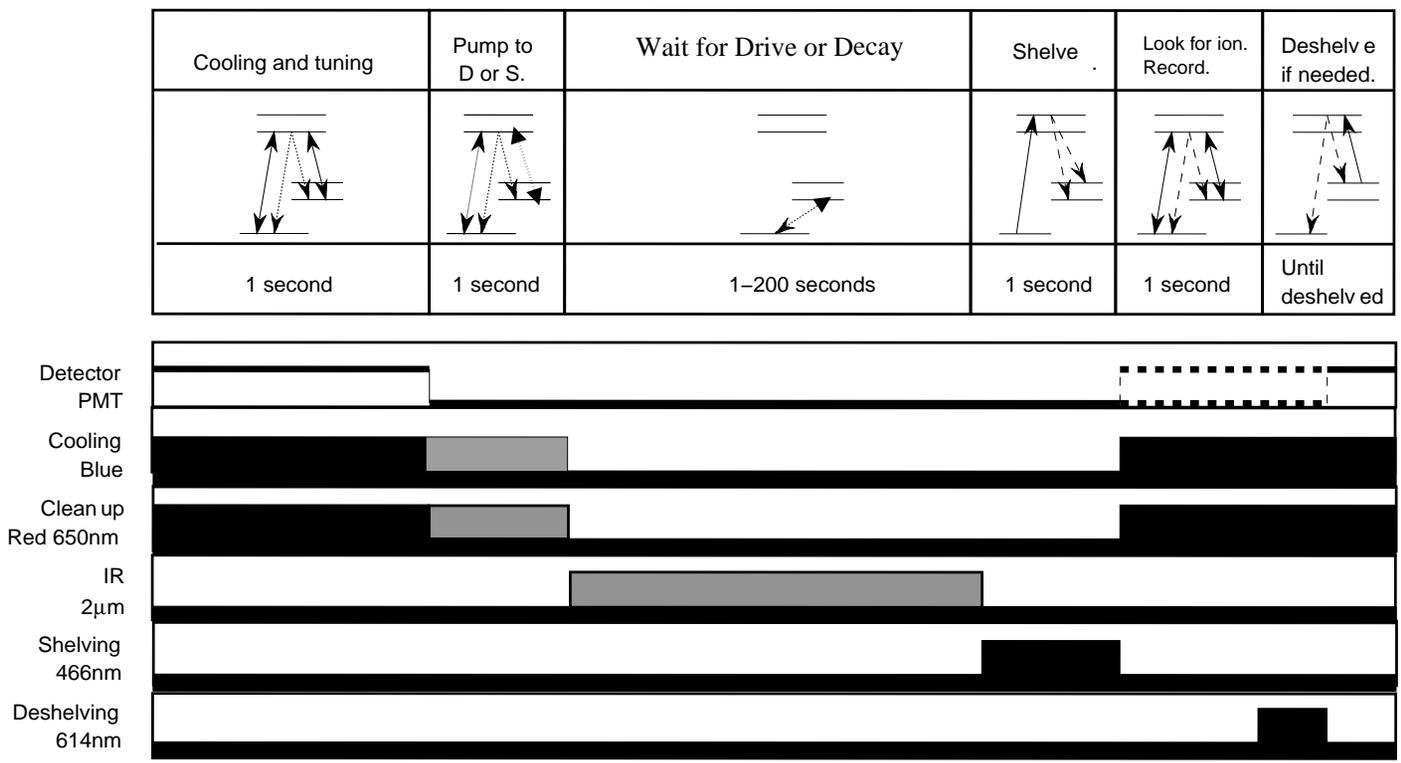}} \par}

\caption{\label{Fig:5D_3/2LifetimeSequence}\protect\( 5D_{3/2}\protect \) lifetime
or quadrupole transition measurement sequence.}
\end{figure}

\begin{figure}
{\par\centering \includegraphics{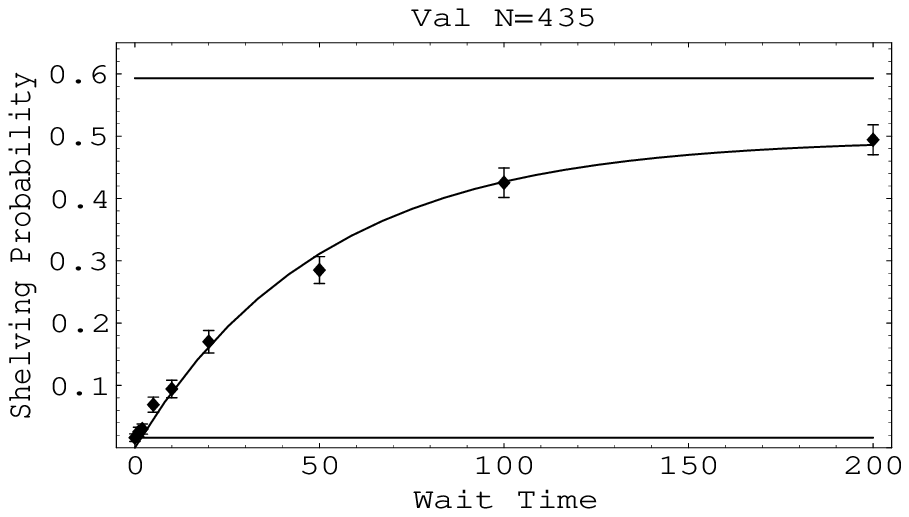} \par}

\caption{\label{Fig:5D3_2LifetimeData}\protect\( 5D_{5/2}\protect \) lifetime data.
Shelving probability as a function of wait time after initially pumping ion
to \protect\( D\protect \) state.}
\end{figure}

The net result is that after this brief shelving step, the ion will likely not
florese if it has decayed from the \( 5D_{3/2} \) state during the wait time,
and likely florese if it has not decayed. The cases are not perfectly correlated
from some practical complications, sec.\ref{Sec:PracticalLimitations}, but
this sequence does yield a shelving probability that increases with an exponential
profile having a time constant of the lifetime of the \( D_{3/2} \) state.
Similarly repeating this sequence for many wait times, fig.\ref{Fig:5D_3/2LifetimeSequence},
yields the shelving probability as a function of time, fig.\ref{Fig:5D3_2LifetimeData},
which provide the \( 5D_{3/2} \) state lifetime.

\subsection{\protect\( S\rightarrow D\protect \) Quadrupole Transition}

These measurements have so far involved a natural transition, they are just
as effective in detecting driven transitions, to measure transition rates or
resonance frequencies, such as that spin flip transitions that must be measured
to determine the PNC splitting. A simple illustration is with the same state
just considered for the \( 6S_{1/2}\rightarrow 5D_{3/2} \) quadrupole transition.
Here even a conventional atomic experiment could not continuously monitor the
state of the transition. The transition probability could be determined as a
function of time by turning on the clean-up beam and measuring the amount of
blue floresence from ion excited to, or remaining in the \( D \) state. The
amount of blue floresence measures the number of atoms that made the transition.
Again for a single ion this would be inefficient since the resulting single
photon that signals the transition is very hard to detect, and again this can
be remedied using shelving to generate instead an almost arbitrarily large number
of photons if the transition was made, and none if it hasn't.

The ion can be started in either state with the cooling beams as in the \( 5D_{3/2} \)
lifetime measurement, pumping to the ground state simply requires the complementary
case of blocking the blue beam. Light applied to the ion resonant with this
transition will result in transitions between the states. A transition can be
detected by attempting to shelve from the ground state. Sufficiently monochromatic
or intense light will result in a harmonic modulation of the probability of
the ion to be in its ground state, yielding an oscillating shelving probability
as a function of time. The more easily arranged effectively broadband excitation
will give a first order rate equation exponential transition profile. Either
can be used to determine the transition rate. 

\begin{figure}
{\par\centering \resizebox*{!}{0.9\textheight}{\includegraphics{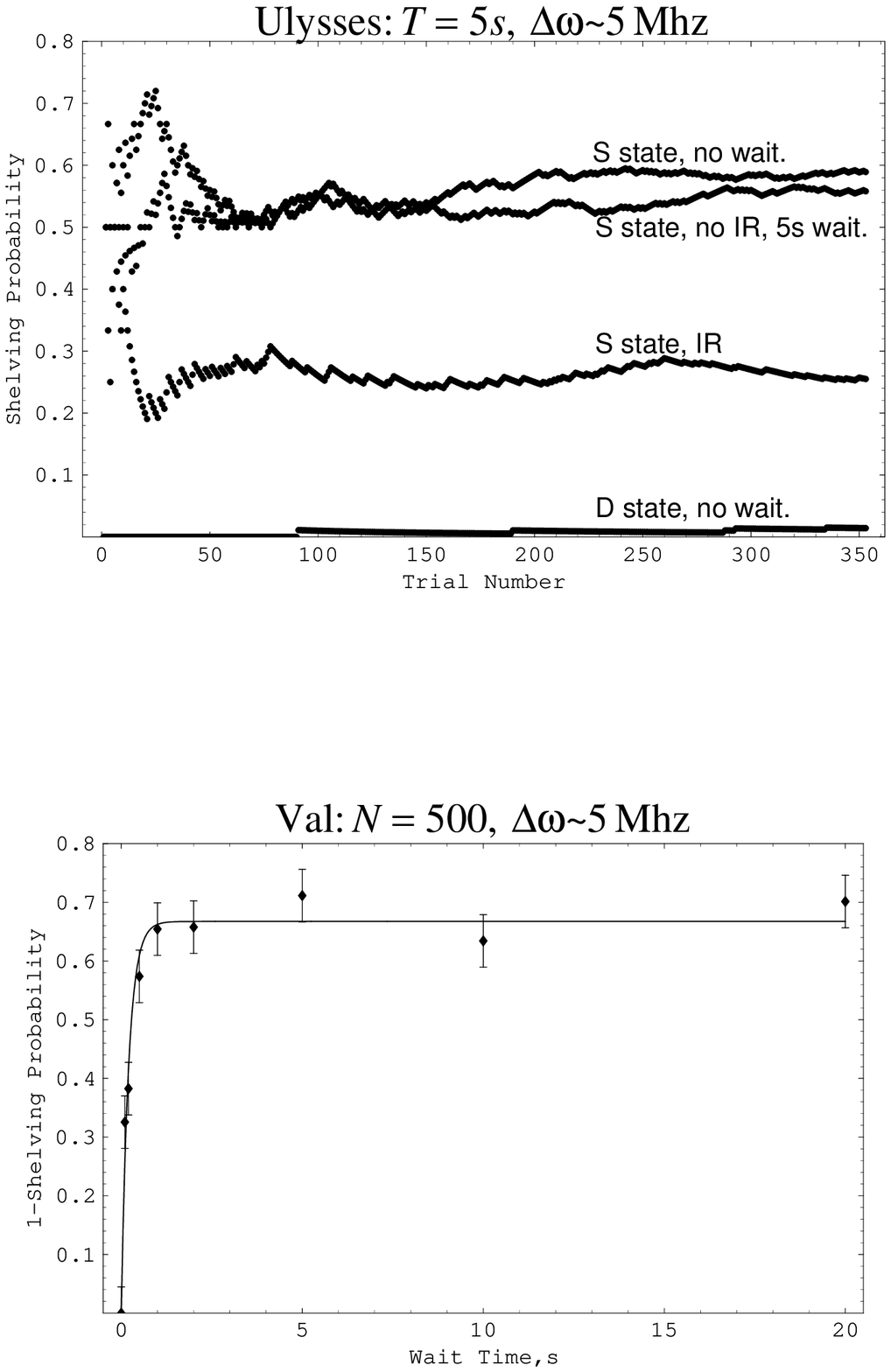}} \par}

\caption{\label{Fig:2umTranstionData}Shelving Probability as a function of tiral number
for initially pumped \protect\( S\protect \) and \protect\( D\protect \) states,
and pumped \protect\( S\protect \) state with and without \protect\( 2\mu m\protect \)
laser for \protect\( 5s\protect \), and shelving probability as a function
of \protect\( 2\mu m\protect \) exposure time for initially pumped \protect\( S\protect \)
state.}
\end{figure}

An example of the results of this kind of measurement is shown in fig.\ref{Fig:2umTranstionData}.
The measurement sequence is similar to fig.\ref{Fig:5D_3/2LifetimeSequence}
for the \( 5D_{3/2} \) lifetime except that the ion is initially pumped to
the ground state with the cleanup beam, and the wait is done while applying
a broadened \( 2\mu m \) laser to the ion to drive the \( S\rightarrow D \)
transition effectively incoherently. For short times, or far from resonance,
the transition probability is small and the ion remains in the ground state
from which it can be shelved so the shelving probability is high. For longer
times near resonance transitions are likely and the resulting shelving probability
is reduced. The shelving probability then simply decays exponentially with a
time constant given by the excitation rate.

\subsection{Efficiency, Stability, Precision}

\label{Sec:PracticalLimitations}

It is easy to understand how these shelving methods qualitatively yield the
desired information about transitions, but there are a few practical considerations
that slightly complicate a perfect interpretation of detected floresence after
the detection sequence as a transition between the set of states under investigation. 

The central feature exploited in these measurements is that an ion in the \( 6S \)
state can be shelved and an ion in the \( 5D_{3/2} \) state can not be. However,
this doesn't imply that an ion in the \( 6S \) state will be shelved, or that
an ion in the \( D \) state will not end up shelved. First the shelving rate
is finite, so that even when starting in the ground state, the ion may not end
up in the shelved state when the shelving lamp is applied simply because it
has not yet made any transitions to the \( P_{3/2} \) state. 

Similarly even if it has made a transition to the \( P \) state the decay may
be back to the ground state, where it may try again, or to the \( 5D_{3/2} \)
state where it is stuck. Then after the shelving lamp is applied for some period
of time, this gives a shelving probability, \( p_{shelving} \), limited by
the shelving rate and the branching ratio of the \( P_{3/2}\rightarrow D_{5/2} \)
decay to the \( P_{3/2}\rightarrow D_{3/2} \) decay of about \( 10:1 \). 

The intensity and exposure time of the shelving lamp can be increased so that
the shelving transition is saturated, eliminating limits from the rate and leaving
the branching ratio limit given \( p_{shelving}\approx 0.9 \), but for this
apparatus, the shelving rate is slow enough that waiting long enough to saturate
the transition slows down data taking too much so a slightly shorter exposure
time is used that typically yields an estimated \( p_{shelving}\approx 0.8 \).

The \( 5D_{3/2} \) state has a finite lifetime so that while attempting to
shelve from the \( S \) state that ion can decay from the \( D \) state and
also be shelved, \( p_{residual} \). This depends on the exposure time and
\( D_{3/2} \) lifetime as well as the shelving rate and branching ratio. Typically
the shelving step takes only \( 0.1s \) and this effect is negligible since
the \( D \) state is very unlikely to have decayed in this time. Typically,
\( P_{res}<0.01 \).

Finally, even if the ion has been shelved, this may not be detected by the cooling
beams either from counting statistics, or because the ion then decays from the
\( D_{5/2} \) state, or is kicked out by the broadband background in the cleanup
laser. This yields a shelved state detection efficiency of \( p_{det}\approx 0.9 \). 

The net result, with the possibilities that the ion may be in the ground state,
\( p_{S} \) and not shelved, \( p_{shelving} \), the ion may be in the \( D \)
state, \( p_{D} \) and then still shelved, \( p_{residual} \) , and the ion
may be shelved but not detected as shelved, \( 1-p_{det} \), gives a floresence
probability, \( p_{fl} \), after this detection sequence of
\[
p_{fl}=p_{det}(p_{shelving}p_{S}+p_{residual}p_{D})\]
With the \( S \) and the \( D \) state the only possibilities, \( p_{S}+p_{D}=1 \).
Again the residual shelving from the \( D \) state is usually negligible so
that,

\begin{eqnarray*}
p_{fl} & = & p_{det}((p_{shelving}-p_{residual})p_{S}+p_{residual})\\
 & \approx  & p_{0}p_{S}\\
p_{0} & \approx  & p_{det}p_{shelving}\sim 0.7-0.8
\end{eqnarray*}
The net detection efficiency for a transition is slightly reduced, but the profile
is unaffected.

These offsets and modifications depend on experimental conditions and so many
be variable and a source of noise or systematic errors in the final measurement.
Precision measurements require a careful understanding of this kind of instability
and their consequences. This kinds of studies were the central focus of earlier
work on this project and the problems and solutions and end results are discussed
in \cite{KristiThesis}.

\chapter{Spin State Manipulation and Detection}

\label{Sec:SpinStuff}

The quantity to be directly measured in this experiment is the PNC induced splitting
of the ground state or \( D \) state magnetic sublevels generated when the
\( 6S_{1/2}\rightarrow 5D_{3/2} \) quadrupole and parity violating dipole couplings
are simultaneously driven with the appropriate phases and beam geometries. One
possible route to determining the size of this splitting is to start the ion
in a particular sublevel, drive the spin flip transitions and measure the transition
probability as a function of interaction time or the frequency of the applied
spin flipping fields to get a precession rate or resonance frequency. This will
require being able to set and detect the initial and final spin state. The same
general procedures are used to manipulate and detect the ion among its different
energy levels in more conventional ion experiments. In most cases these spin
manipulations and measurements can be made with the same methods modified to
make them spin sensitive.

\section{Pumping}

\label{Sec:Pumping}

Manipulating spin states by optical pumping is a familiar idea in vapors. The
same techniques can be used on a single ion with the only difference being in
interpretation of the final result. Rather than pumping resulting in an unequal
distribution among particular spin states of atoms in a large population, it
will determine relative probabilities for the single ion to be in a given spin
state when pumping is completed.

\subsection{Laser Polarization and Magnetic Fields}

Pumping between the \( 6S \) and \( 5D_{3/2} \) energy levels is easily done
with the cooling and cleanup beam. Blocking the laser that drives transitions
out the the desired final state uncouples that state. When the ion decays to
this state is remains there. Similarly pumping can be done among the spin sublevels
simply by uncoupling the desired final state, in this case, by adjusting the
polarization of the beams, and the magnetic field defining the quantization
axis.
\begin{figure}
{\par\centering \includegraphics{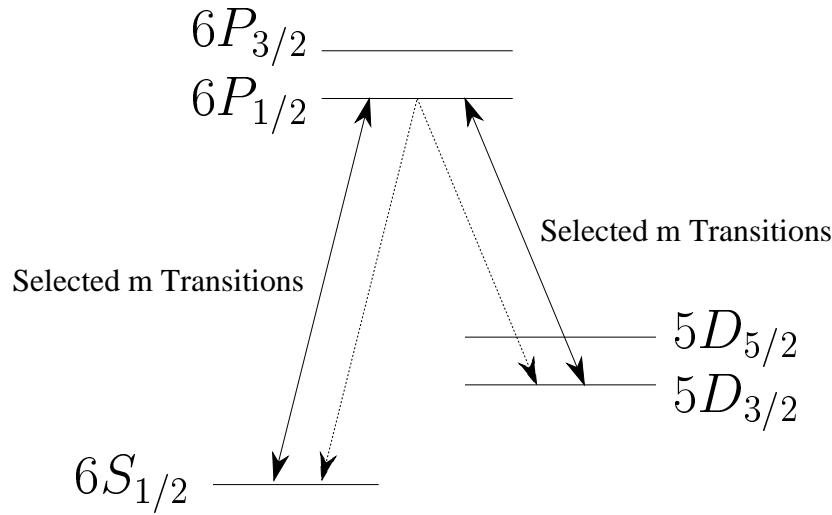} \par}

\caption{\label{fig:SpinStatePumping}Spin state pumping.}
\end{figure}
 Generally the magnetic field is kept fixed, either parallel to the beam axis
or parallel to the linear polarization of the red laser. The applied field is
difficult to change quickly, perhaps on time scales of a few tenths of seconds
at best without a lot of work, giving a significant limitation to the data collection
rate. More importantly, high magnetic field stability is required for the resonance
experiments and changing the direction of the applied field dramatically can
result in long time scale drifts of the resulting field due to nearby ferromagnetic
materials. As a result, in practice, only the laser polarizations are changed
to alter the relative spin couplings.

The polarizations of both laser are independently controlled using a voltage
controlled variable retardation plate that allows adjustment to vertical and
horizontal linear polarization and left or right handed circular polarization.These
particular control units are rather slow. Switching polarization states requires
times on the order of tenths of seconds, also limiting data collection rates.
This is never likely to be the biggest limit, but future improvements may involve
an upgrade to the higher speed and additional flexibility, and complications,
of a pockel cell. 

The red beam also passes through a half wave plate so that the direction of
its linear polarization can be continuously adjusted manually. This additional
freedom is helpful in aligning the magnetic field as explained later, but is
otherwise not necessary. The same ability for the blue laser is not helpful
due to the structure of the \( 1/2\rightarrow 1/2 \), \( S\rightarrow P \)
transition.

The beams are then combined with a non polarizing 50\% beam splitter cube. Afterwards
they pass only through a set of focusing optics and to the trap so the final
polarization of the beams is unmodified other than by imperfections in the lenses
or vacuum windows.

\subsection{Ground State Pumping}

\label{Sec:GroundStatePumping}

\begin{figure}
{\par\centering \includegraphics{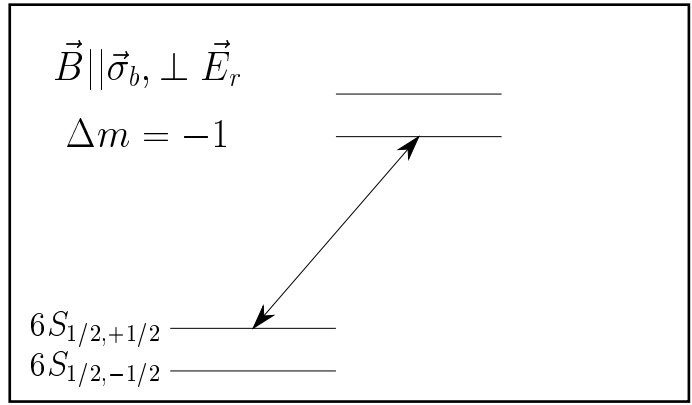} \par}

\caption{\label{Fig:PolarizationDecoupling}Pumping in the Ground State.}
\end{figure}

With a magnetic field parallel to the beam axis, and circularly polarized blue
light used to drive only \( \Delta m=+1 \) transitions, the \( S_{1/2,+1/2} \)
state is uncoupled, fig.\ref{Fig:PolarizationDecoupling}. At the same time
the red beam is linearly polarized to include all \( D \) states.

If the ion starts in the \( S_{+1/2} \) state it will remain there as it can't
be excited by the cooling beams from this state. When the ion starts in the
\( S_{-1/2} \) state it will be driven through the \( P \) and \( D \) states
and at some point decay from the \( P \) state to the \( S_{+1/2} \) where
it will again remain.

With the ion in this uncoupled state, it will no longer scatter blue photons
out of the cooling beam to the PMT as it can no longer be excited to the \( P \)
state by the cooling laser. As a result, the count rate drops to the background
rate and evidence of this spin pumping is available immediately.

Transition and decay rates are on the order of MHz, so that only a few \( \mu  \)s
is required for a transition to the intended \( S_{+1/2} \) state to be very
likely, in a ms it is virtually guaranteed, the fluorescence drops, and the
cooling beams can be shut off. Pumping happens very quickly and is practically
limited by the switching times of polarization controls and beam blocks well
before any limits due to excitation rates, with the end result being the ion
in the desired spin state.

\subsection{Practical Limits to Pumping Efficiencies}

In practice, the excitation rate out of this pumped state can not be made to
be exactly zero because of impure polarizations or imperfect alignment. This
will be considered in full generality shortly, but the immediate consequence
is that the ion is occasionally excited out of this pumped state and this results
in the occasional blue photon above the background as the ion decays from the
\( P \) state. The count rate then is not exactly the background rate, but
is still much reduced, typically to within 20-50 cps of the background, so the
PMT count rate is around 100cps including a 30-60cps background compared to
1000cps while not pumping, a reduction in the floresence rate by a factor of
20 or more. This limit is probably largely due to a less than perfectly circularly
polarized beam, though it is also possible that some other perturbation would
kick the ion out of the pumped state. The latter would require a transition
rate of a few tens of kHz, and is one motivation for studying the spin lifetimes
discussed later that turn out to exclude this possibility.

The non-optimal configuration also yields a smaller pumping efficiency. The
probability to be in the intended destination state during pumping is slightly
less than the ideal 100\%. As shown later, the rates described above correspond
to the ion spending \( >95\% \) of the time in the \( S_{+1/2} \) state. This
is not exceedingly close to ideal, but is, more to the point, far from the 50\%
uniform probability of an unpolarized ion and is completely sufficient for making
spin states detectable and spin transitions visible. The reduced efficiency
does eventually result in a decreased experimental efficiency. Improvements
might be made with better alignment of the magnetic field and slightly more
elaborate control of the polarization to improved the purity of the preferred
polarization state and compensate for possible perturbations to it from the
few remaining optical elements between the ion and where the polarization is
defined. These improvements may eventually be worth considering if they could
be made without excess effort as the yield would be small, though they may also
be desirable for other purposes.

Finally the residual excitation rate out of the pumped state require the pumping
beams to be shut off in a well-defined order. Clearly the red beam must not
be blocked first as the ion may be removed from the ground state in the time
it takes to then block the blue. The beams can not be blocked simultaneously
without considerable effort as the high transition rates result in simultaneous
meaning to be within some small fraction of a microsecond. Even if achieved
this would still yield a small probability for the ion to end up in the \( D \)
state as the ion does spend some time there during pumping. The best method
is then also the easiest, the blue beam must be shut off first followed immediately
by the red beam. This also slightly reduces the final pumping efficiency since
an ion that is in the \( D \) state` when the blue beam is shut off` will be
pumped to the ground state by the red but with no preferred spin orientation
so it enters either ground state spin level with equal probability. This is
a negligible modification since the probability to be in the \( D \) state
to begin with is only a few percent.

\subsection{D State Pumping}

\label{Sec:DStatePumping}

\begin{figure}
{\par\centering \resizebox*{1\textwidth}{!}{\includegraphics{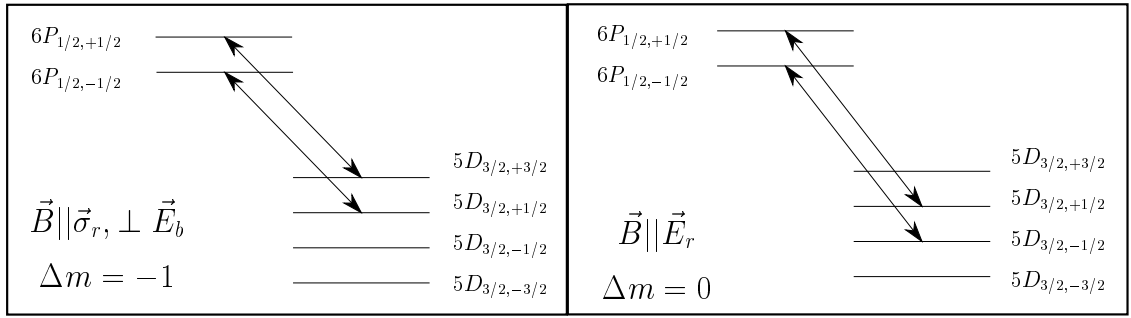}} \par}

\caption{\label{Fig:DStatePumping}D State Pumping}
\end{figure}

With the same magnetic field parallel to the beam axis and circularly polarized
red light with now linearly polarized blue, the \( D_{+3/2} \) and \( D_{+1/2} \)
states are uncoupled and the ion is quickly driven to these spin states. With
linearly polarized red and a magnetic field now parallel to this polarization
axis, only \( \Delta m=0 \) transitions are made on the the \( D \) state
and the ion is driven to the \( D_{\pm 3/2} \) states, fig.\ref{Fig:DStatePumping}.

As for the ground state a pumping signal is immediately visible as reduced fluorescence
because the ion lingers in the uncoupled state rather than cycling through the
\( P \) state and decaying. Also, for the same reasons as in the ground state,
the order that the pumping beam are shut off is important. In this case the
red beam is shut off first, followed closely by the blue.

In these \( D \) state methods the ion is driven to a set of two selected spin
states rather than just one as in the ground state, but the result is still
a highly polarized ion in the sense that the probability distribution among
the various spin states is far from the uniform, unpolarized case. For the case
on linearly polarized red, the probability is evenly distributed between the
\( \pm 3/2 \) states. For circular polarization, as seen in the general analysis
presented later, the result is a 75\% chance to be in the \( +3/2 \) state
and 25\% to be in the \( +1/2 \) state, and in fact for arbitrary polarizations
it turns out the ratio of the probabilities to be in the \( m=\pm 1/2 \) state
to the \( m=\pm 3/2 \) state is \( 3:1 \), though this particular solution
is unstable and very sensitive to the alignment of the pumping beam, sec.\ref{Sec:FinalPumpedDStates}.

In principle, these two procedures could be sequenced to yield selection of
a single spin state. However, this would require changing the magnetic field
between the two cases, which as already mentioned is difficult and undesirable,
or using additional laser beams with different propagation directions, also
a practical nuisance. In addition, the previously discussed limits of polarization
purity and alignment, which prevent any coupling from being made to be exactly
zero, make the final state effectively independent of the initial configuration
except for particular narrow windows of excitation rates or pumping times because
any coupling to a state will eventually result in transitions out of that state.
So, in practice, a single step is used and found to be completely sufficient.

\subsection{General Pumping Analysis }

\label{Sec:PumpingAnalysis} 

For the idealized cases, pumping is easy to understand and simple to analyze.
One beam is used to pump by being set to uncouple a desired set of states, and
the other beam is set to couple to all spin states and act a cleanup by keeping
the ion in the desired spin multiplet and out of all other levels. In practice,
the polarizations can't be set to exactly the desired states, and the alignments
relative to the applied magnetic field can't be made perfectly precise. So states
that are ideally uncoupled, are instead only much more weakly coupled relative
to the other states. 

In this case, the resulting populations must be determined from the relative
rates. The results will be qualitatively quite similar, as the ideal case is
simply the limit of a spurious rate returning to zero. The only difference will
be in the dependence of the final pumped state on the initial pre-pumping configuration.
The precise results require a fully general treatment of the rate equations.
As the final configuration does affect detection efficiencies in any future
experiment, a more accurate and detailed understanding of the end results of
pumping are worthwhile and helpful for planning measurements or evaluating possible
new techniques.

The transitions and lasers are broad relative to the excitations rates used,
10's of MHz compared to a few MHz. So the limit of broadband excitations can
be considered and simple rate equations used to describe the evolution of the
populations. Lingering coherent effects will be ignored as they would require
a considerably more complicated density matrix analysis. This may not be completely
appropriate, but the rate equations will certainly capture most of the spirit
of the resulting behavior if possibly not all the details.

Take \( S_{m} \), \( P_{m} \), \( D_{m} \) to be the probability to be in
the \( m^{th} \) sublevel of the corresponding state. \( R^{R}_{mm'} \) and
\( R^{B}_{mm'} \) can represent the transition rates of the red and blue laser,
and \( \Gamma ^{S,D}_{mm'} \) the decay rate from the \( P \) state to the
\( S \) or \( D \) states. The \( m \) dependence of the decay rates is given
by the square of the appropriate Clebsch-Gordan coefficients, 
\begin{eqnarray*}
\Gamma ^{S,D}_{m'm} & = & \Gamma ^{S,D}\frac{\left| \left\langle \frac{3}{2}m'|1,m'-m;j_{S,D},m\right\rangle \right| ^{2}}{2(3/2)+1}
\end{eqnarray*}
 With the normalization, \( 2j_{P}+1 \), the scalars \( \Gamma ^{S,D} \) can
be properly interpreted simply as the fractional decay widths of the P state
to the S or D manifold and can be written in terms of the total width of the
state \( \Gamma  \) and a branching fraction \( f \). With \( f \) taken
to be the branching fraction to the \( S \) state, the fractional widths are
\( \Gamma ^{S}=f\Gamma  \), \( \Gamma ^{D}=(1-f)\Gamma  \). 

Excitation rates depend on Clebsch-Gordan coefficients in the same way and,
in this case, also the polarization of the lasers.

\begin{eqnarray*}
R^{S,D}_{m,m+s} & = & R_{s}\left| \left\langle \frac{3}{2},m+s|1,s;j_{S,D},m\right\rangle \right| ^{2}
\end{eqnarray*}

\( R_{s} \) gives the total excitation rate for a particular transition and
depends on the direction of the applied electric field and atomic structure
\( R_{s}=\left| eE_{s}\left\langle f\left| \left| D\right| \right| i\right\rangle \right| ^{2}/\Gamma _{Laser} \)
with \( E_{0}=E_{z},E_{\pm 1}=(-E_{x}\pm iE_{y})/\sqrt{2} \). The overall rate,
\( R=\left| eE\left\langle f\left| \left| D\right| \right| i\right\rangle \right| ^{2}/\Gamma _{Laser} \)
can be factored out and the remaining dependence on direction is easiest to
understand in terms of the polarization and propagation direction of the applied
laser. The fully general case is a bit cumbersome but a general analysis yields
the tidy result, \cite{Schacht00},

\begin{eqnarray*}
\left| E_{0}\right| ^{2} & = & E^{2}(\hat{\varepsilon }\cdot \hat{B})^{2}\\
\left| E_{\pm }\right| ^{2} & = & \frac{E^{2}}{2}\left( 1\pm \left( \hat{k}\cdot \hat{B}\right) \sigma -\left( \hat{\varepsilon }\cdot \hat{B}\right) ^{2}\right) 
\end{eqnarray*}
\( \hat{\varepsilon }\cdot \hat{B} \) is more conveniently written as,
\[
\hat{\varepsilon }\cdot \hat{B}=\frac{1}{2}\left( 1+sin(2\alpha )\sqrt{1-\sigma ^{2}}\right) \left( 1-\hat{k}\cdot \hat{B}\right) \]
giving the rates in terms of three completely independent parameters, \( \sigma  \),
\( \hat{k}\cdot \hat{B} \) and \( \alpha  \).

The population of a particular spin state includes a loss from every excitation
out of it and a gain for any stimulated or spontaneous decay into it. The equations
of motion for all states are,
\begin{eqnarray*}
\dot{S}_{m} & = & -R^{B}_{m'm}S_{m}+\left( R^{B}_{m'm}+\Gamma ^{S}_{mm'}\right) P_{m'}\\
\dot{P}_{m} & = & -\left( R^{B}_{m'm}+\Gamma ^{S}_{m'm}\right) P_{m}-\left( R^{R}_{m'm}+\Gamma _{m'm}^{D}\right) P_{m}\\
 & + & R^{B}_{mm'}S_{m'}+R^{R}_{mm'}D_{m'}\\
\dot{D}_{m} & = & -R^{R}_{m'm}D_{m}+\left( R^{B}_{m'm}+\Gamma ^{P}_{mm'}\right) P_{m'}
\end{eqnarray*}

In all cases the source or destination state \( m' \) is summed over. This
allows for a more compact matrix notation. Let \( L_{m} \) be the total loss
rate out of a given state due to absorption or stimulated or spontaneous emission,
\begin{eqnarray*}
L^{S}_{m'm} & \equiv  & \sum _{m''}R^{B}_{m''m}\\
L^{D}_{m'm} & \equiv  & \sum _{m''}R^{R}_{m''m}\\
L_{m'm}^{P} & \equiv  & \sum _{m''}R^{B}_{mm''}+R^{R}_{mm''}+\Gamma ^{S}_{m''m}+\Gamma ^{D}_{m''m}\\
 & = & \Gamma +\sum _{m''}R^{B}_{mm''}+R^{R}_{mm''}
\end{eqnarray*}
 For the \( P \) state, the sum of the decay rate over all possible destinations
gives the lifetime for a particular spin state. The lifetime will be spin independent
so this becomes simply the lifetime of the \( P \) state. 

With these definitions, and moving some indices around for notational tidiness,
the matrix form of the rate equations follows easily,

\begin{eqnarray*}
\dot{S} & = & -L_{S}S+\left( R_{ST}+\Gamma _{S}\right) P\\
\dot{P} & = & -L_{P}P+R_{S}S+R_{D}D\\
\dot{D} & = & -L_{D}D+\left( R_{DT}+\Gamma _{D}\right) P
\end{eqnarray*}

With cooling rates at around a \( MHz \) the populations quickly come to their
asymptotic values, so only these steady state populations are required for analysis
and they are the solutions giving zero time derivatives.

\begin{eqnarray*}
0 & = & -L_{S}S+\left( R_{S}^{T}+\Gamma _{S}\right) P\\
0 & = & -L_{P}P+R_{S}S+R_{D}D\\
0 & = & -L_{D}D+\left( R_{D}^{T}+\Gamma _{D}\right) P
\end{eqnarray*}
Since the \( L_{i} \) are diagonal, and for the \( P \) state has no nonzero
elements on the diagonal because at least the decay rate is non-zero, it is
easy to solve for \( P \), as \( L_{P} \) must be invertible, 
\[
P=L_{P}^{-1}\left( R_{S}S+R_{D}D\right) \]
giving,
\begin{eqnarray*}
L_{S}S & = & \left( R_{S}^{T}+\Gamma _{S}\right) L_{P}^{-1}\left( R_{S}S+R_{D}D\right) \\
L_{D}D & = & \left( R_{D}^{T}+\Gamma _{D}\right) L_{P}^{-1}\left( R_{S}S+R_{D}D\right) 
\end{eqnarray*}
 This is a simple system of 6 coupled equations which are now best dealt with
using explicit algebra or just simply solved numerically. Further manipulations
can result in the solution appearing as an eigenvector but this isn't particularly
illuminating and doesn't make an analytic solution transparent.

An additional simplification is possible if the excitation rates are much slower
than the decay rates the \( R \) can be neglected relative to the \( \Gamma  \).
Losses from the \( P \) state are dominated by decays so \( L_{P} \) becomes
spin independent and \( L_{P}^{-1} \) is given just by \( 1/\Gamma  \) times
an identity, giving

\begin{eqnarray*}
\dot{S} & \approx  & -L_{S}S+f_{S}(R_{S}S+R_{D}D)\\
\dot{D} & \approx  & -L_{D}D+f_{D}(R_{S}S+R_{D}D)
\end{eqnarray*}
With \( f_{i}=\Gamma _{i}/\Gamma  \). For the work presented here, this limit
is generally not valid, the transitions are usually saturated and the excitation
rates are of the same order as the decay rates. In either case, only numerical
solutions will be presented here so the loss of generality provides no advantage
and this form is completely sufficient.

These equations are partly redundant and incomplete. Probability conservation
requires that the time derivatives add to zero, so one of the equations is not
linearly independent. Also the normalization is not fixed as any scalar multiple
of a solution remains a valid solution. Replacing one of the equations with
a normalization condition, that all populations add to one, completely defines
the problem.

\subsection{Pumping Signal and Pumped Populations}

\label{Sec:PumpingSignal}

\subsubsection{Final Populations with Ideal Beams}

\begin{figure}
{\par\centering \resizebox*{1\textwidth}{!}{\includegraphics{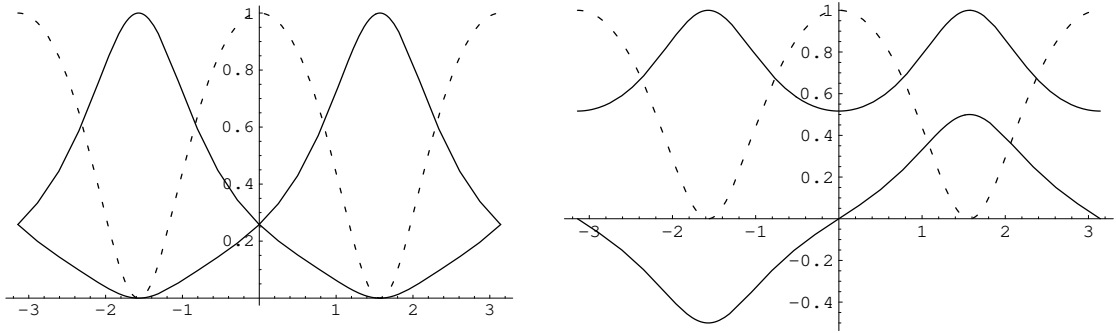}} \par}

\caption{\label{fig:SPumpingSignal}Floresence and ground state populations as a function
of circular polarization of blue cooling laser.}
\end{figure}

\begin{figure}
{\par\centering \includegraphics{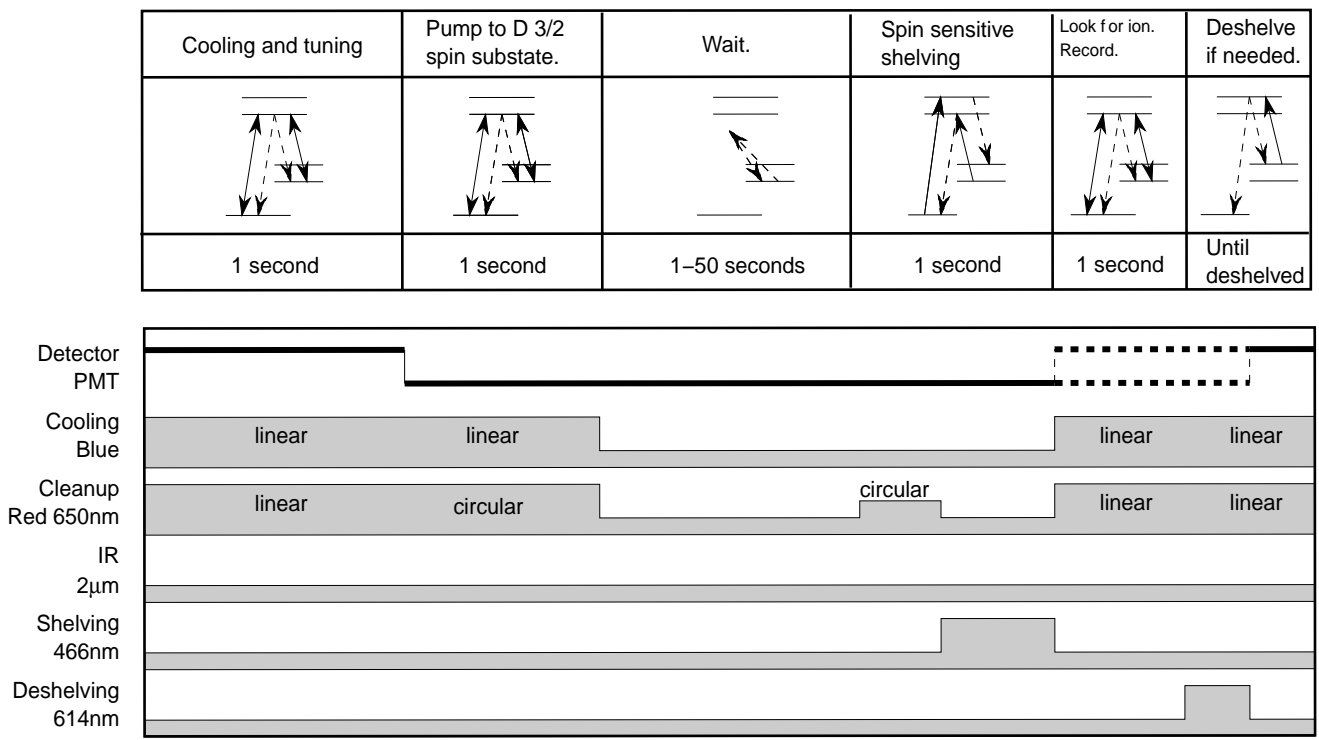} \par}

\caption{\label{fig:DCircPumpLabel}Floresence and \protect\( D_{3/2}\protect \) state
populations as a function of circular polarization of red cooling laser. The
\protect\( D_{\pm 1/2}\protect \) levels turn out to have the largest populations.}
\end{figure}

\begin{figure}
{\par\centering \resizebox*{1\textwidth}{!}{\includegraphics{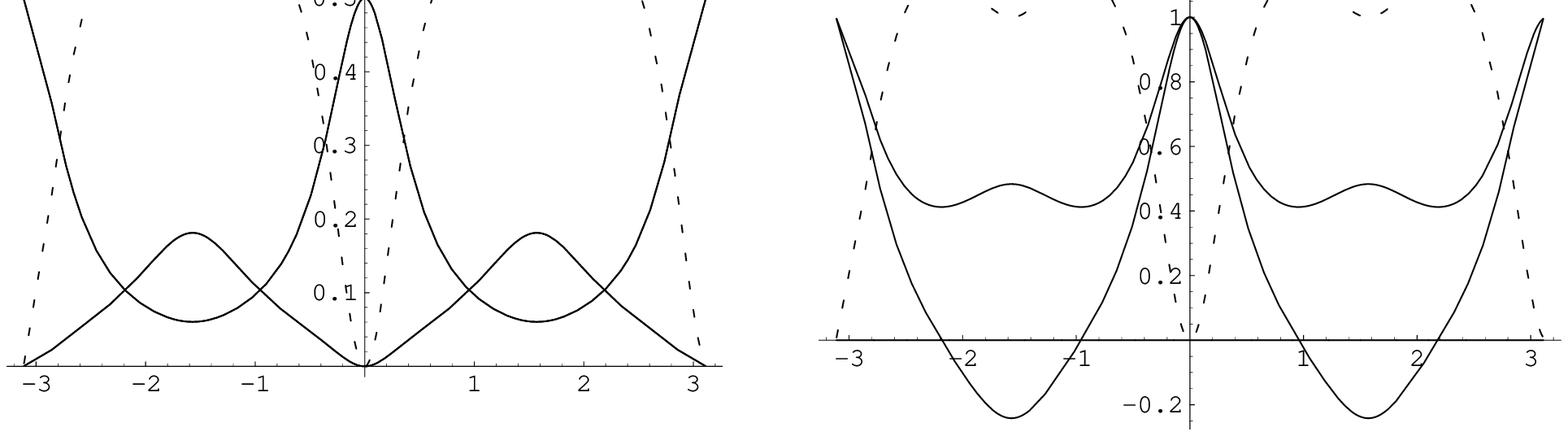}} \par}

\caption{\label{fig:DLinPumpSignal}Floresence and \protect\( D_{3/2}\protect \) state
populations as a function of polarization angle of linearly polarized red cooling
laser.}
\end{figure}

Numerical solutions of the floresence and populations as a function of polarization
in figs.\ref{fig:SPumpingSignal},\ref{fig:DCircPumpLabel},\ref{fig:DLinPumpSignal}.
Scalar, dipole and quadrupole moments of the resulting populations are also
shown, these are helpful in the later analysis and the basis vectors are defined
in that discussion. With properly normalized basis vectors, the components are
given by \( \sigma _{z}^{\left( k\right) T}p \). 


\subsubsection{Polarization Errors in Pumping Beam}

\label{Sec:FinalPumpedDStates}

These plot show that the final states for arbitrary polarizations and so indicate
that they are not particularly sensitive to errors in polarization. However
the assume perfect alignment of the beams relative to the magnetic field and
it is not immediately clear that small perturbations of this alignment do not
yield dramatic differences in the final pumped state. This is easily illustrated
for the \( D \) state. 

Consider circular pumping, where ideally only the \( D_{-3/2} \) and \( D_{-1/2} \)
states are coupled so that the ion ends up in the \( D_{+3/2} \) and \( D_{+1/2} \)
states with relative populations of \( 1:3 \) as discussed previously. Keep
the polarization correct by now alter the alignment of the beam. This gives
the electric field a small \( \hat{z} \) component which will drive the \( \Delta m=0 \)
transition, which then couples the \( D_{+1/2} \) state is now coupled as well
to the \( P_{+1/2} \) state. This might result in the final state being given
by the ion only in the \( D_{+3/2} \) state as it is the only uncoupled state.
But with this misalignment the beam is now actually slightly polarized in the
\( x=y \) plane so in fact \( \Delta m=-1 \) transitions are driven and the
\( D_{+3/2} \) state is not uncoupled. Clearly now the solution depends on
determining the relative rates which requires the correct form of the amplitudes
of all the transitions for arbitrary fields. 

\begin{figure}
{\par\centering \includegraphics{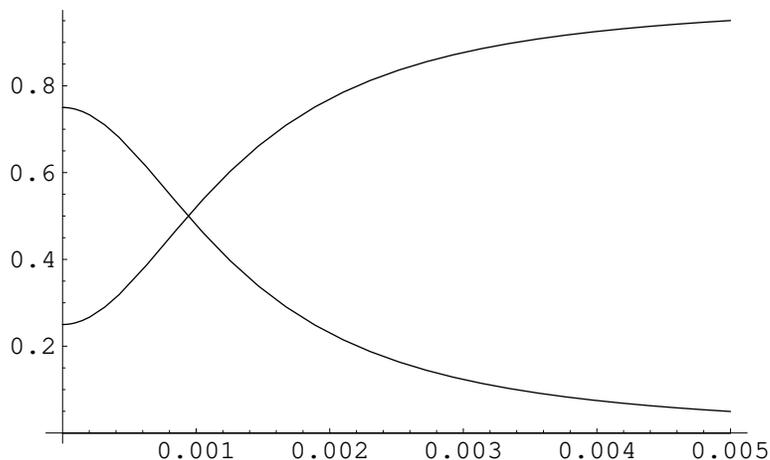} \par}

\caption{\label{Fig:DPumpPopsWithAlignPert}\protect\( 5D_{+3/2}\protect \) and \protect\( 5D_{+1/2}\protect \)
state pumped populations as a function of red laser alignment perturbations,
\protect\( \hat{\varepsilon }_{R}\cdot \hat{B}\protect \).}
\end{figure}

Fig.\ref{Fig:DPumpPopsWithAlignPert} show the populations of the \( 5D_{+3/2} \)
and \( 5D_{+1/2} \) states with \( \sigma _{r}=1 \), \( \alpha _{r}=0 \)
, as a function of \( \hat{\varepsilon }\cdot \hat{B} \) with perfectly aligned
and purely linearly polarized blue. As \( \hat{\varepsilon }\cdot \hat{B}=0 \)
the populations are \( 1:3 \) respectively as previously discussed. 
For even very small perturbations in the alignments the resulting residual \( \Delta m=0 \)
transition rate quickly cleans out the \( 5D_{+1/2} \) state and results in
the final state being instead composed of full population in only the \( 5D_{+3/2} \)
state. This perturbed solutions actually remains very accurately stable until
very large misalignments, \( \hat{\varepsilon }\cdot \hat{B}\approx 0.1 \).
Aligning the pumping beam to a part in \( 10^{4} \) would require excessive
effort and the most likely beam geometry is somewhere in the stable, misaligned
region so for practical applications this perturbed solution should be considered
to be the final pumped state.

The other pumping geometries, linear in the \( D \) state, and circular in
the \( S \), are not similarly sensitive to this same kinds of misalignment,
they remain approximately the same until very large misalignments.

\subsubsection{Polarization Errors in Cleanup Beam}

\begin{figure}
{\par\centering \includegraphics{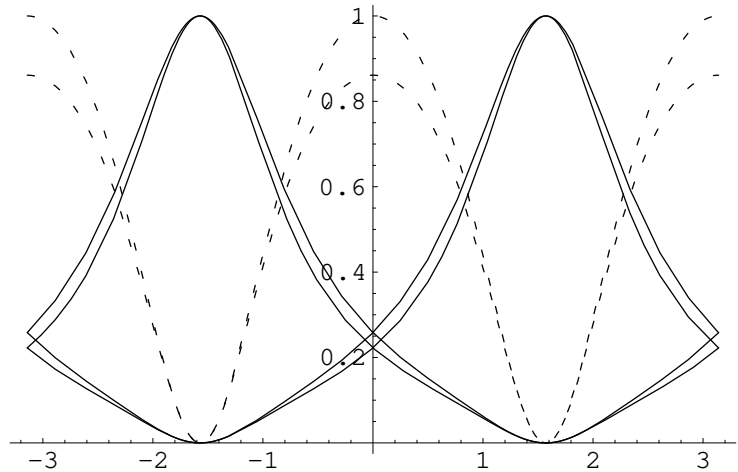} \par}

\caption{\label{Fig:PumpingSignalWithNonIdealCleanupPolarizations}Fluorescence and
Ground State populations and a function of circular polarization of blue cooling
laser for linearly polarized and partially circularly polarized red laser.}
\end{figure}

The data plotted for ideal alignments as a function of pump beam polarization
assumes that the complementary cleanup beam is perfectly set to its ideal polarizations
which equally couple all spin states in the other, unpumped, energy levels.
For these cases this is always a linear polarization so that that any polarization
of the final states is completely determined by the polarization of the pump
beam. This ideal is not realized in practice and residual bits circular polarization
in the cleanup beam alter the populations. This residual polarization will be
small, and it turns out that even it is not the effect on the populations is
small, even far from the region of maximum pumping and, of course, at that maximum
point there is exactly no change since the desired states are exactly uncoupled.\ref{Fig:PumpingSignalWithNonIdealCleanupPolarizations}

\subsubsection{Floresence and Pumped Spin State Populations}

The results are qualitatively intuitive. For polarizations and alignment that
significantly reduce the coupling to any state, the floresence decreases as
the ion gets stuck in that state and can't scatter photons as rapidly. This
reduction in count rate gives a measure of the time spent in the pumped state
by providing an estimate of the ratio of the excitation rate out of the pumped
states \( R \) to the return rate to those same states \( \Gamma  \). At equilibrium
the rates times the occupation probabilities for the set of pumped states, \( p \),
and the set of all other states, \( 1-p \), will be equal, \( pR=\left( 1-p\right) \Gamma  \)
so that for \( R<<\Gamma  \), \( p=1/\left( 1+R/\Gamma \right) \approx 1-R/\Gamma  \).

Consider an ideally uncoupled state or set of states. An ion that is excited
out of this set will be quickly returned to it with some blue photons generated
along the way. For pumping in the ground state at least one blue photon is generated
in the decay back to the uncoupled state, but more could be emitted by repeated
decays to the other spin state. In the \( D \) state it is possible that no
blue photons are generated if the ion immediately decays back to the \( D \)
state, but it is more likely to make a few trips through the ground state first.
In either case there is a well defined average number of blue photons generated
for every residual excitation out an uncoupled state and the floresence is simply
proportional to the sum of the occupation probability times the residual excitation
rate for each uncoupled state \( \sum _{\left\{ m|uncoupled\right\} }R_{m}p_{m} \).
For the ground state there is only one ideally uncoupled state, for the \( D \)
state there are two but their populations are simply related at and near the
region of optimal pumping for deviations in the appropriate polarization, for
linear \( \Delta m=0 \) pumping the populations of the \( m=\pm 3/2 \) states
are equal for arbitrary polarization angles, and for \( \Delta m=+1 \) pumping
the \( m=3/2 \) population is 3 times the \( m=1/2 \) population for arbitrary
circular polarizations. Then total population and excitation rate of any set
of states can be treated as a single state,

\begin{eqnarray*}
R_{m_{1}}p_{m_{1}}+R_{m_{2}}p_{m_{2}} & = & R_{m_{1}}p_{m_{1}}+\alpha R_{m_{2}}p_{m_{1}}\\
 & = & p\left( R_{m_{1}}+\alpha R_{m_{2}}\right) \\
 & = & pR
\end{eqnarray*}

\begin{figure}
{\par\centering \includegraphics{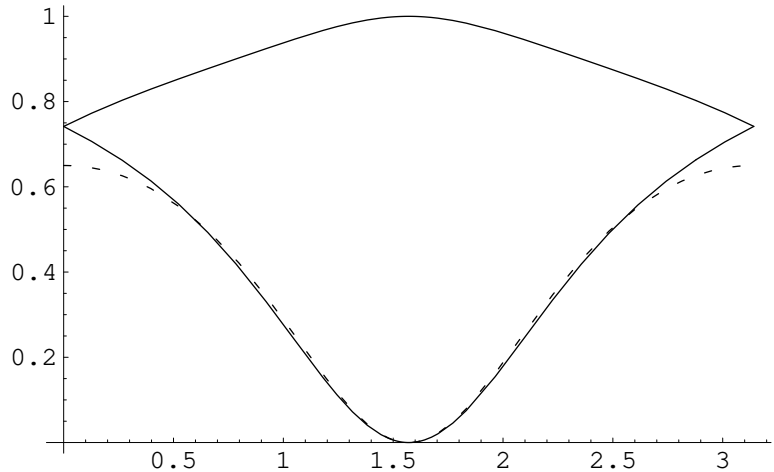} \par}

\caption{\label{Fig:FlouresencePopulations}Floresence and complement of ground state
populations as a function of blue polarization.}
\end{figure}

When the excitation rate out of the ideally uncoupled states is much slower
than the other excitation and decay rates \( p \) will be very close to 1 as
the ion is not frequently excited out of the states. In this limit the floresence
is then just proportional to the total residual excitation rate. As as rate
increases the probability to be in that state decreases but doesn't get far
from 1 until the residual rate approaches the cleanup and decay rates, \( R\approx \Gamma  \).
Here the occupation probabilities of all states are similar and so \( p \)
is still of order 1, the examples above show \( p\approx 0.2 \) where the populations
cross. At this point the floresence is again limited by the decay and cleanup
rates, but for most of the way the floresence is simply proportional to the
residual excitation rate. This turns out to be an even better approximation
in practice because of how the relative populations of the other states increase
far from the exact pumping point. fig.\ref{Fig:FlouresencePopulations} show
that the floresence is approximately proportional to the pumped state populations
even in for relatively large deviations from the ideal pump polarization.

Taking this to be approximately true for all rates, an estimate of the residual
excitation rate at the optimal pumping point, \( R \) relative to the excitation
rate out of the same states when no pumping occurs, \( R_{0} \) which is then
about the same as the decay rate, is given simply by the ratio of the floresence
at the pumping point \( \gamma  \) to the maximum floresence \( \gamma _{0} \),
\( R/R_{0}\approx \gamma /\gamma _{0} \). In the region where no pumping occurs
\( R\approx \Gamma  \), so \( R/\Gamma \approx \gamma /\gamma _{0} \) so that
finally the occupation probability is given simply by \( p\approx 1-\gamma /\gamma _{0} \).
For these measurement \( \gamma /\gamma _{0} \) is typically about \( 1/10 \)
to \( 1/20 \) so that \( p\approx 0.90-0.95 \) as mentioned above.

\subsection{Pumping and Field Alignment}

\label{Sec:PolarizationPumping}

The pumping signal can be used to align the magnetic field relative to the beam
axis and polarizations and optimize polarizations. The applied magnetic field
is not precisely known due to imperfections in the construction or alignment
of the coils generating the fields. There are also offsets from other magnetic
sources which in the case of a local soft ferromagnetic material that can even
depend on the applied field, as the directions of the material's domains are
altered by the applied field or ambient temperature, so that the total field
can be a complicated or even unknown function of the current supplied to the
coils, time and temperature. Nearby probes are little help since unknown sources
giving unknown offsets may also include unknown gradients so that knowing the
field, however precisely, a few centimeters away from the ion provides no information
about the field at the ion. The pumping signal allows the ion to be used as
a probe.

For example, pumping occurs for a magnetic field parallel to the polarization
axis of a linearly polarized red laser. This provides a means of zeroing the
magnetic field along the beam axis. A large field in the general direction of
the red laser polarization axis can be applied. If this field is sufficiently
large enough to define the direction of the resulting total field some reduction
in the floresence due to pumping should be visible. Then, both the angle of
polarization can be adjusted, with a half-wave plate, and the magnetic field
along the beam axis can be adjusted to minimize the floresence. As previously
mentioned, this typically occurs at about 1/10th to 1/20th of the unpumped count
rate, though it is ideally zero. The applied field can be reduced a bit at a
time and the other parameters adjusted again until the applied field is the
desired size. At this point the magnetic field is aligned with the red laser
polarization axis, perpendicular to the beam axis. If particular direction perpendicular
to the beam axis is desired, vertical for example, the polarization direction
can stay fixed and another magnetic field component adjusted instead.

A similar procedure can be used to align the field along the beam axis. With
circularly polarized red or blue pumping will occur when the magnetic field
is parallel to the beam axis. With an initially large applied field the pumping
signal can be used to null the components of the magnetic field perpendicular
to the beam axis. 

When the directions of the applied fields are well known these procedures are
sufficient to determine the magnetic field offset and with that information
the field can be reliably set to any desired direction. More often the direction
of the applied fields are not precisely known. In this case this alignment by
pumping can be done for a variety of directions and field sizes. Eventually
there are enough constraints that the size and direction contributions from
all the coils can be determined. This processes is a bit easier if the total
size of the magnetic field is known, this can be done using the spin resonance
techniques discussed later and significantly reduced the number of field configurations
that must be tested. 

A peculiar feature of the structure of the \( D_{3/2} \) to \( P_{1/2} \)
transition can also be used isolate the offset. For this transition there are
always two linear combinations of \( D \) states that are not coupled by a
red laser with fixed polarization, \ref{Sec:M+NZeros}, they are not necessarily
combinations generated from just spatial rotations though they obviously are
for the cases of linear and circular polarization. If there is no magnetic field
to drive the ion out of these states, the ion gets stuck in these \( D \) states
and pumping occurs just as in the cases already discussed. The floresence drops
and this polarization pumping can be used to locate the point where the magnetic
field is zero by adjusting the applied field until the PMT count rate is minimized.
Some care must be taken to check that this minimum is independent of polarization
so that the field is not really zero but just in a direction suitable for pumping
with the laser polarizations used.

\label{Sec:FieldCoilDependance}

If the field offset is fixed, some combination of these procedures gives enough
information that the resulting field as a function of coil currents can be completely
determined. In practice this provided a model of the applied field that was
able to predict the size of the resulting field to only about \( 10\% \), though
for smaller variations of the applied field agreement was better than a few
percent. This suggests a significant contribution from some ferro-magnetic materials
that would be redirected by the applied field giving an applied field dependent
offset. This is an important observation to consider when trying to understand
the source of larger than expected spin resonance transition widths. In principle,
these contributions could also be identified by a careful study of the resulting
field, but no such attempt was made other than to confirm that the contribution
was large and nonlinear.

The pumping signal could also be used to tune the polarizations of the lasers.
They can be measured independently and are well known outside the chamber, but
the final optics, and in particular the window needed to deliver the lasers
to ion can alter the polarization after any point where it could be measured
before reaching the ion. Any resulting induced circular, or more generally elliptical,
polarization could be compensated for by maximizing the pumping signal, this
time as a function of laser polarization. This is not completely reliable as
there may be continuous flat directions along some combination of polarizations
and field directions that always give pumping, \cite{Schacht00}, so that some
combination of imperfectly polarized beams and a slightly misaligned magnetic
field can also give maximal pumping. As a result the pumping signal can not
necessarily be used to optimize both the magnetic field and laser polarizations,
but some combination of pumping and spin resonance measurements to determine
the resulting magnetic field size can be used to determine the magnetic field
independent of laser polarizations so that the field can then be set reliably
to optimize polarizations. 

With enough work, all these imperfection can be uncoupled and all field and
polarization parameters determined and optimized. In practice, this sort of
precision isn't really important. The most important result of pumping is that
it leaves a spin state that is significantly different than an unpolarized state.
This is achieved in any configuration giving a good pumping signal, even including
the case of pumping as a result of zero magnetic field. The lack of floresence
implies occupation of some uncoupled state and not all the spin states are uncoupled
so the ion can't be unpolarized. Lack of precise knowledge of the magnetic field
and laser polarizations then just results in lack of knowledge of the resulting
pumped state. The pumped state could be a single single spin state defined along
an unknown axis, or some other combination of state not corresponding to a single
rotated state, but it is still well defined, if unknown. It does help to have
a coarse idea of the parameters to keep track of things, as discussed far more
later, maximizing the S/N of the spin detection methods requires that the probed
state be the same as the pumped state and coordinating this is easier when fields
and polarizations are at least coarsely known. But then coarse knowledge is
sufficient and further efforts towards precise knowledge is less important.

\subsection{Final States}

These solutions provide the steady state populations of all of the states during
pumping. If the beams could be shut off instantly and simultaneously, or attenuated
in unison so that their relative intensities stayed fixed, then these would
also be the populations after pumping. General this can't, or isn't done in
practice. The beams are shut off sequentially so that the ion is sure to at
least end up in the desired energy level, if not the ideal spin state. That
is, for pumping in the \( D \) state, the red beam is shut off first, so that
if the ion happened to be in the ground state, it is moved to the \( D \) state.
Similarly for pumping in the \( D \) state. This final pulse from the clean
up beam can alter the relative populations of the pumped level. 

For polarizations having maximal pumping, cases where the floresence has dropped
to zero, this modification will be very small. In these cases the probability
to be in the other state is very small so that the final cleaning out of that
state will not add much extra population to the pumped state, and so not significantly
alter its distribution. For polarizations still yielding significant floresence,
there is a significant population in the other state and the redistribution
of that population to the pumped state during the final cleanup pulse is an
important modification.

It is not immediately obvious that these latter cases are important, but it
will be useful in the measurements discussed later to repeat a particular measurement
for a variety of initial polarizations. For circularly polarized pumping, there
are two fully pumped initial states available. They are generated by using the
two oppositely circularly polarized pumping beams. For linear pumping in the
\( D \) state there is only the single fully pumped state. In either case it
can be useful to have even more initial states. These will be generated by using
non ideal polarizations for the pump beams and then require a careful analysis
of the effects of the final cleanup pulse on the spin distribution of the pumped
state. 

One convenient initial state for reference is an unpolarized state. For the
ground state is it clear how this unpolarized state can be generated. With the
fully pumped states given by a left and right circularly polarized blue beam,
and linearly polarized red, both propagating parallel to the quantization axis,
a simple switch to linearly polarized light now couples both ground state spin
sublevels equally. With this spin symmetry it is obvious that while pumping
the populations of all state in the \( S \) and \( D \) levels are spin sign
independent, and since no state is completely uncoupled, the populations of
all the state in both levels are comparable. A quick numerical result gives,
\begin{eqnarray*}
S & = & \left( \begin{array}{cc}
0.26 & 0.26
\end{array}\right) \\
D & = & \left( \begin{array}{cccc}
0.06 & 0.18 & 0.18 & 0.06
\end{array}\right) 
\end{eqnarray*}
The final pulse is the linearly polarized red cleanup beam. Again the couplings
are spin sign independent, so the leftover population in the \( D \) state
must empty into the \( S \) state symmetrically to yield, 
\[
S=\left( \begin{array}{cc}
0.5 & 0.5
\end{array}\right) \]

Trying to generate and unpolarized \( D \) state in this way is not as clear.
Again the result must be spin sign independent, but the relative populations
of the \( m=\pm 3/2 \) states to the \( m=\pm 1/2 \) states is not immediately
obvious. The correct final state can be determined with the rate equations for
this system,

\begin{eqnarray*}
\dot{S} & = & -L_{S}S+\left( R_{ST}+\Gamma _{S}\right) L_{P}^{-1}(R_{S}S+R_{D}D)\\
\dot{D} & = & -L_{D}D+\left( R_{DT}+\Gamma _{D}\right) L_{P}^{-1}(R_{S}S+R_{D}D)
\end{eqnarray*}
For simplicity consider any stimulated decay rate to be negligible compared
to radiative decays so that, in particular, \( L=\Gamma ^{-1} \). With \( R_{D}=0 \),
implying \( L_{D}=0 \), this gives,

\begin{eqnarray*}
\dot{S} & = & -(L_{S}-f_{S}R_{S})S\\
\dot{D} & = & f_{D}R_{S}S
\end{eqnarray*}
The final \( D \) state populations during the last bit of application of the
cleanup beam are given by the integration,

\[
D(\infty )=D(0)+\int _{0}^{\infty }dt\dot{D}==D(0)+f_{D}R_{S}\int _{0}^{\infty }dtS\]
This can be written formally in term of the solutions of the ground state populations
using the eigenvalues of \( L_{S}-f_{s}R_{s} \). This gives solutions of the
form. 

\[
S=av_{1}e^{-\gamma _{1}t}+bv_{2}e^{-\gamma _{2}t}\]
Which can easily be integrated.

\[
\int _{0}^{\infty }dtS=\frac{av_{1}}{\gamma _{1}}+\frac{bv_{2}}{\gamma _{2}}\]
Giving, formally, the final \( D \) populations,
\[
D(0)+f_{R}R_{S}(\frac{a}{\gamma _{1}}v_{1}+\frac{b}{\gamma _{2}}v_{2})\]

These eigenvalues and eigenvectors can also easily be determined numerically.
For circular pumping in the \( D \) state, using circularly polarized red,
linearly polarized blue, and magnetic field parallel to beam axis, linearly
polarized red light and the final blue cleanup pulse yields the state,

\[
D=\left( \begin{array}{cccc}
0.19 & 0.31 & 0.31 & 0.19
\end{array}\right) \]
This is not precisely unpolarized, the populations are certainly significantly
different from the pumped state, which will turn out to be all that is really
important for detecting transitions. For linearly polarized pumping in the \( D \)
state , using linearly polarized red and blue, and magnetic field perpendicular
to the beams, a circularly polarized red pump beam and final blue cleanup pulse,

\[
D=\left( \begin{array}{cccc}
0.28 & 0.22 & 0.22 & 0.28
\end{array}\right) \]

Solutions for other polarizations for the \( D \) state are given from the
same calculations. The same procedure can be used to determine final ground
state spin distributions for arbitrary pump polarizations, though the analysis
is a bit more complicated as four eigenvectors and eigenvalues are the required.
These few solutions show that a significantly different initial spin state is
available with a simple change of pump beam polarization and approximately unpolarized
states are possible.

\section{Spin State Detection}

While setting the initial spin state of a single ion is relatively straightforward,
determining an unknown final state is more subtle. In a vapor this spin state
can be determined with a probe laser in a variety of ways, but none can be directly
applied to the case of a single ion.

\subsection{Vapor Methods and Single Ions}

With a magnetic field parallel to the propagation direction, resonant probe
light will couple \( \Delta m=\pm 1 \) transitions. As an illustration, consider
spin detection in the ground state. The issues for the \( D \) state are almost
identical, only details of the structure of the transition are different.

\subsubsection{Absorption}

On a \( j=1/2\rightarrow j'=1/2 \) transition, atoms in the \( S_{1/2,+1/2} \)
state will absorb photons of only one helicity, while atoms in the \( S_{1/2,-1/2} \)
state will absorb photons of the opposite helicity. This picture would be slightly
more complicated in more general cases such as a \( j=1/2\rightarrow j'=3/2 \)
transition where both spin states can absorb photons of either chirality, but
in this case the relative rates are different and you simply get atoms in one
particular spin state preferentially, rather than exclusively, absorbing photons
of a particular chirality.

In either case the net effect is qualitatively the same, with an excess of atoms
in one spin state, more photons of one helicity will be absorbed and scattered
out of the beam than of the opposite helicity. With incoming probe light linearly
polarized, the net effect would be an induced elliptical polarization of the
probe after interacting with the atoms by an amount, and direction, dependent
on the relative spin states of the atoms.

A direct application of this method to a single ion could involve using the
blue cooling laser as a probe and looking for changes in its final state as
a result of the spin state of the ion. In this case the reduction of the system
from a macroscopic collection to a single ion is fatal. The resonant interaction
of the light with the ion destroys the spin state of the ion. After absorbing
a photon and being excited to the \( P_{1/2} \) state, the ion can decay to
either spin sublevel in the ground state. The branching fraction for decay to
each sublevel is different, the ratio is given simply by Clebsch-Gordan coefficients
as 2:1, so the final state of the ion, and probability for absorption of a second
photon is slightly different for different initial spin states. For this particular
system, there is also the possibility of a round trip through the \( D_{3/2} \)
state before returning to the \( S \) state further diluting the final dependence
on the initial spin state. As a result, after absorbing even just a few photons,
the information about the initial spin state of the ion is lost and the ion
will absorb photons of either helicity without preference. 

In this way, the final state of the probe beam would be altered by the loss
of at most a few more photons of on helicity than another and the resulting
change in the probe's polarization state would be undetectable.

\subsubsection{Optical Rotation}

A more common method in vapors is to use off resonant light as a probe. Here
absorption is negligible and instead the net effect is easiest to understand
as a difference in the index of refraction for left or right circularly polarized
light propagating through the vapor by an amount dependent on the relative spin
states of the atoms. This will give a different relative phase of the two components,
and appear as a rotation of the plane of polarization of the probe after it
exits the vapor.

The amount of rotation depends linearly on the number of atoms, so again this
method fails for the single ion because of number. A rotation of even a few
full revolutions in a vapor with \( 10^{10} \) or more atoms corresponds to
an undetectably small rotation for a single particle though perhaps it is possible
that variations of this method could be made to work by, for example, putting
the ion and the probe light in a cavity so that the same light could interact
with the ion many times before being detected.

\subsubsection{Fluorescence}

Finally, consider again a resonant probe beam. Now, instead of looking at the
final state of the beam, which is basically the light that didn't interact,
look instead at the scattered probe light. Using a circularly polarized beam
to excite, for example, \( \Delta m=+1 \) transitions, only atoms in the \( S_{-1/2} \)
can be excited to the \( P \) state, and so only those will scatter the probe
light. Atoms in the \( S_{+1/2} \) state will make no contribution to the fluorescence.
The population among spin states then appears as a change in the number of detected
photons from some baseline corresponding to an unpolarized state. An increase
in the number of atoms in the \( S_{-1/2} \) state is an increase in the number
of atoms able to absorb the incoming probe light and so an increase in fluorescence.

A decay from the \( P \) state can be to either ground state sublevel. If the
decay is to the \( -1/2 \) state, another photon can be scattered by the same
process. Eventually the final state will be \( m=+1/2 \) and fluorescence must
stop, so each particle can generate at most a few photons. For a single ion
in a trap these few photons are the entire signal but with the previously mentioned
detection efficiency on the order of a part in a thousand these photons will
most often not be seen and so about a thousand trials are necessary to begin
to be able to distinguish spin states. Detection is possible, but sensitivity
is far too low to be of any practical use, \ref{sec:StateManipulationAndDetection}.

The situation can be slightly modified when considering other transitions for
spin detection such as using the \( P_{3/2} \) state rather than the \( P_{1/2} \).
With probe light again set to excite \( \Delta m=+1 \) transitions, an ion
initially in the \( S_{1/2,+1/2} \) state will be excited to the \( P_{3/2,+3/2} \)
state. From there, if the ion decays back to the ground state it can only go
back to the \( S_{+1/2} \) sublevel, as it can change \( m \) by at most 1.
So in this case the ions starting in the \( S_{+1/2} \) level can scatter photons
as long as the probe beam is on, but now, ions starting in the \( S_{-1/2} \)
level can also. From that level, ions are excited to the \( P_{+1/2} \) state
where they can decay back to either ground state, both of which can again be
excited and in the end the ion will end up in the same loop as when starting
from the \( S_{+1/2} \) state. The excitation rate is slower out of the \( -1/2 \)
state by half, so one spin state may generate a few more photons than the other,
but low detection efficiency also results in this method having low sensitivity
and even this difference is slightly eroded by possible decay to the \( D_{5/2} \)
state from the \( P_{3/2} \) state which results in further loss of information
about the initial spin state.

\subsection{Shelving}

In each case direct application of vapor methods to spin state detection in
a single ion fails because of low sensitivity and detection efficiency and so
something fundamentally different must be considered. At the higher level, and
lower resolution, of distinguishing the state of the ion between different energy
levels, similar difficulties exist and these have been solved with great success
using now well established shelving methods, \ref{sec:StateManipulationAndDetection}.
Modification of these methods to make them spin sensitive finally becomes a
practical means of spin state detection.

\subsection{Shelving Laser}

To generalize these kind of techniques to detecting spin states, the simplest
idea to consider, and most straightforward to analyze, is a modification of
shelving by direct coupling to the shelved state with a 1.76\( \mu  \)m laser.
As with pumping, the laser polarization and magnetic field direction can be
chosen to couple only particular spin states to make transition rates spin dependent.

Consider again \( \Delta m=+1 \) transitions. The structure of the \( j=1/2\rightarrow j'=5/2 \)
transition doesn't leave either state completely uncoupled, but the \( m=+1/2\rightarrow m'=+3/2 \)
is faster then the \( m=-1/2\rightarrow m'=+1/2 \) transition by the ratio
\[
\left| \left\langle \frac{5}{2},\frac{3}{2}|2,1;\frac{1}{2},\frac{1}{2}\right\rangle \right| ^{2}/\left| \left\langle \frac{5}{2},\frac{1}{2}|2,1;\frac{1}{2},-\frac{1}{2}\right\rangle \right| ^{2}=2\]
 This can be exploited in two ways,depending on the nature of the light used
to drive the transition.

\begin{figure}
{\par\centering \resizebox*{1\textwidth}{!}{\includegraphics{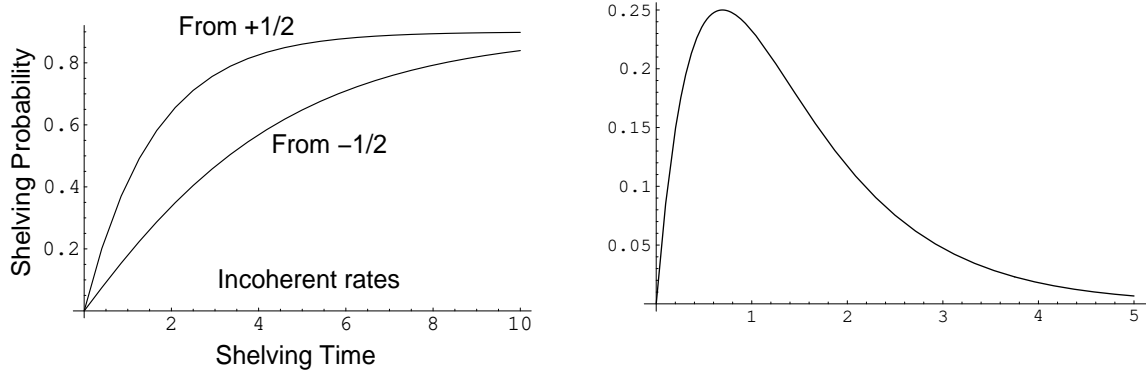}} \par}

\caption{\label{Fig:IncoherantShelvingLaser}Shelving probability as a function of time
for circularly polarized shelving laser.}
\end{figure}

The most conservative assumption is that the laser linewidth is sufficiently
broad, and the transition rate sufficiently slow, that the laser is effectively
broadband and the transition is being driven incoherently. In this case probabilities
evolve as simple first order rate equations. Neglect any spurious couplings
to unintended states, though they would be important at long times and modify
the steady state probabilities by the ratio of g-factors. This then gives simple
exponential excitation profiles with different time constants for the different
spin states, fig.\ref{Fig:IncoherantShelvingLaser}.
 The ground state spin can then be determined by measuring the shelving probability
since for a fixed time, the shelving probability is a function of the initial
spin state. 

Maximizing detection efficiency requires maximizing this difference. For short
times, neither spin state has a large shelving probability so the difference
is small. At long times both state have reached the same steady state value
and the difference is zero. The spin states can only be distinguished in a particular
range of times so the shelving time must be well chosen. With \( P_{\infty } \)
the steady state shelving probability and the rates given by \( \gamma _{1} \)
and \( \gamma _{2} \), the difference is \( \Delta P=P_{1}\left( t\right) -P_{2}\left( t\right) =P_{\infty }\left( e^{-\gamma _{1}t}-e^{-\gamma _{2}t}\right)  \).
This is maximized at the time \( t_{max}=ln\left( \gamma _{1}/\gamma _{2}\right) /\left( \gamma _{1}-\gamma _{2}\right)  \),
and at \( t_{max} \) the difference depends only on the ratio of the rates
\( \alpha =\gamma _{1}/\gamma _{2} \), 
\[
\Delta P_{max}=P_{\infty }\alpha ^{\alpha /\left( 1-\alpha \right) }\left( 1-\alpha \right) \]
Both \( t_{max} \) and \( \Delta P_{max} \) give the same result for \( \gamma _{1}\leftrightarrow \gamma _{2} \),
\( \alpha \leftrightarrow \alpha ^{-1} \) except for the sign of \( \Delta P_{max} \).
Generally the rates aren't known well enough to be able to use the optimal exposure
time. At time \( rt_{max} \) the difference becomes \( P_{\infty }\alpha ^{r\alpha /(1-\alpha )}\left( 1-\alpha ^{r}\right)  \).

\begin{figure}
{\par\centering \includegraphics{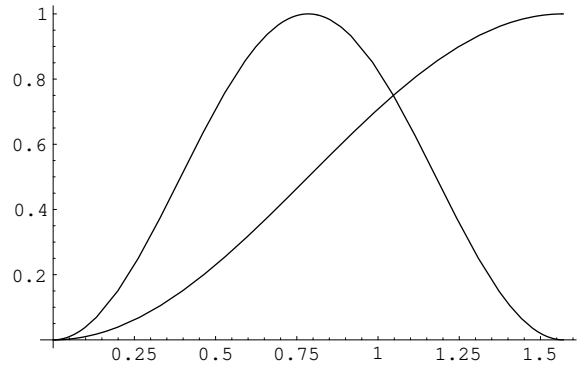} \par}

\caption{\label{Fig:CoherantShelvingLaserSpinProbe}Shelving probability as a function
of time for \protect\( S_{\pm 1/2}\protect \) states with coherant transitions
driven by a circularly polarized shelving laser.}
\end{figure}

This case gives \( P_{\infty }=1/2 \), \( \alpha =2 \), \( \Delta P_{max}=1/8 \)
at \( \tau \equiv Rt=ln2=0.7 \). The difference is small, though certainly
detectable. However it can be greatly improved if the transitions can be driven
coherently. With a narrow laser linewidth and sufficiently high transition rate,
the width of the laser can be neglected and assumed to be monochromatic. For
this case the probability just oscillates between an \( S \) and a \( D \)
state with a frequency given by the square of the matrix element. Since the
couplings from the two spin states are different, the oscillation frequencies
are different and again the states can be distinguished by a difference in the
shelving probability as a function of time. With a ratio of rates of two, or
any integer, the difference can be 100\%. By driving the transition until the
slower transition, here the \( m=+1/2\rightarrow m'=+3/2 \), has completed
a half cycle, the faster transition has completed a whole cycle. A \( \pi  \)
pulse for one is a \( 2\pi  \) pulse for the other and the shelving probability
is the ideal 100\% for one spin state and 0\% for the other, fig.\ref{Fig:CoherantShelvingLaserSpinProbe}.

This can work from the \( D_{3/2} \) state as well, though the energies are
very different. The \( D_{3/2} \) to \( D_{5/2} \) is far in the infrared
at \( 12\mu m \). The additional structure of the \( D_{3/2} \) state makes
things a little more complicated. For \( \Delta m=0 \) transitions, the ratio
of frequencies is,

\[
\left| \left\langle \frac{5}{2},\pm \frac{3}{2}|2,0;\frac{3}{2},\pm \frac{3}{2}\right\rangle \right| ^{2}/\left| \left\langle \frac{5}{2},\pm \frac{1}{2}|2,0;\frac{3}{2},\pm \frac{1}{2}\right\rangle \right| ^{2}=6\]
This gives a 100\% discriminant between \( m=\pm 3/2 \) and \( m=\pm 1/2 \)
at \( t=T/2 \), with \( T \) the period for the slower \( 1/2\rightarrow 1/2 \)
transition. For \( \Delta m=+1 \) transitions there are four rates,
\[
R_{\frac{3}{2},\frac{5}{2}}:R_{\frac{1}{2},\frac{3}{2}}:R_{-\frac{1}{2},\frac{1}{2}}:R_{-\frac{3}{2},-\frac{1}{2}}=30:2:25:27\]

Spin detection using the shelving transition directly in this way yields the
ideal 100\% detection efficiency but with the considerable technical complications
of a very well stabilized, high power laser and carefully timed pulses. Driving
the transition coherently with a laser stabilized to even 100kHz requires a
transition rate of about a MHz. This rate electric field strengths of \( 1000V/cm \)
using, for example, a \( 100mW \)mW laser focussed to a \( 100\mu m \) spot.
This may be pursued if the S/N of other methods proves insufficient.

\subsection{Shelving Lamp}

Shelving through the \( P_{3/2} \) state can also be modified to be sensitive
to spin, again by choosing polarizations and magnetic fields that drive only
\( \Delta m=+1 \) transitions. Here also both ground state spin levels are
coupled and a spin discriminant requires exploiting different relative rates.
In this case, coherent transitions would not be useful. On resonance, with the
transition rate very much faster than the decay rate out of the \( P \) state,
the ion can simply be considered to be in either relevant \( P \) state with
50\% probability, from which it could decay to the shelved \( D_{5/2} \) state.
But the decay rates are spin independent so the ground state sublevels cannot
be distinguished.

As the transition rate slows, the populations evolve less coherently and the
spin states can be distinguished. The largest disparity is when the transition
is completely decoherent. This situation is also the easiest to analyze and
the most appropriate physically since in the current system the shelving transitions
are being driven broadband by a discharge lamp.

\subsubsection{Rate Equations}

The previous analysis used for pumping can be used directly. The structure of
the system is a bit different in that the \( P \) state can decay to either
of two \( D \) states, but this doesn't change the form of the equations. With
\( R^{D}=0 \) in this case, and also assuming that \( \Gamma >>R^{S} \), the
rate equations from sec.\ref{Sec:PumpingAnalysis} become,

\begin{eqnarray*}
\dot{S} & = & -L^{S}S+\Gamma ^{S}P\\
\dot{P} & = & -L^{P}P+R^{S}S\\
\dot{D}_{3/2} & = & \Gamma ^{D_{3/2}}P\\
\dot{D}_{5/2} & = & \Gamma ^{D_{5/2}}P
\end{eqnarray*}

With \( \Gamma >>R \) the \( P \) population evolve quickly to a quasi-independent
steady-state value and we can take \( \dot{P}\approx 0 \). \( L^{P} \) include
only decays so is spin independent, \( L^{P}=\Gamma  \). Take \( R^{S}=R \)
as it is now the only excitation rate in the problem and drop the \( D \) state
terms, they will be given by normalization. This gives \( P=(R/\Gamma )S \)
and \( \dot{S}=\left( -L+(\Gamma ^{S}/\Gamma )R\right) S \). \( \Gamma ^{S} \)
is the matrix describing the relative rates of decays to particular spin states
and \( \Gamma  \) is just the usual total decay rate, so the overall rate drops
out. With \( \Gamma ^{S}_{m'm}=f\Gamma \gamma _{m'm} \) with \( f \) the branching
ratio to the \( S \) state compared to the \( P \) state and \( \gamma  \)
the appropriate matrix of squares of Clebsch-Gordan coefficients the rate equation
for the ground state becomes 
\[
\dot{S}=\left( -L+f\gamma R\right) S\]
\( f \) is well defined but not well known so left as a variable which also
allows for the easy application of this solution to other systems.

This leaves a single uncoupled equation. The \( D \) populations can be derived
from this solution but the spin details aren't necessary for this analysis.
The quantity of interest is the shelving probability, which will be the final
probability to be in the \( D_{5/2} \) state. Summing over spin state for either
\( D \) state,

\begin{eqnarray*}
\sum _{m}\dot{D}^{j}_{m} & = & \sum _{m}\Gamma ^{D_{j}}_{mm'}P_{m'}\\
 & = & f_{j}\Gamma \sum _{m}\gamma _{mm'}\left( R_{m'm''}/\Gamma \right) S_{m''}=f_{j}LS
\end{eqnarray*}

Each \( D \) state evolves with the same structure but a rate given by its
branching ratio. Then at any time, the relative population among the \( D \)
states, which is just that missing from the \( S \) states, is given by these
ratios. In this case this is 90\% to the \( D_{5/2} \). So the evolution of
populations among any of the states is given simply by the solution for the
\( S \) state.

\subsubsection{Solution}

Clearly the ground state spin can not be determined with a linearly polarized
probe beam, so consider pure circular polarization selecting \( \Delta m=+1 \)
transitions. The desired solutions are for the total population of the ground
state as a function of time when starting from a particular spin state. For
this 2 dimensional first-order linear system the general solution is a sum of
exponentials with 2 time constants. 

For this problem one of these solutions is trivial. As previously discussed
for pumping, and ion beginning in the \( S_{+1/2} \) state will be excited
to the \( P_{+3/2} \) state, at the excitation rate \( R \), from which it
can decay to the \( D \) states, and be shelved with 90\% probability, or back
to the ground state with probability \( f\approx .8 \) and so a rate \( fR \).
The decay to the ground state can only be back to the \( S_{+1/2} \) level
for this dipole decay to the ion stays in this cycle until it reaches a \( D \)
state. So the initial state \( s_{+}=\left( \begin{array}{cc}
1 & 0
\end{array}\right)  \) decays with time constant \( R(1-f) \) and the population in the ground state
decays with the simple single exponential.

The solution for an initial state of spin down is less obvious, though straight-forward.
The excitation rate from the \( S_{-1/2} \) state is a factor of 3 slower,
as given by a simple ratio of Clebsch-Gordan coefficients, 
\[
\left\langle \frac{1}{2},\frac{1}{2}|1,1;\frac{3}{2},\frac{3}{2}\right\rangle ^{2}/\left\langle \frac{1}{2},\frac{1}{2}|1,1;\frac{3}{2},\frac{1}{2}\right\rangle ^{2}=3\]
In this case the decay to the ground state can be to either spin sublevel and
a decay to the \( S_{+1/2} \) level is preferred
\[
\left\langle \frac{3}{2},\frac{1}{2}|1,0;\frac{1}{2},\frac{1}{2}\right\rangle ^{2}/\left\langle \frac{3}{2},\frac{3}{2}|1,-1;\frac{1}{2},-\frac{1}{2}\right\rangle ^{2}=2\]
So the branching fraction to the \( S_{+1/2} \) state is 2/3. The initial spin
information is quickly lost as the populations soon evolve identically and both
decay as if starting from the \( S_{+1/2} \) state.

These processes completely describe the evolution of the system. The repopulation
rate of the \( S_{-1/2} \) state is \( 1/3*fR(1/3)=fR/9 \), and of the \( S_{+1/2} \)
state is \( 1/3*fR(2/3)=2fR/9 \). This gives all the terms in the rate equation
coupling matrix.

\[
-L+f\gamma R=R\left( \begin{array}{cc}
-1+f & 2f/9\\
0 & -1/3+f/9
\end{array}\right) \]
The scalar overall excitation rate \( R \) factors out, and with \( \tau =Rt \),
can be scaled away. 

The detailed behavior for this particular problem is easily obtained with a
bit a algebra, but a more general analysis is not much harder and will be useful
for other processes to be considered later. An eigenvalue/eigenvector solution
is the most transparent. There will be two eigenvectors, call them \( v_{+}=\left( \begin{array}{cc}
1 & a_{+}
\end{array}\right) ^{T},v_{-}=\left( \begin{array}{cc}
a_{-} & 1
\end{array}\right) ^{T} \), and two eigenvalues \( \gamma _{\pm } \). The eigenvectors evolve simply
as \( v_{\pm }(t)=v_{\pm }e^{-\gamma _{\pm }\tau } \). The two initial spin
states will ideally be pure spin up and pure spin down, but to allow an easier
analysis of systematic errors involving impure polarization state, misalignment
of axis and incomplete pumping, and for later work with lifetimes and transitions,
it helps to consider the more general case an initial states.

The probabilities can be fully parameterized by a single variable, a convenient
form is \( p=1/2\left( \begin{array}{cc}
1+s & 1-s
\end{array}\right) ^{T} \), or \( p=1/2\left( \begin{array}{cc}
1 & 1
\end{array}\right) ^{T}+s/2\left( \begin{array}{cc}
1 & -1
\end{array}\right) ^{T} \) which is already expanded in terms of the usual spherical multipole moments.
To make things even more efficient, these basis vectors can be derived from
the \( 2\times 2 \) identity matrix and pauli matrices by operating them on
a column vector of \( 1 \)s, with \( 1=\left( \begin{array}{cc}
1 & 1
\end{array}\right) ^{T} \), as \( \left( \begin{array}{cc}
1 & 1
\end{array}\right) ^{T}=1\times 1,\left( \begin{array}{cc}
1 & -1
\end{array}\right) ^{T}=\sigma _{z}\times 1 \). For diagonal matrices, such as the \( 1 \) and \( \sigma _{z} \) used here,
this simply gives a column vector of the diagonal elements of the matrix. The
matrix dimensions will be clear from context and so the \( \times 1 \) can
also be dropped. Then the state is just written as
\[
p=\left( 1/2\right) +\left( s/2\right) \sigma _{z}\]
From this it is clearer that \( s \) is just the spin polarization, \( \left\langle s_{z}\right\rangle =p_{1/2}-p_{-1/2}=s \).

The state can be written in terms of the eigenvectors as,

\[
p=\frac{1}{1-a_{+}a_{-}}\frac{1}{2}\left( \left( \left( 1-a_{-}\right) +s\left( 1+a_{-}\right) \right) v_{+}+\left( \left( 1-a_{+}\right) -s\left( 1+a_{+}\right) \right) v_{-}\right) \]
 This immediately gives the time evolution by \( v_{\pm }\rightarrow v_{\pm }(t) \).
The absolute probability is not as important as its dependence on the initial
spin polarization, so it is convenient to just compare this result to some reference
state with some different \( s \). A natural comparison to the unpolarized
state with \( s=0 \). The \( s \) independent pieces just cancel to give,
\[
\Delta p\left( t\right) =p-p_{unpolarized}=\frac{1}{1-a_{+}a_{-}}\frac{s}{2}\left( \left( 1+a_{-}\right) v_{+}e^{-\gamma _{+}\tau }-\left( 1+a_{+}\right) v_{-}e^{-\gamma _{-}\tau }\right) \]
 The total population in the \( S \) state is just the sum of the populations
in each spin state. Summing the components of the column vectors gives,

\[
\Delta P=\sum _{m}\Delta p_{m}=\frac{s}{2}\frac{\left( 1+a_{+}\right) \left( 1+a_{-}\right) }{1-a_{+}a_{-}}\left( e^{-\gamma _{+}\tau }-e^{-\gamma _{-}\tau }\right) \]
 and the difference is just proportional to the difference of the initial spin
polarizations.

For this problem the rates are given by \( \gamma _{+}=1-f,\gamma _{-}=1/3-f/9 \).
For the eigenvectors, as argued above \( a_{+}=0 \) and it turns out \( a_{-}=-f/\left( 4f-3\right)  \).
This yields,
\[
\Delta P_{Max}=\frac{s}{2}(1-\frac{f}{4f-3})\alpha ^{\alpha /\left( 1-\alpha \right) }\left( 1-\alpha \right) \]
With \( f=3.6/(1+3.6)=0.78 \), this yields \( \Delta P_{Max}=0.12s \) at \( \tau =4.3 \).
This ideal \( \Delta P \) is reduced by the previously discussed branching
ratio to the \( D_{3/2} \) state of about 10\%, which gives a maximum shelving
probability of 90\%, and the shelved state detection efficiency, the probability
that a shelved state is detected as shelved which is less than 1 because the
finite lifetime of the \( D_{5/2} \) state might allow the ion to decay while
checking for shelving. The detection efficiency depends primarily on the time
taken to check for shelving, which in turn is limited by the PMT fluorescence
signal and background. Observations times are generally \( 0.2-1s \), and the
\( D_{5/2} \) state lifetime is about 10s, when the cooling beams are on, giving
detection efficiencies of \( 90-98\% \). These reduce the theoretical maximum
discriminant to at least 80\% of its ideal value, \( \Delta P\approx 0.09s \).
Ideally the initial state can be polarized fully in either of two opposite directions
for \( s=\pm 1 \), and the comparison of shelving probabilities between these
two cases is twice the \( \Delta P \) just given.

\subsubsection{Experiment}

\begin{figure}
{\par\centering \rotatebox{90}{\includegraphics{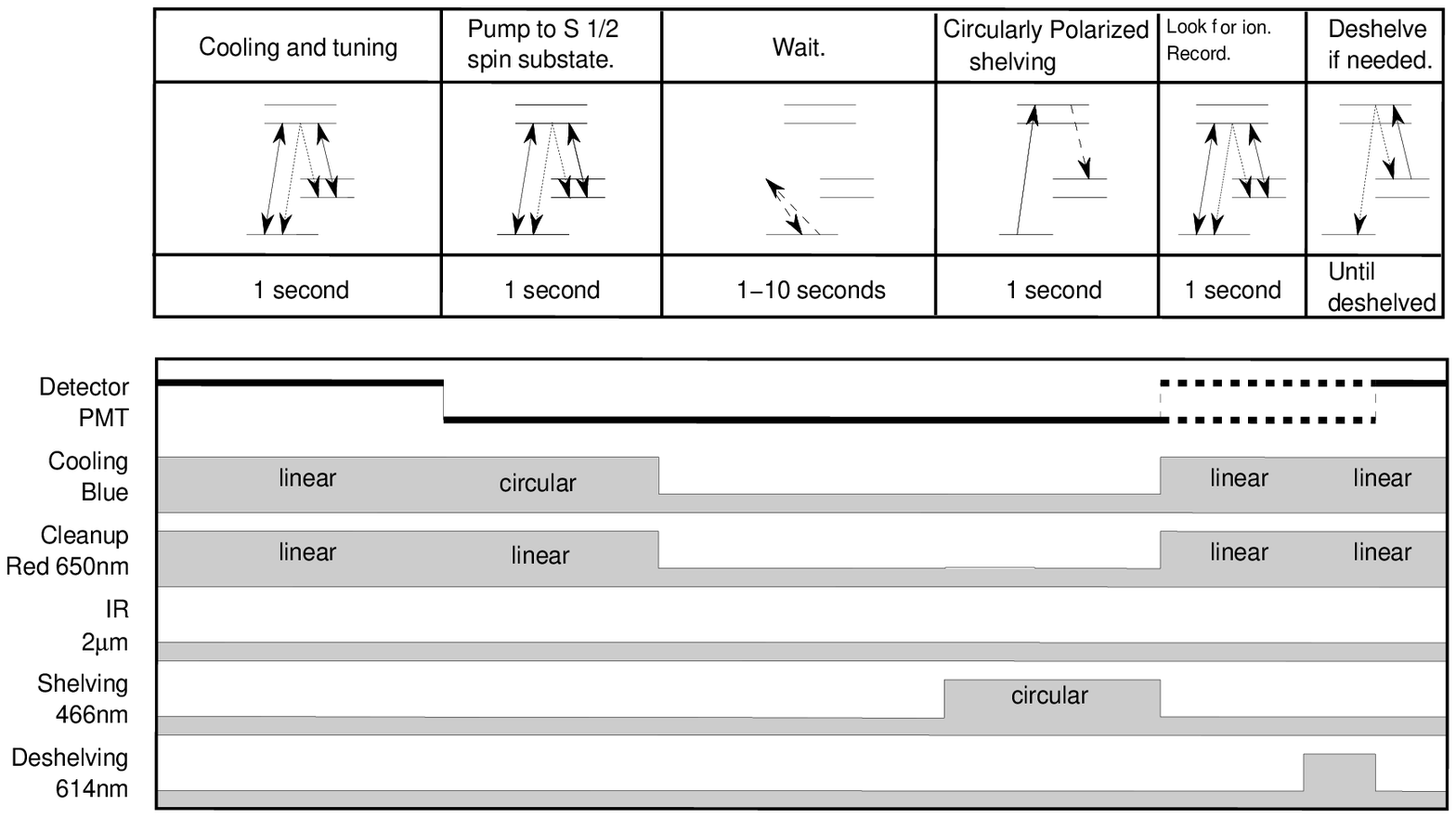}} \par}

\caption{\label{Fig:ChiralShelvingMeasurementSequence}Chiral Shelving Measurement Sequence}
\end{figure}

The resulting difference in shelving probability is smaller than for the techniques
discussed later, but it should be detectable and was the first to be tried since
it is the easiest to implement. A simple two step experiment was used to study
the techniques and determine if it would yield a usable spin discriminant. After
cooling the ion the spin state is set using the pumping methods discussed in
Sec.\ref{Sec:GroundStatePumping} to one of the spin states. Immediately after
these pumping beams are shut off, the circularly polarized shelving lamp is
applied for fixed times of about a second, the shelving lamp is blocked and
the ion's state is determined by again applying the cooling beams with polarization
set for cooling and detection. No fluorescence indicates a shelved ion. The
result is recording and the entire sequence is repeated with a second, different,
initial spin state. Both cases are repeated until the statistical uncertainty
is small enough that a discriminant can be detected, this is typically a few
hundred trials. A detailed diagram of the measurement sequence and timing is
shown in fig.\ref{Fig:ChiralShelvingMeasurementSequence}.
 The sensitivity of this detection procedure to the ions initial spin state
should appear as a difference in shelving probability between the two cases.

The details of this particular implementation of this method require some modifications
to the previous analysis. The shelving lamp was circularly polarized with a
linear polarizer and a wavelength specific quarter-wave plate. Relative angle
alignment was coarse but good providing at least 90\% circular polarization.
More importantly, physical constraints required that the shelving light propagate
at a 45 degree angle relative to the quantization axis define by the pumping
beams. This could be accounted for by modifying the probe polarization in the
previous analysis, but it is easiest to temporarily just rotate the states and
use the previous generalized results for an arbitrary initial state. After rotation
in a plane containing the z axis a spin up state becomes \( \left( \begin{array}{cc}
cos\left( \theta /2\right) e^{-i\phi /2} & sin\left( \theta /2\right) e^{i\phi /2}
\end{array}\right)  \). For 45 degrees, this gives a probability distribution of \( \left( \begin{array}{cc}
cos^{2}\left( \pi /8\right)  & sin^{2}\left( \pi /8\right) 
\end{array}\right)  \) or \( s=1/\sqrt{2}=0.71 \) giving \( \Delta P=0.066 \). Also limitations
in the polarization control of the blue pumping beam allowed for circular polarization
of only one handedness, so the comparison was made between this resulting pumped
state and an unpolarized state generated using linearly polarized blue light
giving rather than completely opposite polarizations and the discriminant yields
only one \( \Delta P \). Finally, since the shelving rate is not precisely
known it is difficult to pick the optimal shelving time. An error in the rate
of a factor of two gives a further reduction in the discriminant by about a
factor of two or more. This makes the signal marginally detectable. 

The actual experiment yielded a discriminant of only about 1.5\%, Table \ref{Tab:D5_2SpinLife}.
It was reproducible and reversible but it is just at the edge of statistical
limits for these \textasciitilde{}thousand trials of \( \sqrt{p\left( 1-p\right) /N}\approx 1.2\% \)
so it is far from a reliable detection. Improvement could be made. The loss
due to the \( D_{3/2} \) branching ratio could be reduced by using the usual
red cleanup beam to return the ion to the ground state but this considerably
complicates the analysis and most of the spin information is lost in the many
decays and excitations back into the probe cycle, and anyway this problem is
far from the biggest limit on spin detection efficiency. More important is to
improve pump polarization, align the shelving lamp with the pump beam axis,
and accurately determine the ideal shelving time. These improvements could be
made with some effort, but such work was put off in favor of pursuing method
with inherently better signals which, in addition, eliminate the sensitive dependence
of the discriminant to the probe rate.

\subsection{Pumping Beams as a Probe}

\subsubsection{Ground State Spin Detection}

The fundamental problem with ground state spin detection through an intermediate
\( P \) state is the high branching ratio for decay back to the ground state
which leads to a loss of the initial spin information. This penalty is particularly
high when using the \( P_{3/2} \) state since the preferred decay from the
\( P_{3/2,+3/2} \) state is to the \( S_{1/2,+1/2} \) state compared to the
\( S_{1/2,-1/2} \) by a factor of 2:1, so that evolution profile of an ion
beginning in the \( S_{-1/2} \) state quickly merges with that of an ion beginning
in the \( S_{+1/2} \) state. This situation is reversed in the case of excitations
to the \( P_{1/2} \) state. Here the preferred decay from the \( P_{1/2,+1/2} \)
state is to the \( S_{1/2,-1/2} \) state, again by a factor of 2:1. Then an
ion starting in the \( S_{1/2,-1/2} \) state, excited to the \( P_{1/2} \)
state by circularly polarized light and not decaying to the \( D_{3/2} \) state
will most likely return to the spin state from which it started where it can
make another attempt for the \( D \) state. This partial preservation of the
initial spin information in the decay delays the dilution of that initial spin
information and allows a much larger discriminant. 

In this case, the \( S_{1/2,+1/2} \) state ideally is completely uncoupled
and the end result is the ion in the \( S \) or \( D_{3/2} \) state depending
on its initial spin information. To complete the process, a conventional, spin
independent, shelving step is added. Then an ion starting in the \( S_{+1/2} \)
state remains there during the spin sensitive probe step and is shelved during
the shelving step, while an ion beginning in the state is possibly moved to
the \( D_{3/2} \) state from which it won't be shelved.

The same beams and polarizations used to pump the ion can be used for this spin
sensitive probe, making implementation rather easy, though they must be highly
attenuated or quickly switched as discussed later. The method is a bit more
roundabout, requiring an extra shelving step, but the ideal discriminant is
twice as large as that for using the \( P_{3/2} \) state. And since the \( S_{+1/2} \)
state is uncoupled, it turns out that it is easier to realize this ideal making
the \( P_{1/2} \) state a far more tractable means of spin detection.

The detailed analysis is the same as for the \( P_{3/2} \) state. The time
evolution is determined by,
\[
-L+f\gamma R=R\left( \begin{array}{cc}
0 & 2f/9\\
0 & -2/3+4f/9
\end{array}\right) \]

Again there is one trivial eigenstate, \( v_{+}=\left( \begin{array}{cc}
1 & 0
\end{array}\right)  \) with \( \gamma _{+}=0 \), and it turns out \( v_{-}=\left( \begin{array}{cc}
f/(2f-3) & 1
\end{array}\right)  \) with, as is clear by inspection, \( \gamma _{-}=2/3-4f/9 \), so \( a_{+}=0,a_{-}=f/(2f-3) \).
Since one of the rates is zero, the maximum discriminant is simply at \( t\rightarrow \infty  \)
and the actual rates aren't important. Substituting this we have,

\begin{eqnarray*}
\Delta P & = & \frac{s}{2}\frac{\left( 1+a_{+}\right) \left( 1+a_{-}\right) }{1-a_{+}a_{-}}
\end{eqnarray*}
which reduces, for this case, to,

\begin{eqnarray*}
\Delta P & = & \frac{s}{2}\left( 1-\frac{f}{3-2f}\right) 
\end{eqnarray*}

Note that this has sensible limits for \( f\rightarrow 0 \) and \( f\rightarrow 1 \).
For \( f=2.85/\left( 2.85+1\right)  \) and the ideal \( s=1 \) this gives
\( \Delta P=0.25 \). For a comparison between oppositely polarized initial
states, \( \Delta s=\pm 1 \), this is a difference of \( 2\Delta P=0.5 \)
in the shelving probability. This ideal value is also slightly reduced by shelving
and detection efficiency as before by a factor of about 0.81 to \( 0.4 \).

The difference is independent of the rates, it is no longer important to accurately
know the probe rates, provided they are applied for a long enough time. So a
reduction of this ideal rate because of an incorrectly chosen exposure time
is no longer a factor. As a practical matter, the rates are still important
to consider. Since the probe beam will not be perfectly polarized the spin up
state will not be completely uncoupled. If the probe beam is applied for too
long, both spin states will be emptied. Therefore exposure times should be chosen
to be not longer than a few \( \left( R\gamma _{-}\right) ^{-1} \) to fully
empty the spin down state and minimize loss from the spin up state. For this
work the pumping beams are also used as probes, as proposed earlier. The pumping
and cooling rates are MHz, so at full power, microsecond switching times would
be necessary. This alone is possible with the right hardware, but it is also
necessary that when the beams are 'off' the rates are really zero on times scales
of a few minutes. This requires attenuation ratios of about \( 10^{6} \), current
devices, like pockel cells, provide around \( 10^{3} \) without too much works.
Alternately, mechanical beam blocks can easily provide this kind of extinction,
but switching times are limited to a few ms. For this work, the pump beams were
attenuated so that their excitation rates were are around a few per second for
use as a probe, and a beam blocks on stepper motor were switched at times on
the order of tenths of a second to provide the ideal exposure.

This method also has the advantage that the probe polarization is the same as
the pump polarization. Recall that in attempting to implement the case of the
circularly polarized shelving lamp physical constraints prevented being able
to align the axis of the shelving lamp with the pumping lasers and this contributed
a loss in the resulting spin discriminant of a factor of 2. In general, maximizing
the spin discriminant requires probing the same state that was pumped which
then requires that pump and probe light have the same polarizations. This requirement
is automatically satisfied for this method since pump and probe light are the
same. This is partly complicated by the fact that two lasers are used for pumping
while one is used for probing, and the pumped state is exactly determined by
both pump lasers. As discussed in the pumping, sec.\ref{Sec:Pumping}, ideally
one laser is linearly polarized to equally couple all of its states and act
as a cleanup while the other is used as the pump to select a particular state
to empty. If the cleanup laser has some residual circular polarization the resulting
pumped state is somewhat altered, though the effect is small, and with a perfectly
polarized pump the shift is zero as the desired state is exactly uncoupled.
As a result the the probed state is almost exactly the pumped state and the
spin discriminant is maximized with no additional effort.

This method works just as well for spin detection in the \( D_{3/2} \) state
and it turns out to have an even larger discriminant. So this technique was
tried first in the \( D \) state to make study and optimization easier and
work on the ground state put off.

\subsubsection{D State Spin Detection}

The same branching ratio that makes spin detection in the ground state difficult
using a \( P \) state, make it very easy to detect the ion's spin in a \( D \)
state. A ion excited with a probe from the \( D_{3/2} \) state to the \( P_{1/2} \),
would most likely decay to the ground state rather than back to the \( D \)
state by a factor of 2.85. With this spin detection is done in one step. This
advantage is partially eroded by the structure of the decay back to the \( D \)
state, as with the the \( S_{1/2}\rightarrow P_{3/2} \) probe but this is less
important and the resulting discriminant is much higher than for the \( S \)
state.

For the \( D \) state there are two probe possibilities that are worth considering.
Like for the ground, \( \Delta m=\pm 1 \) light can be used. In this case,
for \( \Delta m=+1 \) for example, an ion in the \( D_{-3/2} \) or \( D_{-1/2} \)
states will be excited and possibly moved to the ground state, from which it
can then be shelved, and an ion starting in the \( D_{+3/2} \) or \( D_{+1/2} \)
states will remain in the \( D \) state, and not be shelved. Here there is
also the possibility of using a \( \Delta m=0 \) probe. This will result in
the \( D_{\pm 1/2} \) states being emptied, and the \( D_{\pm 3/2} \) states
being untouched.

Again the detailed behavior is given by the same rate equations with the only
difference being the dimension of the problem. Before looking at any particular
case, consider the general structure of the problem. For this 4 dimensional
system there will be 4 eigenvalues, but since the excitation is to a spin \( 1/2 \)
intermediate state, two of these eigenvalues will be zero. This is obvious from
the special polarization cases just discussed, but it is generally true as well.
The evolution of the ground state will then be given by a constant plus two
decaying terms. For long times only the constant terms will contribute. So to
get the probability of remaining in the \( D \) state, only the coefficients
of the eigenvectors with zero eigenvalue in the expansion of the initial state
are required.

The simplest case is the \( \Delta m=0 \) probe,
\[
-L+f\gamma R=R\left( \begin{array}{cccc}
0 & f/6 & 0 & 0\\
0 & -1/3+f/9 & f/18 & 0\\
0 & f/18 & -1/3+f/9 & 0\\
0 & 0 & f/6 & 0
\end{array}\right) \]
In addition to the trivial eigenvectors, written here for convenience as, 
\begin{eqnarray*}
v_{1} & = & \left( \begin{array}{cccc}
1 & 0 & 0 & 1
\end{array}\right) \\
v_{2} & = & \left( \begin{array}{cccc}
1 & 0 & 0 & -1
\end{array}\right) 
\end{eqnarray*}
are 
\begin{eqnarray*}
v_{3} & = & \left( \begin{array}{cccc}
-\frac{3f}{6-f} & 1 & -1 & \frac{3f}{6-f}
\end{array}\right) \\
v_{4} & = & \left( \begin{array}{cccc}
-\frac{f}{2-f} & 1 & 1 & -\frac{f}{2-f}
\end{array}\right) 
\end{eqnarray*}
 An ion starting in the fully linearly pumped \( D \) state, \( p_{pumped}=\left( \begin{array}{cccc}
1 & 0 & 0 & 1
\end{array}\right) /2 \) will remain in the \( D \) state after probing. This should be compared to
a completely unpolarized state \( p_{unpolarized}=\left( \begin{array}{cccc}
1 & 1 & 1 & 1
\end{array}\right) /4=\left( v_{4}+\left( 1+f/\left( 2-f\right) \right) v_{1}\right) /4 \) since the exact complement, \( p=\left( \begin{array}{cccc}
0 & 1 & 1 & 0
\end{array}\right) /2=\left( v_{4}+f/\left( 2-f\right) v_{1}\right) /2 \), can't be generated directly. The \( v_{4} \) state will decay, leaving \( p_{unpol}\left( \infty \right) =\left( 1+f/\left( 2-f\right) \right) v_{1}/2 \),
giving the total \( D \) state population \( P_{unpol}\left( \infty \right) =1/\left( 2-f\right)  \).
Relative to the pumped state, this gives,

\[
\Delta P=1-\frac{1}{2-f}\]
With \( f=1/(2.85+1) \), \( \Delta P\approx 0.43 \). This would turn out to
be \( \Delta P=1-f/(2-f)\approx 0.85 \) if compared to the complement to the
pumped state.

For circularly polarized light with \( \Delta m=+1 \), 
\[
-L+f\gamma R=R\left( \begin{array}{cccc}
0 & 0 & -f/12 & 0\\
0 & 0 & -f/18 & -f/12\\
0 & 0 & 1/6-f/36 & -f/6\\
0 & 0 & 0 & 1/2-f/4
\end{array}\right) \]
The algebra begins to get a bit cumbersome, but the strategy is the same. In
the end, for the fully circularly pumped state, which is in practice, fig.\ref{Sec:FinalPumpedDStates}
,

\begin{eqnarray*}
d_{pumped} & = & \left( \begin{array}{cccc}
1 & 0 & 0 & 0
\end{array}\right) \\
P_{pumped}\left( \infty \right)  & = & 1
\end{eqnarray*}

The complement state can be generated with an oppositely circularly polarized
pump beam giving 
\begin{eqnarray*}
d_{complement} & = & \left( \begin{array}{cccc}
0 & 0 & 0 & 1
\end{array}\right) /4\\
P_{complement}\left( \infty \right)  & = & \frac{f\left( 2+3f\right) }{\left( 2-f\right) \left( 6-f\right) }
\end{eqnarray*}
and a maximum possible discriminant,
\[
P_{pumped}-P_{complement}=1-\frac{f\left( 2+3f\right) }{\left( 2-f\right) \left( 6-f\right) }\approx 0.92\]
This discriminant is larger than for the linearly polarized probe completely
because oppositely polarized states can be compared. Just for reference, an
unpolarized state in the basis would yield 
\[
P_{pumped}-P_{unpolarized}=\Delta P=1-\frac{1}{2-f}\approx 0.43\]
Again, both are slightly reduced by the shelving efficiency and shelved state
detection efficiency.

As with the ground state, a solution for a general initial state would be convenient
for studying transitions and systematic errors. The \( D_{3/2} \) is not quite
as straight forward as three variables are required to parameterize the populations.
An expansion in the spherical tensor basis is still most convenient,for this
dimension 4 problem the basis vectors are 
\begin{eqnarray*}
1 & = & \left( \begin{array}{cccc}
1 & 1 & 1 & 1
\end{array}\right) ^{T}\\
\sigma _{z} & = & \left( \begin{array}{cccc}
3 & 1 & -1 & -3
\end{array}\right) ^{T}\\
\tau _{zz} & \equiv  & \left( \begin{array}{cccc}
1 & -1 & -1 & 1
\end{array}\right) ^{T}\\
\upsilon _{zzz} & \equiv  & \left( \begin{array}{cccc}
1 & -3 & 3 & -1
\end{array}\right) ^{T}
\end{eqnarray*}
The normalizations are not conventional but they aren't important and these
are the tidiest with a minimum of fractions and roots in the end. Then the state
is written as,
\[
p=\frac{1}{4}+s\sigma _{z}+t\tau _{zz}+u\upsilon _{zzz}\]
To get the final occupation probability for this state, again write the basis
vectors in terms of the eigenvector, take the long time limit so any eigenvectors
with nonzero eigenvalue will give zero at long times, and sum over spin states.
Note that this last step also eliminates contributions from eigenvalues whose
components sum to zero, such as \( v_{2} \) from case of linear polarization
as its components were chosen to sum to zero, and in the end only the component
of \( v_{1} \) contributes to the final probability. 

For the case of linear polarization, 
\[
1=v_{4}+\left( 1+\frac{f}{2-f}\right) \left( v_{1}+v_{2}\right) \]
 For long times, \( v_{4} \) gives zero, and summing over spin states, \( v_{1} \)
gives 2 and \( v_{2} \) gives 0, so 
\begin{eqnarray*}
1 & \rightarrow  & 2\left( 1+f/\left( 2-f\right) \right) \\
\tau _{zz} & \rightarrow  & 2\left( 1-\frac{f}{2-f}\right) =4\left( 1-\frac{1}{2-f}\right) 
\end{eqnarray*}
Similarly \( \sigma _{z}\rightarrow 0 \), and \( \upsilon _{zzz}\rightarrow 0 \).
Comparing to the unpolarized \( s=t=u=0 \) case, this yields,
\[
\Delta P=4t\left( 1-\frac{1}{2-f}\right) \]
Giving the tidy result that the final probability depends only on the amplitude
of this quadrupole moment. This is consistent with the previously considered
case \( p=\left( \begin{array}{cccc}
1 & 0 & 0 & 1
\end{array}\right) /2=\left( 1+\tau _{z}\right) /4 \), or \( t=1/4 \), so \( \Delta P=1-1/\left( 2-f\right)  \).

As before the result for circular polarization requires less tidy algebra. The
result becomes,
\[
\Delta P=\frac{4\left( 1-f\right) }{\left( 2-f\right) \left( 6-f\right) }\left( 12s-2tf-u\left( 6-5f\right) \right) \]
For pumped state, which is in practice, \( p_{pumped}=\left( \begin{array}{cccc}
1 & 0 & 0 & 0
\end{array}\right) /4 \), \( s=3/20 \), \( t=1/4 \), \( u=1/20 \), 
\begin{eqnarray*}
\Delta P & = & \frac{4\left( 1-f\right) }{\left( 2-f\right) \left( 6-f\right) }(12\frac{3}{20}-2f\frac{1}{4}-\frac{1}{20}(6-5f))\\
 & = & \frac{1-f}{2-f}=1-\frac{1}{2-f}
\end{eqnarray*}
as before. For the complement to the pumped state \( p_{comp}=\left( \begin{array}{cccc}
0 & 0 & 0 & 1
\end{array}\right) /4 \), \( s=-3/20 \), \( t=1/4 \), \( u=-1/20 \), yielding,

\begin{eqnarray*}
\Delta P & = & \frac{4\left( 1-f\right) }{\left( 2-f\right) \left( 6-f\right) }\left( 12\frac{3}{10}-\frac{1}{10}\left( 6-5f\right) \right) \\
 & = & \frac{4\left( 1-f\right) }{\left( 2-f\right) \left( 6-f\right) }\frac{1}{2}\left( 6+f\right) \\
 & = & 2\frac{\left( 1-f\right) \left( 6+f\right) }{\left( 2-f\right) \left( 6-f\right) }
\end{eqnarray*}

\subsubsection{Experiment}

\begin{figure}
{\par\centering \includegraphics{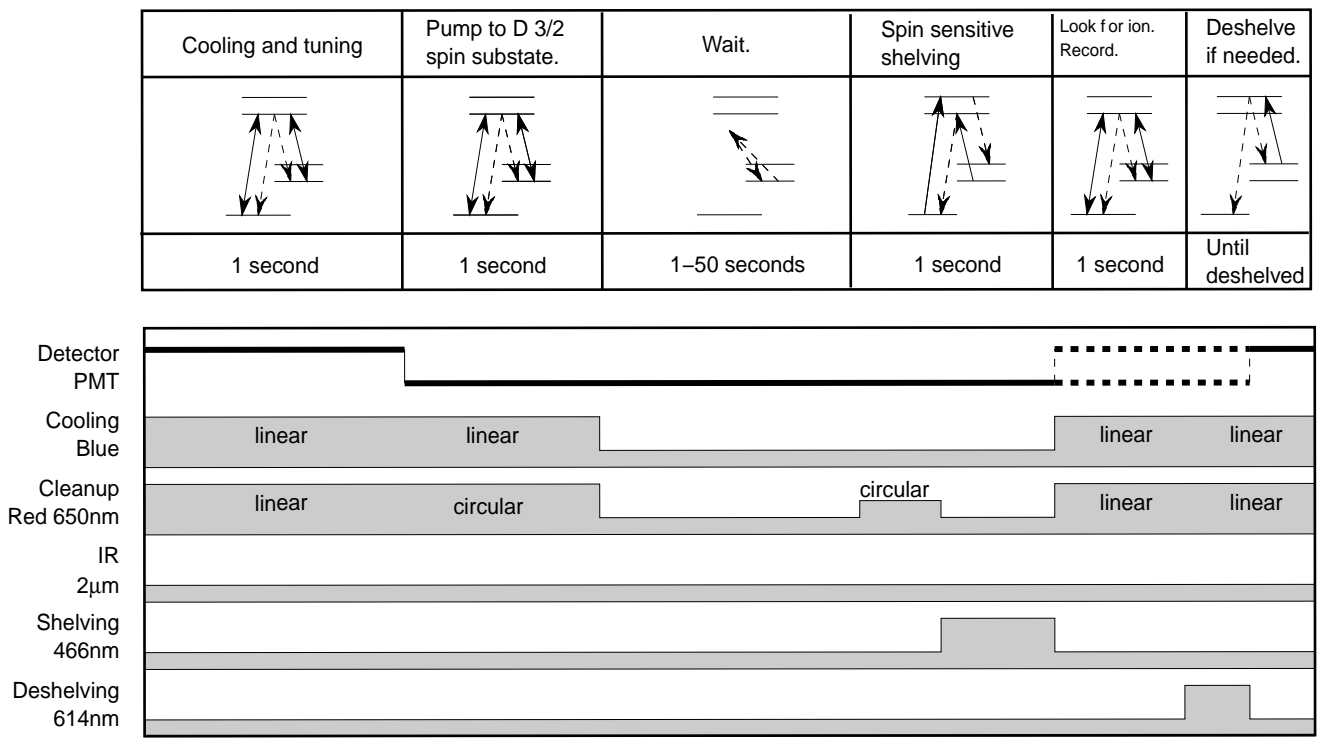} \par}

\caption{\label{Fig:PumpBeamSpinProbeSequence}Pump beam spin probe measurement sequence
for circularly pumped and probed \protect\( 5D_{3/2}\protect \) state.}
\end{figure}

This method was tested in the same way as for chiral shelving for both polarization
schemes. A typical measurement sequence is shown in fig.\ref{Fig:PumpBeamSpinProbeSequence},
for a circularly pumped initial state and circular probe. For other measurements,
the spin pump and probe polarizations are altered, as well as the cleanup beam
polarization if appropriate. A probe beam was supplied simply by mechanically
blocking the brightest part of the red cleanup beam and letting the weaker diffuse
light pass through to the ion. 
First a preparatory step is required to the the probe rate properly. The ion
is set to the \( D \) state with the blue cooling beam, the blue beam is blocked
and the attenuated red beam with polarization set to couple to all the \( D \)
states is applied for a fixed time, at first a few seconds, lately a few tenths
of seconds. The red beam is blocked and a shelving attempt is made. The red
attenuator is adjusted so that the shelving probability shelving probability
after the probe approaches the maximum shelving probability. This guarantees
that the rate and interaction time is long enough to empty a \( D \) spin state
to the ground state, but not so long that all are emptied because of residual
couplings to those state from an imperfectly polarized probe beam or beam axis
or magnetic field alignment. 

When the probe rate is properly set the spin detection test can be made. For
either spin probe scheme the ion was prepared by pumping with one of two polarizations.
For the linearly polarized spin probe, there is a spin polarizing pump polarization,
linear, parallel to an applied magnetic field and and reference pump polarization
which was chosen to be circular, which would yield an unpolarized \( D \) state
for the same magnetic field. For the circularly polarized spin probe, the polarizing
and reference pump polarization were simply left and right circular polarization,
in this case the reference polarization also polarizes the spin state. Immediately
after the state is prepared it is probed with the red beam alone set to the
spin polarizing pump polarization. A shelving attempt is then made, and the
shelving probability as a function of pump polarization is determined. The shelving
polarization should be lower when trying to shelve and probe the pumped states. 

\begin{figure}
{\par\centering \includegraphics{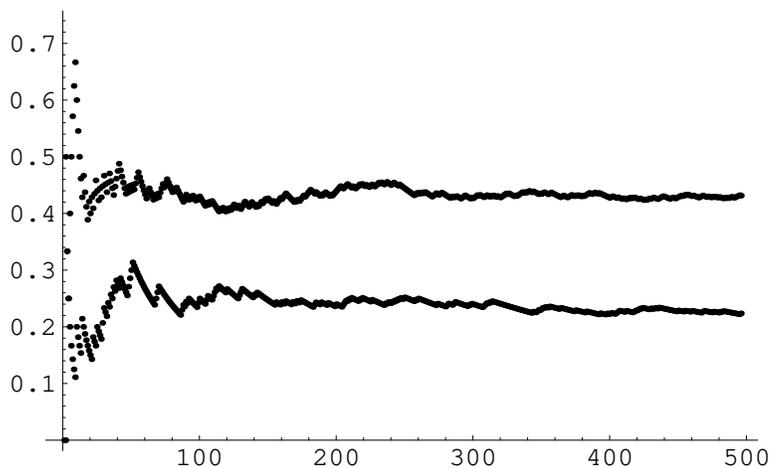} \par}

\caption{\label{Fig:LinearProbeSpinDiscriminant}Shelving probabilitity as a function
of trial number for linear \protect\( 5D_{3/2}\protect \) spin probe for fully
linearly pumped and unpolarized initial states.}
\end{figure}

A typical result is shown in fig.\ref{Fig:LinearProbeSpinDiscriminant}for a
linearly polarized probe.
 The shelving probability is plotted as a function of trial number for two cases.
The pumped line gives the shelving probability after probing a spin polarized
state. The shelving probability is low for this case because the ion tends to
stay in the pumped states, the 

\( D_{\pm 3/2} \) states, so during the probe, which couples only the \( D_{\pm 1/2} \)
states to the \( P \) state, it remains in the \( D \) state and will not
be shelved. The unpumped line shows a higher shelving probability since the
ion more often begins in the \( D_{\pm 1/2} \) states from which the probe
beam can move it to the ground state and then be shelved. There is a difference
in shelving probability of about \( 0.16 \). This dependence on the initial
spin state is apparent even after only a few trials and the initial state can
be determined reliably after about a hundred trials.

\begin{figure}
{\par\centering \includegraphics{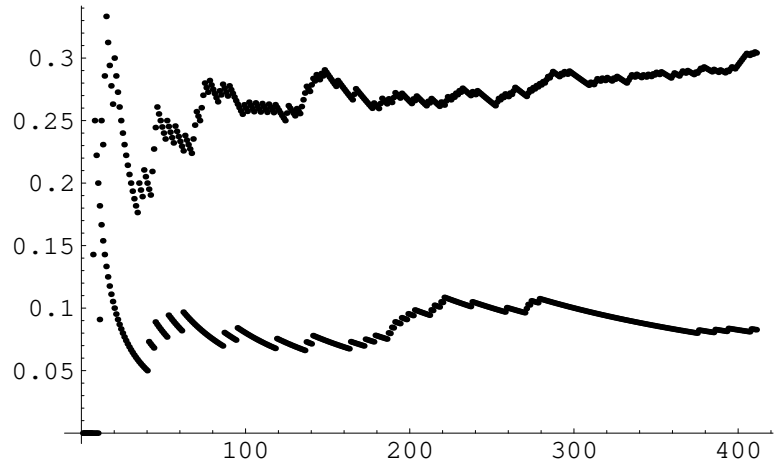} \par}

\caption{\label{Fig:CircularProbeSpinDiscriminant}Shelving probabilitity as a function
of trial number for circularly polarized \protect\( 5D_{3/2}\protect \) spin
probe for fully left and right circularly pumped initial states.}
\end{figure}

An example of results from a circularly polarized spin probe is shown in fig.
\ref{Fig:CircularProbeSpinDiscriminant}. 
In this case the ion is pumped to the \( D_{+m} \) states the \( D_{-m} \)
states and the probe only empties one of these sets of states. When the probe
and pump are the same polarization, the probe is probing an emptied state and
the shelving probability is low. For opposite polarizations the ion starts in
the probed states and the shelving probability is high. Here the difference
is about \( 0.22 \), larger than for the linear polarized probe, as expected,
and again, easily and quickly detected.

For both cases the discriminant is only about 1/4 of its ideal value. As for
the ground state, this is most likely due to imperfectly polarized beams and
misaligned fields. This reduces the difference in the initial spin states of
the ion, reducing the maximum possible discriminant, and also makes the discriminant
probe rate sensitive, which then requires a more careful determination of the
optimal probe time. No optimizations were made, other than an occasional adjustment
of the probe rate using the attenuator, as the signal resolution was completely
sufficient for the studies discussed presently, but when maximizing S/N becomes
important later substantial improvement should easily be available here. However,
as a practical consequence of this neglect, the particular values for the shelving
probability can't easily be predicted, as they are dependent on the details
of the implementation, so they must be remeasured when needed in any particular
experiment. But this is easily done as just described and no great burden.

The reduced discriminant could also be due to a shortened spin lifetime, which
would erode the difference in spin states while probing, and as a result also
reduce the discriminant. This possibility is important to consider as it affect
the sensitivity of a parity measurement as well. Such limits can now be studied
in detail with these spin setting and detection techniques.

\section{Spin Lifetimes}

\label{Sec:SpinLifetimes}

These probe methods provide a spin dependent shelving probability, and with
it, its inverse, a means of determining the initial spin state from the shelving
probability. With only this tool, spin lifetimes can be studied. The parity
experiment requires measuring a shift of about a half Hertz. Generally, this
requires measurement trials of a few seconds or tens of seconds, either to watch
it precess or to drive the transition slowly so that it has a sufficiently narrow
linewidth. This is a relatively long time for atoms, where many processes happen
at megahertz rates, and it is reasonable to worry that there are processes that
could completely scramble the spin well within that time scale such as off axis
fluctuating magnetic fields, or stray DC electric fields, or even the trapping
fields themselves. The details of the ion's environment are not completely known
at the levels which this experiment's sensitivity now requires. The spin needs
to stay put on its own long enough that when it is observed to change it is
only because of an applied interaction.

For the parity experiment there is already a process known to shorten the spin
lifetime. The parity laser will couple the ground state to the \( D_{3/2} \)
state. This \( D \) state has a finite lifetime of around 50\( s \). A given
ground state spin sublevel can make a transition to the other spin state through
an excitation and decay through the \( D_{3/2} \) state. In terms of the new
energy eigenstates this appears as a finite spin lifetime, in this case, for
a transition driven to full saturation the resulting spin lifetime will be twice
the \( D \) state lifetime. At that level this is not a problem, though it
is also the case that the \( D \) state lifetime, and do the resulting spin
lifetime, is shortened by off-resonant couplings of the \( D \) state to other
\( P \) and \( F \) states driven by the intense electric field of the parity
laser. In any case this is all well known, and the concern at present is to
processes driven by unknown environmental perturbations.

A spin lifetime longer than a few tenths of milliseconds is already implied
by existence of pumping signal in the fluorescence. The reduced rate that appears
during pumping is probably determined by pump beam polarization or alignment
errors, but it could, in principle, be limited by a short spin life time for
the pumped state. If the spin lifetimes were very short, microseconds, the ion
wouldn't stay in the uncoupled spin states long enough to significantly reduce
its time in the \( P \) state and so reduce the blue fluorescence produced
by decay to the ground state. Instead here the count rate drops by a factor
of about 100, from megahertz to tens of kilohertz, implying that the the ion
spends at least a few tenths of microseconds in the pumped state.

The appearance of a spin discriminant provides an indication of an even longer
spin lifetime. The polarized probe beam is applied for times of up to a second.
If spin lifetimes were much shorter than this the spin would be unpolarized
before probing was complete and any any dependence on the initial spin state
would disappear since the initial spin state disappears before probing is complete.
The spin discriminant is already less than expected, this could be the result
of an already partially decayed spin state. And even if not, this alone is not
enough to probe lifetimes of many tens of seconds which is the real goal.

These longer times are easily studied with a simple modification of the experiment
used to generate the spin discriminant. After the pump step, instead of immediately
applying the probe beam, wait a fixed time with no light on the ion, then apply
the probe beam and attempt to shelve as usual. For a finite spin lifetime, any
initial spin state will decay to a unpolarized state and eventually the shelving
probability will become independent of the initial spin polarization.

\subsection{Ground State Spin Lifetime}

Consider for now just incoherent perturbations that can be understood as a relaxation,
rather than coherent effects that generate energy shifts. The former reduce
signal to noise and sensitivity while the latter are included when analyzing
systematic errors. Relaxations in the ground state are the easiest to account
for as they have the simplest structure. The fully general case is a dipole
coupling between the two spin states. The result is simple an exponential decay
of the polarization, but to prepare for the more complicated \( j=3/2 \) case
consider a more formal development. The time evolution is given again by a first
order rate equation as \( \dot{S}=MS \) with off diagonal elements given by
Clebsch-Gordan coefficients and the diagonal elements set to conserve probability
by setting them to make rows and columns sum to zero, or \( \sum _{m}M_{mn}=\sum _{n}M_{mn}=0 \),
as this gives \( \sum _{m}\dot{S}_{m}=\sum _{mm\prime }M_{mm\prime }S_{m\prime }=0 \).
For this spin \( 1/2 \) problem, \( M \) is simply,
\[
M=\frac{\Gamma }{2}\left( \begin{array}{cc}
-1 & 1\\
1 & -1
\end{array}\right) \]
with \( \Gamma  \) some rate. 

Again eigenvectors are the most direct path to the solution, here they are \( \left( \begin{array}{cc}
1 & 1
\end{array}\right)  \), \( \left( \begin{array}{cc}
1 & -1
\end{array}\right)  \), or simply \( 1 \) and \( \sigma _{z} \), if as before, these are understood
as vectors derived from the usual matrices operating on a column vector of all
ones giving just the diagonal of the original vector. The eigenvalues are \( 0 \)
and \( \Gamma  \). The \( 1 \) vector with \( 0 \) eigenvalue will always
appear in the general case. It must to conserve probability, and it is clear
mechanically. If all the rows sum to zero than any row operating on a column
vector of \( 1 \)s will give zero. As a result, the elements of all the other
eigenvalues must sum to zero since the \( 1 \) accounts for all the of occupation
probability.

The state, as written before, already appears in this basis as \( P_{S}=\left( 1/2\right) +\left( s/2\right) \sigma _{z} \).
The \( \sigma _{z} \) component decays and the full time evolution is simply
\[
P_{S}=\frac{1}{2}+\frac{s_{0}}{2}e^{-\Gamma t}\sigma _{z}\]
The spin discriminant then also decays with the same time constant,

\begin{eqnarray*}
\Delta P & = & \frac{s_{0}}{2}e^{-\Gamma t}\left( 1-\frac{f}{3-2f}\right) 
\end{eqnarray*}

Experimental studies of spin lifetime in the ground state were limited because
of the low discriminant achieved with the chiral shelving spin probe method.
Data was taken for two wait times, \( 0.5 \)s and \( 5 \)s. The results are
shown in table.\ref{Tab:D5_2SpinLife}. Each case is derived from about 1000
trials giving a statistical uncertainty of 0.012. The shelving probabilities
for a linear or circularly polarized pump beam differ by about 1\( \sigma  \).
The shelving probabilities as a function of initial spin state are basically
independent of time implying spin lifetimes of at least 5 seconds and most likely
10 or more. 

\begin{table}

\caption{\label{Tab:D5_2SpinLife}Shelving probability as a function of pumping polarization
and wait time for circularly polarized ground state pump and probe.}
{\centering \begin{tabular}{|c||c|c|}
\hline 
&
0.5s&
5s\\
\hline 
\hline 
\( P_{\sigma } \)&
0.189&
0.186\\
\hline 
\( P_{\pi } \)&
0.174&
0.176\\
\hline 
\end{tabular}\par}\end{table}

\subsection{D State Spin Lifetime}

A more extensive study of spin lifetimes was done in the \( D_{3/2} \) state
because the much larger spin discriminant obtained made work easier. This four
state system is considerably more complicated, making analysis a bit more involved,
but giving a richer structure that can provide more information about the ion's
environment. As with the ground state, a perpendicular fluctuating magnetic
field could be considered with the same dipole couplings. Another source of
decoherance could be fluctuating static electric fields, or even the applied
high frequency trapping fields. The effects in the ground state would be indistinguishable
from a magnetic field as dipole couplings are the only interactions that can
fit into a spin 1/2 system. In this \( D_{3/2} \) state quadrupole interactions
are allowed, either from direct couplings within the manifold from electric
field gradients, or from dipole couplings to nearby \( P \) states. It is worthwhile
to consider both as the resulting effects on the ion's spin state are different
and could, in principle, be distinguished with a careful study of the spin lifetime
and used to identify the perturbation.

Again the most straightforward solution is to write the eigenvalues of the coupling
in terms of the spherical tensor basis vectors. The spin discriminant is already
written in terms of these parameters so getting the time dependence of these
variables provides a direct solution for the behavior of the experimental observable. 

The dipole interaction turns out to be particularly simple. For this case, again
from Clebsch-Gordan coefficients and probability conservation,
\[
M=-\frac{R}{2}\left( \begin{array}{cccc}
-3 & 3 & 0 & 0\\
3 & -7 & 4 & 0\\
0 & 4 & -7 & 3\\
0 & 0 & 3 & -3
\end{array}\right) \]
The eigenvalues happen to be exactly the basis vectors chosen to parameterize
the states with eigenvalues for \( 1,\sigma ,\tau ,\upsilon  \) of \( 0,R,3R,6R \)
respectively. The times dependence of the state is then just given by exponentially
decaying variables with the appropriate time constants, each moment decays independently
with a different rate,

\[
p=\frac{1}{4}+s_{0}e^{-\tau }\sigma +t_{0}e^{-3\tau }\tau +u_{0}e^{-6\tau }\upsilon \]
Either spin discriminant is similarly given by the same substitutions.

The quadrupole case requires a bit more generality as the \( \Delta m=\pm 1 \)
transitions will be driven at a different rate the the \( \Delta m=\pm 2 \)
transitions. Parameterize these as \( R_{1} \) and \( R_{2} \). This gives,

\[
M=-\frac{1}{2}\left( \begin{array}{cccc}
-R_{1}-R_{2} & R_{1} & R_{2} & 0\\
R_{1} & -R_{1}-R_{2} & 0 & R_{2}\\
R_{2} & 0 & -R_{1}-R_{2} & R_{1}\\
0 & R_{2} & R_{1} & -R_{1}-R_{2}
\end{array}\right) \]
Here \( 1 \) and \( \tau  \) are eigenvectors, with eigenvalues \( 0 \) and
\( R_{1}+R_{2} \). The other eigenvectors are now 
\begin{eqnarray*}
v_{1} & = & \left( \begin{array}{cccc}
1 & -1 & 1 & -1
\end{array}\right) =\left( \sigma +2\upsilon \right) /5\\
v_{2} & = & \left( \begin{array}{cccc}
1 & 1 & -1 & -1
\end{array}\right) =\left( 2\sigma -\upsilon \right) /5
\end{eqnarray*}
 with eigenvalues \( R_{1} \) and \( R_{2} \) respectively. Then \( \sigma =v_{1}+2v_{2} \)
and \( \upsilon =2v_{1}-v_{2} \). This gives,

\begin{eqnarray*}
s & = & \frac{s_{0}+2u_{0}}{5}e^{-R_{1}t}+\frac{4s_{0}-2u_{0}}{5}e^{-R_{2}t}\\
t & = & t_{0}e^{-\left( R_{1}+R_{2}\right) t}\\
u & = & \frac{2s_{0}+4u_{0}}{5}e^{-R_{1}t}+\frac{-2s_{0}+u_{0}}{5}e^{-R_{2}t}
\end{eqnarray*}
To simplify slightly, the initial states that have been considered all have
\( u_{0}=0 \), giving,

\begin{eqnarray*}
s & = & \frac{s_{0}}{5}\left( e^{-R_{1}t}+4e^{-R_{2}t}\right) \\
t & = & t_{0}e^{-\left( R_{1}+R_{2}\right) t}\\
u & = & \frac{2s_{0}}{5}\left( e^{-R_{1}t}-e^{-R_{2}t}\right) 
\end{eqnarray*}

In both the dipole and quadrupole cases the \( \tau  \) component decays with
a single exponential, unfortunately preventing easy identification using a linearly
polarized probe which depends only on this component. A circular probe must
be used, and perhaps with sufficient resolution the difference in the detailed
behavior of the decay could be identified.

\begin{figure}
{\par\centering \resizebox*{1\textwidth}{!}{\includegraphics{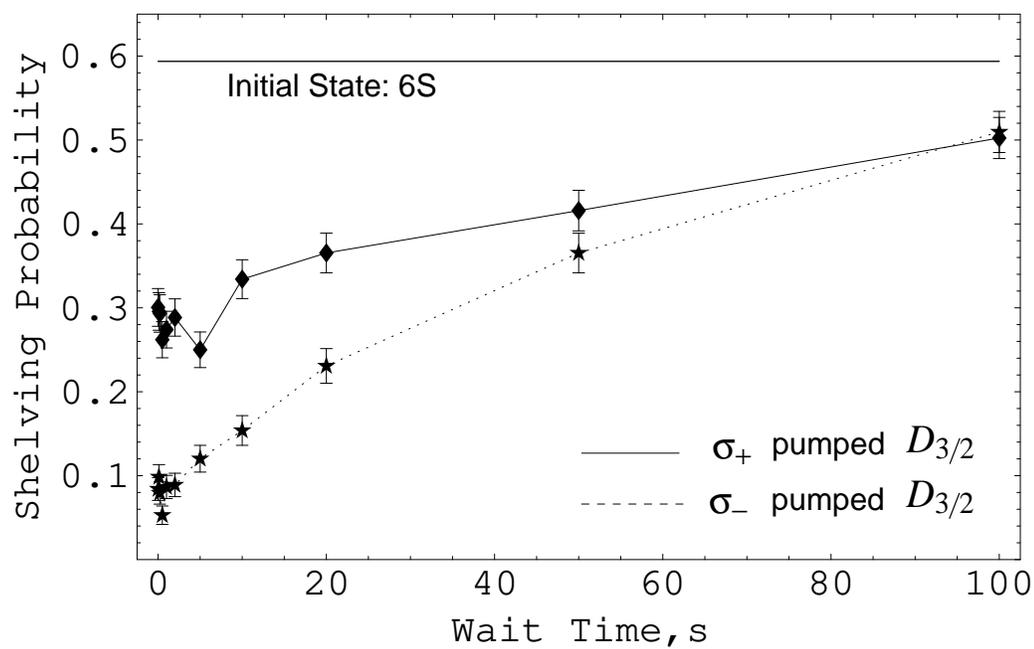}} \par}

\caption{\label{Fig:SpinLifetimeData}\protect\( D_{3/2}\protect \) spin lifetime.
Shelving probability as a function of wait time for linear probe of \protect\( D_{3/2}\protect \)
state with initally linearly pumped and unpolarized states.}
\end{figure}

The experiment done follows a sequence similar to that for all the other spin
measurements. In this case a circularly polarized pump and probe was used. The
ion pumped to one of two initial spin state, here spin up and spin down. Then
followed a wait time ranging from 0.1 to 100 seconds with no applied interactions.
Finally the ion was probed with one particular polarization for both pump cases
and a shelving attempt was made. The final shelving probabilities as a function
of wait time for about 500 trials is shown in fig.\ref{Fig:SpinLifetimeData}. 

The shelving probability increases as a function of time for both initial spin
states, this simply reflects the finite, 50 second, decay lifetime of the \( D_{3/2} \)
state back to the ground state. A decay of the spin state would appear as a
merging of the lines so that the shelving probability becomes independent of
initial spin state. This may be happening at times longer than about 60 seconds,
but the loss of resolutions from the decaying \( D \) state prevents any accurate
determination of a spin lifetime, other than that it is longer than about 50
seconds, reinforcing the bit of evidence from the ground state of long spin
lifetimes. 

This is ideal for future PNC experiments as it shows the S/N will not be limited
by spin decoherance, but difficult for studying spin decoherance. Because the
spin lifetime is so long compared to the \( D \) state lifetime, that is not
very much shorter, the decay profile is completely dominated by the \( D \)
state decay and the contribution from any spin relaxation is difficult to isolate.
This makes identifying a decay mechanism impossible as, effectively, no spin
decay could be detected. 

The primary purpose of this measurement was to determine if the spin lifetime
is long enough to make a parity measurement possible with a reasonable S/N.
Having confirmed this, further work on identifying spin decay mechanisms was
put off. Future work might include simply better statistics, applied broadband
interactions to deliberately scramble the spin, and working again in the ground
state, when an improved discriminant is available, where the infinite manifold
lifetime won't mask the effects of a spin relaxation.

\section{Spin Flip Transitions}

\label{Sec:SpinFlipTransitions}

Instead of simple relaxation from incoherent perturbations, coherent transitions
due to applied interactions can be detected, and this finally provides a means
of measuring the parity light shift. The simplest thing to consider is a harmonic
dipole magnetic field to drive the \( \Delta m=\pm 1 \) transitions. When the
frequency of the applied field coincides with the energy between spin states
the ion will leave the pumped spin state for one that can be shelved after the
spin sensitive probe giving an increase in the shelving probability. Measuring
the change in the position of this resonance frequency when other interactions
are applied, such as the parity lasers, gives a precise measurement of the resulting
energy shift.

As usual, the ground state is most directly relevant, and easiest to analyze,
but initial work was done in the \( D_{3/2} \) state because of it higher S/N
and richer structure, in this case it also have a narrower spin transition line
width making it easier to identify and track down sources of noise.

\subsection{Ground State Spin Transitions}

After making the Rotating-Wave approximation the dynamics of the system can
be described by
\[
H=RJ_{x}+\delta \omega J_{z}=R\left( J_{x}+\gamma J_{z}\right) \]
One subtler aspect of the problem for this application is that after spin pumping
the populations of the spin states are well defined but their relative phases
are unknown. This can be dealt with formally using a density matrix, but it
is easier to see what is going on by explicitly averaging over the initial phases.
Also, for the data presented here, the period of oscillation is much shorter
than the time that the interaction is applied, and the interaction can't be
controlled precisely enough to stop the transitions at a well defined and consistent
phase, the practical result is observation of the time average populations of
the states.

\begin{figure}
{\par\centering \resizebox*{1\textwidth}{!}{\includegraphics{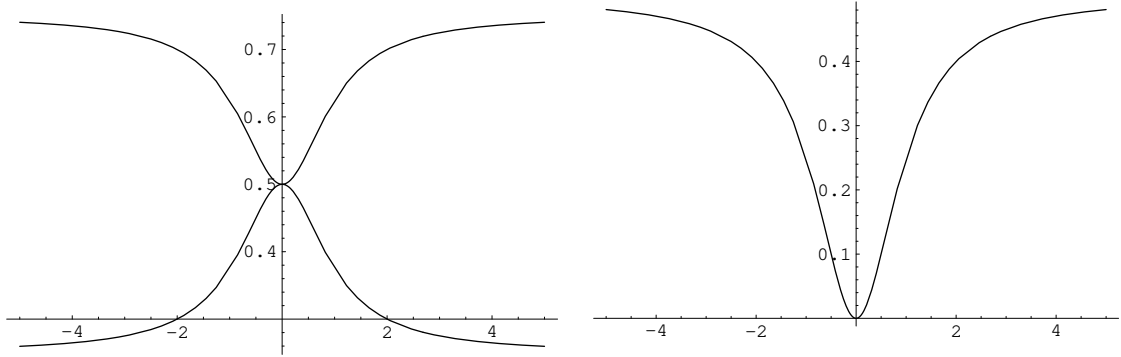}} \par}

\caption{\label{Fig:SSpinTransitions}\protect\( S_{1/2}\protect \) Spin Transitions
populations and dipole moment.}
\end{figure}

The problem can be solved easily with a bit of algebra from either the two dimensional
coupled differential equations, or the operator equations of motion. After suitable
averaging,
\[
\left\langle \psi \right\rangle _{t,\phi }=\left( \begin{array}{c}
\frac{1+s_{0}}{2}+\left( \frac{1}{2}-\frac{1+s_{0}}{2}\right) \frac{1}{1+\gamma ^{2}}\\
\frac{1-s_{0}}{2}+\left( \frac{1}{2}-\frac{1-s_{0}}{2}\right) \frac{1}{1+\gamma ^{2}}
\end{array}\right) \]
The time average populations are equal on resonance, with a Lorentzian frequency
dependence. More simply in terms of the \( s=\left\langle s_{z}\right\rangle  \)
variable used before,
\[
\left\langle s\right\rangle _{t,\phi }=s_{0}\left( 1-\frac{1}{1+\gamma ^{2}}\right) \]
The time average of the spin becomes zero with a Lorentzian lineshape, fig.\ref{Fig:SSpinTransitions}.

This spin \( 1/2 \) problem also has a quick geometric solution. In this basis,
the interaction is like a static magnetic field which the spin simply precesses
around. The time average of the spin will then just be in the direction of the
magnetic field, the initial phase is just the angle of the perpendicular component
of the spin about the \( \hat{z} \) axis and the direction of the time average
is independent of this initial spin direction. The component of the time average
of the spin in the \( \hat{B} \) direction fixed so that \( \left\langle \vec{s}\right\rangle _{t}\cdot \hat{B}=\vec{s}_{0}\cdot \hat{B} \),
as is the time average of the \( \hat{z} \) component of the spin in particular,
\( \left\langle s_{z}\right\rangle _{t}=\vec{s}_{0}\cdot \hat{B}_{z} \). Then
the \( z \) component of the unit vector in the direction of the magnetic field
is \( \hat{B}_{z}=\pm \gamma ^{2}/(1+\gamma ^{2})\hat{z} \) and \( \left\langle s_{z}\right\rangle _{t}=\vec{s}_{0}\cdot \hat{B}_{z}=s_{0}\left( \hat{z}\cdot \hat{B}_{z}\right) =s_{0}\left( 1-1/\left( 1+\gamma ^{2}\right) \right)  \)
as before.

\subsection{\protect\( D_{3/2}\protect \) Spin Transitions}

The spin \( 3/2 \) system is considerably more complicated. An easy geometric
solution does exist but it is less transparent without a lot of development.
The spin expectation value behaves in the same way, but here there are other
degrees of freedom that don't evolve in an obviously trivial way, in the initial
states considered so far there is also a quadrupole component. For now, the
quickest way to a solution is to just treat the general problem formally and
calculate the profiles numerically.

\subsubsection{Resonance Profiles for Dipole Splittings}

With the same static Hamiltonian, the time evolution of the states is given
by, \( \psi \left( t\right) =exp\left( -iHt\right) \psi _{0} \). The population
can be written using spin projection operators \( P_{m} \) which just give
\( P_{m}\left| j,m'\right\rangle =\left| j,m'\right\rangle \delta _{m,m'} \).
With these, 
\begin{eqnarray*}
p_{m} & = & P_{m}\psi ^{\dagger }\psi \\
 & = & P_{m}\psi ^{\dagger }_{0}e^{iHt}P_{m}e^{-iHt}\psi _{0}\\
 & = & P_{m}Tr\left( e^{iHt}P_{m}e^{-iHt}\psi _{0}\psi ^{\dagger }_{0}\right) 
\end{eqnarray*}

The phase average is easy to do in this form, \( \psi _{0}\psi ^{\dagger }_{0} \)
is really just the density matrix, it is real on the diagonal and contains phase
information in the off-diagonal elements, averaging over phases leaves only
the diagonal elements, \( \left\langle \psi _{0}\psi ^{\dagger }_{0}\right\rangle _{\phi }=diag\left( p_{0}\right)  \).
The RHS is just a diagonal matrix, the elements are the initial populations.

The time average can be done in the same way after a rotation in the \( x-z \)
plane by \( \theta =\pi /2-tan^{-1}\gamma  \) with \( R=R_{xz}\left( \theta \right)  \).
This makes the Hamiltonian diagonal, \( H\propto J_{z} \), the magnitude won't
matter here, call it \( \alpha  \). Then 
\begin{eqnarray*}
\left\langle e^{iHt}P_{m}e^{-iHt}\right\rangle _{t} & = & \left\langle R^{T}\left( Re^{iHt}R^{T}\right) RP_{m}R^{T}\left( Re^{-iHt}R^{T}\right) R\right\rangle _{t}\\
 & = & \left\langle R^{T}e^{i\alpha J_{z}t}RP_{m}R^{T}e^{-i\alpha J_{z}t}R\right\rangle _{t}
\end{eqnarray*}
The matrix between the time evolution operators will get a time dependent phase
on off diagonal elements, so again time averaging will leave only the diagonal
elements. This finally yields,
\[
\left\langle p_{m}\right\rangle _{t,\phi }=P_{m}Tr\left( R^{T}diag\left( RP_{m}R\right) Rdiag\left( p_{0}\right) \right) \]

\begin{figure}
{\par\centering \resizebox*{1\textwidth}{!}{\includegraphics{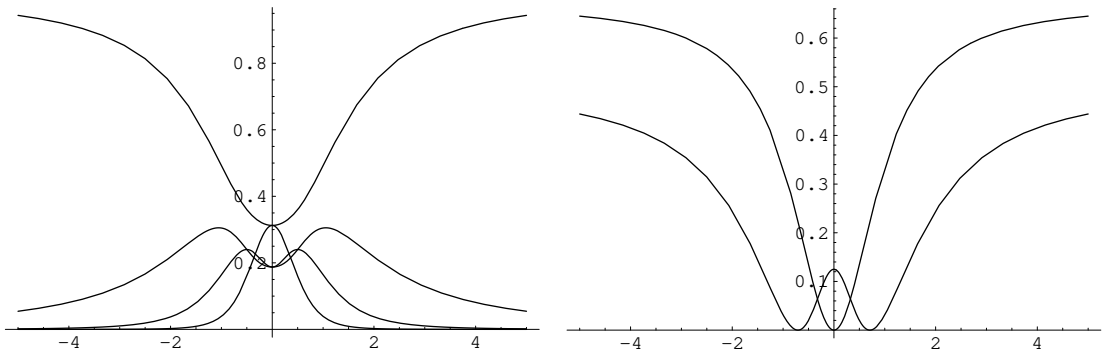}} \par}

\caption{\label{fig:CircPumpDSpinTransitions}Time average populations and moments for
dipole spin flip transition in circularly pumped \protect\( 5D_{3/2}\protect \)
sta}
\end{figure}

\begin{figure}
{\par\centering \resizebox*{1\textwidth}{!}{\includegraphics{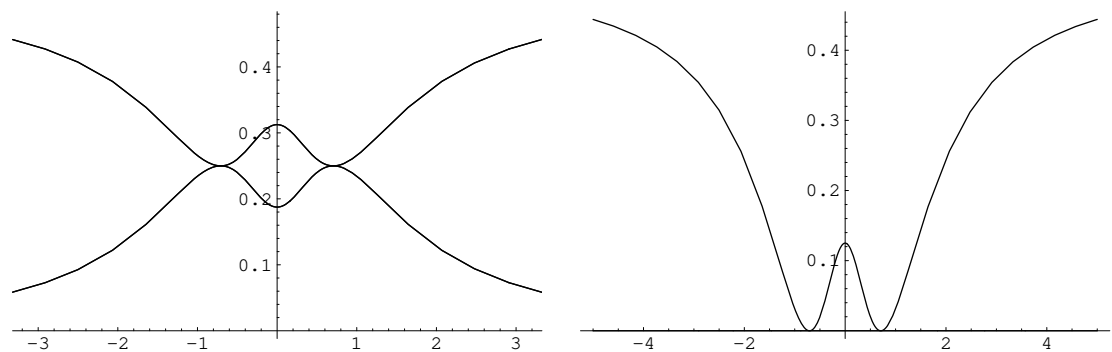}} \par}

\caption{\label{fig:LinPumpDStateSpinTransitions}Time average populations and moments
for dipole spin flip transition in linearly pumped \protect\( 5D_{3/2}\protect \)
state.}
\end{figure}

The results for the populations, and their moments for the two initially pumped
states usually considered are shown in figs.\ref{fig:CircPumpDSpinTransitions}
and \ref{fig:LinPumpDStateSpinTransitions}. The moments are the coefficients
of the normalized spherical basis vectors used previously.
The dipole component behaves simply as it did in the \( j=1/2 \) problem as
expected from the geometric solution, but the profile of the quadrupole component
contains some interesting new structure. The octapole moment conveniently doesn't
contribute, as would be apparent from a geometrical analysis, no octapole moment
exists in the initial configuration and none is generated so its evolution is
trivial.

\subsubsection{Resonance Structure with Quadrupole Shifts}

\label{Sec:ResonancesForQuadrupoleShifts}

A less formal solution to this problem is possible but not generally useful.
This analysis was valid only for dipole splittings of the \( D \) state magnetic
sublevels, the energy separation between each set of adjacent state is equal.
The most interesting case will include quadrupole shifts which change the \( m=\pm 1/2 \)
states independently of the \( m=\pm 3/2 \) states resulting in three different
splittings. The shift will be proportional to \( \tau _{z}=\left( \begin{array}{cccc}
1 & -1 & -1 & 1
\end{array}\right) ^{T} \)so \( \Delta E_{+3/2,+1/2} \) will be shift by same amount as \( \Delta E_{-3/2,-1/2} \).
This generality can be treated exactly with the previous method but that provides
no easy insight. 

The quadrupole shifts of interest will, in all cases, be much larger than the
expected linewidth. In this limit the problem can be treated simply as a collection
of independent two state problems whose full solution has already been given.
This gives, apparently, resonance profiles centered at three different resonance
frequencies, but some care must be taken to consider the amplitudes of these
profiles. Consider the case of a linearly polarized pump and probe the the \( D \)
state. The ion begins in the \( m=\pm 3/2 \) states, if the applied RF frequency
matches either of the \( m=\pm 3/2\rightarrow \pm 1/2 \) transitions the ion
can move to the \( m=\pm 1/2 \) states where it can be emptied by the probe
beam and shelved. The pumped state has equal population in the \( m=\pm 3/2 \)
states, so if these frequencies are different only one of the \( m=\pm 3/2 \)
state will be emptied rather than both, so the probe will empty the \( D \)
state only half as often as before and the ion will be shelved only half as
often. As the same time, if the RF frequency matches the resonance between the
\( m=\pm 1/2 \) states, which is unaltered by the quadrupole shift, the populations
will not be altered, as the initial populations of these states was zero, and
there will be no observable effect on the shelving probability. The net result
is a single peak, splitting into two peaks half as high, shifted by equal amounts
in opposition directions from the original resonance frequency. The \( m=\pm 1/2 \)
transition at the original dipole frequency is not visible.

\begin{figure}
{\par\centering \includegraphics{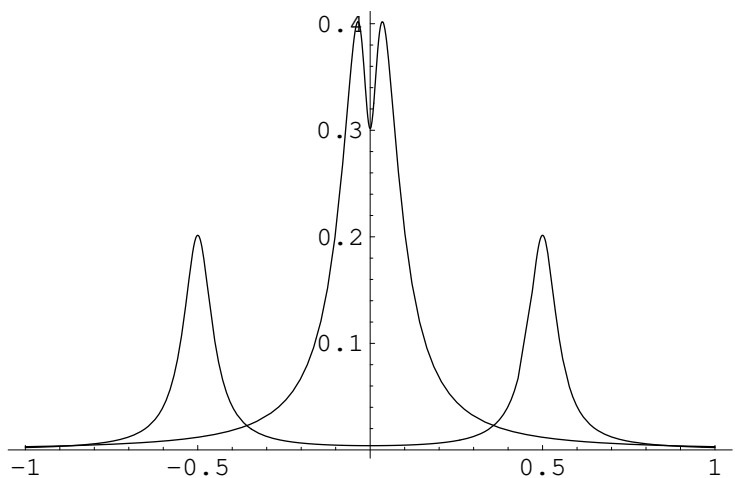} \par}

\caption{\label{Fig:QuadrupoleSplittingLinearProbe}Quadrupole splitting in \protect\( D_{3/2}\protect \)
state with linear probe.}
\end{figure}

The shape and width of these split peaks depends on the dominant broadening
mechanism. For magnetic field noise the lineshape and linewidth should be the
same as before as the energies between the states are fluctuating slightly.
For power broadening only the width depends on the relative rates, the \( m=\pm 3/2\rightarrow \pm 1/2 \)
transition is \( 3/4 \) as fast as the \( m=\pm 1/2 \) transition. The width
of the dipole splitting only profile is some combination of these widths, and
recall has a more complicated lineshape, the split peaks would be simple lorentzians.
For the moment, the most important effect in practice is the stability of the
interaction creating the quadrupole splitting. As its strength changes, the
shift changes and, like the magnetic field noise, this smears out the peaks.
Generally, the split quadrupole peaks can have a completely different shape
and width than the single dipole peak which depends on the broadening mechanism
so only the shifts will be considered here along with a coarse qualitative discussion
of the relative amplitudes. Figure \ref{Fig:QuadrupoleSplittingLinearProbe}
shows the general behavior, the profiles are computed assuming transitions between
spin states having the same width and lineshape, the resulting structure is
due entirely to the effects of the probe beam. Recall that the notch in the
dipole peak it due to the structure of the \( D_{3/2} \) spin transition.

The situation for a circularly polarized pump and probe is similar. The an ideal
probe empties, for example, the \( m=+3/2 \) and \( m=+1/2 \) states. Though
recall that this requires precise alignment of the pumping beams and in practical
cases a residual \( \Delta m=0 \) excitation rate empties the \( m=-1/2 \)
state as well. Then transitions between any of these spin states will not alter
the populations and as a result have no effect on the final shelving probability.
In addition, with the same polarizations, only the \( m=+3/2 \) and \( m=+1/2 \)
states are significantly probed and as a result even transitions from the \( m=-3/2 \)
state to the \( m=-1/2 \) state will not be seen since they just alter the
populations of states that are not probed. With no quadrupole splitting, transitions
are eventually possible to the probed states, and in fact, since the initial
state is fully spin up, after a half period, the state will be fully spin down,
so the resonance is visible with full amplitude. With a quadrupole splitting
making all three energy separations different, the only transitions will be
between pairs of unpopulated states, or pairs of unprobed states and so no resonances
are visible. 

This difficulty was not immediately appreciated when the first spin resonance
experiments were being done in the \( D_{3/2} \) state using circularly polarized
pump and probe beams. Those experiments never detected a spin resonance and
this altered initial pumped state, and naive measurement sequence are the likely
problems.

\begin{figure}
{\par\centering \includegraphics{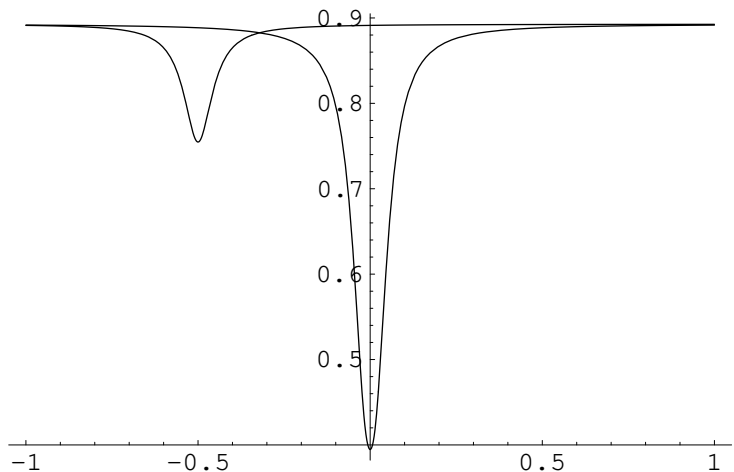} \par}

\caption{\label{Fig:QuadrupoleSplittingCircularProbe}Spin flip transition profile for
quadrupole splitting in \protect\( D_{3/2}\protect \) state with circular pump
and probe.}
\end{figure}

One simple modification can recover some of this. For the transition to alter
the shelving probability the initial populations of the states being considered
must be different and the probe beam must couple to them. The probe beam doesn't
necessarily have to couple to these states with different strengths as the structure
of decay from the \( P \) state back to the \( D \) state also makes the two
states inequivilant. The populations of the \( m=-3/2,-1/2 \) states are already
unequal, so simply switching to a probe beam with the opposite chirality as
the pump beam couples these states and allows them so be emptied differentially,
and the \( m=-3/2\rightarrow -1/2 \) resonance now becomes visible, though
with a reduced amplitude since both states are coupled. The other transitions
are still not detected. The profiles for this case are shown in fig.\ref{Fig:QuadrupoleSplittingCircularProbe}.
The quadrupole peak is relatively small, though not much smaller than the quadrupole
peaks for the linearly probed \( D \) state so detection should not be particularly
harder, and, as argued, the dipole peak is not visible with the quadrupole shift.

\begin{figure}
{\par\centering \includegraphics{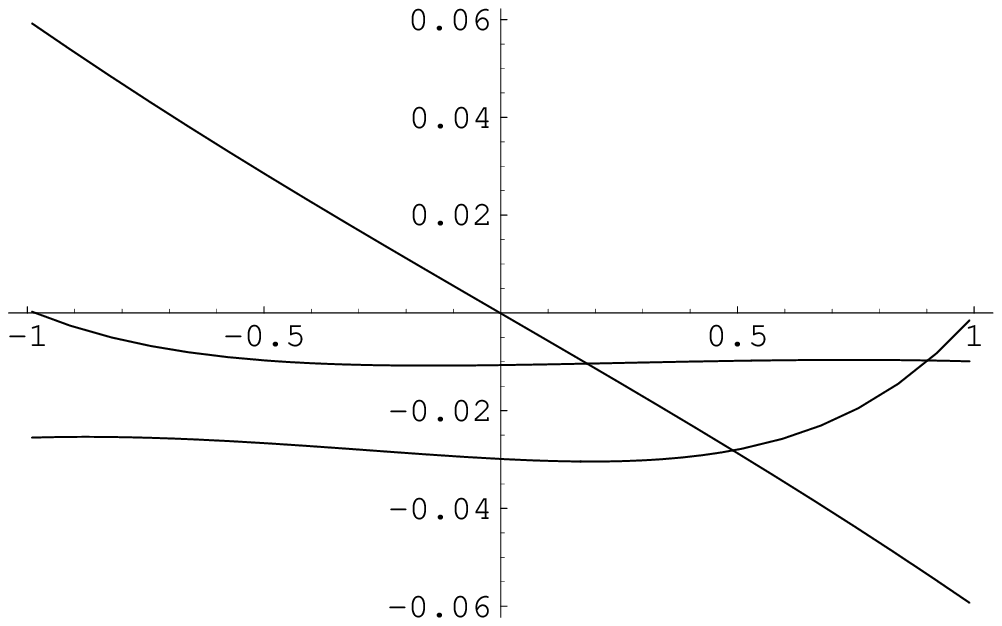} \par}

\caption{\label{Fig:PeakSensitivityForPartialCircularPump}Split flip resonance peak
heights in quadrupole split \protect\( D_{3/2}\protect \) state as a function
of probe circular polarization for partially circularly polarized pump beam.}
\end{figure}

It is not strictly necessary to be able to measure the dipole splitting with
a quadrupole shift. Ideally, the dipole splitting should be unchanged from the
value it had without the quadrupole shift. But, systematic errors that pollute
the quadrupole shift can change the dipole splitting as well and it is useful
to be able to measure both to detect these kinds of possibilities. It is possible
to restore sensitivity to the \( m=-1/2\rightarrow 1/2 \) transition by using
a different initial state by using, for example, a less the perfectly circularly
polarized red pump beam. This then results in a partially populated \( m=-1/2 \)
state so that transitions to the \( m=+1/2 \) state now do result in a change
of the \( D \) state sublevel populations and a change in the shelving probability
after probing. This gives a dipole peak again, but since it also reduces the
difference in population between the \( m=-3/2 \) and \( m=-1/2 \) states,
the quadrupole peak is made further smaller and the height of the quadrupole
peak is further reduced. For example, \( \sigma _{R}=0.5 \), gives the initial
state, \( D=\left( \begin{array}{cccc}
0.22 & 0.41 & 0.22 & 0.15
\end{array}\right)  \) and the transition profile shown in fig.\ref{Fig:PeakSensitivityForPartialCircularPump}.
Note that the directions of the peak are different due to the initial difference
in the initial population. 
\begin{figure}
{\par\centering \includegraphics{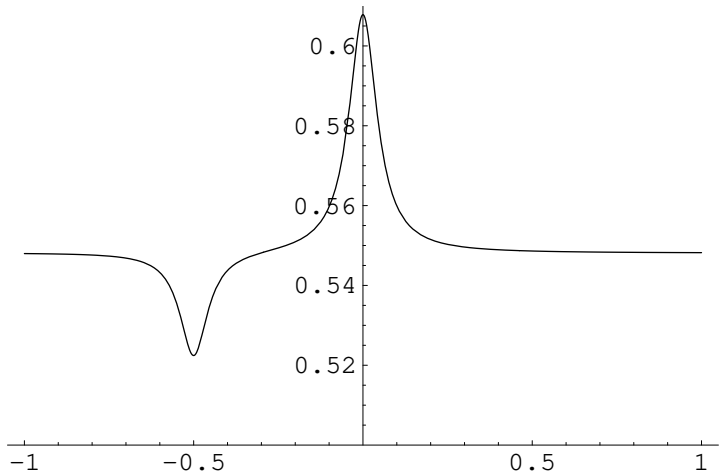} \par}

\caption{\label{Fig:QuadrupoleSplittingPartialCircularPump}Spin Flip transition profile
for quadrupole splitting in \protect\( D_{3/2}\protect \) with partially circularly
polarized pump, \protect\( \sigma _{R}=0.5\protect \) , and circularly polarized
probe. }
\end{figure}

The peaks are now very small and if detectable another slight alteration could
make the second quadrupole peak visible as well. By partially circularly polarizing
the probe beam, sensitivity to population in the \( m=+1/2 \) and \( m=+3/2 \)
states is partly restored and the final shelving probability becomes dependent
on the \( m=+1/2\rightarrow +3/2 \) transitions as well. Fig.\ref{Fig:PeakSensitivityForPartialCircularPump}
show the peak heights as a function of probe polarization for all three peaks
for an initial state pumped with \( \sigma _{R}=-0.5 \). The peak heights are
collectively maximized around \( \sigma _{R}=0.5 \), though still in all cases
very small.

Generally all cases of polarizations altered from their ideal values increase
sensitivity to one transition at the cost of sensitivity to another. A better
method may be to use an alternating series of pump/probe polarizations, or magnetic
field directions. As usual, ideally the magnetic field should be left unchanged
to avoid stability and settling problems. In addition, it will be convenient
to be able to measure shifts in the \( S \) and \( D \) states simultaneously.
Though both quadrupole peaks are visible for a single pump/probe polarization
with a linear probe, this geometry would give no sensitivity to ground state
spin flip transitions. Then the ground state requires a circularly polarized
probe, so to avoid having to alternate magnetic fields, a practical solution
for using a circular probe in the \( D \) state is most desirable.

For this case it is clear that for the second quadrupole peak only changing
polarizations are necessary. If \( \sigma _{R,,pump}=-\sigma _{R,,probe}-1 \)
gives the \( m=-3/2\rightarrow -1/2 \) resonance, then \( \sigma _{R,,pump}=-\sigma _{R,,probe}=1 \)
will give the \( m=3/2\rightarrow 1/2 \) resonance. This reduces sensitivity
since the data collection rate is now reduced by half, but so far all schemes
to make a single polarization sensitive to all transitions reduce peak heights
by more than half, giving an even lower sensitivity overall.

\begin{figure}
{\par\centering \includegraphics{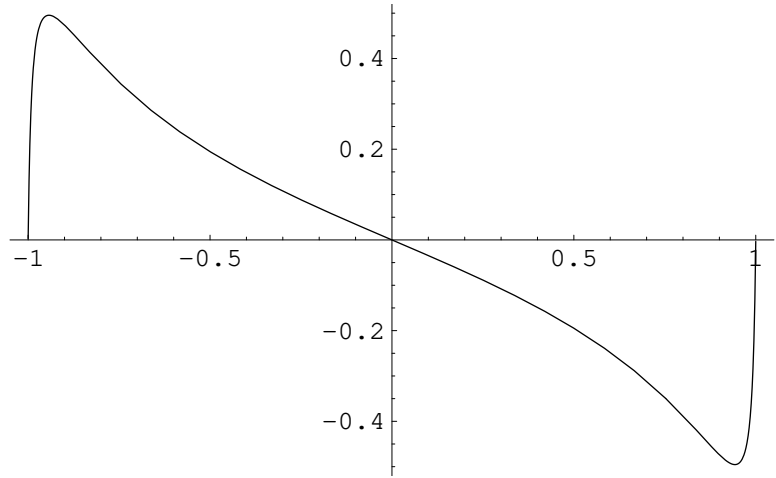} \par}

\caption{\label{Fig:D_pm1/2StateDisparity}Difference in populations of \protect\( D_{1/2}\protect \)
and \protect\( D_{-1/2}\protect \) states as a function of circular polarization
of red pump laser.}
\end{figure}

\begin{figure}
{\par\centering \includegraphics{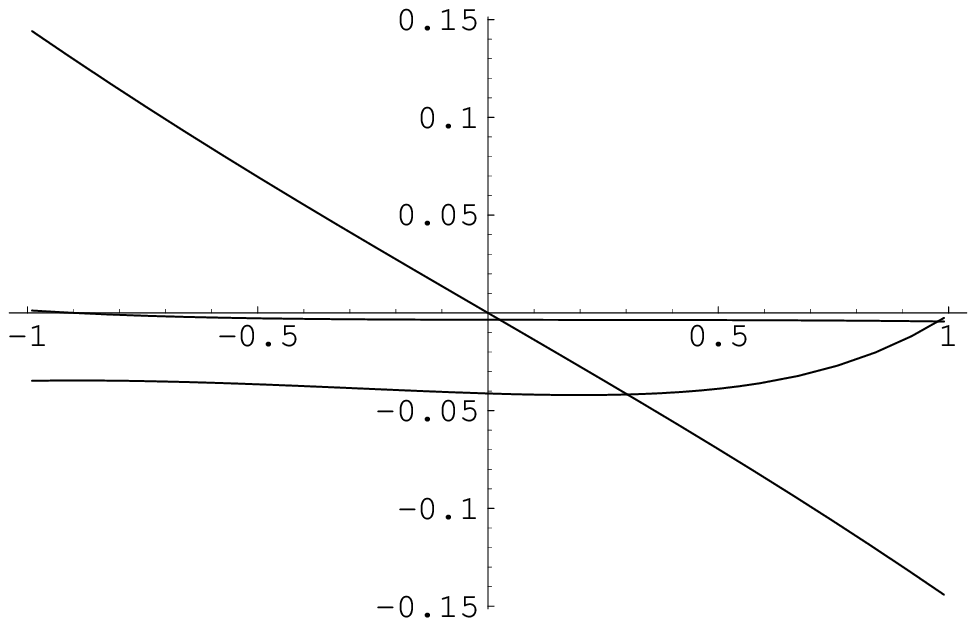} \par}

\caption{\label{Fig:PeakSensitivityForOptimalDipolePump}Size of quadrupole split resonance
peakes as a function of probe polatization for initian state given to maximumize
\protect\( D_{3/2,+1/2}-D_{+3/2,-1,2}\protect \).}
\end{figure}

This allows a means of detecting both quadrupole peaks with reasonable sensitivity,
but in the cases considered so far, the sensitivity to the dipole peak is still
very small. A bit more work yields are reasonable solution. Since the pump and
probe polarizations are independent, it is reasonable that the sensitivity to
this \( m=-1/2\rightarrow 1/2 \) transition will be the largest when the initial
different in populations between the two state is largest, and then the difference
in probe rates is also the largest. Using the solution developed earlier for
the final pumped state for arbitrary pumping polarizations including the effects
of the cleanup pulse, sec.\ref{Sec:FinalPumpedDStates}, a quick calculation,
fig.\ref{Fig:D_pm1/2StateDisparity}, shows a maximum population difference
for \( \sigma _{D}\approx \pm 0.9 \). 
 With this initial state the peak heights as a function of probe polarization
are shown in fig.\ref{Fig:PeakSensitivityForOptimalDipolePump}
The peak height is maximized for a purely circularly polarized probe and the
resulting transition profile gives a relatively small but sufficiently large
dipole peak, again comparable to the quadrupole peak heights with a linear probe,
as well a a bit of one quadrupole peak, \ref{Fig:PeakSensitivityForOptimalDipolePump}.
\begin{figure}
{\par\centering \includegraphics{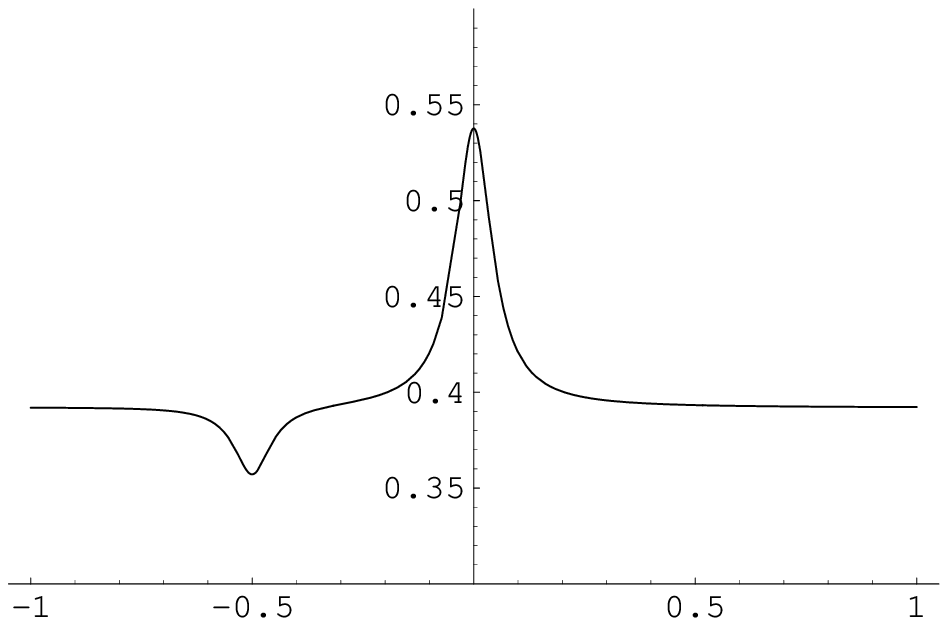} \par}

\caption{\label{Fig:DipoleOptimizedSpinTransitionProfile}Spin flip transition profile
when pump and probe polarizations are optimized for the dipole \protect\( m=-1/2\rightarrow 1/2\protect \)
transition.}
\end{figure}

\subsection{Experiment Implementation}

The measurement sequence follows the same general structure as the other spin
measurements. For spin transitions a perpendicular RF magnetic field is applied
during the period between spin pumping and probing. The field is provided by
a \textasciitilde{}\( 5mm \) loop inside the chamber \textasciitilde{}\( 2cm \)
directly below the ion. The plane of the loop contains the ion so that the resulting
magnetic field is horizontal. The current is provided directly from a function
generator with a 2W output into a 50\( \Omega  \) load providing a few hundred
milliamp. This should give a field of a few milligauss. For this \( j=3/2 \)
system the g-factor is \( \mu =\mu _{B}\left( 2j+s/\left( 2l+1\right) \right) =\left( 4/5\right) \mu _{B}\approx 1.1MHz/G \).
So this field should give transition rates of around a \( kHz \).

The external static applied field ranges from 2-5\( kHz \), giving resonance
frequencies around a few \( mHz \). There are a few practical constraints to
the size of this applied field which could affect the implementation of a future
parity experiment. First, the secular motion of the ion turns out to also be
around a few \( mHz \). It was found that large fields applied near this frequency
results loss of the ion from the trap. Presumably this is due to the small electric
field generated by the oscillating magnetic field coupling to the large charge
of the ion, amplifying the ion secular motion until it escapes the trap. With
large enough fields the ion is lost immediately, even while cooling beams are
simultaneously applied. With small applied fields at these frequencies the ion
will stay in the trap while being cooled but can be lost during the spin interaction
period of the measurement cycle when the only applied interaction is the RF
field. Above about 2-3\( MHz \), or below 1\( Mhz \) the ion trap lifetime
was not so dramatically shortened.

Work was done above this secular motion resonance region. Ideally, lower fields
and resonance frequencies would be better so that the much smaller shifts due
to parity violation can be more easily detected. For applied fields of a few
Gauss, the linewidth would probably be dominated by magnetic field noise, a
desired linewidth of \( 0.1Hz \) to measure a 1\( Hz \) parity shift would
require an applied magnetic field stable to better than a part in \( 10^{7} \),
which is probably intractably difficult. But using lower magnetic fields introduces
another practical complication. As discussed in sec:\ref{Sec:PumpingSignal}
the structure the the \( D_{3/2} \) to \( P_{1/2} \) clean-up transition requires
that there are always two \( D \) states, generally linear combinations of
the \( J_{z} \) eigenstates, that are uncoupled by the clean-up beam. This
polarization pumping is exploited for spin pumping in the \( D \) state, but
it complicates cooling and detection because the ion gets stuck in the \( D \)
state. This is generally not notices because, without explicit attention, there
is always some ambient magnetic field also coupling these spin states which
sufficient strength and appropriate direction that the ion quickly leaves these
pumped spin states. The magnetic field acts to clean out these spin states just
as the red beam is used to clean out the \( D \) state. When the applied magnetic
field is deliberately made small, less than about a Gauss, corresponding to
\( Mhz \) transition rates, the cleaning rate slows enough that fluorescence
is reduced making ion detection difficult. At 1\( kHz \) transition rates the
fluorescence would drop from a few 1000cps to just a few cps.

This could be avoided by using a varying magnetic field, low fields for spin
interactions and large fields for detection. This would generally be very slow
and difficult to do precisely. If initial spin level splittings are around1\( kHz \),
the magnetic field must be stable to better than at least a part in \( 10^{4} \),
which is difficult but reasonable for a static field, but may be too difficult
for a continually changing field if things things like paramagnetism and thermal
time constants begin to cause trouble. A better solution is to modulate the
polarization of the clean-up beam. Since the uncoupled states are polarization
dependent, a changing polarization could, in a sense, clean out a state that
was pumped a short time before. This is easily done at the \( Mhz \) rates
required with a pockel cell, but again the complication was avoided for these
initial studies.

\subsection{Data}

The applied magnetic field was known to about 10\%, \( 0.2-0.3G \), so the
resonance frequency could be initially estimated to only within a few hundred
\( kHz \). To find the resonance the applied RF frequency was swept over this
range in two bins with a linear triangular sweep profile at a rate of \( 1-10Hz \).
The measurement consisted of four cases. Two are the usual spin discriminant
points to provide calibration and reference, these were a spin polarizing and
an unpolarizing pump followed by a 5 second wait with no interactions before
a spin probe sequence. In this case the red pump polarization was linear, parallel
to the applied magnetic field. The other two cases were with the RF magnetic
field applied during the 5 second period after a spin polarizing pump preparation,
the sweep widths were the same in both cases but the center frequencies were
different so that the entire range overlapped the expected position of the resonance. 

\begin{figure}
{\par\centering \includegraphics{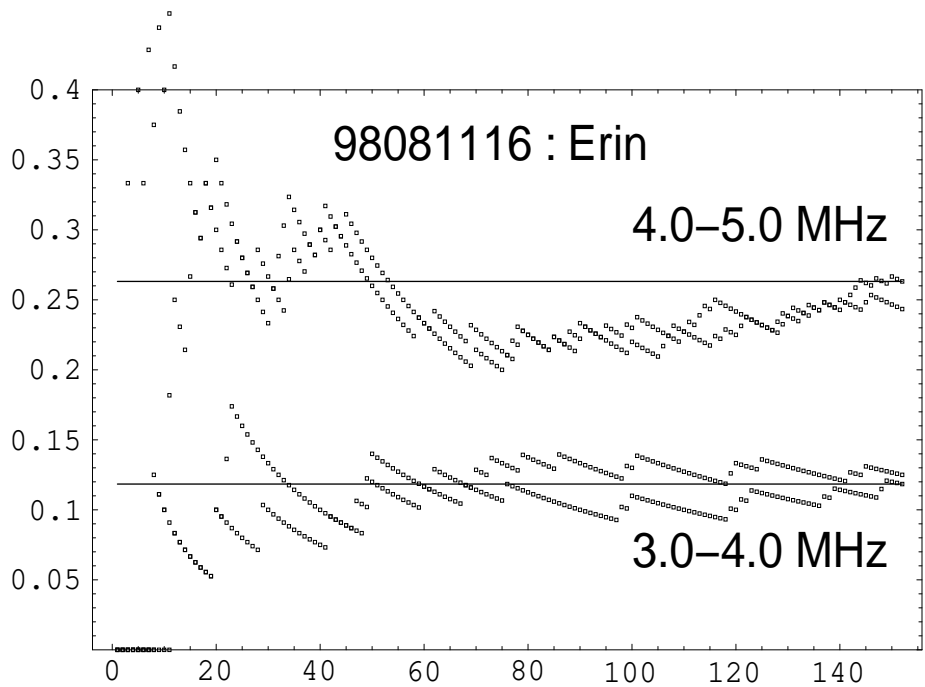} \par}

\caption{\label{Fig:SpinTransitionData}Spin discriminant and Spin Flip Signal.}
\end{figure}

If the frequencies are correct this should result in one range being out of
resonance where the spin will be unaffected and the shelving probability will
be identical to the pumped case with no interaction and the other sweep range
will include the resonance and result in the end in an unpolarized ion so that
the shelving probability will merge with that for the initially unpolarized
case. The data in fig.\ref{Fig:SpinTransitionData} show the initial appearance
of the resonance. The shelving probability as a function of trial number is
shown for each case and it is clear that the difference between on and off resonance
is as well defined as the difference between polarized and unpolarized spins.

This provided the position of the resonance to about \( 100kHz \). The sweep
ranges were halved and the center frequencies moved so that the search range
was then contained in this previously successful region and the entire procedure
repeated. After just a few iterations, the resonance was known to a few \( kHz \).
At this point the sweep range was reduced to just a \( kHz \), or in some cases
to zero, and more center frequencies were added to map out a resonance for this
spin transition and provide information about linewidth, noise sources and stability
and some ideals about the coherence of the transition. 

\begin{figure}
{\par\centering \includegraphics{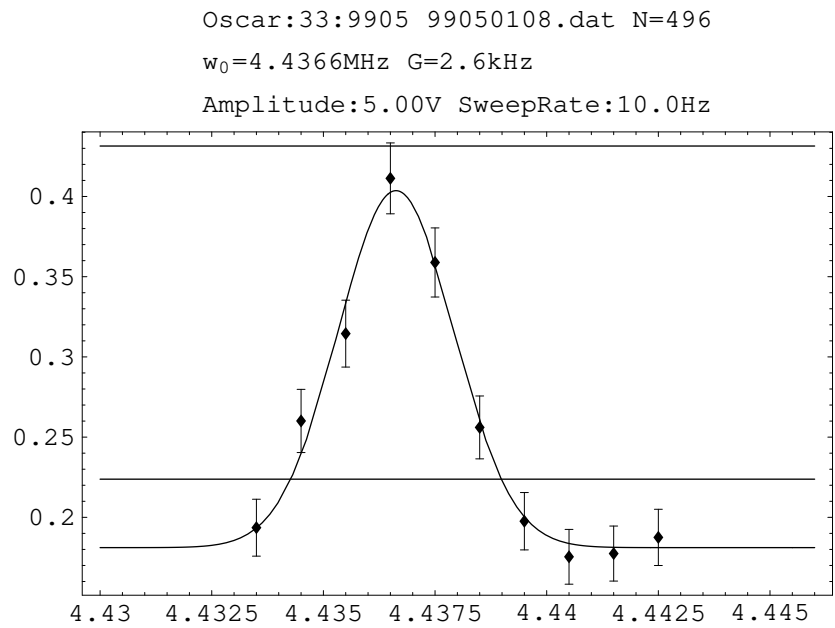} \par}

\caption{\label{Fig:SpinFlipResonance}Spin flip resonance profile.}
\end{figure}

The very large sweep widths were almost certainly wider than the linewidth of
the transition so the applied interaction is effectively broadband and incoherent,
though it could also be treated adiabatically if the transition rate was sufficiently
faster than the sweep rate which was expected but not yet known. Ideally smaller
or zero sweep ranges would result in coherent transition as previously analyzed
but magnetic noise sources or other perturbations could complicated that. An
incoherent excitation would result in a simple dipole relaxation of the spin,
as discussed with spin lifetimes, that would be time and frequency dependent
while coherent excitations should yield the time average provide just calculated
above. After some work eliminating noise sources in the environment and electronics
a very well defined resonance appears as in fig.\ref{Fig:SpinFlipResonance}.
This data shows the shelving probability as a function of frequency for a few
hundred trials at 10 frequencies. The data took about 3 hours to collect.

\subsection{Linewidth, Noise, and Stability}

The linewidth was initially about \( 20kHz \). This turned out to be largely
due to line noise on ovens and filaments in the trap generating small \( 60Hz \)
magnetic fields. With this carefully eliminated the linewidth dropped to \( 5kHz \).
At this point the electronics controlling the current to the coil providing
the magnetic field became important. At this level the field must be stable
to better than a part in \( 10^{3} \), which requires more than a casual effort.
Improvements and testing yielded a current stable to a better than at least
a part in \( 10^{4} \). This finally gave a consistent linewidth of \( 1-2kHz \).
The nature of this limit is still being determined. 

The line shape was found to be independent of the time the RF field was applied
implying the transition is a largely coherent precession, rather than an incoherent
relaxation. The linewidth is consistent with the previously expected transition
rate, but the lineshape is not as expected, Gaussian rather than the Lorentzian
with a central notch, and the width was found to be independent of the amplitude
of the applied field as it was varied by a factor of 50 from its initial value. 

The width is then almost certainly due to magnetic field noise. The system is
completely unshielded, except for shielding on the nearby ion pump to reduce
the field offset the coils must buck out. A few \( mG \) fluctuations is reasonable
from building blower motors or nearby ferromagnetic materials and fluctuating
room temperatures. Eliminating these perturbations with magnetic shielding is
straight-forward and well known. However the external field was independently
monitored and found to be stable to a few tenths of \( mG \), shielding will
certainly be requires for linewidths below \( 0.1kHz \) but clearly there is
another large source of noise.

The lineshape provides some clues. The fit is clearly best to a gaussian, which
suggests a slow random walk. The spin would be completely unpolarized while
in resonance, but a particular frequency would only intermittently be resonant.
The linewidth is independent of time, rather than linearly growing with time
as for an unconstrained random walk, but a confined variation is reasonable
from a ferromagnetic material whose domains are periodically flipping or changing
slightly as the ambient temperature changes. 

Faster fluctuations would involve convolutions of the frequency spectrum of
the perturbation and none should be gaussian. A harmonic frequency fluctuation
spends a large fraction of its period at the ends of its range than the middle,
a triangle or sawtooth variation spends an equal time at all frequencies. The
earlier line noise problem partly illustrated this, in that case the line shape
not well defined, but it was definitely not gaussian, and instead more extended
perhaps as a box convolved with this gaussian fluctuation. 

The ion's motion in the trap could also be the nature of the problem. If there
is a large magnetic field gradient at the position of the ion then the ion's
motion would result in it seeing a continually varying magnetic fields. Here
the frequency of the fluctuations would be at the secular motion or trapping
RF frequencies of about \( 1-2MHz \) and \( 27-30MHz \) respectively. Again
the lineshape is not completely consistent with this kind of variation. Also
the line shape was found to be independent of the power of the trapping RF.
Varying this should change the amplitude of the ion's motion and so change the
range of magnetic fields the ion samples.

An unknown instability in the current controlling electronics could exists.
The fluctuations for this kind of problem should be fractional, proportional
to the applied field. But, similarly, the linewidth was independent of the applied
field. 

There is further evidence in favor of slow field fluctuations from local ferromagnetic
materials. The linewidth is not completely stable, it is time dependent though
not in a reproducible way. The position of the resonance is least stable immediately
after the external magnetic field is applied or changed. It typically takes
a few hours before a profile with a linewidth of \( 1-2kHz \) can be generated,
sometime a day or more is required. This is consistent with some kind of domain
relaxation time constant. To try to pin this down further, the resonant frequency
as a function of time was determined for a long period of time. Data was taken
continuously for more than a day, giving more almost 2000 trials at each RF
frequency. The frequency for each 100 trial subset was calculated and was found
to fluctuate by well over the statistical uncertainty of the average used to
measure the frequency. To test the statistical accuracy, and provide some more
clues about noise timescales, the even and odd trials were averaged separately
and compared. They were found to agree to within statistical accuracy and fluctuate
together outside of that statistical uncertainty. They are also found to periodically
snap together to a significantly different position as if there was a large
discrete change in the applied field.

The fluctuations of the field seem to be a timescales very long compared to
the time of a measurement cycles, and fluctuate and creak in a way suggesting
ferromagnetic contributions. The nonlinear dependence of the field on the current
applied to the magnetic field coils already suggested a large contribution from
ferro-magnetism, sec.\ref{Sec:FieldCoilDependance}. However the source is not
external, as implied by the external monitor. This leaves a nearby source, inside
the chamber, or the chamber itself. A careful inventory turned up many suspects,
and the magnetic properties of the chamber material at these levels is being
studied. Future improvements will require adjustments in the materials used
to build the trap, or possibly internal magnetic shielding, as well as shielding
of external fluctuations.

\subsection{Sensitivity and S/N}

\label{Sec:StatisticsAndSensitivity}

\begin{figure}
{\par\centering \resizebox*{1\textwidth}{!}{\includegraphics{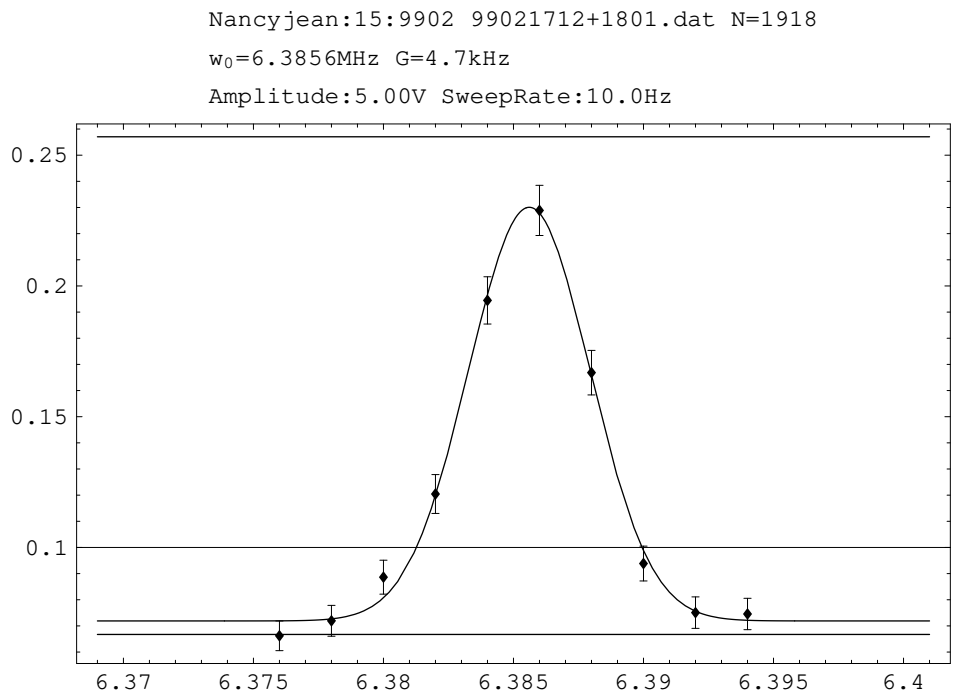}} \par}

\caption{\label{Fig:PrecisionSpinProfile}Spin resonance profile with high S/N.}
\end{figure}

The linewidth reflects technical problems but even in its present state illustrates
the statistical power of these methods. The linewidth of the previous profile
is about \( 2kHz \) but the center frequency is determined to \( 0.5kHz \)
with its few hundred trials at each frequency taking about 3 hours to generate.
This can be further improved simply with better statistics. Fig.\ref{Fig:PrecisionSpinProfile}
shows a profile using about 2000 trials at each frequency which required 40
hours to collect. 

The linewidth is similarly about \( 2kHz \), but here the fit yields a center
frequency accurate to \( 0.1kHz \), or about \( 1/20 \)th of the linewidth.
With shielding, and an interaction time of about 10s the linewidth should be
able to be reduced to \( 0.1Hz \). This would make measurement of a few \( Hz \)
parity shift precise to about \( 0.01Hz \), 1 part in 200 in a day, already
very close to the 1000:1 S/N desired. Possible improvements on this are easily
seen from a careful estimate of the statistical uncertainty in determining a
resonance frequency with this procedure.

The frequency and with had been determined from a fit to the profile, but a
simple mean works just as well, the results agree to within their uncertainties,
and a mean is far simple to analyze, though in practice a fit is better since
using a simple mean requires sampling the entire distribution, an easily asymmetric
sample will skew the mean. The pumped shelving probability provides the offset
and normalization can be computed from the data. Then the mean frequency \( \left\langle \omega \right\rangle  \)
will be given by \( \left\langle \omega \right\rangle =\sum _{i}\Delta p_{i}\omega _{i}/\sum _{i}\Delta p_{i} \),
with \( \Delta p_{i}=p_{i}-p_{0} \). The largest uncertainties in this as an
estimate of the resonance frequency is from the statistical accuracy of the
shelving probabilities, \( \sigma _{p}^{2}=p\left( 1-p\right) /N \), rather
than errors in the frequency which is provided by a high quality arbitrary waveform
generator with a frequency stable, and accurate, to a part in \( 10^{6} \).Then
\( \sigma _{\Delta p}^{2}=\sigma _{p}^{2}+\sigma _{p_{0}}^{2} \). Generally
\( p_{0} \) is known better than any \( p_{i} \) since it is based on more
data points, thought the uncertainties could be about the same order they are
certainly not very much larger so simply estimate the error in \( \Delta p \)
by \( \sigma _{\Delta p}^{2}<\approx 2\sigma _{p}^{2}=2p\left( 1-p\right) /N \).
Further, \( p<1 \) gives \( p^{2}<p \) and an estimate on an upper bound of
\( \sigma _{\Delta p} \) can be made by, \( \sigma _{\Delta p}^{2}<2p/N \).
This turns out to make the final form of the uncertainty in \( \left\langle \omega \right\rangle  \)
particularly simple and intuitive. 

The error in the mean depends on these uncertainties in the shelving probability
through 
\begin{eqnarray*}
\partial \left\langle \omega \right\rangle /\partial \Delta p_{i} & = & \frac{\omega _{i}}{\sum _{j}\Delta p_{j}}-\frac{\sum _{j}\Delta p_{j}\omega _{j}}{\left( \sum _{j}\Delta p_{j}\right) ^{2}}\\
 & = & \frac{1}{\sum _{j}\Delta p_{j}}\left( \omega _{i}-\frac{\sum _{j}\Delta p_{j}\omega _{j}}{\sum _{j}\Delta p_{j}}\right) \\
 & = & \frac{\omega _{i}-\left\langle \omega \right\rangle }{\sum _{j}\Delta p_{j}}
\end{eqnarray*}
 Treating these statistical uncertainties as independent the total uncertainty
in the mean can be written
\begin{eqnarray*}
\delta \left\langle \omega \right\rangle ^{2} & = & \sum _{i}\left( \frac{\partial \left\langle \omega \right\rangle }{\partial \Delta p_{i}}\delta \Delta p_{i}\right) ^{2}\\
 & = & \frac{2}{\sum _{j}\Delta p_{j}}\sum _{i}\frac{\left( \omega _{i}-\left\langle \omega \right\rangle \right) ^{2}}{\sum _{j}\Delta p_{j}}\frac{p_{i}}{N}
\end{eqnarray*}
The sum involving \( \Delta \omega ^{2} \) is simply the width of the distribution
which will be related to \( \Gamma  \) and some geometrical factor \( \alpha  \)
that depends on the shape of the distribution defined by \( \left\langle x^{2}\right\rangle =\alpha ^{2}\Gamma ^{2} \).
For example, for a gaussian distribution \( \alpha ^{2}=\int dxx^{2}e^{-(x/2\Gamma )^{2}}/\int dxe^{-(x/2\Gamma )^{2}}/\Gamma ^{2}=1/8 \),
generally the area should include the portion of the distribution that is sampled.
This gives 
\[
\delta \left\langle \omega \right\rangle \approx 2\alpha \frac{\Gamma }{\sqrt{N}}\frac{1}{\sqrt{\sum \Delta p}}\]
This has a sensible dependence on \( \Gamma  \) and \( N \), and shows that
data taken at points off resonance don't improve the accuracy of the mean as
for those point \( \Delta p=0 \), and more points taken close to resonance
reduces the uncertainty are they contribute another \( \Delta p \). 

With some simple assumptions about the distribution of data points, this uncertainty
can be written in terms of more physical parameters including shelving probabilities,
detection efficiency and spin lifetime. If the frequencies are distributed closely
and evenly over the the resonance than the sum approaches the integral of the
distribution. Not yet including the offset and amplitude, this gives 
\begin{eqnarray*}
\left( \sum \Delta p\Delta \omega \right) /\Delta \omega  & \approx  & \left( \int d\omega p\left( \omega \right) \right) /\Delta \omega \\
 & = & \beta \Gamma /\left( 2\Gamma /N_{\omega }\right) \\
 & = & \beta N_{\omega }/2
\end{eqnarray*}
Again \( \beta  \) is geometrical factor that depends on the distribution,
\( \int d\omega p(\omega )=\beta \Gamma  \), for a gaussian it is \( \beta =\sqrt{\pi }/2 \),
and \( \Delta \omega  \) was written in terms of the total width of the distribution
that is significantly non-zero, about twice the half width, and the number of
frequencies sampled in that range.

The shelving probability ranges between the pumped spin probe shelving probability
and the alternate pumped or unpolarized probability, giving an amplitude of
\( p_{altpump}-p_{pump} \), or simply \( s \) in terms of the variables in
sec.\ref{Sec:SpinLifetimes}. With a finite spin lifetime, \( \Gamma _{s} \),
this discriminant become time dependent and the expression for the uncertainty
in \( \left\langle \omega \right\rangle  \) becomes
\begin{eqnarray*}
\delta \left\langle \omega \right\rangle  & \approx  & 2\alpha \frac{\Gamma }{\sqrt{N}}\frac{1}{\sqrt{s_{0}e^{-\Gamma _{s}t}\beta N_{\omega }/2}}\\
 & = & \left( \frac{2\sqrt{2}\alpha }{\sqrt{\beta }}\right) \frac{\Gamma }{\sqrt{NN_{\omega }}}\frac{e^{\Gamma _{s}t/2}}{\sqrt{s_{0}}}
\end{eqnarray*}
In turn \( s_{0} \) depends on pump and probe laser polarizations, and shelving
and probe rates and times. Currently for the \( D \) state this precisions
is limited by the large spin transition linewidth on the order of a few \( kHz \),likely
due to magnetic field noise. For the ground state this would be larger due to
the larger \( g \) factor, \( 2.8:1 \).

Once the technical parameters are optimized, the statistical limits are improved
simply by increasing the amount of data that is collected, which depends in
terms on the data collection rate, \( 1/\Delta t \), and running time, \( T \)
by \( NN_{\omega }=T/\Delta t \). In terms of these parameters, 
\begin{eqnarray*}
\delta \left\langle \omega \right\rangle  & \approx  & \left( \frac{2\sqrt{2}\alpha }{\sqrt{\beta }}\right) \frac{\Gamma }{\sqrt{T/\Delta t}}\frac{e^{\Gamma _{s}t/2}}{\sqrt{s_{0}}}
\end{eqnarray*}
\( \Delta t \) is not necessarily the observation time, \( t \), due to an
offset for overhead in preparing and analyzing an individual trial. They will
be related by some \( t_{0} \), \( \Delta t=t+t_{0} \). For the current measurements
each trial takes about 2 seconds and sensitivity is largely limited by this
slow data collection rate. Each step in a measurement cycle takes a few tenths
of a second, though most processes occur at \( kHz \) to \( MHz \) rates.
These limits are partly mechanical, beam blocks are chopper wheels on stepper
motors which cannot be switched in much less than 0.1\( s \), the shelving
lamp is controlled by a filter wheel on a stepper motor that requires about
a half second to switch, and the laser polarizations are controlled by an LCD
variable retardation plate with a time constant of around a second. The biggest
limit though is the shelving rate, currently at about a few per second. Maximizing
the shelving transition probability, and in turn the spin discriminant, then
requires shelving times of no less than a half second. The mechanical constraints
can be eased with modified hardware, switching times on the order of several
\( ms \) should be readily possible. The shelving rate could be improved slightly
with more careful alignment and focusing of the shelving lamp, but a dramatic
increase would probably require a different shelving method such as with the
use of a \( 1.76\mu m \) laser.

These improvements would make the data collection limited by the observation
time, \( \Delta t\approx t \). This observation time could be reduced to yield
a higher data collection rate. The transition rate \( R \) must be fast enough
that something happens during the observation time, \( R\sim 1/t \). This faster
rate gives an increasing contribution to the linewidth \( \Gamma _{R}\sim R \)
so that when it is comparable to the linewidth limit given by the magnetic field
environment and other environmental constraints, \( \Gamma _{0} \), the rate
increases like \( 1/t \), so that the uncertainty increases like \( 1/\sqrt{t} \)
and sensitivity decreases. This suggests selecting \( t \) such that \( t\sim 1/\Gamma _{0} \)
giving,

\begin{eqnarray*}
\delta \left\langle \omega \right\rangle  & \approx  & \left( \frac{2\sqrt{2}\alpha }{\sqrt{\beta }}\right) \sqrt{\frac{\Gamma _{0}}{T}}\frac{e^{\Gamma _{s}/2\Gamma _{0}}}{\sqrt{s_{0}}}
\end{eqnarray*}
Ideally this limiting linewidth is due to the finite spin lifetime so that \( \Gamma _{0}=\Gamma _{s} \),
returning to \( \Gamma _{0}=1/t \)

\begin{eqnarray*}
\delta \left\langle \omega \right\rangle  & \approx  & \left( \frac{2\sqrt{2e}\alpha }{\sqrt{\beta }}\right) \frac{1}{\sqrt{Tts_{0}}}
\end{eqnarray*}

\subsection{Other spin resonance experiments}

Once the spin transition linewidth is better controlled a number of other experiments
will be possible. Eventually the linewidth will be determined solely by the
spin flip transition rate. Faster transition rates give broader resonances.
This should be accessible with even a slight improvement of the linewidth as
the transition rate is expected to be a large fraction of a \( kHz \). At this
point the interaction gives effectively complete control over a pure \( j=1/2 \)
or \( j=3/2 \) system, \cite{Schacht00}.

In particular, the details of the spin precession of a single ion will be observable.
The spin flip transition can be applied on resonance for various times and the
transition probability determined as a function of time, yielding oscillations
in the shelving probability corresponding to the ion alternating between spin
up and spin down. This requires careful control of the interaction time, which
can be difficult when the transition rate is \( kHz \) as the period is only
\( ms \), and switching on these time scales is not trivial. The programmable
function generator used in this system to provide the RF spin flip frequency
could be programmed to provide a single short pulse of RF and a number a pulses
applied to the ion. The transition probability as a function of the number of
applied pulses should provide a map of the precession. This will be directly
applicable to the parity measurement if the precession is resolved to improve
sensitivity.

This function generator also provides detailed control of the applied frequency
and the frequency could be changed adiabatically to provide ramps and chirps
that flip the slip exactly independent of the interaction time. The frequency
could also be modulated from monochromatic to effectively broadband to study
the transition from coherent precession to decoherent decay. The limits are
well known, but the intermediate cases are difficult to analyze and the behavior
probably depends on the detailed phase structure of the modulation rather than
just the frequency content. This has less immediate practical applications to
the parity experiment but is an interesting diversion as such studies on this
kind of pure spin system are not otherwise possible.

\section{Light Shifts}

\label{Sec:NonresonantLightShifts}

The intrinsic sensitivity of this spin resonance technique is already very good,
and can be improved significantly with a little effort. The limit now is the
\( kHz \) sized spin resonance linewidth. Considerable work will be required
to improve this to the fraction of a hertz necessary for a parity measurement
though it should be straightforward with careful magnetic shielding and careful
selection of the materials that must be near the ion. Alternately, for much
larger shifts of a few \( kHz \) and more, a \( kHz \) linewidth is perfectly
sufficient and provides a means measuring some quantities with a precision difficult
to achieve with other methods. 

Light shifts can be generated by driving any transition. Measuring the resulting
shift gives the transition matrix element for the states involved if the electric
field of the light driving the transition is known. This amplitude is usually
difficult to determine independently with sufficient precision, but if the same
light happens to also drive another transition whose matrix element is known,
the field can be measured and used to determine the unknown matrix element.
More often, neither matrix element is well known but instead, both light shifts
can be used to determine the ratio of the matrix elements precisely since the
ratio of the lights shifts will be independent of the electric field amplitude.
This doesn't give accurate information about the overall size of the matrix
elements but is just as useful for purposes such as testing atomic structure
calculations.

Two transitions usually can't be driven simultaneously at resonance with a single
applied interaction without a convenient accidental coincidence of transition
energies. Every transition can be driven nonresonantly, but with reasonable
laser powers and spot sizes the resulting electric fields yield shifts on the
order of tens of \( kHz \). Relative to \( MHz \) sized shift on resonance
this is relatively very small and difficult to measure with traditional spectroscopic
methods even with tremendous work on lasers since the natural linewidths of
detectable transitions are on the order of \( MHz \). But the size is just
right for these spin resonance techniques.

With Barium, circularly polarized light could be used to shift a single spin
level in the ground state and two spin levels in the \( D_{3/2} \) state simultaneously.
These shifts would be measured by the change in the spin flip resonance frequency
currently detectable to within about 0.1\( kHz \). 10\( kHz \) sized shifts
known to \( 0.1kHz \) give a \( 0.1\% \) measurement of the ratio. This provides
a tremendously stringent constraint on any atomic structure calculation hoping
to account for the result as it is usually a considerable challenge to achieve
even a \( 1\% \) precision. Precisions of \( 0.1\% \) for atomic calculations
of some important parameters will eventually be required to interpret the results
of a parity measurement, so these light shift ratios are useful to provide an
early benchmark for improving these calculations. The measurement sequence will
also be operationally identical to the parity measurement and so provides a
further test of the techniques and a framework for improving stability, sensitivity
and accuracy, and for simplifying and streamlining the procedure.

\subsection{Tensor Structure of Shifts}

Far off resonance, the rotating wave approximation, which discards the counter-rotating
term, is no longer appropriate as it now contributes at about the same order
as the co-rotating term so both must be included but the contributions can simply
be added if the resulting shifts are small compared to the laser detunings.
Transforming either case to a static frame gives
\[
H=\left( \begin{array}{cccc}
0 & \Omega _{01}^{\dagger } & \Omega _{02}^{\dagger } & \cdots \\
\Omega _{01} & \Delta E_{01}\pm \omega  & \Omega _{12}^{\dagger } & \cdots \\
\Omega _{02} & \Omega _{12} & \Delta E_{02}\pm \omega  & \cdots \\
\vdots  & \vdots  & \vdots  & \ddots 
\end{array}\right) \]
\( \Omega _{ij} \) are matrices giving the coupling between multiplets, for
a dipole interaction
\[
\left( \Omega _{ij}\right) _{m'm}=e\frac{\vec{E}}{2}\left\langle i,m'\left| \vec{r}\right| j,m\right\rangle \]
In general, there is also a magnetic field splitting the magnetic sublevels
of each multiplet, but large detuning this splitting can be neglected so that
the diagonal blocks in \( H \) can be considered to be simple scalars, proportional
to the appropriate identity matrix, to compute the shifts.

Eigenstates corresponding to a particular unperturbed state taken to be \( i=0 \)
can be computed approximately by neglecting the couplings between \( i,j\neq 0 \)
states and are given by the eigenstates, \( \chi _{0m} \), of 
\[
H_{0m}=\sum _{j,\pm \omega }\frac{\Omega _{0j}^{\dagger }\Omega _{0j}}{E_{0m}-\left( \Delta E_{0j}\pm \omega \right) }\]
by
\[
\psi _{0m}=\left( \begin{array}{c}
\chi _{0m}\\
\Omega _{01}\chi _{0}/\left( E_{0m}-\left( \Delta E_{01}\pm \omega \right) \right) \\
\vdots 
\end{array}\right) \]
\( E_{0m} \) is the eigenvalue for a particular spin state in the level being
considered, \( H_{0m}\chi _{0m}=E_{0m}\chi _{0m} \). For resulting energy shift
very small compared to energy separations between levels and the laser frequency
the \( E \) in the denominators can be neglected,
\[
H_{0m}=H_{0}=\sum _{j,\pm \omega }\frac{\Omega _{0j}^{\dagger }\Omega _{0j}}{\Delta E_{0j}\pm \omega }\]
\( H_{0m} \) then becomes \( m \) independent and acts as an effective hamiltonian
completely within a level providing a compact summary of the effects of the
entire interaction, \cite{Schacht00}. An effective \( H_{i} \) can be similarly
constructed for any initial energy level.

The sum is over all states with a dipole coupling to the initial state, for
the \( S_{1/2} \) and \( D_{3/2} \) initial states that will be considered
here the sum will include \( p \) and \( f \) states with \( j=1/2,3/2,5/2 \)
as appropriate. The sum over all spin levels within a particular energy level
is contained in the matrix multiplication of the \( \Omega  \). The energy
denominator does not strongly suppress more highly excited levels, though the
matrix elements do tend to get smaller. The relevant set of state can turn out
to be quite large and all must be included for a precise prediction, but their
general effects are well defined, and can be classified and sorted. This can
be dealt with efficiently using the techniques developed to analyze quadrupole
misalignment systematics, in particular, in term of generalized pauli matrices,
\[
H_{0}=\frac{e^{2}}{4}\sum _{j,\pm \omega }\frac{\left| \left\langle i\left| \left| D\right| \right| j\right\rangle \right| ^{2}}{\Delta E_{0j}\pm \omega }\sigma _{k}^{\dagger }\left( j_{i},j_{j}\right) \sigma _{l}\left( j_{i},j_{j}\right) E_{k}^{*}E_{l}\]
 As before the \( \sigma  \) have simple multiplication rules of the form,
\[
\sigma _{i}^{\dagger }\left( j,j'\right) \sigma _{j}\left( j,j'\right) =s(j,j')\delta _{ij}+id(j,j')\epsilon _{ijk}j_{k}+q(j,j')j_{ij}\]
where \( j_{i} \) and \( j_{ij} \) are the dipole and quadrupole angular momentum
operators given previously. With this normalization the coefficients needed
for the problem considered here are listed in Table.\ref{Tab:PauliMatrixProductCoefficients}. 
\begin{table}

\caption{\label{Tab:PauliMatrixProductCoefficients}Product coefficients}
{\centering \begin{tabular}{|c|c|c|c|}
\hline 
\( j,j' \)&
\( s \)&
\( d \)&
\( q \)\\
\hline 
1/2,1/2&
\( 1/6 \)&
\( 1/3 \)&
\\
\hline 
1/2,3/2&
\( 1/6 \)&
\( -1/6 \)&
\\
\hline 
3/2,1/2&
\( 1/12 \)&
\( 1/12 \)&
\( -1/12 \)\\
\hline 
3/2,3/2&
\( 1/12 \)&
\( 1/30 \)&
\( 1/15 \)\\
\hline 
3/2,5/2&
\( 1/12 \)&
\( -1/20 \)&
\( -1/60 \)\\
\hline 
\end{tabular}\par}\end{table}

Now consider the \( S \) and \( D \) states separately. The contributions
for intermediate states can be separated and the \( \sigma  \) matrices multiplied
giving scalar, vector and quadrupole tensor components. The effective hamiltonian
will then have the general structure,
\begin{eqnarray*}
H/\left( e^{2}/4\right)  & = & \left( a\delta _{ij}+ib\epsilon _{ijk}j_{k}+j_{ij}\right) E_{i}^{*}E_{j}\\
 & = & s\left( \vec{E}^{*}\cdot \vec{E}\right) +id\left( \vec{E}^{*}\times \vec{E}\right) \cdot \vec{j}+qj_{ij}\left( E_{i}^{*}E_{j}\right) \\
H & = & \left( e^{2}/4\right) \left| \vec{E}\right| ^{2}\left( s+d\vec{\sigma }_{L}\cdot \vec{j}+q\left( \epsilon _{i}^{*}\epsilon _{j}\right) j_{ij}\right) \\
 & = & \delta \omega _{s}+\delta \omega _{d}\vec{\sigma }_{L}\cdot \vec{j}+\delta \omega _{q}\left( \epsilon _{i}^{*}\epsilon _{j}\right) j_{ij}
\end{eqnarray*}
The \( \epsilon _{i} \) are the components of the polarization of the light,
then \( \vec{\sigma }_{L}=i\left( \vec{\epsilon }\times \vec{\epsilon }^{*}\right)  \)
is the helicity. \( j_{ij} \) is symmetric so \( \epsilon _{i}^{*}\epsilon _{j} \)
can be replaced by \( \left( \epsilon _{i}^{*}\epsilon _{j}+\epsilon _{j}^{*}\epsilon _{i}\right) /2=Re\left( \epsilon _{i}^{*}\epsilon _{j}\right)  \). 

The coefficients are given by the matrix elements and detunings. Using 
\[
\gamma _{m}^{i}=\sum _{j=j,\pm \omega }\frac{\left| \left\langle i\left| \left| D\right| \right| j\right\rangle \right| ^{2}}{\Delta E_{ij}\pm \omega }\]
The coefficients for the \( S_{1/2} \) state are,

\begin{eqnarray*}
s_{S} & = & \frac{1}{6}\left( \gamma ^{S}_{1/2}+\gamma ^{S}_{3/2}\right) \\
d_{S} & = & \frac{1}{6}\left( 2\gamma ^{S}_{1/2}-\gamma ^{S}_{3/2}\right) \\
q_{S} & = & 0
\end{eqnarray*}

and for the \( D_{3/2} \) state,

\begin{eqnarray*}
s_{D} & = & \frac{1}{12}\left( \gamma ^{D}_{1/2}+\gamma ^{D}_{3/2}+\gamma ^{D}_{5/2}\right) \\
d_{D} & = & \frac{1}{12}\left( \gamma ^{D}_{1/2}+\frac{2}{5}\gamma ^{D}_{3/2}-\frac{3}{5}\gamma ^{D}_{5/2}\right) \\
q_{D} & = & \frac{1}{12}\left( -\gamma ^{D}_{1/2}+\frac{4}{5}\gamma ^{D}_{3/2}-\frac{1}{5}\gamma ^{D}_{5/2}\right) 
\end{eqnarray*}

\begin{figure}
{\par\centering \includegraphics{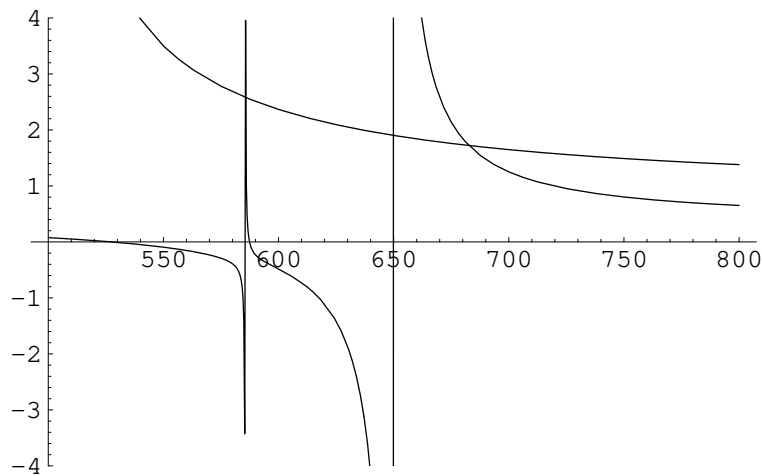} \par}

\caption{\label{Fig:CalculatedLightShifts} Calculated Light Shifts as a function of
frequency in \protect\( kHz/mW\protect \) for a \protect\( 200\mu m\protect \)
diameter beam.}
\end{figure}

The \( s_{i},d_{i},q_{i},\gamma _{j}^{i} \) contain the contributions from
atomic structure and are the points at which to compare to atomic theory. To
estimate the general size of these shifts the wavefunctions are approximated
by coulomb wavefunctions corresponding to the empirically determined energies
for the states. The scalar and dipole shifts, \( s_{i} \) and \( d_{i} \),
in \( kHz/mW \) for the \( S \) and \( D \) state are shown in fig.\ref{Fig:CalculatedLightShifts}
for a laser focussed to a \( 200\mu  \) diameter spot.

The shift is wavelength dependent and the ideal wavelength would shift both
\( S_{1/2} \) and \( D_{3/2} \) spin states by comparable amounts. This might
initially suggest a frequency somewhere between the two resonances to the \( P_{1/2} \)
state. But, in the sum over states, the energy denominator determines the sign.
If \( \omega  \) is chosen such that \( \omega <\Delta E_{ij} \) for all transitions,
that is the frequency is below the resonance frequency of all transitions, all
denominators are positive and the shift is largest, while when some transitions
are driven above resonance the shift is reduces. As a result, in the region
between the \( S_{1/2}\rightarrow P_{1/2} \) and \( D_{3/2}\rightarrow P_{1/2} \)
resonances the \( D_{3/2} \) shifts are actually very small and the shifts
are largest for both near, and on the red side of the \( D_{3/2}\rightarrow P_{1/2} \)
resonance. 

Since the larger ground state \( g \)-factor is larger than the \( D \) state
\( g \)-factor by about a factor of three, the spin resonance linewidth is
about three times larger in the ground state. Then to get a comparable sensitivity
for shifts in both states the ground state shift should be three times larger
than the \( D \) state shift. But this far to red of any ground state resonance
the shift is fairly independent of frequency, and it doesn't help to deliberately
make the \( D \) state shifts smaller without a gain in the size the the ground
state shift. So almost any frequency somewhat to the blue, or far red of the
\( D_{3/2}\rightarrow P_{1/2} \) resonance is suitable. The region very close
to the resonance should be avoided to keep the shifts on the same order of magnitude
so that the same method can be used to measure both, and so that the \( D \)
state shift is not dominated by interactions with the \( P_{1/2} \) state at
the cost of lost information about other states. Frequencies in this region
give shifts on the order of \( 1-3kHz/mW \).

\subsection{Resonances}

This gives the formal structure in a general interaction. The energy shifts
are detected through changes in the spin flip resonance frequencies, so to make
the final connection to an experimental observable the effect of a particular
field configuration on the resonances should be considered.

\subsubsection{Linear Polarization}

Linear polarization parallel to the applied magnetic field gives no splitting
in the ground state but does change the structure of the \( D \) state. In
the general result, \( \sigma =0 \) leaves only the scalar term in the ground
state shift but also keeps the quadrupole shift in the \( D \) state. This
can also be understood as a consequence of simple selection rules. Spherical
symmetry requires that the resulting \( \Delta m=0 \) transitions couple \( \pm m \)
states with equal strengths, but states with different \( \left| m\right|  \)
will be coupled differently. For the ground state there is only one \( \left| m\right|  \),
so there is only one shift and both states move by the same amount. In the \( D \)
state, the \( \left| m\right| =3/2 \) and \( \left| m\right| =1/2 \) states
are coupled differently. For transitions to \( j=3/2 \), or \( j=5/2 \) this
is because the Clebsch-Gordan coefficients differ by more than just a sign between
the \( m=\pm 3/2 \) and the \( m=\pm 1/2 \) states. For transitions to \( j=1/2 \)
states the difference is simply because the \( m=\pm 3/2 \) states are left
uncoupled as there are no corresponding states that satisfy the selection rule.
This results in the \( m=\pm 1/2 \) states being shifted differently than the
\( m=\pm 3/2 \) states.

This polarization isn't useful for light shift ratios as there will be no ground
state splitting to use for normalization, but this case will briefly be considered
first as it is the configuration used for the first detection of these off resonant
dipole light shifts and the general structure of the shift is very simple. Let
the coordinate system be defined by the magnetic field so that \( \hat{z}||\hat{B} \).
For this case the electric field is ideally also completely in the \( \hat{z} \)
direction, \( \vec{E}=E\hat{z} \). The shift is then given simply by,
\[
\Delta H=\frac{e^{2}}{4}\left| \vec{E}\right| ^{2}(s_{D}+q_{D}j_{zz})\]
For \( j=3/2 \), \( J_{zz} \) is \( diag\left( \begin{array}{cccc}
1 & -1 & -1 & 1
\end{array}\right)  \). Again showing that the end result is for the the \( m=\pm 1/2 \) states to
be shifted together away from the \( m=\pm 3/2 \) states. This gives three
different spin resonance frequencies rather than just one. The splitting between
the \( m=\pm 1/2 \) states is unchanged, it stays at the frequency corresponding
to the dipole splitting due to the magnetic field. The energies of the \( m=\pm 3/2 \)
to \( m=\pm 1/2 \) transitions are changed, \( \Delta \omega =\Delta \omega _{+3/2,+1/2}=-\Delta \omega _{-1/2,-3/2}=2q_{D}(e^{2}/4)\left| \vec{E}\right| ^{2} \).
As discussed in Sec.\ref{Sec:ResonancesForQuadrupoleShifts} with spin resonance
profiles, this gives two resonance peaks. If using a linearly polarized pump
and probe, the peaks are due to the \( m=\pm 3/2 \) to \( m=\pm 1/2 \) transitions
and both are shifted in opposite directions from the original position of the
dipole resonance, at \( \omega _{0}\pm \Delta \omega  \), and the \( m=\pm 1/2 \)
transition is not visible. For a circularly polarized probe, the dipole peak
remains and a second peak to one side appears at \noun{}\( \omega _{0}\pm \Delta \omega  \).
Either probe polarization is perfectly as the shifting beams and pump/probe
beams are independent, but since the quantity of interest here is \( q_{D} \)
a linear probe gives the best S/N as the resulting split peaks are separated
by \( 2\Delta E \) rather than \( \Delta E \) as for a circular probe.

\subsubsection{Circular Polarization}

A change in the splitting of the ground state magnetic sublevels requires a
dipole shift, which is given only by circular polarization. First consider perfectly
circularly polarized off-resonant light. Again, let the coordinate system be
defined by the magnetic field with \( \hat{z}||\vec{B} \). Ideally, the light
also propagates exactly along this direction so that \( \vec{\sigma }\propto \hat{k}=\hat{z} \),
so the dipole term is just \( 2dj_{z} \). The polarization vectors are, for
example, \( \vec{\epsilon }_{x}=\hat{x}/\sqrt{2},\vec{\epsilon }_{y}=i\hat{y}/\sqrt{2} \),
which gives \( Re\left( \epsilon _{x}^{*}\epsilon _{y}\right) =0 \) and \( Re\left( \epsilon _{x}^{*}\epsilon _{x}\right) =Re\left( \epsilon _{y}^{*}\epsilon _{y}\right) =1/2 \).
This is particularly convenient as only \( j_{i,j=i} \) have off diagonal terms,
and since \( j_{ij} \) is traceless in the \( i,j \) indices \( j_{xx}+j_{yy}=-j_{zz}=\left( \begin{array}{cccc}
-1 & 1 & 1 & -1
\end{array}\right)  \). The quadrupole shift is just proportional to \( \sum _{i,j}Re\left( \epsilon _{i}^{*}\epsilon _{j}\right) j_{ij}=\left( 1/2\right) \left( j_{xx}+j_{yy}\right) =-\left( 1/2\right) j_{zz} \).
And the total shift is given by
\[
\Delta H=\frac{e^{2}}{4}\left| \vec{E}\right| ^{2}(s+dj_{z}-\left( q/2\right) j_{zz})\]
For the ground state, \( j_{z}=diag\left( \begin{array}{cc}
1/2 & -1/2
\end{array}\right)  \), \( j_{zz}=0 \), so the change in the splitting of the ground state sublevels
is \( \delta \omega ^{S}_{d}=2d_{S}(e^{2}/4)\left| \vec{E}\right| ^{2} \).
For the \( D \) state the \( m=\pm 1/2 \) splitting is changed only by the
dipole term. Here also \( \left( j_{z}\right) _{\pm 1/2,\pm 1/2}=\pm 1/2 \)
so \( \delta \omega _{d}^{D}=2d_{D}(e^{2}/4)\left| \vec{E}\right| ^{2} \).
This term also changes the \( m=\pm 3/2 \) to \( m=\pm 1/2 \) transition energies,
by the same amount since \( \left( j_{z}\right) _{m,m}=\left( j_{z}\right) _{m-1,m-1}=1 \).
The quadrupole term also changes these energies, now in different directions
by \( \delta \omega ^{D}_{q}=\pm q_{D}(e^{2}/4)\left| \vec{E}\right| ^{2} \)
so that the total shift is given by \( \delta \omega ^{D}_{d}\pm \delta \omega _{q}^{D} \). 

Circularly polarized pump/probe light must be used to resolve ground state spin
levels, so circularly polarized light should be used for the \( D \) state
as well to avoid having to switch polarizations and move magnetic fields. The
shifts appear as changes in the spin resonance frequencies, initially \( \omega _{0}^{S,D} \).
The ground state shift moves the single dipole peak by \( \delta \omega ^{S}_{d} \)
and is then measured simply by \( \delta \omega ^{S}_{d}=\omega ^{S}-\omega _{0}^{S} \),
where \( \omega ^{S} \) is the new position of the resonance. The dipole resonance
peak in the \( D \) state is also just moved by the dipole shift so that \( \delta \omega ^{D}_{d}=\omega ^{D}_{d}-\omega ^{D}_{0} \),
\( \omega ^{D}_{d} \) is the position of the dipole peak. The quadrupole peak
appears at \( \omega ^{D}_{q}=\omega ^{D}_{0}+\delta \omega _{d}^{D}\pm \delta \omega _{q}^{D} \).
Only one quadrupole peak appears when using a circularly polarized probe, the
\( \pm  \) depends on the sense of circular polarization used for the probe.
Since \( \omega ^{D}_{d}=\delta \omega ^{D}_{d}+\omega ^{D}_{0} \) the quadrupole
peak is just split from the dipole peak by \( \pm \delta \omega _{q}^{D} \)
and can simply be measured by \( \delta \omega _{q}^{D}=\omega ^{D}_{q}-\omega ^{D}_{d} \).

With this scheme the dipole and quadrupole shifts for both states are independently
determined, and simple functions that depend only on the matrix elements and
the frequency of the applied laser can be made simply by taking ratios. As initially
discussed, the ground state shift can be used to normalized the \( D \) state
shifts, or vise-versa, giving
\begin{eqnarray*}
\delta \omega ^{D}_{d}/\delta \omega ^{S}_{d} & = & d_{D}/d_{S}\\
\delta \omega _{q}^{D}/\delta \omega _{s}^{S} & = & q_{D}/d_{S}
\end{eqnarray*}
Note that the ratio of \( D \) state shifts is also non-trivial,
\[
\delta \omega _{q}^{D}/\delta \omega ^{D}_{d}=q_{D}/d_{D}\]
allowing for useful constrains on atomic matrix elements using only the \( D \)
state. This wasn't possible using a linearly polarized interaction beam as in
that case there was really only one shift and the changes in frequencies were
all exactly correlated.

\subsubsection{Systematics, Polarization and Alignment Independence of Ratios}

The most important feature of these ratios is that they are all independent
of the applied electric field. This eliminates what is generally the largest
uncertainty in these measurements. The size the applied electric field depends
on laser power, and the size and position of the beam at the ion. The power
could be measured and monitored outside the chamber, but that will be altered,
at least, by the window the beam must pass through to reach the ion in the vacuum
chamber. Even if this modification could be exactly accounted for so that power
reaching the ion was exactly known, the even less well known beam size and position,
also modified by the window as well, would prevent precise knowledge of the
electric field at the ion. Since the intended spot size is very small, gradients
are large and the electric field is very sensitive to position.

These ratios eliminate the difficulty of this inability to independently determine
the size of the applied electric field. The field must be large enough to provide
a measurable splitting with sufficient S/N, and it must be stable enough that
the shifted peaks are narrow enough that their positions can be measured with
sufficient S/N, but beyond that the precise magnitude is not important.

Another source of uncertainty are imperfections in the polarization or alignment
of the shifting laser. These results were derived assuming a perfectly circularly
polarized beam, perfectly aligned with the magnetic field. The ratios are still
trivially independent of the overall size of the electric field whatever the
polarization and alignment as the amplitude is simply an overall factor for
all the shifts, but these imperfections can change the structure and relative
sizes of the shifts and the meaning of the ratios. These parameters are difficult
to control and measure with arbitrarily high precision, and even if they could
outside the chamber, they can be altered by windows and local magnetic source
inside the chamber where the resulting induced imperfections can't easily be
measured so it is important to consider the ignorance of these changes on the
resulting ratios. 

Start from the ideal case and consider a small deviation from perfect circular
polarization, \( \delta \sigma  \) so that \( \sigma =1-\delta \sigma  \),
and a small nonzero angle, \( \alpha  \) between \( \vec{k} \) and \( \vec{B} \).
Take the \( \vec{k} \) and \( \vec{B} \) to define the \( x-z \) plane so
that this misalignment is in the \( \hat{x} \) direction. The misalignment
gives the electric field a small component in the \( \hat{z} \) direction proportional
to \( sin\left( \alpha \right) \approx \alpha  \), and reduces the field in
the \( \hat{x} \) direction by \( cos\left( \alpha \right)  \), which is second
order in \( \alpha  \) so it will eventually be neglected. \( \vec{\sigma } \)
gets a small \( \hat{x} \) component proportional to \( sin\left( \alpha \right)  \). 

The general result is a \( 4\times 4 \) matrix, now with off-diagonal terms.
The exact eigenvalues can be obtained with a lot of work, but they are not of
too much use and more insight is gained from a perturbative adjustment to the
ideal case. First consider the effects of these imperfections on the dipole
splittings, in terms of the original shift,
\[
\Delta H_{d}=\delta \omega _{d}\left( 1-\delta \sigma \right) \left( cos\left( \alpha \right) j_{z}+sin\left( \alpha \right) j_{x}\right) \]
This structure is the same for both the ground state and \( D \) state so their
ratios are exactly independent of polarization and alignment, but these shifts
are determined from the resulting changes in spin resonances and that translation
is altered by these errors. With no magnetic field, these shifts would be the
only dipole contribution, not including perturbations from the quadrupole term
as discuss later, so they would be available directly from the \( m=\pm 1/2 \)
splittings. But practical constraints require a magnetic field and it is the
relation shift between these shifts and the direction and size of that applied
field that introduce uncertainties in determining the size of the dipole term
from the same resonances.

\( \Delta H \) alone has eigenvalues of \( m\delta \omega _{d}\left( 1-\sigma \delta \right) /2 \),
for a given \( m \) sublevel, as the misalignment just gives a rotation here
and doesn't change the amplitude of the interaction. This can no longer simply
be added to the magnetic field splitting as the \( j_{x} \) contains off-diagonal
terms so that the eigenvectors of \( \Delta H_{d} \) are not the same as the
eigenvectors of \( j_{z} \) so the whole hamiltonian must be considered,
\begin{eqnarray*}
H_{d} & = & \mu \vec{B}\cdot \vec{J}+\Delta H_{d}\\
 & = & \left( \omega _{0}+\delta \omega _{d}\left( 1-\delta \sigma \right) cos\left( \alpha \right) \right) j_{z}\\
 & + & \delta \omega _{d}\left( 1-\delta \sigma \right) sin\left( \alpha \right) j_{x}\\
 & \equiv  & \vec{\omega }\cdot \vec{j}
\end{eqnarray*}
This is still a vector interaction so the energies are given by the magnitude
of the vector, \( E_{m}=m\left| \vec{\omega }\right|  \), 
\begin{eqnarray*}
\left| \vec{\omega }\right| ^{2} & = & \left( \omega _{0}+\delta \omega _{d}\left( 1-\delta \sigma \right) cos\left( \alpha \right) \right) ^{2}\\
 & + & \left( \delta \omega _{d}\left( 1-\delta \sigma \right) sin\left( \alpha \right) \right) ^{2}\\
\left| \vec{\omega }\right|  & = & \left( \omega _{0}+\delta \omega _{d}\left( 1-\delta \sigma \right) cos\left( \alpha \right) \right) \times \\
 &  & \sqrt{1+\left( \frac{\delta \omega _{d}\left( 1-\delta \sigma \right) sin\left( \alpha \right) }{\omega _{0}+\delta \omega _{d}\left( 1-\delta \sigma \right) cos\left( \alpha \right) }\right) ^{2}}
\end{eqnarray*}
The correction to a simple sum of the eigenvalues is second order in both \( \alpha  \)
and \( \delta \omega _{d} \),
\[
\Delta H_{d}\approx \left( \delta \omega _{d}\left( 1-\delta \sigma \right) cos\left( \alpha \right) +o\left( \alpha \delta \omega _{d}/\omega _{0}\right) ^{2}\right) j_{z}\]
 If the polarization and magnetic field can be made parallel to a part in \( 10^{2} \),
about half a degree, the error is only a part in \( 10^{4} \). Alternately,
if this alignment constraint proves too stringent, the applied magnetic field
can be made large enough that there error is negligible. For an applied field
giving an initial splitting of a few \( MHz \), and off-resonant light generating
shifts of \( 10-100kHz \) , \( \left( \delta \omega _{d}/\omega _{0}\right) ^{2}\approx 10^{-2}-10^{-4} \).
This can make the correction almost completely insensitive to misalignments,
and at least relax alignment constraints to only \( \alpha <10^{-1}\approx 5^{\circ } \)

This can easily be understood geometrically. The dipole shift is a vector interaction
and so can be understood simply as an effective magnetic field, \( \delta \vec{\omega }_{d} \).
The energies are given by the length of the vector sum of this shift and the
magnetic field, \( \mu \vec{B} \). When pointed in the same direction the total
length of the sum just changes by \( \left| \delta \vec{\omega }_{d}\right|  \),
when pointed in different directions the total change is less. The change is
second order for components of the shift perpendicular to the magnetic field
by simple geometry, and linear in shifts parallel to it, so if the perpendicular
components are very much smaller then \( \mu \vec{B} \) their effects on the
length are negligible and the change is just given by the parallel components
of the shift. In effect, the large magnetic field results in only components
of the shift along the field being detected.

These dipole shifts can be determined from the resonance shifts of the \( m=\pm 1/2 \)
transitions as before, so the ratios are approximately insensitive to misalignments
to \( o\left( \alpha \delta \omega _{d}/\omega _{0}\right) ^{2} \) and, in
fact, exactly insensitive to polarization impurities. Both dipole shifts depend
linearly on the circular polarization so that polarization cancels out in the
ratio and the only effect is on the sensitivity as both shifts get smaller.
This is partly confounded by the quadrupole term.

This completely describes the behavior of the dipole terms and as a result the
ground state shifts as it has only a dipole contribution but the \( D \) state
has additional quadrupole structure that must be studied. For this, write the
polarization error as a deviation in the relative phase of \( \hat{\epsilon }_{x} \)
and \( \hat{\epsilon }_{y} \) from perfectly imaginary, \( \hat{\epsilon }_{y}=ie^{i\delta }\hat{y}/\sqrt{2} \).
The misalignment gives, \( \hat{\epsilon }_{x}=cos\left( \alpha \right) \hat{x}/\sqrt{2},\hat{\epsilon }_{z}=sin\left( \alpha \right) /\sqrt{2} \).
With these modifications, the quadrupole term becomes,
\begin{eqnarray*}
Re\left( \epsilon _{i}^{*}\epsilon _{j}\right) j_{ij} & = & \left( cos^{2}\left( \alpha \right) j_{xx}+j_{yy}+sin^{2}\left( \alpha \right) j_{zz}\right) /2\\
 & + & sin\left( \delta \right) cos\left( \alpha \right) j_{xy}+sin\left( \delta \right) sin\left( \alpha \right) j_{yz}\\
 & + & cos\left( \alpha \right) sin\left( \alpha \right) j_{xz}
\end{eqnarray*}

As with the dipole shift, this modifies the diagonal terms, and introduces off
diagonal terms. The diagonal terms are easily dealt with, they come from the
\( j_{i=j} \) terms, \( j_{zz} \) is diagonal, and the diagonals of \( j_{xx} \)
and \( j_{yy} \) are simply related to \( j_{zz} \) as \( j_{xx}^{diag}=j_{yy}^{diag}=-\left( 1/2\right) j_{zz} \).
Then the diagonal part of the quadrupole term is, \( \Delta H_{q}^{diag}=-\left( 1-\left( 3/2\right) sin^{2}\left( \alpha \right) \right) j_{zz}/2 \).
The quadrupole term is primarily sensitive to alignment rather than polarization.

The off-diagonal terms are not so trivially dealt with exactly. With \( j_{xx}+j_{yy}=-j_{zz} \)
so that \( j_{xx}^{off-diag}+j_{yy}^{off-diag}=0 \), the remaining terms are,
\begin{eqnarray*}
Re\left( \epsilon _{i}^{*}\epsilon _{j}\right) j_{ij} & = & \left( cos^{2}\left( \alpha \right) j_{xx}+j_{yy}+sin^{2}\left( \alpha \right) j_{zz}\right) /2\\
 & + & sin\left( \delta \right) cos\left( \alpha \right) j_{xy}+sin\left( \delta \right) sin\left( \alpha \right) j_{yz}\\
 & + & cos\left( \alpha \right) sin\left( \alpha \right) j_{xz}
\end{eqnarray*}

but, also as with the dipole term, it is easily shown that they can be neglected.
Any off diagonal term, \( \omega ^{q}_{mm'} \)will couple states separated
by energies of order a few \( MHz \) with strengths of \( 10-100kHz \). Couplings
between other states complicate an exact result for the resulting shift, but
perturbatively they will be like as with a simple two state problem, \( \delta \omega ^{2}\approx \left( \left( m-m'\right) \omega _{0}\right) ^{2}+\omega ^{q2}_{mm'} \).
This is an adjustment to second order in \( \omega ^{q}_{mm'}/\omega _{0} \),
and the \( \omega ^{q}_{mm'} \) are themselves linear in \( \alpha  \) or
\( \delta  \). So with sufficiently large \( \omega _{0} \) and sufficiently
small \( \alpha ,\delta  \) these off-diagonal terms are negligible, the same
bounds as discussed with the dipole term are sufficient, \( \left( \alpha ,\delta \right) \omega _{q}/\omega _{0}<10^{-2} \). 

For a final observation on the structure of the quadrupole shift, even though
this is a quadrupole coupling, only the diagonal terms leave \( \Delta E_{\pm 1/2} \)
unchanged, the off-diagonal element of this term now can also change this splitting
as the dipole term does. But the error is this same small correction so it doesn't
alter the interpretation of the \( \Delta m=\pm 1/2 \) resonance as the dipole
splitting any more that the misalignment perturbations of the dipole term itself.

The entire shift can then be written as,
\begin{eqnarray*}
\Delta H & = & \delta \omega _{d}\left( 1-\delta \sigma \right) cos\left( \alpha \right) j_{z}-\delta \omega _{q}\left( 1/2\right) \left( 1-\left( 3/2\right) sin^{2}\left( \alpha \right) \right) j_{zz}
\end{eqnarray*}
with corrections to the dipole term to \( o\left( \alpha \omega _{d}/\omega _{0}\right) ^{2} \),
and corrections to both dipole and quadrupole term to \( o\left( \alpha \omega _{q}/\omega _{0}\right) ^{2} \)
and \( o\left( \delta \omega _{q}/\omega _{0}\right) ^{2} \). 

The ratio of the dipole shifts is independent of small alignment and polarization
errors and so is the most promising observable to measure. The ratio of the
quadrupole term to either dipole term is more sensitive to these errors, linear
in \( \delta \sigma  \) and to second order in \( \alpha  \), not additionally
suppressed by \( \left( \omega _{q}/\omega _{0}\right) ^{2} \). This is simply
because, as observed above, the quadrupole term is more sensitive to alignment
than polarization, while the dipole term is highly sensitive to polarization
and equally sensitive to alignment. These terms can be adjusted almost independently
with alignment and polarization, while the dipole terms depend on both in the
same way. A measurement of ratios using the quadrupole shift to a part in \( 10^{3} \),
though very convenient as it could be made using only the \( D \) state, would
require the polarization be accurately known to that precision and the alignment
right to a bit less than a part in \( 10^{2} \), though parts of these errors
can be measured directly and corrected with suitable diagnostics as discussed
shortly.

The corrections all appear as modifications to the sizes of the existing terms.
This gives a tidy result but is somewhat disappointing in practice as it effectively
means that these errors only masquerade as legitimate effects rather than some
anomalous structure. It would be convenient if, for example, the errors generated
an octapole term which simultaneously increases or decreases the \( \pm 3/2\rightarrow \pm 1/2 \)
transition energies which would then show up as a shift in the mean position
of the quadrupole peaks from the position of the dipole peak, otherwise constrained
to be equal. However, the dipole squared structure of the transitions allow
for at most a quadrupole shift and so no such unambiguous error signal is available
and elimination of these systematic problems would generally depend on independently
detecting and minimizing or correcting for these polarization and alignment
imperfections. 

For example, suppression of these errors depends on a large applied magnetic
field so a possible indication of the absence of these problems is the independence
of the ratios of the size of the magnetic field, though it can be difficult
to accurately change the size of the field without also changing its direction. 

As a more promising possibility, the dipole shift is linear in the circular
polarization \( \sigma  \), so that if \( \sigma  \) can be accurately reversed
outside the chamber, and the modification to the polarization resulting from
entering the chamber is well behaved, large problems due to improperly polarized
light would appear as a change in the quadrupole to dipole ratios. The difference
in the two ratios gives the size of the error in polarization. This can then
be used to reduce the imperfection to that required for an accurate dipole ratio,
and when small enough even correct the quadrupole-dipole results, improving
confidence in the dipole ratio and restoring much of the precision to the quadrupole
to dipole ratios. 

Alignment of the applied beams propagation direction relative to the magnetic
field will probably depend on mechanical methods. The pump beams can be fairly
well aligned with the magnetic field using the pumping signal in the floresence,
the shifting laser is then easily made to overlap these pump beams. This alignment
is less easy to systematically change than the polarization, but both the beam
propagation direction, or the magnetic field direction could be slightly, arbitrarily
altered and any change in the ratios interpreted as a measure of systematic
errors due to non-ideal alignment.

\subsection{Data}

Final preparations for a complete precision measurement of these ratios is in
progress and light-shifts due this kind of off-resonant interaction has already
been detected. As usual, initial studies involve the \( D \) state. To make
beginning observations easier, the light shift was made as large as possible
to compensate for uncertainties in the matrix elements and alignment by trying
to guarantee that the shift is significantly larger than the spin resonance
linewidth. 

The shift depends on the laser power and beam size, about \( 100mW \) in about
\( 100\mu  \) any state the shift increases dramatically, so the frequency
was chosen to be near the \( 5D_{3/2}\rightarrow 6P_{1/2} \) resonance. The
frequency can't be arbitrarily close to resonance since the amplitude to be
in the \( P \) state begins to increase as well, this \( P \) state has a
large decay rate to the ground state and the biggest effect quickly becomes
loss of population to the ground state rather than just shifts in the \( D \)
state. This doesn't turn out to be a large restriction. The amplitude to be
in the \( P \) state increases like \( \Gamma ^{2}/\delta \omega ^{2} \),
so the induced loss rate will increase like \( \Gamma ^{3}/\delta \omega ^{2} \).
To keep this rate much less than about \( 1/s \), for example around \( 1/100s \),
and allow for enough RF interaction time to probe the spin states requires \( \delta \omega ^{2}<<\Gamma ^{3}/\left( 1Hz\right)  \).
Decay rates are around \( 1Mhz \), which in turn gives \( \delta \omega >10GHz \)
or \( \delta \lambda >0.1 \)Angstrom.

The shift depends linearly on the applied power so intensity stability is very
important, a 10\% fluctuations in the power of a laser giving a \( 10kHz \)
shifts gives an additional \( 1kHz \) linewidth in the spin resonance profile.
Good stability must be maintained for relatively long times as a measurement
giving sufficient sensitivity requires a few hundred trials which take a few
hours to collect. For this reason a diode laser was initially selected to generate
the shifts. 

Bare diodes with powers of \( 30-50mW \) at \( 555-560nm \) were used. These
would give shifts of only a few tenths of \( kHz \). In addition, the broadband
background included a significant amount of power at the \( D\rightarrow P \)
resonance which resulted in quick loss of the ion from the \( D \) state. This
latter problem was partly resolved with a diffraction grating, and a commercial
dielectric interference filter would eliminate the trouble, but the shifts still
proved to be too small to detect. Higher power diodes are readily available
and improvements could be made in spot size and alignment to increase the shift.
This is a likely path for a final measurement, but a first look favored the
more immediately accessible alternative of a dye laser.

Dye lasers can produce 100's of \( mW \) at a broad range of continuously tunable
wavelengths. The tunability could prove to powerful advantage as mapping the
shifts as a function of wavelength would help isolate and more tightly constrain
contributions to the shift from individual transitions rather than only the
collective effects of all of them. The tradeoff is a relatively enormous, expensive
apparatus compared to a simple diode, and relatively much poorer intensity stability.
Without external help, a dye laser's light intensity will fluctuate by at least
a few percent, and more often 10\% and more, at time scales of a few per second
probably associated with changes in the dye jet. Completely unattended, thermal
and mechanical changes in alignment can change the laser's power by more than
50\% over a few minutes. Managing these difficulties eventually required active
feedback and regular attention to the laser's alignment.

Light from an Argon Ion pumped dye laser was delivered to the ion from a separate
room initially by means of a multimode fiber. This eliminated the chore of hauling
the gear around and prevented it becoming a nearby source of noise, vibration
and background light. This also effectively removed the sensitivity of the focussed
spot's position to fluctuations in the direction and position of the beam exiting
the laser from changes in alignment as the spot position was instead determined
by the position of the fiber output. A polarizing beam splitter cube reflected
the vertically polarized light to the ion to coincide with a vertical magnetic
field. The fiber output is such a strange patchwork collection of intensities
and polarizations that collectively it is effectively unpolarized and so half
of the power is lost to the horizontal polarization.

The horizontal polarization does not necessarily need to be discarded, in this
coordinate system it gives additional equal amounts of left and right circularly
polarized light at an arbitrary, but fixed, phase relative to the vertical polarization.
This could increase the shift, making it more easily visible, but the structure
of the shifts would be different, so this additional complication was avoided.
The discarded power also allowed for an easy means of later actively stabilizing
the output power. A measurement of light shift ratios will require circularly
polarized shifting light, as discussed above, but linear polarization was used
first as it was the easiest to generate and work with and yields a larger S/N
as the dipole peak splits into two quadrupole peaks, both shifted equally from
the initial position of the resonance.

About \( 70-80mW \) of appropriately polarized near-resonant light was available
from the fiber but the resulting beam proved difficult to work with. The focussed
spot was very inhomogeneous and probably would have made it possible for the
ion to sit at a local zero even is the beam was completely contained in, and
filled the trap. Moderate effort also failed to give a spot that would fit entirely
within the trap. This reduces the electric field strength at the center of the
spot, and so reduces the shift, but more importantly in this case, the stray
high intensity light incident on the trap electrode results in a polluted trap
environment that significantly reduces the lifetime of the ion in the trap.
As with the instability created when the applied RF is resonant with the secular
frequency, the ion is generally stable while the off resonant light is applied
if it is also, simultaneously being cooled. During the pumping, probing or interaction
stages when the ion is not being actively cooled, it quickly disappears from
the trap. The trap lifetime is reduced to a few minutes. This is likely due
to the intense light heating the electrode which then outgasses and provides
a local source of grit that collides with and generally heats the ion until
it is knocked out of the trap. It could conceivably also be force on the ion
generated by the off-resonant interaction pushing the ion from the trap, or
some similarly weird effect, but when the intense part of the beam is deliberately
positioned directly on the electrode, the trap lifetime is further reduced,
now even while cooling, reinforcing the former interpretation of problems due
to heating of the electrode. Similarly reducing the overall power reaching the
trap restored the lifetime, not eliminating other explanations, but at least
consistent with the idea of local heating.

The was most easily dealt with by replacing the multi-mode fiber with a single
mode preserving fiber. The price is some loss of intensity as reduced input
coupling efficiency reduced the total output power to to \( 60-80mW \). The
fiber was not polarization preserving, but unlike the multimode fiber the final
polarization was well defined, but arbitrary and variable. The well defined
polarization allowed for minimizing the amount of light lost as horizontal polarization.
Coarse manipulation of the fiber position resulted in less than about \( 10-20\% \)
of the light being discarded, but the output polarization is very sensitive
to the positioning of the fiber and changes with thermal and mechanical fluctuations,
on timescales of a few minutes. As the polarization changes, the power reflected
by the polarizing beam-splitter cube also changes which provides another source
of long term variability in the electric field intensity at the ion. Strictly,
the multimode fiber output polarization also changes with positioning, and so
the relative strengths of different regions of the beam after the beam splitter
changes, also affecting the resulting field at the ion, but the contributions
from all parts of the beam tends to reduce the total change, and in any case
the collective output polarizations of the multimode fiber seemed to be generally
much less sensitive to changes in its position before the output. 

The resulting, almost perfectly gaussian beam was very easy to manipulate and
focus. The beam size could be made much smaller than the \( \sim 140\mu m \)
trap opening with little effort using a single lens. A small spot maximizes
the electric field strength at the center of the spot, but the precise location
of the ion within the trap electrode is not well known and could easily be missed
by a small, tightly focussed spot, so for initial alignment the beam was made
as large as was possible while still being almost completely contained within
the trap electrode so that some part of the beam would be guaranteed to overlap
the ion.

\begin{figure}
{\par\centering \resizebox*{1\textwidth}{!}{\includegraphics{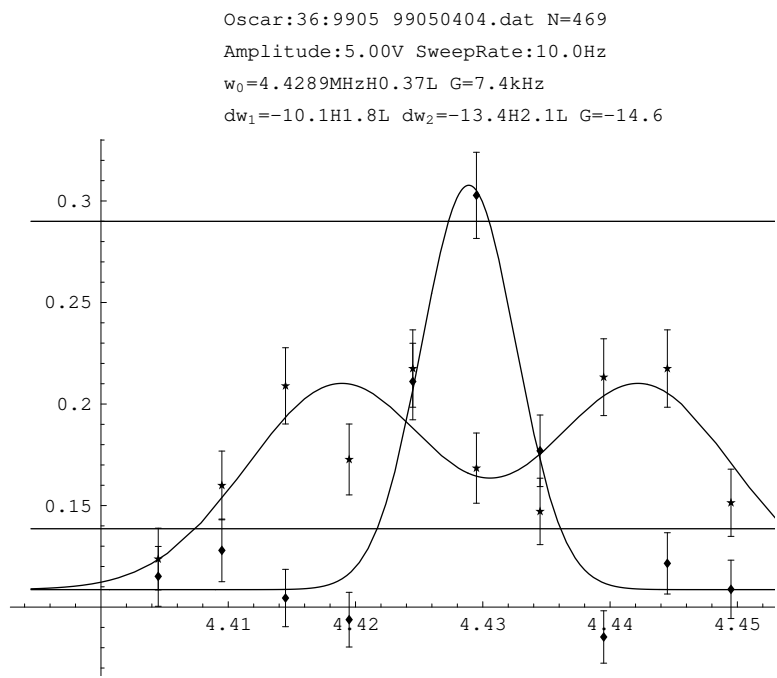}} \par}

\caption{\label{Fig:SmallLightShift} Initially detected quadrupole light shift in \protect\( 5D_{3/2}\protect \)
state.}
\end{figure}

The measurement sequence then consisted of two alternating spin resonance profile
trials where the shelving probability of a pumped, RF driven, and probed state
is measured as a function of the applied RF frequency. The trials were identical
in state preparation, interaction and probing and differed only in the addition
of the application of the intense off-resonant dye laser light during alternate
trials. The wavelength initially used was about \( 655nm \). With this improved
configuration the shift was visible almost immediately. The first evidence appeared
as a significant broadening compared to the dipole peak and partial resolution
of two peaks. The centers of the new peaks were approximately equally shifted
from the dipole peak's original position, fig.\ref{Fig:SmallLightShift}, but
the widths were significantly larger than the usual width of the dipole peak
which was perfectly reasonable given the large intensity fluctuations of the
dye laser.

Further improvements required managing these fluctuations by stabilizing the
intensity. A pockel cell was inserted into the beam path to manipulate the polarization.
For convenience, this was placed immediately at the output of the laser before
the fiber rather than after where it would have taken more work to direct the
beam through the cell. A small fraction of the beam going to the ion, following
the polarizing beam splitter cube, was monitored and the voltage applied to
the pockel cell adjusted to keep this intensity fixed, compensating for changes
in the total laser power as well as changes in the output polarization from
the fiber due to thermal and mechanical fluctuations. The gave a field strength
stable to better than \( 1\% \), and typically as good as \( 0.1\% \) on time
scales from a few \( ms \) to a few tens of minutes. The only attention required
was then an occasional adjustment to the dye laser alignment to restore maximum
power before being reduced below the usable operating range of the feedback
loop.

\begin{figure}
{\par\centering \resizebox*{1\textwidth}{!}{\includegraphics{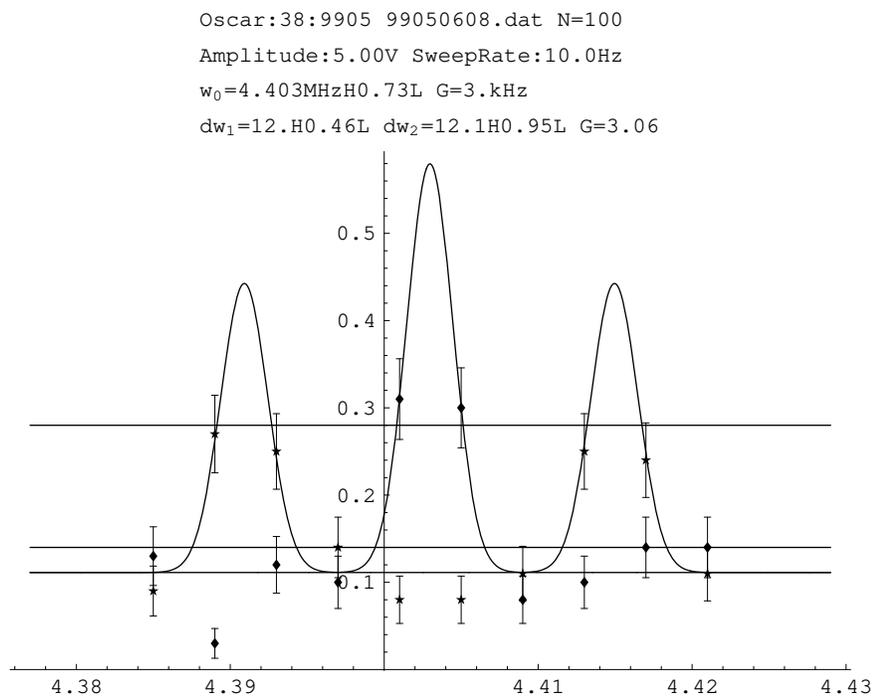}} \par}

\caption{\label{Fig:GoodLargeLightShift}Quadrupole splitting of \protect\( D_{3/2}\protect \)
state with intensity stabilized shifting laser.}
\end{figure}

With this significantly improved performance, the input alignment was adjusted
slightly and the wavelength further reduced to \( 653nm \), closer to resonance
with the \noun{\( P \)} state. This gave clearly resolved quadrupole peaks,
significantly shifted from the position of the initial dipole resonance, fig.\ref{Fig:GoodLargeLightShift}.
The widths of the individual peaks was comparable to the dipole width indicating
that the dominant broadening mechanism was now the same as that affecting the
dipole resonance, rather than fluctuations of the laser power. The density,
and distribution, of frequencies sampled was not sufficient to clearly define
the shape of the split peaks. There are only two peaks per point, so the preside
widths and heights are somewhat ambiguous. In particular it is hard to verify
that the peak height are related to each other and the height of the dipole
peak as expected, but certainly the presence, and general size of the shift
is unambiguous and well defined.

\begin{figure}
{\par\centering \resizebox*{1\textwidth}{!}{\includegraphics{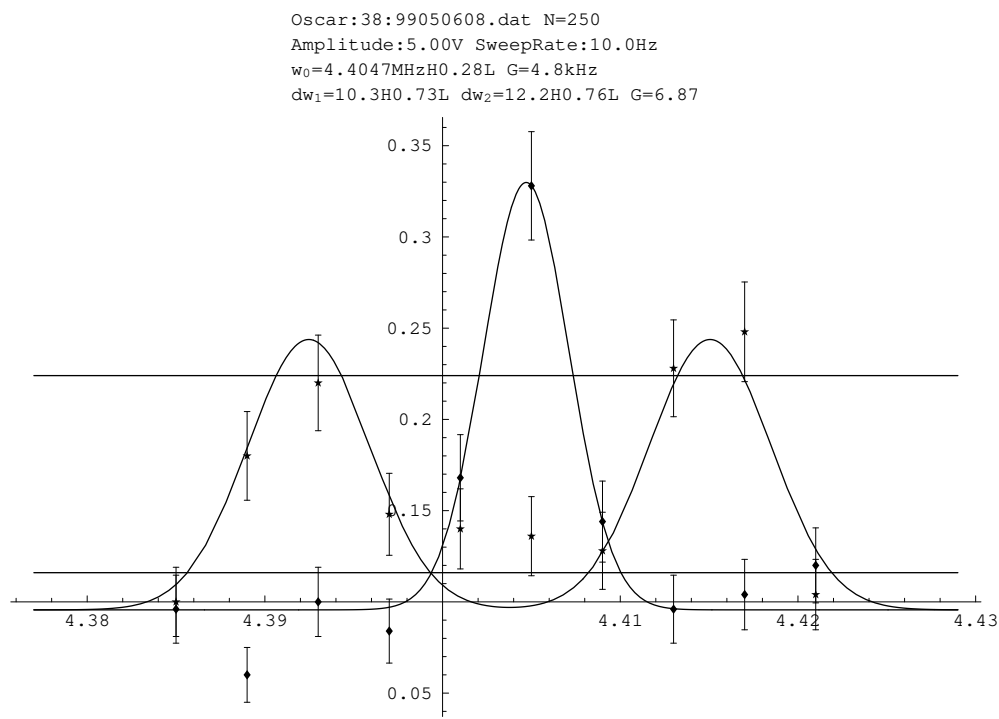}} \par}

\caption{\label{Fig:BadLargeLightShift}Broadening of peaks due to intensity instability
of dye laser.}
\end{figure}

The plot shown is based on only the first 100 trials collected for each case
when the dye laser was being closely monitored. After a few hours, when attention
to, and patience with the laser waned, the intensity stabilization feedback
loop would drop out of lock for several minutes at a time before being noticed
and restored and the quadrupole peaks began to broaden again, fig,\ref{Fig:BadLargeLightShift}.

\section{Outstanding Issues and Future Projects}

Detection of this off-resonance light shift provides direct evidence that these
methods can be used for a measurement of the parity violation light shift. Further
measurements will be taken to improve the precision of these measurements and
determine the \( S-D \) light shift ratios. Besides the parity measurement,
other experiments mentioned to study the details of the spin flip transition
would be interesting.

The statistical dependence of the sensitivity is already sufficient for the
measurement and further improvements can be made. Overall sensitivity is limited
by a broad spin flip resonance linewidth. Studies suggest magnetic field noise
as the source of the width and the system can be sufficiently well shielded,
and cleanly constructed to reduce this linewidth to the required level.

Two other technical details remain before initial measurements can begin. A
sufficiently narrow and powerful \( 2.05\mu m \) laser to drive the parity
transition is not yet available. The current laser produced only \( 5mW \),
has short times drifts around a few \( 100kHz \) and long time drifts over
many \( MHz \). The frequency can be actively stabilized and implementing this
locking is in progress. However, the power must ultimately by around \( 100mW \)
. This will require at least the construction of a different laser based on
the same technology as the existing version, \cite{KristiThesis}, or possible
a completely different kind of system. Such a system is not commercially available
and likely to be the most difficult practical obstacle to further progress. 

Also a system must be implemented to produce \( 2\mu m \) standing waves containing
the ion that can be precisely and stably positioned. An in situ mirror mount
based on the zero thermal expansion glass zerodur and a piezoelectric cylinder
was constructed so that the final note relative to the ion can be stably positioned,
\cite{KristiThesis}. As an insulator, the mount turned out to collect enough
charge that the resulting electric field prevented trapping ions. A copper coated
version worked, but a dielectric coated mirror caused similar problems. A gold
coated mirror with a conductor shielded piezo has not yet been successful.

Once the mirror has been implemented the stability of the standing waves can
be studied using the cleanup beam. An ion in the node of the beam will get stuck
in the \( D_{3/2} \) state and not florese. Position stability can be studied
by measuring the position of the zero as a function of time. If the position
is stable for at least many tens of seconds, the mirror can simply be periodically
repositioned and a means of doing so quickly and precisely must be determined.
A less stable result could possibly require the use of continuous active feedback.

Also with a mirror some of the quadrupole misalignment systematics can be studied,
even with the existing laser if properly stabilized. Quadrupole shifts in \( S \)
and \( D \) states as a function of beam geometry and polarization can be determined
to verify that they behave as expected as method for detecting, minimizing or
correcting them can be determined.

With these pieces a measurement of the PNC shift in one isotope should be possible
with a week or so of data collection. The time will depend on what systematic
problems are turned up and what measurements must be done in parallel with the
parity measurement to correct them. From there further work would be done do
reduce systematics and improve precision and to repeat the measurement on other
isotopes. These later results should follow quickly once the procedure is optimized.
It could take less than a year from the acquisition of a usable \( 2\mu m \)
laser and working mirror to precise results on number of isotopes. Interpretation
as corrections to the Standard Model will require accurate atomic structure
results.

\appendix

\vita{M. Schacht

\begin{lyxlist}{00.00.0000}
\item [B.S~Physics ]University of Wisconsin-Madison 1992
\item [B.S~Math ]University of Wisconsin-Madison 1992
\item [B.S~Astronomy ]University of Wisconsin-Madison 1992
\item [M.S~Physics ]University of Washington 1996
\end{lyxlist}
}

\end{document}